\documentclass[a4paper,11pt]{book}
\usepackage{lmodern}
\usepackage{graphicx}
\usepackage{siunitx}
\usepackage{amsmath}
\usepackage{authblk}
\usepackage{cancel}

\usepackage{geometry}
\usepackage{hyperref}
\hypersetup{
pdftitle={The Numerics of VMEC++},
pdfsubject={The Numerics of VMEC++},
pdfauthor={Jonathan Schilling, Proxima Fusion, Munich, Germany},
pdfkeywords={Variational Moments Equilibrium Code, Magnetohydrodynamics, Spectral Condensation, Free-Boundary}
}

\usepackage{fancyhdr}
\fancyhfoffset[L]{1cm} % left extra length
\fancyhfoffset[R]{1cm} % right extra length
\def\eqn#1{Eqn.~(\ref{eqn:#1})}

\usepackage{amssymb}

\def\naturalsWithZero{\mathbb{N}_0}
\def\integers{\mathbb{Z}}
\def\realnumbers{\mathbb{R}}
\def\complexnumbers{\mathbb{C}}

\usepackage{bm}

\usepackage{mathtools}

\usepackage{xcolor}

\usepackage{algorithm}
\usepackage{algpseudocode}

\def\realpart{\mathfrak{Re}}
\def\imagpart{\mathfrak{Im}}

\usepackage{placeins}

\newcommand{\code} {\texttt}

\usepackage{nicefrac}
\newcommand{\half} {\nicefrac{1}{2}}

\usepackage[style=phys,
            backend=biber,
            articletitle = true,
            biblabel = brackets,
            maxbibnames=9,
            maxcitenames=2,
            chaptertitle = true,
            pageranges = true,
            doi = false,
            url = true
            ]{biblatex}
\bibliography{main.bib}

\title{The Numerics of VMEC++}
\author[1]{Jonathan~Schilling\footnote{Corresponding author. E-mail: \href{mailto:jons@proximafusion.com}{jons@proximafusion.com}}}
\affil[1]{Proxima~Fusion, Munich, Germany}

\begin{document}

% This environment creates a standalone title page
\begin{titlepage}

% We center the content on the page
\centering

% We include the Froxima Fusion logo, specifying a suitable width for the image
\includegraphics[width=0.7\textwidth]{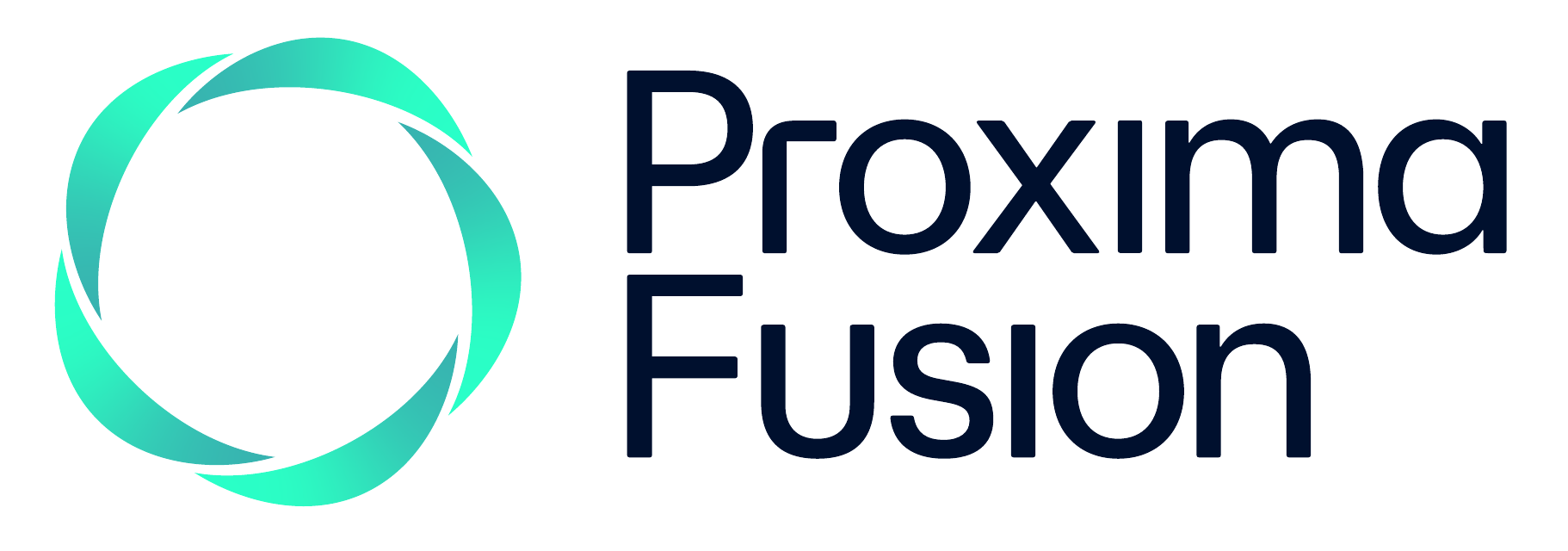}

% We add vertical space after the logo
\vspace{2cm}

% We specify the title in a large, bold font
{\Huge \textbf{The Numerics of VMEC++}}

% We add additional vertical space
\vspace{1cm}

% We specify the author name
{\Large \textsc{Jonathan Schilling}}

\vspace{0.3cm}

{\large Proxima Fusion, Munich, Germany}

% We push the date to the bottom of the page
\vfill

% We automatically generate today's date
{\large \today}

\end{titlepage}

\tableofcontents

\chapter{Introduction}
The Variational Moments Equilibrium Code~(VMEC)~\cite{hirshman_whitson_1983}
is the main computational workhorse used in the stellarator community
to compute the three-dimensional geometry of
and magnetic field in the plasma.
VMEC computes the plasma equilibrium in a stellarator
under the assumption of ideal magneto-hydrodynamics~(MHD).
It therefore makes the explicit assumption
of the existence of nested flux surfaces.
It is usually argued that any stellarator plasma
relevant for a fusion power plant
needs to possess nested flux surface in its core anyway for sufficiently good performance,
despite the existence of nested flux surfaces not being guaranteed in a stellarator.
On the other hand,
the explicit assumption about the topology of the plasma,
i.e., the existence of nested flux surfaces,
allows for a rather straightforward inverse formulation
of the numerical problem that VMEC solves.

Starting out from an initial guess for the flux surface geometry,
the residual MHD force components are computed in real space, i.e.,
on a grid in radial, poloidal and toroidal magnetic flux coordinates,
and transformed back into Fourier space.
The Fourier coefficients of the forces are then used to move
the flux surface geometry (represented by cylindrical coordinates~$R$ and~$Z$)
and stream function~$\lambda$ towards force balance
using an accelerated descent method.
The computation is converged when the force residuals for $R$, $Z$ and $\lambda$
have decayed below a user-provided threshold.

VMEC is designed to allow computation of the ideal MHD equilibrium
for both symmetric and asymmetric two- and three-dimensional configurations,
which is configured by selection of which subset of Fourier harmonics to include in the computation.
VMEC still is the most used 3D ideal-MHD equilibrium code these days,
despite the initial version having been developed in the early 1980s.
This is due to its speed, robustness and relative simplicity in use.

VMEC can be used in two operational modes,
called fixed-boundary and free-boundary.
The fixed-boundary mode solves the ideal MHD force balance
for a given, fixed, plasma boundary geometry and fixed radial profiles
of pressure and either rotational transform or enclosed toroidal current
on a number of flux surfaces inside the plasma boundary.
The free-boundary mode extends the fixed-boundary operation
by allowing the plasma boundary to move into force balance
with an external magnetic field produced by coils.
This adds computational complexity and hence,
free-boundary computations are usually a factor of 2 to 10 slower
than the corresponding fixed-boundary computations.
On the other hand, free-boundary operation of VMEC
is a closer approximation of the real plasma equilibrium in an actually-built stellarator.

The free-boundary force contribution on the plasma boundary
is computed by the Neumann Solver for Toroidal Regions (NESTOR)~\cite{merkel_1986, merkel_1988}.
Free-boundary VMEC,
i.e., the amalgamation of VMEC and NESTOR,
was initially called NEMEC~\cite{hirshman_vanrij_merkel_1986}
and the respective branch is still maintained under that name
at the Max-Planck-Institute for Plasma Physics in Garching, Germany.
The latest version of NEMEC features an updated implementation of NESTOR,
which is based on a current potential formulation~\cite{merkel2015linearmhdstabilitystudies}
rather than an earlier magnetic scalar potential formulation,
which is used in the main branch of VMEC
maintained by the Oak Ridge National Laboratory~(ORNL).
It is noted that, when running VMEC in free-boundary mode,
a so-called \texttt{mgrid} file is required.
This file stores a cache of the magnetic field contributions
for unit currents in each independent circuit
in the main coil system of the experiment to be modeled.
In case of W7-X, e.g., there are seven circuits: all 10 copies of each of the five non-planar coil type wired in series, respectively,
and similar circuits of each 10 copies of the two planar coil types.
The VMEC input file then only needs to contain the respective circuit currents
and the total magnetic field produced by all main coils
is then computed by linear superposition of the response tables
stored in the \texttt{mgrid} file, weighted by the user-provided circuit currents.

\section*{Related Publications}
In the following, a complete (to the author's knowledge) list
of publications and, where applicable, corresponding internal reports of ORNL
(with sometimes additional details in comparison with the published journal articles)
related to the numerical details implemented in VMEC is provided:
\begin{enumerate}
 \item ``Curvilinear coordinates for magnetic confinement geometries''~\cite{ornl_tm_8393_hirshman_1982}.
       This technical report introduces the notation used by S. P. Hirshman
       in the following publications related to VMEC.
 \item ``Steepest-descent moment method for three-dimensional magnetohydrodynamic equilibria''~\cite{hirshman_whitson_1983};
       the corresponding ORNL report is~\cite{ornl_tm_8861_hirshman_whitson_1983}.
       This is the first article on the basic numerical approach that later became fixed-boundary VMEC.
       A first approach to regularizing the poloidal parameterization is included in an appendix.
 \item ``Accelerated Convergence of the Steepest Descent Method for Magnetohydrodynamic Equilibria''~\cite{handy_hirshman_1985};
       the corresponding ORNL report is~\cite{ornl_tm_9133_handy_hirshman_1984}.
       This is an early attempt at improving the convergence of the initial method published in Ref.~\cite{hirshman_whitson_1983}.
       No further details about this branch of VMEC are known to the author.
 \item ``Optimized Fourier representations for three‐dimensional magnetic surfaces''~\cite{hirshman_meier_1985};
       the corresponding ORNL report is~\cite{ornl_tm_9300_hirshman_meier_1984}.
       The spectral width~$\langle M \rangle$ is introduced as a metric
       to quantify the compactness of a Fourier representation of the geometry of flux surfaces.
       This forms the basis for the first implementation of the DESCUR curve fitting tool,
       which was written as an interface between vacuum flux surfaces obtained from Poincare plots
       and the Fourier representation of the plasma boundary required by VMEC.
       Moreover, the approach to spectral condensation introduced in this article
       is the one still mainly in use in the main branch of VMEC,
       albeit with slight modifications (see below).
 \item ``A convergent spectral representation for three-dimensional inverse magnetohydrodynamic equilibria''~\cite{hirshman_weitzner_1985};
       the corresponding ORNL report is~\cite{ornl_tm_9302_hirshman_weitzner_1984}.
       This is an alternative approach to spectral condensation,
       which apparently was not followed further.
 \item ``MOMCON: A spectral code for obtaining three-dimensional magnetohydrodynamic equilibria''~\cite{hirshman_lee_1986};
       the corresponding ORNL report is~\cite{ornl_tm_9531_hirshman_lee_1985}.
       This is a further iteration of the numerical implementation presented in Ref.~\cite{hirshman_whitson_1983}.
       The code which later becomes VMEC got its first name: MOMCON.
       Further details on spectral condensation are included.
       The ORNL technical report lists subroutine names,
       some of which have survived into the latest state of the main Fortran VMEC implementation.
 \item ``ORMEC: A three-dimensional MHD spectral inverse equilibrium code''~\cite{hirshman_hogan_1986};
       the corresponding ORNL report is~\cite{ornl_tm_9547_hirshman_hogan_1986}.
       This article introduces the enclosed toroidal current constraint into VMEC.
       A rebranding is performed to a new name: ORMEC.
 \item ``An Integral Equation Technique for the Exterior and Interior Neumann Problem in Toroidal Regions''~\cite{merkel_1986}.
       This is the main reference for NESTOR,
       which computes the free-boundary force contribution by solving Laplace's equation
       with a Neumann boundary condition on the plasma boundary.
 \item ``Three-dimensional free boundary calculations using a spectral Green's function method''~\cite{hirshman_vanrij_merkel_1986}.
       This is the main reference for free-boundary VMEC,
       where the coupling between fixed-boundary VMEC and NESTOR is presented.
       It is also the article where VMEC gets its final name.
 \item ``Evaluation of magnetic coordinates in finite-beta, asymmetric plasmas''~\cite{ornl_tm_10222_beasly_rome_attenberger_hirshman_1987}.
       This report presents a first attempt at computing Boozer coordinates
       from an ideal MHD equilibrium computed by MOMCON or VMEC.
 \item ``Some Practial Considerations Involving Spectral Representations of 3D Plasma Equilibria''~\cite{attenberger_1987};
       the corresponding ORNL report is~\cite{ornl_tm_10412_attenberger_houlberg_hirshman_1987}.
       This article introduces several important post-processing methods
       commonly used on VMEC equilibria: an inverse coordinate transform
       from real space back into VMEC's flux coordinates,
       evaluation of the cylindrical components of the magnetic field,
       and tracking a line-of-sight through the plasma.
 \item ``Optimum Fourier representations for stellarator magnetic flux surfaces''~\cite{lee_1988}.
       This paper introduces an alternative approach to computing an ideal MHD equilibrium
       in the VMEC representation by reconstructing the Fourier coefficients of~$\lambda$
       as well as the rotational transform profile from the plasma geometry.
       While not directly related, this paper turned out to be quite valuable
       as a cross-check of the understanding of the other articles related to VMEC.
 \item ``Applications of the Neumann Problem to Stellarators: Magnetic Surfaces, Coils, Free-Boundary Equilibrium, Magnetic Diagnostics''~\cite{merkel_1988}.
       This is a further iteration of the NESTOR publication by Merkel.
       It mainly corrects some spelling mistakes of the original NESTOR article~\cite{merkel_1986}.
 \item ``Improved Radial Differencing for Three-Dimensional Magnetohydrodynamic Equilibrium Calculations''~\cite{hirshman_schwenn_nuehrenberg_1990}.
       This article introduces regularization factors for the odd-$m$ Fourier coefficients
       in order to increase the numerical accuracy of radial derivatives approximated by finite differences.
 \item ``Preconditioned Descent Algorithm for Rapid Calculations of Magnetohydrodynamic Equilibria''~\cite{hirshman_betancourt_1991}.
       A radial preconditioner for the residual MHD forces on $R$ and $Z$ is introduced,
       which strongly accelerates the convergence of VMEC.
       A corresponding preconditioner for the force on~$\lambda$ is found in Ref.~\cite{betancourt_1988_betas},
       related to the BETAS code.
 \item ``Calculation of $\beta_I$ and $l_i$ for three‐dimensional plasma configurations''~\cite{hirshman_1993}.
       This article goes into the details of some of the post-processing computations
       found in the various output files produced by VMEC.
 \item ``Explicit spectrally optimized Fourier series for nested magnetic surfaces''~\cite{hirshman_breslau_1998}.
       This is yet another attempt at spectral condensation.
       It is the basis for the latest version of the DESCUR curve fitting tool (see above)
       and this representation was also found to be implemented as an option in VMEC,
       but not used in practise.
 \item ``BCYCLIC: A parallel block tridiagonal matrix cyclic solver''~\cite{hirshman_etal_2010}.
       This article introduces a parallel block-tridiagonal matrix solver
       used in the context of a two-dimensional preconditioner in VMEC
       as well as in PARVMEC.
 \item ``PARVMEC: An Efficient, Scalable Implementation of the Variational Moments Equilibrium Code''~\cite{seal_hirshman_2016}.
       PARVMEC is the latest implementation of VMEC, which makes use of distributed-memory
       parallelization using the Message Passing Interface~(MPI)
       in order to speed up high-resolution VMEC computations
       by distributing the workload across multiple computers.
\end{enumerate}
The computation of the Mercier ideal-MHD stability criterion
implemented in VMEC is described in Ref.~\cite{beta_2}.

It is noted that before the implementation of PARVMEC,
several other attempts had been made to parallelize VMEC:
\begin{enumerate}
  \item ``Parallelization Strategies for the VMEC Program''~\cite{romero_1998}.
  \item ``Partial Parallelization of VMEC System''~\cite{msc_zhou}.
  \item ``Data-task parallelism for the VMEC program''~\cite{romero_2001}.
  \item ``Optimization strategy for the VMEC stellarator equilibrium code''~\cite{merz_2013}.
        This is the basic blueprint according to which PARVMEC was then implemented.
\end{enumerate}
Several earlier versions of VMEC were found in the context of this work.
The earliest among those is labelled VMEC\_9011.
The numerical suffix is interpreted as indicating a release date in November of 1990.
It is noted that already this rather early version
includes almost all features of the present-day fixed-boundary VMEC,
and only few minor details have been changed regarding the core numerical approach
to computing the MHD equilibrium since that version.

\FloatBarrier

\chapter{Know your tools}

\FloatBarrier
\section{Normal Derivative of a Scalar Field}
The normal derivative of scalar field $\Phi$ is:
\begin{equation}
  \partial \Phi / \partial \mathbf{n} = \nabla \Phi \cdot \mathbf{n} \, .
\end{equation}

\FloatBarrier
\section{Elliptic Partial Differential Equations}
Let $\Omega$ be some domain.
The following names are defined:
\begin{itemize}
  \item Laplace's equation: $\Delta \Phi = 0$ in $\Omega$ \\
  \item Poisson's equation: $\Delta \Phi = f ~(f\neq0)$ in $\Omega$ \\
  \item Dirichlet boundary condition: $\forall \mathbf{x} \in \partial\Omega:          \Phi(\mathbf{x})                       = f(\mathbf{x})$ \\
  \item Neumann boundary condition:   $\forall \mathbf{x} \in \partial\Omega: \partial \Phi(\mathbf{x}) / \partial \mathbf{n} = f(\mathbf{x})$ ($\mathbf{n}$: surface normal vector)
\end{itemize}

\FloatBarrier
\section{Green's Second Identity}
Consider a region $\mathcal{D}$ with a surface $\partial\mathcal{D}$.
Assume $\Delta \Phi = 0$ in $\mathcal{D}$, which means that $\Phi$ fulfills Laplace's equation.
Green's function is
\begin{equation}
  G(\mathbf{x}, \mathbf{x}') = \frac{1}{|\mathbf{x}-\mathbf{x}'|} \label{eqn:GreensFunction}
\end{equation}
with
\begin{equation*}
  \Delta G(\mathbf{x}, \mathbf{x}') = -4 \pi \delta(\mathbf{x}-\mathbf{x}') \, .
\end{equation*}
Then:
\begin{align*}
  \forall \mathbf{x} \in \mathcal{D}: & \int\limits_\mathcal{D} \left[ \Phi(\mathbf{x}') \Delta G(\mathbf{x},\mathbf{x}')
                                - \Delta \Phi(\mathbf{x}') G(\mathbf{x},\mathbf{x}') \right] \,\mathrm{d}V' \\
  ~ & = \int\limits_\mathcal{D} \Phi(\mathbf{x}') \Delta G(\mathbf{x},\mathbf{x}') \,\mathrm{d}V' \\
  ~ & = -4 \pi \Phi(\mathbf{x}) \\
  ~ & = \int\limits_{\partial\mathcal{D}} \left(\frac{\partial G(\mathbf{x}, \mathbf{x}')}{\partial \mathbf{n}'} \Phi(\mathbf{x}')
            - G(\mathbf{x}, \mathbf{x}') \frac{\partial \Phi(\mathbf{x}')}{\partial \mathbf{n}'}     \right)  \,\mathrm{d}S'
\end{align*}
and re-arranging leads to:
\begin{equation}
  \forall \mathbf{x} \in \mathcal{D} :
  \Phi(\mathbf{x}) =
  - \frac{1}{4 \pi} \int\limits_{\partial\mathcal{D}} \frac{\partial G(\mathbf{x}, \mathbf{x}')}{\partial \mathbf{n}'} \Phi(\mathbf{x}') \,\mathrm{d}S'
  + \frac{1}{4 \pi} \int\limits_{\partial\mathcal{D}} G(\mathbf{x}, \mathbf{x}') \frac{\partial \Phi(\mathbf{x}')}{\partial \mathbf{n}'} \,\mathrm{d}S' \, .
  \label{eqn:GreensIdentityOffSurface}
\end{equation}
The factor $4 \pi$ originates from the solid angle subtended by any point $\mathbf{x}$.
For evaluation points on the surface, the solid angle reduces to $2 \pi$ and Green's second identity can be formulated as follows~\cite{martensen_potentialtheorie, roy_2007}:
\begin{equation}
  \forall \mathbf{x} \in \partial\mathcal{D}:
  \Phi(\mathbf{x}) =
  - \frac{1}{2 \pi} \int\limits_{\partial\mathcal{D}} \frac{\partial G(\mathbf{x}, \mathbf{x}')}{\partial \mathbf{n}'} \Phi(\mathbf{x}') \,\mathrm{d}S'
  + \frac{1}{2 \pi} \int\limits_{\partial\mathcal{D}} G(\mathbf{x}, \mathbf{x}') \frac{\partial \Phi(\mathbf{x}')}{\partial \mathbf{n}'} \,\mathrm{d}S' \, .
  \label{eqn:GreensIdentityOnSurface}
\end{equation}
These are the two formulations of Green's second identity used by Merkel in NESTOR~\cite{merkel_1986}.
Note that the integrands in the on-surface evaluation typically feature a singularity at $\mathbf{x} = \mathbf{x}'$
due to Green's function $G$, which requires special treatment.

\FloatBarrier
\section{Trigonometry}
The derivative of $\tan(.)$ is:
\begin{equation}
  \frac{\mathrm{d}}{\mathrm{d}x} \tan(x) = \frac{1}{\cos^2(x)} \label{eqn:tan_derivative}
\end{equation}
Euler's identity:
\begin{equation}
  \exp(i \varphi) = \cos(\varphi) + i \sin(\varphi) \label{eqn:eulers_identity}
\end{equation}
Area Hyperbolic Sine for $x \in \realnumbers$ with $x>0$:
\begin{equation}
  \mathrm{arsinh}\,(x) = \ln \left( x + \sqrt{x^2 + 1} \right) \label{eqn:arsinh}
\end{equation}
Area Hyperbolic Tangent for $x \in \realnumbers$ with $|x|<1$:
\begin{equation}
  \mathrm{artanh}(x) = \frac{1}{2} \log\left(\frac{1 + x}{1 - x} \right)
\end{equation}
Identities:
\begin{align}
  \sin(\theta) \cos(\theta)       =&\, \frac{1}{2} \left[ \sin(\theta + \theta) + \sin(\theta - \theta) \right] = \frac{1}{2} \sin(2 \theta) \label{eqn:trigid_1} \\
  \cos^2(\theta) - \sin^2(\theta) =&\, \cos(2 \theta) \label{eqn:trigid_2} \\
  1 - \cos(\theta)                =&\, 2 \sin^2(\theta/2)
\end{align}

\FloatBarrier
\section{Series}
The Binomial series:
\begin{equation}
  (s-t)^l = \sum\limits_{n=0}^{l} (-1)^{n} {\genfrac(){0pt}{0}{l}{n}} s^{l-n} t^{n}
          = \sum\limits_{n=0}^{l} (-1)^{n} \frac{l!}{n!\, (l-n)!} s^{l-n} t^{n}
\end{equation}
An inverse power series:
\begin{equation}
  \frac{1}{(1-x)^{l+1}} = \sum\limits_{k=0}^{+\infty} {\genfrac(){0pt}{0}{k+l}{k}} x^k
                        = \sum\limits_{k=0}^{+\infty} \frac{(k+l)!}{k!\, l!} x^k
\end{equation}
The geometric series for $z \in \complexnumbers$ with $|z|<1$:
\begin{equation}
  \sum_{n=0}^{+\infty} z^n = \frac{1}{1-z} \label{eqn:geometric_series}
\end{equation}

\FloatBarrier
\section{Generating Functions}
Sequences of numbers appear for example as coefficients of the Fourier transform of a given function.
Generalized analytical expressions for the set of coefficients can sometimes be obtained
by introducing a so-called generating function where the given coefficients
are interpreted as coefficients of a power series.
Consider for example the Fourier coefficients $I_{m,n}$,
for which the generating function $I(s,t)$ can be introduced:
\begin{equation}
  I(s,t) = \sum_{m=0}^{+\infty} \sum_{n=0}^{+\infty} I_{m,n} s^m t^n \, .
\end{equation}
Standard mathematical tools can now be applied to the function $I(s,t)$
instead of having to work with the coefficients $I_{m,n}$ individually.
Note that the parameters $s$ and $t$ are completely artificial and thus,
any reformulation of the $I_{m,n}$ is valid as long as it holds for
at least \underline{a single} value of $s$ and $t$~\cite{discrete_mathematics}.

\FloatBarrier
\section{Orthogonality of the Fourier Basis}
For $k,l \in \naturalsWithZero$ it holds:
\begin{align}
  \frac{1}{2 \pi} \int_{0}^{2 \pi} \sin(k x) & \cos(l x) \,\mathrm{d}x = 0 \nonumber \\
  \frac{1}{2 \pi} \int_{0}^{2 \pi} \sin(k x) & \sin(l x) \,\mathrm{d}x = \begin{cases}
                                                                           0                         & k=l=0 \\
                                                                           \frac{1}{2} \delta_{k, l} & \textrm{ else}
                                                                         \end{cases}   \label{eqn:ortho_fourier_basis}\\
  \frac{1}{2 \pi} \int_{0}^{2 \pi} \cos(k x) & \cos(l x) \,\mathrm{d}x = \begin{cases}
                                                                           1                         & k=l=0 \\
                                                                           \frac{1}{2} \delta_{k, l} & \textrm{ else}
                                                                         \end{cases} \nonumber
\end{align}

\FloatBarrier
\section{Forward and Inverse Fourier Transform}
Throughout this book, the following definitions will be used to distinguish
between a forward Fourier transform and an inverse Fourier transform.
This is based on Sec.~2.2~(Fourier series) in Ref.~\cite{boyd_spectral_methods}.
We consider truncated Fourier series (only up to some maximum mode number~$M$)
of some $2 \pi$-periodic function~$f$ ($f(\theta) = f(\theta + 2 \pi)$).
An inverse Fourier transform is the evaluation of the Fourier series
to obtain the (real-valued) representation of the function
at given real-space sampling points~$\theta$
encoded by a set of Fourier coefficients:
\begin{equation}
  f(\theta) = \hat{f}_0^\mathrm{cos} + \sum_{m=1}^{M} \left[ \hat{f}_m^\mathrm{cos} \cos(m \theta) + \hat{f}_m^\mathrm{sin} \sin(m \theta) \right] \, .
\end{equation}
A forward Fourier transform is the process of obtaining those Fourier coefficients
from given (real-valued) function values~$f(\theta)$:
\begin{align}
  \hat{f}_0^\mathrm{cos} =&\, \frac{1}{2 \pi} \int\limits_{0}^{2 \pi} f(\theta)                \,\mathrm{d}\theta                            \nonumber \\
  \hat{f}_m^\mathrm{cos} =&\, \frac{1}{  \pi} \int\limits_{0}^{2 \pi} f(\theta) \cos(m \theta) \,\mathrm{d}\theta \quad \textrm{ for } m > 0 \label{eqn:fc_def} \\
  \hat{f}_m^\mathrm{sin} =&\, \frac{1}{  \pi} \int\limits_{0}^{2 \pi} f(\theta) \sin(m \theta) \,\mathrm{d}\theta \quad \textrm{ for } m > 0 \nonumber \, .
\end{align}

\FloatBarrier
\subsection{Use of Parity}
If the function $f$ to be Fourier-transformed has definite parity (even or odd),
the integrals in~\eqn{fc_def} can be evaluated on half of the interval.
In a discrete Fourier transform implementation, this is equivalent to a factor of 2
improvement in memory requirements and computational work reduction.
Suppose the function $f$ to Fourier-transform has even (cosine-like) parity.
The function $f$ is then also called to be symmetric.
It then holds:
\begin{equation}
  f(-x) = f(x) \, .
\end{equation}
It follows:
\begin{align}
 \hat{f}_0^\mathrm{cos}
  =&\, \frac{1}{2 \pi} \left[ \int\limits_{0}^{\pi} f(\theta) \,\mathrm{d}\theta + \int\limits_{\pi}^{2 \pi} f(\theta) \,\mathrm{d}\theta \right]
  =    \frac{1}{2 \pi} \Biggl[ \int\limits_{0}^{\pi} f(\theta) \,\mathrm{d}\theta + \int\limits_{0}^{\pi} \underbrace{f(2 \pi - \theta)}_{=f(- \theta)} \,\mathrm{d}\theta \Biggr] \nonumber \\
  =&\, \frac{1}{2 \pi} \int\limits_{0}^{\pi} \Bigl[  f(\theta) + \underbrace{f(- \theta)}_{=f(\theta)} \Bigr] \,\mathrm{d}\theta
  =    \frac{\bcancel{2}}{\bcancel{2} \pi} \int\limits_{0}^{\pi} f(\theta) \,\mathrm{d}\theta
  =    \frac{1}{\pi} \int\limits_{0}^{\pi} f(\theta) \,\mathrm{d}\theta
\end{align}
where the $2 \pi$-periodicity of $f$ has been used to find $f(2 \pi - \theta) = f(-\theta)$.
Similarly, we find for $m>0$:
\begin{align}
 \hat{f}_m^\mathrm{cos}
  =&\, \frac{1}{\pi} \int\limits_{0}^{\pi} \Bigl[ f(  \theta)                           \cos(  m \theta)
                                    + \underbrace{f(- \theta)}_{=f(\theta)} \underbrace{\cos(- m \theta)}_{=\cos(m \theta)} \Bigr] \,\mathrm{d}\theta \nonumber \\
  =&\, \frac{2}{\pi} \int\limits_{0}^{\pi} f(\theta) \cos(m \theta) \,\mathrm{d}\theta
\end{align}
as well as:
\begin{align}
 \hat{f}_m^\mathrm{sin}
  =&\, \frac{1}{\pi} \int\limits_{0}^{\pi} \Bigl[ f(  \theta)                           \sin(  m \theta)
                                    + \underbrace{f(- \theta)}_{=f(\theta)} \underbrace{\sin(- m \theta)}_{=-\sin(m \theta)} \Bigr] \,\mathrm{d}\theta \nonumber \\
  =&\, 0 \, .
\end{align}
Thus, a function with even parity has only cosine Fourier coefficients.

Suppose now the function $f$ to Fourier-transform has odd (sine-like) parity.
The function $f$ is then also called to be antisymmetric.
It then holds:
\begin{equation}
  f(-x) = -f(x) \, .
\end{equation}
It follows:
\begin{align}
 \hat{f}_0^\mathrm{cos}
  =&\, \frac{1}{2 \pi} \int\limits_{0}^{\pi} \Bigl[  f(\theta) + \underbrace{f(- \theta)}_{=-f(\theta)} \Bigr] \,\mathrm{d}\theta
  =    0
\end{align}
as well as for $m>0$:
\begin{align}
 \hat{f}_m^\mathrm{cos}
  =&\, \frac{1}{\pi} \left[   \int\limits_{  0}^{ \pi} f(\theta) \cos(m \theta) \,\mathrm{d}\theta
                            + \int\limits_{\pi}^{2\pi} f(\theta) \cos(m \theta) \,\mathrm{d}\theta \right] \nonumber \\
  =&\, \frac{1}{\pi} \Biggl[  \int\limits_{0}^{\pi} f(\theta) \cos(m \theta) \,\mathrm{d}\theta
                            + \int\limits_{0}^{\pi} \underbrace{f(2 \pi - \theta)}_{=f(-\theta)} \underbrace{\cos\left(m (2 \pi - \theta)\right)}_{=\cos(- m \theta)} \,\mathrm{d}\theta \Biggr] \nonumber \\
  =&\, \frac{1}{\pi} \int\limits_{0}^{\pi} \Bigl[ f(  \theta)                           \cos(  m \theta)
                                    + \underbrace{f(- \theta)}_{=-f(\theta)} \underbrace{\cos(- m \theta)}_{=\cos(m \theta)} \Bigr] \,\mathrm{d}\theta \nonumber \\
  =&\, 0 \, .
\end{align}
However, we find:
\begin{align}
 \hat{f}_m^\mathrm{sin}
  =&\, \frac{1}{\pi} \int\limits_{0}^{\pi} \Bigl[ f(  \theta)                            \sin(  m \theta)
                                    + \underbrace{f(- \theta)}_{=-f(\theta)} \underbrace{\sin(- m \theta)}_{=-\sin(m \theta)} \Bigr] \,\mathrm{d}\theta \nonumber \\
  =&\, \frac{2}{\pi} \int\limits_{0}^{\pi} f(\theta) \sin(m \theta) \,\mathrm{d}\theta \, .
\end{align}
An overview is given in the following table:
\begin{table}[htbp]
\centering
\begin{tabular}{l|l|l}
           ~              & \multicolumn{2}{c}{parity of function $f$} \\
         term             & even                                                                                                & odd                          \\
 \hline
 $\hat{f}_0^\mathrm{cos}$ & $\hat{f}_0^\mathrm{cos} = \frac{1}{\pi} \int_{0}^{\pi} f(\theta)                \,\mathrm{d}\theta$ & $\hat{f}_0^\mathrm{cos} = 0$ \\
 $\hat{f}_m^\mathrm{cos}$ & $\hat{f}_m^\mathrm{cos} = \frac{2}{\pi} \int_{0}^{\pi} f(\theta) \cos(m \theta) \,\mathrm{d}\theta$ & $\hat{f}_m^\mathrm{cos} = 0$ \\
 $\hat{f}_m^\mathrm{sin}$ & $\hat{f}_m^\mathrm{sin} = 0$ & $\hat{f}_m^\mathrm{sin} = \frac{2}{\pi} \int_{0}^{\pi} f(\theta) \sin(m \theta) \,\mathrm{d}\theta$
\end{tabular}
\end{table}

The use of poloidal symmetric can be summarized as follows.
Consider the integral over the poloidal coordinate~$\theta$ of an even-parity, $2\pi$-periodic function~$f$.
Then $f$ has the following properties:
\begin{align}
  f(\theta + 2 \pi) =&\, f(\theta) \\
  f(- \theta      ) =&\, f(\theta) \\
  \Leftrightarrow
  f(2 \pi - \theta) =&\, f(\theta) \, .
\end{align}
The following two integrals are then equivalent:
\begin{align}
  \frac{1}{2 \pi} \int\limits_0^{2 \pi} f(\theta) \,\mathrm{d}\theta \approx&\, \frac{1}{2 \pi} \sum\limits_{l=0}^{n_{\theta,1}-1} f\left( 2 \pi \frac{l}{n_{\theta,1}  } \right) \Delta \theta
    = \frac{1}{n_{\theta,1}  } \sum\limits_{l=0}^{n_{\theta,1}-1} f\left( 2 \pi \frac{l}{n_{\theta,1}  } \right) \label{eqn:theta_int_full} \\
  \frac{1}{  \pi} \int\limits_0^{  \pi} f(\theta) \,\mathrm{d}\theta \approx&\, \frac{1}{n_{\theta,2}-1} \sum\limits_{l=0}^{n_{\theta,2}-1} f\left(   \pi \frac{l}{n_{\theta,2}-1} \right)
  \begin{cases}
    1/2 &: l=0 \textrm{ or } l=n_{\theta,2}-1 \\
    1   &: \textrm{else.}
  \end{cases}
\end{align}
In the first discretized variant, the function is evaluated at regular increments over the open interval~$[0,2 \pi[$.
In the second variant, the function is evaluated at regular increments over the closed interval~$[0, \pi]$.
The analytical equivalence can be seen as follows:
\begin{align}
  ~&\, \frac{1}{2 \pi} \int\limits_0^{2 \pi} f(\theta) \,\mathrm{d}\theta
  = \frac{1}{2 \pi} \left[ \int\limits_0^{\pi} f(\theta) \,\mathrm{d}\theta + \int\limits_\pi^{2\pi} f(\theta) \,\mathrm{d}\theta \right]
  = \frac{1}{2 \pi} \left[ \int\limits_0^{\pi} f(\theta) \,\mathrm{d}\theta + \int\limits_0^{\pi} f(2 \pi - \theta) \,\mathrm{d}\theta \right] \nonumber \\
  =&\, \frac{1}{2 \pi} \left[ \int\limits_0^{\pi} f(\theta) \,\mathrm{d}\theta + \int\limits_0^{\pi} f(\theta) \,\mathrm{d}\theta \right]
  = \frac{\bcancel{2}}{\bcancel{2} \pi} \int\limits_0^{\pi} f(\theta) \,\mathrm{d}\theta
  = \frac{1}{  \pi} \int\limits_0^{  \pi} f(\theta) \,\mathrm{d}\theta \, .
\end{align}

\begin{figure}[htbp]
  \centering
  \includegraphics{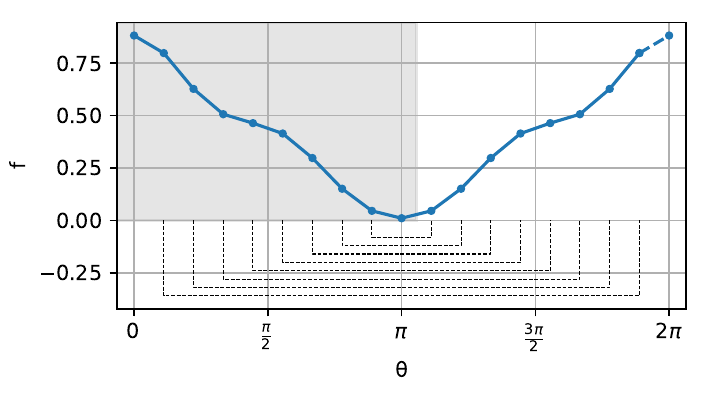}
  \caption{Demo of the reduced integration interval for integrals over an even-parity function.
  The grey area marks those points that are actually included in the discrete summation.
  The dashed lines at the bottom link points of equal value in the full interval
  and their counterparts in the half interval.}
  \label{fig:even_parity_demo}
\end{figure}

The symmetry-aware reduction to half the poloidal interval can be understood also in the discrete formuation.
We start with the first term in~\eqn{theta_int_full} and consider the symmetry depicted in Fig.~(\ref{fig:even_parity_demo}).
This leads to:
\begin{align}
  ~&\, \frac{1}{2 \pi} \Delta \theta \sum\limits_{l=0}^{n_{\theta,1}-1} f\left( 2 \pi \frac{l}{n_{\theta,1}  } \right)
  =    \frac{1}{2 \pi} \Delta \theta \Biggl[ f(0) + f(\pi) + 2 \sum\limits_{l=1}^{n_{\theta,2}-2} f\Bigl( \bcancel{2} \pi \underbrace{\frac{l}{n_{\theta,1}}}_{=l/[\bcancel{2}(n_{\theta,2}-1)]} \Bigr) \Biggr] \nonumber \\
  =&\, \frac{1}{2 \bcancel{\pi}} \frac{\bcancel{\pi}}{n_{\theta,2}-1}
       \left[ f(0) + f(\pi) + 2 \sum\limits_{l=1}^{n_{\theta,2}-2} f \left( \pi \frac{l}{n_{\theta,2}-1} \right) \right] \nonumber \\
  =&\, \frac{1}{n_{\theta,2}-1} \left[ \frac{1}{2} f(0) + \frac{1}{2} f(\pi) + \sum\limits_{l=1}^{n_{\theta,2}-2} f \left( \pi \frac{l}{n_{\theta,2}-1} \right) \right] \nonumber \\
  =&\, \frac{1}{n_{\theta,2}-1} \sum\limits_{l=0}^{n_{\theta,2}-1} f \left( \pi \frac{l}{n_{\theta,2}-1} \right) \begin{cases}
                                                                                                                   1/2 &: l=0 \textrm{ or } l=n_{\theta,2}-1 \\
                                                                                                                   1   &: \textrm{else.}
                                                                                                                 \end{cases}
\end{align}

% src/poloidal_Fourier_basis.ipynb

Summarizing:
\begin{itemize}
  \item If the normalization factor is $1/n_{\theta,1}$, the inner points need to be weighted by $2$ and only the endpoints of the half-interval are weighted by $1$.
  \item If the normalization factor is $1/(n_{\theta,2}-1)$, the inner points need to be weighted by $1$ and only the endpoints need to be weighted by $1/2$.
\end{itemize}

\FloatBarrier
\newpage
\section{Midpoint vs. Trapezoidal} \label{sec:midpoint_vs_trapz}
The midpoint rule and the trapezoidal rule are used to approximate the one-dimensional integral
of a function~$f$ over a finite interval~$[a,b]$.
The integrand has to be evaluated at the centers of the intervals for the midpoint rule:
\begin{equation}
  A_\mathrm{mid} = \Delta x \left[   f \left(a + \frac{1}{2} \Delta x \right)
                                   + f \left(a + \frac{3}{2} \Delta x \right)
                                   + ...
                                   + f \left(b - \frac{1}{2} \Delta x \right) \right] \, .
\end{equation}
The error estimate for the midpoint rule is:
\begin{equation}
  \left|\int\limits_a^b f(x) \,\mathrm{d}x - A_\mathrm{mid} \right|
  \leq
  \frac{M_2 (b-a)^3}{24 n^2}
\end{equation}
where $M_2$ is the maximum of the absolute value of $f''(x)$ over the interval $[a,b]$.

The integrand has to be evaluated at the borders of the intervals for the trapezoidal rule:
\begin{equation}
  A_\mathrm{trapz} = \Delta x \left[   1/2 f(a)
                                     +     f(a +   \Delta x)
                                     +     f(a + 2 \Delta x)
                                     + ...
                                     +     f(b -   \Delta x)
                                     + 1/2 f(b)\right] \, .
\end{equation}
The error estimate for the trapezoidal rule is:
\begin{equation}
  \left|\int\limits_a^b f(x) \,\mathrm{d}x - A_\mathrm{trapz} \right|
  \leq
  \frac{M_2 (b-a)^3}{12 n^2}
\end{equation}
where $M_2$ is the maximum of the absolute value of $f''(x)$ over the interval $[a,b]$.

This info is quoted from \texttt{https://en.wikipedia.org/wiki/Riemann\_sum}.

\chapter{Toroidal Coordinate System}
\begin{figure}[htbp]
  \centering
  \includegraphics[width=0.9\textwidth]{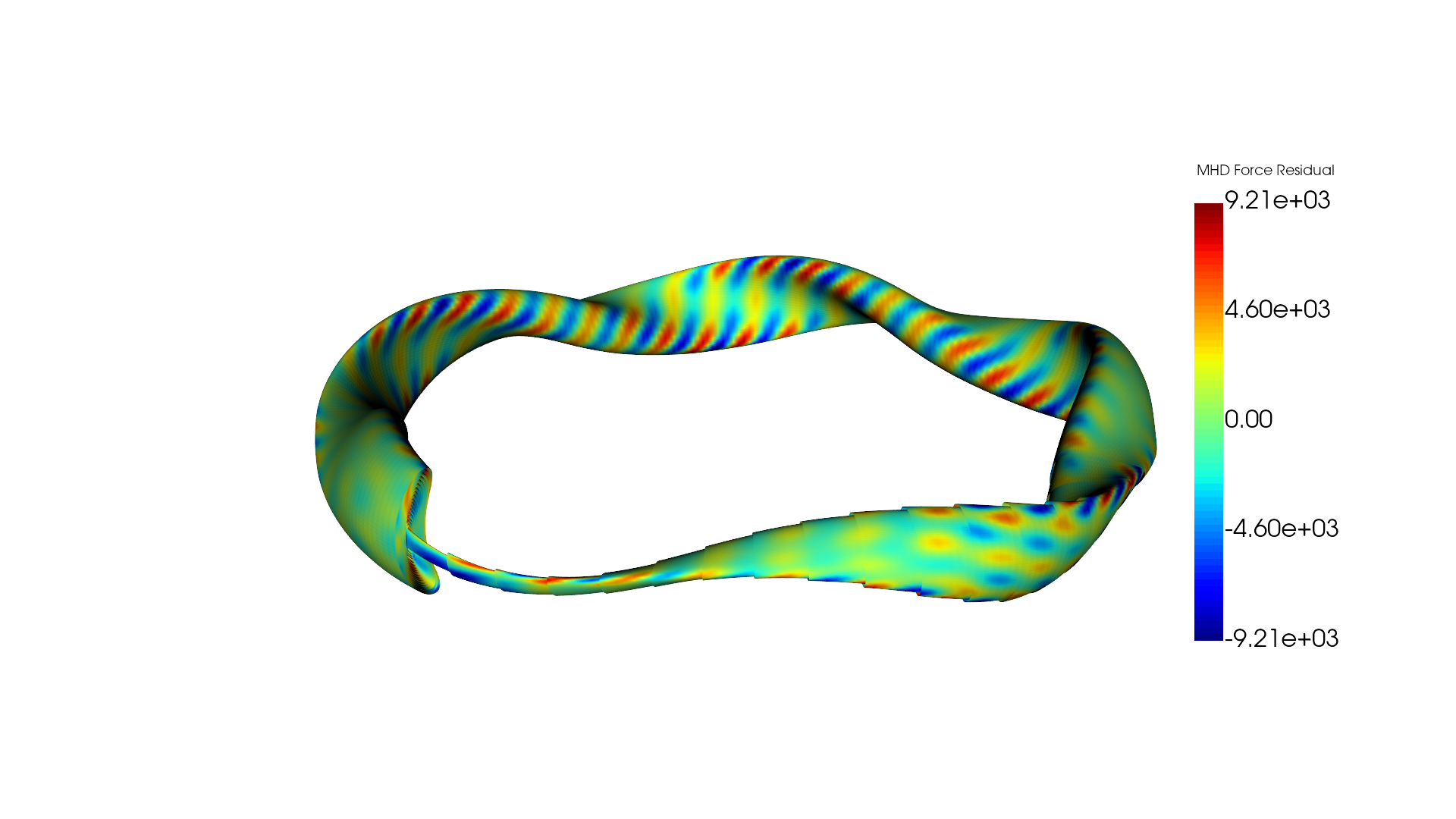}
  \caption{Flux surface geometry and MHD force residual color-coded on it,
           of the standard magnetic configuration of the Wendelstein 7-X stellarator.
           Some of the flux surface are shown as cut-open to see the inner surfaces.}
  \label{fig:w7x_ref_1}
\end{figure}

\begin{figure}[htbp]
  \centering
  \includegraphics[width=0.7\textwidth]{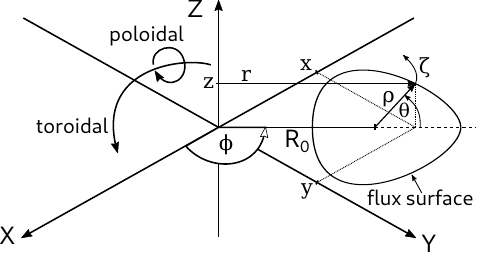}
  \caption{Coordinate systems in use in this work.
           A poloidal cut through a flux surface at the toroidal angle $\phi$ is shown.
           A point on the (torus-like) flux surface can be uniquely specified
           either by its Cartesian coordinates $(x,y,z)$,
           by its cylindrical coordinates $(r, \phi, z)$ or
           by its toroidal coordinates $(\rho,\theta,\phi)$.
           $R_0$ is the major radius of the toroidal coordinate system at $\phi$.}
  \label{fig:torcoords}
\end{figure}

The flux surface geometry of the standard magnetic configuration of the Wendelstein 7-X~(\mbox{W7-X}) stellarator is shown in Fig.~\ref{fig:w7x_ref_1}
as an example for the type of geometry discussed for in this chapter.
\FloatBarrier
In Fig.~\ref{fig:torcoords}, the toroidal coordinate system in use is shown.
Note that a point on a flux surface is described uniquely by three sets of coordinates:
$(x,y,z)$ in Cartesian coordinates, $(r, \phi, z)$ in cylindrical coordinates and $(\rho, \theta, \phi)$ in toroidal coordinates.
All three coordinate systems are orthogonal. The Cartesian and the cylindrical coordinate systems are right-handed
and the choice of the poloidal angle as illustrated implies that the toroidal coordinate system is left-handed.

\FloatBarrier
\section{Inverse Coordinate Representation}
\begin{figure}[htbp]
  \centering
  \includegraphics{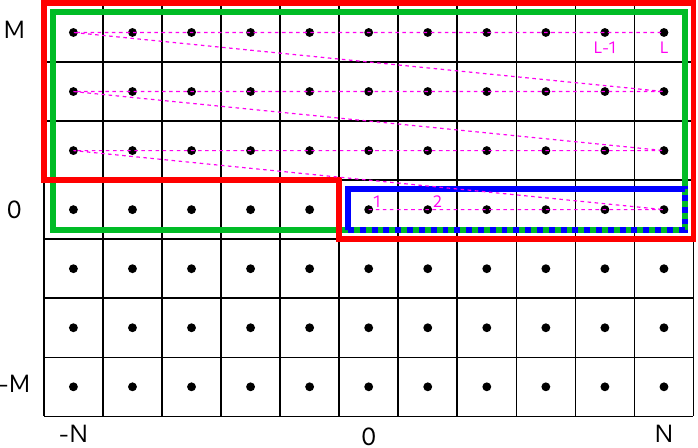}
  \caption{Full mode number spectrum used in \eqn{R2} and \eqn{Z2} (black grid)
           for $M=3$ and $N=5$.
           The horizontal axis enumerates toroidal mode numbers $-N \leq n \leq N$
           and the vertical axis enumerated poloidal mode numbers $-M \leq m \leq M$.
           Single Fourier modes are represented by black dots in the grid cells.
           %The dashed lines at $m=0$ and $n=0$ shall guide the eye.
           The green rectangle contains the modes with $m \geq 0$.
           The blue rectangle contains the modes with $m=0$ and $n \geq 0$.
           The red polygon includes all mode number combinations required to uniquely define a real-space quantity, e.g., $R$ or $Z$.
           The pink dashed line and the pink numbers indicate the linear arrangement of the Fourier harmonics
           used to vectorize the numerical implementation.}
  \label{fig:mode_numbers}
\end{figure}

A toroidal surface like this is conveniently parameterized by two angle-like coordinates.
The long way around the torus (along its major circumference) is called the toroidal direction
and the short way around the torus (wristband-like) is called the poloidal direction (cf Fig.~\ref{fig:torcoords}).
The toroidal coordinate is $\zeta$ and the poloidal coordinate is $\theta$.
This torus is embedded in a cylindrical coordinate system,
where the $z$-axis is aligned with the major axis of the torus.
The cylindrical angle $\phi$ can be identified with the toroidal coordinate~$\zeta$ of the torus,
although usually one introduces an integer scaling factor to exploit toroidal symmetry (see below).
Note that this works only for singly-linked tori, whereas knotted configurations need a special treatment~\cite{hudson_knotatrons}.
This allows to specify the surface geometry using two-dimensional discrete Fourier series in $\theta$ and $\zeta$
for the coordinates $R$ and $Z$ of a surface:
\begin{align}
  R(\theta, \zeta) =&\, \realpart \left( \sum_{m=-\infty}^{+\infty} \sum_{n=-\infty}^{+\infty} \hat{R}_{m,n} \,e^{-i (m \theta + n \zeta)} \right) \nonumber \\
  Z(\theta, \zeta) =&\, \realpart \left( \sum_{m=-\infty}^{+\infty} \sum_{n=-\infty}^{+\infty} \hat{Z}_{m,n} \,e^{-i (m \theta + n \zeta)} \right) \nonumber
\end{align}
where $\theta$ is the poloidal angle-like coordinate, $m$ and $n$ are (integer) poloidal and toroidal mode numbers, respectively,
and $\hat{R}_{m,n}$, $\hat{Z}_{m,n}$ are the (complex-valued) Fourier coefficients of $R$ and $Z$, respectively.
In VMEC, the toroidal coordinates $(\rho, \theta, \zeta)$ form a left-handed coordinate system
due to the chosen direction for the poloidal coordinate $\theta$ depicted in Fig.~\ref{fig:torcoords}.
The sign of $\theta$ is therefore inverted in the Fourier representation:
\begin{align}
  R(\theta, \zeta) =&\, \realpart \left( \sum_{m=-\infty}^{+\infty} \sum_{n=-\infty}^{+\infty} \hat{R}_{m,n} \,e^{-i (m (-\theta) + n \zeta)} \right)
                   =    \realpart \left( \sum_{m=-\infty}^{+\infty} \sum_{n=-\infty}^{+\infty} \hat{R}_{m,n} \,e^{ i (m   \theta  - n \zeta)} \right) \label{eqn:R} \\
  Z(\theta, \zeta) =&\, \realpart \left( \sum_{m=-\infty}^{+\infty} \sum_{n=-\infty}^{+\infty} \hat{Z}_{m,n} \,e^{-i (m (-\theta) + n \zeta)} \right)
                   =    \realpart \left( \sum_{m=-\infty}^{+\infty} \sum_{n=-\infty}^{+\infty} \hat{Z}_{m,n} \,e^{ i (m   \theta  - n \zeta)} \right) \label{eqn:Z} \, .
\end{align}
Note that the specification of the surface geometry in \eqn{R} and \eqn{Z} is somewhat inconvenient to implement numerically:
it needs complex arithmetics and it features infinite sums over the mode numbers.
These issues are dealt with by two simplicifications outlined in the following.
First, the complex Fourier coefficients and complex basis functions are split into real and imaginary parts and rearranged.
This is illustrated exemplarly for $R$:
\begin{align}
  R(\theta, \zeta) =&\, \realpart \left( \sum_{m=-\infty}^{+\infty} \sum_{n=-\infty}^{+\infty}
                       \hat{R}_{m,n} \,e^{ i (m   \theta  - n \zeta)} \right)        \nonumber                 \\
                   =&\, \realpart \left( \sum_{m=-\infty}^{+\infty} \sum_{n=-\infty}^{+\infty}
                       \left( \hat{R}_{m,n}^\mathrm{Re} + i \cdot \hat{R}_{m,n}^\mathrm{Im} \right) \cdot
                       \left(  \cos(m \theta  - n \zeta) + i \cdot  \sin(m \theta  - n \zeta) \right) \right) \label{eqn:cplxR}
\end{align}
with $\hat{R}_{m,n}^\mathrm{Re} = \realpart \left( \hat{R}_{m,n} \right)$
and  $\hat{R}_{m,n}^\mathrm{Im} = \imagpart \left( \hat{R}_{m,n} \right)$.
Now consider the complex multiplication in \eqn{cplxR} separately:
\begin{align}
  ~& \left( \hat{R}_{m,n}^\mathrm{Re} + i \cdot \hat{R}_{m,n}^\mathrm{Im} \right) \cdot
     \left(  \cos(m \theta  - n \zeta) + i \cdot  \sin(m \theta  - n \zeta) \right) \nonumber \\
  =& \phantom{+ i \cdot}~~ \left( \hat{R}_{m,n}^\mathrm{Re} \cos(m \theta - n \zeta) - \hat{R}_{m,n}^\mathrm{Im} \sin(m \theta - n \zeta) \right) \nonumber \\
  ~&          + i \cdot    \left( \hat{R}_{m,n}^\mathrm{Re} \sin(m \theta - n \zeta) + \hat{R}_{m,n}^\mathrm{Im} \cos(m \theta - n \zeta) \right)
\end{align}
Only the real part of above result is required to express $R$ and the imaginary part is ignored.
The Fourier coefficients for $R$ are renamed after their (now real-valued) Fourier basis functions:
\begin{align}
  \hat{R}_{m,n}^\mathrm{cos} \equiv& \phantom{-}~ \hat{R}_{m,n}^\mathrm{Re} \\
  \hat{R}_{m,n}^\mathrm{sin} \equiv&          -   \hat{R}_{m,n}^\mathrm{Im}
\end{align}
and equivalently for $Z$:
\begin{align}
  \hat{Z}_{m,n}^\mathrm{cos} \equiv& \phantom{-}~ \hat{Z}_{m,n}^\mathrm{Re} \\
  \hat{Z}_{m,n}^\mathrm{sin} \equiv&          -   \hat{Z}_{m,n}^\mathrm{Im} \, .
\end{align}
This results in the following representation, which only involves real arithmetics now:
\begin{align}
  R(\theta, \zeta) =& \sum_{m=-\infty}^{+\infty} \sum_{n=-\infty}^{+\infty} \left( \hat{R}_{m,n}^\mathrm{cos} \cos(m \theta - n \zeta) + \hat{R}_{m,n}^\mathrm{sin} \sin(m \theta - n \zeta) \right) \nonumber \\
  Z(\theta, \zeta) =& \sum_{m=-\infty}^{+\infty} \sum_{n=-\infty}^{+\infty} \left( \hat{Z}_{m,n}^\mathrm{cos} \cos(m \theta - n \zeta) + \hat{Z}_{m,n}^\mathrm{sin} \sin(m \theta - n \zeta) \right) \nonumber \, .
\end{align}
Second, the Fourier spectrum of $R$ and $Z$ is truncated at some finite maximim mode numbers $M$ and $N$:
\begin{align}
  R(\theta, \zeta) =& \sum_{m=-M}^{M} \sum_{n=-N}^{N}
                       \left( \hat{R}_{m,n}^\mathrm{cos} \cos(m \theta - n \zeta) + \hat{R}_{m,n}^\mathrm{sin} \sin(m \theta - n \zeta) \right) \label{eqn:R2} \\
  Z(\theta, \zeta) =& \sum_{m=-M}^{M} \sum_{n=-N}^{N}
                       \left( \hat{Z}_{m,n}^\mathrm{cos} \cos(m \theta - n \zeta) + \hat{Z}_{m,n}^\mathrm{sin} \sin(m \theta - n \zeta) \right) \label{eqn:Z2} \, .
\end{align}
A sketch of the layout of the full mode spectrum used in \eqn{R2} and \eqn{Z2} is shown in Fig.~\ref{fig:mode_numbers}.
Redundant information among the Fourier harmonics $\hat{R}_{m,n}$ and $\hat{Z}_{m,n}$ is eliminated
by taking into account only modes with $0 \leq m \leq M$ and $|n| \leq N$,
which still fully define the (truncated) Fourier representation, as is shown in the following.
For any $(\theta, \zeta) \in \realnumbers$ and given mode numbers $(m,n)$ with $m < 0$ it holds:
\begin{align}
    \cos(m \theta - n \zeta)
 =& \cos\left(- ( m  \theta -   n  \zeta)\right) \nonumber \\
 =& \cos\left(  (-m) \theta - (-n) \zeta \right) \nonumber \\
 =& \cos\left(  m' \theta - n' \zeta \right)
\end{align}
and similarly:
\begin{align}
      \sin(m \theta - n \zeta)
 =& - \sin\left(- ( m  \theta -   n  \zeta)\right) \nonumber \\
 =& - \sin\left(  (-m) \theta - (-n) \zeta \right) \nonumber \\
 =& - \sin\left(  m' \theta - n' \zeta \right)
\end{align}
with $m' = -m$ and $n' = -n$ where now $m' > 0$.
Thus, the Fourier coefficients for $m<0$ can either be included in the Fourier series for $R$ and $Z$
by weighting them with their own basis functions
or they can equivalently be included in the Fourier coefficients for $m>0$ as follows:
\begin{align}
 \tilde{R}_{m,n}^\mathrm{cos} =& \hat{R}_{m,n}^\mathrm{cos} + \hat{R}_{-m,-n}^\mathrm{cos} \\
 \tilde{R}_{m,n}^\mathrm{sin} =& \hat{R}_{m,n}^\mathrm{sin} - \hat{R}_{-m,-n}^\mathrm{sin}
\end{align}
and analogously for $Z$:
\begin{align}
 \tilde{Z}_{m,n}^\mathrm{cos} =& \hat{Z}_{m,n}^\mathrm{cos} + \hat{Z}_{-m,-n}^\mathrm{cos} \\
 \tilde{Z}_{m,n}^\mathrm{sin} =& \hat{Z}_{m,n}^\mathrm{sin} - \hat{Z}_{-m,-n}^\mathrm{sin}
\end{align}
for $0 \leq m \leq M$ and $-N \leq n \leq N$.
Using these combined Fourier coefficients
    $\tilde{R}_{m,n}^\mathrm{cos}$, $\tilde{R}_{m,n}^\mathrm{sin}$
and $\tilde{Z}_{m,n}^\mathrm{cos}$, $\tilde{Z}_{m,n}^\mathrm{sin}$,
the number of summands in \eqn{R2} and \eqn{Z2} can already be almost halved:
\begin{align}
  R(\theta, \zeta) =& \sum_{m=0}^{M} \sum_{n=-N}^{N}
                       \left( \tilde{R}_{m,n}^\mathrm{cos} \cos(m \theta - n \zeta) + \tilde{R}_{m,n}^\mathrm{sin} \sin(m \theta - n \zeta) \right) \nonumber \\
  Z(\theta, \zeta) =& \sum_{m=0}^{M} \sum_{n=-N}^{N}
                       \left( \tilde{Z}_{m,n}^\mathrm{cos} \cos(m \theta - n \zeta) + \tilde{Z}_{m,n}^\mathrm{sin} \sin(m \theta - n \zeta) \right) \nonumber
\end{align}
which is equivalent to the green rectangle in Fig.~\ref{fig:mode_numbers}.
One further possible way of compression becomes evident when considering the Fourier basis for $m=0$.
Note that:
\begin{align}
 \cos(\underbrace{m \theta}_{=0} -n \zeta) =& \phantom{-}~ \cos(n \zeta) \label{eqn:m0cos} \\
 \sin(\underbrace{m \theta}_{=0} -n \zeta) =&          -   \sin(n \zeta) \label{eqn:m0sin} \, .
\end{align}
Thus, the Fourier coefficients for $m=0$ and $n<0$ can be combined with the ones for $m=0$ and $n>0$ as follows:
\begin{align}
 \tilde{\tilde{R}}_{0,n}^\mathrm{cos} \equiv& \phantom{-}~ \tilde{R}_{0,n}^\mathrm{cos} + \tilde{R}_{0,-n}^\mathrm{cos} \\
 \tilde{\tilde{R}}_{0,n}^\mathrm{sin} \equiv&          -   \tilde{R}_{0,n}^\mathrm{sin} + \tilde{R}_{0,-n}^\mathrm{sin}
\end{align}
where yet another sign change was performed for $\tilde{\tilde{R}}_{0,n}^\mathrm{sin}$ to cancel
the minus in the right-hand side of \eqn{m0sin}.
Analogous expressions follow for $Z$:
\begin{align}
 \tilde{\tilde{Z}}_{0,n}^\mathrm{cos} \equiv& \phantom{-}~ \tilde{Z}_{0,n}^\mathrm{cos} + \tilde{Z}_{0,-n}^\mathrm{cos} \\
 \tilde{\tilde{Z}}_{0,n}^\mathrm{sin} \equiv&          -   \tilde{Z}_{0,n}^\mathrm{sin} + \tilde{Z}_{0,-n}^\mathrm{sin} \, .
\end{align}
In the end, the two-dimensional Fourier series for $R$ and $Z$ can be written as follows:
\begin{align}
    R(\theta, \zeta)
 =& \phantom{+ \sum_{m=1}^{M}}~~\, \sum_{n=0}^{N}
      \left( \tilde{\tilde{R}}_{0,n}^\mathrm{cos} \cos(n \zeta) + \tilde{\tilde{R}}_{0,n}^\mathrm{sin} \sin(n \zeta) \right) \nonumber \\
 ~&          + \sum_{m=1}^{M}   \sum_{n=-N}^{N}
      \left( \tilde{R}_{m,n}^\mathrm{cos} \cos(m \theta - n \zeta) + \tilde{R}_{m,n}^\mathrm{sin} \sin(m \theta - n \zeta) \right) \nonumber \\
 =& \phantom{+}\, \sum_{n=0}^{N}                 \tilde{\tilde{R}}_{0,n}^\mathrm{cos} \cos(n \zeta)
             +    \sum_{m=1}^{M} \sum_{n=-N}^{N} \tilde{R}_{m,n}^\mathrm{cos}         \cos(m \theta - n \zeta) \nonumber \\
 ~&          +    \sum_{n=1}^{N}                 \tilde{\tilde{R}}_{0,n}^\mathrm{sin} \sin(n \zeta)
             +    \sum_{m=1}^{M} \sum_{n=-N}^{N} \tilde{R}_{m,n}^\mathrm{sin}         \sin(m \theta - n \zeta) \label{eqn:fullFourierR} \\
    Z(\theta, \zeta)
 =& \phantom{+ \sum_{m=1}^{M}}~~\, \sum_{n=0}^{N}
      \left( \tilde{\tilde{Z}}_{0,n}^\mathrm{cos} \cos(n \zeta) + \tilde{\tilde{Z}}_{0,n}^\mathrm{sin} \sin(n \zeta) \right) \nonumber \\
 ~&          + \sum_{m=1}^{M}   \sum_{n=-N}^{N}
      \left( \tilde{Z}_{m,n}^\mathrm{cos} \cos(m \theta - n \zeta) + \tilde{Z}_{m,n}^\mathrm{sin} \sin(m \theta - n \zeta) \right) \nonumber \\
 =& \phantom{+}\, \sum_{n=0}^{N}                 \tilde{\tilde{Z}}_{0,n}^\mathrm{cos} \cos(n \zeta)
             +    \sum_{m=1}^{M} \sum_{n=-N}^{N} \tilde{Z}_{m,n}^\mathrm{cos}         \cos(m \theta - n \zeta) \nonumber \\
 ~&          +    \sum_{n=1}^{N}                 \tilde{\tilde{Z}}_{0,n}^\mathrm{sin} \sin(n \zeta)
             +    \sum_{m=1}^{M} \sum_{n=-N}^{N} \tilde{Z}_{m,n}^\mathrm{sin}         \sin(m \theta - n \zeta) \label{eqn:fullFourierZ} \, .
\end{align}
Note that the $(0,0)$-Fourier coefficients for sine basis functions are redundant,
since $\sin(0\, \theta + 0\, \zeta) = 0$ for all values of $\theta$ and $\zeta$.
Thus, no matter what the value of $\tilde{\tilde{R}}_{0,0}^\mathrm{sin}$ or $\tilde{\tilde{Z}}_{0,0}^\mathrm{sin}$ would be,
their contributions would not enter into $R$ or $Z$, respectively, anyway.

\FloatBarrier
\newpage
\section{Symmetry Properties}
The surface geometry depicted in Fig.~\ref{fig:w7x_ref_1} features a five-fold symmetry in the toroidal direction
which can be taken into account to further reduce the number of Fourier coefficients required to describe the surface geometry.
The toroidal periodicity of the surface is denoted $n_\mathrm{fp}$ ($=5$ for W7-X).
Note that for a (axisymmetric) Tokamak, $n_\mathrm{fp}=1$ and $N=0$ are sufficient to describe the toroidal direction.
The cylindrical angle $\phi$ is now brought into correspondence with the toroidal coordinate $\zeta$
by the following relation:
\begin{equation}
  \zeta = n_\mathrm{fp} \cdot \phi \label{eqn:zeta} \, .
\end{equation}
Frequently, the Fourier representation of the surface geometry is written in terms as $\phi$ instead of $\zeta$:
\begin{align}
    R(\theta, \phi)
 =& \phantom{+ \sum_{m=1}^{M}}~~\, \sum_{n=0}^{N}
      \left( \tilde{\tilde{R}}_{0,n}^\mathrm{cos} \cos(n n_\mathrm{fp} \phi) + \tilde{\tilde{R}}_{0,n}^\mathrm{sin} \sin(n n_\mathrm{fp} \phi) \right) \nonumber \\
 ~&          + \sum_{m=1}^{M}   \sum_{n=-N}^{N}
      \left( \tilde{R}_{m,n}^\mathrm{cos} \cos(m \theta - n n_\mathrm{fp} \phi) + \tilde{R}_{m,n}^\mathrm{sin} \sin(m \theta - n n_\mathrm{fp} \phi) \right) \label{eqn:fullFourierRPhi} \\
    Z(\theta, \phi)
 =& \phantom{+ \sum_{m=1}^{M}}~~\, \sum_{n=0}^{N}
      \left( \tilde{\tilde{Z}}_{0,n}^\mathrm{cos} \cos(n n_\mathrm{fp} \phi) + \tilde{\tilde{Z}}_{0,n}^\mathrm{sin} \sin(n n_\mathrm{fp} \phi) \right) \nonumber \\
 ~&          + \sum_{m=1}^{M}   \sum_{n=-N}^{N}
      \left( \tilde{Z}_{m,n}^\mathrm{cos} \cos(m \theta - n n_\mathrm{fp} \phi) + \tilde{Z}_{m,n}^\mathrm{sin} \sin(m \theta - n n_\mathrm{fp} \phi) \right) \label{eqn:fullFourierZPhi} \, .
\end{align}
The $\zeta$ coordinate is also called toroidal angle per module,
since it advances by $2 \pi$ when going from one toroidal module to the next one.
Toroidal symmetry is expressed in mathematical form as follows:
\begin{align}
  R(\theta, \zeta) =& R \left( \theta, \zeta + 2 \pi k \right) \\
  Z(\theta, \zeta) =& Z \left( \theta, \zeta + 2 \pi k \right)
\end{align}
and equivalently for $\phi$:
\begin{align}
  R(\theta, \phi) =& R \left( \theta, \phi + \frac{2 \pi k}{n_\mathrm{fp}} \right) \\
  Z(\theta, \phi) =& Z \left( \theta, \phi + \frac{2 \pi k}{n_\mathrm{fp}} \right)
\end{align}
for $k \in \integers$.
Note that while $\phi$ takes on values between $0$ and $2 \pi$ for a full toroidal turn,
the $\zeta$-coordinates takes on values between $0$ and $n_\mathrm{fp} \cdot (2 \pi)$ for a full toroidal turn
around the whole machine.

One further symmetry property, commonly referred to as stellarator symmetry,
can be taken into account (if applicable) to reduce the number of Fourier coefficients.
It is observed that each of the $n_\mathrm{fp}$ toroidal modules is flip-symmetric about the axis in the $(x,y)$-plane at half its toroidal angle
as well as the axis in the $(x,y)$-plane at its toroidal boundaries.
This symmetry property is expressed as follows:
\begin{align}
  R(\theta, \zeta) =& \phantom{-}~ R(-\theta, -\zeta) \\
  Z(\theta, \zeta) =&          -   Z(-\theta, -\zeta) \, .
\end{align}
In order to recover this symmetry property in the Fourier representation of $R$ and $Z$,
it is required to only take into account Fourier coefficients for $R$ and $Z$ with appropriate parity.
For $R$, which needs to retain even parity, these are all Fourier coefficients weighted by cosine basis functions.
For $Z$, which needs to retain  odd parity, these are all Fourier coefficients weighted by sine basis functions.
Therefore, a stellarator-symmetric surface can be expressed as follows:
\begin{align}
    R(\theta, \zeta)
 =&                  \sum_{n= 0}^{N} \tilde{\tilde{R}}_{0,n}^\mathrm{cos}            \cos(n \zeta)
    + \sum_{m=1}^{M} \sum_{n=-N}^{N}         \tilde{R}_{m,n}^\mathrm{cos} \cos(m \theta - n \zeta) \label{eqn:stellsymR} \\
    Z(\theta, \zeta)
 =&                  \sum_{n= 0}^{N} \tilde{\tilde{Z}}_{0,n}^\mathrm{sin}            \sin(n \zeta)
    + \sum_{m=1}^{M} \sum_{n=-N}^{N}         \tilde{Z}_{m,n}^\mathrm{sin} \sin(m \theta - n \zeta) \label{eqn:stellsymZ}
\end{align}
and similarly for $\phi$:
\begin{align}
    R(\theta, \phi)
 =&                  \sum_{n= 0}^{N} \tilde{\tilde{R}}_{0,n}^\mathrm{cos}            \cos(n n_\mathrm{fp} \phi)
    + \sum_{m=1}^{M} \sum_{n=-N}^{N}         \tilde{R}_{m,n}^\mathrm{cos} \cos(m \theta - n n_\mathrm{fp} \phi) \label{eqn:stellsymRPhi} \\
    Z(\theta, \phi)
 =&                  \sum_{n= 0}^{N} \tilde{\tilde{Z}}_{0,n}^\mathrm{sin}            \sin(n n_\mathrm{fp} \phi)
    + \sum_{m=1}^{M} \sum_{n=-N}^{N}         \tilde{Z}_{m,n}^\mathrm{sin} \sin(m \theta - n n_\mathrm{fp} \phi) \label{eqn:stellsymZPhi} \, .
\end{align}
The terms only required when no stellarator symmetry can be assumed are marked {\color{blue}blue} from here on.

\FloatBarrier
\section{Vectorized Formulation of Fourier Series} \label{sec:torcoords_numbering}
It is evident that the Fourier summation for $R$ and $Z$ in \eqn{fullFourierRPhi} and \eqn{fullFourierZPhi}
and also for the stellarator-symmetric case in \eqn{stellsymRPhi} and \eqn{stellsymZPhi}
is rather complicated in terms of which coefficients to take into account and which not.
Efficiently implementing these expressions numerically would be rather difficult.
In order to fix this, the Fourier coefficients are arranged along one linear dimension
and appropriate mode number arrays are introduced.
The total number of mode number combinations to take into account is:
\begin{equation}
 K = (N + 1) + M (2N + 1) \, .
\end{equation}
The order in which the mode numbers and the Fourier coefficients appear in the linear arragement is shown in Tab.~\ref{tab:xm_xn_contents}.
\begin{table}[htb]
  \centering
  \begin{tabular}{c|c|c|c|c|c|c}
    index $k$ & $m_k \equiv m$ & $n_k \equiv n_\mathrm{fp} n$ &               $R_k^\mathrm{cos}$       &               $Z_k^\mathrm{sin}$       & {\color{blue}               $R_k^\mathrm{sin}$       } & {\color{blue}               $Z_k^\mathrm{cos}$       } \\
    \hline %  &                &                              &                                        &                                        & {\color{blue}                                        } & {\color{blue}                                        } \\
    1         & 0              & 0                            & $\tilde{\tilde{R}}_{0,0}^\mathrm{cos}$ & $\tilde{\tilde{Z}}_{0,0}^\mathrm{sin}$ & {\color{blue} $\tilde{\tilde{R}}_{0,0}^\mathrm{sin}$ } & {\color{blue} $\tilde{\tilde{Z}}_{0,0}^\mathrm{cos}$ } \\
    ...       & 0              & ...                          & ...                                    & ...                                    & {\color{blue} ...                                    } & {\color{blue} ...                                    } \\
    $N+1$     & 0              & $  n_\mathrm{fp} N$          & $\tilde{\tilde{R}}_{0,N}^\mathrm{cos}$ & $\tilde{\tilde{Z}}_{0,N}^\mathrm{sin}$ & {\color{blue} $\tilde{\tilde{R}}_{0,N}^\mathrm{sin}$ } & {\color{blue} $\tilde{\tilde{Z}}_{0,N}^\mathrm{cos}$ } \\
    \hline %  &                &                              &                                        &                                        & {\color{blue}                                        } & {\color{blue}                                        } \\
    $N+2$     & 1              & $- n_\mathrm{fp} N$          &        $\tilde{R}_{1,-N}^\mathrm{cos}$ &        $\tilde{Z}_{1,-N}^\mathrm{sin}$ & {\color{blue}        $\tilde{R}_{1,-N}^\mathrm{sin}$ } & {\color{blue}        $\tilde{Z}_{1,-N}^\mathrm{cos}$ } \\
    ...       & 1              & ...                          & ...                                    & ...                                    & {\color{blue} ...                                    } & {\color{blue} ...                                    } \\
    $2N+2$    & 1              & 0                            &        $\tilde{R}_{1, 0}^\mathrm{cos}$ &        $\tilde{Z}_{1, 0}^\mathrm{sin}$ & {\color{blue}        $\tilde{R}_{1, 0}^\mathrm{sin}$ } & {\color{blue}        $\tilde{Z}_{1, 0}^\mathrm{cos}$ } \\
    ...       & 1              & ...                          & ...                                    & ...                                    & {\color{blue} ...                                    } & {\color{blue} ...                                    } \\
    $3N+2$    & 1              & $  n_\mathrm{fp} N$          &        $\tilde{R}_{1, N}^\mathrm{cos}$ &        $\tilde{Z}_{1, N}^\mathrm{sin}$ & {\color{blue}        $\tilde{R}_{1, N}^\mathrm{sin}$ } & {\color{blue}        $\tilde{Z}_{1, N}^\mathrm{cos}$ } \\
    \hline %  &                &                              &                                        &                                        & {\color{blue}                                        } & {\color{blue}                                        } \\
    ...       & ...            & ...                          & ...                                    & ...                                    & {\color{blue} ...                                    } & {\color{blue} ...                                    } \\
    \hline %  &                &                              &                                        &                                        & {\color{blue}                                        } & {\color{blue}                                        } \\
    ...       & $M$            & $- n_\mathrm{fp} N$          &        $\tilde{R}_{M,-N}^\mathrm{cos}$ &        $\tilde{Z}_{M,-N}^\mathrm{sin}$ & {\color{blue}        $\tilde{R}_{M,-N}^\mathrm{sin}$ } & {\color{blue}        $\tilde{Z}_{M,-N}^\mathrm{cos}$ } \\
    ...       & ...            & ...                          & ...                                    & ...                                    & {\color{blue} ...                                    } & {\color{blue} ...                                    } \\
    $K$       & $M$            & $  n_\mathrm{fp} N$          &        $\tilde{R}_{M, N}^\mathrm{cos}$ &        $\tilde{Z}_{M, N}^\mathrm{sin}$ & {\color{blue}        $\tilde{R}_{M, N}^\mathrm{sin}$ } & {\color{blue}        $\tilde{Z}_{M, N}^\mathrm{cos}$ } \\
  \end{tabular}
  \caption{Mode number combinations used for the truncated Fourier series representation of flux surface coordinates
  in VMEC. For $m=0$, only positive $n$ are needed to support a complete truncated Fourier basis.
  Note that the number of toroidal periods $n_\mathrm{fp}$ has already been included in the $n_k$.}
  \label{tab:xm_xn_contents}
\end{table}

Using this linear arragement of mode numbers, the Fourier series for $R$ and $Z$ can be reformulated as follows:
\begin{align}
    R(\theta, \phi) =& \sum_{k=1}^{K}   R_k^\mathrm{cos} \cos(m_k \theta - n_k \phi)
                        {\color{blue} + R_k^\mathrm{sin} \sin(m_k \theta - n_k \phi) } \label{eqn:linearFourierRPhi} \\
    Z(\theta, \phi) =& \sum_{k=1}^{K}   Z_k^\mathrm{sin} \sin(m_k \theta - n_k \phi)
                        {\color{blue} + Z_k^\mathrm{cos} \cos(m_k \theta - n_k \phi) } \label{eqn:linearFourierZPhi} \, .
\end{align}

\FloatBarrier
\section{Normalized Coordinates}
In numerical implementations is it often convenient to work with normalized coordinates,
which do not range from $0$ to some arbitrary real number, but to $1$.
The normalized coordinates for parameterizing flux surfaces in a stellarator
are commonly called $u$ for the poloidal direction and $v$ for the toroidal direction.
They are defined as follows:
\begin{align}
  u \equiv& \frac{\theta}{2 \pi} \\
  v \equiv& \frac{\zeta}{2 \pi}
\end{align}
which results in $0 \leq u \leq 1$ and $0 \leq v \leq n_\mathrm{fp}$.
For convenience, note that
\begin{equation}
  \phi = \frac{2 \pi}{n_\mathrm{fp}} v \label{eqn:phiFromV} \, .
\end{equation}

\FloatBarrier
\section{Tangential Derivatives} \label{sec:tangential_derivatives}
The Fourier representation of $R$ and $Z$ makes it almost trivial to compute tangential derivatives of $R$ and $Z$,
that is, derivatives with respect to $\theta$ and $\phi$.
First consider the first-order derivatives with respect to $\theta$:
\begin{align}
     \frac{\partial R}{\partial \theta}
  =& \sum_{k=1}^{K}   m_k  R_k^\mathrm{cos} (- \sin) (m_k \theta - n_k \phi) {\color{blue} + m_k R_k^\mathrm{sin}   \cos (m_k \theta - n_k \phi) } \nonumber \\
  =& \sum_{k=1}^{K}  -m_k  R_k^\mathrm{cos}    \sin  (m_k \theta - n_k \phi) {\color{blue} + m_k R_k^\mathrm{sin}   \cos (m_k \theta - n_k \phi) } \label{eqn:dRdt} \\
     \frac{\partial Z}{\partial \theta}
  =& \sum_{k=1}^{K}   m_k  Z_k^\mathrm{sin}    \cos  (m_k \theta - n_k \phi) {\color{blue} + m_k Z_k^\mathrm{cos} (-\sin)(m_k \theta - n_k \phi) } \nonumber \\
  =& \sum_{k=1}^{K}   m_k  Z_k^\mathrm{sin}    \cos  (m_k \theta - n_k \phi) {\color{blue} - m_k Z_k^\mathrm{cos}   \sin (m_k \theta - n_k \phi) } \label{eqn:dZdt} \, .
\end{align}
Next consider the first-order derivatives with respect to $\phi$:
\begin{align}
     \frac{\partial R}{\partial \phi}
  =& \sum_{k=1}^{K} (-n_k) R_k^\mathrm{cos} (- \sin) (m_k \theta - n_k \phi) {\color{blue} - n_k R_k^\mathrm{sin}   \cos (m_k \theta - n_k \phi) } \nonumber \\
  =& \sum_{k=1}^{K}   n_k  R_k^\mathrm{cos}    \sin  (m_k \theta - n_k \phi) {\color{blue} - n_k R_k^\mathrm{sin}   \cos (m_k \theta - n_k \phi) } \label{eqn:dRdp} \\
     \frac{\partial Z}{\partial \phi}
  =& \sum_{k=1}^{K} (-n_k) Z_k^\mathrm{sin}    \cos  (m_k \theta - n_k \phi) {\color{blue} - n_k Z_k^\mathrm{cos} (-\sin)(m_k \theta - n_k \phi) } \nonumber \\
  =& \sum_{k=1}^{K}  -n_k  Z_k^\mathrm{sin}    \cos  (m_k \theta - n_k \phi) {\color{blue} + n_k Z_k^\mathrm{cos}   \sin (m_k \theta - n_k \phi) } \label{eqn:dZdp} \, .
\end{align}
The second-order derivatives are equally simple to compute.
We start by considering the second-order derivatives with respect to $\theta$:
\begin{align}
     \frac{\partial^2 R}{\partial \theta^2}
  =& \sum_{k=1}^{K}   m_k(-m_k)  R_k^\mathrm{cos}   \cos (m_k \theta - n_k \phi) {\color{blue} + m_k   m_k  R_k^\mathrm{sin} (-\sin)(m_k \theta - n_k \phi) } \nonumber \\
  =& \sum_{k=1}^{K}  -m_k^2      R_k^\mathrm{cos}   \cos (m_k \theta - n_k \phi) {\color{blue} - m_k^2      R_k^\mathrm{sin}   \sin (m_k \theta - n_k \phi) } \label{eqn:d2Rdt2} \\
     \frac{\partial^2 Z}{\partial \theta^2}
  =& \sum_{k=1}^{K}   m_k m_k    Z_k^\mathrm{sin} (-\sin)(m_k \theta - n_k \phi) {\color{blue} + m_k (-m_k) Z_k^\mathrm{cos}   \cos (m_k \theta - n_k \phi) } \nonumber \\
  =& \sum_{k=1}^{K}  -m_k^2      Z_k^\mathrm{sin}   \sin (m_k \theta - n_k \phi) {\color{blue} - m_k^2      Z_k^\mathrm{cos}   \cos (m_k \theta - n_k \phi) } \label{eqn:d2Zdt2} \, .
\end{align}
We continue by considering the second-order mixed derivatives with respect to $\theta$ and $\phi$:
\begin{align}
     \frac{\partial^2 R}{\partial \theta \partial \phi}
  =& \sum_{k=1}^{K} (-n_k)(-m_k) R_k^\mathrm{cos}   \cos (m_k \theta - n_k \phi) {\color{blue} + (-n_k) m_k  R_k^\mathrm{sin} (-\sin)(m_k \theta - n_k \phi) } \nonumber \\
  =& \sum_{k=1}^{K}   m_k   n_k  R_k^\mathrm{cos}   \cos (m_k \theta - n_k \phi) {\color{blue} +   m_k   n_k  R_k^\mathrm{sin}   \sin (m_k \theta - n_k \phi) } \label{eqn:d2Rdtdp} \\
     \frac{\partial^2 Z}{\partial \theta \partial \phi}
  =& \sum_{k=1}^{K} (-n_k) m_k   Z_k^\mathrm{sin} (-\sin)(m_k \theta - n_k \phi) {\color{blue} + (-n_k)(-m_k) Z_k^\mathrm{cos}   \cos (m_k \theta - n_k \phi) } \nonumber \\
  =& \sum_{k=1}^{K}   m_k  n_k   Z_k^\mathrm{sin}   \sin (m_k \theta - n_k \phi) {\color{blue} +   m_k   n_k  Z_k^\mathrm{cos}   \cos (m_k \theta - n_k \phi) } \label{eqn:d2Zdtdp} \, .
\end{align}
Finally, the second-order derivatives with respect to $\phi$ are:
\begin{align}
     \frac{\partial^2 R}{\partial \phi^2}
  =& \sum_{k=1}^{K} (-n_k) n_k  R_k^\mathrm{cos}    \cos  (m_k \theta - n_k \phi) {\color{blue} - (-n_k) n_k R_k^\mathrm{sin} (-\sin) (m_k \theta - n_k \phi) } \nonumber \\
  =& \sum_{k=1}^{K}  -n_k^2     R_k^\mathrm{cos}    \cos  (m_k \theta - n_k \phi) {\color{blue} -   n_k^2    R_k^\mathrm{sin}   \sin  (m_k \theta - n_k \phi) } \label{eqn:d2Rdp2} \\
     \frac{\partial^2 Z}{\partial \phi^2}
  =& \sum_{k=1}^{K}-(-n_k) n_k  Z_k^\mathrm{sin}  (-\sin) (m_k \theta - n_k \phi) {\color{blue} + (-n_k) n_k Z_k^\mathrm{cos}   \cos (m_k \theta - n_k \phi) } \nonumber \\
  =& \sum_{k=1}^{K}  -n_k^2     Z_k^\mathrm{sin}    \sin  (m_k \theta - n_k \phi) {\color{blue} -   n_k^2    Z_k^\mathrm{cos}   \cos (m_k \theta - n_k \phi) } \label{eqn:d2Zdp2} \, .
\end{align}

\FloatBarrier
\section{Toroidal Rotation of the Coordinate System}
The whole toroidal coordinate system can be rotated in the toroidal direction by simple changes in the Fourier coefficients alone.
Consider, e.g., the representation of the cylindrical $R$ coordinate:
\begin{equation}
  R(\theta, \phi) = \sum_{k=1}^{K}   R_k^\mathrm{cos} \cos(m_k \theta - n_k \phi)
                     {\color{blue} + R_k^\mathrm{sin} \sin(m_k \theta - n_k \phi) } \, .
\end{equation}
Suppose a shift in the toroidal angle is to be done according to $\phi \rightarrow \phi + \Delta \phi$.
Then, the cylindrical $R$ coordinate can be evaluated as follows:
\begin{align}
 R(\theta, \phi + \Delta \phi) =& \sum_{k=1}^{K}   R_k^\mathrm{cos} \cos(m_k \theta - n_k (\phi +     \Delta \phi))
                                   {\color{blue} + R_k^\mathrm{sin} \sin(m_k \theta - n_k (\phi +     \Delta \phi)) } \nonumber \\
                               =& \sum_{k=1}^{K}   R_k^\mathrm{cos} \cos(m_k \theta - n_k  \phi - n_k \Delta \phi)
                                   {\color{blue} + R_k^\mathrm{sin} \sin(m_k \theta - n_k  \phi - n_k \Delta \phi) } \, .
\end{align}
The Fourier basis functions can be split up according to the addition theorems of trigonometry:
\begin{align}
    \cos(m_k \theta - n_k  \phi - n_k \Delta \phi)
 =& \cos(m_k \theta - n_k  \phi) \cos(- n_k \Delta \phi) - \sin(m_k \theta - n_k  \phi) \sin(- n_k \Delta \phi) \nonumber \\
 =& \cos(m_k \theta - n_k  \phi) \cos(  n_k \Delta \phi) + \sin(m_k \theta - n_k  \phi) \sin(  n_k \Delta \phi) \\
    \sin(m_k \theta - n_k  \phi - n_k \Delta \phi)
 =& \sin(m_k \theta - n_k  \phi) \cos(- n_k \Delta \phi) + \cos(m_k \theta - n_k  \phi) \sin(- n_k \Delta \phi) \nonumber \\
 =& \sin(m_k \theta - n_k  \phi) \cos(  n_k \Delta \phi) - \cos(m_k \theta - n_k  \phi) \sin(  n_k \Delta \phi) \, .
\end{align}
The terms with the same two-dimensional Fourier basis terms are collected and re-arranged:
\begin{align}
  R(\theta, \phi + \Delta \phi)
 =&          \sum_{k=1}^{K} \Biggl\{          \phantom{+} \cos(m_k \theta - n_k \phi)   \left[                 R_k^\mathrm{cos} \cos(n_k \Delta \phi)
                                                                                               {\color{blue} - R_k^\mathrm{sin} \sin(n_k \Delta \phi) } \right] \nonumber \\
 ~& \phantom{\sum_{k=1}^{K} \Biggl\{ } ~ {\color{blue} +  \sin(m_k \theta - n_k \phi) } \left[                 R_k^\mathrm{cos} \sin(n_k \Delta \phi)
                                                                                               {\color{blue} + R_k^\mathrm{sin} \cos(n_k \Delta \phi) } \right] \Biggr\} \, .
\end{align}
New Fourier coefficients $R_k^{\mathrm{c},\Delta \phi}$, $R_k^{\mathrm{s},\Delta \phi}$ are now defined, which incorporate the toroidal offset:
\begin{align}
  R_k^{\mathrm{c},\Delta \phi} =&\, R_k^\mathrm{cos} \cos(n_k \Delta \phi) {\color{blue} - R_k^\mathrm{sin} \sin(n_k \Delta \phi) } \label{eqn:rkcosrot} \\
  R_k^{\mathrm{s},\Delta \phi} =&\, R_k^\mathrm{cos} \sin(n_k \Delta \phi) {\color{blue} + R_k^\mathrm{sin} \cos(n_k \Delta \phi) } \label{eqn:rksinrot} \, .
\end{align}
Using these rotated coefficients, the Fourier series representation for $R(\theta, \phi + \Delta \phi)$ reads as follows:
\begin{equation}
  R(\theta, \phi + \Delta \phi) = \sum_{k=1}^{K}   R_k^{\mathrm{c},\Delta \phi} \cos(m_k \theta - n_k \phi)
                                                 + R_k^{\mathrm{s},\Delta \phi} \sin(m_k \theta - n_k \phi) \, .
\end{equation}
It should be emphasized that while the original (non-rotated) Fourier series for $R$
might have implied stellarator symmetry by leaving out the ${\color{blue} R_k^\mathrm{sin} }$ terms (highlighted in blue),
the rotated Fourier coefficients are generally no longer stellarator-symmetric.
This can be seen from the fact that there are contributions from the stellarator-symmetric (and thus always present) terms $R_k^\mathrm{cos}$
to the non-stellarator-symmetric rotated Fourier coefficients $R_k^{\mathrm{s},\Delta \phi}$ in \eqn{rksinrot}.

\FloatBarrier
\newpage
\section{Cartesian Components, Unit Vectors and Surface Normal}
The transformation from cylindrical to Cartesian coordinates is done as follows:
\begin{equation}
  \mathbf{x}
  = \left( \begin{array}{c} x \\ y \\ z \end{array} \right)
  = \left( \begin{array}{c}
                        R \cos(\phi) \\
                        R \sin(\phi) \\
                        Z
                      \end{array} \right) \label{eqn:cyl2cart}
\end{equation}
where $\mathbf{x}$ is the Cartesian equivalent of $(R, \phi, Z)$.
The cylindrical components of the differential line element $\mathrm{d}\mathbf{x}$ is:
\begin{equation}
  \mathrm{d}\mathbf{x} = \mathrm{d}R \, \hat{\mathbf{e}}_R + R \, \mathrm{d}\phi \, \hat{\mathbf{e}}_\phi + \mathrm{d}Z \, \hat{\mathbf{e}}_Z \label{eqn:dx_cyl}
\end{equation}
with the Cartesian components of the cylindrical unit vectors as follows:
\begin{equation}
  \hat{\mathbf{e}}_R    = \left( \begin{array}{c}  \cos(\phi) \\ \sin(\phi) \\ 0 \end{array} \right), ~
  \hat{\mathbf{e}}_\phi = \left( \begin{array}{c} -\sin(\phi) \\ \cos(\phi) \\ 0 \end{array} \right), ~
  \hat{\mathbf{e}}_Z    = \left( \begin{array}{c}      0      \\     0      \\ 1 \end{array} \right) \, .
\end{equation}
The only dependence of the cylindrical unit vectors on the cylindrical coordinates
is through $\phi$ in $\hat{\mathbf{e}}_R$ and in $\hat{\mathbf{e}}_\phi$:
\begin{align}
  \frac{\partial \hat{\mathbf{e}}_R   }{\partial \phi} =& \left( \begin{array}{c} -\sin(\phi) \\  \cos(\phi) \\ 0 \end{array} \right) =  \hat{\mathbf{e}}_\phi \\
  \frac{\partial \hat{\mathbf{e}}_\phi}{\partial \phi} =& \left( \begin{array}{c} -\cos(\phi) \\ -\sin(\phi) \\ 0 \end{array} \right) = -\hat{\mathbf{e}}_R \, .
\end{align}
These expressions are required when computing derivatives with respect to $\phi$ of \eqn{cyl2cart}.
The Cartesian components~$(N^X, N^Y, N^Z)$ of a vector~$\mathbf{N}$ specified via its cylindrical components~$(N^R, N^\varphi, N^Z)$ are as follows:
\begin{equation}
 \left( \begin{matrix}
    N^X \\
    N^Y \\
    N^Z
 \end{matrix} \right)
 =
 \left( \begin{matrix}
   N^R \cos(\phi) - N^\varphi  \sin(\phi) \\
   N^R \sin(\phi) + N^\varphi  \cos(\phi) \\
   N^Z
 \end{matrix} \right) \, .
\end{equation}
With a little bit of impreciseness in notation when using \eqn{dx_cyl}, the Cartesian components of the tangential derivatives can be written as follows:
\begin{align}
  \frac{\partial \mathbf{x}}{\partial \theta} =&                 \frac{\partial R}{\partial \theta} \, \hat{\mathbf{e}}_R
                                                 + R \underbrace{\frac{\partial \phi}{\partial \theta}}_{=0} \, \hat{\mathbf{e}}_\phi
                                                 +               \frac{\partial Z}{\partial \theta} \, \hat{\mathbf{e}}_Z
                                              = \left( \begin{aligned}
                                                  \frac{\partial R}{\partial \theta} & \cos(\phi)  \\
                                                  \frac{\partial R}{\partial \theta} & \sin(\phi)  \\
                                                  \frac{\partial Z}{\partial \theta} & ~
                                                \end{aligned} \right) \label{eqn:dXdt} \\
  \frac{\partial \mathbf{x}}{\partial \phi}   =&                 \frac{\partial R   }{\partial \phi} \, \hat{\mathbf{e}}_R
                                                 + R \underbrace{\frac{\partial \phi}{\partial \phi}}_{=1} \, \hat{\mathbf{e}}_\phi
                                                 +               \frac{\partial Z   }{\partial \phi} \, \hat{\mathbf{e}}_Z
                                              = \left( \begin{aligned}
                                                  \frac{\partial R}{\partial \phi} \cos(\phi) -& R \sin(\phi) \\
                                                  \frac{\partial R}{\partial \phi} \sin(\phi) +& R \cos(\phi) \\
                                                  \frac{\partial Z}{\partial \phi}             & ~
                                                \end{aligned} \right) \label{eqn:dXdp} \, .
\end{align}
Note that with \eqn{zeta}, it follows:
\begin{equation}
    \frac{\partial \mathbf{x}}{\partial \zeta}
  = \frac{\partial \mathbf{x}}{\partial \phi } \underbrace{\frac{\partial \phi}{\partial \zeta}}_{=1/n_\mathrm{fp}}
  = \frac{1}{n_\mathrm{fp}} \frac{\partial \mathbf{x}}{\partial \phi} \label{eqn:dXdz} \, .
\end{equation}
The surface normal vector~$\mathbf{N}$ is defined as:
\begin{equation}
  \mathbf{N} = - \frac{\partial \mathbf{x}}{\partial \theta} \times \frac{\partial \mathbf{x}}{\partial \phi} \label{eqn:surfN}
\end{equation}
and the unit vector normal to the surface~$\mathbf{n}$ is then:
\begin{equation}
    \mathbf{n}
  = \frac{\mathbf{N}}{|\mathbf{N}|}
  = - \frac{       \frac{\partial \mathbf{x}}{\partial \theta} \times \frac{\partial \mathbf{x}}{\partial \phi}        }
           {\left| \frac{\partial \mathbf{x}}{\partial \theta} \times \frac{\partial \mathbf{x}}{\partial \phi} \right|} \, .
\end{equation}
The cross product in \eqn{surfN} can be evaluated using the cylindrical components of
the tangential derivatives of $\mathbf{x}$ from \eqn{dXdt} and \eqn{dXdp}:
\begin{equation}
 \mathbf{N} = \left( \begin{aligned} N^R \\ N^\phi \\ N^Z \end{aligned} \right)
            = - \left( \begin{aligned}
                                   0                 \cdot \frac{\partial Z}{\partial \phi  } &- \frac{\partial Z}{\partial \theta} \cdot                  R               \\
                  \frac{\partial Z}{\partial \theta} \cdot \frac{\partial R}{\partial \phi  } &- \frac{\partial R}{\partial \theta} \cdot \frac{\partial Z}{\partial \phi} \\
                  \frac{\partial R}{\partial \theta} \cdot                  R                 &-                  0                 \cdot \frac{\partial R}{\partial \phi}
                \end{aligned} \right)
            =   \left( \begin{aligned}
                                                       R                                      &\cdot                 \frac{\partial Z}{\partial \theta}                    \\
                  \frac{\partial R}{\partial \theta} \cdot \frac{\partial Z}{\partial \phi  } &-     \frac{\partial Z}{\partial \theta} \cdot \frac{\partial R}{\partial \phi} \\
                                                      -R                                      &\cdot                 \frac{\partial R}{\partial \theta}
                \end{aligned} \right) \label{eqn:surfNComponents} \, .
\end{equation}
The second-order tangential derivatives of the Cartesian position vector $\mathbf{x}$ are a little bit involved,
as will become evident in the following.
We start with the second-order derivative in poloidal direction:
\begin{align}
       \frac{\partial^2 \mathbf{x}}{\partial \theta^2}
  =&   \frac{\partial}{\partial \theta} \left( \frac{\partial R}{\partial \theta} \, \hat{\mathbf{e}}_R \right)
     + \frac{\partial}{\partial \theta} \left( \frac{\partial Z}{\partial \theta} \, \hat{\mathbf{e}}_Z \right) \nonumber \\
  =&   \frac{\partial^2 R}{\partial \theta^2} \, \hat{\mathbf{e}}_R
     + \frac{\partial^2 Z}{\partial \theta^2} \, \hat{\mathbf{e}}_Z \label{eqn:d2Xdt2} \, .
\end{align}
The mixed second-order derivative can be computed in two ways and it will be shown that both lead to the same result.
We start by computing the derivative of \eqn{dXdt} with respect to $\phi$:
\begin{align}
      \frac{\partial}{\partial \phi} \left( \frac{\partial \mathbf{x}}{\partial \theta} \right)
 =&   \frac{\partial}{\partial \phi} \left( \frac{\partial R}{\partial \theta} \, \hat{\mathbf{e}}_R \right)
    + \frac{\partial}{\partial \phi} \left( \frac{\partial Z}{\partial \theta} \, \hat{\mathbf{e}}_Z \right) \nonumber \\
 =&   \frac{\partial^2 R}{\partial \theta \partial \phi} \, \hat{\mathbf{e}}_R
    + \frac{\partial R}{\partial \theta} \underbrace{\frac{\partial \hat{\mathbf{e}}_R}{\partial \phi}}_{=\hat{\mathbf{e}}_\phi}
    + \frac{\partial^2 Z}{\partial \theta \partial \phi} \, \hat{\mathbf{e}}_Z
    + \frac{\partial Z}{\partial \theta} \underbrace{\frac{\partial \hat{\mathbf{e}}_Z}{\partial \phi}}_{=0} \nonumber \\
 =&   \frac{\partial^2 R}{\partial \theta \partial \phi} \, \hat{\mathbf{e}}_R
    + \frac{\partial R}{\partial \theta} \, \hat{\mathbf{e}}_\phi
    + \frac{\partial^2 Z}{\partial \theta \partial \phi} \, \hat{\mathbf{e}}_Z \label{eqn:d2Xdtdz} \, .
\end{align}
For comparison, we also compute the derivative of \eqn{dXdp} with respect to $\theta$:
\begin{align}
      \frac{\partial}{\partial \theta} \left( \frac{\partial \mathbf{x}}{\partial \phi} \right)
 =&   \frac{\partial}{\partial \theta} \left( \frac{\partial R}{\partial \phi  } \, \hat{\mathbf{e}}_R    \right)
    + \frac{\partial}{\partial \theta} \left(                R                   \, \hat{\mathbf{e}}_\phi \right)
    + \frac{\partial}{\partial \theta} \left( \frac{\partial Z}{\partial \phi  } \, \hat{\mathbf{e}}_Z    \right) \nonumber \\
 =&   \frac{\partial^2 R}{\partial \phi \partial \theta} \, \hat{\mathbf{e}}_R
    + \frac{\partial R}{\partial \theta}                 \, \hat{\mathbf{e}}_\phi
    + \frac{\partial^2 Z}{\partial \phi \partial \theta} \, \hat{\mathbf{e}}_Z  \label{eqn:d2Xdzdt} \, .
\end{align}
\eqn{d2Xdtdz} and \eqn{d2Xdzdt} result in the same expression and therefore,
the second-order mixed derivative can be written as:
\begin{equation}
      \frac{\partial^2 \mathbf{x}}{\partial \theta \partial \phi}
  =   \frac{\partial^2         R }{\partial \theta \partial \phi} \, \hat{\mathbf{e}}_R
    + \frac{\partial           R }{\partial \theta              } \, \hat{\mathbf{e}}_\phi
    + \frac{\partial^2         Z }{\partial \theta \partial \phi} \, \hat{\mathbf{e}}_Z  \label{eqn:d2Xdtz} \, .
\end{equation}
It remains to compute the second-order derivative in toroidal direction:
\begin{align}
      \frac{\partial^2 \mathbf{x}}{\partial \phi^2}
 =&   \frac{\partial}{\partial \phi} \left( \frac{\partial R}{\partial \phi} \, \hat{\mathbf{e}}_R    \right)
    + \frac{\partial}{\partial \phi} \left(                R                 \, \hat{\mathbf{e}}_\phi \right)
    + \frac{\partial}{\partial \phi} \left( \frac{\partial Z}{\partial \phi} \, \hat{\mathbf{e}}_Z    \right) \nonumber \\
 =&   \frac{\partial^2 R }{\partial \phi^2} \, \hat{\mathbf{e}}_R
    + \frac{\partial   R }{\partial \phi  } \underbrace{\frac{\partial \hat{\mathbf{e}}_R}{\partial \phi}}_{=\hat{\mathbf{e}}_\phi}
    + \frac{\partial   R }{\partial \phi  } \,\hat{\mathbf{e}}_\phi
    +                  R                    \underbrace{\frac{\partial \hat{\mathbf{e}}_\phi}{\partial \phi}}_{=-\hat{\mathbf{e}}_R}
    + \frac{\partial^2 Z }{\partial \phi^2} \, \hat{\mathbf{e}}_Z
    + \frac{\partial   Z }{\partial \phi  } \underbrace{\frac{\partial \hat{\mathbf{e}}_Z}{\partial \phi}}_{=0} \nonumber \\
 =&   \left( \frac{\partial^2 R}{\partial \phi^2} - R \right) \, \hat{\mathbf{e}}_R
    +   2    \frac{\partial   R}{\partial \phi  }             \, \hat{\mathbf{e}}_\phi
    + \frac{\partial^2 Z }{\partial \phi^2} \, \hat{\mathbf{e}}_Z \label{eqn:d2Xdz2} \, .
\end{align}

\FloatBarrier
\section{Implementation Notes}
The mapping between quantities used in this chapter and the corresponding variables in the Fortran VMEC implementation
is given in Tab.~\ref{tab:fourier_vars}.
\begin{table}[htb]
  \centering
  \begin{tabular}[t]{ c | c }
    quantity & variable \\
    \hline
    $(M+1)$            & \code{mpol} \\
    $N$                & \code{ntor} \\
    $K$                & \code{mnmax} \\
    $n_\mathrm{fp}$    & \code{nfp} \\
    $m_k$              & \code{xm} \\
    $n_k$              & \code{xn} \\
    $R_k^\mathrm{cos}$ & \code{rmnc} \\
    $Z_k^\mathrm{sin}$ & \code{zmns} \\
    $R_k^\mathrm{sin}$ & \code{rmns} \\
    $Z_k^\mathrm{cos}$ & \code{zmnc} \\
 \end{tabular}
 \caption{Mapping of variables occuring in this chapter to the corresponding Fortran variables in VMEC.}
 \label{tab:fourier_vars}
\end{table}

\chapter{Spectral Condensation}
When given a real-space representation of a flux surface, e.g., from field line tracing,
the question arises how to transform this set of points into Fourier coefficients suitable for use in VMEC.
The toroidal angle $\zeta$ is fixed by definition to be the cylindrical angle.
Tangential freedom remains in the definition of the poloidal angle,
which can be exploited to make the most economic use of the available, finite, Fourier coefficients used in the computation.
This process is called spectral condensation~\cite{hirshman_meier_1985}.

For the moment, the focus is on one-dimensional Fourier series
that describe the shape $(x,y)$ of a single flux surface in one poloidal cut
parameterized by the poloidal coordinate $\theta$:
\begin{align}
  x(\theta) =&\, \sum\limits_{m=0}^{m_\textrm{max}} x_m \cos(m \theta) \label{eqn:x_of_theta} \\
  y(\theta) =&\, \sum\limits_{m=0}^{m_\textrm{max}} y_m \sin(m \theta) \label{eqn:y_of_theta}
\end{align}
with Fourier coefficients $x_m, y_m$ for $x, y$, respectively, up to some maximum poloidal mode number $m_\textrm{max}$.
Note that above series assume stellarator symmetry.
The extension to asymmetric boundaries can be performed by including additional coefficients and basis functions
as described in the previous chapter. This is omitted here for a clearer focus on the subject of spectral condensation.

\FloatBarrier
\section{Spectral Width}
The power spectrum $S_p(m)$ of this description of the surface is defined as:
\begin{equation}
 S_p(m) = m^p (x_m^2 + y_m^2)
\end{equation}
where $p \geq 0$.
Then, a $q$-moment of $S_p(m)$ is introduced and this leads to the spectral width $M$:
\begin{equation}
 M(p,q) = \frac{\sum\limits_{m=1}^{m_\textrm{max}} m^q S_p(m)}
               {\sum\limits_{m=1}^{m_\textrm{max}}     S_p(m)}
        = \frac{\sum\limits_{m=1}^{m_\textrm{max}} m^{p+q} (x_m^2 + y_m^2)}
               {\sum\limits_{m=1}^{m_\textrm{max}} m^p     (x_m^2 + y_m^2)} \, . \label{eqn:M}
\end{equation}
The quantity $M$ measures the spectral extent of the Fourier series that describes the flux surface shape via $x_m$ and $y_m$ for $q>0$.

A given set of real-space points $(x_i, y_i)$ for $i=1, 2, ..., n_\theta$ is to be described by a spectrally condensed Fourier series.
This implies that not only the corresponding poloidal coordinates $\theta_i$ must be inferred,
but also the Fourier coefficients $x_m, y_m$ themselves are to be varied to yield a description of the surface
that at the same time is spectrally condensed with respect to the Fourier coefficients and simultaneously fits the given real-space points well.
Changes in the definition of the poloidal angle thus not only imply a change in the Fourier coefficients
but also in the values of the $\theta_i$ that correspond to prescribed real-space locations.
However, the changes in the Fourier coefficients must comply with the requirement that only the poloidal coordinate changes.
This implies that variations $\delta u$ in the poloidal coordinate must translate into purely tangential variations of the coodinates:
\begin{align}
 \delta x =&\, \frac{\partial x}{\partial \theta} \delta u \label{eqn:delta_x} \\
 \delta y =&\, \frac{\partial y}{\partial \theta} \delta u \label{eqn:delta_y} \, .
\end{align}
The corresponding changes in the Fourier coefficients that describe the surface is then obtained by Fourier-transforming the variations $\delta x$ and $\delta y$:
\begin{align}
 \delta x_m =&\, \oint \cos(m \theta) \underbrace{\frac{\partial x}{\partial \theta} \delta u}_{= \delta x} \,\mathrm{d}\theta \label{eqn:delta_x_m} \\
 \delta y_m =&\, \oint \sin(m \theta) \underbrace{\frac{\partial y}{\partial \theta} \delta u}_{= \delta y} \,\mathrm{d}\theta \label{eqn:delta_y_m} \, .
\end{align}
The changes in each of the Fourier coefficients $\delta x_m$ and $\delta y_m$ lead to a corresponding change $\delta M$ in the spectral width $M$:
\begin{equation}
 \delta M = \sum\limits_{m=1}^{m_\textrm{max}} \left[
    \frac{\partial M}{\partial x_m} \delta x_m
  + \frac{\partial M}{\partial y_m} \delta y_m \right] \, . \label{eqn:delta_M}
\end{equation}
For a fixed $m$, the partial derivatives of $M$ are obtained via the quotient rule, slightly re-arranged from the usual form:
\begin{equation}
 \left(\frac{u}{v}\right)' = \frac{u'v - u v'}{v^2} = \frac{1}{v}\left(u' - \frac{u}{v} \, v'\right)
\end{equation}
for a ratio of two functions $u$ and $v$ corresponding to the numerator and the denominator in $M$ here.
Application of this to the partial derivatives of $M$ leads to:
\begin{align}
 \frac{\partial M}{\partial x_m} =&\, \frac{1}{\sum\limits_{\tilde{m}=1}^{m_\textrm{max}} S_p(\tilde{m})}
     \Biggl( 2 m^{p+q} x_m - \underbrace{\frac{\sum\limits_{\tilde{m}=1}^{m_\textrm{max}} \tilde{m}^q S_p(\tilde{m})}
                                              {\sum\limits_{\tilde{m}=1}^{m_\textrm{max}}             S_p(\tilde{m})}}_{= M} 2 m^p x_m \Biggr)
                                 = \frac{2 m^p (m^q - M) x_m}{\sum\limits_{\tilde{m}=1}^{m_\textrm{max}} S_p(\tilde{m})} \label{eqn:dMdxm} \\
 \frac{\partial M}{\partial y_m} =&\, \frac{1}{\sum\limits_{\tilde{m}=1}^{m_\textrm{max}} S_p(\tilde{m})}
     \Biggl( 2 m^{p+q} y_m - \underbrace{\frac{\sum\limits_{\tilde{m}=1}^{m_\textrm{max}} \tilde{m}^q S_p(\tilde{m})}
                                              {\sum\limits_{\tilde{m}=1}^{m_\textrm{max}}             S_p(\tilde{m})}}_{= M} 2 m^p y_m \Biggr)
                                 = \frac{2 m^p (m^q - M) y_m}{\sum\limits_{\tilde{m}=1}^{m_\textrm{max}} S_p(\tilde{m})} \label{eqn:dMdym} \, .
\end{align}
These expressions can be inserted into the variation of $M$ in \eqn{delta_M}.
For convenience, the common denominator in \eqn{dMdxm} and \eqn{dMdym} is pulled to the front:
\begin{align}
 \delta M \sum\limits_{m=1}^{m_\textrm{max}} S_p(m)
 =&\, 2 \sum\limits_{m=1}^{m_\textrm{max}} \left[
      \left( m^p(m^q - M) x_m \delta x_m \right)
    + \left( m^p(m^q - M) y_m \delta y_m \right)
 \right] \, .
\end{align}
A common factor $f(m)$ is identified:
\begin{equation}
 f(m) = m^p (m^q - M) \, .
\end{equation}
Now the variations in the Fourier coefficients from \eqn{delta_x_m} and \eqn{delta_y_m} can be inserted:
\begin{equation}
 \delta M \sum\limits_{m=1}^{m_\textrm{max}} S_p(m)
 = 2 \sum\limits_{m=1}^{m_\textrm{max}} \left[
     f(m) x_m \oint \cos(m \theta) \frac{\partial x}{\partial \theta} \delta u \,\mathrm{d}\theta \\
   + f(m) y_m \oint \sin(m \theta) \frac{\partial y}{\partial \theta} \delta u \,\mathrm{d}\theta
 \right] \, .
\end{equation}
The integrals and the sums can be swapped:
\begin{equation}
 \delta M \sum\limits_{m=1}^{m_\textrm{max}} S_p(m)
 = 2 \oint \left[
      \sum\limits_{m=1}^{m_\textrm{max}} f(m) x_m \cos(m \theta) \frac{\partial x}{\partial \theta}
    + \sum\limits_{m=1}^{m_\textrm{max}} f(m) y_m \sin(m \theta) \frac{\partial y}{\partial \theta}
 \right] \delta u \,\mathrm{d}\theta \, .
\end{equation}
A few short-hand notations are introduced:
\begin{align}
 X(\theta) =&\, \sum\limits_{m=1}^{m_\textrm{max}} f(m) x_m \cos(m \theta) \\
 Y(\theta) =&\, \sum\limits_{m=1}^{m_\textrm{max}} f(m) y_m \sin(m \theta) \\
 I(\theta) =&\, X(\theta) \frac{\partial x}{\partial \theta} + Y(\theta) \frac{\partial y}{\partial \theta}
\end{align}
to finally arrive at:
\begin{equation}
 \delta M \sum\limits_{m=1}^{m_\textrm{max}} S_p(m)
 = 2 \oint I(\theta) \delta u \,\mathrm{d}\theta \, . \label{eqn:var_M}
\end{equation}

\FloatBarrier
\section{Curve Fitting}
Recall that the aim of this is to obtain a spectrally condensed Fourier representation
of a set of points $(x_i, y_i)$ using Fourier series for $x(\theta)$ and $y(\theta)$ as given in \eqn{x_of_theta} and \eqn{y_of_theta}
and a set of poloidal coordinates $\theta_i$ for each of the $i$ points to be represented.
For any trial set of parameters $\boldsymbol{\xi} = \{x_m, y_m, \theta_i | m=0,1,...,m_\textrm{max}, i=1, 2, ..., n_\theta \}$,
the fit to the given points can be quantified by the sum of squared deviations, also called the \textit{curve energy}:
\begin{align}
 W_c =&\, \frac{1}{2} \sum\limits_{i=1}^{n_\theta}
          \left\{ \left[ x(\theta_i) - x_i \right]^2 + \left[ y(\theta_i) - y_i \right]^2 \right\} \nonumber \\
     =&\, \frac{1}{2} \sum\limits_{i=1}^{n_\theta} \left\{
          \left[ \left(\sum\limits_{m=0}^{m_\textrm{max}} x_m \cos(m \theta_i) \right) - x_i \right]^2
        + \left[ \left(\sum\limits_{m=0}^{m_\textrm{max}} y_m \sin(m \theta_i) \right) - y_i \right]^2 \right\} \, . \label{eqn:Wc}
\end{align}
Corresponding forces on the parameters can be derived when $W_c$ is thought of as an energy:
\begin{equation}
 \mathbf{F} = - \nabla W_c = - \left\{
   \frac{\partial W_c}{\partial x_m},
   \frac{\partial W_c}{\partial y_m},
   \frac{\partial W_c}{\partial \theta_i} \right\} \, .
\end{equation}
The force components on the Fourier coefficients are:
\begin{align}
 F^x_m =&\, - \frac{\partial W_c}{\partial x_m} = - \sum\limits_{i=1}^{n_\theta} \left( x(\theta_i) - x_i \right)
   \underbrace{\cos(m \theta_i)}_{=\partial x(\theta_i) / \partial x_m} \\
 F^y_m =&\, - \frac{\partial W_c}{\partial y_m} = - \sum\limits_{i=1}^{n_\theta} \left( y(\theta_i) - y_i \right)
   \underbrace{\sin(m \theta_i)}_{=\partial y(\theta_i) / \partial y_m}
\end{align}
and the force components on the $\theta_i$ are:
\begin{equation}
   F^\theta_i
 = - \frac{\partial W_c}{\partial \theta_i}
 = - \left[  \left( x(\theta_i) - x_i \right) \frac{\partial x}{\partial \theta}(\theta_i)
           + \left( y(\theta_i) - y_i \right) \frac{\partial y}{\partial \theta}(\theta_i) \right] \, .
\end{equation}
The Jacobian matrix of this system is a $(3 \times 3)$-block matrix $\partial \mathbf{F} / \partial \boldsymbol{\xi}$ with the following layout:
\begin{align}
  \frac{\partial \mathbf{F}}{\partial \boldsymbol{\xi}}
  = \begin{pmatrix}
      \dfrac{\partial F^x_m     }{\partial x_{m'}} & \dfrac{\partial F^x_m     }{\partial y_{m'}} & \dfrac{\partial F^x_m     }{\partial \theta_i} \\[2ex]
      \dfrac{\partial F^y_m     }{\partial x_{m'}} & \dfrac{\partial F^y_m     }{\partial y_{m'}} & \dfrac{\partial F^y_m     }{\partial \theta_i} \\[2ex]
      \dfrac{\partial F^\theta_i}{\partial x_m   } & \dfrac{\partial F^\theta_i}{\partial y_m   } & \dfrac{\partial F^\theta_i}{\partial \theta_i}
    \end{pmatrix} \, .
\end{align}
It consists of the following elements:
\begin{align}
 \frac{\partial F^x_m}{\partial x_{m'}} =&\, - \sum\limits_{i=1}^{n_\theta} \cos(m' \theta_i) \cos(m \theta_i) \label{eqn:dFxmdxm}\\
 \frac{\partial F^x_m}{\partial y_{m'}} =&\, 0 \\
 \frac{\partial F^x_m}{\partial \theta_i} =&\, - \left[
     \frac{\partial x(\theta_i)}{\partial \theta_i} \cos(m \theta_i)
   - \left( x(\theta_i) - x_i \right) m \sin(m \theta_i) \right] \\
 \frac{\partial F^y_m}{\partial x_{m'}} =&\, 0 \\
 \frac{\partial F^y_m}{\partial y_{m'}} =&\, - \sum\limits_{i=1}^{n_\theta} \sin(m' \theta_i) \sin(m \theta_i) \label{eqn:dFymdym} \\
 \frac{\partial F^y_m}{\partial \theta_i} =&\, - \left[
     \frac{\partial y(\theta_i)}{\partial \theta_i} \sin(m \theta_i)
   + \left( y(\theta_i) - y_i \right) m \cos(m \theta_i) \right] \\
 \frac{\partial F^\theta_i}{\partial x_m} =&\, \frac{\partial F^x_m}{\partial \theta_i} \\
 \frac{\partial F^\theta_i}{\partial y_m} =&\, \frac{\partial F^y_m}{\partial \theta_i} \\
 \frac{\partial F^\theta_i}{\partial \theta_i} =&\, - \left[
     \left( \frac{\partial x(\theta_i)}{\partial \theta_i} \right)^2 + \left( x(\theta_i) - x_i \right) \frac{\partial^2 x(\theta_i)}{\partial \theta_i^2}
   + \left( \frac{\partial y(\theta_i)}{\partial \theta_i} \right)^2 + \left( y(\theta_i) - y_i \right) \frac{\partial^2 y(\theta_i)}{\partial \theta_i^2}
   \right] \label{eqn:dFthetadTheta}
\end{align}
where $m, m' \in [0, m_\textrm{max}]$ and $i, i' \in [1, n_\theta]$.
A sharp look reveals that the Jacobian of the curve fit forces is symmetric and relatively sparse;
note that for $i' \neq i$, $\partial F^\theta_i/\partial \theta_{i'} = 0$.

\begin{figure}[h]
  \centering
  \begin{minipage}[b]{0.49\textwidth}
    \includegraphics[width=\textwidth]{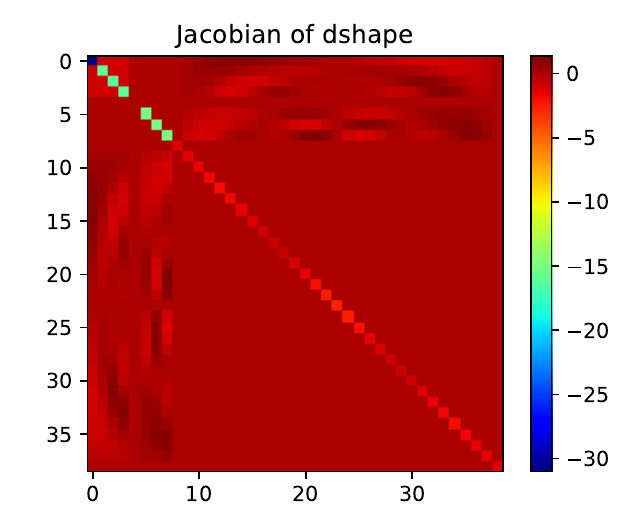}
  \end{minipage}
  \hfill
  \begin{minipage}[b]{0.49\textwidth}
    \includegraphics[width=\textwidth]{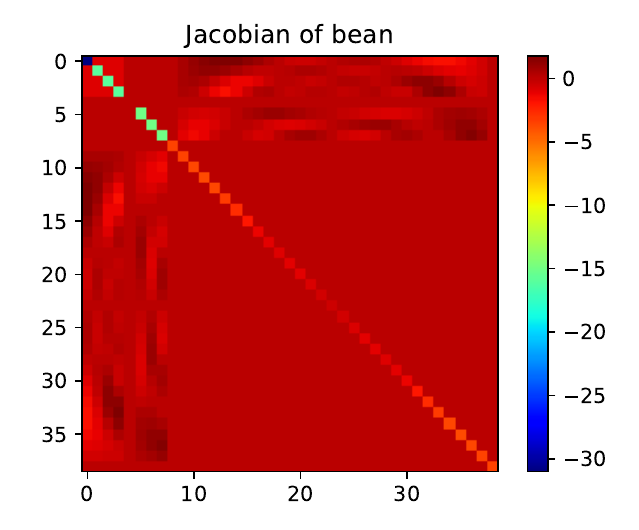}
  \end{minipage}
  \caption{Curve fit Jacobian matrix of a D-shaped and a bean-shaped boundary.}
  \label{fig:boundary_jacobians}
\end{figure}

Two exemplary cases are investigated in the following.
They are the D-shaped and the bean-shaped examples from the DESCUR article~\cite{hirshman_meier_1985}.
The Jacobian matrix for each of these cases is depicted in Fig.~\ref{fig:boundary_jacobians}.
It is evident that these matrices are approximately diagonal.
This is true for the upper left $(2 \times 2)$-block matrix (the forces on the Fourier coefficients due to changes in the Fourier coefficients)
since the orthogonality of the Fourier basis (see \eqn{ortho_fourier_basis}) ensures that all terms in \eqn{dFxmdxm} for $m \neq m'$
and all terms in \eqn{dFymdym} for $m \neq m'$ are approximately zero.
Also, as mentioned above, the lower right block is diagonal by definition, as seen in \eqn{dFthetadTheta}.

According to Ref.~\cite{gill_murray_wright}, the maximum eigenvalue of a real, symmetric matrix $\mathbf{A}$ can be computed as:
\begin{equation}
 \lambda_\textrm{max} = \max_{x \neq 0} \frac{x^T \mathbf{A} x}{x^T x} = \max_{||x|| = 1} x^T \mathbf{A} x \, .
\end{equation}
Any real symmetric $(n \times n)$ matrix $\mathbf{A}$ can be diagonalized into a form
\begin{equation}
  \mathbf{A} = \mathbf{U}^T \mathbf{\Lambda} \mathbf{U}
\end{equation}
where $\mathbf{U}$ is an orthonormal matrix and $\mathbf{\Lambda} = \mathrm{diag}\left( \lambda_1, \lambda_2, ..., \lambda_n \right)$ is the diagonal matrix
of eigenvalues $\lambda_i$ of $\mathbf{A}$.

\begin{figure}[h]
  \centering
  \begin{minipage}[b]{0.49\textwidth}
    \includegraphics[width=0.74\textwidth]{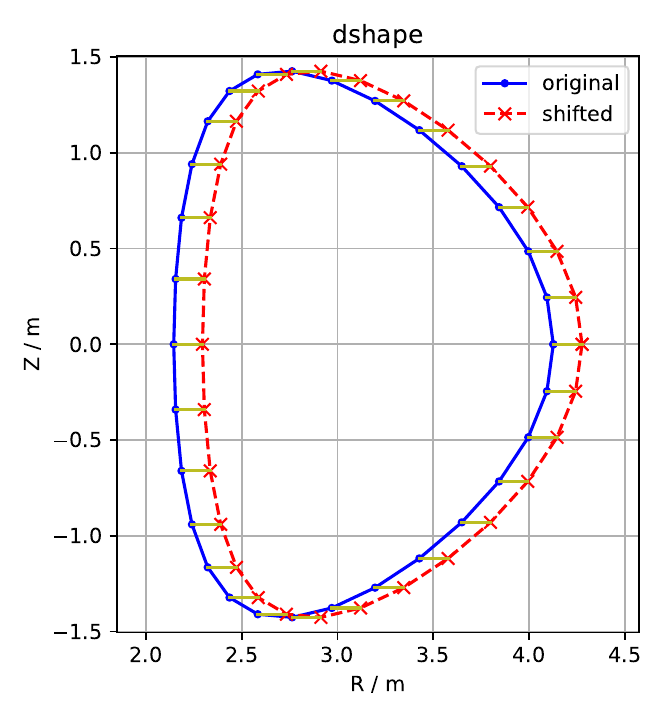}
  \end{minipage}
  \hfill
  \begin{minipage}[b]{0.49\textwidth}
    \includegraphics[width=0.8\textwidth]{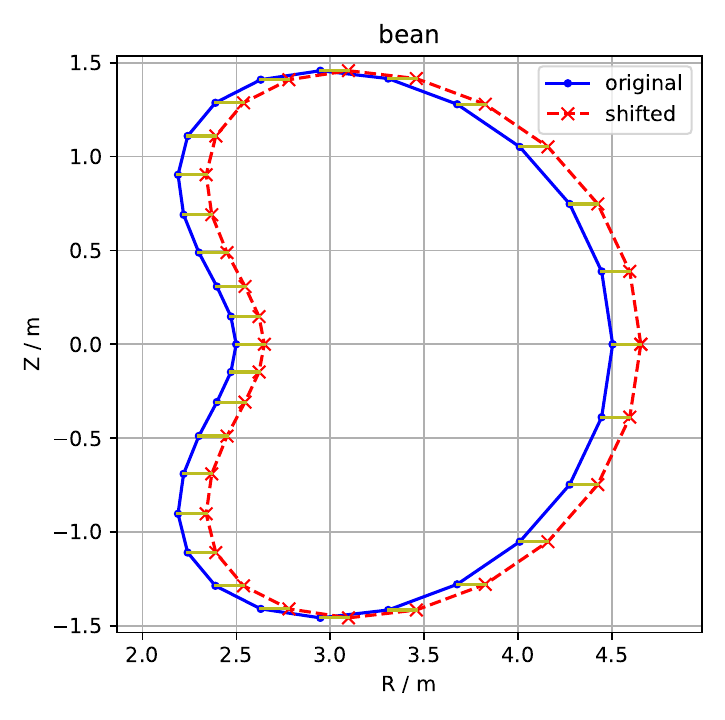}
  \end{minipage}
  \caption{A D-shaped and a bean-shaped boundary in the original shape and for a radial shift of 5~\% of their major radius.
           The yellow lines visualize fictional rubber bands between the original points and the shifted model points.}
  \label{fig:rubber_bands_jac}
\end{figure}

The Jacobian matrix of the curve-fitting problem is approximately diagonal.
Therefore, the maximum eigenvalue of the Jacobian matrix approximately corresponds to the maximum diagonal element in this matrix.
The maximum element of the Jacobian matrix is located at the $(0,0)$ index and corresponds to $\partial F^x_0 / \partial x_0$, leading to:
\begin{equation}
  \lambda_\textrm{max} = \left| \frac{\partial F^x_0}{\partial x_0} \right|
  = \sum\limits_{i=1}^{n_\theta} \underbrace{\cos(0 \theta_i)}_{=1} \underbrace{\cos(0 \theta_i)}_{=1} = n_\theta \, . \label{eqn:max_eigenvalue_jac}
\end{equation}
This can be understood also intuitively.
The eigenvalues of the Jacobian matrix correspond to the changes in the ``net'' force of the curve fit problem
when any of the free parameters is changed.
Any change in $x_0$ corresponds to a radial shift of \underline{all} points simultaneously.
The curve-fit forces can be visualized by fictional rubber bands between the points to fit $(x_i, y_i)$ and the model points $(x(\theta_i), y(\theta_i))$.
For the two demo cases mentioned above, this is illustrated in Fig.~\ref{fig:rubber_bands_jac}.
None of the other free parameters has such a global effect on the forces between the original points and the model points
and thus, for each rubber band (with assumed unit spring constant) a unit restoring force acts on $x_0$, for a total of $n_\theta$
unit force contributions to $F^x_0$ and (since this part of the model is linear) also to $\partial F^x_0 / \partial x_0$.
An eigendecomposition of the two Jacobian matrices was computed to verify the simplifying assumptions above.
The maximum eigenvalue of the Jacobian in the D-shaped case was $\lambda_\textrm{max, D} = 31.72$
and the maximum eigenvalue of the Jacobian in the bean-shaped case was $\lambda_\textrm{max, bean} = 32.31$.
These two numerical values have been obtained for $n_\theta = 31$ individual points around the circumference of the shapes
and are deemed sufficiently close to the approximated value in \eqn{max_eigenvalue_jac}.

\FloatBarrier
\section{Spectral Condensation Forces}
The energy functional is augmented to include an additional term related to the spectral width,
leading to the Lagrangian of the problem:
\begin{equation}
  W = W_c + \epsilon M
\end{equation}
where $\epsilon$ is a Lagrange multiplier yet to be determined.
The variation in energy due to a tangential displacement~$\delta u$ is:
\begin{equation}
  \delta W = \delta W_c + \epsilon \delta M
\end{equation}
and it follows with $\delta M$ from \eqn{var_M}:
\begin{equation}
  \delta W = \delta W_c + \frac{2 \epsilon}{\sum\limits_{m=1}^{m_\textrm{max}} S_p(m)} \oint I(\theta) \delta u \,\mathrm{d}\theta \, .
\end{equation}
The Lagrange multiplier~$\epsilon$ is rescaled according to:
\begin{equation}
  \epsilon_m = \begin{cases}
                 \epsilon / \sum\limits_{\tilde{m}=1}^{m_\textrm{max}} S_p(\tilde{m})  & m > 0 \\
                 0                                                                     & m = 0
               \end{cases}
\end{equation}
which leads to:
\begin{equation}
  \delta W = \delta W_c + 2 \epsilon_m \oint I(\theta) \delta u \,\mathrm{d}\theta \, .
\end{equation}

\FloatBarrier
\section{A Constrained Optimization Problem}
The task of finding a spectrally condensed Fourier series to describe a plane curve
represented by (randomly sampled) points along the curve can be treated as a constrained optimization problem:
\begin{align}
  \min M(\hat{x}_m, \hat{y}_m) \\
  \textrm{s.t. } W_\mathrm{c} (\hat{x}_m, \hat{y}_m, \theta_i) = 0
\end{align}
The non-linear objective function to be minimized is chosen to be the spectral width from \eqn{M}.
Note that the optimization routine is fed with the value of $M-1$, since $M$ cannot get lower than 1
and it is deemed useful to incorporate this information,
since the path taken by the optimizer in parameter space might depend on the actual value of the objective function.
The non-linear equality constraint is that the curve energy $W_\mathrm{c}$ from (\eqn{Wc}) vanishes.

The problem of finding the Fourier coefficients $\hat{x}_m$ and $\hat{y}_m$ as well as the discrete
tangential parameter values $\theta_i$ has large amounts of local minima, since for tiny changes in any of the Fourier coefficients
all $\theta_i$ would have to change in an ordered manner simultaneously.
This can (to an extent) be compensated by preconditioning, but the basic problem remains.

Here, a Fourier parameterization is chosen for the modulation $\lambda$ of the tangential parameter,
which is added to a linearly-increasing fixed background coordinate $\theta$:
\begin{align}
 \theta_i =& 2 \pi \frac{i}{N-1} \\
 \Omega_i =& \theta_i + \lambda(\theta_i)
\end{align}
for $i = 0, 1, ..., (N-1)$ with
\begin{equation}
 \lambda(\theta) = \sum_{m=1}^{m^\lambda} \hat{\lambda}_m \sin(m \theta) \, .
\end{equation}
The number of Fourier coefficients for $\lambda$, $m^\lambda$, is chosen as $m^\lambda = \lfloor \frac{N}{2} \rfloor$,
so that the full amount of information contained in the $\theta_i$ can be representated by the $\hat{\lambda}_m$.
Here, stellarator (equivalent to up-down) symmetry is assumed.

The Fourier series of $x$ and $y$ are then evaluated at the parameter $\Omega$:
\begin{align}
 x(\theta_i) =& \sum_{m=0}^{m^\mathrm{XY}} \hat{x}_m \cos(m \Omega_i) = \sum_{m=0}^{m^\mathrm{XY}} \hat{x}_m \cos(m \left[\theta_i + \lambda(\theta_i) \right]) \\
 y(\theta_i) =& \sum_{m=1}^{m^\mathrm{XY}} \hat{y}_m \sin(m \Omega_i) = \sum_{m=1}^{m^\mathrm{XY}} \hat{y}_m \sin(m \left[\theta_i + \lambda(\theta_i) \right]) \, .
\end{align}
The number of Fourier coefficients for $x$ and $y$, $m^\mathrm{XY}$, is generally chosen much smaller than $m^\lambda$.
Typical values are defined by the Fourier resolution used in a subsequently used MHD equilibrium code,
where the Fourier coefficients of the curve geometry are used to specify the shape of flux surfaces.
For VMEC, this would correspond to $m^\mathrm{XY} \lesssim 12$.

Analytical first- and second-order derivatives are now derived for both the objective function and the constraint.
They will be used by the non-linear optimization algorithm later on.

It turns out nicely that the Hessian matrices are symmetric.
This is due to the symmetry of both the constraint and the objective wrt. swapping $x$ and $y$.

The free parameters in this model are then $\hat{x}_m$, $\hat{y}_m$ and $\hat{\lambda}_m$.
The objective function $M$ only depends on $\hat{x}_m$ and $\hat{y}_m$ and its gradient is thus given fully by \eqn{dMdxm} and \eqn{dMdym}.
Here and below, $m>0$ and $m'>0$ are assumed.
The constraint function $W_\mathrm{c}$ depends on all free parameters and its gradient is given below:
\begin{align}
  \frac{\partial W_c}{\partial x(\Omega_i)}          =&\, x(\Omega_i) - x_i \\
  \frac{\partial W_c}{\partial y(\Omega_i)}          =&\, y(\Omega_i) - y_i \, ,
\end{align}
going on:
\begin{align}
  \frac{\partial x(\Omega_i)}{\partial \hat{x}_m}    =&\, \cos( m \Omega_i ) \\
  \frac{\partial x(\Omega_i)}{\partial \hat{y}_m}    =&\, 0                  \\
  \frac{\partial y(\Omega_i)}{\partial \hat{x}_m}    =&\, 0                  \\
  \frac{\partial y(\Omega_i)}{\partial \hat{y}_m}    =&\, \sin( m \Omega_i )
\end{align}
and furthermore:
\begin{align}
  \frac{\partial x(\Omega_i)}{\partial \Omega_i}     =&\,          -    \sum_{m=1}^{m^\mathrm{XY}} \hat{x}_m m \sin(m \Omega_i) \\
  \frac{\partial y(\Omega_i)}{\partial \Omega_i}     =&\, \phantom{-}\, \sum_{m=1}^{m^\mathrm{XY}} \hat{y}_m m \cos(m \Omega_i) \\
  \frac{\partial^2 x(\Omega_i)}{\partial \Omega_i^2} =&\, - \sum_{m=1}^{m^\mathrm{XY}} \hat{x}_m m^2 \cos(m \Omega_i) \\
  \frac{\partial^2 y(\Omega_i)}{\partial \Omega_i^2} =&\, - \sum_{m=1}^{m^\mathrm{XY}} \hat{y}_m m^2 \sin(m \Omega_i) \\
  \frac{\partial \Omega_i}{\partial \hat{\lambda}_m} =&\, \sin( m \theta_i ) \, .
\end{align}
Some second-order derivatives are useful later on:
\begin{align}
  \frac{\partial^2 W_c}{\partial \hat{x}_m \partial x(\Omega_i)}
  =&\, \frac{\partial x(\Omega_i)}{\partial \hat{x}_m}
  =    \cos( m \Omega_i ) \\
  \frac{\partial^2 W_c}{\partial \hat{y}_m \partial y(\Omega_i)}
  =&\, \frac{\partial y(\Omega_i)}{\partial \hat{y}_m}
  =    \sin( m \Omega_i )
\end{align}
It follows:
\begin{align}
  \frac{\partial x(\Omega_i)}{\partial \hat{\lambda}_m} =&\,          -    \left( \sum_{\tilde{m}=1}^{m^\mathrm{XY}} \hat{x}_{\tilde{m}} \tilde{m} \sin(\tilde{m} \Omega_i) \right) \sin( m \theta_i ) \\
  \frac{\partial y(\Omega_i)}{\partial \hat{\lambda}_m} =&\, \phantom{-}\, \left( \sum_{\tilde{m}=1}^{m^\mathrm{XY}} \hat{y}_{\tilde{m}} \tilde{m} \cos(\tilde{m} \Omega_i) \right) \sin( m \theta_i )
\end{align}
as well as:
\begin{align}
  \frac{\partial^2 x(\Omega_i)}{\partial \hat{x}_{m'} \partial \hat{\lambda}_m}
  =&\, \frac{\partial}{\partial \hat{x}_{m'}} \left[ - \left( \sum_{\tilde{m}=1}^{m^\mathrm{XY}} \hat{x}_{\tilde{m}} \tilde{m} \sin(\tilde{m} \Omega_i) \right) \sin( m \theta_i )  \right]
  = - m' \sin(m' \Omega_i) \sin( m \theta_i ) \\
  \frac{\partial^2 x(\Omega_i)}{\partial \hat{y}_{m'} \partial \hat{\lambda}_m}
  =&\, \frac{\partial}{\partial \hat{y}_{m'}} \left[ - \left( \sum_{\tilde{m}=1}^{m^\mathrm{XY}} \hat{x}_{\tilde{m}} \tilde{m} \sin(\tilde{m} \Omega_i) \right) \sin( m \theta_i )  \right]
  = 0 \\
  \frac{\partial^2 y(\Omega_i)}{\partial \hat{x}_{m'} \partial \hat{\lambda}_m}
  =&\, \frac{\partial}{\partial \hat{x}_{m'}} \left[\phantom{-}\, \left( \sum_{\tilde{m}=1}^{m^\mathrm{XY}} \hat{y}_{\tilde{m}} \tilde{m} \cos(\tilde{m} \Omega_i) \right) \sin( m \theta_i )  \right]
  = 0 \\
  \frac{\partial^2 y(\Omega_i)}{\partial \hat{y}_{m'} \partial \hat{\lambda}_m}
  =&\, \frac{\partial}{\partial \hat{y}_{m'}} \left[\phantom{-}\, \left( \sum_{\tilde{m}=1}^{m^\mathrm{XY}} \hat{y}_{\tilde{m}} \tilde{m} \cos(\tilde{m} \Omega_i) \right) \sin( m \theta_i )  \right]
  = \phantom{-}  m' \cos(m' \Omega_i) \sin( m \theta_i ) \\
  \frac{\partial^2 x(\Omega_i)}{\partial \hat{\lambda}_{m'} \partial \hat{\lambda}_m}
  =&\, \frac{\partial^2 x(\Omega_i)}{\partial \Omega_i^2} \frac{\partial \Omega_i}{\partial \hat{\lambda}_{m'}} \frac{\partial \Omega_i}{\partial \hat{\lambda}_m}
  = - \left( \sum_{\tilde{m}=1}^{m^\mathrm{XY}} \hat{x}_{\tilde{m}} {\tilde{m}}^2 \cos(\tilde{m} \Omega_i) \right) \sin( m' \theta_i ) \sin( m \theta_i ) \\
  \frac{\partial^2 y(\Omega_i)}{\partial \hat{\lambda}_{m'} \partial \hat{\lambda}_m}
  =&\, \frac{\partial^2 y(\Omega_i)}{\partial \Omega_i^2} \frac{\partial \Omega_i}{\partial \hat{\lambda}_{m'}} \frac{\partial \Omega_i}{\partial \hat{\lambda}_m}
  = - \left( \sum_{\tilde{m}=1}^{m^\mathrm{XY}} \hat{y}_{\tilde{m}} {\tilde{m}}^2 \sin(\tilde{m} \Omega_i) \right) \sin( m' \theta_i ) \sin( m \theta_i )
\end{align}
From above parts, the full gradient vector of the constraint can be assembled:
\begin{align}
 \frac{\partial W_c}{\partial \hat{x}_m}
  =&\, \sum_{i=0}^{N-1} \frac{\partial W_c}{\partial x(\Omega_i)} \frac{\partial x(\Omega_i)}{\partial \hat{x}_m}
  =    \sum_{i=0}^{N-1} \left( x(\Omega_i) - x_i \right) \cos( m \Omega_i ) \\
 \frac{\partial W_c}{\partial \hat{y}_m}
  =&\, \sum_{i=0}^{N-1} \frac{\partial W_c}{\partial y(\Omega_i)} \frac{\partial y(\Omega_i)}{\partial \hat{y}_m}
  =    \sum_{i=0}^{N-1} \left( y(\Omega_i) - y_i \right) \sin( m \Omega_i ) \\
\frac{\partial W_c}{\partial \hat{\lambda}_m}
 =&\, \sum_{i=0}^{N-1} \left(  \frac{\partial W_c}{\partial x(\Omega_i)} \frac{\partial x(\Omega_i)}{\partial \Omega_i}
                             + \frac{\partial W_c}{\partial y(\Omega_i)} \frac{\partial y(\Omega_i)}{\partial \Omega_i}
                       \right) \frac{\partial \Omega_i}{\partial \hat{\lambda}_m} \nonumber \\
~=&\,         \sum_{i=0}^{N-1} \Biggl[  - \left( x(\Omega_i) - x_i \right) \left( \sum_{\tilde m =1}^{m^\mathrm{XY}} \hat{x}_{\tilde{m}} \tilde{m} \sin(\tilde{m} \Omega_i) \right) \nonumber \\
~&\, \phantom{\sum_{i=0}^{N-1} \Biggl[} + \left( y(\Omega_i) - y_i \right) \left( \sum_{\tilde m =1}^{m^\mathrm{XY}} \hat{y}_{\tilde{m}} \tilde{m} \cos(\tilde{m} \Omega_i) \right) \Biggr] \sin( m \theta_i )
\end{align}
The problem at hand is fully analytic and therefore, analytical second-order derivatives of the objective as well as the constraint are derived next.
First the second-order derivatives of the objective function $M$ are considered:
\begin{align}
  \frac{\partial^2 M}{\partial \hat{x}_{m'} \partial \hat{x}_m}
  =&\, \frac{\partial}{\partial \hat{x}_{m'}} \left( \frac{\partial M}{\partial \hat{x}_m} \right)
  =    \frac{\partial}{\partial \hat{x}_{m'}} \left( \frac{2 m^p (m^q - M) \hat{x}_m}{\sum\limits_{\tilde{m}=1}^{m^\textrm{XY}} \tilde{m}^p (\hat{x}_{\tilde{m}}^2 + \hat{y}_{\tilde{m}}^2)} \right) \nonumber \\
  =&\, \frac{1}{\sum\limits_{\tilde{m}=1}^{m^\textrm{XY}} \tilde{m}^p (\hat{x}_{\tilde{m}}^2 + \hat{y}_{\tilde{m}}^2)}
       \left[                                        \frac{\partial}{\partial \hat{x}_{m'}} \Bigl( 2 m^p (m^q - M) \hat{x}_m                                                                             \Bigr)
             - \frac{\partial M}{\partial \hat{x}_m} \frac{\partial}{\partial \hat{x}_{m'}} \left( \sum\limits_{\tilde{m}=1}^{m^\textrm{XY}} \tilde{m}^p (\hat{x}_{\tilde{m}}^2 + \hat{y}_{\tilde{m}}^2) \right)
       \right] \nonumber \\
  =&\, - \frac{2}{\sum\limits_{\tilde{m}=1}^{m^\textrm{XY}} \tilde{m}^p (\hat{x}_{\tilde{m}}^2 + \hat{y}_{\tilde{m}}^2)}
         \left[ m^p \hat{x}_{m} \frac{\partial M}{\partial \hat{x}_{m'}} + {m'}^p \hat{x}_{m'} \frac{\partial M}{\partial \hat{x}_m} - m^p (m^q - M) \delta_{m m'} \right]
\end{align}
Furthermore:
\begin{align}
  \frac{\partial^2 M}{\partial \hat{y}_{m'} \partial \hat{x}_m}
  =&\, \frac{\partial}{\partial \hat{y}_{m'}} \left( \frac{\partial M}{\partial \hat{x}_m} \right)
  =    \frac{\partial}{\partial \hat{y}_{m'}} \left( \frac{2 m^p (m^q - M) \hat{x}_m}{\sum\limits_{\tilde{m}=1}^{m^\textrm{XY}} \tilde{m}^p (\hat{x}_{\tilde{m}}^2 + \hat{y}_{\tilde{m}}^2)} \right) \nonumber \\
  =&\, \frac{1}{\sum\limits_{\tilde{m}=1}^{m^\textrm{XY}} \tilde{m}^p (\hat{x}_{\tilde{m}}^2 + \hat{y}_{\tilde{m}}^2)}
       \left[- 2 m^p \hat{x}_m \frac{\partial M}{\partial \hat{y}_{m'}}
             - \frac{\partial M}{\partial \hat{x}_m} \frac{\partial}{\partial \hat{y}_{m'}} \left( \sum\limits_{\tilde{m}=1}^{m^\textrm{XY}} \tilde{m}^p (\hat{x}_{\tilde{m}}^2 + \hat{y}_{\tilde{m}}^2) \right)
       \right] \nonumber \\
  =&\, - \frac{2}{\sum\limits_{\tilde{m}=1}^{m^\textrm{XY}} \tilde{m}^p (\hat{x}_{\tilde{m}}^2 + \hat{y}_{\tilde{m}}^2)}
         \left[ m^p \hat{x}_{m} \frac{\partial M}{\partial \hat{y}_{m'}} + {m'}^p \hat{y}_{m'} \frac{\partial M}{\partial \hat{x}_m} \right]
\end{align}
Finally:
\begin{align}
  \frac{\partial^2 M}{\partial \hat{y}_{m'} \partial \hat{y}_m}
  =&\, \frac{\partial}{\partial \hat{y}_{m'}} \left( \frac{\partial M}{\partial \hat{y}_m} \right)
  =    \frac{\partial}{\partial \hat{y}_{m'}} \left( \frac{2 m^p (m^q - M) \hat{y}_m}{\sum\limits_{\tilde{m}=1}^{m^\textrm{XY}} \tilde{m}^p (\hat{x}_{\tilde{m}}^2 + \hat{y}_{\tilde{m}}^2)} \right) \nonumber \\
  =&\, \frac{1}{\sum\limits_{\tilde{m}=1}^{m^\textrm{XY}} \tilde{m}^p (\hat{x}_{\tilde{m}}^2 + \hat{y}_{\tilde{m}}^2)}
       \left[                                        \frac{\partial}{\partial \hat{y}_{m'}} \Bigl( 2 m^p (m^q - M) \hat{y}_m                                                                             \Bigr)
             - \frac{\partial M}{\partial \hat{y}_m} \frac{\partial}{\partial \hat{y}_{m'}} \left( \sum\limits_{\tilde{m}=1}^{m^\textrm{XY}} \tilde{m}^p (\hat{x}_{\tilde{m}}^2 + \hat{y}_{\tilde{m}}^2) \right)
       \right] \nonumber \\
  =&\, - \frac{2}{\sum\limits_{\tilde{m}=1}^{m^\textrm{XY}} \tilde{m}^p (\hat{x}_{\tilde{m}}^2 + \hat{y}_{\tilde{m}}^2)}
         \left[ m^p \hat{y}_{m} \frac{\partial M}{\partial \hat{y}_{m'}} + {m'}^p \hat{y}_{m'} \frac{\partial M}{\partial \hat{y}_m} - m^p (m^q - M) \delta_{m m'} \right]
\end{align}
The second-order derivatives of the constraint are considered next.
The second-order derivatives with respect to $\hat{x}_m$ and $\hat{y}_m$ are similar to the corresponding components of the curve-fit Jacobian:
\begin{align}
  \frac{\partial^2 W_\mathrm{c}}{\partial \hat{x}_{m'} \partial \hat{x}_m}
  =&\, \frac{\partial}{\partial \hat{x}_{m'}} \left( \frac{\partial W_\mathrm{c}}{\partial \hat{x}_m} \right)
  = \sum_{i=0}^{N-1} \cos( m' \Omega_i ) \cos( m \Omega_i ) \\
  \frac{\partial^2 W_\mathrm{c}}{\partial \hat{y}_{m'} \partial \hat{x}_m}
  =&\, 0 \\
  \frac{\partial^2 W_\mathrm{c}}{\partial \hat{x}_{m'} \partial \hat{y}_m}
  =&\, 0 \\
  \frac{\partial^2 W_\mathrm{c}}{\partial \hat{y}_{m'} \partial \hat{y}_m}
  =&\, \frac{\partial}{\partial \hat{y}_{m'}} \left( \frac{\partial W_\mathrm{c}}{\partial \hat{y}_m} \right)
  = \sum_{i=0}^{N-1} \sin( m' \Omega_i ) \sin( m \Omega_i )
\end{align}
Furthermore:
\begin{align}
  \frac{\partial^2 W_\mathrm{c}}{\partial \hat{x}_{m'} \partial \hat{\lambda}_m}
  =&\, \sum_{i=0}^{N-1} \frac{\partial}{\partial \hat{x}_{m'}}
                        \left(  \frac{\partial W_c}{\partial x(\Omega_i)} \frac{\partial x(\Omega_i)}{\partial \hat{\lambda}_m}
                              + \frac{\partial W_c}{\partial y(\Omega_i)} \frac{\partial y(\Omega_i)}{\partial \hat{\lambda}_m}
                       \right) \nonumber \\
  =&\, \sum_{i=0}^{N-1} \Biggl(\phantom{+}\,  \frac{\partial^2 W_c}{\partial \hat{x}_{m'} \partial x(\Omega_i)} \frac{\partial   x(\Omega_i)}{                      \partial \hat{\lambda}_m}
                               + \frac{\partial   W_c}{                      \partial x(\Omega_i)} \frac{\partial^2 x(\Omega_i)}{\partial \hat{x}_{m'} \partial \hat{\lambda}_m} \nonumber \\
~ &\, \phantom{\sum_{i=0}^{N-1} \Biggl(}\,  + \underbrace{\frac{\partial^2 W_c}{\partial \hat{x}_{m'} \partial y(\Omega_i)}}_{=0} \frac{\partial   y(\Omega_i)}{                      \partial \hat{\lambda}_m}
                               + \frac{\partial   W_c}{                      \partial y(\Omega_i)} \underbrace{\frac{\partial^2 y(\Omega_i)}{\partial \hat{x}_{m'} \partial \hat{\lambda}_m}}_{=0}
                        \Biggr) \nonumber \\
  =&\, \sum_{i=0}^{N-1} \left(  \cos(m' \Omega_i) \frac{\partial x(\Omega_i)}{\partial \Omega_i}
                              - (x(\Omega_i) - x_i) m' \sin(m' \Omega_i) \right) \sin( m \theta_i ) \label{eqn:d2WcdXmdLm}
\end{align}
as well as:
\begin{align}
  \frac{\partial^2 W_\mathrm{c}}{\partial \hat{y}_{m'} \partial \hat{\lambda}_m}
  =&\, \sum_{i=0}^{N-1} \frac{\partial}{\partial \hat{y}_{m'}}
                        \left(  \frac{\partial W_c}{\partial x(\Omega_i)} \frac{\partial x(\Omega_i)}{\partial \hat{\lambda}_m}
                              + \frac{\partial W_c}{\partial y(\Omega_i)} \frac{\partial y(\Omega_i)}{\partial \hat{\lambda}_m}
                       \right) \nonumber \\
  =&\, \sum_{i=0}^{N-1} \Biggl(\phantom{+}\,  \underbrace{\frac{\partial^2 W_c}{\partial \hat{y}_{m'} \partial x(\Omega_i)}}_{=0} \frac{\partial   x(\Omega_i)}{                      \partial \hat{\lambda}_m}
                               + \frac{\partial   W_c}{                      \partial x(\Omega_i)} \underbrace{\frac{\partial^2 x(\Omega_i)}{\partial \hat{y}_{m'} \partial \hat{\lambda}_m}}_{=0} \nonumber \\
~ &\, \phantom{\sum_{i=0}^{N-1} \Biggl(}\,  + \frac{\partial^2 W_c}{\partial \hat{y}_{m'} \partial y(\Omega_i)} \frac{\partial   y(\Omega_i)}{                      \partial \hat{\lambda}_m}
                               + \frac{\partial   W_c}{                      \partial y(\Omega_i)} \frac{\partial^2 y(\Omega_i)}{\partial \hat{y}_{m'} \partial \hat{\lambda}_m}
                        \Biggr) \nonumber \\
  =&\, \sum_{i=0}^{N-1} \left(  \sin(m' \Omega_i) \frac{\partial y(\Omega_i)}{\partial \Omega_i}
                                 + (y(\Omega_i) - y_i) m' \cos(m' \Omega_i) \right) \sin( m \theta_i ) \label{eqn:d2WcdYmdLm}
\end{align}
The second-order derivatives with respect to $\hat{\lambda}_m$ follow:
\begin{align}
  \frac{\partial^2 W_\mathrm{c}}{\partial \hat{\lambda}_{m'} \partial \hat{x}_m}
  =&\, \sum_{i=0}^{N-1} \frac{\partial}{\partial \hat{\lambda}_{m'}} \Bigl[ \left( x(\Omega_i) -  x_i \right) \cos( m \Omega_i ) \Bigr] \nonumber \\
  =&\, \sum_{i=0}^{N-1} \left[  \frac{\partial x(\Omega_i)}{\partial \hat{\lambda}_{m'}} \cos( m \Omega_i )
                               - \left( x(\Omega_i) - x_i \right) m \sin(m \Omega_i) \sin(m' \theta_i) \right] \nonumber \\
  =&\, \sum_{i=0}^{N-1} \left[   \cos( m \Omega_i ) \frac{\partial x(\Omega_i)}{\partial \Omega_i}
                               - \left( x(\Omega_i) - x_i \right) m \sin(m \Omega_i) \right] \sin(m' \theta_i) \label{eqn:d2WcdLmdXm} \\
  \frac{\partial^2 W_\mathrm{c}}{\partial \hat{\lambda}_{m'} \partial \hat{y}_m}
  =&\, \sum_{i=0}^{N-1} \frac{\partial}{\partial \hat{\lambda}_{m'}} \Bigl[ \left( y(\Omega_i) - y_i \right) \sin( m \Omega_i ) \Bigr] \nonumber \\
  =&\, \sum_{i=0}^{N-1} \left[  \frac{\partial y(\Omega_i)}{\partial \hat{\lambda}_{m'}} \sin( m \Omega_i )
                               + \left( y(\Omega_i) - y_i \right) m \cos(m \Omega_i) \sin(m' \theta_i) \right] \nonumber \\
  =&\, \sum_{i=0}^{N-1} \left[   \sin( m \Omega_i ) \frac{\partial y(\Omega_i)}{\partial \Omega_i}
                               + \left( y(\Omega_i) - y_i \right) m \cos(m \Omega_i) \right] \sin(m' \theta_i) \label{eqn:d2WcdLmdYm}
\end{align}
It is (expectedly) observed that \eqn{d2WcdXmdLm} equals \eqn{d2WcdLmdXm} and \eqn{d2WcdYmdLm} equals \eqn{d2WcdLmdYm} under exchange of $m$ and $m'$.
Finally:
\begin{align}
  \frac{\partial^2 W_\mathrm{c}}{\partial \hat{\lambda}_{m'} \partial \hat{\lambda}_m}
  =&\, \sum_{i=0}^{N-1} \frac{\partial}{\partial \hat{\lambda}_{m'}}
                        \left(  \frac{\partial W_c}{\partial x(\Omega_i)} \frac{\partial x(\Omega_i)}{\partial \hat{\lambda}_m}
                              + \frac{\partial W_c}{\partial y(\Omega_i)} \frac{\partial y(\Omega_i)}{\partial \hat{\lambda}_m}
                        \right) \nonumber \\
  =&\,          \sum_{i=0}^{N-1} \Biggl(   \phantom{+}\, \frac{\partial^2 W_c}{\partial \hat{\lambda}_{m'} \partial x(\Omega_i)} \frac{\partial   x(\Omega_i)}{                            \partial \hat{\lambda}_m}
                                                    +    \frac{\partial   W_c}{                            \partial x(\Omega_i)} \frac{\partial^2 x(\Omega_i)}{\partial \hat{\lambda}_{m'} \partial \hat{\lambda}_m} \nonumber \\
 ~ &\, \phantom{\sum_{i=0}^{N-1} \Biggl(}\,         +    \frac{\partial^2 W_c}{\partial \hat{\lambda}_{m'} \partial y(\Omega_i)} \frac{\partial   y(\Omega_i)}{                            \partial \hat{\lambda}_m}
                                                    +    \frac{\partial   W_c}{                            \partial y(\Omega_i)} \frac{\partial^2 y(\Omega_i)}{\partial \hat{\lambda}_{m'} \partial \hat{\lambda}_m}
                                 \Biggr) \nonumber \\
  =&\,          \sum_{i=0}^{N-1} \Biggl(   \phantom{+}\, \frac{\partial x(\Omega_i)}{\partial \hat{\lambda}_{m'}} \frac{\partial x(\Omega_i)}{\partial \hat{\lambda}_m}
                                                    +    (x(\Omega_i) - x_i) \frac{\partial^2 x(\Omega_i)}{\partial \hat{\lambda}_{m'} \partial \hat{\lambda}_m} \nonumber \\
 ~ &\, \phantom{\sum_{i=0}^{N-1} \Biggl(}\,         +    \frac{\partial y(\Omega_i)}{\partial \hat{\lambda}_{m'}} \frac{\partial y(\Omega_i)}{\partial \hat{\lambda}_m}
                                                    +    (y(\Omega_i) - y_i) \frac{\partial^2 y(\Omega_i)}{\partial \hat{\lambda}_{m'} \partial \hat{\lambda}_m}
                                 \Biggr) \nonumber \\
  =&\,          \sum_{i=0}^{N-1} \Biggl[
      \phantom{+}\, \left( \frac{\partial x(\Omega_i)}{\partial \Omega_i} \right)^2 + \left( x(\Omega_i) - x_i \right) \frac{\partial^2 x(\Omega_i)}{\partial \Omega_i^2} \nonumber \\
  ~&\, \phantom{\sum_{i=0}^{N-1} \Biggl[}
               +    \left( \frac{\partial y(\Omega_i)}{\partial \Omega_i} \right)^2 + \left( y(\Omega_i) - y_i \right) \frac{\partial^2 y(\Omega_i)}{\partial \Omega_i^2} \Biggr]
       \sin(m' \theta_i) \sin(m \theta_i)
\end{align}

\FloatBarrier
\newpage
\section{Direct Equal-Arclength Parameterization}
Kuhl and Giardina~\cite{kuhl_giardina_1982} present a method to directly and robustly compute equal-arclength parameterized
Fourier coefficients for a given closed curve in the plane.
A Python implementation of their method is available in Ref.~\cite{pyefd}.
This method has applications in two areas of Stellarator research.
First, it can be used to obtain a Fourier description of Poincare data from magnetic field line tracing.
Second, it can be used in  a 3D MHD equilibrium code to define a unique (although not optimal; see below)
poloidal angle-like coordinate/parameter.

The input data for this method is a set of points $\{(x_i, y_i)|i=0,1,..., (N-1)\}$ along a closed curve in the plane.
We define $x_N \equiv x_0$ and $y_N \equiv y_0$ for convenience.
If the geometry to be used as input is already a Fourier series, it needs to be evaluated at consecutive values
of the tangential parameter, e.g., by an inverse FFT. This would imply evaluations at equal increments of the tangential parameter,
which is not a strict necessity for the method under discussion here.
The tangential parameter $t$ is introduced as the arclength along the curve, which can be computed as follows
assuming piecewise-linear segments connecting the points given as input.

Define for $i = 0, 1, ..., (N-1)$:
\begin{align}
  \Delta x_i =&\, x_{i+1} - x_i \\
  \Delta y_i =&\, y_{i+1} - y_i \\
  \Delta t_i =&\, \sqrt{ \Delta x_i^2 + \Delta y_i^2 } \\
         t_i =&\, \begin{cases}
                    0                    &: i=0 \\
                    t_{i-1} + \Delta t_i &: \textrm{ else} \, .
                  \end{cases}
\end{align}
The total arclength or circumference of the curve $T$ is then given by:
\begin{equation}
  T = t_N = t_{N-1} + \Delta t_{N-1} \, .
\end{equation}
An angle-like tangential parameter $\theta$ can then be introduced by:
\begin{equation}
  \theta = \frac{2 \pi t}{T}
\end{equation}
for disrete values $\theta_i$:
\begin{equation}
  \theta_i = \frac{2 \pi}{T} t_i \, .
\end{equation}
The points along the curve can be reconstructed from the discrete differentials $\Delta x_i$, $\Delta y_i$
and the first point $(x_0, y_0)$:
\begin{align}
  x_i =& x_0 + \sum_{k=0}^{i-1} \Delta x_k \\
  y_i =& y_0 + \sum_{k=0}^{i-1} \Delta y_k \, .
\end{align}
A piecewise-linear continuous description of the curve geometry is formulated as follows:
\begin{align}
  x(t) =&\, x_i + (t - t_i) \frac{x_{i+1} - x_i}{t_{i+1} - t_i} \nonumber \\
       =&\, x_i + (t - t_i) \frac{\Delta x_i}{\Delta t_i} \label{eqn:efd_pl_x} \\
  y(t) =&\, y_i + (t - t_i) \frac{\Delta y_i}{\Delta t_i} \, .
\end{align}
for $t_i \leq t < t_{i+1}$ and $0 \leq i < N$.
The targeted truncated Fourier description of this curve geometry is formulated as follows:
\begin{align}
  x(t) =&\, A_0 + \sum_{m=1}^{m^\mathrm{XY}} \left[   a_m \cos\left( \frac{2 \pi}{T} m t \right)
                                                    + b_m \sin\left( \frac{2 \pi}{T} m t \right) \right] \label{eqn:efd_fs_x} \\
  y(t) =&\, C_0 + \sum_{m=1}^{m^\mathrm{XY}} \left[   c_m \cos\left( \frac{2 \pi}{T} m t \right)
                                                    + d_m \sin\left( \frac{2 \pi}{T} m t \right) \right]
\end{align}
where
\begin{align}
  A_0 =&\, \frac{1}{T} \int\limits_{0}^{T} x(t) \mathrm{d}t \\
  C_0 =&\, \frac{1}{T} \int\limits_{0}^{T} y(t) \mathrm{d}t \\
  a_m =&\, \frac{2}{T} \int\limits_{0}^{T} x(t) \cos\left( \frac{2 \pi}{T} m t \right) \mathrm{d}t \label{eqn:efd_a_m} \\
  b_m =&\, \frac{2}{T} \int\limits_{0}^{T} x(t) \sin\left( \frac{2 \pi}{T} m t \right) \mathrm{d}t \\
  c_m =&\, \frac{2}{T} \int\limits_{0}^{T} y(t) \cos\left( \frac{2 \pi}{T} m t \right) \mathrm{d}t \\
  d_m =&\, \frac{2}{T} \int\limits_{0}^{T} y(t) \sin\left( \frac{2 \pi}{T} m t \right) \mathrm{d}t \, .
\end{align}
Several approaches are possible for the numerical evaluation of the Fourier integrals listed above
to obtain the Fourier coefficients $A_0$, $C_0$ and $a_m$, $b_m$, $c_m$ and $d_m$
up to a maximum poloidal mode number $m^\mathrm{XY}$.

\subsection{Trapezoidal Quadrature}
Direct application of the trapezoidal rule to above Fourier integrals leads to:
\begin{align}
  A_0 =&\, \frac{1}{T} \int\limits_{0}^{T} x(t) \mathrm{d}t \nonumber \\
      \approx&\, \frac{1}{T} \sum_{i=0}^{N-1} \frac{x_i + x_{i+1}}{2} \delta t_i \nonumber \\
      =&\, \frac{1}{2 T} \sum_{i=0}^{N-1} (x_i + x_{i+1}) \delta t_i
\end{align}
and
\begin{align}
  a_m =&\, \frac{2}{T} \int\limits_{0}^{T} x(t) \cos\left( \frac{2 \pi}{T} m t \right) \mathrm{d}t \nonumber \\
      \approx&\, \frac{\bcancel{2}}{T} \sum_{i=0}^{N-1} \frac{1}{\bcancel{2}} \left[   x_i     \cos\left( \frac{2 \pi}{T} m t_i     \right)
                                                                                     + x_{i+1} \cos\left( \frac{2 \pi}{T} m t_{i+1} \right) \right] \Delta t_i \nonumber \\
      =&\, \frac{1}{T} \sum_{i=0}^{N-1} \left[   x_i     \cos\left( \frac{2 \pi}{T} m t_i     \right)
                                               + x_{i+1} \cos\left( \frac{2 \pi}{T} m t_{i+1} \right) \right] \Delta t_i
\end{align}
as well as
\begin{align}
  b_m =&\, \frac{2}{T} \int\limits_{0}^{T} x(t) \sin\left( \frac{2 \pi}{T} m t \right) \mathrm{d}t \nonumber \\
      \approx&\, \frac{\bcancel{2}}{T} \sum_{i=0}^{N-1} \frac{1}{\bcancel{2}} \left[   x_i     \sin\left( \frac{2 \pi}{T} m t_i     \right)
                                                                                     + x_{i+1} \sin\left( \frac{2 \pi}{T} m t_{i+1} \right) \right] \Delta t_i \nonumber \\
      =&\, \frac{1}{T} \sum_{i=0}^{N-1} \left[   x_i     \sin\left( \frac{2 \pi}{T} m t_i     \right)
                                               + x_{i+1} \sin\left( \frac{2 \pi}{T} m t_{i+1} \right) \right] \Delta t_i \, .
\end{align}
The remaining coefficients $C_0$, $c_m$ and $d_m$ are obtained by replacing $x$ with $y$ in above expressions, respectively.
The computational cost for this method is comparably low.
A significant factor of oversampling is required to achieve an accurate match with the original point data
from the Fourier representation obtained using trapezoidal quadrature.

\subsection{Analytical Integration over Piecewise-Linear Segments}
This approach is presented by Kuhl and Giardina in their article~\cite{kuhl_giardina_1982}.
The key idea is to evaluate the Fourier integrals for $A_0$, ..., $d_m$ analytically for assumed piecewise-linear segments connecting the given input points.
It might seem counterintuitive to explicitly put straight segments into the model when trying to fit curved geometries.
(Note that Kuhl and Giardina applied this method to chain codes, where linear segments are desired.)
When operating close to the Nyquist limit, there are not enough modes present to force the Fourier representation straight in between sample points.
When operating way below the Nyquist limit (i.e., more points than needed for a given number of Fourier coefficients according to the sampling theorem),
the implied line segments are much shorter than the wavelength of the highest mode to be reconstructed, so no artificial straightening happens.
It is only possible to obtain straight segments from the Fourier representation when reconstructing way more Fourier modes thatn the Nyquist theorem
would allow for based on the given number of input points. This is desired and intended by Kuhl and Giardina for their application
of representing chain codes, but not a useful modus operandi in the context of this work.

There are (at least) two different ways to derive the results in Ref.~\cite{kuhl_giardina_1982}.
The first one is to express both $x(t)$ and $\partial x/\partial t$ as Fourier series
and perform analytical integration in the Fourier integrals for $\partial x/\partial t$, then equate coefficients.
This is the way chosen by Kuhl and Giardina originally.
It is repeated here due to typographic errors in the original article.

A new Fourier series is introduced for $\partial x/\partial t$:
\begin{equation}
  \frac{\partial x}{\partial t}(t) = \sum_{m=1}^{m^\mathrm{XY}} \left[ \alpha_m \cos\left( \frac{2 \pi}{T} m t \right)
                                                                      + \beta_m \sin\left( \frac{2 \pi}{T} m t \right) \right]
\end{equation}
with
\begin{align}
  \alpha_m =&\, \frac{2}{T} \int\limits_0^T \frac{\partial x}{\partial t}(t) \cos\left( \frac{2 \pi}{T} m t \right) \mathrm{d}t \\
   \beta_m =&\, \frac{2}{T} \int\limits_0^T \frac{\partial x}{\partial t}(t) \sin\left( \frac{2 \pi}{T} m t \right) \mathrm{d}t \, .
\end{align}
The tangential derivative $\partial x/\partial t$ is piecewise constant for assumed straight segments between given points.
The integrals for $\alpha_m$ and $\beta_m$ can therefore be evaluated as follows:
\begin{align}
  \alpha_m =&\, \frac{2}{T} \sum_{i=0}^{N-1} \frac{\Delta x_i}{\Delta t_i} \int\limits_{t_i}^{t_{i+1}} \cos\left( \frac{2 \pi}{T} m t \right) \mathrm{d}t \nonumber \\
           =&\, \frac{\bcancel{2}}{\cancel{T}} {\color{blue} \frac{\cancel{T}}{\bcancel{2} \pi m} }
                \sum_{i=0}^{N-1} \frac{\Delta x_i}{\Delta t_i} \left[ \sin\left( \frac{2 \pi}{T} m t_{i+1} \right) - \sin\left( \frac{2 \pi}{T} m t_i \right) \right] \nonumber \\
           =&\, \frac{1}{\pi m} \sum_{i=0}^{N-1} \frac{\Delta x_i}{\Delta t_i} \left[ \sin\left( \frac{2 \pi}{T} m t_{i+1} \right) - \sin\left( \frac{2 \pi}{T} m t_i \right) \right] \\
   \beta_m =&\, \frac{2}{T} \sum_{i=0}^{N-1} \frac{\Delta x_i}{\Delta t_i} \int\limits_{t_i}^{t_{i+1}} \sin\left( \frac{2 \pi}{T} m t \right) \mathrm{d}t \nonumber \\
           =&\, - \frac{\bcancel{2}}{\cancel{T}} {\color{blue} \frac{\cancel{T}}{\bcancel{2} \pi m} }
                \sum_{i=0}^{N-1} \frac{\Delta x_i}{\Delta t_i} \left[ \cos\left( \frac{2 \pi}{T} m t_{i+1} \right) - \cos\left( \frac{2 \pi}{T} m t_i \right) \right] \nonumber \\
           =&\, \frac{-1}{\pi m} \sum_{i=0}^{N-1} \frac{\Delta x_i}{\Delta t_i} \left[ \cos\left( \frac{2 \pi}{T} m t_{i+1} \right) - \cos\left( \frac{2 \pi}{T} m t_i \right) \right] \, .
\end{align}
The terms marked blue are missing in the original article, although their final results are correct.
Note that $\partial x/\partial t$ can also be obtained by differentiation of the Fourier series for $x(t)$ from \eqn{efd_fs_x}:
\begin{align}
  \frac{\partial x}{\partial t}(t) =&\, \frac{\partial}{\partial t} \left\{
    A_0 + \sum_{m=1}^{m^\mathrm{XY}} \left[   a_m \cos\left( \frac{2 \pi}{T} m t \right)
                                            + b_m \sin\left( \frac{2 \pi}{T} m t \right) \right] \right\} \\
     =&\, \sum_{m=1}^{m^\mathrm{XY}} \left[ - a_m \frac{2 \pi m}{T} \sin\left( \frac{2 \pi}{T} m t \right)
                                            + b_m \frac{2 \pi m}{T} \cos\left( \frac{2 \pi}{T} m t \right) \right] \, .
\end{align}
Equating the two representations of $\partial x/\partial t$ we obtain:
\begin{align}
  - a_m \frac{2 \pi m}{T} =  \beta_m \Leftrightarrow&\, a_m =          -    \frac{T}{2 \pi m}  \beta_m \\
    b_m \frac{2 \pi m}{T} = \alpha_m \Leftrightarrow&\, b_m = \phantom{-}\, \frac{T}{2 \pi m} \alpha_m \, .
\end{align}
The coefficients $a_m$, $b_m$ can now be obtained by inserting the explicit forms of $\alpha_m$, $\beta_m$:
\begin{align}
  a_m =&\, \bcancel{-} \frac{T}{2 \pi m} \frac{\bcancel{-}1}{\pi m} \sum_{i=0}^{N-1} \frac{\Delta x_i}{\Delta t_i} \left[ \cos\left( \frac{2 \pi}{T} m t_{i+1} \right) - \cos\left( \frac{2 \pi}{T} m t_i \right) \right] \nonumber \\
      =&\, \frac{T}{2 \pi^2 m^2} \sum_{i=0}^{N-1} \frac{\Delta x_i}{\Delta t_i} \left[ \cos\left( \frac{2 \pi}{T} m t_{i+1} \right) - \cos\left( \frac{2 \pi}{T} m t_i \right) \right] \\
  b_m =&\, \frac{T}{2 \pi m} \frac{1}{\pi m} \sum_{i=0}^{N-1} \frac{\Delta x_i}{\Delta t_i} \left[ \sin\left( \frac{2 \pi}{T} m t_{i+1} \right) - \sin\left( \frac{2 \pi}{T} m t_i \right) \right] \nonumber \\
      =&\, \frac{T}{2 \pi^2 m^2} \sum_{i=0}^{N-1} \frac{\Delta x_i}{\Delta t_i} \left[ \sin\left( \frac{2 \pi}{T} m t_{i+1} \right) - \sin\left( \frac{2 \pi}{T} m t_i \right) \right] \, .
\end{align}
These are the coefficients also published by Kuhl and Giardina.
Analogously, expressions for $c_m$ and $d_m$ can be obtained by replacing $x$ with $y$ in above derivations for $a_m$ and $b_m$, respectively.

The second way to obtain these coefficients is by inserting the piecewise-linear representation for $x(t)$ from Eqn.~\eqn{efd_pl_x}
into the Fourier integrals for $a_m$, $b_m$ and evaluating these integrals directly.
This is presented here for reference and as an independent check of the original results.
We consider the integral for $a_m$ from Eqn.~\eqn{efd_a_m} here and split it up at the input data points:
\begin{align}
  a_m =&\, \frac{2}{T} \sum_{i=0}^{N-1} \left[ \int\limits_{t_i}^{t_{i+1}} x(t) \cos\left( \frac{2 \pi}{T} m t \right) \mathrm{d}t \right] \nonumber \\
      =&\, \frac{2}{T} \sum_{i=0}^{N-1} \left[ \int\limits_{t_i}^{t_{i+1}}
             \left( x_i + (t - t_i) \frac{\Delta x_i}{\Delta t_i} \right)
             \cos\left( \frac{2 \pi}{T} m t \right) \mathrm{d}t \right] \nonumber \\
      =&\, \frac{2}{T} \sum_{i=0}^{N-1} \left\{
               \left[ \left( x_i - t_i \frac{\Delta x_i}{\Delta t_i} \right) \int\limits_{t_i}^{t_{i+1}} \cos\left( \frac{2 \pi}{T} m t \right) \mathrm{d}t \right]
             + \left[ \frac{\Delta x_i}{\Delta t_i} \int\limits_{t_i}^{t_{i+1}} t \cos\left( \frac{2 \pi}{T} m t \right) \mathrm{d}t \right] \right\} \nonumber \\
      =&\, \frac{2}{T} \sum_{i=0}^{N-1} \Biggl\{
           \phantom{+}\, \left( x_i - t_i \frac{\Delta x_i}{\Delta t_i} \right) \frac{T}{2 \pi m} \left[ \sin\left( \frac{2 \pi}{T} m t_{i+1} \right) - \sin\left( \frac{2 \pi}{T} m t_i \right) \right] \nonumber \\
      ~&\, \phantom{\frac{2}{T} \sum_{i=0}^{N-1} \Biggl\{}\,
             + \frac{\Delta x_i}{\Delta t_i} \int\limits_{t_i}^{t_{i+1}} t \cos\left( \frac{2 \pi}{T} m t \right) \mathrm{d}t \Biggr\}
\end{align}
Integration by parts is employed to solve the second integral in above expression:
\begin{align}
  ~&\, \int\limits_{t_i}^{t_{i+1}} t \cdot \cos\left( \frac{2 \pi}{T} m t \right) \mathrm{d}t \nonumber \\
  =&\,   \left[ t \cdot \frac{T}{2 \pi m} \sin\left( \frac{2 \pi}{T} m t \right) \right]_{t_i}^{t_{i+1}}
       - \int\limits_{t_i}^{t_{i+1}} 1 \cdot \frac{T}{2 \pi m} \sin\left( \frac{2 \pi}{T} m t \right) \mathrm{d}t \nonumber \\
  =&\,   \frac{T}{2 \pi m} \left[ t_{i+1} \sin\left( \frac{2 \pi}{T} m t_{i+1} \right) - t_i \sin\left( \frac{2 \pi}{T} m t_i \right) \right]
       - \frac{T}{2 \pi m} \underbrace{\int\limits_{t_i}^{t_{i+1}} \sin\left( \frac{2 \pi}{T} m t \right) \mathrm{d}t}_{
                                     = \frac{T}{2 \pi m} \left[ - \cos\left( \frac{2 \pi}{T} m t \right) \right]_{t_i}^{t_{i+1}}} \nonumber \\
  =&\,         \frac{T}{2 \pi m}          \left[ t_{i+1} \sin\left( \frac{2 \pi}{T} m t_{i+1} \right) - t_i \sin\left( \frac{2 \pi}{T} m t_i \right) \right] \nonumber \\
  ~&\, + \left(\frac{T}{2 \pi m}\right)^2 \left[         \cos\left( \frac{2 \pi}{T} m t_{i+1} \right) -     \cos\left( \frac{2 \pi}{T} m t_i \right) \right] \, .
\end{align}
This can now be inserted into the expression for $a_m$:
\begin{align}
  a_m =&\, \frac{\bcancel{2}}{\cancel{T}} \sum_{i=0}^{N-1} \Biggl\{
           \phantom{+}\, \left( x_i - t_i \frac{\Delta x_i}{\Delta t_i} \right) \frac{\cancel{T}}{\bcancel{2} \pi m} \left[ \sin\left( \frac{2 \pi}{T} m t_{i+1} \right) - \sin\left( \frac{2 \pi}{T} m t_i \right) \right] \nonumber \\
      ~&\, \phantom{\frac{2}{T} \sum_{i=0}^{N-1} \Biggl\{}\,
             + \frac{\Delta x_i}{\Delta t_i} \frac{\cancel{T}}{\bcancel{2} \pi m} \left[ t_{i+1} \sin\left( \frac{2 \pi}{T} m t_{i+1} \right) - t_i \sin\left( \frac{2 \pi}{T} m t_i \right) \right] \nonumber \\
      ~&\, \phantom{\frac{2}{T} \sum_{i=0}^{N-1} \Biggl\{}\,
             + \frac{\Delta x_i}{\Delta t_i} \frac{\cancel{T}}{\bcancel{2} \pi m} \frac{T}{2 \pi m} \left[ \cos\left( \frac{2 \pi}{T} m t_{i+1} \right) - \cos\left( \frac{2 \pi}{T} m t_i \right) \right] \Biggr\} \nonumber \\
      =&\, \frac{T}{2 \pi^2 m^2} \sum_{i=0}^{N-1} \frac{\Delta x_i}{\Delta t_i} \left[ \cos\left( \frac{2 \pi}{T} m t_{i+1} \right) - \cos\left( \frac{2 \pi}{T} m t_i \right) \right] \nonumber \\
      ~&\, + \frac{1}{\pi m} \sum_{i=0}^{N-1} \Biggl\{ \left( x_i - t_i \frac{\Delta x_i}{\Delta t_i} \right) \left[ \sin\left( \frac{2 \pi}{T} m t_{i+1} \right) - \sin\left( \frac{2 \pi}{T} m t_i \right) \right] \nonumber \\
      ~&\, \phantom{+ \frac{1}{\pi m} \sum_{i=0}^{N-1} \Biggl\{}\, + \frac{\Delta x_i}{\Delta t_i} \left[ t_{i+1} \sin\left( \frac{2 \pi}{T} m t_{i+1} \right) - t_i \sin\left( \frac{2 \pi}{T} m t_i \right) \right]
           \Biggr\} \, .
\end{align}
The first sum in above intermediate result is the desired expression for $a_m$.
The remaining sum is now shown to vanish:
\begin{align}
  ~&\,          \sum_{i=0}^{N-1} \Biggl\{      \left( x_i - t_i \frac{\Delta x_i}{\Delta t_i} \right) \left[ \sin\left( \frac{2 \pi}{T} m t_{i+1} \right) - \sin\left( \frac{2 \pi}{T} m t_i \right) \right] \nonumber \\
  ~&\, \phantom{\sum_{i=0}^{N-1} \Biggl\{}\, + \frac{\Delta x_i}{\Delta t_i} \left[ t_{i+1} \sin\left( \frac{2 \pi}{T} m t_{i+1} \right) - t_i \sin\left( \frac{2 \pi}{T} m t_i \right) \right] \Biggr\} \nonumber \\
  =&\,          \sum_{i=0}^{N-1} \Biggl\{  x_i \left[ \sin\left( \frac{2 \pi}{T} m t_{i+1} \right) - \sin\left( \frac{2 \pi}{T} m t_i \right) \right] \nonumber \\
  ~&\, \phantom{\sum_{i=0}^{N-1} \Biggl\{}\, + \frac{\Delta x_i}{\Delta t_i} \left[         - t_i     \sin\left( \frac{2 \pi}{T} m t_{i+1} \right)
                                                                                    \cancel{+ t_i     \sin\left( \frac{2 \pi}{T} m t_i     \right)}
                                                                                            + t_{i+1} \sin\left( \frac{2 \pi}{T} m t_{i+1} \right)
                                                                                    \cancel{- t_i     \sin\left( \frac{2 \pi}{T} m t_i     \right)} \right] \Biggr\} \nonumber \\
  =&\, \sum_{i=0}^{N-1} \Biggl\{  x_i \left[ \sin\left( \frac{2 \pi}{T} m t_{i+1} \right) - \sin\left( \frac{2 \pi}{T} m t_i \right) \right]
                               + \frac{\Delta x_i}{\cancel{\Delta t_i}} \cancel{(t_{i+1} - t_i)} \sin\left( \frac{2 \pi}{T} m t_{i+1} \right) \Biggr\} \nonumber \\
  =&\, \sum_{i=0}^{N-1} \Biggl\{ \cancel{  x_i     \sin\left( \frac{2 \pi}{T} m t_{i+1} \right)}
                            \underbrace{- x_i     \sin\left( \frac{2 \pi}{T} m t_i     \right)
                                        + x_{i+1} \sin\left( \frac{2 \pi}{T} m t_{i+1} \right)}_\textrm{cancels along the sum}
                                \cancel{- x_i     \sin\left( \frac{2 \pi}{T} m t_{i+1} \right)} \Biggr\} \nonumber \\
  =&\, 0 \, .
\end{align}
Thus it remains for $a_m$:
\begin{equation}
  a_m = \frac{T}{2 \pi^2 m^2} \sum_{i=0}^{N-1} \frac{\Delta x_i}{\Delta t_i} \left[ \cos\left( \frac{2 \pi}{T} m t_{i+1} \right) - \cos\left( \frac{2 \pi}{T} m t_i \right) \right]
\end{equation}
which is the desired result put forward by Kuhl and Giardina, obtained here using an alternative approach.
Similar results can be obtained for $b_m$, ..., $d_m$.

It remains to derive expressions for the DC-coefficients $A_0$ and $C_0$.
Kuhl and Giardina provide expressions for these, but do not go further into detail how to derive them.
Here, expressions for $A_0$ and $C_0$ are obtained again by inserting the piecewise-linear expressions for $x(t)$ and $y(t)$ into the Fourier integrals.
We start by considering $A_0$:
\begin{align}
  A_0 =&\, \frac{1}{T} \sum_{i=0}^{N-1} \left[ \int\limits_{t_i}^{t_{i+1}} x(t) \mathrm{d}t \right]
      =    \frac{1}{T} \sum_{i=0}^{N-1} \left[ \int\limits_{t_i}^{t_{i+1}} \left( x_i + (t - t_i) \frac{\Delta x_i}{\Delta t_i} \right) \mathrm{d}t \right] \nonumber \\
      =&\, \frac{1}{T} \sum_{i=0}^{N-1} \Biggl\{ \left( x_i - t_i \frac{\Delta x_i}{\Delta t_i} \right) \underbrace{\int\limits_{t_i}^{t_{i+1}} \mathrm{d}t}_{= t_{i+1} - t_i = \Delta t_i}
                                                + \frac{\Delta x_i}{\Delta t_i} \underbrace{\int\limits_{t_i}^{t_{i+1}} t\, \mathrm{d}t}_{= \frac{1}{2} \left( t_{i+1}^2 - t_i^2 \right)} \Biggr\} \nonumber \\
      =&\, \frac{1}{T} \sum_{i=0}^{N-1} \left[ \frac{\Delta x_i}{2 \Delta t_i} \left( t_{i+1}^2 - t_i^2 \right) + \left(x_i - t_i \frac{\Delta x_i}{\Delta t_i} \right) \Delta t_i \right] \nonumber \\
      =&\, \frac{1}{T} \sum_{i=0}^{N-1} \left[ \frac{\Delta x_i}{2 \Delta t_i} \left( t_{i+1}^2 - t_i^2 \right) + x_i \Delta t_i - t_i \Delta x_i \right] \, .
\end{align}
Similarly, it follows:
\begin{equation}
  C_0 = \frac{1}{T} \sum_{i=0}^{N-1} \left[ \frac{\Delta y_i}{2 \Delta t_i} \left( t_{i+1}^2 - t_i^2 \right) + y_i \Delta t_i - t_i \Delta y_i \right] \, .
\end{equation}
Note that:
\begin{align}
  x_i =&\, x_0 + \sum_{k=0}^{i-1} \Delta x_k \\
  y_i =&\, y_0 + \sum_{k=0}^{i-1} \Delta y_k \\
  t_i =&\,       \sum_{k=0}^{i-1} \Delta t_k \, .
\end{align}
A shorthand notation ${\tilde{\xi}}_i$ is introduced:
\begin{align}
  {\tilde{\xi}}_i \equiv&\, x_i - t_i \frac{\Delta x_i}{\Delta t_i} \\
    =&\, x_0 + \sum_{k=0}^{i-1} \Delta x_k - \frac{\Delta x_i}{\Delta t_i} \sum_{k=0}^{i-1} \Delta t_k \, .
\end{align}
Insert this into the expression for $A_0$:
\begin{align}
  A_0 =&\, \frac{1}{T} \sum_{i=0}^{N-1} \Bigl[ \frac{\Delta x_i}{2 \Delta t_i} \left( t_{i+1}^2 - t_i^2 \right)
    + x_0 \Delta t_i + \underbrace{\left(\sum_{k=0}^{i-1} \Delta x_k - \frac{\Delta x_i}{\Delta t_i} \sum_{k=0}^{i-1} \Delta t_k \right)}_{\equiv \xi_i} (t_{i+1} - t_i) \Bigr] \nonumber \\
      =&\, x_0 \frac{1}{\cancel{T}} \underbrace{\sum_{i=0}^{N-1} \Delta t_i}_{= \cancel{T}}
           + \underbrace{\frac{1}{T} \sum_{i=0}^{N-1} \left[ \frac{\Delta x_i}{2 \Delta t_i} \left( t_{i+1}^2 - t_i^2 \right) + \xi_i (t_{i+1} - t_i) \right]}_{A_0 \textrm{ as presented by Kuhl and Giardina}} \, .
\end{align}
Thus we recover the expression for $A_0$ presented by Kuhl and Giardina and their $C_0$ can be obtained by replacing $x$ with $y$ and $\xi$ with $\delta$ in the expression for $A_0$, respectively.
Their $A_0$ does not contain the offset of $x_0$, which can be understood by considering that they only work with the line segments between the points and not the point positions directly.
Formulated differently, in their formulation $x_0 = 0$ and $y_0 = 0$ hold by construction and their $A_0$ and $C_0$ are therefore referenced to the starting position of the curve.

\subsection{Side Remark: Adaptive Quadrature}
The equal-arclength Fourier coefficients can be numerically computed from a curve description that can be continuously evaluated using the methodology described in this section.
The key idea is to formulate the differential arclength $\mathrm{d}l$ as a function of the tangential parameter $t$ and use available adaptive quadrature routines
to compute the total and incremental arclength to the desired numerical accuracy.
Depending on how the model function for the curve is formulated, interpolation might be required in order to evaluate the differential arclength in between given input data points.
Two types of interpolation methods should yield reasonable results:
linear interpolation to match the model assumptions made by Kuhl and Giardina (although this would defeat the purpose of the analytical integrals in their method)
as well as sinc interpolation~\cite{boyd_spectral_methods}, which is typically used to interpolate randomly sampled periodic time series data.
If the curve to be described by equal-arclength Fourier coefficients is already described by a Fourier series (or some other representation that permits exact evaluation
at arbitrary parameter values), no interpolation is required. It is duely noted however that generally the evaluation of Fourier series at non-uniformly distributed locations
(as occurs in, e.g., Gauss-Kronrod quadrature) precludes the use of FFT methods, making this method computationally expensive in comparison with the methods presented in the previous sections.
It is therefore not considered here in more detail.

\FloatBarrier
\section{Unique Poloidal Origin}
Two approaches are laid out in the following
to determine a unique origin of the poloidal coordinate.

\subsection{First Harmonic Phasor}
For a vertically asymmetric curve, a unique origin of the tangential parameter remains to be determined.
Here, we follow the method outlined on p.~251 of Ref.~\cite{kuhl_giardina_1982}.
The contributions to the geometry from the $m=1$-mode are denoted $x_1$ and $y_1$:
\begin{align}
  x_1 =&\, a_1 \cos(\theta) + b_1 \sin(\theta) \\
  y_1 =&\, c_1 \cos(\theta) + d_1 \sin(\theta) \, .
\end{align}
The first harmonic phasor (magnitude of the $m=1$ contribution) $E$ is given by:
\begin{equation}
  E = \sqrt{x_1^2 + y_1^2} \, .
\end{equation}
The criterion to determine a unique tangential origin $\theta_1$ suggested by Kuhl and Giardina is:
\begin{equation}
  \frac{\partial E}{\partial \theta}(\theta_1) = 0 \, .
\end{equation}
The tangential derivative of $E$ is computed using the chain rule:
\begin{align}
  \frac{\partial   E}{\partial \theta} =&\, \frac{\partial E}{\partial x_1} \frac{\partial x_1}{\partial \theta} + \frac{\partial E}{\partial y_1} \frac{\partial y_1}{\partial \theta} \\
  \frac{\partial   E}{\partial    x_1} =&\, \frac{x_1}{\sqrt{x_1^2 + y_1^2}} \\
  \frac{\partial   E}{\partial    y_1} =&\, \frac{y_1}{\sqrt{x_1^2 + y_1^2}} \\
  \frac{\partial x_1}{\partial \theta} =&\, - a_1 \sin(\theta) + b_1 \cos(\theta) \\
  \frac{\partial y_1}{\partial \theta} =&\, - c_1 \sin(\theta) + d_1 \cos(\theta)
\end{align}
leading to:
\begin{align}
  \frac{\partial E}{\partial \theta}
  =&\, \phantom{+}\, \frac{\left( a_1 \cos(\theta) + b_1 \sin(\theta) \right) \cdot \left( - a_1 \sin(\theta) + b_1 \cos(\theta) \right)}
                          {\sqrt{\left( a_1 \cos(\theta) + b_1 \sin(\theta) \right)^2 + \left( c_1 \cos(\theta) + d_1 \sin(\theta) \right)^2}} \nonumber \\
  ~&\,          +    \frac{\left( c_1 \cos(\theta) + d_1 \sin(\theta) \right) \cdot \left( - c_1 \sin(\theta) + d_1 \cos(\theta) \right)}
                          {\sqrt{\left( a_1 \cos(\theta) + b_1 \sin(\theta) \right)^2 + \left( c_1 \cos(\theta) + d_1 \sin(\theta) \right)^2}} \nonumber \\
  =&\, \frac{\left(-a_1^2 + b_1^2 - c_1^2 + d_1^2 \right) \sin(\theta) \cos(\theta) + (a_1 b_1 + c_1 d_1) \left( \cos^2(\theta) - \sin^2(\theta) \right)}
            {\sqrt{\left( a_1 \cos(\theta) + b_1 \sin(\theta) \right)^2 + \left( c_1 \cos(\theta) + d_1 \sin(\theta) \right)^2}} \nonumber \\
  =&\, \frac{\left(-a_1^2 + b_1^2 - c_1^2 + d_1^2 \right) \frac{1}{2} \sin(2 \theta) + (a_1 b_1 + c_1 d_1) \cos(2 \theta)}
            {\sqrt{\left( a_1 \cos(\theta) + b_1 \sin(\theta) \right)^2 + \left( c_1 \cos(\theta) + d_1 \sin(\theta) \right)^2}}
\end{align}
where trigonomic identities~\eqn{trigid_1} and~\eqn{trigid_2} were used.
The denominator of $\partial E/\partial \theta$ is positive for non-trivial Fourier coefficients $a_1$, ..., $d_1$.
Thus, it is sufficient to investigate the zero in the numerator:
\begin{align}
                 0 =&\, \frac{1}{2} \left(b_1^2 + d_1^2 -a_1^2 - c_1^2 \right) \sin(2 \theta_1) + (a_1 b_1 + c_1 d_1) \cos(2 \theta_1) \nonumber \\
 \Leftrightarrow 0 =&\, \frac{\sin(2 \theta_1)}{\cos(2 \theta_1)} + \frac{2 (a_1 b_1 + c_1 d_1)}{b_1^2 + d_1^2 - a_1^2 - c_1^2} \nonumber \\
 \Leftrightarrow \tan(2 \theta_1) =&\, \frac{2(a_1 b_1 + c_1 d_1)}{a_1^2 + c_1^2 - b_1^2 - d_1^2 } \nonumber \\
 \Leftrightarrow \theta_1 =&\, \frac{1}{2} \arctan \left(\frac{2(a_1 b_1 + c_1 d_1)}{a_1^2 + c_1^2 - b_1^2 - d_1^2} \right) \, .
\end{align}
This is the result presented by Kuhl and Giardina.
The tangential coordinate system then has to be shifted by $\theta_1$ by re-scaling the Fourier coefficients $a_m$, ..., $d_m$.
In the symmetric case, $b_m=0$ and $c_m=0$ for all $m$ under consideration, leading to:
\begin{align}
  \theta_1 =&\, \frac{1}{2} \arctan \left(\frac{2(a_1 \cdot 0 + 0 \cdot d_1)}{a_1^2 - d_1^2} \right) \nonumber \\
           =&\, \frac{1}{2} \arctan \left( 0 \right) = 0 \, .
\end{align}
As expected, above criterion does not induce a shift in the tangential coordinate if the curve is inherently symmetric.
(The tangential parameter is implicitly well-defined in this case.)

\subsection{3-Parameter-Ellipse}
\begin{figure}[htbp]
 \centering
 \includegraphics{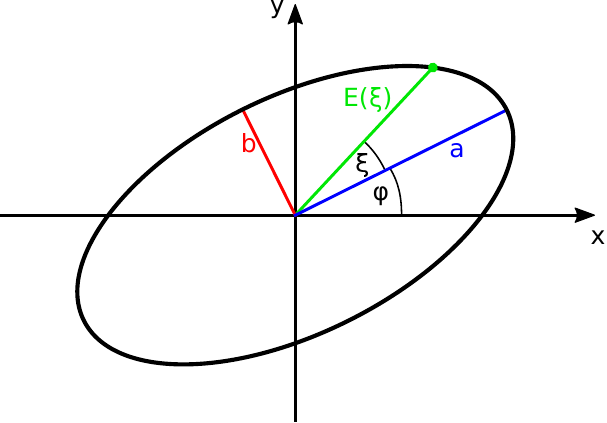}
 \caption{A general ellipse can be described by the lengths of its two half-axes~$a$ and~$b$
 and two angles, namely the rotation angle of the major half-axis~$\varphi$
 and the tangential origin~$\xi$.}
 \label{fig:ellipse_sketch}
\end{figure}

The general Fourier series for an arbitrary flux surface geometry (described by $M$ Fourier harmonics) is:
\begin{align}
 x(\theta) =&\, \hat{x}_0 + \sum\limits_{m=1}^{m^\mathrm{XY}} \left[ \hat{x}_m^\mathrm{c} \cos(m \theta) + \hat{x}_m^\mathrm{s} \sin(m \theta) \right] \\
 y(\theta) =&\, \hat{y}_0 + \sum\limits_{m=1}^{m^\mathrm{XY}} \left[ \hat{y}_m^\mathrm{c} \cos(m \theta) + \hat{y}_m^\mathrm{s} \sin(m \theta) \right] \, .
\end{align}
The contributions~$(x_m(\theta), y_m(\theta))$ for each mode number~$m$ are ellipses:
\begin{align}
 x_m(\theta) =&\, \hat{x}_m^\mathrm{c} \cos(m \theta) + \hat{x}_m^\mathrm{s} \sin(m \theta) \label{eqn:x_m_contrib} \\
 y_m(\theta) =&\, \hat{y}_m^\mathrm{c} \cos(m \theta) + \hat{y}_m^\mathrm{s} \sin(m \theta) \label{eqn:y_m_contrib}
\end{align}
and the surface geometry can be expressed this way as a superposition of ellipses:
\begin{align}
 x(\theta) =&\, \hat{x}_0 + \sum\limits_{m=1}^{m^\mathrm{XY}} x_m(\theta) \\
 y(\theta) =&\, \hat{y}_0 + \sum\limits_{m=1}^{m^\mathrm{XY}} y_m(\theta) \, .
\end{align}
We clearly see that an ellipse can be uniquely described by its four Fourier coefficients.
In above example, these would be~$\{\hat{x}_m^\mathrm{c}, \hat{x}_m^\mathrm{s}, \hat{y}_m^\mathrm{c}, \hat{y}_m^\mathrm{s}\}$.
A general ellipse is shown in Fig~\ref{fig:ellipse_sketch}.
Alternatively, any ellipse can also be described by the lengths of its two half-axes~$a$ and~$b$
and two angles~$\varphi$ and~$\xi$ (see Fig.~\ref{fig:ellipse_sketch}).
It turns out that yet another parameterization
in terms of three Fourier coefficients~$\hat{c}$, $\hat{s}$ and~$\hat{r}$ plus one angle~$\Delta \theta$ is possible.
We focus only on the $m=1$-contribution for now:
\begin{align}
 x_1(\theta) =&\, \hat{c} \cos(\theta + \Delta \theta) + \hat{r} \sin(\theta + \Delta \theta) \label{eqn:ellipse_4_x} \\
 y_1(\theta) =&\, \hat{s} \sin(\theta + \Delta \theta) + \hat{r} \cos(\theta + \Delta \theta) \label{eqn:ellipse_4_y} \, .
\end{align}
The shape of the ellipse in realspace is completely determined by three parameters,
which in terms of Fig.~\ref{fig:ellipse_sketch} would be the set~$\{a, b, \varphi\}$.
The fourth parameter sets the origin for the tangential coordinate.
Varying it only shifts points back and forth along the circumference of the ellipse,
but does not change its general appearance.
It should be noted that the poloidal origin is the same for all contributions
to the Fourier representation of the surface.
Consequently, fixing the poloidal origin for any one of the various-$m$ contributions
implicitly defines the poloidal origin also for all other-$m$ contributions.
This is put to use when finding a spectrally-condensed Fourier representation
for a given surface geometry.
Here, the $m=1$ - contribution is reparameterized according to~\eqn{ellipse_4_x} and~\eqn{ellipse_4_y}
with the tangential offset~$\Delta \theta$ assumed to be zero,
which leads to an implicit determination of the tangential origin:
\begin{align}
 x_1(\theta) =&\, \hat{c} \cos(\theta) + \hat{r} \sin(\theta) \label{eqn:ellipse_3_x} \\
 y_1(\theta) =&\, \hat{s} \sin(\theta) + \hat{r} \cos(\theta) \label{eqn:ellipse_3_y} \, .
\end{align}

Summarizing, the set of Fourier coefficients to describe a given surface geometry
is reduced by one element by assuming~$\hat{x}_1^\mathrm{s} = \hat{y}_1^\mathrm{c}$
and by this constraint on the $m=1$ - contribution,
the location of the tangential origin is implictly defined.
All other-$m$ contributions retain their full set of four Fourier coefficients.

It might be useful for one or the other purpose to know the $\{\hat{c}, \hat{s}, \hat{r}, \Delta \theta \}$-representation
of an ellipse specified in terms of the regular four
Fourier coefficients~$\{\hat{x}_1^\mathrm{c}, \hat{x}_1^\mathrm{s}, \hat{y}_1^\mathrm{c}, \hat{y}_1^\mathrm{s}\}$.
This is done next.
We start by considering the representation of an ellipse by~\eqn{ellipse_4_x} and~\eqn{ellipse_4_y}:
\begin{align}
                ~&\, \begin{cases}
                       x_1(\theta) =&\, \hat{c} \cos(\theta + \Delta \theta) + \hat{r} \sin(\theta + \Delta \theta) \\
                       y_1(\theta) =&\, \hat{s} \sin(\theta + \Delta \theta) + \hat{r} \cos(\theta + \Delta \theta)
                     \end{cases} \nonumber \\
 \Leftrightarrow &\, \begin{cases}
                       x_1(\theta) =&\,   \hat{c} \left[ \cos(\theta) \cos(\Delta \theta) - \sin(\theta) \sin(\Delta \theta) \right]
                                        + \hat{r} \left[ \sin(\theta) \cos(\Delta \theta) + \cos(\theta) \sin(\Delta \theta) \right]  \\
                       y_1(\theta) =&\,   \hat{s} \left[ \sin(\theta) \cos(\Delta \theta) + \cos(\theta) \sin(\Delta \theta) \right]
                                        + \hat{r} \left[ \cos(\theta) \cos(\Delta \theta) - \sin(\theta) \sin(\Delta \theta) \right]
                     \end{cases} \nonumber \\
 \Leftrightarrow &\, \begin{cases}
                       x_1(\theta) =&\,   \cos(\theta) \left[             \hat{c} \cos(\Delta \theta) + \hat{r} \sin(\Delta \theta) \right]
                                        + \sin(\theta) \left[          -  \hat{c} \sin(\Delta \theta) + \hat{r} \cos(\Delta \theta) \right] \\
                       y_1(\theta) =&\,   \sin(\theta) \left[             \hat{s} \cos(\Delta \theta) - \hat{r} \sin(\Delta \theta) \right]
                                        + \cos(\theta) \left[ \phantom{-} \hat{s} \sin(\Delta \theta) + \hat{r} \cos(\Delta \theta) \right]
                     \end{cases}
\end{align}
Equating coefficients with~\eqn{x_m_contrib} and~\eqn{y_m_contrib}, we find:
\begin{align}
 \hat{x}_1^\mathrm{c} =&\, \phantom{-}~ \hat{c} \cos(\Delta \theta) + \hat{r} \sin(\Delta \theta) \\
 \hat{x}_1^\mathrm{s} =&\,          -   \hat{c} \sin(\Delta \theta) + \hat{r} \cos(\Delta \theta) \\
 \hat{y}_1^\mathrm{c} =&\, \phantom{-}~ \hat{r} \cos(\Delta \theta) + \hat{s} \sin(\Delta \theta) \\
 \hat{y}_1^\mathrm{s} =&\,          -   \hat{r} \sin(\Delta \theta) + \hat{s} \cos(\Delta \theta)
\end{align}
or equally in matrix form:
\begin{align}
 \begin{pmatrix}
  \hat{x}_1^\mathrm{c} \\
  \hat{x}_1^\mathrm{s}
 \end{pmatrix}
 =&\,
 \mathbf{R}(\Delta \theta)
 \begin{pmatrix}
  \hat{c} \\
  \hat{r}
 \end{pmatrix} \\
 \begin{pmatrix}
  \hat{y}_1^\mathrm{c} \\
  \hat{y}_1^\mathrm{s}
 \end{pmatrix}
 =&\,
 \mathbf{R}(\Delta \theta)
 \begin{pmatrix}
  \hat{r} \\
  \hat{s}
 \end{pmatrix}
\end{align}
with the rotation matrix~$\mathbf{R}(\Delta \theta)$:
\begin{equation}
 \mathbf{R}(\Delta \theta)
 = \begin{pmatrix}
    \phantom{-}\, \cos(\Delta \theta) & \sin(\Delta \theta) \\
             -    \sin(\Delta \theta) & \cos(\Delta \theta)
   \end{pmatrix} \, .
\end{equation}
The rotation matrix is readily inverted by observing:
\begin{equation}
 \mathbf{R}^{-1}(\Delta \theta) = \mathbf{R}(-\Delta \theta) \, ,
\end{equation}
leading to:
\begin{equation}
 \mathbf{R}^{-1}(\Delta \theta)
 = \begin{pmatrix}
    \cos(\Delta \theta) &          -    \sin(\Delta \theta) \\
    \sin(\Delta \theta) & \phantom{-}\, \cos(\Delta \theta)
   \end{pmatrix} \, .
\end{equation}
This allows us to obtain the three-parameter representation for the ellipse
given the standard four Fourier coefficients and assuming $\Delta \theta$ is known already:
\begin{align}
 \begin{pmatrix}
  \hat{c} \\
  \hat{r}
 \end{pmatrix}
 =&\,
 \mathbf{R}^{-1}(\Delta \theta)
 \begin{pmatrix}
  \hat{x}_1^\mathrm{c} \\
  \hat{x}_1^\mathrm{s}
 \end{pmatrix} \label{eqn:cr_from_x1cs} \\
 \begin{pmatrix}
  \hat{r} \\
  \hat{s}
 \end{pmatrix}
 =&\,
 \mathbf{R}^{-1}(\Delta \theta)
 \begin{pmatrix}
  \hat{y}_1^\mathrm{c} \\
  \hat{y}_1^\mathrm{s}
 \end{pmatrix} \label{eqn:rs_from_y1cs} \, .
\end{align}
We note that this way we have two expressions for the shared Fourier coefficient~$\hat{r}$:
\begin{align}
 \hat{r} =&\, \hat{x}_1^\mathrm{c} \sin(\Delta \theta) + \hat{x}_1^\mathrm{s} \cos(\Delta \theta) \\
 \hat{r} =&\, \hat{y}_1^\mathrm{c} \cos(\Delta \theta) - \hat{y}_1^\mathrm{s} \sin(\Delta \theta) \, .
\end{align}
These two expressions can be equated and solved for the so-far unknown $\Delta \theta$:
\begin{align}
                 \hat{x}_1^\mathrm{c} \sin(\Delta \theta) + \hat{x}_1^\mathrm{s} \cos(\Delta \theta)
            =&\, \hat{y}_1^\mathrm{c} \cos(\Delta \theta) - \hat{y}_1^\mathrm{s} \sin(\Delta \theta) \nonumber \\
 \Leftrightarrow \sin(\Delta \theta) \left[ \hat{x}_1^\mathrm{c} + \hat{y}_1^\mathrm{s} \right]
            +&\, \cos(\Delta \theta) \left[ \hat{x}_1^\mathrm{s} - \hat{y}_1^\mathrm{c} \right] = 0 \nonumber \\
 \Leftrightarrow \frac{\sin(\Delta \theta)}{\cos(\Delta \theta)}
            =&\, \frac{\hat{y}_1^\mathrm{c} - \hat{x}_1^\mathrm{s}}{\hat{x}_1^\mathrm{c} + \hat{y}_1^\mathrm{s}} \nonumber \\
 \Leftrightarrow \Delta \theta
            =&\, \arctan\left( \frac{\hat{y}_1^\mathrm{c} - \hat{x}_1^\mathrm{s}}{\hat{x}_1^\mathrm{c} + \hat{y}_1^\mathrm{s}} \right) \label{eqn:delta_theta_ellipse_3} \, .
\end{align}
A possible division by zero can be circumvented by evaluating \eqn{delta_theta_ellipse_3} using the \texttt{arctan2} function:
\begin{equation}
 \boxed{\Delta \theta = \texttt{arctan2}\left(\hat{y}_1^\mathrm{c} - \hat{x}_1^\mathrm{s}, \hat{x}_1^\mathrm{c} + \hat{y}_1^\mathrm{s}\right)} \, .
\end{equation}
The other coefficients~$\hat{c}$, $\hat{s}$ and~$\hat{r}$ can then be obtained
from the original four Fourier coefficients
using \eqn{cr_from_x1cs} and \eqn{rs_from_y1cs}.
There is one last ambiguity.
The tangential origin displacement~$\Delta \theta$ is only unique up to a shift by $\pi$.
Thus, the following two representations yield geometrically equivalent ellipses:
\begin{equation}
 \{ \hat{c}, \hat{s}, \hat{r}, \Delta \theta \} \textrm{ and } \{ -\hat{c}, -\hat{s}, -\hat{r}, \Delta \theta + \pi \} \, .
\end{equation}
They differ by the poloidal origin being displaced to the opposite side of the ellipse,
similar to the illustration in Fig.~8c in Ref.~\cite{kuhl_giardina_1982}.

\chapter{VMEC}

\FloatBarrier
\section{Initial Guess for Boundary}
This section goes into the detail of how the (initial guess for) the geometry
of the last closed flux surface, which is the computational boundary in VMEC,
is specified in the input file and how this translates into the VMEC-internal array format.
In case of a fixed-boundary VMEC run, the boundary is taken as specified in the input file.
In case of a free-boundary VMEC run, the boundary specified in the input file
is taken as an initial guess and it evolves along with the geometry of the other flux surfaces.
The boundary geometry is specified using the arrays \code{rbc} and \code{zbs}
in case of a (stellarator-)symmetric case. For an asymmetric case,
additional arrays \code{rbs} and \code{zbc} can be specified.
First, consider an axisymmetric case. The configuration has no toroidal variation
and therefore, only a single Fourier mode in the toroidal direction,
namely the DC offset component at $n=0$, is required.
The magnetic axis has no poloidal variation and thus, only a single Fourier mode,
namely the DC offset at $m=0$ is required to describe its geometry.
If furthermore up/down symmetry is assumed in the axisymmetric case,
the axis has to be located at the $z=0$ plane and its geometry is then specified
by a single number, namely its radial position specified by $\hat{R}^\mathrm{cc}_{0,0}$.

\FloatBarrier
\section{Magnetic Field Representation}
The contravariant magnetic field components in VMEC are defined as follows:
\begin{align}
  B^\theta =&\, \frac{\Phi'}{\sqrt{g}} \left( \iota - \lambda_\zeta  \right)
           =    \frac{1}{\sqrt{g}} \left( \chi' - \Phi' \lambda_\zeta  \right) \label{eqn:bsuptheta} \\
  B^\zeta  =&\, \frac{\Phi'}{\sqrt{g}} \left(   1   + \lambda_\theta \right) \label{eqn:bsupzeta} \, .
\end{align}
The covariant magnetic field components are then computed using the contravariant magnetic field components and the metric elements:
\begin{align}
  B_\theta =&\, B^\theta g_{\theta \theta} + B^\zeta g_{\theta \zeta } \label{eqn:bsubtheta_from_contra} \\
  B_\zeta  =&\, B^\theta g_{\theta \zeta } + B^\zeta g_{\zeta  \zeta } \label{eqn:bsubzeta_from_contra}  \, .
\end{align}

\FloatBarrier
\section{Radial Discretization} \label{sec:radial_discretization}
In VMEC, the radial direction is discretized into a finite number (usually denoted $n_\mathrm{s}$) of flux surfaces.
Physical quantities are represented either at radial knots on the flux surfaces (full-grid)
or half-way (in toroidal flux-space) in between these (half-grid).
The radial coordinate is the normalized toridal flux:
\begin{equation}
 s_i = \frac{\Phi_i}{\Phi_\mathrm{LCFS}} \textrm{ with } i = 0, 1, ..., (n_\mathrm{s}-1)
\end{equation}
where $\Phi_\mathrm{LCFS}$ is the toroidal magnetic flux enclosed by the last closed flux surface
and $\Phi_i$ is the toroidal magnetic flux enclosed by the $i$-th flux surface.
The positions of the radial knots are usually chosen equally-distributed in $s$, leading to:
\begin{equation}
 s_i = \frac{i}{n_\mathrm{s}-1} \textrm{ with } i = 0, 1, ..., (n_\mathrm{s}-1) \, .
\end{equation}
Furthermore, the physical quantities represented as Fourier series are split
into contributions from their even poloidal harmonics and from their odd poloidal harmonics,
were a factor of $\sqrt{s}$ is factored out of the odd poloidal harmonics.
A physical quantity $f$ is then represented as follows:
\begin{equation}
 f(s) = f_\mathrm{e}(s) + \sqrt{s} f_\mathrm{o}(s)
\end{equation}
where $f_\mathrm{e}(s)$ is the contribution from even-$m$ harmonics
and $f_\mathrm{o}(s)$ is the contribution from odd-$m$ harmonics.
Then it follows for the radial derivative of $f$:
\begin{align}
 \frac{\partial f}{\partial s}
  ~ &= \frac{\partial f_\mathrm{e}}{\partial s} + \frac{\partial}{\partial s} \left( \sqrt{s} f_\mathrm{o} \right) \nonumber \\
  ~ &= \frac{\partial f_\mathrm{e}}{\partial s} + \sqrt{s} \frac{\partial f_\mathrm{o}}{\partial s} + \frac{\partial \sqrt{s}}{\partial s} f_\mathrm{o} \nonumber \\
  ~ &= \frac{\partial f_\mathrm{e}}{\partial s} + \sqrt{s} \frac{\partial f_\mathrm{o}}{\partial s} + \frac{1}{2 \sqrt{s}} f_\mathrm{o} \, .
\end{align}
Assume now that $f$ is given on the full grid
and the radial derivative is to be computed at a half-grid position $s_{i+1/2}$.
The two neighboring full-grid positions are then $s_i$ and $s_{i+1}$.
The radial derivatives of the parity-separated contributions are computed by finite differencing separately:
\begin{align}
 \frac{\partial f_\mathrm{e}}{\partial s} \left( s_{i+1/2} \right)
   &= \left[ f_\mathrm{e}(s_{i+1}) - f_\mathrm{e}(s_i) \right] / \Delta s
    = \left( n_\mathrm{s} - 1 \right) \left[ f_\mathrm{e}(s_{i+1}) - f_\mathrm{e}(s_i) \right] \\
 \frac{\partial f_\mathrm{o}}{\partial s} \left( s_{i+1/2} \right)
   &= \left[ f_\mathrm{o}(s_{i+1}) - f_\mathrm{o}(s_i) \right] / \Delta s
    = \left( n_\mathrm{s} - 1 \right) \left[ f_\mathrm{o}(s_{i+1}) - f_\mathrm{o}(s_i) \right]
\end{align}
with $\Delta s = 1 / \left( n_\mathrm{s} - 1 \right)$.
The odd contribution at the half-grid location $f_\mathrm{o}\left( s_{i+1/2} \right)$ is computed
by averaging the values at the neighboring full-grid points:
\begin{equation}
 f_\mathrm{o}\left( s_{i+1/2} \right) = \frac{1}{2} \left[ f_\mathrm{o}(s_{i+1}) + f_\mathrm{o}(s_i) \right]
\end{equation}
Putting above results together, we arrive at the following expression for the radial derivative of $f$ at $s_{i+1/2}$:
\begin{align}
   \frac{\partial f}{\partial s} \left( s_{i+1/2} \right)
 =& \phantom{+}
   \left( n_\mathrm{s} - 1 \right)
   \left[                    \left( f_\mathrm{e}(s_{i+1}) - f_\mathrm{e}(s_i) \right)
          + \sqrt{s_{i+1/2}} \left( f_\mathrm{o}(s_{i+1}) - f_\mathrm{o}(s_i) \right) \right] \nonumber \\
 ~& + \frac{1}{4 \sqrt{s_{i+1/2}}} \left[ f_\mathrm{o}(s_{i+1}) + f_\mathrm{o}(s_i) \right] \, .
\end{align}

% TODO: reverse way is used in Compute_Currents:
% The covariant magnetic field components are on the half-grid
% and the current coefficients are to be computed on the full grid.

If products of two quantities that are defined on the full-grid
are to be interpolated onto the half-grid, the so-called product interpolation rule has to be employed~\cite{hirshman_schwenn_nuehrenberg_1990}:
\begin{equation}
  (X Y)_{j-1/2} = \frac{1}{2} \left[ (X Y)_j + (X Y)_{j-1} \right] \, . \label{eqn:product_interp}
\end{equation}

\begin{figure}[htbp]
  \centering
  \includegraphics[width=\textwidth]{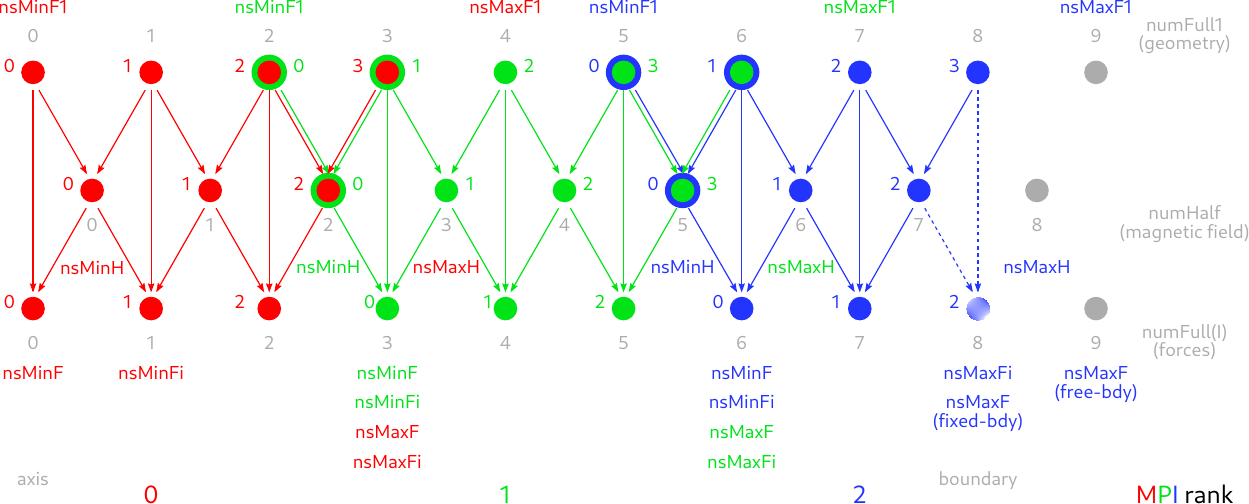}
  \caption{Radial partitioning scheme used in VMEC and associated MPI partitioning (colors).
           The particular example shown is for 3 MPI ranks and $\texttt{ns}=9$ flux surfaces.}
  \label{fig:vmec_radial_grids}
\end{figure}

\FloatBarrier
\section{Radial Profiles}
The distribution of flux surfaces along the radial coordinate is parameterized in VMEC via the $a^\phi$ user input polynomial,
which determines the mapping from the normalized radial coordinate $s$ (on which the flux surfaces are distributed in equal increments) to the normalized toroidal flux~$\phi$.
Thus, $a^\phi$ does not carry any physical unit.
It is specified using $n$ coefficients $a^\phi_0$, ..., $a^\phi_{n-1}$ (\code{aphi} in the input namelist).
The evaluation of the derivative of the polynomial $\phi'(s)$ at a radial location $s$ is performed using a recurrence relation starting at $i=n-1$:
\begin{equation}
  {\phi'}_{(i-1)}(s)
    \leftarrow \begin{cases}
                 n \,a^\phi_{n-1}                        &\textrm{: } i = {n-1} \\
                 s {\phi'}_{(i)}(s) + (i+1) \,a^\phi_{i} &\textrm{: else}
               \end{cases}
\end{equation}
where ${\phi'}_{(i-1)}(s)$ denotes the value of $\phi'(s)$ in the $i$-th iteration and the recurrence stops at $i=0$.
Thus, $a^\phi(s)$ specifies $\phi(s)$, but the first coefficient $a^\phi_0$ is for the term linear in $s$:
\begin{align}
              \phi(s)  =&\, a^\phi_0 s + a^\phi_1 s^2 + ... \\
  \Rightarrow \phi'(s) =&\, a^\phi_0 + 2 a^\phi_1 s + ... \, .
\end{align}
No offset is possible in $\phi(s)$ with this parameterization (and therefore always $\phi(0)=0$).
$\phi(s)$ gets computed by radially integrating ${\phi'}(s)$ using trapezoidal quadrature:
\begin{equation}
  \phi(s) = \int\limits_0^s {\phi'}(s') \,\mathrm{d}s'
    \approx \sum\limits_{i=0}^{N-1} {\phi'}({s_i}') \Delta {s_i}' \begin{cases}
                                                                    \half &\textrm{: } i=0 \textrm{ or } i=(N-1) \\
                                                                    1     &\textrm{: else}
                                                                  \end{cases}
\end{equation}
with $\Delta {s_i}' = s/N$, ${s_i}' = i \Delta {s_i}'$ and $N=100$ fixed.
The normalized poloidal flux (differential) is computed using the normalized toroidal flux (differential) and the user-provided rotational transform profile~$p_\iota(\phi_s)$.
Here is how to evaluate the normalized poloidal flux differential $\mathcal{X}'(s)$:
\begin{align}
           \phi_s =&\, \mathrm{min}\,(\phi(s, 1) \\
  \mathcal{X}'(s) =&\, p_\iota(\phi_s) {\phi'}(s) \, .
\end{align}
$\mathcal{X}(s)$ gets computed as well by radially integrating ${\mathcal{X}'}(s)$ using trapezoidal quadrature:
\begin{equation}
  \mathcal{X}(s) = \int\limits_0^s {\mathcal{X}'}(s') \,\mathrm{d}s'
    \approx \sum\limits_{i=0}^{N-1} {\mathcal{X}'}({s_i}') \Delta {s_i}' \begin{cases}
                                                                    \half &\textrm{: } i=0 \textrm{ or } i=(N-1) \\
                                                                    1     &\textrm{: else} \, .
                                                                  \end{cases}
\end{equation}
The toroidal flux enclosed by the LCFS is specified in Wb in the user input variable \code{phiedge} (denoted $\Phi_\mathrm{edge}$ here).
It is incorporated here as $\Phi_\mathrm{max}$:
\begin{equation}
  \Phi_\mathrm{max} = \frac{\code{signgs}}{2 \pi} \Phi_\mathrm{edge}
    \begin{cases}
      1/\phi(1) &\textrm{: } \phi(1) \neq 0 \\
      1         &\textrm{: else.}
    \end{cases}
\end{equation}
Note that $\phi(1)=0$ can happen, e.g., for $a^\phi_0=1$ and $a^\phi_1 = -1$.
Since $\phi(0)=0$ is fixed, $\phi(1)=0$ implies either $\phi(s)=0$ for all $s\in[0,1]$
or $\phi(s)=0$ is not monotone on $[0,1]$.
This occurs for example if an equilibrium is to be computed for a reversed-field pinch~\cite{2010_terranova, 2011_momo}.
For a Tokamak and a Stellarator, $\phi(1)=0$ is not a valid modus operandi.
The poloidal flux enclosed by the LCFS is denoted $\mathcal{X}_\mathrm{max}$:
\begin{equation}
  \mathcal{X}_\mathrm{max} = \Phi_\mathrm{max}
    \begin{cases}
      1/\mathcal{X}(1) &\textrm{: } \mathcal{X}(1) \neq 0 \\
      1                &\textrm{: else.}
    \end{cases}
\end{equation}
With the introduction of $a^\phi$, we now have three levels of parameterizations for the radial coordinate in terms of the toroidal flux:
\begin{enumerate}
  \item $s$ is the lowest level, where an equally-spaced grid ranges from $0$ to $1$ in $n_s$ steps (and the half-grid at points in between). \\
  \item $\phi(s)$ is the normalized toroidal flux from $\phi(0)=0$ at the axis to $\phi(1)$ at the LCFS.
        This profile is parameterized by the $a^\phi$ polynomial. It does not have a physical unit.
        Its value is (in the use further down) limited to $1$ and then
        used to evaluate the radial profile functions for mass, rotational transform and net toroidal current. \\
  \item $\Phi(s)$ is the physical toroidal magnetic flux enclosed in a given flux surface.
        For a Stellarator or a Tokamak we have:
        \begin{equation}
          \Phi(s) = \Phi(\phi(s)) = \frac{\code{signgs}}{2 \pi} \Phi_\mathrm{edge} \frac{\phi(s)}{\phi(1)} \, .
        \end{equation}
\end{enumerate}
The profiles of $\Phi'$, $\chi'$ and $\iota$ as well as toroidal current $I^\zeta$ are initialized on the half-grid as follows:
\begin{align}
  \Phi'_{j-\half} =&\, \Phi_\mathrm{max} \phi'(s_{j-\half}) \\
  \chi'_{j-\half} =&\, \Phi_\mathrm{max} \mathcal{X}'(s_{j-\half}) \\
  \iota_{j-\half} =&\, p_\iota    (\phi_{j-\half}) \\
I^\zeta_{j-\half} =&\, p_{I^\zeta}(\phi_{j-\half})
\end{align}
with $\phi_{j-\half} = \mathrm{min}\,(\phi(s_{j-\half}, 1)$ and $j \in \{ 1, 2, ..., (n_s-1) \}$.
The toroidal flux differential is extrapolated to the first half-grid position outside the LCFS:
\begin{equation}
  \phi_{n_s+\half} = 2 \phi_{n_s-\half} - \phi_{n_s-\nicefrac{3}{2}} \, .
\end{equation}
A scaling factor for $\lambda$ is derived from the toroidal flux differential profile:
\begin{equation}
  \lambda_\mathrm{scale} = \sqrt{\frac{1}{n_s-1} \sum\limits_{j=1}^{n_s-1} \left( \Phi'_{j-\half} \right)^2} \, . \label{eqn:lamscale}
\end{equation}
This scaling factor must not be zero, which would otherwise imply $\Phi'(s)=0$ everywhere.
It is noted that this quantity (called \texttt{lamscale} in the code)
is the RMS value of the radial profile of $\Phi'$.
The geometry of the LCFS is checked for compatibility with the assumed sign of the Jacobian compiled into VMEC (\code{signgs} or \code{isigng}).
If the sign of the $\theta$ coordinate needs to be flipped, this change also has to be incorporated into the profiles of $\iota$ and $\chi$:
\begin{align}
  \iota_{j-\half} \leftarrow&\, -\iota_{j-\half} \\
  \chi'_{j-\half} \leftarrow&\, -\chi'_{j-\half} \, .
\end{align}
The profiles of $\Phi'$, $\chi'$ and $\iota$ are initialized on the full-grid as follows:
\begin{align}
  \Phi'_{j} =&\, \Phi_\mathrm{max} \phi'(s_{j}) \\
  \chi'_{j} =&\, \Phi_\mathrm{max} \mathcal{X}'(s_{j}) \\
  \iota_{j} =&\, p_\iota    (\phi_{j})
\end{align}
with $\phi_{j} = \mathrm{min}\,(\phi(s_{j}, 1)$ and $j \in \{ 0, 1, ..., (n_s-1) \}$.

Note that the user-provided profiles for mass, rotational transform and net toroidal current $p_m$, $p_\iota$ and $p_{I^\zeta}$, respectively,
are always evaluated at a normalized toroidal flux coordinate ranging from $0$ at the magnetic axis to $1$ at the LCFS.
Thus, the re-distribution of flux surfaces using the $a^\phi$ polynomial
can be performed without changing the assumed coordinate of the prescribed profiles.

The Jacobian $\sqrt{g}$ is available on the radial half-grid.
This allows to compute the differential volume profile $V'(s)$
where $V$ is the volume enclosed by a given flux surface and the prime denotes a derivative with respect to $s$ (see Eqn.~(4.9.10) in Ref.~\cite{dHaseleer}):
\begin{equation}
  V'(s) = \frac{\mathrm{d}V}{\mathrm{d}s}(s) = \int\limits_0^{2 \pi} \int\limits_0^{2 \pi} \sqrt{g}(s, \theta, \zeta) \,\mathrm{d}\theta \,\mathrm{d}\zeta \, .
\end{equation}
In VMEC, the sign of the Jacobian $\sqrt{g}$ is usually negative and thus, the absolute value of $\sqrt{g}$ should be taken
for computing the (differential) plasma volume; $\sqrt{g}$ is to be regarded as a symbol and throughts should be prevented to wander into considering imaginary volumes.
In VMEC, the differential volume is computed on the half-grid as follows:
\begin{align}
        ~&\, \frac{1}{(2 \pi)^2} V'(s_{j-\half}) = \code{vp}(s_{j-\half}) \nonumber \\
        =&\, \frac{1}{(2 \pi)^2}\int\limits_0^{2 \pi} \int\limits_0^{2 \pi} |\sqrt{g}(s_{j-\half}, \theta, \zeta)| \,\mathrm{d}\theta \,\mathrm{d}\zeta \nonumber \\
  \approx&\, \frac{1}{(2 \pi)^2}\sum\limits_{k=0}^{n_\theta - 1} \sum\limits_{l=0}^{n_\zeta - 1} |\sqrt{g}(s_{j-\half}, \theta_k, \zeta_l)| \Delta \theta \Delta \zeta \nonumber \\
        =&\, \frac{1}{\bcancel{(2 \pi)^2}} \frac{\bcancel{2 \pi}}{n_\theta} \frac{\bcancel{2 \pi}}{n_\zeta}
               \sum\limits_{k=0}^{n_\theta - 1} \sum\limits_{l=0}^{n_\zeta - 1} |\sqrt{g}(s_{j-\half}, \theta_k, \zeta_l)| \nonumber \\
        =&\, \frac{1}{n_\theta n_\zeta} \sum\limits_{k=0}^{n_\theta - 1} \sum\limits_{l=0}^{n_\zeta - 1} |\sqrt{g}(s_{j-\half}, \theta_k, \zeta_l)|
\end{align}
with the following discretization in the tangential coordinates:
\begin{align}
  \Delta \theta =&\, 2 \pi / n_\theta \\
  \Delta \zeta  =&\, 2 \pi / n_\zeta  \\
  \theta_k      =&\, k \Delta \theta \\
  \zeta_l       =&\, l \Delta \zeta \, .
\end{align}
The total plasma volume is computed as follows:
\begin{align}
  V       =&\, \int\limits_0^1 V'(s) \,\mathrm{d}s \\
    \approx&\, \frac{(2 \pi)^2}{n_s-1} \sum\limits_{j=1}^{n_s - 1} \frac{V'(s_{j-\half})}{(2 \pi)^2} \\
          =&\, \frac{(2 \pi)^2}{n_s-1} \sum\limits_{j=1}^{n_s - 1} \code{vp}(s_{j-\half}) \, . \label{eqn:voli}
\end{align}
In the first iteration, the initial plasma volume is computed using~\eqn{voli} and stored in~\code{voli}.
The mass profile $m(s)$ is specified in the user input as a parameterized function $\code{pmass}(s)$,
a scaling factor \code{pres\_scale} and the adiabatic index $\gamma$:
\begin{equation}
  \mu_0 m(s_{j-\half}) = \code{mass}((s_{j-\half}))
   = \mu_0 \cdot \code{pres\_scale} \cdot \code{pmass}(s_{j-\half}) \cdot \left( |\code{vpnorm}| \cdot \code{r00} \right)^\gamma
\end{equation}
where
\begin{equation}
  \code{vpnorm} = \code{signgs} \cdot \frac{\Phi_\mathrm{LCFS}}{2 \pi} \cdot \Phi'(s_{j-\half})
\end{equation}
and $\code{r00}$ is the $R_{00}$ component of the LCFS as specified in the user input.
Given the mass profile $m(s)$, the pressure profile $p(s)$ is computed on the half-grid:
\begin{align}
  \mu_0 p(s_{j-\half}) = \code{pres}(s_{j-\half})
   =&\, \frac{\mu_0 m(s_{j-\half})}{\left( V'(s_{j-\half}) / (2 \pi)^2 \right)^\gamma} \nonumber \\
   =&\, \frac{\code{mass}(s_{j-\half})}{\left(\code{vp}(s_{j-\half})\right)^\gamma} \, .
\end{align}
For $\gamma = 0$, the pressure profile is prescribed:
\begin{equation}
  \mu_0 p(s_{j-\half}) = \mu_0 m(s_{j-\half}) = \mu_0 \cdot \code{pres\_scale} \cdot \code{pmass}(s_{j-\half}) \, .
\end{equation}
Two typical use cases are found:
\begin{itemize}
  \item $\code{pmass}(s)$ specified a normalized pressure profile with $p(0) = 1$ at the magnetic axis and $p(1) = 0$ at the LCFS.
        \code{pres\_scale} then scales this profile to a given physical pressure, e.g., 10\,kPa.
  \item $\code{pmass}(s)$ specified a physical pressure profile in Pa, e.g., from measurements and
        \code{pres\_scale} scales this profile by a small factor (say $0.5 \leq \code{pres\_scale} \leq 2$)
        to study the influence of a global change in plasma pressure on some subsequent physical quantity
        obtained from the equilibrium output.
\end{itemize}

\FloatBarrier
\section{Real-Space Grid Setup}
For up/down-symmetric two-dimensional configurations
and for Stellarator-symmetric three-dimensional configurations,
use can be made of the parity of the basis functions
to reduce the computational work by a factor of ca.~2.
This is done by computing quantities only on half the poloidal domain.
The choice to make use of parity in the poloidal direction
vs. the toroidal direction is gouverned by the fact
that this way, also axisymmetric configurations can make use of this symmetry.

\begin{figure}[htbp]
  \centering
  \includegraphics{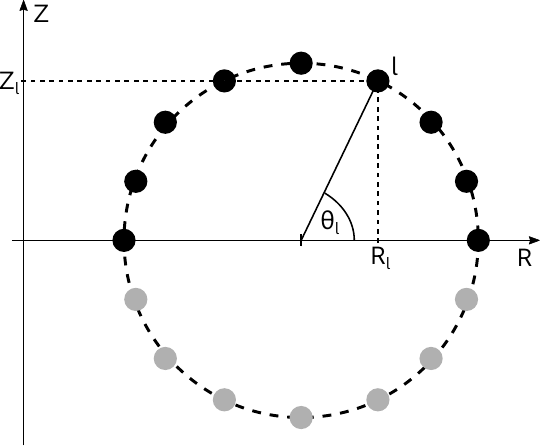}
  \caption{Poloidal symmetry can be used to reduce the grid from the full $[0, 2 \pi[$ interval
  (black and grey dots) to only the symmetric half $[0, \pi]$ (black dots only).
  The $l$-th poloidal grid point at $(R_l,Z_l)$ is indicated with its poloidal coordinate~$\theta_l$.
  Here, $n_{\theta,1} = 16$ and $n_{\theta,2} = 9$.}
  \label{fig:poloidal_symmetry}
\end{figure}

The number of poloidal grid points needs to be even for this to work out.
The user-prescribed number of poloidal grid points, \texttt{ntheta},
is rounded down to the nearest even integer value:
\begin{equation}
  n_{\theta, 1} = 2 \left\lfloor \frac{\texttt{ntheta}}{2} \right\rfloor \, .
\end{equation}
If \texttt{ntheta} is not given in the input file,
the minimum value required to support the prescribed number of poloidal modes \texttt{mpol} is used.
In case of a symmetric computation,
the poloidal grid is reduced to $n_{\theta,2}$ grid points with
\begin{equation}
  n_{\theta,2} = \frac{n_{\theta, 1}}{2} + 1 \, .
\end{equation}
This is illustrated in Fig.~\ref{fig:poloidal_symmetry}.
If the full poloidal range is used and thus
\begin{equation}
  \theta_l = \frac{2 \pi}{n_{\theta,1}} l \textrm{ for } l = 0, 1, ..., (n_{\theta,1}-1) \, .
\end{equation}
In case use is made of symmetry, this leads to
\begin{align}
    \theta_l = \frac{2 \pi}{n_{\theta,1}} l
  =&\, \frac{\bcancel{2} \pi}{\bcancel{2}(n_{\theta,2}-1)} l \nonumber \\
  =&\, \frac{\pi}{n_{\theta,2}-1} l \textrm{ for } l = 0, 1, ..., (n_{\theta,2}-1) \, .
\end{align}
The equations to be solved are formulated using~$\phi$ as the toroidal coordinate.
Assuming discrete toroidal translation invariance over $n_\mathrm{fp}$~field periods,
quantities only need to be evaluated on the interval $\phi \in [0, 2\pi/n_\mathrm{fp}[$.
This is equivalent to evaluating them in terms of the toroidal angle per module,~$\zeta$,
on the interval~$\zeta \in [0, 2\pi[$.
Note however that toroidal derivatives are still to be taken wrt.~$\phi$
and thus a factor $n \cdot n_\mathrm{fp}$ appears in front of the Fourier coefficients.
The number of toroidal grid points, $n_\zeta$, to be used in the computation
is specified in the input variable~\texttt{nzeta}.
The discrete grid in toroidal direction consists of $n_\zeta$ points:
\begin{equation}
  \zeta_k = \frac{2 \pi}{n_\zeta} k \textrm{ for } k = 0, 1, ..., (n_\zeta-1) \, .
\end{equation}
This corresponds to a grid over one field period in the cylindrical coordinate~$\phi$:
\begin{equation}
  \phi_k = \frac{2 \pi}{n_\mathrm{fp} n_\zeta} k \textrm{ for } k = 0, 1, ..., (n_\zeta-1) \, ,
\end{equation}
leading to~$\phi_k \in [0, 2 \pi / n_\mathrm{fp}[$.

In a nutshell:
The basis function argument is~$(m \theta - n n_\mathrm{fp} \phi)$.
Evaluation of functions and their tangential derivatives is only ever done in the first toroidal period:
$\phi \in [0, 2\pi/n_\mathrm{fp}[ \Rightarrow \zeta \in [0, 2\pi[$.
Therefore, the argument $(m \theta - n \zeta)$ is used in the evaluation of the Fourier basis functions.
Toroidal derivatives are to be taken with respect to $\phi$.
The Fourier representation allows to do this analytically and a factor~$-n n_\mathrm{fp}$ comes out of the basis function.

\FloatBarrier
\section{Fourier Representation} \label{sec:vmec_fourier_transforms}
\begin{figure}[htbp]
  \centering
  \includegraphics{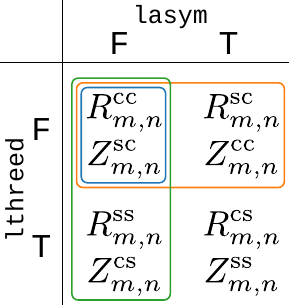}
  \caption{Coefficients to take into account for a given set of symmetry flags $\texttt{lasym}$ and $\texttt{lasym}$.
  The blue   box (top left corner) contains the coefficients required for  symmetric   two-dimensional configurations.
  The orange box (top row)         contains the coefficients required for asymmetric   two-dimensional configurations.
  The green  box (left column)     contains the coefficients required for  symmetric three-dimensional configurations.
  All coefficients                                       are required for asymmetric three-dimensional configurations.
  $R$ can be seen as a placeholder for all quantities with cosine parity.
  $Z$ can be seen as a placeholder for all quantities with   sine parity.}
  \label{fig:4_term_DFT_coeffs}
\end{figure}

The quantites $R$, $Z$ and $\lambda$ are the state variables in VMEC that describe the solution.
These quantites are represented by two-dimensional Fourier series in $\theta$ and $\zeta$ on each of the flux surfaces.
VMEC is designed to allow computation of the MHD equilibrium for both symmetric and asymmetric two- and three-dimensional configurations.
This freedom comes at a cost.
Multi-dimensional Fourier transforms with a complex-valued Fourier basis
can easily be evaluated as a separable product over the dimensions:
\begin{equation}
  \exp(i (m \theta - n \zeta)) = \exp(i m \theta) \exp(i n \zeta) \, .
\end{equation}
The real-valued two-dimensional Fourier basis functions used in VMEC
are split up into two contributing separable products of basis functions to be added up:
\begin{align}
  \cos(m \theta - n \zeta) =& \cos(m \theta) \cos(n \zeta) + \sin(m \theta) \sin(n \zeta) \\
  \sin(m \theta - n \zeta) =& \sin(m \theta) \cos(n \zeta) - \cos(m \theta) \sin(n \zeta) \, .
\end{align}
These expressions are inserted into the Fourier representations for $R$, $Z$ and $\lambda$.
Therefore, consider the following for $X \in \{R, Z, \lambda\}$:
\begin{align}
     X(\theta, \zeta)
  =&          \sum_{m=0}^M \sum_{n=-N}^N \left[   \hat{X}_{m,n}^\mathrm{cos} \cos(m \theta - n \zeta)
                                                + \hat{X}_{m,n}^\mathrm{sin} \sin(m \theta - n \zeta) \right] \nonumber \\
  =&          \sum_{m=0}^M \sum_{n=-N}^N \Biggl[ \phantom{+}\, \hat{X}_{m,n}^\mathrm{cos} \Bigl( \cos(m \theta) \cos(n \zeta) + \sin(m \theta) \sin(n \zeta) \Bigr) \nonumber \\
  ~& \phantom{\sum_{m=0}^M \sum_{n=-N}^N \Biggl[}\,       +    \hat{X}_{m,n}^\mathrm{sin} \Bigl( \sin(m \theta) \cos(n \zeta) - \cos(m \theta) \sin(n \zeta) \Bigr) \Biggr] \, .
\end{align}
The range of toroidal mode numbers~$n$ required to describe the solution
can be limited to non-negative values by taking the parity of the Fourier basis functions into account:
\begin{align}
  X(\theta, \zeta)
  =& \sum_{m=0}^M \sum_{n=0}^N \Biggl\{
     \cos(m \theta) \Biggl[ \phantom{-}\,
                            \left(   \hat{X}_{m,n}^\mathrm{cos} \cos(n \zeta)
                                  + \begin{cases}
                                     0                                          &: n = 0 \\
                                     \hat{X}_{m,-n}^\mathrm{cos} \cos(-n \zeta) &: n > 0
                                   \end{cases} \Biggr\} \right) \nonumber \\
  ~& \phantom{\sum_{m=0}^M \sum_{n=0}^N \Biggl\{}\,
     \phantom{\cos(m \theta) \Biggl[}\,
                           -\left(    \hat{X}_{m,n}^\mathrm{sin} \sin(n \zeta)
                                  + \begin{cases}
                                      0                                          & : n = 0 \\
                                      \hat{X}_{m,-n}^\mathrm{sin} \sin(-n \zeta) & : n > 0
                                    \end{cases} \Biggr\} \right) \Biggr] \nonumber \\
  ~& \phantom{\sum_{m=0}^M \sum_{n=0}^N \Biggl\{}\,
     \sin(m \theta) \Biggl[ \phantom{+}\,
                            \left(   \hat{X}_{m,n}^\mathrm{cos} \sin(n \zeta)
                                  + \begin{cases}
                                     0                                          & : n = 0 \\
                                     \hat{X}_{m,-n}^\mathrm{cos} \sin(-n \zeta) & : n > 0
                                   \end{cases} \Biggr\} \right) \nonumber \\
  ~& \phantom{\sum_{m=0}^M \sum_{n=0}^N \Biggl\{}\,
     \phantom{\sin(m \theta) \Biggl[}\,
                           +\left(    \hat{X}_{m,n}^\mathrm{sin} \cos(n \zeta)
                                  + \begin{cases}
                                      0                                          & : n = 0 \\
                                      \hat{X}_{m,-n}^\mathrm{sin} \cos(-n \zeta) & : n > 0
                                    \end{cases} \Biggr\} \right) \Biggr] \Biggr\}
\end{align}
Collecting terms, we end up with:
\begin{align}
  X(\theta, \zeta)
  =&          \sum_{m=0}^M \sum_{n=0}^N \,\Biggl[ \phantom{+}\, \hat{X}_{m,n}^\mathrm{cc} \cos(m \theta) \cos(n \zeta)
                                                           +    \hat{X}_{m,n}^\mathrm{ss} \sin(m \theta) \sin(n \zeta) \nonumber \\
  ~& \phantom{\sum_{m=0}^M \sum_{n=0}^N \,\Biggl[}\,       +    \hat{X}_{m,n}^\mathrm{sc} \sin(m \theta) \cos(n \zeta)
                                                           +    \hat{X}_{m,n}^\mathrm{cs} \cos(m \theta) \sin(n \zeta) \Biggr] \label{eqn:four_term_dft}
\end{align}
with
\begin{align}
  \hat{X}_{m,n}^\mathrm{cc}
  =& \begin{cases}
       \hat{X}_{m,n}^\mathrm{cos}                               &: n = 0 \\
       \hat{X}_{m,n}^\mathrm{cos} + \hat{X}_{m,-n}^\mathrm{cos} &: n > 0
     \end{cases} \\
  \hat{X}_{m,n}^\mathrm{ss}
  =& \begin{cases}
       0                                                        &: n = 0 \textrm{ or } m = 0 \\
       \hat{X}_{m,n}^\mathrm{cos} - \hat{X}_{m,-n}^\mathrm{cos} &: n > 0, m > 0
     \end{cases} \\
  \hat{X}_{m,n}^\mathrm{sc}
  =& \begin{cases}
       0                                                        &: \phantom{n = 0,\,} m = 0 \\
       \hat{X}_{m,n}^\mathrm{sin}                               &: n = 0, m > 0 \\
       \hat{X}_{m,n}^\mathrm{sin} + \hat{X}_{m,-n}^\mathrm{sin} &: n > 0, m > 0
     \end{cases} \\
  \hat{X}_{m,n}^\mathrm{cs}
  =& \begin{cases}
       0                                                         &: n = 0 \\
       -\hat{X}_{m,n}^\mathrm{sin} + \hat{X}_{m,-n}^\mathrm{sin} &: n > 0
     \end{cases} \, .
\end{align}
In the following, only $R$ and $Z$ are considered,
since $R$ has cosine parity and $Z$ and $\lambda$ both have sine parity.
Thus, the symmetry properties of $Z$ are directly applicable
to the Fourier representation of $\lambda$ as well.
In case of an up-down symmetric two-dimensional (symmetric Tokamak) configuration,
only terms from $\hat{R}^\mathrm{cos}$ and $\hat{Z}^\mathrm{sin}$ with $n=0$ contribute:
\begin{align}
  R(\theta) =& \sum_{m=0}^M \hat{R}^\mathrm{cc}_m \cos(m \theta) \\
  Z(\theta) =& \sum_{m=1}^M \hat{Z}^\mathrm{sc}_m \sin(m \theta) \, .
\end{align}
In case of an up-down-asymmetric two-dimensional (asymmetric Tokamak) configuration,
all terms with $n=0$ contribute:
\begin{align}
  R(\theta) =& \sum_{m=0}^M \left[ \hat{R}^\mathrm{cc}_m \cos(m \theta) + \hat{R}^\mathrm{sc}_m \sin(m \theta) \right] \\
  Z(\theta) =& \sum_{m=1}^M \left[ \hat{Z}^\mathrm{cc}_m \cos(m \theta) + \hat{Z}^\mathrm{sc}_m \sin(m \theta) \right] \, .
\end{align}
In case of a symmetric three-dimensional (Stellarator-symmetric) configuration,
only $\hat{R}^\mathrm{cos}$ and $\hat{Z}^\mathrm{sin}$ contribute:
\begin{align}
  R(\theta, \zeta) =& \sum_{m=0}^M \sum_{n=0}^N \left[  \hat{R}^\mathrm{cc}_{m,n} \cos(m \theta) \cos(n \zeta)
                                                      + \hat{R}^\mathrm{ss}_{m,n} \sin(m \theta) \sin(n \zeta) \right] \\
  Z(\theta, \zeta) =& \sum_{m=0}^M \sum_{n=0}^N \left[  \hat{Z}^\mathrm{sc}_{m,n} \sin(m \theta) \cos(n \zeta)
                                                      + \hat{Z}^\mathrm{cs}_{m,n} \cos(m \theta) \sin(n \zeta) \right] \, .
\end{align}
The full set of Fourier coefficients in~\eqn{four_term_dft} is only required
for an asymmetric three-dimensional configuration.
The number of coefficient types $\texttt{ntmax}$ that contribute to the state quantities is given by
\begin{equation}
  \texttt{ntmax} = 2^{\texttt{ithreed} + \texttt{iasym}}
\end{equation}
where $\texttt{ithreed}=1$ for a three-dimensional configuration and $\texttt{ithreed}=0$ otherwise
and $\texttt{iasym}=1$ in case of an up-down asymmetric configuration (at $\zeta=0$) and $\texttt{iasym}=0$ else.
The multi-dimensional Fourier transforms implemented in FFTW only cover separable products of transforms along the individual dimensions.
This is compatible with the decomposition of the two-dimensional Fourier series for the flux surface geometry
into real-even and real-odd transforms in VMEC.
The coefficients to take into account for a given set of symmetry flags
is depicted in Fig.~\ref{fig:4_term_DFT_coeffs}.

\begin{figure}[h]
  \centering
  \begin{minipage}[b]{0.49\textwidth}
    \includegraphics{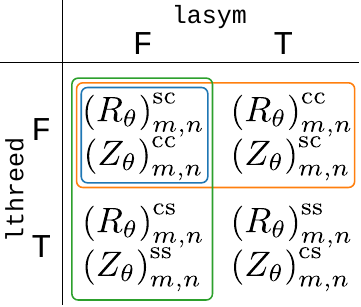}
  \end{minipage}
  \hfill
  \begin{minipage}[b]{0.49\textwidth}
    \includegraphics{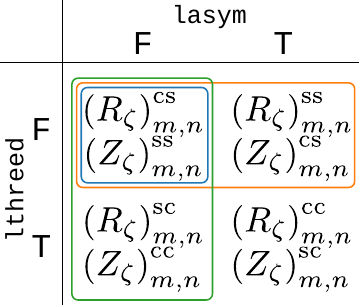}
  \end{minipage}
  \caption{Coefficients to take into account for a given set of symmetry flags $\texttt{lasym}$ and $\texttt{lasym}$
  when evaluating the tangential derivatives in poloidal (left panel) and toroidal (right panel) directions.
  The blue   box (top left corner) contains the coefficients required for  symmetric   two-dimensional configurations.
  The orange box (top row)         contains the coefficients required for asymmetric   two-dimensional configurations.
  The green  box (left column)     contains the coefficients required for  symmetric three-dimensional configurations.
  All coefficients                                       are required for asymmetric three-dimensional configurations.}
  \label{fig:4_term_DFT_derivatives}
\end{figure}

The tangential derivatives of the state variables in VMEC are computed analytically
as laid out in Sec.~\ref{sec:tangential_derivatives}.
For the four-term separable-product Fourier basis described above,
the tangential derivatives are computed as follows.
The poloidal derivative $U(\theta,\zeta)$ of a state variable $X$ is given by:
\begin{align}
  U(\theta,\zeta) = \frac{\partial X}{\partial \theta} (\theta, \zeta)
  =&          \sum_{m=0}^M \sum_{n=0}^N \Biggl[    - m \hat{X}_{m,n}^\mathrm{cc} \sin(m \theta) \cos(n \zeta)
                                                   + m \hat{X}_{m,n}^\mathrm{ss} \cos(m \theta) \sin(n \zeta) \nonumber \\
  ~& \phantom{\sum_{m=0}^M \sum_{n=0}^N \Biggl[}\, + m \hat{X}_{m,n}^\mathrm{sc} \cos(m \theta) \cos(n \zeta)
                                                   - m \hat{X}_{m,n}^\mathrm{cs} \sin(m \theta) \sin(n \zeta) \Biggr]
\end{align}
which can be reformulated as
\begin{align}
  U(\theta,\zeta)
  =&          \sum_{m=0}^M \sum_{n=0}^N \Biggl[    \phantom{+}\, \hat{U}_{m,n}^\mathrm{cc} \cos(m \theta) \cos(n \zeta)
                                                            +    \hat{U}_{m,n}^\mathrm{ss} \sin(m \theta) \sin(n \zeta) \nonumber \\
  ~& \phantom{\sum_{m=0}^M \sum_{n=0}^N \Biggl[}\,          +    \hat{U}_{m,n}^\mathrm{sc} \sin(m \theta) \cos(n \zeta)
                                                            +    \hat{U}_{m,n}^\mathrm{cs} \cos(m \theta) \sin(n \zeta) \Biggr]
\end{align}
with
\begin{align}
  \hat{U}_{m,n}^\mathrm{cc} &= + m \hat{X}_{m,n}^\mathrm{sc} \\
  \hat{U}_{m,n}^\mathrm{ss} &= - m \hat{X}_{m,n}^\mathrm{cs} \\
  \hat{U}_{m,n}^\mathrm{sc} &= - m \hat{X}_{m,n}^\mathrm{cc} \\
  \hat{U}_{m,n}^\mathrm{cs} &= + m \hat{X}_{m,n}^\mathrm{ss} \, .
\end{align}
The toroidal derivative $V(\theta,\zeta)$ of a state variable $X$ is given by:
\begin{align}
  ~&\, V(\theta,\zeta) = \frac{\partial X}{\partial \phi} \underbrace{\frac{\partial \phi}{\partial \zeta}}_{= n_\mathrm{fp}} (\theta, \zeta) \nonumber \\
  =&\,          \sum_{m=0}^M \sum_{n=0}^N \Biggl[    - n n_\mathrm{fp} \hat{X}_{m,n}^\mathrm{cc} \cos(m \theta) \sin(n \zeta)
                                                     + n n_\mathrm{fp} \hat{X}_{m,n}^\mathrm{ss} \sin(m \theta) \cos(n \zeta) \nonumber \\
  ~&\, \phantom{\sum_{m=0}^M \sum_{n=0}^N \Biggl[}\, - n n_\mathrm{fp} \hat{X}_{m,n}^\mathrm{sc} \sin(m \theta) \sin(n \zeta)
                                                     + n n_\mathrm{fp} \hat{X}_{m,n}^\mathrm{cs} \cos(m \theta) \cos(n \zeta) \Biggr]
\end{align}
which can be reformulated as
\begin{align}
  V(\theta,\zeta)
  =&          \sum_{m=0}^M \sum_{n=0}^N \Biggl[    \phantom{+}\, \hat{V}_{m,n}^\mathrm{cc} \cos(m \theta) \cos(n \zeta)
                                                            +    \hat{V}_{m,n}^\mathrm{ss} \sin(m \theta) \sin(n \zeta) \nonumber \\
  ~& \phantom{\sum_{m=0}^M \sum_{n=0}^N \Biggl[}\,          +    \hat{V}_{m,n}^\mathrm{sc} \sin(m \theta) \cos(n \zeta)
                                                            +    \hat{V}_{m,n}^\mathrm{cs} \cos(m \theta) \sin(n \zeta) \Biggr]
\end{align}
with
\begin{align}
  \hat{V}_{m,n}^\mathrm{cc} &= + n n_\mathrm{fp} \hat{X}_{m,n}^\mathrm{cs} \\
  \hat{V}_{m,n}^\mathrm{ss} &= - n n_\mathrm{fp} \hat{X}_{m,n}^\mathrm{sc} \\
  \hat{V}_{m,n}^\mathrm{sc} &= + n n_\mathrm{fp} \hat{X}_{m,n}^\mathrm{ss} \\
  \hat{V}_{m,n}^\mathrm{cs} &= - n n_\mathrm{fp} \hat{X}_{m,n}^\mathrm{cc} \, .
\end{align}
The symmetry properties of the tangential derivatives are shown in Fig.~\ref{fig:4_term_DFT_derivatives}.
Above inverse Fourier transforms are evaluated in the symmetric case only on half of the poloidal interval.
This gives the real-space representation of the quantities, which due to the definite-parity basis functions then also have a definite parity.
This is implemented in the Fortran implementation of VMEC in the subroutine~\texttt{totzsps}.
In the asymmetric case, the same routine is used to compute the contribution to the quantities
with that parity those quantities would have in the symmetric case.
Additionally, a second routine is used to compute the other-parity contributions to the quantities,
also only on half the poloidal interval. This is done in~\texttt{totzspa}.
These contributions are combined in the asymmetric case to yield the asymmetric quantites on the full interval.
This is a two-step process and it is implemented in~\texttt{symrzl}.
First, the ``extended'' interval~$]\pi,2\pi[$ needs to be filled,
since the original data (present in $[0,\pi]$) must not be overwritten.
Then, the first half of the interval ($[0,\pi]$) is overwritten in-place
with the combination of the symmetric and anti-symmetric contributions.

\begin{figure}[htbp]
  \centering
  \includegraphics{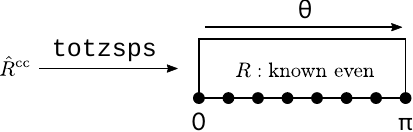}
  \caption{The real-space $R$~coordinate is known to have even parity in the symmetric case.
  It is computed from the Fourier coefficients~$\hat{R}^\mathrm{cc}$.}
  \label{fig:R_sym}
\end{figure}

\begin{figure}[htbp]
  \centering
  \includegraphics{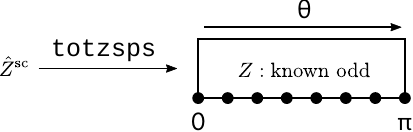}
  \caption{The real-space $Z$~coordinate is known to have odd parity in the symmetric case.
  It is computed from the Fourier coefficients~$\hat{Z}^\mathrm{sc}$.}
  \label{fig:Z_sym}
\end{figure}

\begin{figure}[htbp]
  \centering
  \includegraphics{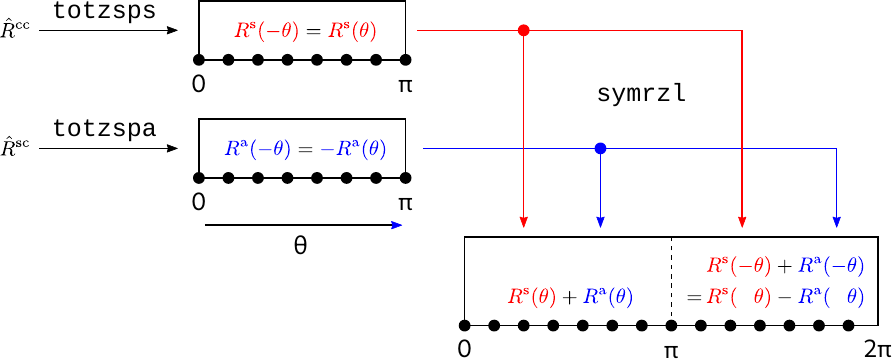}
  \caption{The real-space $R$~coordinate is constructed from its even-parity and odd-parity components in the general case.
  The even-parity component is the even~$R$ of the symmetric case, computed from the Fourier coefficients~$\hat{R}^\mathrm{cc}$.
  The  odd-parity component is computed additionally in the asymmetric case from the Fourier coefficients~$\hat{R}^\mathrm{sc}$.
  These components are combined using \texttt{symrzl} to yield the general~$R$ on the full poloidal domain.}
  \label{fig:R_asym}
\end{figure}

\begin{figure}[htbp]
  \centering
  \includegraphics{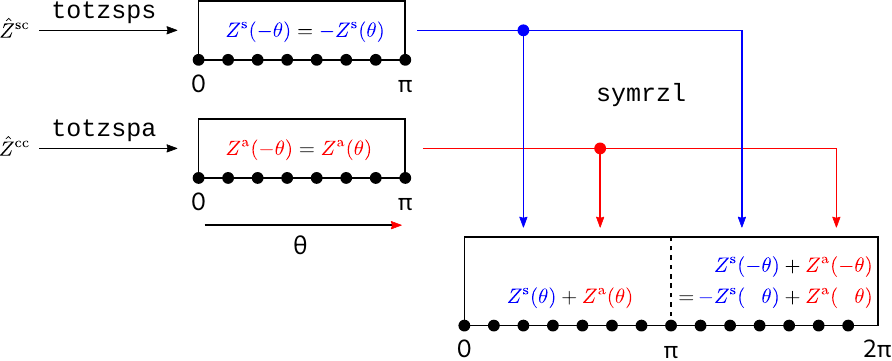}
  \caption{The real-space $Z$~coordinate is constructed from its even-parity and odd-parity components in the general case.
  The  odd-parity component is the odd~$Z$ of the symmetric case, computed from the Fourier coefficients~$\hat{Z}^\mathrm{sc}$.
  The even-parity component is computed additionally in the asymmetric case from the Fourier coefficients~$\hat{Z}^\mathrm{cc}$.
  These components are combined using \texttt{symrzl} to yield the general~$Z$ on the full poloidal domain.}
  \label{fig:Z_asym}
\end{figure}

\FloatBarrier
\section{Jacobian} \label{sec:jacobian}
The Jacobian $\sqrt{g}$ is computed on the radial half-grid.
The Jacobian is decomposed into a toroidal factor $R$
and the two-dimensional Jacobian in the poloidal plane $\tau$:
\begin{align}
                  \sqrt{g} =&\, R \tau \\
  \textrm{with} \quad \tau =&\, R_\theta Z_s - R_s Z_\theta \, .
\end{align}
This is Eqn.~(17) with $\tau = G$ in Ref.~\cite{hirshman_whitson_1983}.
The terms appearing in above expression must be combined from contributions due to even-$m$ and odd-$m$ Fourier harmonics~\cite{hirshman_schwenn_nuehrenberg_1990}:
\begin{equation}
  X(s) = X^\mathrm{e}(s) + \sqrt{s} X^\mathrm{o}(s) \label{eqn:even_odd_comb}
\end{equation}
for $X \in \{ R, Z, R_\theta, Z_\theta, ... \}$.
The geometry of the flux surfaces $R$, $Z$ and its tangential derivatives are available on the full grid.
The geometry and the tangential derivatives are interpolated/averaged onto intermediate half-grid points.
Radial derivatives are computed using finite differences over neighboring full-grid points.
This leads to the following for $R$ and for the tangential derivatives:
\begin{align}
  R(s_{j-\half})        =&\, \frac{1}{2} \left[ R(s_j) + R(s_{j-1}) \right] \\
                        =&\, \frac{1}{2} \left[                      \left( R^\mathrm{e}(s_j) + R^\mathrm{e}(s_{j-1}) \right)
                                                + \sqrt{s_{j-\half}} \left( R^\mathrm{o}(s_j) + R^\mathrm{o}(s_{j-1}) \right) \right] \label{eqn:r_half} \\
  R_\theta(s_{j-\half}) =&\, \frac{1}{2} \left[ R_\theta(s_j) + R_\theta(s_{j-1}) \right] \\
                        =&\, \frac{1}{2} \left[                      \left( R_\theta^\mathrm{e}(s_j) + R_\theta^\mathrm{e}(s_{j-1}) \right)
                                                + \sqrt{s_{j-\half}} \left( R_\theta^\mathrm{o}(s_j) + R_\theta^\mathrm{o}(s_{j-1}) \right) \right] \\
  Z_\theta(s_{j-\half}) =&\, \frac{1}{2} \left[ Z_\theta(s_j) + Z_\theta(s_{j-1}) \right] \\
                        =&\, \frac{1}{2} \left[                      \left( Z_\theta^\mathrm{e}(s_j) + Z_\theta^\mathrm{e}(s_{j-1}) \right)
                                                + \sqrt{s_{j-\half}} \left( Z_\theta^\mathrm{o}(s_j) + Z_\theta^\mathrm{o}(s_{j-1}) \right) \right] \, .
\end{align}
The chain rule needs to be taken into account when computing the radial derivatives (see Sec.~\ref{sec:radial_discretization}).
The radial derivatives can be decomposed into two parts:
\begin{align}
  R_s(s_{j-\half}) =&\, {\tilde{R_s}}(s_{j-\half}) + \frac{1}{2 \sqrt{s_{j-\half}}} R^\mathrm{o}(s_{j-\half}) \\
  Z_s(s_{j-\half}) =&\, {\tilde{Z_s}}(s_{j-\half}) + \frac{1}{2 \sqrt{s_{j-\half}}} Z^\mathrm{o}(s_{j-\half}) \, .
\end{align}
The first terms in above expression are computed using finite differences as follows:
\begin{align}
  {\tilde{R_s}}(s_{j-\half}) =&\, \frac{1}{\Delta s} \left[ R(s_j) - R(s_{j-1}) \right] \\
                             =&\, \frac{1}{\Delta s} \left[                      \left( R^\mathrm{e}(s_j) - R^\mathrm{e}(s_{j-1}) \right)
                                                            + \sqrt{s_{j-\half}} \left( R^\mathrm{o}(s_j) - R^\mathrm{o}(s_{j-1}) \right) \right] \\
  {\tilde{Z_s}}(s_{j-\half}) =&\, \frac{1}{\Delta s} \left[ Z(s_j) - Z(s_{j-1}) \right] \\
                             =&\, \frac{1}{\Delta s} \left[                      \left( Z^\mathrm{e}(s_j) - Z^\mathrm{e}(s_{j-1}) \right)
                                                            + \sqrt{s_{j-\half}} \left( Z^\mathrm{o}(s_j) - Z^\mathrm{o}(s_{j-1}) \right) \right] \, .
\end{align}
Inserting the radial derivatives into the expression for $\tau$, we arrive at:
\begin{align}
  \tau(s_{j-\half}) =&\, \phantom{-}\, R_\theta(s_{j-\half}) \left[ {\tilde{Z_s}}(s_{j-\half}) + \frac{1}{2 \sqrt{s_{j-\half}}} Z^\mathrm{o}(s_{j-\half}) \right] \nonumber \\
                    ~&\,          -    \left[ {\tilde{R_s}}(s_{j-\half}) + \frac{1}{2 \sqrt{s_{j-\half}}} R^\mathrm{o}(s_{j-\half}) \right] Z_\theta(s_{j-\half}) \nonumber \\
                    =&\,   R_\theta(s_{j-\half}) {\tilde{Z_s}}(s_{j-\half}) - {\tilde{R_s}}(s_{j-\half}) Z_\theta(s_{j-\half}) \nonumber \\
                    ~&\, + \frac{1}{2 \sqrt{s_{j-\half}}} \left[ R_\theta(s_{j-\half}) Z^\mathrm{o}(s_{j-\half}) - R^\mathrm{o}(s_{j-\half}) Z_\theta(s_{j-\half}) \right] \label{eqn:tau_almost}
\end{align}
The last term has to be expanded further:
\begin{align}
   R_\theta(s_{j-\half}) Z^\mathrm{o}(s_{j-\half})
  =&\, \left[ R_\theta^\mathrm{e}(s_{j-\half}) + \sqrt{s_{j-\half}} R_\theta^\mathrm{o}(s_{j-\half}) \right] Z^\mathrm{o}(s_{j-\half}) \nonumber \\
  =&\, R_\theta^\mathrm{e}(s_{j-\half}) Z^\mathrm{o}(s_{j-\half}) + \sqrt{s_{j-\half}} R_\theta^\mathrm{o}(s_{j-\half}) Z^\mathrm{o}(s_{j-\half}) \\
   R^\mathrm{o}(s_{j-\half}) Z_\theta(s_{j-\half})
  =&\, R^\mathrm{o}(s_{j-\half}) \left[ Z_\theta^\mathrm{e}(s_{j-\half}) + \sqrt{s_{j-\half}} Z_\theta^\mathrm{o}(s_{j-\half}) \right] \nonumber \\
  =&\, R^\mathrm{o}(s_{j-\half}) Z_\theta^\mathrm{e}(s_{j-\half}) + \sqrt{s_{j-\half}} R^\mathrm{o}(s_{j-\half}) Z_\theta^\mathrm{o}(s_{j-\half})
\end{align}
Inserting this into the expression for $\tau$ from \eqn{tau_almost} leads to:
\begin{align}
  \tau(s_{j-\half}) =&\,   R_\theta(s_{j-\half}) {\tilde{Z_s}}(s_{j-\half}) - {\tilde{R_s}}(s_{j-\half}) Z_\theta(s_{j-\half}) \nonumber \\
                    ~&\, + \frac{1}{2 \sqrt{s_{j-\half}}} \Bigl\{
                      \phantom{-}\,  R_\theta^\mathrm{e}(s_{j-\half}) Z^\mathrm{o}(s_{j-\half}) + \sqrt{s_{j-\half}} R_\theta^\mathrm{o}(s_{j-\half}) Z^\mathrm{o}(s_{j-\half}) \nonumber \\
                    ~&\, \phantom{+    \frac{1}{2 \sqrt{s_{j-\half}}} \Bigl\{}\,
                               - R^\mathrm{o}(s_{j-\half}) Z_\theta^\mathrm{e}(s_{j-\half}) - \sqrt{s_{j-\half}} R^\mathrm{o}(s_{j-\half}) Z_\theta^\mathrm{o}(s_{j-\half}) \Bigr\} \nonumber \\
                    =&\,   R_\theta(s_{j-\half}) {\tilde{Z_s}}(s_{j-\half}) - {\tilde{R_s}}(s_{j-\half}) Z_\theta(s_{j-\half}) \nonumber \\
                    ~&\, + \frac{1}{2 \sqrt{s_{j-\half}}} \left[ R_\theta^\mathrm{e}(s_{j-\half}) Z^\mathrm{o}(s_{j-\half}) - R^\mathrm{o}(s_{j-\half}) Z_\theta^\mathrm{e}(s_{j-\half}) \right] \nonumber \\
                    ~&\, + \frac{1}{2}                    \left[ R_\theta^\mathrm{o}(s_{j-\half}) Z^\mathrm{o}(s_{j-\half}) - R^\mathrm{o}(s_{j-\half}) Z_\theta^\mathrm{o}(s_{j-\half}) \right]
\end{align}
Here, the product interpolation rule~\eqn{product_interp} has to be employed:
\begin{align}
  R_\theta^\mathrm{e}(s_{j-\half}) Z^\mathrm{o}(s_{j-\half}) =&\, \frac{1}{2} \Bigl[ R_\theta^\mathrm{e}(s_{j}) Z^\mathrm{o}(s_{j}) + R_\theta^\mathrm{e}(s_{j-1}) Z^\mathrm{o}(s_{j-1}) \Bigr] \\
  R^\mathrm{o}(s_{j-\half}) Z_\theta^\mathrm{e}(s_{j-\half}) =&\, \frac{1}{2} \Bigl[ R^\mathrm{o}(s_{j}) Z_\theta^\mathrm{e}(s_{j}) + R^\mathrm{o}(s_{j-1}) Z_\theta^\mathrm{e}(s_{j-1}) \Bigr] \\
  R_\theta^\mathrm{o}(s_{j-\half}) Z^\mathrm{o}(s_{j-\half}) =&\, \frac{1}{2} \Bigl[ R_\theta^\mathrm{o}(s_{j}) Z^\mathrm{o}(s_{j}) + R_\theta^\mathrm{o}(s_{j-1}) Z^\mathrm{o}(s_{j-1}) \Bigr] \\
  R^\mathrm{o}(s_{j-\half}) Z_\theta^\mathrm{o}(s_{j-\half}) =&\, \frac{1}{2} \Bigl[ R^\mathrm{o}(s_{j}) Z_\theta^\mathrm{o}(s_{j}) + R^\mathrm{o}(s_{j-1}) Z_\theta^\mathrm{o}(s_{j-1}) \Bigr] \, .
\end{align}
Finally, $\tau$ is assembled as follows:
\begin{align}
  \tau(s_{j-\half}) =&\,   R_\theta(s_{j-\half}) {\tilde{Z_s}}(s_{j-\half}) - {\tilde{R_s}}(s_{j-\half}) Z_\theta(s_{j-\half}) \nonumber \\
                    ~&\, + \frac{1}{4 \sqrt{s_{j-\half}}} \Bigl[ \phantom{-}\, R_\theta^\mathrm{e}(s_{j}) Z^\mathrm{o}(s_{j}) + R_\theta^\mathrm{e}(s_{j-1}) Z^\mathrm{o}(s_{j-1}) \nonumber \\
           ~&\, \phantom{+ \frac{1}{4 \sqrt{s_{j-\half}}} \Bigl[}\,       -    R^\mathrm{o}(s_{j}) Z_\theta^\mathrm{e}(s_{j}) - R^\mathrm{o}(s_{j-1}) Z_\theta^\mathrm{e}(s_{j-1}) \Bigr] \nonumber \\
                    ~&\, + \frac{1}{4}                    \Bigl[ \phantom{-}\, R_\theta^\mathrm{o}(s_{j}) Z^\mathrm{o}(s_{j}) + R_\theta^\mathrm{o}(s_{j-1}) Z^\mathrm{o}(s_{j-1}) \nonumber \\
           ~&\, \phantom{+ \frac{1}{4}                    \Bigl[}\,       -    R^\mathrm{o}(s_{j}) Z_\theta^\mathrm{o}(s_{j}) - R^\mathrm{o}(s_{j-1}) Z_\theta^\mathrm{o}(s_{j-1}) \Bigr] \label{eqn:tau_final}
\end{align}
The Jacobian on the half-grid can then be assembled without further due:
\begin{equation}
  \sqrt{g}(s_{j-\half}) = R(s_{j-\half}) \tau(s_{j-\half})
\end{equation}
with $R(s_{j-\half})$ from \eqn{r_half} and $\tau(s_{j-\half})$ from \eqn{tau_final}.

\FloatBarrier
\section{Metric Elements} \label{sec:metric_elements}
The metric elements $g_{\alpha \beta}$ with $(\alpha, \beta) \in \{ \theta, \zeta \}$ are given by:
\begin{equation}
  g_{\alpha \beta} = R_\alpha R_\beta + Z_\alpha Z_\beta + \delta_{\alpha \zeta} \delta_{\beta \zeta} R^2 \, . \label{eqn:metric_elements_abstract}
\end{equation}
Concretely, this leads to~\cite{geiger_calc_current_in_prout}:
\begin{align}
  g_{\theta \theta} =&\, \left( \frac{\partial R}{\partial \theta} \right)^2 + \left( \frac{\partial Z}{\partial \theta} \right)^2 \\
  g_{\theta \zeta } =&\, \frac{\partial R}{\partial \theta} \frac{\partial R}{\partial \zeta} + \frac{\partial Z}{\partial \theta} \frac{\partial Z}{\partial \zeta} \\
  g_{\zeta  \zeta } =&\, \left( \frac{\partial R}{\partial \zeta } \right)^2 + \left( \frac{\partial Z}{\partial \zeta } \right)^2 + R^2 \, .
\end{align}
The individual terms in above expressions are composed of contributions from even-$m$ and odd-$m$ Fourier harmonics as noted in~\eqn{even_odd_comb}.
Inserting this decomposition into the abstract metric element expression from~\eqn{metric_elements_abstract} leads to:
\begin{align}
  g_{\alpha \beta} =&\, \phantom{+}\, \left( R_\alpha^\mathrm{e} + \sqrt{s} R_\alpha^\mathrm{o} \right) \left( R_\beta^\mathrm{e} + \sqrt{s} R_\beta^\mathrm{o} \right) \nonumber \\
                   ~&\,          +    \left( Z_\alpha^\mathrm{e} + \sqrt{s} Z_\alpha^\mathrm{o} \right) \left( Z_\beta^\mathrm{e} + \sqrt{s} Z_\beta^\mathrm{o} \right) \nonumber \\
                   ~&\,          +    \delta_{\alpha \zeta} \delta_{\beta \zeta} \left( R^\mathrm{e} + \sqrt{s} R^\mathrm{o} \right) \left( R^\mathrm{e} + \sqrt{s} R^\mathrm{o} \right) \nonumber \\
                   =&\, \phantom{+ \sqrt{s}}~ \left(   R_\alpha^\mathrm{e} R_\beta^\mathrm{e} + Z_\alpha^\mathrm{e} Z_\beta^\mathrm{e}
                                                     + \delta_{\alpha \zeta} \delta_{\beta \zeta} R^\mathrm{e} R^\mathrm{e} \right) \nonumber \\
                   ~&\,          + \sqrt{s}   \left(   R_\alpha^\mathrm{e} R_\beta^\mathrm{o} + R_\alpha^\mathrm{o} R_\beta^\mathrm{e}
                                                     + Z_\alpha^\mathrm{e} Z_\beta^\mathrm{o} + Z_\alpha^\mathrm{o} Z_\beta^\mathrm{e}
                                                     + 2 \delta_{\alpha \zeta} \delta_{\beta \zeta} R^\mathrm{e} R^\mathrm{o} \right) \nonumber \\
                   ~&\,          + ~~s~       \left(   R_\alpha^\mathrm{o} R_\beta^\mathrm{o} + Z_\alpha^\mathrm{o} Z_\beta^\mathrm{o}
                                                     + \delta_{\alpha \zeta} \delta_{\beta \zeta} R^\mathrm{o} R^\mathrm{o} \right) \, . \label{eqn:metric_elements_evenodd}
\end{align}
Direct evaluation of this intermediate result at $s_{j-\half}$ leads to Eqn.~(10) in Ref.~\cite{hirshman_schwenn_nuehrenberg_1990}
(up to the omitted term $\delta_{\alpha \zeta} \delta_{\beta \zeta} R^2$):
\begin{align}
  g_{\alpha \beta}^{j-\half} =&\, \phantom{+ \sqrt{s_{j-\half}}}~ \left(   R_\alpha^\mathrm{e} R_\beta^\mathrm{e} + Z_\alpha^\mathrm{e} Z_\beta^\mathrm{e}
                                                                         + \delta_{\alpha \zeta} \delta_{\beta \zeta} R^\mathrm{e} R^\mathrm{e} \right)^{j-\half} \nonumber \\
                             ~&\,          + \sqrt{s_{j-\half}}   \left(   R_\alpha^\mathrm{e} R_\beta^\mathrm{o} + R_\alpha^\mathrm{o} R_\beta^\mathrm{e}
                                                                         + Z_\alpha^\mathrm{e} Z_\beta^\mathrm{o} + Z_\alpha^\mathrm{o} Z_\beta^\mathrm{e}
                                                                         + 2 \delta_{\alpha \zeta} \delta_{\beta \zeta} R^\mathrm{e} R^\mathrm{o} \right)^{j-\half} \nonumber \\
                             ~&\,          + ~~s_{j-\half}~       \left(   R_\alpha^\mathrm{o} R_\beta^\mathrm{o} + Z_\alpha^\mathrm{o} Z_\beta^\mathrm{o}
                                                                         + \delta_{\alpha \zeta} \delta_{\beta \zeta} R^\mathrm{o} R^\mathrm{o} \right)^{j-\half} \, . \label{eqn:metric_elements_10}
\end{align}
What is implemented in VMEC is different.
It suggests the following interpretation.
Suppose the metric elements $g_{\alpha \beta}$ can also be decomposed into even-$m$ and odd-$m$ contributions:
\begin{equation}
  g_{\alpha \beta} = g_{\alpha \beta}^\mathrm{e} + \sqrt{s} \, g_{\alpha \beta}^\mathrm{o} \, .
\end{equation}
Then, interpolation onto the half-grid would be done separately:
\begin{equation}
  g_{\alpha \beta}^{j-\half} = g_{\alpha \beta}^{\mathrm{e}, j-\half} + \sqrt{s_{j-\half}} \, g_{\alpha \beta}^{\mathrm{o}, j-\half} \, .
\end{equation}
The odd-$m$ contributions to the metric elements are odd functions of the poloidal coordinate.
Thus, the radial regularization factors~$ \sqrt{s}$ need to be factored out prior to interpolation onto the half-grid.
However, in the metric elements, products of two odd-$m$-terms appear (see, e.g., the last group in \eqn{metric_elements_evenodd}).
The product of two odd functions is an even function again.
The interpolation of this function onto the half-grid therefore should happen without factoring out the $\sqrt{s}$ normalization factors.
However, all odd-$m$ terms in the current VMEC implementation have them factored in already.
Therefore, one needs to cancel the factor $s = \sqrt{s} \sqrt{s}$ (where each $\sqrt{s}$ stems from one of the odd-$m$ factors)
before performing the radial interpolation onto the half-grid.
Concretely, this leads to:
\begin{align}
  g_{\alpha \beta}^{j-\half} =&\, \left[     \left(   R_\alpha^\mathrm{e} R_\beta^\mathrm{e} + Z_\alpha^\mathrm{e} Z_\beta^\mathrm{e}
                                                    + \delta_{\alpha \zeta} \delta_{\beta \zeta} R^\mathrm{e} R^\mathrm{e} \right)
                                         + s \left(   R_\alpha^\mathrm{o} R_\beta^\mathrm{o} + Z_\alpha^\mathrm{o} Z_\beta^\mathrm{o}
                                                    + \delta_{\alpha \zeta} \delta_{\beta \zeta} R^\mathrm{o} R^\mathrm{o} \right) \right]^{j-\half} \nonumber \\
                             ~&\, + \sqrt{s_{j-\half}}   \left(   R_\alpha^\mathrm{e} R_\beta^\mathrm{o} + R_\alpha^\mathrm{o} R_\beta^\mathrm{e}
                                                                + Z_\alpha^\mathrm{e} Z_\beta^\mathrm{o} + Z_\alpha^\mathrm{o} Z_\beta^\mathrm{e}
                                                                + 2 \delta_{\alpha \zeta} \delta_{\beta \zeta} R^\mathrm{e} R^\mathrm{o} \right)^{j-\half} \, .
\end{align}
Application of the product interpolation rule~\eqn{product_interp} leads to:
\begin{align}
  g_{\alpha \beta}^{j-\half} =&\,          \frac{1}{2} \Biggl[    \phantom{+ s_{j-1}}~  \left(   R_\alpha^\mathrm{e} R_\beta^\mathrm{e} + Z_\alpha^\mathrm{e} Z_\beta^\mathrm{e}
                                                                                               + \delta_{\alpha \zeta} \delta_{\beta \zeta} R^\mathrm{e} R^\mathrm{e} \right)^j     \nonumber \\
                             ~&\, \phantom{\frac{1}{2} \Biggl[}\, + \phantom{s_{j-1}}   \left(   R_\alpha^\mathrm{e} R_\beta^\mathrm{e} + Z_\alpha^\mathrm{e} Z_\beta^\mathrm{e}
                                                                                               + \delta_{\alpha \zeta} \delta_{\beta \zeta} R^\mathrm{e} R^\mathrm{e} \right)^{j-1} \nonumber \\
                             ~&\, \phantom{\frac{1}{2} \Biggl[}\, +          s_j ~~\,   \left(   R_\alpha^\mathrm{o} R_\beta^\mathrm{o} + Z_\alpha^\mathrm{o} Z_\beta^\mathrm{o}
                                                                                               + \delta_{\alpha \zeta} \delta_{\beta \zeta} R^\mathrm{o} R^\mathrm{o} \right)^j     \nonumber \\
                             ~&\, \phantom{\frac{1}{2} \Biggl[}\, +          s_{j-1}    \left(   R_\alpha^\mathrm{o} R_\beta^\mathrm{o} + Z_\alpha^\mathrm{o} Z_\beta^\mathrm{o}
                                                                                               + \delta_{\alpha \zeta} \delta_{\beta \zeta} R^\mathrm{o} R^\mathrm{o} \right)^{j-1} \Biggr] \nonumber \\
                             ~&\,          + \frac{1}{2} \sqrt{s_{j-\half}} \Biggl[ \phantom{+}\, \left(   R_\alpha^\mathrm{e} R_\beta^\mathrm{o} + R_\alpha^\mathrm{o} R_\beta^\mathrm{e}
                                                                                                         + Z_\alpha^\mathrm{e} Z_\beta^\mathrm{o} + Z_\alpha^\mathrm{o} Z_\beta^\mathrm{e}
                                                                                                         + 2 \delta_{\alpha \zeta} \delta_{\beta \zeta} R^\mathrm{e} R^\mathrm{o} \right)^j \nonumber \\
                             ~&\, \phantom{+ \frac{1}{2} \sqrt{s_{j-\half}} \Biggl[}\,       +    \left(   R_\alpha^\mathrm{e} R_\beta^\mathrm{o} + R_\alpha^\mathrm{o} R_\beta^\mathrm{e}
                                                                                                         + Z_\alpha^\mathrm{e} Z_\beta^\mathrm{o} + Z_\alpha^\mathrm{o} Z_\beta^\mathrm{e}
                                                                                                         + 2 \delta_{\alpha \zeta} \delta_{\beta \zeta} R^\mathrm{e} R^\mathrm{o} \right)^{j-1} \Biggr]
\end{align}
Concretely:
\begin{align}
  g_{\theta \theta}^{j-\half} =&\, \phantom{+}~ \frac{1}{2} \left[           \left( R_\theta^\mathrm{e} R_\theta^\mathrm{e} + Z_\theta^\mathrm{e} Z_\theta^\mathrm{e} \right)^j
                                                                   +         \left( R_\theta^\mathrm{e} R_\theta^\mathrm{e} + Z_\theta^\mathrm{e} Z_\theta^\mathrm{e} \right)^{j-1} \right] \nonumber \\
                              ~&\,          +   \frac{1}{2} \left[   s_j     \left( R_\theta^\mathrm{o} R_\theta^\mathrm{o} + Z_\theta^\mathrm{o} Z_\theta^\mathrm{o} \right)^j
                                                                   + s_{j-1} \left( R_\theta^\mathrm{o} R_\theta^\mathrm{o} + Z_\theta^\mathrm{o} Z_\theta^\mathrm{o} \right)^{j-1} \right] \nonumber \\
                              ~&\,          + \sqrt{s_{j-\half}} \left[   \left( R_\theta^\mathrm{e} R_\theta^\mathrm{o} + Z_\theta^\mathrm{e} Z_\theta^\mathrm{o} \right)^j
                                                                        + \left( R_\theta^\mathrm{e} R_\theta^\mathrm{o} + Z_\theta^\mathrm{e} Z_\theta^\mathrm{o} \right)^{j-1} \right] \label{eqn:g_theta_theta} \\
  g_{\theta \zeta }^{j-\half} =&\, \phantom{+}~ \frac{1}{2} \left[           \left( R_\theta^\mathrm{e} R_\zeta^\mathrm{e} + Z_\theta^\mathrm{e} Z_\zeta^\mathrm{e} \right)^j
                                                                   +         \left( R_\theta^\mathrm{e} R_\zeta^\mathrm{e} + Z_\theta^\mathrm{e} Z_\zeta^\mathrm{e} \right)^{j-1} \right] \nonumber \\
                              ~&\,          +   \frac{1}{2} \left[   s_j     \left( R_\theta^\mathrm{o} R_\zeta^\mathrm{o} + Z_\theta^\mathrm{o} Z_\zeta^\mathrm{o} \right)^j
                                                                   + s_{j-1} \left( R_\theta^\mathrm{o} R_\zeta^\mathrm{o} + Z_\theta^\mathrm{o} Z_\zeta^\mathrm{o} \right)^{j-1} \right] \nonumber \\
                              ~&\, + \frac{1}{2} \sqrt{s_{j-\half}} \Bigl[ \phantom{+}\, \left(   R_\theta^\mathrm{e} R_\zeta^\mathrm{o} + R_\theta^\mathrm{o} R_\zeta^\mathrm{e}
                                                                                                + Z_\theta^\mathrm{e} Z_\zeta^\mathrm{o} + Z_\theta^\mathrm{o} Z_\zeta^\mathrm{e} \right)^j \nonumber \\
                              ~&\, \phantom{+ \frac{1}{2} \sqrt{s_{j-\half}} \Bigl[}\, + \left(   R_\theta^\mathrm{e} R_\zeta^\mathrm{o} + R_\theta^\mathrm{o} R_\zeta^\mathrm{e}
                                                                                                + Z_\theta^\mathrm{e} Z_\zeta^\mathrm{o} + Z_\theta^\mathrm{o} Z_\zeta^\mathrm{e} \right)^{j-1} \Bigr] \label{eqn:g_theta_zeta} \\
  g_{\zeta  \zeta }^{j-\half} =&\, \phantom{+}~ \frac{1}{2} \left[           \left( R_\zeta^\mathrm{e} R_\zeta^\mathrm{e} + Z_\zeta^\mathrm{e} Z_\zeta^\mathrm{e} + R^\mathrm{e} R^\mathrm{e} \right)^j
                                                                   +         \left( R_\zeta^\mathrm{e} R_\zeta^\mathrm{e} + Z_\zeta^\mathrm{e} Z_\zeta^\mathrm{e} + R^\mathrm{e} R^\mathrm{e} \right)^{j-1} \right] \nonumber \\
                              ~&\,          +   \frac{1}{2} \left[   s_j     \left( R_\zeta^\mathrm{o} R_\zeta^\mathrm{o} + Z_\zeta^\mathrm{o} Z_\zeta^\mathrm{o} + R^\mathrm{o} R^\mathrm{o} \right)^j
                                                                   + s_{j-1} \left( R_\zeta^\mathrm{o} R_\zeta^\mathrm{o} + Z_\zeta^\mathrm{o} Z_\zeta^\mathrm{o} + R^\mathrm{o} R^\mathrm{o} \right)^{j-1} \right] \nonumber \\
                              ~&\,          + \sqrt{s_{j-\half}} \left[   \left( R_\zeta^\mathrm{e} R_\zeta^\mathrm{o} + Z_\zeta^\mathrm{e} Z_\zeta^\mathrm{o} + R^\mathrm{e} R^\mathrm{o} \right)^j
                                                                        + \left( R_\zeta^\mathrm{e} R_\zeta^\mathrm{o} + Z_\zeta^\mathrm{e} Z_\zeta^\mathrm{o} + R^\mathrm{e} R^\mathrm{o} \right)^{j-1} \right] \, . \label{eqn:g_zeta_zeta}
\end{align}
This is what is actually implemented in VMEC.
A change to the variant in~\eqn{metric_elements_10} was tried,
but VMEC does not converge as fast as with the metric elements computed from~\eqn{g_theta_theta} to~\eqn{g_zeta_zeta}.

\FloatBarrier
\section{Constraint of Toroidal Current Profile}
The net toroidal current density is given by~\cite{hirshman_hogan_1986}:
\begin{equation}
  \langle j_\zeta \rangle = \int\limits_0^{2 \pi} \int\limits_0^{2 \pi} \underbrace{\left( \mathbf{j} \cdot \nabla \zeta \right)}_{=j_\zeta} \sqrt{g} \,\mathrm{d}\theta \,\mathrm{d}\zeta
                          = \frac{1}{(2 \pi)^2} \frac{\partial \langle B_\theta \rangle}{\partial s}
\end{equation}
with
\begin{equation}
  \langle B_\theta \rangle = \frac{1}{(2 \pi)^2} \int\limits_0^{2 \pi} \int\limits_0^{2 \pi} B_\theta \,\mathrm{d}\theta \,\mathrm{d}\zeta \, . \label{eqn:bsubu_bar}
\end{equation}
The net toroidal current enclosed by a given flux surface is $I_\zeta(s)$ with:
\begin{equation}
  \frac{\mu_0}{2 \pi} I_\zeta = \langle B_\theta \rangle \, . \label{eqn:curtor_bsubtheta}
\end{equation}
\eqn{bsubu_bar} is now inserted into \eqn{curtor_bsubtheta}:
\begin{equation}
  \frac{\mu_0}{2 \pi} I_\zeta = \frac{1}{(2 \pi)^2} \iint B_\theta \,\mathrm{d}\theta \,\mathrm{d}\zeta \, .
\end{equation}
The covariant magnetic field component $B_\theta$ is expressed using~\eqn{bsubtheta_from_contra}:
\begin{equation}
  \frac{\mu_0}{2 \pi} I_\zeta
  =
  \frac{1}{(2 \pi)^2} \iint \left( B^\theta g_{\theta \theta} + B^\zeta g_{\theta \zeta } \right) \,\mathrm{d}\theta \,\mathrm{d}\zeta \, . \label{eqn:constr_current_start}
\end{equation}
Inserting the explicit forms of the contravariant magnetic field components from~\eqn{bsuptheta} and~\eqn{bsupzeta},
it follows with~\eqn{constr_current_start}:
\begin{equation}
  \frac{\mu_0}{2 \pi} I_\zeta
  =
  \frac{1}{(2 \pi)^2}
  \iint \left[  \frac{  1  }{\sqrt{g}} \left( \chi' - \Phi' \lambda_\zeta  \right) g_{\theta \theta}
              + \frac{\Phi'}{\sqrt{g}} \left(   1   +       \lambda_\theta \right) g_{\theta \zeta } \right] \,\mathrm{d}\theta \,\mathrm{d}\zeta \, . \label{eqn:izeta_constraint}
\end{equation}
In \eqn{izeta_constraint}, $I_\zeta$, $\Phi'$, $\lambda$ and the geometric quantities $g_{\theta \theta}$, $g_{\theta \zeta}$ and $\sqrt{g}$ are given.
Thus, it is rearranged to yield the $\chi'$ profile compatible with the prescribed current profile:
\begin{align}
 \chi' \frac{1}{(2 \pi)^2} \iint \frac{g_{\theta \theta}}{\sqrt{g}} \,\mathrm{d}\theta \,\mathrm{d}\zeta
  =&\,
   \frac{\mu_0}{2 \pi} I_\zeta
 - \frac{1}{(2 \pi)^2}
   \iint \left[  \left(    - \frac{\Phi' \lambda_\zeta  }{\sqrt{g}} \right) g_{\theta \theta}
               + \left( \frac{\Phi' (1 + \lambda_\theta)}{\sqrt{g}} \right) g_{\theta \zeta } \right] \,\mathrm{d}\theta \,\mathrm{d}\zeta \\
\Leftrightarrow
 \chi'
  =&\,
  \frac{\frac{\mu_0}{2 \pi} I_\zeta - \frac{1}{(2 \pi)^2} \iint \left[  \left(    - \frac{\Phi' \lambda_\zeta  }{\sqrt{g}} \right) g_{\theta \theta}
                                                                      + \left( \frac{\Phi' (1 + \lambda_\theta)}{\sqrt{g}} \right) g_{\theta \zeta } \right] \,\mathrm{d}\theta \,\mathrm{d}\zeta}
       {\frac{1}{(2 \pi)^2} \iint \frac{g_{\theta \theta}}{\sqrt{g}} \,\mathrm{d}\theta \,\mathrm{d}\zeta} \, . \label{eqn:chip_from_current_constraint}
\end{align}
The rotational transform profile is then computed via:
\begin{equation}
 \iota = \frac{\chi'}{\Phi'} \, . \label{eqn:iota_from_fluxes}
\end{equation}
Eqn.~(11) in Ref.~\cite{hirshman_hogan_1986} is recovered for $I_\zeta = 0$, where above approach is termed the ``Zero-Current Algorithm''.
In VMEC, \eqn{chip_from_current_constraint} and \eqn{iota_from_fluxes} are evaluated at each time step and on every flux surface
to obtain the $\iota$ profile consistent with a prescribed net toroidal current profile.

In VMEC, $\lambda$ is one of the state variables that (together with the radial profiles and a few constants) completely define the MHD equilibrium.
In the following, the program flow is described that starts from $\lambda$ and the metric elements and results in the $\iota$ profile
consistent with the prescribed net toroidal current profile.
The factor $\Phi'$ is included in the $\lambda$ state variable since version~8.47 of VMEC.
It is normalized to the root-mean-square (RMS) value of the toroidal flux derivative (\texttt{lamscale}, see \eqn{lamscale}).
This was necessary for application of VMEC to RFPs, according to the release notes.
The tangential derivatives of $\lambda$ are computed first on the full-grid via an inverse DFT.
Contributions from even-$m$ and odd-$m$ Fourier harmonics are kept separately.
The normalization factor $\lambda_\textrm{scale} (= \code{lamscale})$ is multiplied in again
to retrieve the physically relevant quantities:
\begin{align}
  \code{lu}^\mathrm{e} \leftarrow&\, \lambda_\textrm{scale}             \left[\phantom{-} \frac{\Phi'}{\lambda_\textrm{scale}} \left( \frac{\partial \lambda}{\partial \theta} \right)^\mathrm{e} \right] \\
  \code{lu}^\mathrm{o} \leftarrow&\, \lambda_\textrm{scale}             \left[\phantom{-} \frac{\Phi'}{\lambda_\textrm{scale}} \left( \frac{\partial \lambda}{\partial \theta} \right)^\mathrm{o} \right] \\
  \code{lv}^\mathrm{e} \leftarrow&\, \lambda_\textrm{scale}             \left[         -  \frac{\Phi'}{\lambda_\textrm{scale}} \left( \frac{\partial \lambda}{\partial \zeta } \right)^\mathrm{e} \right] \\
  \code{lv}^\mathrm{o} \leftarrow&\, \lambda_\textrm{scale} \underbrace{\left[         -  \frac{\Phi'}{\lambda_\textrm{scale}} \left( \frac{\partial \lambda}{\partial \zeta } \right)^\mathrm{o} \right]}_{\textrm{from inv-DFTs}} \, .
\end{align}
In the next step, $\Phi'$ is added to $\code{lu}^\mathrm{e}$ in order to arrive at:
\begin{equation}
 \code{lu}^\mathrm{e} \leftarrow \code{lu}^\mathrm{e} + \Phi' = \Phi' \left( 1 + \left(\frac{\partial \lambda}{\partial \theta} \right)^\mathrm{e} \right) \, . \label{eqn:lu_e_full}
\end{equation}
All of this is still on the full-grid.
The offset~(= $(m=0)$ - contribution) of $1$ is only present in the even-$m$ term of \code{lu}, since the $(m=0)$ - contribution has even parity.
Above computations happen on the full-grid.
It is now required to interpolate these contributions onto the half-grid
in order to compute the contravariant magnetic field components,
where also the Jacobian and radial derivatives (later required for the MHD forces) are available.
Additionally, a factor of $1/\sqrt{g}$ is included:
\begin{align}
  \code{bsupu}^{j-\half}
 \leftarrow&\,
  \frac{1}{\sqrt{g}} \Biggl[                      \frac{1}{2} \left( \code{lv}^{\mathrm{e}, j-1} + \code{lv}^{\mathrm{e}, j} \right)
                             + \sqrt{s_{j-\half}} \frac{1}{2} \left( \code{lv}^{\mathrm{o}, j-1} + \code{lv}^{\mathrm{o}, j} \right) \Biggr] \nonumber \\
    \approx&\, \left( - \frac{\Phi' \lambda_\zeta}{\sqrt{g}} \right)^{j-\half} \label{eqn:lu_half} \\
  \code{bsupv}^{j-\half}
 \leftarrow&\,
  \frac{1}{\sqrt{g}} \Biggl[                      \frac{1}{2} \left( \code{lu}^{\mathrm{e}, j-1} + \code{lu}^{\mathrm{e}, j} \right)
                             + \sqrt{s_{j-\half}} \frac{1}{2} \left( \code{lu}^{\mathrm{o}, j-1} + \code{lu}^{\mathrm{o}, j} \right) \Biggr] \nonumber \\
    \approx&\, \left( \frac{\Phi' \left(1 + \lambda_\theta\right)}{\sqrt{g}} \right)^{j-\half} = B_\zeta^{j-\half} \label{eqn:lv_half} \, .
\end{align}
If the rotational transform profile is prescribed in the user input, it is used below and the next step is omitted.
Otherwise, the next step is to compute the $\iota$ profile using~\eqn{chip_from_current_constraint} and~\eqn{iota_from_fluxes}.
The surface-averaging integrals are replaced by discrete sums for trapezoidal quadrature
where $A$ denotes the quantitiy to integrate and $A_{kl} = A(\theta_k, \zeta_l)$:
\begin{align}
  \frac{1}{(2 \pi)^2} \int\limits_0^{2 \pi} \int\limits_0^{2 \pi} A(\theta, \zeta) \,\mathrm{d}\theta \,\mathrm{d}\zeta
  \approx&\, \frac{1}{(2 \pi)^2} \sum\limits_{k=0}^{n_\theta - 1} \sum\limits_{l=0}^{n_\zeta - 1} A(\theta_k, \zeta_l) \,\Delta \theta \,\Delta \zeta \nonumber \\
        =&\, \frac{1}{n_\theta n_\zeta} \sum\limits_{k=0}^{n_\theta - 1} \sum\limits_{l=0}^{n_\zeta - 1} A(\theta_k, \zeta_l)  \nonumber \\
        =&\, \frac{1}{n_\theta n_\zeta} \sum\limits_{k=0}^{n_\theta - 1} \sum\limits_{l=0}^{n_\zeta - 1} A_{kl} \, .
\end{align}
This leads to:
\begin{align}
  \chi'
  =&\,
  \frac{\frac{\mu_0}{2 \pi} I_\zeta -
        \frac{1}{n_\theta n_\zeta} \sum\limits_{k=0}^{n_\theta - 1} \sum\limits_{l=0}^{n_\zeta - 1}
                                     \left[  \left(    - \frac{\Phi' \lambda_\zeta  }{\sqrt{g}} \right) g_{\theta \theta}
                                           + \left( \frac{\Phi' (1 + \lambda_\theta)}{\sqrt{g}} \right) g_{\theta \zeta } \right]_{kl}}
       {\frac{1}{n_\theta n_\zeta} \sum\limits_{k=0}^{n_\theta - 1} \sum\limits_{l=0}^{n_\zeta - 1}
                                     \left[ \frac{g_{\theta \theta}}{\sqrt{g}} \right]_{kl}} \, . \label{eqn:iota_from_current_discrete}
\end{align}
The discrete profile of the enclosed toroidal current is given by:
\begin{equation}
  \code{jv}(s_{j-\half}) = \frac{\mu_0}{2 \pi} \cdot \code{signgs} \cdot \underbrace{I_\zeta(s_{j-\half})}_{= \code{curtor} \cdot \code{pcurr}(s_{j-\half}) / \code{pcurr}(1) } \, .
\end{equation}
Using the quantities from~\eqn{lu_half} and~\eqn{lv_half},
the $\chi'$ profile can now be computed discretely on the half-grid:
\begin{equation}
  \boxed{
  \chi'^{j-\half} = \frac{ \code{jv}(s_{j-\half})
                          - \frac{1}{n_\theta n_\zeta} \sum\limits_{k=0}^{n_\theta - 1} \sum\limits_{l=0}^{n_\zeta - 1}
                            \left[   \left( - \frac{\Phi'           \lambda_\zeta        }{\sqrt{g}} \right)^{j-\half} g_{\theta \theta}^{j-\half}
                                   + \left(   \frac{\Phi' \left(1 + \lambda_\theta\right)}{\sqrt{g}} \right)^{j-\half} g_{\theta \zeta }^{j-\half} \right]_{kl}}
                         {  \frac{1}{n_\theta n_\zeta} \sum\limits_{k=0}^{n_\theta - 1} \sum\limits_{l=0}^{n_\zeta - 1}
                            \left[ g_{\theta \theta}^{j-\half} / \left( \sqrt{g} \right)^{j-\half} \right]_{kl}} } \, . \label{eqn:chip_half}
\end{equation}
Using the $\iota$ profile from user input in case of a constrained-$\iota$ case
(and \eqn{iota_from_fluxes} re-arranged to get $\chi'$)
or from~\eqn{chip_half} in case of a constrained-current case,
the contravariant magnetic field component $B^\theta(s_{j-\half})$ can be computed:
\begin{align}
  \code{bsupu}^{j-\half} \leftarrow&\, \underbrace{\left( - \frac{\Phi' \lambda_\zeta}{\sqrt{g}} \right)^{j-\half}}_{\code{bsupu}^{j-\half}} + \left( \frac{\chi'}{\sqrt{g}} \right)^{j-\half} \nonumber \\
                               =&\, \left( \frac{1}{\sqrt{g}} \left( \chi' - \Phi' \lambda_\zeta \right) \right)^{j-\half} = B^\theta(s_{j-\half}) \, .
\end{align}
At this point, the computation of the contravariant magnetic field components is done.

\FloatBarrier
\section{Covariant Magnetic Field Components} \label{sec:bcov}
The covariant magnetic field components~$B_\theta$ and $B_\zeta$ are now computed
from the contravariant magnetic field components~$B^\theta$ and $B^\zeta$
via~\eqn{bsubtheta_from_contra} and~\eqn{bsubzeta_from_contra}:
\begin{align}
  B_\theta(s_{j-\half}) = \code{bsubu}^{j-\half} =&\,          \,   \code{bsupu}^{j-\half}                            \, g_{\theta \theta}^{j-\half} +             \code{bsupv}^{j-\half}                           \, g_{\theta \zeta}^{j-\half} \\
  B_\zeta (s_{j-\half}) = \code{bsubv}^{j-\half} =&\,   \underbrace{\code{bsupu}^{j-\half}}_{= B^\theta(s_{j-\half})}    g_{\theta \zeta }^{j-\half} + \underbrace{\code{bsupv}^{j-\half}}_{= B^\zeta(s_{j-\half})}    g_{\zeta  \zeta}^{j-\half} \, .
\end{align}
In the case of an axisymmetric run, $g_{\theta \zeta } = 0$ and $g_{\zeta  \zeta} = R^2$,
which can be used to simplify these equations.

\FloatBarrier
\section{Magnetic and Thermal Energies}
The magnetic pressure in the plasma volume can be computed now:
\begin{equation}
  \mu_0 \frac{|B|^2}{2 \mu_0}(s_{j-\half}) = \code{bsq}^{j-\half} = \frac{1}{2} \left( B^\theta B_\theta + B^\zeta B_\zeta \right)^{j-\half} \, .
\end{equation}
The thermal energy $W_\mathrm{th}$ is given by:
\begin{align}
  \frac{\mu_0}{(2 \pi)^2} W_\mathrm{th} =&\, \frac{\mu_0}{(2 \pi)^2} \int\limits_\mathrm{V} p \,\mathrm{d}V = \mu_0 \int\limits_0^1 p(s) \frac{V'(s)}{(2 \pi)^2} \,\mathrm{d}s \nonumber \\
                 =&\, \int\limits_0^1 \frac{\mu_0 m(s)}{\left(V'(s)\right)^\gamma} \frac{V'(s)}{(2 \pi)^2} \,\mathrm{d}s
                 =  \frac{1}{(2 \pi)^2}  \int\limits_0^1 \frac{\mu_0 m(s)}{\left(V'(s)\right)^{\gamma - 1}} \,\mathrm{d}s \nonumber \\
           \approx&\, \sum\limits_{j=1}^{n_s-1} \mu_0 p(s_{j-\half}) \frac{V'(s_{j-\half})}{(2 \pi)^2} \Delta s \nonumber \\
                 =&\, \frac{1}{n_s-1} \sum\limits_{j=1}^{n_s-1} \mu_0 p(s_{j-\half}) \frac{V'(s_{j-\half})}{(2 \pi)^2} \nonumber \\
                 =&\, \frac{1}{n_s-1} \sum\limits_{j=1}^{n_s-1} \code{pres}(s_{j-\half}) \, \code{vp}(s_{j-\half}) = \code{wp} \, .
\end{align}
where V denotes the total plasma volume.
The magnetic energy $W_B$ is computed from the magnetic pressure as follows:
\begin{align}
  \frac{\mu_0}{(2 \pi)^2} W_B =&\, \frac{\mu_0}{(2 \pi)^2} \int\limits_0^{1} \int\limits_0^{2 \pi} \int\limits_0^{2 \pi}
                   \frac{|B|^2}{2 \mu_0} \sqrt{g} \,\mathrm{d}\theta \,\mathrm{d}\zeta \,\mathrm{d}s \nonumber \\
      \approx&\, \frac{1}{(2 \pi)^2} \sum\limits_{j=1}^{n_s-1} \sum\limits_{k=0}^{n_\theta - 1} \sum\limits_{l=0}^{n_\zeta - 1}
                   \left( \frac{|B|^2}{2} \right)^{j-\half}_{kl} \sqrt{g}^{j-\half}_{kl}
                   \,\Delta s \,\Delta \theta \,\Delta \zeta \nonumber \\
            =&\, \frac{1}{\bcancel{(2 \pi)^2}} \frac{1}{n_s-1} \frac{\bcancel{2 \pi}}{n_\theta} \frac{\bcancel{2 \pi}}{n_\zeta}
                 \sum\limits_{j=1}^{n_s-1} \sum\limits_{k=0}^{n_\theta - 1} \sum\limits_{l=0}^{n_\zeta - 1}
                   \left( \frac{|B|^2}{2} \right)^{j-\half}_{kl} \sqrt{g}^{j-\half}_{kl} \nonumber \\
            =&\, \frac{1}{n_s-1} \frac{1}{n_\theta n_\zeta}
                 \sum\limits_{j=1}^{n_s-1} \sum\limits_{k=0}^{n_\theta - 1} \sum\limits_{l=0}^{n_\zeta - 1}
                   \left( \frac{|B|^2}{2} \right)^{j-\half}_{kl} \left(\sqrt{g}\right)^{j-\half}_{kl} = \code{wb} \, .
\end{align}
This allows to compute the MHD energy as follows:
\begin{align}
 \frac{\mu_0}{(2 \pi)^2} W_\mathrm{MHD}
             =&\, \frac{\mu_0}{(2 \pi)^2} \int\limits_\mathrm{V} \left( \frac{B^2}{2 \mu_0} + \frac{p}{\gamma - 1} \right) \,\mathrm{d}V \nonumber \\
             =&\, \frac{\mu_0}{(2 \pi)^2} \left( W_B + \frac{1}{\gamma - 1} W_\mathrm{th} \right) \\
       \approx&\, \code{wb} + \code{wp} / (\gamma - 1) = \code{w0} \nonumber \, .
\end{align}
Note that in VMEC, the MHD energy is only a diagnostic quantity that gets printed to the screen.
In particular, its evolution along the iterations does not influence the algorithm.
Using this way to compute the MHD energy, it becomes trivial to also compute the volume-averaged plasma beta:
\begin{equation}
  \langle \beta \rangle = \frac{W_\mathrm{th}}{W_B} = \left\langle \frac{p}{B^2 / 2 \mu_0} \right\rangle \, .
\end{equation}

\FloatBarrier
\section{Radial Force Balance}
The evaluation of the average radial force balance is described here.
The enclosed poloidal and toroidal currents are computed via surface integrals on the half-grid
using the already-computed (see Sec.~\ref{sec:bcov}) covariant magnetic field components on the half-grid:
\begin{align}
  \frac{1}{(2 \pi)^2}  I^\zeta(s_{j-\half}) =&\, \frac{1}{(2 \pi)^2} \iint B_\theta(s_{j-\half}, \theta, \zeta) \,\mathrm{d}\theta \,\mathrm{d}\zeta \nonumber \\
                            \approx&\, \frac{1}{n_\theta n_\zeta} \sum\limits_{k,l} \left[ B_\theta(s_{j-\half}) \right]^{k,l} \\
  \frac{1}{(2 \pi)^2} I^\theta(s_{j-\half}) =&\, \frac{1}{(2 \pi)^2} \iint  B_\zeta(s_{j-\half}, \theta, \zeta) \,\mathrm{d}\theta \,\mathrm{d}\zeta \nonumber \\
                            \approx&\, \frac{1}{n_\theta n_\zeta} \sum\limits_{k,l} \left[  B_\zeta(s_{j-\half}) \right]^{k,l} \, .
\end{align}
The radial derivatives required to evaluate the radial force balance on the full-grid
are now computed from the half-grid quantites:
\begin{align}
  \frac{1}{(2 \pi)^2}  j^\zeta(s_j) =&\, \phantom{-}~ \frac{\sqrt{g}}{|\sqrt{g}|} \frac{1}{\Delta s} \left[ \frac{ I^\zeta(s_{j+\half})}{(2 \pi)^2} - \frac{ I^\zeta(s_{j-\half})}{(2 \pi)^2} \right] \\
  \frac{1}{(2 \pi)^2} j^\theta(s_j) =&\,          -   \frac{\sqrt{g}}{|\sqrt{g}|} \frac{1}{\Delta s} \left[ \frac{I^\theta(s_{j+\half})}{(2 \pi)^2} - \frac{I^\theta(s_{j-\half})}{(2 \pi)^2} \right] \\
  \mu_0 p'(s_j) =&\, \frac{1}{\Delta s} \left[ \mu_0 p(s_{j+\half}) - \mu_0 p(s_{j-\half}) \right] \, .
\end{align}
The differential volume is interpolated onto the full-grid:
\begin{equation}
  \frac{V'(s_j)}{(2 \pi)^2} = \frac{1}{2} \left[ \frac{V'(s_{j+\half})}{(2 \pi)^2} + \frac{V'(s_{j-\half})}{(2 \pi)^2} \right] \, .
\end{equation}
This allows to compute the average radial force imbalance~$F(s_j)$ as follows:
\begin{align}
  F(s_j) =&\,   \left[   \chi'(s_j) \frac{ j^\zeta(s_j)}{\cancel{(2 \pi)^2}}
                       - \Phi'(s_j) \frac{j^\theta(s_j)}{\cancel{(2 \pi)^2}} \right] \frac{\cancel{(2 \pi)^2}}{V'(s_j)}
              + \mu_0 p'(s_j) \nonumber \\
         =&\,   \frac{1}{V'(s_j)} \left[ \chi' j^\zeta - \Phi' j^\theta + \mu_0 p' V' \right]^j
\end{align}
where the final expression in above equation resembles Eqn.~(33b) in Ref.~\cite{hirshman_whitson_1983}.
Note that $F(s_j)$ has a unit of $\mu_0 \cdot \textrm{ Pa}$ in this formulation.

\FloatBarrier
\section{MHD Forces}
The conservative expressions for the cylindrical MHD force components~$F^R$ and~$F^Z$
are given in Eqn.~(18a) and~(18b) in the main VMEC article~\cite{hirshman_whitson_1983}.
They are:
\begin{align}
 F^R =&\, \phantom{-}\, \frac{\partial}{\partial s} \left( Z_\theta P \right) + \tau \left( \frac{P}{R} - \frac{1}{\mu_0} (R B^\zeta)^2 \right)
          - \frac{\partial}{\partial \theta} \Bigl(\phantom{-}\, Z_s P - \frac{1}{\mu_0} B^\theta b_R \Bigr)
          + \frac{\partial}{\partial \zeta}  \Bigl( \frac{1}{\mu_0} B^\zeta b_R \Bigr) \nonumber \\
 F^Z =&\, \underbrace{-\frac{\partial}{\partial s} \left( R_\theta P \right) \phantom{+ G \left( \frac{P}{R} - \frac{1}{\mu_0} (R B^\zeta)^2 \right)}}_{A}\,
          - \, \frac{\partial}{\partial \theta} \Bigl( \underbrace{- R_s P - \frac{1}{\mu_0} B^\theta b_Z}_{B} \Bigr)
          + \frac{\partial}{\partial \zeta}  \Bigl( \underbrace{\frac{1}{\mu_0} B^\zeta b_Z }_{C} \Bigr) \nonumber \, .
\end{align}
Note that $\tau$ is called $G$ in Ref.~\cite{hirshman_whitson_1983}.
The force on $\lambda$ is given in Eqn.~(14b):
\begin{equation}
 F^\lambda = \Phi' |\sqrt{g}| F_\beta
\end{equation}
with $F_\beta$ from Eqn.~(4e):
\begin{equation}
 F_\beta = \frac{1}{\mu_0 \sqrt{g}} \left( \frac{\partial B_\zeta}{\partial \theta} - \frac{\partial B_\theta}{\partial \zeta} \right) \, .
\end{equation}
The MHD forces in VMEC are organized as follows:
\begin{align}
  F^R       =&\,          A^R    - \frac{\partial B^R      }{\partial \theta} + \frac{\partial C^R      }{\partial \phi} \nonumber \\
  F^Z       =&\,          A^Z    - \frac{\partial B^Z      }{\partial \theta} + \frac{\partial C^Z      }{\partial \phi} \label{eqn:mhd_forces} \\
  F^\lambda =&\, \phantom{A^Z}\, - \frac{\partial B^\lambda}{\partial \theta} + \frac{\partial C^\lambda}{\partial \phi} \nonumber
\end{align}
with
\begin{align}
 A^R =&\, \phantom{-}~ \frac{\partial}{\partial s} \left( Z_\theta P \right) + \tau \left( \frac{P}{R} - \frac{1}{\mu_0} (R B^\zeta)^2 \right) \label{eqn:f_a_r} \\
 A^Z =&\,          -   \frac{\partial}{\partial s} \left( R_\theta P \right)                                                                   \label{eqn:f_a_z} \\
-B^R =&\, \frac{1}{\mu_0} B^\theta b_R - Z_s P                                                                                                 \label{eqn:f_b_r} \\
-B^Z =&\, \frac{1}{\mu_0} B^\theta b_Z + R_s P                                                                                                 \label{eqn:f_b_z} \\
 C^R =&\, \frac{1}{\mu_0}  B^\zeta b_R                                                                                                         \label{eqn:f_c_r} \\
 C^Z =&\, \frac{1}{\mu_0}  B^\zeta b_Z                                                                                                         \label{eqn:f_c_z}
\end{align}
and
\begin{align}
 B^\lambda =&\, - \frac{\Phi'}{\mu_0} \frac{|\sqrt{g}|}{\sqrt{g}} B_\zeta  \label{eqn:blmn_start} \\
 C^\lambda =&\, - \frac{\Phi'}{\mu_0} \frac{|\sqrt{g}|}{\sqrt{g}} B_\theta \label{eqn:clmn_start} \, .
\end{align}
It is noted that $|\sqrt{g}| / \sqrt{g} = \code{signgs}$.
The following shorthand notations are used in above equations:
\begin{align}
 P   =&\, R \left( \frac{|B|^2}{2 \mu_0} + p \right) \label{eqn:r_ptot} \\
 b_R =&\, b^\theta R_\theta + b^\zeta R_\zeta \label{eqn:smallbr} \\
 b_Z =&\, b^\theta Z_\theta + b^\zeta Z_\zeta \\
 b^\theta =&\, \sqrt{g} B^\theta \\
 b^\zeta  =&\, \sqrt{g} B^\zeta \label{eqn:smallbzeta}
\end{align}
which implies
\begin{align}
 b_R =&\, \sqrt{g} \left( B^\theta R_\theta + B^\zeta R_\zeta \right) \\
 b_Z =&\, \sqrt{g} \left( B^\theta Z_\theta + B^\zeta Z_\zeta \right) \, .
\end{align}

\subsection{Forces on Lambda}
%%%%%%% lambda forces %%%%%%%%%
The forces on $\lambda$ are considered first.
These are essentially the covariant magnetic field components.
These have been computed on the half-grid in Sec.~\ref{sec:bcov}.
Some special care is needed when performing the interpolation to the full-grid,
where they are needed to act on the full-grid $\lambda$ state variable.
A hybrid half-grid/full-grid scheme for $B_\zeta$ is implemented in VMEC, which is described here.
Starting from \eqn{blmn_start} and \eqn{clmn_start}, it is noted that the scalar factors are included in the state variable $\lambda$
and hence the source terms for the forces on $\lambda$ are the (negative) covariant magnetic field components.
\begin{align}
  \code{blmn\_e} = \mu_0 \, \code{signgs} \, \left(\frac{B^\lambda}{\Phi'} \right)^j =&\, - B^j_\zeta  \label{eqn:blmn_e} \\
  \code{clmn\_e} = \mu_0 \, \code{signgs} \, \left(\frac{C^\lambda}{\Phi'} \right)^j =&\, - B^j_\theta \label{eqn:clmn_e} \, .
\end{align}
The forces on the odd-$m$ harmonics of $\lambda$ need to get scaled by $\sqrt(s)$:
\begin{align}
  \code{blmn\_o} = \mu_0 \, \code{signgs} \, \sqrt{s_j} \left(\frac{B^\lambda}{\Phi'} \right)^{\mathrm{o},j} =&\, \sqrt{s_j} \left( - B^j_\zeta  \right) \label{eqn:blmn_o} \\
  \code{clmn\_o} = \mu_0 \, \code{signgs} \, \sqrt{s_j} \left(\frac{C^\lambda}{\Phi'} \right)^{\mathrm{o},j} =&\, \sqrt{s_j} \left( - B^j_\theta \right) \label{eqn:clmn_o} \, .
\end{align}
The terms in \eqn{blmn_e} to \eqn{clmn_o} are the $\lambda$ forces taken into account in VMEC.
The only terms required to be computed for the $\lambda$ forces are $B^j_\zeta \equiv B_\zeta(s_j)$ and $B^j_\theta \equiv B_\theta(s_j)$,
which are the covariant magnetic field components on the full-grid.
Then, the odd-$m$ $\lambda$ forces are obtained from the even-$m$ components by scaling:
\begin{align}
  \code{blmn\_o} =&\, \sqrt{s_j} \,\code{blmn\_e} \\
  \code{clmn\_o} =&\, \sqrt{s_j} \,\code{clmn\_e} \, .
\end{align}
The computation of the full-grid covariant magnetic field components follows now.
The component $B_\theta$ is simply interpolated onto the full-grid:
\begin{equation}
  B_\theta^j = \frac{1}{2} \left[ B_\theta(s_{j+\half}) + B_\theta(s_{j-\half}) \right] \, .
\end{equation}
A hybrid scheme is used for $B_\zeta$:
\begin{equation}
  B_\zeta^j =              \epsilon_\mathrm{blend}(s_j)                          \tilde{B_\zeta}^j
              + \left[ 1 - \epsilon_\mathrm{blend}(s_j) \right] \frac{1}{2} \left[ B_\zeta^{j+\half} + B_\zeta^{j-\half} \right]
\end{equation}
with
\begin{equation}
  \epsilon_\mathrm{blend}(s_j) = 2 p_\mathrm{damp} (1 - s_j)
\end{equation}
and $\tilde{B_\zeta}^j$ computed as follows:
\begin{align}
  \tilde{B_\zeta}^j =&\, \left( B^\theta g_{\theta \zeta } \right)^j + \left( B^\zeta g_{\zeta  \zeta } \right)^j \, .
\end{align}
Here, the product interpolation rule is applied again:
\begin{equation}
  \left( B^\theta g_{\theta \zeta } \right)^j
 = \frac{1}{2} \left[ \left( B^\theta g_{\theta \zeta } \right)^{j+\half} + \left( B^\theta g_{\theta \zeta } \right)^{j-\half} \right] \, .
\end{equation}
The quantity $(B^\zeta g_{\zeta  \zeta })^j$ is computed by re-arranging terms before interpolation onto the full-grid.
Recall \eqn{lu_e_full} and it follows:
\begin{align}
  (B^\zeta g_{\zeta  \zeta } )^j
  =&\, \phantom{+}~
         \frac{1}{2} \left[ \phantom{\sqrt{s_{j+\half}}} \left( \frac{g_{\zeta \zeta}}{\sqrt{g}} \right)^{j+\half} + \phantom{\sqrt{s_{j-\half}}} \left( \frac{g_{\zeta \zeta}}{\sqrt{g}} \right)^{j-\half} \right]
         \left[ \Phi' \left( 1 + \lambda_\theta^\mathrm{e} \right) \right]^j \nonumber \\
  ~&\, +
         \frac{1}{2} \left[ \sqrt{s_{j+\half}} \left( \frac{g_{\zeta \zeta}}{\sqrt{g}} \right)^{j+\half} + \sqrt{s_{j-\half}} \left( \frac{g_{\zeta \zeta}}{\sqrt{g}} \right)^{j-\half} \right]
         \left[ \Phi' \lambda_\theta^\mathrm{o} \right]^j \, .
\end{align}
This concludes the derivation of the current hybrid formulation used to compute the $\lambda$-forces in VMEC.
It remains to compute the forces on $R$ and $Z$ as listed in \eqn{f_a_r} to \eqn{f_c_z}.

\subsection{A-Forces on R and Z}
%%%%%%% A^{R,Z} forces %%%%%%%%%
The $A^R$ force component consists of two contributions:
\begin{equation}
  A^R =   \underbrace{\frac{\partial}{\partial s} \left( Z_\theta P \right)}_{A^{R,1}}
        + \underbrace{\tau \left( \frac{P}{R} - \frac{1}{\mu_0} \left( R B^\zeta \right)^2 \right)}_{A^{R,2}}
\end{equation}
First consider $A^{R,1}$:
\begin{equation}
  A^{R,1}(s_j) = A^{R,1,\mathrm{e}}(s_j) + \sqrt{s_j} A^{R,1,\mathrm{o}} (s_j) \, .
\end{equation}
The product differencing rule is employed to compute the even-$m$ component of $A^{R,1}$:
\begin{equation}
  A^{R,1,\mathrm{e}} (s_j) = \frac{1}{\Delta s} \left[ \left( Z_\theta P \right)^{j+\half} - \left( Z_\theta P \right)^{j-\half} \right] \, .
\end{equation}
For the odd-$m$ component, the product rule is used backwards:
\begin{align}
  \frac{\partial}{\partial s} \left( \sqrt{s} Z_\theta P \right) =&\, \sqrt{s} \frac{\partial}{\partial s} \left( Z_\theta P \right) + \frac{1}{2 \sqrt{s}} Z_\theta P \nonumber \\
  \Leftrightarrow
  \sqrt{s} A^{R,1,\mathrm{o}} =
  \sqrt{s} \frac{\partial}{\partial s} \left( Z_\theta P \right) =&\, \frac{\partial}{\partial s} \left( \sqrt{s} Z_\theta P \right) - \frac{1}{2 \sqrt{s}} Z_\theta P \, .
\end{align}
It follows for the odd-$m$ component of $A^{R,1}$:
\begin{align}
  \sqrt{s_j} A^{R,1,\mathrm{o}} (s_j)
  =&\, \phantom{-}~ \frac{1}{\Delta s} \left[ \sqrt{s_{j+\half}} \left( Z_\theta P \right)^{j+\half} - \sqrt{s_{j-\half}} \left( Z_\theta P \right)^{j-\half} \right] \nonumber \\
  ~&\,          -   \left[ \frac{1}{2 \sqrt{s}} \left( Z_\theta^\mathrm{e} + \sqrt{s} Z_\theta^\mathrm{o} \right) P \right]^j
\end{align}
with
\begin{equation}
  \left[ \frac{1}{2 \sqrt{s}} \left( Z_\theta^\mathrm{e} + \sqrt{s} Z_\theta^\mathrm{o} \right) P \right]^j
  = \frac{1}{2} Z_\theta^\mathrm{e}(s_j) \left(\frac{P}{\sqrt{s}}\right)^j + \frac{1}{2} Z_\theta^\mathrm{o}(s_j) P(s_j) \, .
\end{equation}
In here, use the following expressions for interpolating $P/\sqrt{s}$ and $P$ to the full-grid:
\begin{align}
  \left(\frac{P}{\sqrt{s}}\right)^j \approx&\, \frac{1}{2} \left[ \left( \frac{P}{\sqrt{s}} \right)^{j+\half} + \left( \frac{P}{\sqrt{s}} \right)^{j-\half} \right] \label{eqn:psqrts_interp} \\
  P(s_j)  \approx&\, \frac{1}{2} \left[ P\left(s_{j+\half}\right) + P\left(s_{j-\half}\right) \right] \label{eqn:p_interp} \, .
\end{align}
This leads to the final expression for $\sqrt{s} A^{R,1,\mathrm{o}}$:
\begin{align}
  \sqrt{s_j} A^{R,1,\mathrm{o}}(s_j)
  =&\, \phantom{-}~ \frac{1}{\Delta s} \left[ \sqrt{s_{j+\half}} \left( Z_\theta P \right)^{j+\half} - \sqrt{s_{j-\half}} \left( Z_\theta P \right)^{j-\half} \right] \nonumber \\
  ~&\,          -   \frac{1}{4} Z_\theta^\mathrm{e}(s_j) \left[ \left( \frac{P}{\sqrt{s}} \right)^{j+\half} + \left( \frac{P}{\sqrt{s}} \right)^{j-\half} \right] \nonumber \\
  ~&\,          -   \frac{1}{4} Z_\theta^\mathrm{o}(s_j) \left[ P \left(s_{j+\half}\right) + P\left(s_{j-\half}\right) \right] \, .
\end{align}
It remains to consider the computation of $A^{R,2}$ for $A^R$.
Recall the definition of $P$ from \eqn{r_ptot} and it follows:
\begin{align}
  A^{R,2}
  =&\, \tau \left( \frac{P}{R} - \frac{1}{\mu_0} \left( R B^\zeta \right)^2 \right) \nonumber \\
  =&\, \tau \left( \frac{|B|^2}{2 \mu_0} + p \right) - \frac{1}{\mu_0} \tau R^2 {B^\zeta}^2 \nonumber \\
  =&\, \tau p_\mathrm{tot} - \frac{1}{\mu_0} R \underbrace{\tau R}_{=\sqrt{g}} B^\zeta B^\zeta \nonumber \\
  =&\, \tau p_\mathrm{tot} - \frac{1}{\mu_0} R \sqrt{g} B^\zeta B^\zeta
\end{align}
with
\begin{align}
  p_\mathrm{tot} =&\, \left( \frac{|B|^2}{2 \mu_0} + p \right) \, .
\end{align}
The product differencing rule is applied to interpolate the product $\tau p_\mathrm{tot}$ onto the full-grid:
\begin{equation}
  \left[ \tau p_\mathrm{tot} \right]^j = \frac{1}{2} \left[ \left( \tau p_\mathrm{tot} \right)^{j+\half} + \left( \tau p_\mathrm{tot} \right)^{j-\half} \right] \, .
\end{equation}
Recall that $R = R^\mathrm{e} + \sqrt{s} R^\mathrm{o}$ and it follows for the second term in $A^{R,2}$:
\begin{equation}
  \left[ R \sqrt{g} B^\zeta B^\zeta \right]^j
  = \left[ \sqrt{g} B^\zeta B^\zeta \right]^j R^\mathrm{e}(s_j) + \left[ \sqrt{s} \sqrt{g} B^\zeta B^\zeta \right]^j R^\mathrm{o}(s_j)
\end{equation}
with
\begin{equation}
  \left[ \sqrt{g} B^\zeta B^\zeta \right]^j
  \approx
  \frac{1}{2} \left[ \left( \sqrt{g} B^\zeta B^\zeta \right)^{j+\half} + \left( \sqrt{g} B^\zeta B^\zeta \right)^{j-\half} \right]
\end{equation}
and
\begin{equation}
  \left[ \sqrt{s} \sqrt{g} B^\zeta B^\zeta \right]^j
  \approx
  \frac{1}{2} \left[ \sqrt{s_{j+\half}} \left( \sqrt{g} B^\zeta B^\zeta \right)^{j+\half} + \sqrt{s_{j-\half}} \left( \sqrt{g} B^\zeta B^\zeta \right)^{j-\half} \right] \, .
\end{equation}
Thus, the full expression for $A^{R,2,\mathrm{e}}$
% ({\color{red} and ignoring a missing factor of $\mu_0$ for now...})
reads:
\begin{align}
  A^{R,2,\mathrm{e}} (s_j)
  =&\, \phantom{-}~ \frac{1}{2} \left[ \left( \tau p_\mathrm{tot} \right)^{j+\half} + \left( \tau p_\mathrm{tot} \right)^{j-\half} \right] \nonumber \\
  ~&\,          -   \frac{1}{2} \left[   \phantom{\sqrt{s_{j+\half}}} \left( \sqrt{g} B^\zeta B^\zeta \right)^{j+\half}
                                       + \phantom{\sqrt{s_{j-\half}}} \left( \sqrt{g} B^\zeta B^\zeta \right)^{j-\half} \right] R^\mathrm{e}(s_j) \nonumber \\
  ~&\,          -   \frac{1}{2} \left[            \sqrt{s_{j+\half}}  \left( \sqrt{g} B^\zeta B^\zeta \right)^{j+\half}
                                       +          \sqrt{s_{j-\half}}  \left( \sqrt{g} B^\zeta B^\zeta \right)^{j-\half} \right] R^\mathrm{o}(s_j) \, .
\end{align}
An additional factor of $\sqrt{s}$ has to be included for $\sqrt{s} A^{R,2,\mathrm{o}}$ and it follows:
\begin{align}
  \sqrt{s_j} A^{R,2,\mathrm{o}} (s_j)
  =&\, \phantom{-}~ \frac{1}{2}               \left[            \sqrt{s_{j+\half}}  \left( \tau p_\mathrm{tot} \right)^{j+\half}
                                                     +          \sqrt{s_{j-\half}}  \left( \tau p_\mathrm{tot} \right)^{j-\half} \right] \nonumber \\
  ~&\,          -   \frac{1}{2} \phantom{s_j} \left[            \sqrt{s_{j+\half}}  \left( \sqrt{g} B^\zeta B^\zeta \right)^{j+\half}
                                                     +          \sqrt{s_{j-\half}}  \left( \sqrt{g} B^\zeta B^\zeta \right)^{j-\half} \right] R^\mathrm{e}(s_j) \nonumber \\
  ~&\,          -   \frac{1}{2}          s_j  \left[   \phantom{\sqrt{s_{j+\half}}} \left( \sqrt{g} B^\zeta B^\zeta \right)^{j+\half}
                                                     + \phantom{\sqrt{s_{j-\half}}} \left( \sqrt{g} B^\zeta B^\zeta \right)^{j-\half} \right] R^\mathrm{o}(s_j) \, .
\end{align}
Putting things together, we arrive at the following final expressions for the components of $A^R$:
\begin{align}
  A^{R,\mathrm{e}} (s_j)
  =&\, \phantom{+}~ \frac{1}{\Delta s} \left[ \left( Z_\theta P \right)^{j+\half} - \left( Z_\theta P \right)^{j-\half} \right] \nonumber \\
  ~&\,          +   \frac{1}{2} \left[ \left( \tau p_\mathrm{tot} \right)^{j+\half} + \left( \tau p_\mathrm{tot} \right)^{j-\half} \right] \nonumber \\
  ~&\,          -   \frac{1}{2} \left[   \phantom{\sqrt{s_{j+\half}}} \left( \sqrt{g} B^\zeta B^\zeta \right)^{j+\half}
                                       + \phantom{\sqrt{s_{j-\half}}} \left( \sqrt{g} B^\zeta B^\zeta \right)^{j-\half} \right] R^\mathrm{e}(s_j) \nonumber \\
  ~&\,          -   \frac{1}{2} \left[            \sqrt{s_{j+\half}}  \left( \sqrt{g} B^\zeta B^\zeta \right)^{j+\half}
                                       +          \sqrt{s_{j-\half}}  \left( \sqrt{g} B^\zeta B^\zeta \right)^{j-\half} \right] R^\mathrm{o}(s_j) \label{eqn:armn_e}
\end{align}
and
\begin{align}
  \sqrt{s_j} A^{R,\mathrm{o}} (s_j)
  =&\, \phantom{+}~ \frac{1}{\Delta s} \left[ \sqrt{s_{j+\half}} \left( Z_\theta P \right)^{j+\half} - \sqrt{s_{j-\half}} \left( Z_\theta P \right)^{j-\half} \right] \nonumber \\
  ~&\,          -   \frac{1}{4} Z_\theta^\mathrm{e}(s_j) \left[ \left( \frac{P}{\sqrt{s}} \right)^{j+\half} + \left( \frac{P}{\sqrt{s}} \right)^{j-\half} \right] \nonumber \\
  ~&\,          -   \frac{1}{4} Z_\theta^\mathrm{o}(s_j) \left[ P \left(s_{j+\half}\right) + P\left(s_{j-\half}\right) \right] \nonumber \\
  ~&\,          +   \frac{1}{2}               \left[            \sqrt{s_{j+\half}}  \left( \tau p_\mathrm{tot} \right)^{j+\half}
                                                     +          \sqrt{s_{j-\half}}  \left( \tau p_\mathrm{tot} \right)^{j-\half} \right] \nonumber \\
  ~&\,          -   \frac{1}{2} \phantom{s_j} \left[            \sqrt{s_{j+\half}}  \left( \sqrt{g} B^\zeta B^\zeta \right)^{j+\half}
                                                     +          \sqrt{s_{j-\half}}  \left( \sqrt{g} B^\zeta B^\zeta \right)^{j-\half} \right] R^\mathrm{e}(s_j) \nonumber \\
  ~&\,          -   \frac{1}{2}          s_j  \left[   \phantom{\sqrt{s_{j+\half}}} \left( \sqrt{g} B^\zeta B^\zeta \right)^{j+\half}
                                                     + \phantom{\sqrt{s_{j-\half}}} \left( \sqrt{g} B^\zeta B^\zeta \right)^{j-\half} \right] R^\mathrm{o}(s_j) \label{eqn:armn_0} \, .
\end{align}
The $A^Z$ force component consists of two components for the even/odd-$m$ Fourier harmonics:
\begin{equation}
  A^Z(s_j) = A^{Z,\mathrm{e}}(s_j) + \sqrt{s_j} A^{Z,\mathrm{o}} (s_j) \, .
\end{equation}
The product differencing rule is employed to compute the even-$m$ component of $A^Z$:
\begin{equation}
  A^{Z,\mathrm{e}} (s_j) = - \frac{1}{\Delta s} \left[ \left( R_\theta P \right)^{j+\half} - \left( R_\theta P \right)^{j-\half} \right] \, .
\end{equation}
For the odd-$m$ component, the product rule is used backwards:
\begin{align}
  \frac{\partial}{\partial s} \left( \sqrt{s} R_\theta P \right) =&\, \sqrt{s} \frac{\partial}{\partial s} \left( R_\theta P \right) + \frac{1}{2 \sqrt{s}} R_\theta P \nonumber \\
  \Leftrightarrow
  \sqrt{s} A^{Z,\mathrm{o}} =
  - \sqrt{s} \frac{\partial}{\partial s} \left( R_\theta P \right) =&\, - \frac{\partial}{\partial s} \left( \sqrt{s} R_\theta P \right) + \frac{1}{2 \sqrt{s}} R_\theta P \, .
\end{align}
It follows for the odd-$m$ component of $A^Z$:
\begin{align}
  \sqrt{s_j} A^{Z,\mathrm{o}} (s_j)
  =&\, - \frac{1}{\Delta s} \left[ \sqrt{s_{j+\half}} \left( R_\theta P \right)^{j+\half} - \sqrt{s_{j-\half}} \left( R_\theta P \right)^{j-\half} \right] \nonumber \\
  ~&\, + \left[ \frac{1}{2 \sqrt{s}} \left( R_\theta^\mathrm{e} + \sqrt{s} R_\theta^\mathrm{o} \right) P \right]^j
\end{align}
with
\begin{equation}
  \left[ \frac{1}{2 \sqrt{s}} \left( R_\theta^\mathrm{e} + \sqrt{s} R_\theta^\mathrm{o} \right) P \right]^j
  = \frac{1}{2} R_\theta^\mathrm{e}(s_j) \left(\frac{P}{\sqrt{s}}\right)^j + \frac{1}{2} R_\theta^\mathrm{o}(s_j) P(s_j) \, .
\end{equation}
The interpolated values of $P/\sqrt{s}$ and $P$ from \eqn{psqrts_interp} and \eqn{p_interp} are used here again.
This leads to the final expression for $\sqrt{s} A^{Z,\mathrm{o}}$:
\begin{align}
  \sqrt{s_j} A^{Z,\mathrm{o}}(s_j)
  =&\, - \frac{1}{\Delta s} \left[ \sqrt{s_{j+\half}} \left( R_\theta P \right)^{j+\half} - \sqrt{s_{j-\half}} \left( R_\theta P \right)^{j-\half} \right] \nonumber \\
  ~&\, + \frac{1}{4} R_\theta^\mathrm{e}(s_j) \left[ \left( \frac{P}{\sqrt{s}} \right)^{j+\half} + \left( \frac{P}{\sqrt{s}} \right)^{j-\half} \right] \nonumber \\
  ~&\, + \frac{1}{4} R_\theta^\mathrm{o}(s_j) \left[ P \left(s_{j+\half}\right) + P\left(s_{j-\half}\right) \right] \, .
\end{align}

\subsection{B-Forces on R and Z}
%%%%%%% B^{R,Z} forces %%%%%%%%%
The force component $B^R$ is computed as follows, starting from \eqn{f_b_r}
and inserting the expression from \eqn{r_ptot} to \eqn{smallbzeta}:
\begin{align}
  B^R =&\, Z_s P - \frac{1}{\mu_0} B^\theta \left[          b^\theta          R_\theta +          b^\zeta          R_\zeta \right] \nonumber \\
      =&\, Z_s P - \frac{1}{\mu_0} B^\theta \left[ \sqrt{g} B^\theta          R_\theta + \sqrt{g} B^\zeta          R_\zeta \right] \nonumber \\
      =&\, Z_s P - \frac{1}{\mu_0}          \left[ \sqrt{g} B^\theta B^\theta R_\theta + \sqrt{g} B^\theta B^\zeta R_\zeta \right] \, .
\end{align}
Recall that since $Z = Z^\mathrm{e} + \sqrt{s} Z^\mathrm{o}$, it follows:
\begin{equation}
  Z_s = \frac{\partial Z}{\partial s} = \frac{\partial Z^\mathrm{e}}{\partial s} + \sqrt{s} \frac{\partial Z^\mathrm{o}}{\partial s} + \frac{1}{2 \sqrt{s}} Z^o \, .
\end{equation}
For brevity we introduce (see Sec. \ref{sec:jacobian}):
\begin{equation}
  \tilde{Z}_s \equiv \frac{\partial Z^\mathrm{e}}{\partial s} + \sqrt{s} \frac{\partial Z^\mathrm{o}}{\partial s} = \code{dZdSHalf} \, ,
\end{equation}
leading to:
\begin{equation}
  Z_s = \tilde{Z}_s + \frac{1}{2 \sqrt{s}} Z^o \, .
\end{equation}
Two force components are required: one acting on the even-$m$ Fourier coefficients of the geomentry
and one acting on the odd-$m$ Fourier coefficients:
\begin{equation}
  B^R = B^{R,\mathrm{e}} + \sqrt{s} B^{R,\mathrm{o}}
\end{equation}
The first term ($Z_s P$) in $B^{R,\mathrm{e}}$ is implemented as follows:
\begin{equation}
  \left( Z_s P \right) (s_j)
  =   \frac{1}{2} \left[ \left( \tilde{Z}_s P \right)^{j+\half} + \left( \tilde{Z}_s P \right)^{j-\half} \right]
    + Z^o(s_j) \left[ \frac{P^{j+\half}}{4 \sqrt{s_{j+\half}}}  + \frac{P^{j-\half}}{4 \sqrt{s_{j-\half}}}  \right] \, ,
\end{equation}
where the last term is built using interpolation:
\begin{equation}
  \left( \frac{1}{2 \sqrt{s}} Z^o P \right) (s_j)
  = \frac{1}{2} Z^o(s_j) \left(\frac{P}{\sqrt{s}}\right)^j
  = \frac{1}{2} Z^o(s_j) \frac{1}{2} \left[ \left(\frac{P}{\sqrt{s}}\right)^{j+\half} + \left(\frac{P}{\sqrt{s}}\right)^{j-\half} \right] \, .
\end{equation}
The second term in $B^{R,\mathrm{e}}$ is implemented as follows:
\begin{align}
  - \left( B^\theta b_R \right) (s_j)
  =&\, \phantom{-}~ \frac{1}{2} \left[   \phantom{\sqrt{s_{j+\half}}} \left( \sqrt{g} B^\theta B^\theta \right)^{j+\half}
                                       + \phantom{\sqrt{s_{j-\half}}} \left( \sqrt{g} B^\theta B^\theta \right)^{j-\half} \right] R_\theta^\mathrm{e}(s_j) \nonumber \\
  ~&\,          -   \frac{1}{2} \left[            \sqrt{s_{j+\half}}  \left( \sqrt{g} B^\theta B^\theta \right)^{j+\half}
                                       +          \sqrt{s_{j-\half}}  \left( \sqrt{g} B^\theta B^\theta \right)^{j-\half} \right] R_\theta^\mathrm{o}(s_j) \nonumber \\
  ~&\,          -   \frac{1}{2} \left[   \phantom{\sqrt{s_{j+\half}}} \left( \sqrt{g} B^\theta  B^\zeta \right)^{j+\half}
                                       + \phantom{\sqrt{s_{j-\half}}} \left( \sqrt{g} B^\theta  B^\zeta \right)^{j-\half} \right]  R_\zeta^\mathrm{e}(s_j) \nonumber \\
  ~&\,          -   \frac{1}{2} \left[            \sqrt{s_{j+\half}}  \left( \sqrt{g} B^\theta  B^\zeta \right)^{j+\half}
                                       +          \sqrt{s_{j-\half}}  \left( \sqrt{g} B^\theta  B^\zeta \right)^{j-\half} \right]  R_\zeta^\mathrm{o}(s_j) \, .
\end{align}
Putting things together
% ({\color{red} and ignoring a missing factor of $\mu_0$ for now...})
, the full expression for $B^{R,\mathrm{e}}$ reads:
\begin{align}
  B^{R,\mathrm{e}} (s_j)
  =&\, \phantom{-}~ \frac{1}{2} \left[ \left( \tilde{Z}_s P \right)^{j+\half} + \left( \tilde{Z}_s P \right)^{j-\half} \right]
    + Z^o(s_j) \left[ \frac{P^{j+\half}}{4 \sqrt{s_{j+\half}}}  + \frac{P^{j-\half}}{4 \sqrt{s_{j-\half}}}  \right] \nonumber \\
  ~&\, - \frac{1}{2} \left[   \phantom{\sqrt{s_{j+\half}}} \left( \sqrt{g} B^\theta B^\theta \right)^{j+\half}
                            + \phantom{\sqrt{s_{j-\half}}} \left( \sqrt{g} B^\theta B^\theta \right)^{j-\half} \right] R_\theta^\mathrm{e}(s_j) \nonumber \\
  ~&\, - \frac{1}{2} \left[            \sqrt{s_{j+\half}}  \left( \sqrt{g} B^\theta B^\theta \right)^{j+\half}
                            +          \sqrt{s_{j-\half}}  \left( \sqrt{g} B^\theta B^\theta \right)^{j-\half} \right] R_\theta^\mathrm{o}(s_j) \nonumber \\
  ~&\, - \frac{1}{2} \left[   \phantom{\sqrt{s_{j+\half}}} \left( \sqrt{g} B^\theta  B^\zeta \right)^{j+\half}
                            + \phantom{\sqrt{s_{j-\half}}} \left( \sqrt{g} B^\theta  B^\zeta \right)^{j-\half} \right]  R_\zeta^\mathrm{e}(s_j) \nonumber \\
  ~&\, - \frac{1}{2} \left[            \sqrt{s_{j+\half}}  \left( \sqrt{g} B^\theta  B^\zeta \right)^{j+\half}
                            +          \sqrt{s_{j-\half}}  \left( \sqrt{g} B^\theta  B^\zeta \right)^{j-\half} \right]  R_\zeta^\mathrm{o}(s_j) \label{eqn:brmn_e} \, .
\end{align}
The last two terms are only taken into account in case of a three-dimensional configuration,
since $R_\zeta$ is zero in axisymmetric (or two-dimensional) configurations.
The odd-$m$ $B^R$ force component includes an additional factor of $\sqrt{s}$
and thus reads:
\begin{align}
  \sqrt{s_j} B^{R,\mathrm{o}} (s_j)
  =&\, \phantom{-}~ \frac{1}{2} \left[ \sqrt{s_{j+\half}} \left( \tilde{Z}_s P \right)^{j+\half} + \sqrt{s_{j-\half}} \left( \tilde{Z}_s P \right)^{j-\half} \right]
    + \frac{1}{4} Z^o(s_j) \left[ P^{j+\half} + P^{j-\half} \right] \nonumber \\
  ~&\, - \frac{1}{2} \phantom{s_j} \left[            \sqrt{s_{j+\half}}  \left( \sqrt{g} B^\theta B^\theta \right)^{j+\half}
                                          +          \sqrt{s_{j-\half}}  \left( \sqrt{g} B^\theta B^\theta \right)^{j-\half} \right] R_\theta^\mathrm{e}(s_j) \nonumber \\
  ~&\, - \frac{1}{2}          s_j  \left[   \phantom{\sqrt{s_{j+\half}}} \left( \sqrt{g} B^\theta B^\theta \right)^{j+\half}
                                          + \phantom{\sqrt{s_{j-\half}}} \left( \sqrt{g} B^\theta B^\theta \right)^{j-\half} \right] R_\theta^\mathrm{o}(s_j) \nonumber \\
  ~&\, - \frac{1}{2} \phantom{s_j} \left[            \sqrt{s_{j+\half}}  \left( \sqrt{g} B^\theta  B^\zeta \right)^{j+\half}
                                          +          \sqrt{s_{j-\half}}  \left( \sqrt{g} B^\theta  B^\zeta \right)^{j-\half} \right]  R_\zeta^\mathrm{e}(s_j) \nonumber \\
  ~&\, - \frac{1}{2}          s_j  \left[   \phantom{\sqrt{s_{j+\half}}} \left( \sqrt{g} B^\theta  B^\zeta \right)^{j+\half}
                                          + \phantom{\sqrt{s_{j-\half}}} \left( \sqrt{g} B^\theta  B^\zeta \right)^{j-\half} \right]  R_\zeta^\mathrm{o}(s_j) \label{eqn:brmn_o} \, .
\end{align}
Note that in second term in the first line, the odd-$m$ weighting factor $\sqrt{s}$ cancelled with the $1/\sqrt{s}$ from the chain rule
of the radial derivative.
The expressions for the components of $B^Z$ are found by replacing
$R_\theta \rightarrow Z_\theta$, $R_\zeta \rightarrow Z_\zeta$ and $Z_s \rightarrow R_s$, $Z^\mathrm{o} \rightarrow R^\mathrm{o}$
in the expressions for the components of $B^R$.
The derivation starts from
\begin{equation}
  B^Z = -R_s P - \frac{1}{\mu_0} B^\theta b_Z \, .
\end{equation}
After analogous operations as done above for $B^R$, we arrive at the following forms for the $B^Z$ force components.
The even-$m$ term thus reads:
\begin{align}
  B^{Z,\mathrm{e}} (s_j)
  =&\, - \frac{1}{2} \left[ \left( \tilde{R}_s P \right)^{j+\half} + \left( \tilde{R}_s P \right)^{j-\half} \right]
       - R^o(s_j) \left[ \frac{P^{j+\half}}{4 \sqrt{s_{j+\half}}}  + \frac{P^{j-\half}}{4 \sqrt{s_{j-\half}}}  \right] \nonumber \\
  ~&\, - \frac{1}{2} \left[   \phantom{\sqrt{s_{j+\half}}} \left( \sqrt{g} B^\theta B^\theta \right)^{j+\half}
                            + \phantom{\sqrt{s_{j-\half}}} \left( \sqrt{g} B^\theta B^\theta \right)^{j-\half} \right] Z_\theta^\mathrm{e}(s_j) \nonumber \\
  ~&\, - \frac{1}{2} \left[            \sqrt{s_{j+\half}}  \left( \sqrt{g} B^\theta B^\theta \right)^{j+\half}
                            +          \sqrt{s_{j-\half}}  \left( \sqrt{g} B^\theta B^\theta \right)^{j-\half} \right] Z_\theta^\mathrm{o}(s_j) \nonumber \\
  ~&\, - \frac{1}{2} \left[   \phantom{\sqrt{s_{j+\half}}} \left( \sqrt{g} B^\theta  B^\zeta \right)^{j+\half}
                            + \phantom{\sqrt{s_{j-\half}}} \left( \sqrt{g} B^\theta  B^\zeta \right)^{j-\half} \right]  Z_\zeta^\mathrm{e}(s_j) \nonumber \\
  ~&\, - \frac{1}{2} \left[            \sqrt{s_{j+\half}}  \left( \sqrt{g} B^\theta  B^\zeta \right)^{j+\half}
                            +          \sqrt{s_{j-\half}}  \left( \sqrt{g} B^\theta  B^\zeta \right)^{j-\half} \right]  Z_\zeta^\mathrm{o}(s_j) \label{eqn:bzmn_e}
\end{align}
and the odd-$m$ term reads:
\begin{align}
  \sqrt{s_j} B^{Z,\mathrm{o}} (s_j)
  =&\, - \frac{1}{2} \left[ \sqrt{s_{j+\half}} \left( \tilde{R}_s P \right)^{j+\half} + \sqrt{s_{j-\half}} \left( \tilde{R}_s P \right)^{j-\half} \right]
       - \frac{1}{4} R^o(s_j) \left[ P^{j+\half} + P^{j-\half} \right] \nonumber \\
  ~&\, - \frac{1}{2} \phantom{s_j} \left[            \sqrt{s_{j+\half}}  \left( \sqrt{g} B^\theta B^\theta \right)^{j+\half}
                                          +          \sqrt{s_{j-\half}}  \left( \sqrt{g} B^\theta B^\theta \right)^{j-\half} \right] Z_\theta^\mathrm{e}(s_j) \nonumber \\
  ~&\, - \frac{1}{2}          s_j  \left[   \phantom{\sqrt{s_{j+\half}}} \left( \sqrt{g} B^\theta B^\theta \right)^{j+\half}
                                          + \phantom{\sqrt{s_{j-\half}}} \left( \sqrt{g} B^\theta B^\theta \right)^{j-\half} \right] Z_\theta^\mathrm{o}(s_j) \nonumber \\
  ~&\, - \frac{1}{2} \phantom{s_j} \left[            \sqrt{s_{j+\half}}  \left( \sqrt{g} B^\theta  B^\zeta \right)^{j+\half}
                                          +          \sqrt{s_{j-\half}}  \left( \sqrt{g} B^\theta  B^\zeta \right)^{j-\half} \right]  Z_\zeta^\mathrm{e}(s_j) \nonumber \\
  ~&\, - \frac{1}{2}          s_j  \left[   \phantom{\sqrt{s_{j+\half}}} \left( \sqrt{g} B^\theta  B^\zeta \right)^{j+\half}
                                          + \phantom{\sqrt{s_{j-\half}}} \left( \sqrt{g} B^\theta  B^\zeta \right)^{j-\half} \right]  Z_\zeta^\mathrm{o}(s_j) \label{eqn:bzmn_o} \, .
\end{align}

\subsection{C-Forces on R and Z}
%%%%%%% C^{R,Z} forces %%%%%%%%%
The force component $C^R$ is computed as follows, starting from \eqn{f_c_r}
and inserting the expression from \eqn{smallbr} to \eqn{smallbzeta}:
\begin{align}
  C^R =&\, \frac{1}{\mu_0} B^\zeta \left[          b^\theta         R_\theta +          b^\zeta         R_\zeta \right] \nonumber \\
      =&\, \frac{1}{\mu_0} B^\zeta \left[ \sqrt{g} B^\theta         R_\theta + \sqrt{g} B^\zeta         R_\zeta \right] \nonumber \\
      =&\, \frac{1}{\mu_0}         \left[ \sqrt{g} B^\theta B^\zeta R_\theta + \sqrt{g} B^\zeta B^\zeta R_\zeta \right] \, .
\end{align}
The tangential derivatives are defined as contributions from even-$m$ and odd-$m$ Fourier harmonics:
\begin{align}
  R_\theta = R_\theta^\mathrm{e} + \sqrt{s} R_\theta^\mathrm{o} \nonumber \\
   R_\zeta =  R_\zeta^\mathrm{e} + \sqrt{s}  R_\zeta^\mathrm{o} \, .
\end{align}
Two force components are required: one acting on the even-$m$ Fourier coefficients of the geomentry
and one acting on the odd-$m$ Fourier coefficients:
\begin{equation}
  C^R = C^{R,\mathrm{e}} + \sqrt{s} C^{R,\mathrm{o}}
\end{equation}
with
\begin{align}
  C^{R,\mathrm{e}} =&\,                     \left( \sqrt{g} B^\theta  B^\zeta \right) R_\theta^\mathrm{e}
                                 +          \left( \sqrt{g}  B^\zeta  B^\zeta \right)  R_\zeta^\mathrm{e}
                                 + \sqrt{s} \left( \sqrt{g}  B^\zeta  B^\zeta \right)  R_\zeta^\mathrm{o}
                                 + \sqrt{s} \left( \sqrt{g} B^\theta  B^\zeta \right) R_\theta^\mathrm{o} \nonumber \\
  \sqrt{s} C^{R,\mathrm{o}} =&\,   \sqrt{s} \left( \sqrt{g} B^\theta  B^\zeta \right) R_\theta^\mathrm{e}
                                 + \sqrt{s} \left( \sqrt{g}  B^\zeta  B^\zeta \right)  R_\zeta^\mathrm{e}
                                 +       s  \left( \sqrt{g}  B^\zeta  B^\zeta \right)  R_\zeta^\mathrm{o}
                                 +       s  \left( \sqrt{g} B^\theta  B^\zeta \right) R_\theta^\mathrm{o} \, .
\end{align}
The force components are required on the full-grid, but the magnetic field components and the Jacobian are only available on the half-grid.
Radial interpolation is thus needed and leads to the following implementation of the $C^R$ force components:
\begin{align}
  C^{R,\mathrm{e}} (s_j)
  =&\, \phantom{+}~ \frac{1}{2} \left[   \phantom{\sqrt{s_{j+\half}}} \left( \sqrt{g} B^\theta  B^\zeta \right)^{j+\half}
                                       + \phantom{\sqrt{s_{j-\half}}} \left( \sqrt{g} B^\theta  B^\zeta \right)^{j-\half} \right] R_\theta^\mathrm{e}(s_j) \nonumber \\
  ~&\,          +   \frac{1}{2} \left[            \sqrt{s_{j+\half}}  \left( \sqrt{g} B^\theta  B^\zeta \right)^{j+\half}
                                       +          \sqrt{s_{j-\half}}  \left( \sqrt{g} B^\theta  B^\zeta \right)^{j-\half} \right] R_\theta^\mathrm{o}(s_j) \nonumber \\
  ~&\,          +   \frac{1}{2} \left[   \phantom{\sqrt{s_{j+\half}}} \left( \sqrt{g}  B^\zeta  B^\zeta \right)^{j+\half}
                                       + \phantom{\sqrt{s_{j-\half}}} \left( \sqrt{g}  B^\zeta  B^\zeta \right)^{j-\half} \right]  R_\zeta^\mathrm{e}(s_j) \nonumber \\
  ~&\,          +   \frac{1}{2} \left[            \sqrt{s_{j+\half}}  \left( \sqrt{g}  B^\zeta  B^\zeta \right)^{j+\half}
                                       +          \sqrt{s_{j-\half}}  \left( \sqrt{g}  B^\zeta  B^\zeta \right)^{j-\half} \right]  R_\zeta^\mathrm{o}(s_j) \label{eqn:crmn_e}
\end{align}
and
\begin{align}
  \sqrt{s_j} C^{R,\mathrm{o}} (s_j)
  =&\, \phantom{+}~ \frac{1}{2} \phantom{s_j} \left[            \sqrt{s_{j+\half}}  \left( \sqrt{g} B^\theta  B^\zeta \right)^{j+\half}
                                                     +          \sqrt{s_{j-\half}}  \left( \sqrt{g} B^\theta  B^\zeta \right)^{j-\half} \right] R_\theta^\mathrm{e}(s_j) \nonumber \\
  ~&\,          +   \frac{1}{2}          s_j  \left[   \phantom{\sqrt{s_{j+\half}}} \left( \sqrt{g} B^\theta  B^\zeta \right)^{j+\half}
                                                     + \phantom{\sqrt{s_{j-\half}}} \left( \sqrt{g} B^\theta  B^\zeta \right)^{j-\half} \right] R_\theta^\mathrm{o}(s_j) \nonumber \\
  ~&\,          +   \frac{1}{2} \phantom{s_j} \left[            \sqrt{s_{j+\half}}  \left( \sqrt{g}  B^\zeta  B^\zeta \right)^{j+\half}
                                                     +          \sqrt{s_{j-\half}}  \left( \sqrt{g}  B^\zeta  B^\zeta \right)^{j-\half} \right]  R_\zeta^\mathrm{e}(s_j) \nonumber \\
  ~&\,          +   \frac{1}{2}          s_j  \left[   \phantom{\sqrt{s_{j+\half}}} \left( \sqrt{g}  B^\zeta  B^\zeta \right)^{j+\half}
                                                     + \phantom{\sqrt{s_{j-\half}}} \left( \sqrt{g}  B^\zeta  B^\zeta \right)^{j-\half} \right]  R_\zeta^\mathrm{o}(s_j) \label{eqn:crmn_o} \, .
\end{align}
This type of weighted radial interpolation was already found in Sec.~\ref{sec:metric_elements}.
The force component $C^Z$ is computed as follows, starting from \eqn{f_c_z}:
\begin{align}
  C^Z =&\, \frac{1}{\mu_0} B^\zeta \left[ b^\theta Z_\theta + b^\zeta Z_\zeta \right] \nonumber \\
      =&\, \frac{1}{\mu_0} B^\zeta \left[ \sqrt{g} B^\theta Z_\theta + \sqrt{g} B^\zeta Z_\zeta \right] \nonumber \\
      =&\, \frac{1}{\mu_0} \sqrt{g} B^\zeta \left[  B^\theta Z_\theta + B^\zeta Z_\zeta \right] \, .
\end{align}
As all force components, $C^Z$ as well has separate contributions for the even-$m$ and the odd-$m$ Fourier harmonics:
\begin{equation}
  C^Z = C^{Z,\mathrm{e}} + \sqrt{s} C^{Z,\mathrm{o}} \, .
\end{equation}
The final expressions are found by the replacements $R_\theta \rightarrow Z_\theta$ and $R_\zeta \rightarrow Z_\zeta$
in \eqn{crmn_e} and \eqn{crmn_o}:
\begin{align}
  C^{Z,\mathrm{e}} (s_j)
  =&\, \phantom{+}~ \frac{1}{2} \left[   \phantom{\sqrt{s_{j+\half}}} \left( \sqrt{g} B^\theta  B^\zeta \right)^{j+\half}
                                       + \phantom{\sqrt{s_{j-\half}}} \left( \sqrt{g} B^\theta  B^\zeta \right)^{j-\half} \right] Z_\theta^\mathrm{e}(s_j) \nonumber \\
  ~&\,          +   \frac{1}{2} \left[            \sqrt{s_{j+\half}}  \left( \sqrt{g} B^\theta  B^\zeta \right)^{j+\half}
                                       +          \sqrt{s_{j-\half}}  \left( \sqrt{g} B^\theta  B^\zeta \right)^{j-\half} \right] Z_\theta^\mathrm{o}(s_j) \nonumber \\
  ~&\,          +   \frac{1}{2} \left[   \phantom{\sqrt{s_{j+\half}}} \left( \sqrt{g}  B^\zeta  B^\zeta \right)^{j+\half}
                                       + \phantom{\sqrt{s_{j-\half}}} \left( \sqrt{g}  B^\zeta  B^\zeta \right)^{j-\half} \right]  Z_\zeta^\mathrm{e}(s_j) \nonumber \\
  ~&\,          +   \frac{1}{2} \left[            \sqrt{s_{j+\half}}  \left( \sqrt{g}  B^\zeta  B^\zeta \right)^{j+\half}
                                       +          \sqrt{s_{j-\half}}  \left( \sqrt{g}  B^\zeta  B^\zeta \right)^{j-\half} \right]  Z_\zeta^\mathrm{o}(s_j) \label{eqn:czmn_e}
\end{align}
as well as
\begin{align}
  \sqrt{s_j} C^{Z,\mathrm{o}} (s_j)
  =&\, \phantom{+}~ \frac{1}{2} \phantom{s_j} \left[            \sqrt{s_{j+\half}}  \left( \sqrt{g} B^\theta  B^\zeta \right)^{j+\half}
                                                     +          \sqrt{s_{j-\half}}  \left( \sqrt{g} B^\theta  B^\zeta \right)^{j-\half} \right] Z_\theta^\mathrm{e}(s_j) \nonumber \\
  ~&\,          +   \frac{1}{2}          s_j  \left[   \phantom{\sqrt{s_{j+\half}}} \left( \sqrt{g} B^\theta  B^\zeta \right)^{j+\half}
                                                     + \phantom{\sqrt{s_{j-\half}}} \left( \sqrt{g} B^\theta  B^\zeta \right)^{j-\half} \right] Z_\theta^\mathrm{o}(s_j) \nonumber \\
  ~&\,          +   \frac{1}{2} \phantom{s_j} \left[            \sqrt{s_{j+\half}}  \left( \sqrt{g}  B^\zeta  B^\zeta \right)^{j+\half}
                                                     +          \sqrt{s_{j-\half}}  \left( \sqrt{g}  B^\zeta  B^\zeta \right)^{j-\half} \right]  Z_\zeta^\mathrm{e}(s_j) \nonumber \\
  ~&\,          +   \frac{1}{2}          s_j  \left[   \phantom{\sqrt{s_{j+\half}}} \left( \sqrt{g}  B^\zeta  B^\zeta \right)^{j+\half}
                                                     + \phantom{\sqrt{s_{j-\half}}} \left( \sqrt{g}  B^\zeta  B^\zeta \right)^{j-\half} \right]  Z_\zeta^\mathrm{o}(s_j) \label{eqn:czmn_o} \, .
\end{align}
Note that the $C$ force components are only needed when computing three-dimensional equilibria,
i.e., they do not enter in an axisymmetric calculation.

\FloatBarrier
\newpage
\section{Radial Preconditioner}
The highest-order radial derivatives in the MHD force terms
are used to define a one-dimensional radial derivatives.
In order to identify these terms,
one starts from the first terms of the MHD forces
defined in Eqn.~(18a) and~(18b) in Ref.~\cite{hirshman_whitson_1983}:
\begin{align}
 F_R =&\,  \frac{\partial}{\partial \rho} \left(Z_\theta P \right) + ... \\
 F_Z =&\, -\frac{\partial}{\partial \rho} \left(R_\theta P \right) + ...
\end{align}
with
\begin{equation}
 P = R \left( p + \frac{|B|^2}{2 \mu_0} \right) \, .
\end{equation}
The product rule is used to derive:
\begin{align}
 F_R =&\,  \frac{\partial Z_\theta}{\partial \rho} P + Z_\theta \frac{\partial P}{\partial \rho} + ... \\
 F_Z =&\, -\frac{\partial R_\theta}{\partial \rho} P - R_\theta \frac{\partial P}{\partial \rho} + ... \, .
\end{align}
The first terms in these expressions
do not contribute second-order radial derivatives
and are thus omitted in the following.
The radial derivate of~$P$ is considered next:
\begin{align}
 \frac{\partial P}{\partial \rho}
 =&\, \frac{\partial}{\partial \rho} \left[ R \left( p + \frac{|B|^2}{2 \mu_0} \right) \right] \nonumber \\
 =&\,   \frac{\partial R}{\partial \rho} \left( p + \frac{|B|^2}{2 \mu_0} \right)
      + R \frac{\partial}{\partial \rho} \left( p + \frac{|B|^2}{2 \mu_0} \right) \nonumber \\
 =&\,   \frac{\partial R}{\partial \rho} \left( p + \frac{|B|^2}{2 \mu_0} \right)
      + R \left(  \frac{\partial p}{\partial \rho}
                + \frac{1}{2 \mu_0} \frac{\partial (|B|^2)}{\partial \rho} \right) \nonumber \\
 =&\, \frac{R}{2 \mu_0} \frac{\partial (|B|^2)}{\partial \rho} + ... \, .
\end{align}
Now the magnetic pressure~$|B|^2$ is considered again:
\begin{align}
 |B|^2 =&\, B^\theta B_\theta + B^\zeta B_\zeta \\
 =&\,   (B^\theta)^2     g_{\theta \theta}
      + 2 B^\theta B^\zeta g_{\theta \zeta}
      + (B^\zeta)^2     g_{\zeta \zeta}
\end{align}
The metric elements appearing here
only contain derivatives in the tangential directions
and thus do not contribute second-order derivatives in the radial direction.
It follows:
\begin{align}
 \frac{\partial}{\partial \rho} (B^\theta)^2
 =&\, 2 B^\theta \frac{\partial B^\theta}{\partial \rho} \nonumber \\
 =&\, 2 B^\theta \frac{\partial}{\partial \rho} \left[
        \frac{\Phi'}{\sqrt{g}} \left( \iota - \lambda_\zeta  \right) \right] \nonumber \\
 =&\, 2 B^\theta \Phi' \left( \iota - \lambda_\zeta  \right)
      \frac{\partial}{\partial \rho} \left( \frac{1}{\sqrt{g}} \right) + ... \nonumber \\
 =&\, 2 (B^\theta)^2 \sqrt{g}
      \frac{\partial}{\partial \rho} \left( \sqrt{g} \right)^{-1} + ... \nonumber \\
 =&\, 2 (B^\theta)^2 \sqrt{g} \frac{1}{(\sqrt{g})^2}
      \frac{\partial \sqrt{g}}{\partial \rho} + ... \nonumber \\
 =&\, 2 \frac{(B^\theta)^2}{\sqrt{g}}
      \frac{\partial \sqrt{g}}{\partial \rho} + ... \, .
\end{align}
In this expression, we only need to care about
the radial derivative of $\sqrt{g}$
and use the form from Eqn~(17a) and~(17b) of Ref.~\cite{hirshman_whitson_1983}
(although for consistency with other articles, we use~$\tau$ for~$G$:
\begin{align}
 \sqrt{g} =&\, R \tau \nonumber \\
 \textrm{with } \tau =&\, R_\theta Z_\rho - R_\rho Z_\theta \, .
\end{align}
Now it follows:
\begin{align}
 \frac{\partial \sqrt{g}}{\partial \rho}
 =&\, R_\rho \tau + R \frac{\partial \tau}{\partial \rho} \nonumber \\
 =&\, R \frac{\partial}{\partial \rho}
      \left( R_\theta Z_\rho - R_\rho Z_\theta \right) + ... \nonumber \\
 =&\, R \left[   R_\theta \frac{\partial^2 Z}{\partial \rho^2}
               - Z_\theta \frac{\partial^2 R}{\partial \rho^2} \right] + ... \, .
\end{align}
Here, finally the second-order derivatives appear.
Going back to the radial derivative of the magnetic pressure:
\begin{align}
 \frac{\partial}{\partial \rho} (B^\theta B^\zeta)
 =&\, B^\theta \frac{\partial B^\zeta }{\partial \rho}
     + B^\zeta \frac{\partial B^\theta}{\partial \rho} \nonumber \\
 =&\, \left(  B^\theta \frac{B^\zeta }{\sqrt{g}}
            + B^\zeta  \frac{B^\theta}{\sqrt{g}} \right)
            \frac{\partial \sqrt{g}}{\partial \rho} \nonumber \\
 =&\, 2 \frac{B^\theta B^\zeta}{\sqrt{g}}
        \frac{\partial \sqrt{g}}{\partial \rho}
\end{align}
as well as
\begin{align}
 \frac{\partial}{\partial \rho} (B^\zeta)^2
 =&\, 2 B^\zeta \frac{\partial B^\zeta}{\partial \rho} \nonumber \\
 =&\, 2 \frac{(B^\zeta)^2}{\sqrt{g}}
       \frac{\partial \sqrt{g}}{\partial \rho} \, .
\end{align}
Putting things back together, we arrive at:
\begin{align}
 \frac{\partial (|B|^2)}{\partial \rho}
 =&\,     \frac{\partial (B^\theta)^2      }{\partial \rho} g_{\theta \theta}
      + 2 \frac{\partial (B^\theta B^\zeta)}{\partial \rho} g_{\theta \zeta }
      +   \frac{\partial (B^\zeta)^2       }{\partial \rho} g_{\zeta  \zeta } + ... \nonumber \\
 =&\, \frac{2 |B|^2}{\sqrt{g}} \frac{\partial \sqrt{g}}{\partial \rho} + ... \, .
\end{align}
Now using the radial derivative of~$\sqrt{g}$,
we arrive at:
\begin{align}
 \frac{\partial (|B|^2)}{\partial \rho}
 =&\, \frac{2 |B|^2}{\sqrt{g} / R}
      \left[   R_\theta \frac{\partial^2 Z}{\partial \rho^2}
             - Z_\theta \frac{\partial^2 R}{\partial \rho^2} \right] + ... \, .
\end{align}
It follows:
\begin{align}
 \frac{\partial P}{\partial \rho}
 =&\, \frac{R}{2 \mu_0}
      \frac{2 |B|^2}{\sqrt{g} / R}
      \left[   R_\theta \frac{\partial^2 Z}{\partial \rho^2}
             - Z_\theta \frac{\partial^2 R}{\partial \rho^2} \right] + ... \nonumber \\
 =&\, \frac{R}{\mu_0}
      \frac{|B|^2}{\tau}
      \left[   R_\theta \frac{\partial^2 Z}{\partial \rho^2}
             - Z_\theta \frac{\partial^2 R}{\partial \rho^2} \right] + ... \, .
\end{align}
It is useful to introduce a shorthand notation here:
\begin{equation}
 d_0 \equiv \frac{R |B|^2}{\mu_0 \tau} \, ,
\end{equation}
which allows to write:
\begin{align}
 \frac{\partial P}{\partial \rho}
 =&\, d_0
      \left[   R_\theta \frac{\partial^2 Z}{\partial \rho^2}
             - Z_\theta \frac{\partial^2 R}{\partial \rho^2} \right] + ... \, .
\end{align}
Putting this back into the expressions
for the $R$ and $Z$ MHD forces allows to write:
\begin{align}
 F_R =&\,  d_0 Z_\theta \left[   R_\theta \frac{\partial^2 Z}{\partial \rho^2}
                               - Z_\theta \frac{\partial^2 R}{\partial \rho^2} \right] + ... \\
 F_Z =&\, -d_0 R_\theta \left[   R_\theta \frac{\partial^2 Z}{\partial \rho^2}
                               - Z_\theta \frac{\partial^2 R}{\partial \rho^2} \right] + ...
\end{align}
and equivalently:
\begin{align}
 F_R =&\, - Z_\theta^2        d_0 \frac{\partial^2 R}{\partial \rho^2}
          + R_\theta Z_\theta d_0 \frac{\partial^2 Z}{\partial \rho^2} + ... \\
 F_Z =&\,   R_\theta Z_\theta d_0 \frac{\partial^2 R}{\partial \rho^2}
          - R_\theta^2        d_0 \frac{\partial^2 Z}{\partial \rho^2} + ...  \, .
\end{align}
This allows to introduce diffusion coefficients~$D_{ij}$ as follows:
\begin{align}
 D_{RR} =&\, Z_\theta^2        d_0 \\
 D_{RZ} =&\, R_\theta Z_\theta d_0 \\
 D_{ZZ} =&\, R_\theta^2        d_0
\end{align}
and in turn formulate the MHD forces as follows:
\begin{align}
 F_R =&\, - D_{RR} \frac{\partial^2 R}{\partial \rho^2}
          + D_{RZ} \frac{\partial^2 Z}{\partial \rho^2} + ... \\
 F_Z =&\, \phantom{-}\,  D_{RZ} \frac{\partial^2 R}{\partial \rho^2}
          - D_{ZZ} \frac{\partial^2 Z}{\partial \rho^2} + ...  \, .
\end{align}
The diagonal elements~$D_{RR}$ and $D_{ZZ}$
can now be used to formulate a radial preconditioner.
Denoting the magnetic pressure as~$P_B = |B|^2 / (2 \mu_0)$,
we can write:
\begin{align}
 D_{RR}
 =&\, Z_\theta^2 d_0
 =    Z_\theta^2 \frac{R |B|^2}{\mu_0 \tau}
 =    2 \frac{Z_\theta^2 R |B|^2}{2 \mu_0 \sqrt{g}/R} \nonumber \\
 =&\, 2 \frac{(Z_\theta R)^2 P_B}{\sqrt{g}}
 =    2 \frac{Z_\theta^2 R P_B}{\tau}
\end{align}
and
\begin{align}
 D_{ZZ}
 =&\, R_\theta^2 d_0
 =    R_\theta^2 \frac{R |B|^2}{\mu_0 \tau}
 =    2 \frac{R_\theta^2 R |B|^2}{2 \mu_0 \sqrt{g}/R} \nonumber \\
 =&\, 2 \frac{(R_\theta R)^2 P_B}{\sqrt{g}}
 =    2 \frac{R_\theta^2 R P_B}{\tau} \, .
\end{align}

A similar excercise can now be done to identify
the highest-order poloidal and toroidal derivatives.

The derivation starts with considering
the second-order poloidal derivatives in the MHD forces:
\begin{align}
 F_R =&\, -\frac{\partial}{\partial \theta} \left(Z_\rho P \right)
          + \frac{1}{\mu_0} \frac{\partial}{\partial \theta} \left( B^\theta b_R \right)
          + ... \\
 F_Z =&\,  \frac{\partial}{\partial \theta} \left(R_\rho P \right)
          + \frac{1}{\mu_0} \frac{\partial}{\partial \theta} \left( B^\theta b_Z \right)
          + ...
\end{align}
with the cylindrical magnetic field components~$b_R$ and~$b_Z$:
\begin{align}
 b_R =&\, b^\theta R_\theta + b^\zeta R_\zeta \\
 b_Z =&\, b^\theta Z_\theta + b^\zeta Z_\zeta
\end{align}
with
\begin{align}
 b^\theta =&\, \sqrt{g} B^\theta \\
 b^\zeta  =&\, \sqrt{g} B^\zeta  \, .
\end{align}
Putting this together, we arrive at:
\begin{align}
 F_R =&\, -\frac{\partial}{\partial \theta} \left(Z_\rho P \right)
          + \frac{1}{\mu_0} \frac{\partial}{\partial \theta} \left[
            B^\theta \sqrt{g} \left( B^\theta R_\theta + B^\zeta R_\zeta \right) \right]
          + ... \\
 F_Z =&\,  \frac{\partial}{\partial \theta} \left(R_\rho P \right)
          + \frac{1}{\mu_0} \frac{\partial}{\partial \theta} \left[
             B^\theta \sqrt{g} \left( B^\theta Z_\theta + B^\zeta Z_\zeta \right) \right]
          + ... \, .
\end{align}
In the first terms, only the poloidal derivatives of $P$
contribute second-order poloidal derivatives
and we can thus write:
\begin{align}
 F_R =&\, - Z_\rho \frac{\partial P}{\partial \theta}
          + \frac{1}{\mu_0} \frac{\partial}{\partial \theta}
            \left[ \sqrt{g} \left(   \left( B^\theta \right)^2 R_\theta
                                   + B^\theta B^\zeta          R_\zeta  \right) \right]
          + ... \\
 F_Z =&\,   R_\rho \frac{\partial P}{\partial \theta}
          + \frac{1}{\mu_0} \frac{\partial}{\partial \theta}
            \left[ \sqrt{g} \left(   \left( B^\theta \right)^2 Z_\theta
                                   + B^\theta B^\zeta          Z_\zeta  \right) \right]
          + ... \, .
\end{align}
Next, consider the poloidal derivative of~$P$:
\begin{align}
 \frac{\partial P}{\partial \theta}
 =&\, \frac{\partial}{\partial \theta}
        \left[ R \left( p + \frac{|B|^2}{2 \mu_0} \right) \right]
 = R \frac{\partial}{\partial \theta} \left( p + \frac{|B|^2}{2 \mu_0} \right) + ... \nonumber \\
 =&\, \frac{R}{2 \mu_0} \frac{\partial \left( |B|^2 \right)}{\partial \theta} + ... \nonumber \\
\end{align}
Thus, the poloidal derivative of the magnetic pressure needs to be considered next:
\begin{align}
 \frac{\partial \left( |B|^2 \right)}{\partial \theta}
 =&\, \frac{\partial}{\partial \theta} \left[
        B^\theta B_\theta + B^\zeta B_\zeta
      \right] \nonumber \\
 =&\, \frac{\partial}{\partial \theta} \left[
        (B^\theta)^2       g_{\theta \theta}
      + 2 B^\theta B^\zeta g_{\theta \zeta}
      + (B^\zeta)^2        g_{\zeta \zeta}
      \right] \, .
\end{align}
We continue term by term here:
\begin{align}
 \frac{\partial}{\partial \theta} \left( (B^\theta)^2 g_{\theta \theta} \right)
 =&\,   \frac{\partial (B^\theta)^2}{\partial \theta} g_{\theta \theta}
      + (B^\theta)^2 \frac{\partial g_{\theta \theta}}{\partial \theta} \nonumber \\
 \frac{\partial}{\partial \theta} \left( B^\theta B^\zeta g_{\theta \zeta} \right)
 =&\,   \frac{\partial (B^\theta B^\zeta)}{\partial \theta} g_{\theta \zeta}
      + B^\theta B^\zeta \frac{\partial g_{\theta \zeta}}{\partial \theta} \nonumber \\
 \frac{\partial}{\partial \theta} \left( (B^\zeta)^2 g_{\zeta \zeta} \right)
 =&\,   \frac{\partial (B^\zeta)^2}{\partial \theta} g_{\zeta \zeta}
      + (B^\zeta)^2 \frac{\partial g_{\zeta \zeta}}{\partial \theta} \, .
\end{align}
The required derivatives in here are:
\begin{align}
 \frac{\partial (B^\theta)^2}{\partial \theta}
 =&\, 2 B^\theta \frac{\partial B^\theta}{\partial \theta} \nonumber \\
 =&\, 2 B^\theta \frac{\partial}{\partial \theta}
      \left[ \frac{\Phi'}{\sqrt{g}} \left( \iota - \lambda_\zeta  \right) \right] \nonumber \\
 =&\, \frac{2 \left( B^\theta \right)^2}{\sqrt{g}}
        \frac{\partial \sqrt{g}}{\partial \theta} + ...
\end{align}
and in there
\begin{align}
 \frac{\partial \sqrt{g}}{\partial \theta}
 =&\, R_\theta \tau + R \frac{\partial \tau}{\partial \theta} \nonumber \\
 =&\, R \frac{\partial}{\partial \theta} \left( R_\theta Z_\rho - R_\rho Z_\theta \right) + ... \nonumber \\
 =&\, R \left(   Z_\rho \frac{\partial^2 R}{\partial \theta^2}
               - R_\rho \frac{\partial^2 Z}{\partial \theta^2} \right) + ... \, .
\end{align}
The poloidal derivative of~$g_{\theta \theta}$
remains to be done next:
\begin{align}
 \frac{\partial g_{\theta \theta}}{\partial \theta}
 =&\, \frac{\partial}{\partial \theta}
        \left[ R_\theta^2 + Z_\theta^2 \right] \nonumber \\
 =&\, 2 \left(   R_\theta \frac{\partial^2 R}{\partial \theta^2}
               + Z_\theta \frac{\partial^2 Z}{\partial \theta^2} \right) \, .
\end{align}
It follows by putting things back together:
\begin{align}
 \frac{\partial}{\partial \theta} \left( (B^\theta)^2 g_{\theta \theta} \right)
 =&\, \frac{2 \left( B^\theta \right)^2}{\sqrt{g}}
        R \left(   Z_\rho \frac{\partial^2 R}{\partial \theta^2}
                 - R_\rho \frac{\partial^2 Z}{\partial \theta^2} \right)
      +
      2 (B^\theta)^2 \left(   R_\theta \frac{\partial^2 R}{\partial \theta^2}
                            + Z_\theta \frac{\partial^2 Z}{\partial \theta^2} \right) + ... \nonumber \\
 =&\, 2 \left( B^\theta \right)^2  \left\{
          \left[
            \frac{R Z_\rho}{\sqrt{g}} + R_\theta
          \right] \frac{\partial^2 R}{\partial \theta^2}
        + \left[
            \frac{- R R_\rho}{\sqrt{g}} + Z_\theta
          \right] \frac{\partial^2 Z}{\partial \theta^2}
      \right\} + ... \, .
\end{align}
The poloidal derivative of~$B^\theta B^\zeta g_{\theta \zeta}$
needs to be considered next.
% TODO
Then, the poloidal derivative of~$\left(B^\zeta \right)^2 g_{\zeta \zeta}$
needs to be considered.
% TODO

\FloatBarrier
\section{Fourier Transforms of Forces}
The spectral condensation forces are added to the MHD forces.
This happens in Fourier space. Therefore, the forces need to be Fourier-transformed first.
This also facilitates the tangential derivatives in~\eqn{mhd_forces}.
A generic Fourier ansatz is used for the forces~$F^X$ on the state variable~$X$:
\begin{align}
  F^X(\theta, \zeta)
  =&          \sum_{m=0}^M \sum_{n=0}^N \,\Biggl[ \phantom{+}\, \hat{F}_{m,n}^{X,\mathrm{cc}} \cos(m \theta) \cos(n \zeta)
                                                           +    \hat{F}_{m,n}^{X,\mathrm{ss}} \sin(m \theta) \sin(n \zeta) \nonumber \\
  ~& \phantom{\sum_{m=0}^M \sum_{n=0}^N \,\Biggl[}\,       +    \hat{F}_{m,n}^{X,\mathrm{sc}} \sin(m \theta) \cos(n \zeta)
                                                           +    \hat{F}_{m,n}^{X,\mathrm{cs}} \cos(m \theta) \sin(n \zeta) \Biggr] \, .
\end{align}
A similar Fourier ansatz is used for the consistuents of the effective forces,
leading to the corresponding sets of Fourier coefficients for each of the following force components:
\begin{equation}
  F^X \in \left\{ A^R, B^R, C^R, A^Z, B^Z, C^Z, B^\lambda, C^\lambda, F^{R^\mathrm{con}}, F^{Z^\mathrm{con}} \right\} \label{eqn:all_forces} \, .
\end{equation}
Inserting the Fourier representations of the force contributions into~\eqn{mhd_forces} leads to:
\begin{align}
  \hat{F}_{m,n}^{X,\mathrm{cc}} =&\, \hat{A}_{m,n}^{X,\mathrm{cc}} + f(m) \hat{F}_{m,n}^{X^\mathrm{con},\mathrm{cc}} - m \hat{B}_{m,n}^{X,\mathrm{sc}} + n n_\mathrm{fp} \hat{C}_{m,n}^{X,\mathrm{cs}} \label{eqn:f_x_cc} \\
  \hat{F}_{m,n}^{X,\mathrm{ss}} =&\, \hat{A}_{m,n}^{X,\mathrm{ss}} + f(m) \hat{F}_{m,n}^{X^\mathrm{con},\mathrm{ss}} + m \hat{B}_{m,n}^{X,\mathrm{cs}} - n n_\mathrm{fp} \hat{C}_{m,n}^{X,\mathrm{sc}} \\
  \hat{F}_{m,n}^{X,\mathrm{sc}} =&\, \hat{A}_{m,n}^{X,\mathrm{sc}} + f(m) \hat{F}_{m,n}^{X^\mathrm{con},\mathrm{sc}} + m \hat{B}_{m,n}^{X,\mathrm{cc}} + n n_\mathrm{fp} \hat{C}_{m,n}^{X,\mathrm{ss}} \\
  \hat{F}_{m,n}^{X,\mathrm{cs}} =&\, \hat{A}_{m,n}^{X,\mathrm{cs}} + f(m) \hat{F}_{m,n}^{X^\mathrm{con},\mathrm{cs}} - m \hat{B}_{m,n}^{X,\mathrm{ss}} - n n_\mathrm{fp} \hat{C}_{m,n}^{X,\mathrm{cc}} \label{eqn:f_x_cs} \, .
\end{align}
Here, the parity of the tangential derivatives (see Sec.~\ref{sec:vmec_fourier_transforms})
has been used to determine the signs of the contributions.
The negative sign of the $B$-forces has been incorporated as well (see~\eqn{mhd_forces}).
The contributions from the $C$-forces to the effective forces are zero if axisymmetry is assumed.
The constraint force contributions are only present in the effective forces on $R$ and $Z$.
The force components listed in~\eqn{all_forces} are evaluated on a regular grid on each flux surface.
Each force component is Fourier-transformed individually.
The Fourier coefficients of the forces are then combined according to the experession in~\eqn{f_x_cc} to~\eqn{f_x_cs}.

The Fourier coefficients of the forces are computed by discretizing the corresponding Fourier integrals,
leading to discrete Fourier transforms.
In the general case, these are two-dimensional surface integrals.
In the axisymmetric case, the integral along the toroidal direction drops out
and only one-dimensional integrals along the poloidal direction have to be computed.
The Fourier integrals along the poloidal direction are computed over half the domain from contributions of different parity.
In the general/asymmetric case, one could also compute the Fourier integrals of the general functions
(not split up into definite-parity contributions) over the full domain.
In the symmetric case, only the non-vanishing contributions with parity compatible with the symmetry assumption are considered.
In the asymmetric case, the real-space forces are evaluated on the full poloidal interval.
The quantities to be transformed are then decomposed into two contributions with opposite parity,
which are only needed over half the poloidal interval.
This is done in a separate preprocessing step (\texttt{symforce}).
Those contributions with the same parity as would be only present in the symmetric case
are handled by the same Fourier transform code that would be used in the symmetric case alone (\texttt{totzsps}).
The contributions with the other parity (which only enter in case of an asymmetric run)
are handled by a different code segment (\texttt{totzspa}).
The general idea is also discussed in Theorem~26 of Ref.~\cite{boyd_spectral_methods}.

Here, the two-dimensional symmetric case is considered first.
We start with $\hat{F}^{X,\mathrm{cos}}_0$:
\begin{align}
  \hat{F}^{X,\mathrm{cos}}_0
 =&\, \frac{1}{\pi} \int\limits_0^{\pi} F^{X}(\theta) \,\mathrm{d}\theta
 \approx \frac{1}{\pi} \sum\limits_{l=0}^{n_{\theta,2}-1} F^{X}(\theta_l) \Delta\theta \begin{cases}
                                                                                         1/2 &: l=0 \textrm{ or } l=n_{\theta,2}-1 \\
                                                                                         1   &: \textrm{else}
                                                                                       \end{cases} \nonumber \\
 =&\, \frac{1}{\bcancel{\pi}} \frac{\bcancel{\pi}}{n_{\theta,2}-1} \sum\limits_{l=0}^{n_{\theta,2}-1} F^{X}(\theta_l) \begin{cases}
                                                                                                                        1/2 &: l=0 \textrm{ or } l=n_{\theta,2}-1 \\
                                                                                                                        1   &: \textrm{else}
                                                                                                                      \end{cases}
\end{align}
with $\Delta\theta = 2 \pi / n_{\theta,1} = \pi/(n_{\theta,2}-1)$ and $\theta_l = \Delta\theta \,l$.
Note that the standard trapezoidal quadrature rule has been used to discretize the Fourier integral (see also Sec.~\ref{sec:midpoint_vs_trapz}).

\begin{figure}[htbp]
  \centering
  \includegraphics{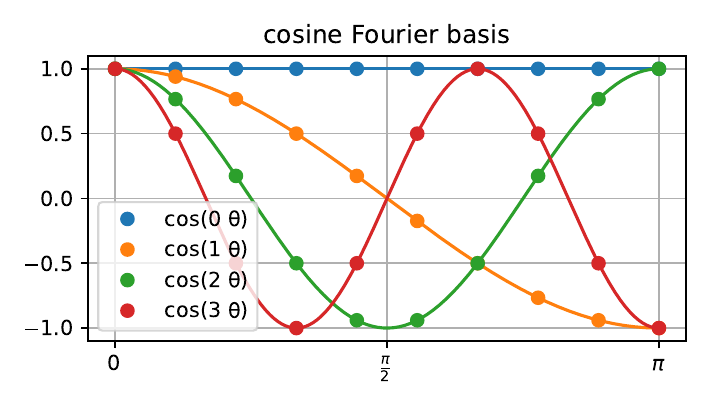}
  \caption{The first four poloidal cosine Fourier basis functions for $n_{\theta,2} = 10$.}
  \label{fig:poloidal_Fourier_basis}
\end{figure}

Similarly, it follows for $\hat{F}^{X,\mathrm{cos}}_m$ with $m>0$:
\begin{align}
 \hat{F}^{X,\mathrm{cos}}_m
  =&\, \frac{2}{\pi} \int\limits_0^{\pi} F^{X}(\theta) \cos(m \theta) \,\mathrm{d}\theta
  \approx \frac{2}{\pi} \sum\limits_{l=0}^{n_{\theta,2}-1} F^{X}(\theta_l) \cos(m \theta_l) \Delta\theta \begin{cases}
                                                                                                           1/2 &: l=0 \textrm{ or } l=n_{\theta,2}-1 \\
                                                                                                           1   &: \textrm{else}
                                                                                                         \end{cases} \nonumber \\
  =&\, \frac{2}{\bcancel{\pi}} \frac{\bcancel{\pi}}{n_{\theta,2}-1} \sum\limits_{l=0}^{n_{\theta,2}-1} F^{X}(\theta_l) \cos(m \theta_l) \begin{cases}
                                                                                                                                          1/2 &: l=0 \textrm{ or } l=n_{\theta,2}-1 \\
                                                                                                                                          1   &: \textrm{else}
                                                                                                                                        \end{cases} \, .
\end{align}
Note that $\theta_0 = 0$ and $\theta_{n_{\theta,2}-1} = \pi$. Thus:
\begin{align}
  \cos(m \theta_0)                =&\, \cos(0) = 1 \\
  \cos(m \theta_{n_{\theta,2}-1}) =&\, \cos(m \pi) = (-1)^m \, .
\end{align}
It follows for $\hat{F}^{X,\mathrm{cos}}_m$:
\begin{align}
 \hat{F}^{X,\mathrm{cos}}_m
  =&\, \frac{1}{n_{\theta,2}-1} \left[f(0) + (-1)^m f(\pi) + 2 \sum\limits_{l=1}^{n_{\theta,2}-2} F^{X}(\theta_l) \cos(m \theta_l) \right] \nonumber \\
  =&\, \frac{1}{n_{\theta,2}-1} \left[f(0) + (-1)^m f(\pi) + 2 \sum\limits_{l=1}^{n_{\theta,2}-2} F^{X}(\theta_l) \cos\left(\frac{\pi}{n_{\theta,2}-1} \,m \,l\right) \right] \label{eqn:f_m_cos} \, .
\end{align}
The basis functions are depicted in Fig.~\ref{fig:poloidal_Fourier_basis}.
% src/poloidal_Fourier_basis.ipynb

In the example of Fig.~\ref{fig:poloidal_symmetry}, the points at $\theta=0$ and at $\theta=\pi$ appear only once
for the even-parity quantity~$R$ in the example and all other value for $R$ of the dots appear twice,
as correctly accounted for by the factors in \eqn{f_m_cos}.

We consider $\hat{F}^{X,\mathrm{sin}}_m$ with $m>0$ next:
\begin{align}
 \hat{F}^{X,\mathrm{sin}}_m
  =&\, \frac{2}{\pi} \int\limits_0^{\pi} F^{X}(\theta) \sin(m \theta) \,\mathrm{d}\theta
  \approx \frac{2}{\pi} \sum\limits_{l=0}^{n_{\theta,2}-1} F^{X}(\theta_l) \sin(m \theta_l) \Delta\theta \begin{cases}
                                                                                                           1/2 &: l=0 \textrm{ or } l=n_{\theta,2}-1 \\
                                                                                                           1   &: \textrm{else}
                                                                                                         \end{cases} \nonumber \\
  =&\, \frac{2}{\bcancel{\pi}} \frac{\bcancel{\pi}}{n_{\theta,2}-1} \sum\limits_{l=1}^{n_{\theta,2}-2} F^{X}(\theta_l) \sin(m \theta_l)
  =    \frac{1}{n_{\theta,2}-1} 2 \sum\limits_{l=0}^{n-1} F^{X}(\theta_l) \sin\left(\frac{\pi}{n+1} \,m \,(l+1) \right)
\end{align}
for $n=n_{\theta,2}-2$.
The first and the last terms in the sum dropped out since $\sin(m 0) = 0$ and $\sin(m \pi) = 0$ at those indices, respectively.
Then, an index shift of $l \rightarrow l-1$ was performed in the sum.

The Fourier transforms are computed in the Fortran implementation of VMEC using
a setup of the Fourier transforms as a matrix-vector multiplication.
Two-dimensional arrays \texttt{cosmui} and \texttt{sinmui} are defined with:
\begin{align}
  \texttt{cosmui}(l,m) =&\, \frac{\texttt{mscale}(m)}{n_\zeta (n_{\theta,2}-1)} \cos\left(\frac{\pi}{n_{\theta,1}} \,m \,l\right) \begin{cases}
                                                                                                                          1/2 &: l=0 \textrm{ or } l=n_{\theta,2}-1 \\
                                                                                                                          1   &: \textrm{else}
                                                                                                                        \end{cases} \\
  \texttt{sinmui}(l,m) =&\, \frac{\texttt{mscale}(m)}{n_\zeta (n_{\theta,2}-1)} \sin\left(\frac{\pi}{n_{\theta,1}} \,m \,l\right)
\end{align}
(in case of \texttt{sinmui}, no attempt is made to trapezoidally-scale the first and last entries along the $l$-dimension, since those are zero anyway)
and a one-dimensional array~\texttt{mscale} with:
\begin{equation}
  \texttt{mscale}(m) = \begin{cases}
                               1  &: m=0           \\
                         \sqrt{2} &: \textrm{else}
                       \end{cases}
\end{equation}
for $l=0,1, ..., (n_{\theta,2}-1)$ and $m=0,1, ..., M$.
The normalization factor is called~\texttt{dnorm} in the code:
\begin{equation}
  \texttt{dnorm} = \frac{1}{n_\zeta (n_{\theta,2}-1)} \, .
\end{equation}
It also contains the normalization factor for the toroidal Fourier integrals (see below).
The computation of $\hat{F}^{X,\mathrm{cos}}_m$ for $m\geq0$ can then be formulated as follows:
\begin{equation}
  \hat{F}^{X,\mathrm{cos}}_m = \sum\limits_{l=0}^{n_{\theta,2}-1} F^{X}(\theta_l) \, \texttt{cosmui}(l,m) \, .
\end{equation}
Note that a factor of $\sqrt{2}$ is missing for $m>0$ to recover \eqn{f_m_cos}.
In turn, if the Fourier transforms are computed using, e.g., FFTW (which implements above transform as \texttt{REDFT00}),
a factor of $1/\sqrt{2}$ has to be applied to the results from that FFT in order to recover the custom VMEC scaling.
Similarly, the computation of $\hat{F}^{X,\mathrm{sin}}_m$ is done as follows:
\begin{equation}
  \hat{F}^{X,\mathrm{sin}}_m = \sum\limits_{l=0}^{n_{\theta,2}-1} F^{X}(\theta_l) \, \texttt{sinmui}(l,m) \, .
\end{equation}
This can be replaced by FFTW's \texttt{RODFT00}, where the first and last points have to be left out explicitly,
the Fourier coefficients are shifted by 1 along the $m$-direction and a factor of $1/\sqrt{2}$ has to be multiplied in as well.
%
% {\color{red} TODO: figure out why exactly mscale, nscale have those particular forms, i.e.,
%  where is made use of the explicit orthogonality with constant factor 0.5?}
%
The poloidal derivatives are implemented using additional matrices:
\begin{align}
  \texttt{cosmumi}(l,m) =&\, (\phantom{-} m) \, \texttt{cosmui}(l,m) \\
  \texttt{sinmumi}(l,m) =&\, (         -  m) \, \texttt{sinmui}(l,m) \, .
\end{align}

\begin{figure}[htbp]
  \centering
  \includegraphics{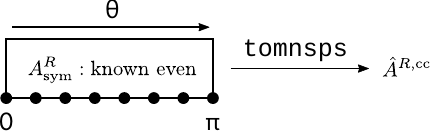}
  \caption{Assembly of the Fourier coefficients~$\hat{A}^{R,\mathrm{cc}}$
    from the $A^R$~real-space force component in the symmetric case.}
  \label{fig:A_R_sym}
\end{figure}

\begin{figure}[htbp]
  \centering
  \includegraphics{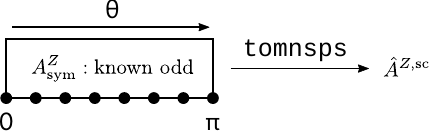}
  \caption{Assembly of the Fourier coefficients~$\hat{A}^{Z,\mathrm{sc}}$
    from the $A^Z$~real-space force component in the symmetric case.}
  \label{fig:A_Z_sym}
\end{figure}

\begin{figure}[htbp]
  \centering
  \includegraphics{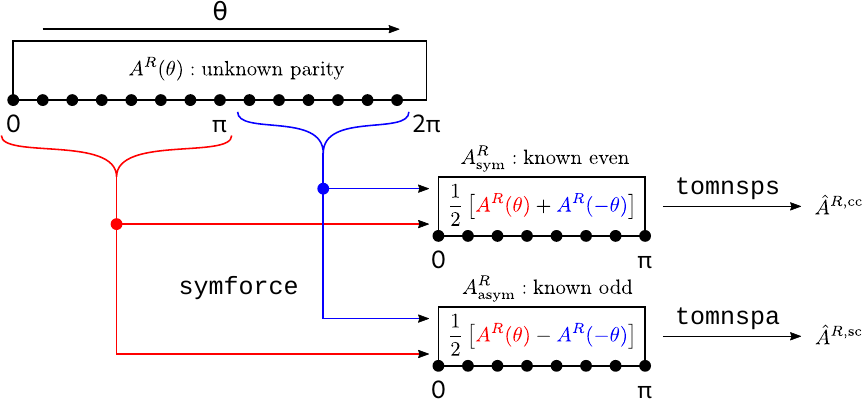}
  \caption{Assembly of the Fourier coefficients~$\hat{A}^{R,\mathrm{cc}}$ and~$\hat{A}^{R,\mathrm{sc}}$
    from the $A^R$~real-space force component in the asymmetric/general case.}
  \label{fig:A_R_asym}
\end{figure}

\begin{figure}[htbp]
  \centering
  \includegraphics{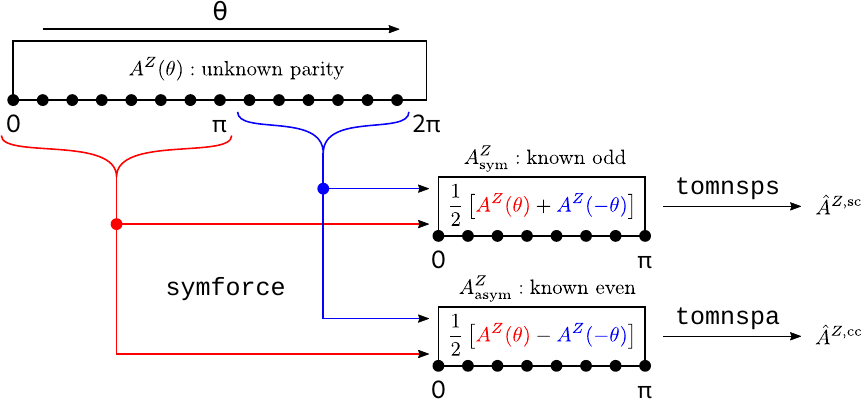}
  \caption{Assembly of the Fourier coefficients~$\hat{A}^{Z,\mathrm{sc}}$ and~$\hat{A}^{Z,\mathrm{cc}}$
    from the $A^Z$~real-space force component in the asymmetric/general case.}
  \label{fig:A_Z_asym}
\end{figure}

\FloatBarrier
Now we turn to the three-dimensional (Stellarator-)symmetric case.
The toroidal Fourier transform can be ignored in the two-dimensional case, since $n_\zeta=1$ and $n=0$,
leading to basis functions $\cos(n\zeta_k)=1$ and $\sin(n\zeta_k)=0$ for the only toroidal grid point at $k=0$.
This effectively disables/removes the toroidal Fourier transform from the setup.
Stellarator symmetry manifests itself as the following parity property
for $R$ and $Z$ of the flux surface geometry under reversal
of both the poloidal $\theta$~coordinate and the toroidal $\zeta$~coordinate:
\begin{align}
  R(-\theta, -\zeta) =&\, \phantom{-}~ R(\theta, \zeta) \\
  Z(-\theta, -\zeta) =&\,          -   Z(\theta, \zeta) \, .
\end{align}
Any other quantity with even (odd) parity has the same symmetry property as $R$ ($Z$) in above equations.

The toroidal grid points (on which the toroidal Fourier transform integrals are discretized)
are distributed evenly in the toroidal direction over the first field period.
The toroidal coordinate~$\zeta$ spans $[0,2\pi[$ over one field period.
This corresponds to $\phi \in [0,2\pi/n_\mathrm{fp}[$ in the underlying cylindrical coordinate system
(recall $\zeta = n_\mathrm{fp} \phi \Leftrightarrow \phi = \zeta / n_\mathrm{fp}$).
It follows for some quantity~$C$ with even parity:
\begin{align}
    C(\zeta) = \sum_{n=0}^{N} \hat{C}_n \cos(n \zeta)
  \Leftrightarrow&\,
    C(\phi ) = \sum_{n=0}^{N} \hat{C}_n \cos(n \,n_\mathrm{fp} \phi) \nonumber \\
    \Rightarrow
    \frac{\partial C}{\partial \zeta} (\zeta) = \sum_{n=0}^{N} (-n              ) \hat{C}_n \sin(n               \zeta)
  \Leftrightarrow&\,
    \frac{\partial C}{\partial \phi } (\phi ) = \sum_{n=0}^{N} (-n \,n_\mathrm{fp}) \hat{C}_n \sin(n \,n_\mathrm{fp} \phi ) \, ,
\end{align}
since
\begin{equation}
  \frac{\partial C}{\partial \phi} = \frac{\partial C}{\partial \zeta} \underbrace{\frac{\partial \zeta}{\partial \phi}}_{=\,n_\mathrm{fp}} (\phi) \, .
\end{equation}

Generally, the argument of the Fourier basis functions is $(m \theta - n \,n_\mathrm{fp} \zeta)$.
Evaluation is done only in the first field period: $\phi \in [0,2\pi/n_\mathrm{fp}[ \Rightarrow \zeta \in [0,2\pi[$.
The argument $(m \theta - n \zeta)$ is used for evaluating the basis functions
on a regular grid in $\zeta$, since no explicit factors of $n_\mathrm{fp}$ occur then.
Toroidal derivatives are to be taken with respect to $\phi$ though,
so the factor to be used in the Fourier sums still is $n \,n_\mathrm{fp}$.

Consider first the Fourier transform of some $2 \pi$-periodic function $f$ of the toroidal coordinate $\zeta$.
Its Fourier coefficients are computed as follows:
\begin{align}
  \hat{f}^\mathrm{cos}_0
  =&\, \frac{1}{2 \pi} \int\limits_0^{2\pi} f(\zeta) \,\mathrm{d}\zeta
    \approx \frac{1}{2 \pi} \sum\limits_{k=0}^{n_\zeta-1} f(\zeta_k) \Delta \zeta
  = \frac{1}{\bcancel{2 \pi}} \frac{\bcancel{2 \pi}}{n_\zeta} \sum\limits_{k=0}^{n_\zeta-1} f(\zeta_k)
  = \frac{1}{n_\zeta} \sum\limits_{k=0}^{n_\zeta-1} f\left(2 \pi \frac{k}{n_\zeta}\right) \\
  \hat{f}^\mathrm{cos}_n
  =&\, \frac{1}{\pi} \int\limits_0^{2\pi} f(\zeta) \cos(n \zeta) \,\mathrm{d}\zeta
    \approx \frac{1}{\pi} \sum\limits_{k=0}^{n_\zeta-1} f(\zeta_k) \cos(n \zeta_k) \Delta \zeta
  = \frac{1}{\bcancel{\pi}} \frac{2 \bcancel{\pi}}{n_\zeta} \sum\limits_{k=0}^{n_\zeta-1} f(\zeta_k) \cos(n \zeta_k) \nonumber \\
  =&\, \frac{2}{n_\zeta} \sum\limits_{k=0}^{n_\zeta-1} f\left(2 \pi \frac{k}{n_\zeta}\right) \cos\left(2 \pi n \frac{k}{n_\zeta}\right) \\
  \hat{f}^\mathrm{sin}_n
  =&\, \frac{1}{\pi} \int\limits_0^{2\pi} f(\zeta) \sin(n \zeta) \,\mathrm{d}\zeta
    \approx \frac{1}{\pi} \sum\limits_{k=0}^{n_\zeta-1} f(\zeta_k) \sin(n \zeta_k) \Delta \zeta
  = \frac{1}{\bcancel{\pi}} \frac{2 \bcancel{\pi}}{n_\zeta} \sum\limits_{k=0}^{n_\zeta-1} f(\zeta_k) \sin(n \zeta_k) \nonumber \\
  =&\, \frac{2}{n_\zeta} \sum\limits_{k=0}^{n_\zeta-1} f\left(2 \pi \frac{k}{n_\zeta}\right) \sin\left(2 \pi n \frac{k}{n_\zeta}\right)
\end{align}
with
\begin{equation}
  \Delta \zeta = \frac{2 \pi}{n_\zeta}
  \textrm{ and }
  \zeta_k = k \, \Delta \zeta = 2 \pi \frac{k}{n_\zeta} \, .
\end{equation}
The Fourier basis functions in the toroidal direction are pre-evaluated
on a regular grid in the toroidal coordinate:
\begin{align}
  \texttt{cosnv}(k,n) =&\, \texttt{nscale}(n) \cos\left(\frac{2 \pi}{n_\zeta} \,n \,k\right) \\
  \texttt{sinnv}(k,n) =&\, \texttt{nscale}(n) \sin\left(\frac{2 \pi}{n_\zeta} \,n \,k\right)
\end{align}
with
\begin{equation}
  \texttt{nscale}(n) = \begin{cases}
                               1  &: n=0           \\
                         \sqrt{2} &: \textrm{else}
                       \end{cases}
\end{equation}
for $k=0,1, ..., (n_\zeta-1)$ and $n=0,1, ..., N$.
Furthermore, the toroidal derivatives are implemented with separate matrices:
\begin{align}
  \texttt{cosnvn}(k,n) =&\, \texttt{nscale}(n) (\phantom{-} n \,n_\mathrm{fp}) \cos\left(\frac{2 \pi}{n_\zeta} \,n \,k\right) \\
  \texttt{sinnvn}(k,n) =&\, \texttt{nscale}(n) (         -  n \,n_\mathrm{fp}) \sin\left(\frac{2 \pi}{n_\zeta} \,n \,k\right) \, .
\end{align}
The computation of the toroidal Fourier integrals $\hat{F}^{X,\mathrm{cos}}_n$ for $n\geq0$ can then be formulated as follows:
\begin{equation}
  \hat{F}^{X,\mathrm{cos}}_n = \sum\limits_{k=0}^{n_\zeta-1} F^{X}(\zeta_k) \, \texttt{cosnv}(k,n) \, .
\end{equation}
Note that also here a factor of $\sqrt{2}$ is missing for $n>0$ to recover the usual Fourier integrals.
Similarly, the computation of $\hat{F}^{X,\mathrm{sin}}_n$ is done as follows:
\begin{equation}
  \hat{F}^{X,\mathrm{sin}}_n = \sum\limits_{k=0}^{n_\zeta-1} F^{X}(\zeta_k) \, \texttt{sinnv}(k,n) \, .
\end{equation}
The general two-dimensional Fourier integrals are assembled
as products of the corresponding one-dimensional transforms:
\begin{align}
  \hat{F}^{X,\mathrm{cc}}_{mn} =&\, \sum\limits_{k=0}^{n_\zeta-1} \sum\limits_{l=0}^{n_{\theta,2}-1} F^{X}(\zeta_k, \theta_l) \, \texttt{cosmui}(l,m) \, \texttt{cosnv}(k,n) \\
  \hat{F}^{X,\mathrm{ss}}_{mn} =&\, \sum\limits_{k=0}^{n_\zeta-1} \sum\limits_{l=0}^{n_{\theta,2}-1} F^{X}(\zeta_k, \theta_l) \, \texttt{sinmui}(l,m) \, \texttt{sinnv}(k,n) \\
  \hat{F}^{X,\mathrm{sc}}_{mn} =&\, \sum\limits_{k=0}^{n_\zeta-1} \sum\limits_{l=0}^{n_{\theta,2}-1} F^{X}(\zeta_k, \theta_l) \, \texttt{sinmui}(l,m) \, \texttt{cosnv}(k,n) \\
  \hat{F}^{X,\mathrm{cs}}_{mn} =&\, \sum\limits_{k=0}^{n_\zeta-1} \sum\limits_{l=0}^{n_{\theta,2}-1} F^{X}(\zeta_k, \theta_l) \, \texttt{cosmui}(l,m) \, \texttt{sinnv}(k,n) \, .
\end{align}
The tangential derivatives are now also easy to assemble.
First consider the poloidal derivatives:
\begin{align}
  \hat{B}^{X,\mathrm{cc}}_{mn} =&\, \sum\limits_{k=0}^{n_\zeta-1} \sum\limits_{l=0}^{n_{\theta,2}-1} B^{X}(\zeta_k, \theta_l) \, \texttt{cosmumi}(l,m) \, \texttt{cosnv}(k,n) \\
  \hat{B}^{X,\mathrm{ss}}_{mn} =&\, \sum\limits_{k=0}^{n_\zeta-1} \sum\limits_{l=0}^{n_{\theta,2}-1} B^{X}(\zeta_k, \theta_l) \, \texttt{sinmumi}(l,m) \, \texttt{sinnv}(k,n) \\
  \hat{B}^{X,\mathrm{sc}}_{mn} =&\, \sum\limits_{k=0}^{n_\zeta-1} \sum\limits_{l=0}^{n_{\theta,2}-1} B^{X}(\zeta_k, \theta_l) \, \texttt{sinmumi}(l,m) \, \texttt{cosnv}(k,n) \\
  \hat{B}^{X,\mathrm{cs}}_{mn} =&\, \sum\limits_{k=0}^{n_\zeta-1} \sum\limits_{l=0}^{n_{\theta,2}-1} B^{X}(\zeta_k, \theta_l) \, \texttt{cosmumi}(l,m) \, \texttt{sinnv}(k,n) \, .
\end{align}
Next consider the toroidal derivatives:
\begin{align}
  \hat{C}^{X,\mathrm{cc}}_{mn} =&\, \sum\limits_{k=0}^{n_\zeta-1} \sum\limits_{l=0}^{n_{\theta,2}-1} C^{X}(\zeta_k, \theta_l) \, \texttt{cosmui}(l,m) \, \texttt{cosnvn}(k,n) \\
  \hat{C}^{X,\mathrm{ss}}_{mn} =&\, \sum\limits_{k=0}^{n_\zeta-1} \sum\limits_{l=0}^{n_{\theta,2}-1} C^{X}(\zeta_k, \theta_l) \, \texttt{sinmui}(l,m) \, \texttt{sinnvn}(k,n) \\
  \hat{C}^{X,\mathrm{sc}}_{mn} =&\, \sum\limits_{k=0}^{n_\zeta-1} \sum\limits_{l=0}^{n_{\theta,2}-1} C^{X}(\zeta_k, \theta_l) \, \texttt{sinmui}(l,m) \, \texttt{cosnvn}(k,n) \\
  \hat{C}^{X,\mathrm{cs}}_{mn} =&\, \sum\limits_{k=0}^{n_\zeta-1} \sum\limits_{l=0}^{n_{\theta,2}-1} C^{X}(\zeta_k, \theta_l) \, \texttt{cosmui}(l,m) \, \texttt{sinnvn}(k,n) \, .
\end{align}
This completes the Fourier transforms of the force components in VMEC.

% TODO: residue() and all the rest until convergence

\section{Output Quantities}
After VMEC has converged, the output quantities are computed.
These are the arrays etc. that eventually end up in the output files.

The \texttt{wout} file is always written.
Also, the logging part of the \texttt{threed1} file is always written.
All other output files are only written if VMEC ended because ...
\begin{enumerate}
 \item ... it converged to the desired force tolerance, or because
 \item ... the number of allowed iterations was exceeded
\end{enumerate}
and are omitted if VMEC failed for any other reason.

The interface between the equilibrium computation and the output quantities
is a little bit weird.
At the beginning of the \texttt{fileout} subroutine that controls the computation
of the output quantities as well as the creation of the output files,
the ideal MHD forward model within VMEC (\texttt{funct3d}) is run again,
but with the global flag \texttt{iequi} set to 1 now (instead of the regular value of 0).
This is required, since many arrays are re-used within VMEC during the iterations
and this would overwrite quantities that are still needed for the output quantities.

In case of a bad Jacobian, the evaluation of the forward model
is terminated after the Jacobian has been computed.
This means that the geometry in the wout file will refer to the intersecting flux surfaces
that triggered the Jacobian error,
but the magnetic field components etc. will refer to the previous iteration.
% TODO: check this!

In summary, if VMEC did not converge properly, don't use the output quantities for anything!

The final forward model evaluation stops
before updating the radial preconditioner.

\subsection{Preparation of the internal data}
\begin{figure}[htbp]
 \centering
 \includegraphics{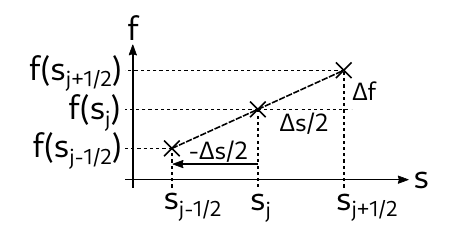}
 \caption{Linear interpolation scheme used for mesh blending of $B_\zeta$.}
 \label{fig:bsubv_mesh_blending}
\end{figure}

The covariant magnetic field component~$B_\zeta$ is averaged back onto the half-grid.
The scheme used for that is illustrated in Fig.~\ref{fig:bsubv_mesh_blending}.
Assume that a function~$f$ is given on a full-grid point $s_j$ and the outward half-grid point~$s_{j+\half}$.
Then, linear extrapolation can be used to compute $f$ at $s_{j-\half}$:
\begin{equation}
 f(s_{j-\half}) \approx f(s_j) + (- \Delta s / 2) \frac{\partial f}{\partial s}\vert_{s=s_j} \label{eqn:linear_extrap} \, .
\end{equation}
The radial derivative of $f$ can be approximated using finite differences:
\begin{equation}
 \frac{\partial f}{\partial s}\vert_{s=s_j}
 \approx
 \frac{f(s_{j+\half}) - f(s_j)}{\Delta s / 2} \, .
\end{equation}
Putting this back into \eqn{linear_extrap} leads to:
\begin{align}
 f(s_{j-\half}) \approx&\, f(s_j) - \bcancel{\Delta s / 2} \frac{f(s_{j+\half}) - f(s_j)}{\bcancel{\Delta s / 2}} \nonumber \\
   ~                =&\, 2 f(s_j) - f(s_{j+\half}) \, .
\end{align}
This can be computed conveniently in-place when starting the radial loop out at the LCFS.

In the next step, the poloidal current is restored by adjusting the offset in~$B_\zeta$.
The poloidal current~$I_\textrm{pol}$ is defined as the average value of $B_\zeta$ over a flux surface:
\begin{equation}
 I_\textrm{pol} = \langle B_\zeta \rangle \, .
\end{equation}
After above mesh blending, the poloidal current is corrected back to its original value,
which was computed during the iterations:
\begin{align}
 B_\zeta \leftarrow&\, B_\zeta - \left[ \langle B_\zeta \rangle - I_\textrm{pol}^\textrm{ref} \right] \nonumber \\
     ~            =&\, B_\zeta + \underbrace{\left[ I_\textrm{pol}^\textrm{ref} - \langle B_\zeta \rangle \right]}_{=\texttt{curpol\_temp}} \, ,
\end{align}
where $I_\textrm{pol}^\textrm{ref}$ is the targeted poloidal current.

The toroidal magnetic flux profile is re-computed with the correct normalization and sign
by quadrature from its radial derivative:
\begin{equation}
 \Phi(s_j) = \texttt{signgs} \cdot 2 \pi \cdot \frac{1}{\Delta s} \sum\limits_{j'=0}^{j} \Phi'(s_{j'})
\end{equation}

\subsection{Computation of $B_s$}
The covariant component~$B_s$ of the magnetic field is computed now
based on the contravariant magnetic field components on the half-grid:
\begin{equation}
 B_s(s_{j+\half}, \zeta_k, \theta_l) = \left[B^\theta g_{s \theta} + B^\zeta g_{s \zeta} \right](s_{j+\half}, \zeta_k, \theta_l) \, .
\end{equation}
The metric elements $g_{s \theta}$ and $g_{s \zeta}$ are computed as follows:
\begin{align}
 g_{s \theta} =&\, R_s R_\theta + Z_s Z_\theta \\
 g_{s \zeta}  =&\, R_s R_\zeta  + Z_s Z_\zeta  \, .
\end{align}
The radial derivatives are computed based on the quantities $\tilde{R_s}$ and $\tilde{Z_s}$
that are already available from the computation of the Jacobian:
\begin{align}
 R_s(s_{j+\half}) =&\, \tilde{R_s}(s_{j+\half}) + \frac{1}{2 \sqrt{s_{j+\half}}} R^\textrm{o}(s_{j+\half}) \\
 Z_s(s_{j+\half}) =&\, \tilde{Z_s}(s_{j+\half}) + \frac{1}{2 \sqrt{s_{j+\half}}} Z^\textrm{o}(s_{j+\half})
\end{align}
with
\begin{align}
 R^\textrm{o}(s_{j+\half}) =&\, \frac{1}{2} \left[ R^\textrm{o}(s_{j+1}) + R^\textrm{o}(s_j) \right] \\
 Z^\textrm{o}(s_{j+\half}) =&\, \frac{1}{2} \left[ Z^\textrm{o}(s_{j+1}) + Z^\textrm{o}(s_j) \right] \, .
\end{align}
The derivatives in the toroidal direction, $R_\zeta$ and $Z_\zeta$,
are taken as zero in the case of an axisymmetric equilibrium calculation.
Otherwise, they are interpolated from the even-$m$ and odd-$m$ contributions
that are available on the full-grid from the inverse-DFTs:
\begin{align}
 R_\zeta(s_{j+\half}) =&\, R_\zeta^\textrm{e}(s_{j+\half}) + \sqrt{s_{j+\half}} R_\zeta^\textrm{o}(s_{j+\half}) \\
 Z_\zeta(s_{j+\half}) =&\, Z_\zeta^\textrm{e}(s_{j+\half}) + \sqrt{s_{j+\half}} Z_\zeta^\textrm{o}(s_{j+\half}) \, .
\end{align}
with
\begin{align}
 R_\zeta^\textrm{e}(s_{j+\half}) =&\, \frac{1}{2} \left[ R_\zeta^\textrm{e}(s_{j+1}) + R_\zeta^\textrm{e}(s_j) \right] \\
 R_\zeta^\textrm{o}(s_{j+\half}) =&\, \frac{1}{2} \left[ R_\zeta^\textrm{o}(s_{j+1}) + R_\zeta^\textrm{o}(s_j) \right] \\
 Z_\zeta^\textrm{e}(s_{j+\half}) =&\, \frac{1}{2} \left[ Z_\zeta^\textrm{e}(s_{j+1}) + Z_\zeta^\textrm{e}(s_j) \right] \\
 Z_\zeta^\textrm{o}(s_{j+\half}) =&\, \frac{1}{2} \left[ Z_\zeta^\textrm{o}(s_{j+1}) + Z_\zeta^\textrm{o}(s_j) \right] \, .
\end{align}
Furthermore, the cylindrical components of the magnetic field
are computed on the half-grid:
\begin{align}
 B^R       =&\, B^\theta R_\theta + B^\zeta R_\zeta \\
 B^\varphi =&\, R B^\zeta \\
 B^Z       =&\, B^\theta Z_\theta + B^\zeta Z_\zeta \, .
\end{align}

% \subsection{Low-Pass-Filtering of Covariant Magnetic Field Components}
% TODO

\FloatBarrier
\subsection{Correction of $B_s$}
The radial part of the ideal MHD force balance equation is given in Eqn.~(4d)
of Ref.~\cite{hirshman_whitson_1983}:
\begin{equation}
 F_s = \frac{1}{\mu_0} \left(
     B^\theta  \frac{\partial B_\theta}{\partial s}
   + B^\zeta   \frac{\partial B_\zeta }{\partial s}
   - \mathbf{B} \cdot \nabla B_s
 \right) + p' \, . \label{eqn:f_rho}
\end{equation}
The dot product $\mathbf{B} \cdot \nabla B_s$ has to be expressed
using contravariant components for $\mathbf{B}$,
because $B_s$ is a covariant component:
\begin{align}
 \mathbf{B} \cdot \nabla B_s
 =&\, \begin{pmatrix}
    0 \\
    B^\theta \\
    B^\zeta
   \end{pmatrix}
   \cdot
   \begin{pmatrix}
    \partial B_s / \partial s \\
    \partial B_s / \partial \theta \\
    \partial B_s / \partial \zeta
   \end{pmatrix} \\
 =&\,  B^\theta \frac{\partial B_s}{\partial \theta}
     + B^\zeta \frac{\partial B_s}{\partial \zeta}
\end{align}
where we have already factored in the assumption of nested flux surfaces,
which implies $B^s = 0$.
Using the assumption of an equilibrium being in force balance,
which implies $F_s = 0$,
we can solve \eqn{f_rho} for the terms depending on $B_s$:
\begin{equation}
   B^\theta \frac{\partial B_s}{\partial \theta}
 + B^\zeta \frac{\partial B_s}{\partial \zeta}
 =   B^\theta  \frac{\partial B_\theta}{\partial s}
   + B^\zeta   \frac{\partial B_\zeta }{\partial s}
   + \mu_0 p' \label{eqn:radial_force_balance}
\end{equation}
We now use a Fourier representation for $B_s$,
which has odd parity in case of assuming stellarator symmetry:
\begin{equation}
 B_s(\theta, \zeta) =
 \sum\limits_{\texttt{mn}=0}^{\texttt{mnmax}-1}
   \hat{B}^\textrm{sin}_{s,\texttt{mn}}
   \sin(m_\texttt{mn} \theta - n_\texttt{mn} \zeta) \, .
\end{equation}
This allows to use analytical expressions
for the tangential derivatives of $B_s$:
\begin{align}
 \frac{\partial B_s}{\partial \theta} =&\,
 \sum\limits_{\texttt{mn}=0}^{\texttt{mnmax}-1}
   m \hat{B}^\textrm{sin}_{s,\texttt{mn}}
   \cos(m_\texttt{mn} \theta - n_\texttt{mn} \zeta) \\
 \frac{\partial B_s}{\partial \zeta} =&\,
 \sum\limits_{\texttt{mn}=0}^{\texttt{mnmax}-1}
   -n \hat{B}^\textrm{sin}_{s,\texttt{mn}}
   \cos(m_\texttt{mn} \theta - n_\texttt{mn} \zeta) \, .
\end{align}
We want to solve \eqn{radial_force_balance} for the Fourier coefficients of $B_s$
on all real-space grid points $\texttt{kl}$ at the same time.
This can be facilitated by formulating \eqn{radial_force_balance}
as a linear system of equations ($\mathbf{A} \mathbf{x} = \mathbf{b}$):
\begin{align}
  \underbrace{
  \left(
    (  m_\texttt{mn} B^\theta_\texttt{kl}
     - n_\texttt{mn} B^\zeta_\texttt{kl} )
     \cos(  m_\texttt{mn} \theta_\texttt{kl}
          - n_\texttt{mn} \zeta_\texttt{kl})
  \right)
  }_{\equiv \mathbf{A}}
  \underbrace{
    \hat{B}^\textrm{sin}_{s,\texttt{mn}}
  }_{\equiv \mathbf{x}}
  =
  \underbrace{
    \left(
       B^\theta  \frac{\partial B_\theta}{\partial s}
     + B^\zeta   \frac{\partial B_\zeta }{\partial s}
     + \mu_0 p'
    \right)_\texttt{kl}
  }_{\equiv \mathbf{b}}
\end{align}
for all grid points indexed by \texttt{kl}.
Summation over \texttt{mn} is implied.

Typically, the real-space resolution
(number of grid points in poloidal and toroidal direction)
is larger than the number of Fourier coefficients,
which implies that the linear system of equations
is over-constrained and hence needs to be
solved in a least-squares sense.

\subsection{Mercier Stability}

The contribution due to magnetic shear is computed as follows:
\begin{align}
 D_\textrm{shear} = (\iota')^2 / 4 \, .
\end{align}
The contribution due to the plasma current is computed as follows:
\begin{align}
 D_\textrm{curr} = -\iota' (t_{jb} - I_\textrm{tor}' t_{bb}) \, .
\end{align}
The contribution due to the magnetic well is computed as follows:
\begin{align}
 D_\textrm{well} = p' (V'' - p' t_{pp}) t_{bb} \, .
\end{align}
The contribution due to geodesic curvature is computed as follows:
\begin{align}
 D_\textrm{geod} = t_{jb}^2 - t_{bb} t_{jj} \, .
\end{align}
The final value for Mercier's stability criterion is then added up from the four contributions:
\begin{align}
 D_\textrm{Merc} = D_\textrm{shear} + D_\textrm{curr} + D_\textrm{well} + D_\textrm{geod}\, .
\end{align}

\chapter{NESTOR}
The Neumann Solver for Toroidal Regions~(NESTOR) provides the free-boundary contribution to the energy functional in VMEC~\cite{merkel_1986, hirshman_vanrij_merkel_1986}.

\FloatBarrier
\section{Problem statement}
\begin{figure}[htbp]
  \centering
  \includegraphics[width=0.3\textwidth]{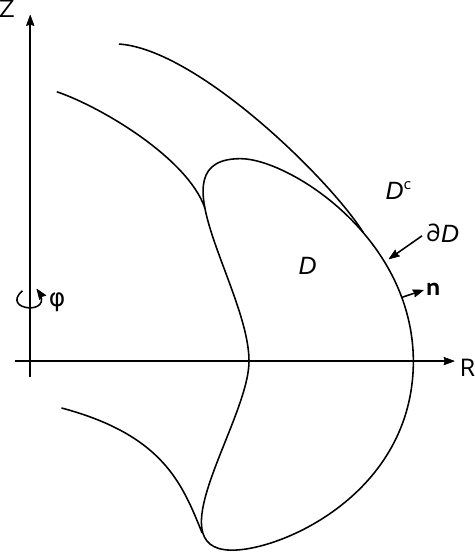}
  \caption{The interior region $\mathcal{D}$ (for the plasma) and the exterior region $\mathcal{D}^\mathrm{c}$ (for the coils) are separated by the boundary $\partial\mathcal{D}$.
           The outward-directed (exterior) unit normal vector is denoted $\mathbf{n}$.}
  \label{fig:00_regions}
\end{figure}

The basic geometry is depicted in Fig.~\ref{fig:00_regions}.

The magnetic field in this situation is denoted $\mathbf{B}_0$.
It is generated by currents $\mathbf{j}$ in $\mathcal{D} \cup \mathcal{D}^\mathrm{c}$:
\begin{equation}
  \nabla \times \mathbf{B}_0 = \mu_0 \mathbf{j}, ~ \nabla \cdot \mathbf{B}_0 = 0 \, .
\end{equation}
A magnetic scalar potential $\Phi$ can be found,
such that the corresponding magnetic field $\mathbf{B} = \nabla \Phi$ cancels the normal component of $\mathbf{B}_0$ on $\partial\mathcal{D}$:
\begin{equation}
  \left( \mathbf{B}_0 + \nabla \Phi \right) \cdot \mathbf{n} = 0 \, .
\end{equation}
The magnetic scalar potential $\Phi$ solves Laplace's equation
with the given Neumann boundary condition:
\begin{align}
  \forall \mathbf{x} \in \mathcal{D}:         &~ \Delta \Phi(\mathbf{x}) = 0 \, , \label{eqn:laplacePhi} \\
  \forall \mathbf{x} \in \partial\mathcal{D}: &~ \frac{\partial \Phi(\mathbf{x})}{\partial \mathbf{n}} = - \mathbf{B}_0(\mathbf{x}) \cdot \mathbf{n} \, . \label{eqn:bnorm}
\end{align}
Green's second identity for on-surface evaluations~(\ref{eqn:GreensIdentityOnSurface}) is used
to convert Laplace's equation for~$\Phi$ in \eqn{laplacePhi} into an integral equation:
\begin{equation}
  \forall \mathbf{x} \in \partial \mathcal{D}: ~
  \Phi(\mathbf{x}) + \frac{1}{2 \pi} \int\limits_{\partial\mathcal{D}} \frac{\partial G(\mathbf{x},\mathbf{x}')}{\partial \mathbf{n}'}                \Phi(\mathbf{x}')                        \mathrm{d}S'
                   = \frac{1}{2 \pi} \int\limits_{\partial\mathcal{D}}                G(\mathbf{x},\mathbf{x}')                        \frac{\partial \Phi(\mathbf{x}')}{\partial \mathbf{n}'} \mathrm{d}S' \, .
  \label{eqn:laplaceIntegral}
\end{equation}
The right hand side of \eqn{laplaceIntegral} is given by the normal component of the magnetic field (see \eqn{bnorm})
and acts as a source term here. The outward-directed unit normal vector at location~$\mathbf{x}'$ is denoted~$\mathbf{n}'$
and $G(\mathbf{x},\mathbf{x}')$ is Green's function as in \eqn{GreensFunction}.

\FloatBarrier
\section{Parity of the Magnetic Scalar Potential}
Let $\mathbf{B}_0$ be the total magnetic field on the plasma boundary of a stellarator.
Using the NESTOR~\cite{merkel_1986} code, a scalar magnetic potential~$\Phi$ can be determined
such that the corresponding solenoidal magnetic field~$\mathbf{B} = \nabla \Phi$
cancels the normal component of the total magnetic field on the boundary:
\begin{equation}
  \left(\mathbf{B}_0 + \nabla \Phi \right) \cdot \mathbf{n} = 0 \, ,
\end{equation}
where $\mathbf{n}$ is the surface normal vector.
The current density representing the coils and the plasma is assumed to be stellarator-symmetric,
which implies that the total magnetic field~$\mathbf{B}_0$ must be stellarator-symmetric as well~\cite{dewar_hudson_1998_stellarator_symmetry}.
Thus, it is concluded that the magnetic scalar potential~$\Phi$ and the solenoidal magnetic field~$\mathbf{B}$ derived from
it are stellarator-symmetric as well.

This note adresses the partity of the magnetic scalar potential
in order to reason whether to use $\sin(m\theta-n\zeta)$~(odd parity) or $\cos(m\theta-n\zeta)$~(even parity) Fourier basis functions for the stellarator-symmetric part of $\Phi$.

The magnetic field is expressed in cylindrical components:
\begin{equation}
  \mathbf{B} = \mathbf{B}(\rho, \varphi, z) = \left[ B_\rho, B_\varphi, B_z \right] \, .
\end{equation}
The relation of $\mathbf{B}$ to the scalar magnetic potential~$\Phi$ is as follows:
\begin{equation}
    \mathbf{B} = \nabla \Phi
  =                  \frac{\partial \Phi}{\partial \rho}    \hat{\mathbf{e}}_\rho
    + \frac{1}{\rho} \frac{\partial \Phi}{\partial \varphi} \hat{\mathbf{e}}_\varphi
    +                \frac{\partial \Phi}{\partial z}       \hat{\mathbf{e}}_z \, . \label{eqn:comp_B}
\end{equation}
Stellarator symmetry of $\mathbf{B}$ implies the following relation for the components of $\mathbf{B}$
under the symmetry operation $I_0$:
\begin{equation}
  I_0 \left[ B_\rho, B_\varphi, B_z \right] = \left[ -B_\rho, B_\varphi, B_z \right] \, . \label{eqn:vec_symm}
\end{equation}
Regarding derivatives, the following holds for an arbitrary scalar function $f$ and in particular for the scalar magnetic potential $\Phi$~\cite{dewar_hudson_1998_stellarator_symmetry}:
\begin{equation}
        \left[ \frac{\partial}{\partial \rho},   \frac{\partial}{\partial \varphi},   \frac{\partial}{\partial z} \right] I_0 f
  = I_0 \left[ \frac{\partial}{\partial \rho}, - \frac{\partial}{\partial \varphi}, - \frac{\partial}{\partial z} \right]     f \, . \label{eqn:der_symm}
\end{equation}
The symmetry operation $I_0$ is applied to $\mathbf{B}$ and Eqn.~(\ref{eqn:der_symm}) is applied (note that $I_0 \rho = \rho$):
\begin{align}
    I_0 \mathbf{B}
  =& I_0 \left( \nabla \Phi \right)
  =  I_0 \left[ \frac{\partial \Phi}{\partial \rho}, \frac{1}{\rho} \frac{\partial \Phi}{\partial \varphi}, \frac{\partial \Phi}{\partial z} \right]
  =  \left[ \frac{\partial}{\partial \rho}, - \frac{1}{\rho}\frac{\partial}{\partial \varphi}, - \frac{\partial}{\partial z} \right] I_0 \Phi \nonumber \\
  =& \left[ \frac{\partial}{\partial \rho} \left( I_0 \Phi \right), - \frac{1}{\rho}\frac{\partial}{\partial \varphi} \left( I_0 \Phi \right), - \frac{\partial}{\partial z} \left( I_0 \Phi \right) \right] \nonumber \\
  =& \left[ - \frac{\partial}{\partial \rho} \left(- I_0 \Phi \right), \frac{1}{\rho}\frac{\partial}{\partial \varphi} \left(- I_0 \Phi \right), \frac{\partial}{\partial z} \left(- I_0 \Phi \right) \right] \, .
\end{align}
Component-wise comparison with Eqn.~(\ref{eqn:comp_B}) and use of the symmetry property~(\ref{eqn:vec_symm}) leads to:
\begin{align}
  B_\rho    =& \frac{\partial \Phi}{\partial \rho} = -I_0 B_\rho = \frac{\partial}{\partial \rho} \left( -I_0 \Phi \right) \Rightarrow \Phi = -I_0 \Phi \Leftrightarrow I_0 \Phi = - \Phi \\
  B_\varphi =& \frac{1}{\rho}\frac{\partial \Phi}{\partial \varphi} = I_0 B_\varphi = \frac{1}{\rho} \frac{\partial}{\partial \varphi} \left(- I_0 \Phi \right) \Rightarrow I_0 \Phi = - \Phi \\
  B_z       =& \frac{\partial \Phi}{\partial z} = I_0 B_z = \frac{\partial}{\partial z} \left( -I_0 \Phi \right) \Rightarrow I_0 \Phi = - \Phi \, .
\end{align}
Concluding, $I_0 \Phi = - \Phi$ is consistent with $\mathbf{B}$ being stellarator-symmetric when $\mathbf{B} = \nabla \Phi$
and this implies that $\Phi$ has \underline{odd} parity and can thus be represented using only a $\sin(m\theta-n\zeta)$ Fourier basis.

\FloatBarrier
\section{Analytical formulation}

\FloatBarrier
\subsection{Coordinate System}
The toroidal surface $\partial\mathcal{D}$ is doubly periodic.
It is thus natural to introduce normalized angle-like coordinates
$0\leq u \leq 1$ and $0 \leq v \leq 1$ for the poloidal and toroidal directions, respectively.
The cylindrical coordinates $(r,\varphi,z)$ on the surface $\partial\mathcal{D}$ are then parameterized with two-dimensional Fourier series as follows:
\begin{align}
  r(u,v)     =& \sum\limits_{m=-m_\mathrm{b}}^{m_\mathrm{b}} \sum\limits_{n=-n_\mathrm{b}}^{n_\mathrm{b}}
    \hat{r}_{m,n} \exp\left( 2 \pi i (m u + n v) \right) \, , \,  \hat{r}_{m,n}^* = \hat{r}_{-m,-n} \, , \nonumber \\
  z(u,v)     =& \sum\limits_{m=-m_\mathrm{b}}^{m_\mathrm{b}} \sum\limits_{n=-n_\mathrm{b}}^{n_\mathrm{b}}
    \hat{z}_{m,n} \exp\left( 2 \pi i (m u + n v) \right) \, , \,  \hat{z}_{m,n}^* = \hat{z}_{-m,-n} \, , \\
  \varphi(v) =& \frac{2 \pi}{n_\mathrm{p}} v \, . \nonumber
\end{align}
The cartesian coordinates of points on the surface in a module~$l$ of a $n_\mathrm{p}$-fold toroidally symmetric
configuration, e.g., a stellarator, then read:
\begin{equation}
  \mathbf{x}^{(l)}(u,v) =
    \begin{pmatrix}
      r(u,v) \cdot \cos\left( \frac{2 \pi}{n_\mathrm{p}} (l-1+v) \right) \\
      r(u,v) \cdot \sin\left( \frac{2 \pi}{n_\mathrm{p}} (l-1+v) \right) \\
      z(u,v)
    \end{pmatrix} ,
    \, l = 1, ..., n_\mathrm{p}
    \label{eqn:def_x}
\end{equation}
and the tangential derivatives are denoted $\mathbf{x}_u \equiv \partial \mathbf{x}/\partial u$ and $\mathbf{x}_v \equiv \partial \mathbf{x}/\partial v$
and a shorthand notation is introduced for the first toroidal period as $\mathbf{x} = \mathbf{x}^{(1)}$.
The notation $\mathbf{x}' = \mathbf{x}(u',v')$ with the definition of $\mathbf{x}$ from \eqn{def_x} is used
and leads to
\begin{equation}
  \mathbf{x}'_{u'} = \frac{\partial \mathbf{x}'(u',v')}{\partial u'} = \left. \frac{\partial \mathbf{x}}{\partial u} \right\rvert_{\substack{u=u'\\v=v'}}
  \, \mathrm{and} ~~
  \mathbf{x}'_{v'} = \frac{\partial \mathbf{x}'(u',v')}{\partial v'} = \left. \frac{\partial \mathbf{x}}{\partial v} \right\rvert_{\substack{u=u'\\v=v'}} \, .
\end{equation}

A change of variables is now performed in \eqn{laplaceIntegral} to the normalized coordinates $u$ and $v$.
The Jacobian of the transformation is (according to Eqn.~(2.5.48a) in Ref.~\cite{dHaseleer}):
\begin{equation}
  \mathrm{d}S' = |\mathbf{x}'_{u'} \times \mathbf{x}'_{v'}| \,\mathrm{d}u' \,\mathrm{d}v' \, . \label{eqn:surfaceJacobian}
\end{equation}
The normal vector $\mathbf{n}'$ can be expressed in terms of the tangential derivatives (after Eqn.~(2.3) in Ref.~\cite{merkel_1987}):
\begin{equation}
  \mathbf{n}' = \frac{\mathbf{x}'_{u'} \times \mathbf{x}'_{v'}}{|\mathbf{x}'_{u'} \times \mathbf{x}'_{v'}|} \, . \label{eqn:normalVector}
\end{equation}
The normal derivative of Green's function is:
\begin{equation}
    \frac{\partial G(\mathbf{x},\mathbf{x}')}{\partial \mathbf{n}'}
  = \nabla G(\mathbf{x},\mathbf{x}') \cdot \mathbf{n}'
  = \frac{\mathbf{x}-\mathbf{x}'}{|\mathbf{x}-\mathbf{x}'|^3} \cdot \frac{\mathbf{x}'_{u'} \times \mathbf{x}'_{v'}}{|\mathbf{x}'_{u'} \times \mathbf{x}'_{v'}|} \, . \label{eqn:normalGreensFunction}
\end{equation}

\FloatBarrier
\subsection{Formulation of Laplace's Equation}
Inserting \eqn{normalVector} into \eqn{bnorm} and this as well as \eqn{surfaceJacobian} and~(\ref{eqn:normalGreensFunction}) into \eqn{laplaceIntegral}, one obtains:
\begin{equation}
  \forall (u,v) \in [0,1]^2: ~
    \Phi(u,v)
  + \int\limits_0^1 \int\limits_0^1
      g(u,v,u',v')
      \Phi(u', v')
      \,\mathrm{d}u' \,\mathrm{d}v'
  = \int\limits_0^1 \int\limits_0^1
      h(u,v,u',v')
      \,\mathrm{d}u' \,\mathrm{d}v' \, .
  \label{eqn:laplaceIntegralUV}
\end{equation}
with
\begin{align}
    g(u,v,u',v')
  =& \, \frac{1}{2 \pi}
    \sum\limits_{l=1}^{n_\mathrm{p}}
    \frac{\mathbf{x}-\mathbf{x}'^{(l)}}{|\mathbf{x}-\mathbf{x}'^{(l)}|^3} \cdot \frac{\mathbf{x}'^{(l)}_{u'} \times \mathbf{x}'^{(l)}_{v'}}{ \bcancel{|\mathbf{x}'^{(l)}_{u'} \times \mathbf{x}'^{(l)}_{v'}|} }
    \bcancel{|\mathbf{x}'^{(l)}_{u'} \times \mathbf{x}'^{(l)}_{v'}|} \nonumber \\
  =& \, \frac{1}{2 \pi}
    \sum\limits_{l=1}^{n_\mathrm{p}}
    \frac{ \left( \mathbf{x}-\mathbf{x}'^{(l)} \right) \cdot \left( \mathbf{x}'^{(l)}_{u'} \times \mathbf{x}'^{(l)}_{v'} \right) }{|\mathbf{x}-\mathbf{x}'^{(l)}|^3} \label{eqn:def_g}
\end{align}
and
\begin{align}
    h(u,v,u',v')
  =& \frac{1}{2 \pi}
    \sum\limits_{l=1}^{n_\mathrm{p}} \frac{1}{|\mathbf{x}-\mathbf{x}'^{(l)}|}
    \left( - \mathbf{B}_0(\mathbf{x}') \cdot \frac{\mathbf{x}'^{(l)}_{u'} \times \mathbf{x}'^{(l)}_{v'}}{ \bcancel{|\mathbf{x}'^{(l)}_{u'} \times \mathbf{x}'^{(l)}_{v'}|} } \right)
    \bcancel{|\mathbf{x}'^{(l)}_{u'} \times \mathbf{x}'^{(l)}_{v'}|} \nonumber \\
  =&\frac{1}{2 \pi}
    \sum\limits_{l=1}^{n_\mathrm{p}}
    \frac{ - \mathbf{B}_0(\mathbf{x}') \cdot \left( \mathbf{x}'^{(l)}_{u'} \times \mathbf{x}'^{(l)}_{v'} \right) }{|\mathbf{x}-\mathbf{x}'^{(l)}|} \, . \label{eqn:def_h}
\end{align}

\FloatBarrier
\subsection{Transformation to Fourier Space}
The magnetic scalar potential $\Phi$ is a doubly-periodic function in $u$ and $v$ and can be represented as a Fourier series:
\begin{equation}
    \Phi(u,v)
  = \sum\limits_{m=-\infty}^{+\infty} \sum\limits_{n=-\infty}^{+\infty}
    \hat{\Phi}_{m,n} e^{2 \pi i (m u + n v)} \label{eqn:fourierPhi}
\end{equation}
with
\begin{equation}
 \hat{\Phi}^*_{m,n}=\hat{\Phi}_{-m,-n} \, .
\end{equation}
This representation of $\Phi(u,v)$ is now inserted into \eqn{laplaceIntegralUV}
and it holds for all $(u,v)$:
\begin{align}
  ~& \sum\limits_{m,n}
    \hat{\Phi}_{m,n} e^{2 \pi i (m u + n v)}
  + \int\limits_0^1 \int\limits_0^1
      g(u,v,u',v')
      \left(\sum\limits_{m',n'} \hat{\Phi}_{m',n'} e^{2 \pi i (m' u' + n' v')} \right)
      \,\mathrm{d}u' \,\mathrm{d}v' \nonumber \\
  =& \int\limits_0^1 \int\limits_0^1
      h(u,v,u',v')
      \,\mathrm{d}u' \,\mathrm{d}v' \, .
\end{align}
The summation over $m', n'$ and the corresponding Fourier coefficients $\hat{\Phi}_{m',n'}$ can be pulled outside of the integral:
\begin{align}
  ~& \sum\limits_{m,n}
    \hat{\Phi}_{m,n} e^{2 \pi i (m u + n v)}
  + \sum\limits_{m',n'}
    \left(
      \int\limits_0^1 \int\limits_0^1
        g(u,v,u',v') e^{2 \pi i (m' u' + n' v')}
      \,\mathrm{d}u' \,\mathrm{d}v' \right)
    \hat{\Phi}_{m',n'} \nonumber \\
 =& \int\limits_0^1 \int\limits_0^1
      h(u,v,u',v')
      \,\mathrm{d}u' \,\mathrm{d}v' \, . \label{eqn:laplace_only_Phi_Fourier}
\end{align}
The two integrals appearing here can be considered as functions of $u$ and $v$
and are expressed as Fourier series as well:
\begin{align}
  \hat{g}_{m',n'}(u, v)
  \equiv&\, \int\limits_0^1 \int\limits_0^1
         g(u,v,u',v') e^{2 \pi i (m' u' + n' v')}
         \,\mathrm{d}u' \,\mathrm{d}v' \nonumber \\
  =&\, \sum\limits_{m,n} \hat{g}_{m,n,m',n'} e^{2 \pi i (m u + n v)} \label{eqn:fourier_g}
\end{align}
with
\begin{align}
     \hat{g}_{m,n,m',n'}
  \equiv&\, \int\limits_0^1 \int\limits_0^1
         \hat{g}_{m',n'}(u, v)
         e^{-2 \pi i(m u + n v)}
       \,\mathrm{d}u \,\mathrm{d}v \nonumber \\
  =& \int\limits_0^1 \int\limits_0^1
       \left(
         \int\limits_0^1 \int\limits_0^1
           g(u,v,u',v') e^{2 \pi i (m' u' + n' v')}
           \,\mathrm{d}u' \,\mathrm{d}v'
       \right)
       e^{-2 \pi i(m u + n v)}
       \,\mathrm{d}u \,\mathrm{d}v
\end{align}
as well as
\begin{align}
  \tilde{h}(u,v)
  \equiv&\, \int\limits_0^1 \int\limits_0^1
         h(u,v,u',v')
         \,\mathrm{d}u' \,\mathrm{d}v' \nonumber \\
  =&\, \sum\limits_{m,n} \hat{h}_{m,n} e^{2 \pi i (m u + n v)} \label{eqn:fourier_h}
\end{align}
with
\begin{align}
     \hat{h}_{m,n}
  \equiv&\, \int\limits_0^1 \int\limits_0^1
         \tilde{h}(u,v)
         e^{-2 \pi i(m u + n v)}
       \,\mathrm{d}u \,\mathrm{d}v \nonumber \\
  =& \int\limits_0^1 \int\limits_0^1
       \left(
         \int\limits_0^1 \int\limits_0^1
           h(u,v,u',v')
           \,\mathrm{d}u' \,\mathrm{d}v'
       \right)
       e^{-2 \pi i(m u + n v)}
       \,\mathrm{d}u \,\mathrm{d}v \, .
\end{align}
These Fourier integrals can be reordered a little bit, leading to:
\begin{align}
     \hat{g}_{m,n,m',n'}
  =& \int\limits_0^1 \int\limits_0^1
     \int\limits_0^1 \int\limits_0^1
       g(u,v,u',v')
       e^{2 \pi i (m' u' + n' v' - m u - n v)}
       \,\mathrm{d}u \,\mathrm{d}v
       \,\mathrm{d}u' \,\mathrm{d}v' \label{eqn:ft_g} \\
     \hat{h}_{m,n}
  =& \int\limits_0^1 \int\limits_0^1
     \int\limits_0^1 \int\limits_0^1
       h(u,v,u',v')
       e^{-2 \pi i(m u + n v)}
       \,\mathrm{d}u \,\mathrm{d}v
       \,\mathrm{d}u' \,\mathrm{d}v' \label{eqn:ft_h} \, .
\end{align}
The Fourier representations of the integrals over $g$ and $h$ from \eqn{fourier_g} and \eqn{fourier_h}
can be inserted back into Laplace's equation~(\ref{eqn:laplace_only_Phi_Fourier}):
\begin{equation}
    \sum\limits_{m,n}
    \hat{\Phi}_{m,n} e^{2 \pi i (m u + n v)}
  + \sum\limits_{m',n'}
    \left( \sum\limits_{m,n} \hat{g}_{m,n,m',n'} e^{2 \pi i (m u + n v)} \right)
    \hat{\Phi}_{m',n'}
  = \sum\limits_{m,n} \hat{h}_{m,n} e^{2 \pi i (m u + n v)} \, .
\end{equation}
The summations in the middle of above equation can be swapped:
\begin{equation}
    \sum\limits_{m,n}
      \hat{\Phi}_{m,n}
    e^{2 \pi i (m u + n v)}
  + \sum\limits_{m,n}
      \sum\limits_{m',n'}
      \hat{g}_{m,n,m',n'}
      \hat{\Phi}_{m',n'}
    e^{2 \pi i (m u + n v)}
  = \sum\limits_{m,n}
      \hat{h}_{m,n}
    e^{2 \pi i (m u + n v)} \, .
\end{equation}
This equation holds for all $(u,v) \in [0,1]^2$.
The orthogonality of the Fourier basis therefore allows to consider only the Fourier coefficients.
This leads to the following infinite set of linear equations:
\begin{align}
  \forall\, (m,n) \in [-\infty,+\infty]^2 :& \,
    \hat{\Phi}_{m,n}
  + \sum\limits_{m'=-\infty}^{+\infty} \sum\limits_{n'=-\infty}^{+\infty}
      \hat{g}_{m,n,m',n'} \hat{\Phi}_{m',n'}
  = \hat{h}_{m,n} \\
  \Leftrightarrow
  \forall\, (m,n) \in [-\infty,+\infty]^2 :& \,
  \sum\limits_{m'=-\infty}^{+\infty} \sum\limits_{n'=-\infty}^{+\infty}
    \left( \hat{g}_{m,n,m',n'} + \delta_{m,m'} \delta_{n,n'} \right) \hat{\Phi}_{m',n'}
  = \hat{h}_{m,n} \label{eqn:linsys_inf} \, .
\end{align}
A matrix~$\mathbf{A}$ and a right-hand side~$\mathbf{b}$
can be defined:
\begin{align}
 (\mathbf{A})_{m,n,m',n'} =&\, \hat{g}_{m,n,m',n'} + \delta_{m,m'} \delta_{n,n'} \\
 (\mathbf{b})_{m,n}       =&\, \hat{h}_{m,n}
\end{align}
as well as a solution vector~$\mathbf{x}$:
\begin{equation}
 (\mathbf{x})_{m',n'} = \hat{\Phi}_{m',n'}
\end{equation}
in order to cast \eqn{linsys_inf} into a linear system of equations:
\begin{equation}
 \mathbf{A} \mathbf{x} = \mathbf{b} \, .
\end{equation}
The Fourier resolution is truncated at some maximum Fourier mode numbers $M$, $N$
in order to arrive at a finite-dimensional linear system of equations,
which then can be solved by standard linear algebra methods,
e.g., from \texttt{LAPACK}.

The remainder of this chapter
deals with the computation of the matrix elements
$\hat{g}_{m,n,m',n'}$ and $\hat{h}_{m,n}$
as well as the computation of the vacuum magnetic pressure
from the obtained solution vector $\hat{\Phi}_{m',n'}$.

\FloatBarrier
\subsection{Cancellation of Singularities using a Subtraction Method}
The kernel functions $g$ and $h$ are singular if simultaneously $u=u'$ and $v=v'$.
Their Fourier transforms in \eqn{ft_g} and \eqn{ft_h} thus cannot be evaluated directly
using standard discrete Fourier transform methods.
The trick is now to subtract a function with the same singular behaviour as the function to be transformed
which can be analytically Fourier-transformed.
A Taylor expansion of $|\mathbf{x}-\mathbf{x}'|$ around fixed values of $(u',v')$ is considered.
\begin{equation}
  |\mathbf{x}-\mathbf{x}'| = \sqrt{ (\mathbf{x}-\mathbf{x}') \cdot (\mathbf{x}-\mathbf{x}') } \, . \label{eqn:dist_x_xprime}
\end{equation}
First look at $\mathbf{x}-\mathbf{x}'$:
\begin{align}
  \mathbf{x}-\mathbf{x}'
     =   &    \underbrace{ \left. (\mathbf{x}-\mathbf{x}') \right\rvert_{\substack{u=u'\\v=v'}} }_{
               \begin{aligned}=&\, \mathbf{x}'-\mathbf{x}' \\=&\, 0 \end{aligned} }
           + \underbrace{ \left. \frac{\partial(\mathbf{x}-\mathbf{x}')}{\partial u} \right\rvert_{\substack{u=u'\\v=v'}} }_{
               \begin{aligned}= \underbrace{\left. \frac{\partial \mathbf{x} }{\partial u}\right\rvert_{\substack{u=u'\\v=v'}}}_{=\mathbf{x}'_{u'}}
                              - \underbrace{\left. \frac{\partial \mathbf{x}'}{\partial u}\right\rvert_{\substack{u=u'\\v=v'}}}_{= 0} \end{aligned}} (u-u')
           + \underbrace{ \left. \frac{\partial(\mathbf{x}-\mathbf{x}')}{\partial v} \right\rvert_{\substack{u=u'\\v=v'}} }_{
               \begin{aligned}= \underbrace{\left. \frac{\partial \mathbf{x} }{\partial v}\right\rvert_{\substack{u=u'\\v=v'}}}_{=\mathbf{x}'_{v'}}
                              - \underbrace{\left. \frac{\partial \mathbf{x}'}{\partial v}\right\rvert_{\substack{u=u'\\v=v'}}}_{= 0} \end{aligned}} (v-v') \nonumber \\
    ~ &\,+ \frac{1}{2} \left[ \mathbf{x}'_{u' u'} (u-u')^2 + 2\, \mathbf{x}'_{u' v'} (u-u') (v-v') + \mathbf{x}'_{v' v'} (v-v')^2 \right] + \, ... \nonumber \\
  \approx&\, \phantom{+}~\underbrace{\mathbf{x}'_{u'} (u-u') + \mathbf{x}'_{v'} (v-v')}_{\textrm{first-order contribution}} \label{eqn:dist_1stOrder} \\
     ~   &\,          +  \underbrace{\frac{1}{2} \left[ \mathbf{x}'_{u' u'} (u-u')^2 + 2\, \mathbf{x}'_{u' v'} (u-u') (v-v') + \mathbf{x}'_{v' v'} (v-v')^2 \right]}_{\textrm{second-order contribution}} \label{eqn:dist_2ndOrder}
\end{align}
due to neglection of higher-order terms.
Inserting the first-order approximation from \eqn{dist_1stOrder} into \eqn{dist_x_xprime} leads to:
\begin{align}
  |\mathbf{x}-\mathbf{x}'|
  \approx &\, \sqrt{ \left[ \mathbf{x}'_{u'} (u-u') + \mathbf{x}'_{v'} (v-v') \right]^2 } \nonumber \\
      =   &\, \sqrt{ {\mathbf{x}'}^2_{u'} (u-u')^2 + 2 \mathbf{x}'_{u'} \mathbf{x}'_{v'} (u-u') (v-v') + {\mathbf{x}'}^2_{v'} (v-v')^2 } \, . \label{eqn:approx_dist}
\end{align}
Only the contribution to $h$ from the toroidal period in which $\mathbf{x}$ is located is singular.
This is equivalent to only considering the first period ($l=1$), since the numbering of toroidal periods is arbitrary.
The approximation from \eqn{approx_dist} is inserted
into the contribution from the first toroidal period to $h$ from \eqn{def_h}:
\begin{equation}
  h(u-u',v-v',u',v')
  \approx\, \frac{1}{2 \pi}
  \frac{- \mathbf{B}_0(\mathbf{x}') \cdot \left( \mathbf{x}'_{u'} \times \mathbf{x}'_{v'} \right)}
       {\left( {\mathbf{x}'}^2_{u'} (u-u')^2 + 2 \mathbf{x}'_{u'} \mathbf{x}'_{v'} (u-u') (v-v') + {\mathbf{x}'}^2_{v'} (v-v')^2 \right)^{1/2}} \, . \label{eqn:approx_h}
\end{equation}
Similarly, only the contribution to $g$ from the first toroidal period ($l=1$) is singular.
The first- and second-order contributions of $\mathbf{x}-\mathbf{x}'$ from \eqn{dist_1stOrder} and \eqn{dist_2ndOrder}
and the first-order approximation of $|\mathbf{x}-\mathbf{x}'|$ from \eqn{approx_dist}
are inserted into the contribution from the first toroidal period to $g$ in \eqn{def_g}:
\begin{align}
  ~ &\, 2 \pi g(u,v,u',v') \nonumber \\
  \approx &\, \phantom{+}~\frac{  \left[ \mathbf{x}'_{u'} (u-u') + \mathbf{x}'_{v'} (v-v') \right] \cdot \left( \mathbf{x}'_{u'} \times \mathbf{x}'_{v'} \right)}
                               {  \left[ {\mathbf{x}'}^2_{u'} (u-u')^2 + 2 \mathbf{x}'_{u'} \mathbf{x}'_{v'} (u-u') (v-v') + {\mathbf{x}'}^2_{v'} (v-v')^2 \right]^{3/2}} \label{eqn:g_1stOrder} \\
     ~    &\,          +  \frac{  \left[ \mathbf{x}'_{u' u'} (u-u')^2 + 2\, \mathbf{x}'_{u' v'} (u-u') (v-v') + \mathbf{x}'_{v' v'} (v-v')^2 \right] \cdot \left( \mathbf{x}'_{u'} \times \mathbf{x}'_{v'} \right)}
                               {2 \left[ {\mathbf{x}'}^2_{u'} (u-u')^2 + 2 \mathbf{x}'_{u'} \mathbf{x}'_{v'} (u-u') (v-v') + {\mathbf{x}'}^2_{v'} (v-v')^2 \right]^{3/2}} \label{eqn:g_2ndOrder}
\end{align}
Note that the first-order contribution from \eqn{g_1stOrder} to the approximation of $g$ vanishes, since
\begin{align*}
  \mathbf{x}'_{u'} \cdot& \left( \mathbf{x}'_{u'} \times \mathbf{x}'_{v'} \right) = 0 ~ \mathrm{and}\\
  \mathbf{x}'_{v'} \cdot& \left( \mathbf{x}'_{u'} \times \mathbf{x}'_{v'} \right) = 0 \, .
\end{align*}
Therefore, the approximation of $g$ can be reduced to:
\begin{equation}
  g(u-u',v-v',u',v') \approx \frac{1}{2 \pi}
    \frac{A (u-u')^2 + 2 B (u-u') (v-v') + C (v-v')^2}
         {\left[ {\mathbf{x}'}^2_{u'} (u-u')^2 + 2 \mathbf{x}'_{u'} \mathbf{x}'_{v'} (u-u') (v-v') + {\mathbf{x}'}^2_{v'} (v-v')^2 \right]^{3/2}}
  \label{eqn:approx_g}
\end{equation}
with
\begin{align}
  A \equiv& \frac{1}{2} \mathbf{x}'_{u' u'} \cdot \left( \mathbf{x}'_{u'} \times \mathbf{x}'_{v'} \right) \, , \nonumber \\
  B \equiv& \frac{1}{2} \mathbf{x}'_{u' v'} \cdot \left( \mathbf{x}'_{u'} \times \mathbf{x}'_{v'} \right) ~ \mathrm{and} \label{eqn:def_ABC}\\
  C \equiv& \frac{1}{2} \mathbf{x}'_{v' v'} \cdot \left( \mathbf{x}'_{u'} \times \mathbf{x}'_{v'} \right) \, . \nonumber
\end{align}
The derivatives with respect to normalized coordinates are transformed into derivatives
with respect to the actual coordinates by application of the chain rule:
\begin{align}
  A =& \frac{1}{2} \frac{\partial^2 \mathbf{x}'}{\partial {u'}^2} \cdot \mathbf{N}'
    =  \frac{1}{2} \underbrace{ \left( \frac{\partial \theta'}{\partial u'} \right)^2}_{=(2\pi)^2} \frac{\partial^2 \mathbf{x}'}{\partial {\theta'}^2} \cdot \mathbf{N}'
    =  (2\pi)^2 \underbrace{\left( \frac{1}{2} \frac{\partial^2 \mathbf{x}'}{\partial {\theta'}^2} \cdot \mathbf{N}' \right)}_{\equiv a_{u,u}} \\
 2B =& \bcancel{2} \frac{1}{\bcancel{2}} \frac{\partial^2 \mathbf{x}'}{\partial u' \partial v'} \cdot \mathbf{N}'
    =  \underbrace{ \frac{\partial \theta'}{\partial u'} \frac{\partial \zeta'}{\partial v'} }_{=(2\pi)^2}
       \underbrace{ \frac{\partial \phi'}{\partial \zeta'} }_{=1/n_\mathrm{fp}}
       \frac{\partial^2 \mathbf{x}'}{\partial \theta' \partial \phi'} \cdot \mathbf{N}'
    =  (2\pi)^2 \underbrace{\left(\frac{1}{n_\mathrm{fp}} \frac{\partial^2 \mathbf{x}'}{\partial \theta' \partial \phi'} \cdot \mathbf{N}' \right)}_{\equiv a_{u,v}} \\
  C =& \frac{1}{2} \frac{\partial^2 \mathbf{x}'}{\partial {v'}^2} \cdot \mathbf{N}'
    =  \frac{1}{2}
       \underbrace{ \left( \frac{\partial \zeta'}{\partial v'} \right)^2}_{=(2\pi)^2}
       \underbrace{ \left( \frac{\partial \phi'}{\partial \zeta'}\right)^2 }_{=1/n_\mathrm{fp}^2}\frac{\partial^2 \mathbf{x}'}{\partial {\phi'}^2} \cdot \mathbf{N}'
    =  (2\pi)^2 \underbrace{\left( \frac{1}{2} \frac{1}{n_\mathrm{fp}^2} \frac{\partial^2 \mathbf{x}'}{\partial {\phi'}^2} \cdot \mathbf{N}' \right)}_{\equiv a_{v,v}}
\end{align}
The components of the surface normal vector are given in \eqn{surfNComponents}
and the second derivatives of the surface geometry are given in \eqn{d2Xdt2}, \eqn{d2Xdtz} and \eqn{d2Xdz2}.
The terms $a_{u,u}$, $a_{u,v}$ and $a_{v,v}$ thus can be evaluated in cylindrical components:
\begin{align}
  a_{u,u} =& \frac{1}{2}
               \left(  \frac{\partial^2 R'}{\partial {\theta'}^2} {N^R}'
                     + \frac{\partial^2 Z'}{\partial {\theta'}^2} {N^Z}' \right) \label{eqn:auu} \\
  a_{u,v} =& \frac{1}{n_\mathrm{fp}}
               \left(  \frac{\partial^2 R'}{\partial \theta' \partial \phi'} {N^R}'
                     + \frac{\partial   R'}{\partial \theta'               } {N^\phi}'
                     + \frac{\partial^2 Z'}{\partial \theta' \partial \phi'} {N^Z}'      \right) \label{eqn:auv} \\
  a_{v,v} =& \frac{1}{2} \frac{1}{n_\mathrm{fp}^2}
               \left[  \left( \frac{\partial^2 R'}{\partial {\phi'}^2} - R' \right) {N^R}'
                     + 2      \frac{\partial   R'}{\partial \phi'  }                {N^\phi}'
                     +        \frac{\partial^2 Z'}{\partial {\phi'}^2}              {N^Z}'    \right] \label{eqn:avv} \, .
\end{align}
Shorthand notations are also introduced for the (squared and mixed) tangential derivatives:
\begin{align}
  a \equiv&\, {\mathbf{x}'}^2_{u'} \nonumber \\
  b \equiv&\,  \mathbf{x}'_{u'} \mathbf{x}'_{v'} \label{eqn:def_abc} \\
  c \equiv&\, {\mathbf{x}'}^2_{v'} \nonumber \, .
\end{align}
It follows:
\begin{align}
  a =&       \Biggl( \frac{\partial \mathbf{x}'}{\partial \theta'} \underbrace{\frac{\partial \theta'}{\partial u'}}_{=2\pi} \Biggr)
       \cdot \Biggl( \frac{\partial \mathbf{x}'}{\partial \theta'} \underbrace{\frac{\partial \theta'}{\partial u'}}_{=2\pi} \Biggr)
    =  (2 \pi)^2 \underbrace{\left( \frac{\partial \mathbf{x}'}{\partial \theta'} \cdot \frac{\partial \mathbf{x}'}{\partial \theta'} \right)}_{=g_{u,u}^\mathrm{b}} \\
 2b =& 2     \Biggl( \frac{\partial \mathbf{x}'}{\partial \theta'} \underbrace{\frac{\partial \theta'}{\partial u'}}_{=2\pi} \Biggr)
       \cdot \Biggl( \frac{\partial \mathbf{x}'}{\partial \phi'} \underbrace{\frac{\partial \phi'}{\partial \zeta'}}_{=1/n_\mathrm{fp}} \underbrace{\frac{\partial \zeta'}{\partial v'}}_{=2\pi} \Biggr)
    =  (2 \pi)^2 \underbrace{\left(\frac{2}{n_\mathrm{fp}} \frac{\partial \mathbf{x}'}{\partial \theta'} \cdot \frac{\partial \mathbf{x}'}{\partial \phi'} \right)}_{=g_{u,v}^\mathrm{b}} \\
  c =&       \Biggl( \frac{\partial \mathbf{x}'}{\partial \phi'} \underbrace{\frac{\partial \phi'}{\partial \zeta'}}_{=1/n_\mathrm{fp}} \underbrace{\frac{\partial \zeta'}{\partial v'}}_{=2\pi} \Biggr)
       \cdot \Biggl( \frac{\partial \mathbf{x}'}{\partial \phi'} \underbrace{\frac{\partial \phi'}{\partial \zeta'}}_{=1/n_\mathrm{fp}} \underbrace{\frac{\partial \zeta'}{\partial v'}}_{=2\pi} \Biggr)
    =  (2 \pi)^2 \underbrace{\left(\frac{1}{n_\mathrm{fp}^2} \frac{\partial \mathbf{x}'}{\partial \phi'} \cdot \frac{\partial \mathbf{x}'}{\partial \phi'} \right)}_{=g_{v,v}^\mathrm{b}} \, .
\end{align}
The cylindrical components of the first-order derivatives can be gotten from \eqn{dXdt} and \eqn{dXdp} and it follows:
\begin{align}
  g_{u,u}^\mathrm{b} =&       \left( \frac{\partial R'}{\partial \theta'}, 0, \frac{\partial Z'}{\partial \theta'} \right)
                        \cdot \left( \frac{\partial R'}{\partial \theta'}, 0, \frac{\partial Z'}{\partial \theta'} \right)
                     =  \left(\frac{\partial R'}{\partial \theta'} \right)^2 + \left( \frac{\partial Z'}{\partial \theta'} \right)^2 \label{eqn:guu_b} \\
  g_{u,v}^\mathrm{b} =& \frac{2}{n_\mathrm{fp}}
                              \left( \frac{\partial R'}{\partial \theta'}, 0 , \frac{\partial Z'}{\partial \theta'} \right)
                        \cdot \left( \frac{\partial R'}{\partial \phi'  }, R', \frac{\partial Z'}{\partial \phi'  } \right)
                     = \frac{2}{n_\mathrm{fp}} \left(  \frac{\partial R'}{\partial \theta'} \frac{\partial R'}{\partial \phi'}
                                                     + \frac{\partial Z'}{\partial \theta'} \frac{\partial Z'}{\partial \phi'} \right) \label{eqn:guv_b} \\
  g_{v,v}^\mathrm{b} =& \frac{1}{n_\mathrm{fp}^2} \left( \frac{\partial R'}{\partial \phi'}, R', \frac{\partial Z'}{\partial \phi'} \right)
                        \cdot                     \left( \frac{\partial R'}{\partial \phi'}, R', \frac{\partial Z'}{\partial \phi'} \right)
                     =  \frac{1}{n_\mathrm{fp}^2} \left[  \left( \frac{\partial R'}{\partial \phi'} \right)^2
                                                        + {R'}^2
                                                        + \left( \frac{\partial Z'}{\partial \phi'} \right)^2 \right] \label{eqn:gvv_b} \, .
\end{align}
The inhomogenity due to the normal component of the external magnetic field is denoted as
\begin{align}
  F(u',v') =&\, - \mathbf{B}_0(\mathbf{x}') \cdot \left( \mathbf{x}'_{u'} \times \mathbf{x}'_{v'} \right) \, .
\end{align}
The only dependance of \eqn{approx_h} on $u$ and $v$ is now in the terms $(u-u')$ and $(v-v')$.
In order to find a periodic approximation of $h$, it is thus required to find a periodic approximation of $(u-u')$ and $(v-v')$.
The normalized variables $u$ and $v$ are periodic over the interval $[0, 1]$ and thus $(u-u')$ spans the range $[-1, 1]$
for all possible combinations of $u$ and $u'$.
Consider the Taylor expansion of $\tan(x)$:
\begin{align}
              \tan(    x) =&\,     x \underbrace{+ \frac{1}{3} x^3 + \, ...}_{\mathrm{neglected}} \nonumber \\
  \Rightarrow \tan(\pi x) \approx &\,          \pi           x  \nonumber \\
  \Leftrightarrow      x  \approx &\, \frac{1}{\pi} \tan(\pi x) \nonumber \\
  \Rightarrow      (u-u') \approx &\, \frac{1}{\pi} \tan\left(\pi (u-u')\right) \label{eqn:tan_u} \\
  \Rightarrow      (v-v') \approx &\, \frac{1}{\pi} \tan\left(\pi (v-v')\right) \label{eqn:tan_v}
\end{align}
which is periodic over the desired interval and spans a corresponding interval of $[-\infty, +\infty]$.
Insertion of the approximations for $(u-u')$ from \eqn{tan_u} and for $(v-v')$ from \eqn{tan_v} into
the approximation of $h$ from \eqn{approx_h} leads to a doubly-periodic function $h^\mathrm{sing}$ which approximates $h$
locally around $(u',v')$ and has the same singular behaviour as $h$:
\begin{align}
  ~& h^\mathrm{sing} (u-u',v-v',u',v') \nonumber \\
  =& \frac{F(u',v')}
          {2 \left[ a \tan^2\left(\pi (u-u')\right) + 2 b \tan\left(\pi (u-u') \right) \tan\left(\pi (v-v') \right) + c \tan^2\left(\pi (v-v') \right) \right]^{1/2}} \, . \label{eqn:h_sing}
\end{align}
Similarly, the approximations for $(u-u')$ from \eqn{tan_u} and for $(v-v')$ from \eqn{tan_v}
are inserted into the approximation of $g$ from \eqn{approx_g}, leading to a doubly-periodic function $g^\mathrm{sing}$ which approximates $g$
locally around $(u',v')$ and has the same singular behaviour as $g$:
\begin{align}
  ~& g^\mathrm{sing} (u-u',v-v',u',v') \nonumber \\
  =& \frac{A \tan^2\left(\pi (u-u')\right) + 2 B \tan\left(\pi (u-u') \right) \tan\left(\pi (v-v') \right) + C \tan^2\left(\pi (v-v') \right)}
          {2 \left[ a \tan^2\left(\pi (u-u')\right) + 2 b \tan\left(\pi (u-u') \right) \tan\left(\pi (v-v') \right) + c \tan^2\left(\pi (v-v') \right) \right]^{3/2}} \, . \label{eqn:g_sing}
\end{align}
The regularization is first applied to the source term elements~$\hat{h}_{m,n}$ from \eqn{ft_h}:
\begin{align}
  \hat{h}_{m,n}
  =& \int\limits_0^1 \int\limits_0^1 \int\limits_0^1 \int\limits_0^1
      h(u,v,u',v') e^{-2 \pi i (m u + n v)}
      \,\mathrm{d}u \,\mathrm{d}v \,\mathrm{d}u' \,\mathrm{d}v' \nonumber \\
  =& \phantom{+}~\int\limits_0^1 \int\limits_0^1 \int\limits_0^1 \int\limits_0^1
                   h^\mathrm{reg} (u,v,u',v') e^{-2 \pi i (m u + n v)}
                 \,\mathrm{d}u \,\mathrm{d}v \,\mathrm{d}u' \,\mathrm{d}v'  \\
  ~&          +  \int\limits_0^1 \int\limits_0^1 \phantom{\int\limits_0^1 \int\limits_0^1}
                   \,\tilde{h}^\mathrm{an}_{m,n} (u',v') \phantom{u,v} e^{-2 \pi i (m u' + n v')}
                 \phantom{\mathrm{d}u \,\mathrm{d}v} \,\mathrm{d}u' \,\mathrm{d}v'
\end{align}
with
\begin{align}
  h^\mathrm{reg} (u,v,u',v')
  = h&(u,v,u',v') - h^\mathrm{sing} (u-u',v-v',u',v') \, , \\
  \tilde{h}^\mathrm{an}_{m,n} (u',v')
  = \int\limits_0^1 \int\limits_0^1&
      h^\mathrm{sing} (u-u',v-v',u',v') e^{-2 \pi i \left[m (u-u') + n (v-v')\right]}
    \,\mathrm{d}u \,\mathrm{d}v \, . \label{eqn:singular_ft_h}
\end{align}
This allows to introduce Fourier coefficients
$\hat{h}^\textrm{reg}_{m,n}$ and $\hat{h}^\textrm{an}_{m,n}$ with
\begin{equation}
 \hat{h}_{m,n} = \hat{h}^\textrm{reg}_{m,n} + \hat{h}^\textrm{an}_{m,n}
\end{equation}
as follows:
\begin{align}
 \hat{h}^\textrm{reg}_{m,n}
 =&\, \int\limits_0^1 \int\limits_0^1 \int\limits_0^1 \int\limits_0^1
        h^\textrm{reg} (u,v,u',v') e^{-2 \pi i (m u + n v)}
        \,\mathrm{d}u \,\mathrm{d}v \,\mathrm{d}u' \,\mathrm{d}v' \label{eqn:reg_ft_h} \\
 =&\, \int\limits_0^1 \int\limits_0^1
        \tilde{h}^\textrm{reg}_{m,n}(u,v) e^{-2 \pi i (m u + n v)}
        \,\mathrm{d}u \,\mathrm{d}v \label{eqn:ft_h_reg}
\end{align}
with
\begin{align}
 \tilde{h}^\textrm{reg}_{m,n}(u,v)
 \equiv&\, \int\limits_0^1 \int\limits_0^1
             h^\textrm{reg} (u,v,u',v')
             \,\mathrm{d}u' \,\mathrm{d}v'
\end{align}
as well as
\begin{align}
 \hat{h}^\textrm{an}_{m,n}
 =&\, \int\limits_0^1 \int\limits_0^1
                   \,\tilde{h}^\mathrm{an}_{m,n} (u',v') e^{-2 \pi i (m u' + n v')}
                 \,\mathrm{d}u' \,\mathrm{d}v' \, . \label{eqn:ft_h_an}
\end{align}
The regularization of the Fourier transform of $g$ from \eqn{ft_g} is considered next:
\begin{align}
  ~& \hat{g}_{m,n,m',n'} \nonumber \\
  =& \int\limits_0^1 \int\limits_0^1 \int\limits_0^1 \int\limits_0^1
       g(u,v,u',v') e^{-2 \pi i (m u + n v)} e^{2 \pi i (m' u' + n' v')}
     \,\mathrm{d}u \,\mathrm{d}v \,\mathrm{d}u' \,\mathrm{d}v' \nonumber \\
  =& \phantom{+}~\int\limits_0^1 \int\limits_0^1 \int\limits_0^1 \int\limits_0^1
                   g^\mathrm{reg}(u,v,u',v') e^{-2 \pi i (m u + n v)} e^{2 \pi i (m' u' + n' v')}
                 \,\mathrm{d}u \,\mathrm{d}v \,\mathrm{d}u' \,\mathrm{d}v' \\
  ~&          +  \int\limits_0^1 \int\limits_0^1 \phantom{\int\limits_0^1 \int\limits_0^1}
                   \,\tilde{g}^\mathrm{an}_{m,n} (u',v') \phantom{u,v} e^{-2 \pi i (m u' + n v')} e^{2 \pi i (m' u' + n' v')}
                 \phantom{\mathrm{d}u \,\mathrm{d}v} \,\mathrm{d}u' \,\mathrm{d}v'
\end{align}
with
\begin{align}
  g^\mathrm{reg}(u,v,u',v'&)
  = g(u,v,u',v') - g^\mathrm{sing} (u-u',v-v',u',v') \, , \\
  \tilde{g}^\mathrm{an}_{m,n} (u',v')
  =& \int\limits_0^1 \int\limits_0^1
       g^\mathrm{sing} (u-u',v-v',u',v') e^{-2 \pi i (m u + n v)} e^{2 \pi i (m u' + n v')}
     \,\mathrm{d}u \,\mathrm{d}v \nonumber \\
  =& \int\limits_0^1 \int\limits_0^1
       g^\mathrm{sing} (u-u',v-v',u',v') e^{-2 \pi i \left[m (u-u') + n (v-v')\right]}
     \,\mathrm{d}u \,\mathrm{d}v \label{eqn:singular_ft_g} \, .
\end{align}
This allows to introduce Fourier coefficients
$\hat{g}^\textrm{reg}_{m,n,m',n'}$ and $\hat{g}^\textrm{an}_{m,n,m',n'}$ with
\begin{equation}
 \hat{g}_{m,n,m',n'} = \hat{g}^\textrm{reg}_{m,n,m',n'} + \hat{g}^\textrm{an}_{m,n,m',n'}
\end{equation}
as follows:
\begin{align}
 \hat{g}^\textrm{reg}_{m,n,m',n'}
 =&\, \int\limits_0^1 \int\limits_0^1 \int\limits_0^1 \int\limits_0^1
        g^\mathrm{reg}(u,v,u',v') e^{-2 \pi i (m u + n v)} e^{2 \pi i (m' u' + n' v')}
        \,\mathrm{d}u \,\mathrm{d}v \,\mathrm{d}u' \,\mathrm{d}v' \label{eqn:reg_ft_g} \\
 =&\, \int\limits_0^1 \int\limits_0^1
        \tilde{g}^\textrm{reg}_{m,n}(u',v') e^{2 \pi i (m' u' + n' v')}
        \,\mathrm{d}u' \,\mathrm{d}v'
\end{align}
with
\begin{align}
 \tilde{g}^\textrm{reg}_{m,n}(u',v')
 \equiv&\, \int\limits_0^1 \int\limits_0^1
             g^\textrm{reg}(u,v,u',v') e^{-2 \pi i (m u + n v)}
             \,\mathrm{d}u \,\mathrm{d}v
\end{align}
as well as
\begin{align}
 \hat{g}^\textrm{an}_{m,n,m',n'}
 =&\, \int\limits_0^1 \int\limits_0^1
        \tilde{g}^\mathrm{an}_{m,n}(u',v') e^{-2 \pi i (m u' + n v')} e^{2 \pi i (m' u' + n' v')}
        \,\mathrm{d}u' \,\mathrm{d}v' \label{eqn:ft_g_an} \\
 =&\, \int\limits_0^1 \int\limits_0^1
        \tilde{\tilde{g}}^\mathrm{an}_{m,n}(u',v') e^{2 \pi i (m' u' + n' v')}
        \,\mathrm{d}u' \,\mathrm{d}v'
\end{align}
with
\begin{equation}
 \tilde{\tilde{g}}^\mathrm{an}_{m,n}(u',v')
 = \tilde{g}^\mathrm{an}_{m,n}(u',v') e^{-2 \pi i (m u' + n v')} \label{eqn:tt_g} \, .
\end{equation}
Looking back, it is noted that the matrix elements $\hat{g}_{m,n,m',n'}$
can be computed as follows:
\begin{align}
 \hat{g}_{m,n,m',n'}
 =&\, \int\limits_0^1 \int\limits_0^1
        \left[
            \tilde{g}^\textrm{reg}_{m,n}(u',v')
          + \tilde{\tilde{g}}^\mathrm{an}_{m,n}(u',v')
        \right] e^{2 \pi i (m' u' + n' v')}
        \,\mathrm{d}u' \,\mathrm{d}v' \nonumber \\
 =&\, \int\limits_0^1 \int\limits_0^1
        \tilde{g}_{m,n}(u',v') e^{2 \pi i (m' u' + n' v')}
        \,\mathrm{d}u' \,\mathrm{d}v' \label{eqn:ft_g_total}
\end{align}
with
\begin{equation}
 \tilde{g}_{m,n}(u',v') = \tilde{g}^\textrm{reg}_{m,n}(u',v')
          + \tilde{\tilde{g}}^\mathrm{an}_{m,n}(u',v') \, .
\end{equation}
The Fourier transform kernel in \eqn{ft_g_total}
is similar to the kernel in \eqn{ft_h_reg}
(since the quantites are independent, one transform can be over $(m' u' + n' v')$
and simultaneously the other transform can be over $(m u + n v)$)
and this again motivates to compute these two Fourier transforms
in the same loop in an implementation.

The integrands in \eqn{reg_ft_h} and \eqn{reg_ft_g} are now finite and standard numerical Fourier transform methods can be applied.
The Fourier transform of the equivalently-singular functions in \eqn{singular_ft_h} and \eqn{singular_ft_g} are calculated analytically in the following.
The Fourier kernel is in both cases
\begin{equation}
    e^{-2 \pi i \left[m (u-u') + n (v-v')\right]}
  = \left[  e^{2 \pi i \left[m (u-u') + n (v-v')\right]} \right]^{*} \, . \label{eqn:fourier_kernel}
\end{equation}
The functions $h^\mathrm{sing}$ and $g^\mathrm{sing}$ are both real-valued.
The integrals in \eqn{singular_ft_h} and \eqn{singular_ft_g} span the whole domain of the integrands,
since the Fourier kernel in \eqn{fourier_kernel} is $2\pi$-periodic and the periodicity in $h^\mathrm{sing}$ and $g^\mathrm{sing}$
comes from $\tan(.)$, which is $\pi$-periodic.
Thus, the shift of the integration variables introduced by $(u-u')$ and $(v-v')$
does not influence the results of the integrals and can be omitted:
\begin{align}
  \tilde{h}^\mathrm{an}_{m,n}(u',v')
  =&\, \frac{F(u',v')}{2}
     \int\limits_0^1 \int\limits_0^1
       \frac{e^{-2 \pi i (m u + n v)}}
            {\left[ a \tan^2(\pi u) + 2 b \tan(\pi u) \tan(\pi v) + c \tan^2(\pi v) \right]^{1/2}}
     \,\mathrm{d}u \,\mathrm{d}v \\
  \tilde{g}^\mathrm{an}_{m,n}(u',v')
  =&\, \int\limits_0^1 \int\limits_0^1
       \frac{\left[A \tan^2(\pi u) + 2 B \tan(\pi u) \tan(\pi v) + C \tan^2(\pi v) \right] e^{-2 \pi i (m u + n v)}}
            {\left[a \tan^2(\pi u) + 2 b \tan(\pi u) \tan(\pi v) + c \tan^2(\pi v) \right]^{1/2}                   }
     \,\mathrm{d}u \,\mathrm{d}v \, .
\end{align}
Note however that there is still the dependence on $u'$ and $v'$ in the factors $a$, $b$ and $c$ as well as $A$, $B$ and $C$.
Analytical coefficients $I_{m,n}(u',v')$ can be identified:
\begin{equation}
  I_{m,n}(u',v')
  = \pi \int\limits_0^1 \int\limits_0^1
       \frac{e^{2 \pi i (m u + n v)}}
            {\left[ a \tan^2(\pi u) + 2 b \tan(\pi u) \tan(\pi v) + c \tan^2(\pi v) \right]^{1/2}}
     \,\mathrm{d}u \,\mathrm{d}v \label{eqn:def_Imn}
\end{equation}
which relate to the $\tilde{h}^\mathrm{an}_{m,n}$ as follows:
\begin{equation}
  \tilde{h}^\mathrm{an}_{m,n} (u',v')
  = \frac{F(u',v')}{2 \pi} I_{m,n}^{*}(u',v') \, .
\end{equation}
This implies that the integral in \eqn{ft_h_an} is then:
\begin{equation}
 \hat{h}^\textrm{an}_{m,n} =
 \frac{1}{2 \pi}
 \int\limits_0^1 \int\limits_0^1
   F(u',v') I_{m,n}^{*}(u',v') e^{-2 \pi i (m u' + n v')}
   \,\mathrm{d}u' \,\mathrm{d}v' \label{eqn:imn_ft} \, .
\end{equation}
Note that
\begin{align}
  \frac{\partial I_{m,n}}{\partial a}
  =& -\pi \int\limits_0^1 \int\limits_0^1
       \frac{           \tan^2(\pi u)                                                 e^{2 \pi i (m u + n v)}}
            {2 \left[ a \tan^2(\pi u) + 2 b \tan(\pi u) \tan(\pi v) + c \tan^2(\pi v) \right]^{3/2}          }
     \,\mathrm{d}u \,\mathrm{d}v \nonumber \\
  \frac{\partial I_{m,n}}{\partial b}
  =& -\pi \int\limits_0^1 \int\limits_0^1
       \frac{                           2   \tan(\pi u) \tan(\pi v)                   e^{2 \pi i (m u + n v)}}
            {2 \left[ a \tan^2(\pi u) + 2 b \tan(\pi u) \tan(\pi v) + c \tan^2(\pi v) \right]^{3/2}          }
     \,\mathrm{d}u \,\mathrm{d}v \\
  \frac{\partial I_{m,n}}{\partial c}
  =& -\pi \int\limits_0^1 \int\limits_0^1
       \frac{                                                           \tan^2(\pi v) e^{2 \pi i (m u + n v)}}
            {2 \left[ a \tan^2(\pi u) + 2 b \tan(\pi u) \tan(\pi v) + c \tan^2(\pi v) \right]^{3/2}          }
     \,\mathrm{d}u \,\mathrm{d}v \nonumber
\end{align}
which can be grouped together as
\begin{equation}
  K_{m,n}(u',v') =
    -2 \left( A \frac{\partial}{\partial a}
             +B \frac{\partial}{\partial b}
             +C \frac{\partial}{\partial c} \right) I_{m,n}(u',v') \, . \label{eqn:Kmn}
\end{equation}
This results in:
\begin{equation}
  K_{m,n}(u',v') =
  \pi \int\limits_0^1 \int\limits_0^1
        \frac{\left[A \tan^2(\pi u) + 2 B \tan(\pi u) \tan(\pi v) + C \tan^2(\pi v) \right] e^{2 \pi i (m u + n v)}}
             {\left[a \tan^2(\pi u) + 2 b \tan(\pi u) \tan(\pi v) + c \tan^2(\pi v) \right]^{3/2}                  }
        \,\mathrm{d}u \,\mathrm{d}v
\end{equation}
and the $K_{m,n}(u',v')$ can be related to $\tilde{g}^\mathrm{an}_{m,n}(u',v')$ as follows:
\begin{equation}
  \tilde{g}^\mathrm{an}_{m,n}(u',v')
  = K_{m,n}^{*}(u',v') \, .
\end{equation}
Inserting this into \eqn{tt_g} leads to:
\begin{equation}
 \tilde{\tilde{g}}^\mathrm{an}_{m,n}(u',v')
 = K_{m,n}^{*}(u',v') e^{-2 \pi i (m u' + n v')} \label{eqn:g_t_an} \, .
\end{equation}
It is noted that the Fourier kernel in \eqn{g_t_an}
is equal to the Fourier kernel in \eqn{imn_ft}.
This allows for computing these products
in the same loop in the code.

\FloatBarrier
\subsection{Analytical Fourier Transform of Singular Coefficients}
A generating function $I$ is introduced to actually compute the $I_{m,n}$:
\begin{equation}
  I(s,t) = \sum\limits_{m=0}^{+\infty}
      \sum\limits_{n=0}^{+\infty}
        I_{m,n} s^m t^n \label{eqn:I_gf} \, .
\end{equation}
This essentially interprets the Fourier coefficients $I_{m,n}$
as coefficients of power series in two artificially-introduced variables $s$ and $t$.
The $I_{m,n}$ from \eqn{def_Imn} are inserted into the generating function:
\begin{align}
  I(s,t)
  =& \sum\limits_{m=0}^{+\infty}
     \sum\limits_{n=0}^{+\infty}
       \left(
         \pi \int\limits_0^1 \int\limits_0^1
         \frac{e^{2 \pi i (m u + n v)}}
              {\left[ a \tan^2(\pi u) + 2 b \tan(\pi u) \tan(\pi v) + c \tan^2(\pi v) \right]^{1/2}}
         \,\mathrm{d}u \,\mathrm{d}v
       \right)
       s^m t^n \\
  =& \int\limits_0^1 \int\limits_0^1
       \sum\limits_{m=0}^{+\infty}
       \sum\limits_{n=0}^{+\infty}
         \frac{\pi e^{2 \pi i (m u + n v)} s^m t^n}
              {\left[ a \tan^2(\pi u) + 2 b \tan(\pi u) \tan(\pi v) + c \tan^2(\pi v) \right]^{1/2}}
         \,\mathrm{d}u \,\mathrm{d}v \\
  =& \int\limits_0^1 \int\limits_0^1
       \frac{\pi}
            {\left[ a \tan^2(\pi u) + 2 b \tan(\pi u) \tan(\pi v) + c \tan^2(\pi v) \right]^{1/2}} \nonumber \\
  ~& \phantom{\int\limits_0^1 \int\limits_0^1}~
     \left(
         \sum\limits_{m=0}^{+\infty}
         \sum\limits_{n=0}^{+\infty}
           e^{2 \pi i (m u + n v)} s^m t^n
       \right)
       \,\mathrm{d}u \,\mathrm{d}v \\
  =& \int\limits_0^1 \int\limits_0^1
       \frac{\pi}
            {\left[ a \tan^2(\pi u) + 2 b \tan(\pi u) \tan(\pi v) + c \tan^2(\pi v) \right]^{1/2}} \nonumber \\
  ~& \phantom{\int\limits_0^1 \int\limits_0^1}~
     \left( \sum\limits_{m=0}^{+\infty} \left( s e^{2 \pi i u} \right)^m \right)
     \left( \sum\limits_{n=0}^{+\infty} \left( t e^{2 \pi i v} \right)^n \right)
     \,\mathrm{d}u \,\mathrm{d}v \, . \label{eqn:def_I}
\end{align}
This expression can be simplified by recognizing
\begin{align}
 z_1 =& s e^{2 \pi i u} \\
 z_2 =& t e^{2 \pi i v} \, .
\end{align}
The infinite sums in \eqn{def_I} are geometric series (as defined in \eqn{geometric_series}):
\begin{align}
    \sum\limits_{m=0}^{+\infty} z_1^m
  = \frac{1}{1 - s e^{2 \pi i u}} \, , \label{eqn:geom_series_s} \\
    \sum\limits_{n=0}^{+\infty} z_2^n
  = \frac{1}{1 - t e^{2 \pi i v}} \, . \label{eqn:geom_series_t}
\end{align}
Above sums only converge for $|z_1|<1$ and $|z_2|<1$, respectively.
This implies $|s|<1$ and $|t|<1$, since for $u\in\realnumbers$, $|\exp \left(2 \pi i u \right)|=1$
and for $v\in\realnumbers$, $|\exp \left(2 \pi i v \right)|=1$.
$s$ and $t$ are parameters of the generating function and therefore, these assumptions on $s$ and $t$ are allowed~\cite{discrete_mathematics}.
The expressions from \eqn{geom_series_s} and \eqn{geom_series_t} can be inserted into $I$ from \eqn{def_I} as follows:
\begin{equation}
  I(s,t) = \int\limits_0^1 \int\limits_0^1
       \frac{\pi \,\mathrm{d}u \,\mathrm{d}v}
            {\left(1 - s \, e^{2 \pi i u} \right)
             \left(1 - t \, e^{2 \pi i v} \right)
             \left[ a \tan^2(\pi u) + 2 b \tan(\pi u) \tan(\pi v) + c \tan^2(\pi v) \right]^{1/2} } \, . \label{eqn:int_I}
\end{equation}
A change of variables in performed to new variables $y$ and $r$ as defined by:
\begin{align}
  y  :=& \tan(\pi u) \, , \\
  ry :=& \tan(\pi v) \Leftrightarrow r = \frac{1}{y} \tan(\pi v) \, .
\end{align}
The corresponding Jacobian elements are:
\begin{align}
  \mathrm{d}y =& \frac{\pi\,\mathrm{d}u}{  \cos^2(\pi u)} \\
  \mathrm{d}r =& \frac{\pi\,\mathrm{d}v}{y \cos^2(\pi v)}
\end{align}
where a little bit of trigonometry from \eqn{tan_derivative} was used.
Also, variables $\alpha$ and $\beta$ are introduced:
\begin{align}
  \alpha := \frac{1-s}{1+s} \, , \\
  \beta  := \frac{1-t}{1+t} \, .
\end{align}
Note that the requirements on $s$ and $t$ from \eqn{geom_series_s} and \eqn{geom_series_t}
imply $\alpha>0$ and $\beta>0$.
Consider the square root in the denominator of the integrand in \eqn{int_I}:
\begin{align}
  ~& a \tan^2(\pi u) + 2 b \tan(\pi u) \tan(\pi v) + c \tan^2(\pi v) \nonumber \\
  =& a y^2 + 2 b r y^2 + c r^2 y^2 \nonumber \\
  =& y^2 (a + 2 b r + c r^2) \, .
\end{align}
This can be already inserted into \eqn{int_I}:
\begin{equation}
  I(s,t) = \int\limits_0^1 \int\limits_0^1
       \frac{\pi \,\mathrm{d}u \,\mathrm{d}v}
            {\left(1 - s \, e^{2 \pi i u} \right)
             \left(1 - t \, e^{2 \pi i v} \right)
             y \sqrt{ a + 2 b r + c r^2 } } \, . \label{eqn:int_I_reduced}
\end{equation}
Now the denominator of parts of the denominator in \eqn{int_I} is considered:
\begin{align}
  ~& 1 - s\, e^{2 \pi i u} \nonumber \\
  =& \underbrace{e^{-i \pi u} \,e^{i \pi u}}_{=1} - s \,\underbrace{e^{i \pi u} \,e^{i \pi u}}_{=e^{2 \pi i u}} \nonumber \\
  =& \, \phantom{i} \left[e^{-i \pi u} - s \,e^{i \pi u} \right]         \,e^{i \pi u} \nonumber \\
  =& \,          i  \left[e^{-i \pi u} - s \,e^{i \pi u} \right] \left[-i\,e^{i \pi u} \right] \nonumber \\
  =& -i \Bigl[ - \underbrace{\left( \cos(-\pi u) + i \sin(- \pi i) \right)}_{= \cos(\pi u) - i \sin(\pi u)}
              + s \left( \cos(\pi u) + i \sin(\pi i) \right) \Bigr]
     \underbrace{\left[-i \left( \cos(\pi u) + i \sin(\pi i) \right) \right]}_{=\sin(\pi u) - i \cos(\pi u)} \nonumber \\
  =& -i \Bigl[ \cos(\pi u) \underbrace{(-1+s)}_{=-(1-s)} + \sin(\pi u) \underbrace{(i+s\,i)}_{=i(1+s)} \Bigr]
     \left( \sin(\pi u) - i \cos(\pi u) \right) \nonumber \\
  =& -i \left[ i(1+s) \sin(\pi u) -(1-s) \cos(\pi u) \right] \left( \sin(\pi u) - i \cos(\pi u) \right) \nonumber \\
  =& -i \cos^2(\pi u)
     \Biggl[ i(1+s) \underbrace{\frac{\sin(\pi u)}{\cos(\pi u)}}_{
     \begin{aligned}
       =& \tan(\pi u) \\
       =& \,y
     \end{aligned}} -(1-s) \Biggr]
     \Biggl[        \underbrace{\frac{\sin(\pi u)}{\cos(\pi u)}}_{
     \begin{aligned}
       =& \tan(\pi u) \\
       =& \,y
     \end{aligned}} - i \Biggr] \nonumber \\
  =& -i \cos^2(\pi u) \left( i(1+s) y - (1-s) \right) (y-i) \nonumber \\
  =& -i \cos^2(\pi u) (1+s) \Biggl[ i y - \underbrace{\frac{1-s}{1+s}}_{= \alpha} \Biggr] (y-i) \nonumber \\
  =& -i \cos^2(\pi u) (1+s) ( i y - \alpha ) (y-i) \, .
\end{align}
Going on with the inverse of above expression, which actually occurs in \eqn{int_I}:
\begin{align}
  \frac{1}{ 1 - s\, e^{2 \pi i u}}
  =&  \frac{i}{\cos^2(\pi u) (1+s)} \frac{1}{(i y - \alpha) (y-i)}
  =  \frac{i}{\cos^2(\pi u)} \frac{\frac{1}{1+s} \,\Bigl\}=\frac{1}{2}\frac{1+s+1-s}{1+s}}{(i y - \alpha) (y-i)} \nonumber \\
  =& \frac{i}{2 \cos^2(\pi u)} \frac{\frac{1+s}{1+s} + \frac{1-s}{1+s}}{(i y - \alpha) (y-i)}
  =  \frac{i}{2 \cos^2(\pi u)} \frac{1 + \alpha}{(i y - \alpha) (y-i)} \nonumber \\
  =& \frac{i}{2 \cos^2(\pi u)} \frac{1 + \alpha}{i (y + i \alpha) (y-i)}
  =  \frac{i}{2 \cos^2(\pi u)} \frac{-i (1 + \alpha)}{(y + i \alpha) (y-i)} \nonumber \\
  =& \frac{i}{2 \cos^2(\pi u)} \frac{y - i - y - i \alpha}{(y + i \alpha) (y-i)}
  =  \frac{i}{2 \cos^2(\pi u)} \frac{(y - i) - (y + i \alpha)}{(y + i \alpha) (y-i)} \nonumber \\
  =& \frac{i}{2 \cos^2(\pi u)} \left( \frac{1}{y + i \alpha} - \frac{1}{y-i} \right) \, .
\end{align}
Analogously it follows by replacing
$s \rightarrow t$,
$u \rightarrow v$,
$y \rightarrow ry$ and
$\alpha \rightarrow \beta$ in above equations:
\begin{equation}
    \frac{1}{ 1 - t\, e^{2 \pi i v}}
  = \frac{i}{2 \cos^2(\pi v)} \left( \frac{1}{ry + i \beta} - \frac{1}{ry-i} \right) \, .
\end{equation}
Inserting these expressions into \eqn{int_I_reduced} results in:
\begin{equation}
  I(s,t) = \frac{-1}{4 \pi}
  \int\limits_0^1 \int\limits_0^1
  \underbrace{\frac{\pi \,\mathrm{d}u}{  \cos^2(\pi u)}}_{= \mathrm{d}y}
  \underbrace{\frac{\pi \,\mathrm{d}v}{y \cos^2(\pi v)}}_{= \mathrm{d}r}
  \left( \frac{1}{y + i \alpha} - \frac{1}{y-i} \right)
  \left( \frac{1}{ry + i \beta} - \frac{1}{ry-i} \right)
  \frac{1}{\sqrt{ a + 2 b r + c r^2 }} \, . \label{eqn:int_I_to_transform}
\end{equation}
A change of variables implies the need to adjust the integration limits.
Note that
\begin{equation}
  \tan(0) = \tan(\pi) = 0 \, ,
\end{equation}
but $\tan(\pi u)$ spans the range $[-\infty,+\infty]$ for $u \in [0,1]$
and so the integrals in \eqn{int_I_to_transform} have to be taken from $-\infty$ to $+\infty$:
\begin{equation}
  I(s,t) = \frac{-1}{4 \pi}
  \int\limits_{-\infty}^{+\infty}
    \left[
    \int\limits_{-\infty}^{+\infty}
      \left( \frac{1}{y + i \alpha} - \frac{1}{y-i} \right)
      \left( \frac{1}{ry + i \beta} - \frac{1}{ry-i} \right)
      \,\mathrm{d}y
    \right]
  \frac{1}{\sqrt{ a + 2 b r + c r^2 }}
  \,\mathrm{d}r
  \, . \label{eqn:int_I_transformed}
\end{equation}
The integral over $y$ in \eqn{int_I_transformed} is performed first.
For this, look at its integrand:
\begin{align}
  ~& \left( \frac{1}{y + i \alpha} - \frac{1}{y-i} \right) \left( \frac{1}{ry + i \beta} - \frac{1}{ry-i} \right) \nonumber \\
  =&   \underbrace{ \Bigl[ (y + i \alpha) (r y + i \beta) \Bigr]^{-1} }_{(1)}
     - \underbrace{ \Bigl[ (y + i \alpha) (r y - i      ) \Bigr]^{-1} }_{(2)}
     - \underbrace{ \Bigl[ (y - i       ) (r y + i \beta) \Bigr]^{-1} }_{(3)}
     + \underbrace{ \Bigl[ (y - i       ) (r y - i      ) \Bigr]^{-1} }_{(4)} \, . \nonumber
\end{align}
The four summands in above expression can be integrated separately due to the linearity of the integral.
An elegant way of performing these $y$-integrals is based on application of the residue theorem.
First, note that the integrands $(1)$ to $(4)$ are actually composed of four individual factors:
\begin{equation}
 \frac{1}{ y+i \alpha},
 \frac{1}{ y-i},
 \frac{1}{ry+i \beta},
 \frac{1}{ry-i} \, . \nonumber
\end{equation}
For $\alpha, \beta, r \in \realnumbers$, these factors become singular when inserting (``analytic continuation'' from $\realnumbers$ to $\complexnumbers$)
certain complex values for $y$. The values $y \in \complexnumbers$ for which above factors become singular
are called poles and are denoted $z_1, ..., z_4$:
\begin{align}
 \frac{1}{ y+i \alpha} \Rightarrow& z_1 = -i \alpha          \nonumber \\
 \frac{1}{ y-i}        \Rightarrow& z_2 =  i                 \nonumber \\
 \frac{1}{ry+i \beta}  \Rightarrow& z_3 = -i \frac{\beta}{r} \nonumber \\
 \frac{1}{ry-i}        \Rightarrow& z_4 =  \frac{i}{r}       \nonumber \, .
\end{align}
The integrands $(1)$ to $(4)$ are composed of above factors and feature pairs of poles $z_1, ..., z_4$:
\begin{align}
 (1) \Rightarrow& (z_1, z_3) \nonumber \\
 (2) \Rightarrow& (z_1, z_4) \nonumber \\
 (3) \Rightarrow& (z_2, z_3) \nonumber \\
 (4) \Rightarrow& (z_2, z_4) \nonumber \, .
\end{align}
The residue theorem allows to replace the $y$-integral along the whole real axis ($-\infty ... +\infty$)
with a closed contour integral denoted $\mathcal{C}$ in the complex plane:
\begin{figure}[htbp]
 \centering
 \includegraphics[width=7cm]{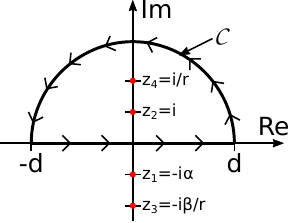}
 \caption{Closed integration contour $\mathcal{C}$ (thick black line) for application of the residue theorem
          and location of the poles $z_1, ..., z_4$ for $\alpha>0, \beta>0, r>0$ (red dots).}
\end{figure}
In order to recover the integral along the whole real axis,
the parameter $d$ of the integration contour $\mathcal{C}$ is taken to the limit $d \rightarrow +\infty$.
The order of the integrands $(1)$ to $(4)$ is $-2$ and thus the contribution from the great semicircle
to the contour integral vanishes, leaving only the contribution from the real axis,
which is equivalent (in the limit $d \rightarrow +\infty$) to the original $y$-integral.
The residue theorem on the other hand allows to express the contour integral
with the sum of the residuals (, i.e., poles) \underline{within} the contour $\mathcal{C}$.
Depending on the signs of $\alpha$, $\beta$ and $r$,
the poles of a given integrand $(1)$ to $(4)$ have $\imagpart(z_k)>0$ or $\imagpart(z_k)<0$ ($k=1,...,4$).
The contour $\mathcal{C}$ (parameterized by $d$) has to be taken large enough to include all poles $z_k$ with $\imagpart(z_k)>0$.
The calculation of the $y$ integrals over $(1)$ to $(4)$ then reduces to computation of the residuals of $z_1, ..., z_4$,
assuming that they are located within $\mathcal{C}$, and then summing them up with cases
for the various combinations of signs of $\alpha$, $\beta$ and $r$.
We start with $(1)$:
\begin{equation}
  \int\limits_{-\infty}^{+\infty}
    (1) \,\mathrm{d}y
= \oint_\mathcal{C}
    f_1(z) \,\mathrm{d}z \label{eqn:int_f1}
\end{equation}
with
\begin{align}
  f_1(z) =& \frac{1}{z + i \alpha}          \cdot \frac{1}{r z + i \beta} \\
         =& \frac{1}{z + i \frac{\beta}{r}} \cdot \frac{1}{r(z + i \alpha)} \, .
\end{align}
The residues of $f_1$ are:
\begin{align}
 \mathrm{res}_{z_1} f_1 =& \frac{1}{r(-i \alpha) + i \beta}
                        =  \frac{1}{-i \alpha r + i \beta}
                        =  \frac{1}{i(\beta - \alpha r)}
                        \textrm{ for } \alpha < 0 \Rightarrow \imagpart(-i\alpha) > 0 \\
 \mathrm{res}_{z_3} f_1 =& \frac{1}{r(-i \frac{\beta}{r} + i \alpha)}
                        =  \frac{1}{-i \beta + i \alpha r}
                        =  \frac{-1}{i(\beta - \alpha r)}
                        \textrm{ for } r>0, \beta< 0 \textrm{ or } r<0, \beta>0 \, .
\end{align}
The integral in \eqn{int_f1} can be evaluated:
\begin{equation}
   \oint_\mathcal{C} f_1(z) \,\mathrm{d}z
 = 2 \pi i \left[ \left\{ \begin{aligned}
                   \frac{1}{i(\beta-\alpha r)} :&\, \alpha<0 \\
                   0                           :&\, \alpha>0
                  \end{aligned} \right\}
                 +\left\{ \begin{aligned}
                   \frac{-1}{i(\beta-\alpha r)} :&\, \left\{\begin{aligned}
                                                       r>0,&\, \beta<0 \textrm{ or} \\
                                                       r<0,&\, \beta>0
                                                      \end{aligned} \right. \\
                   0                           :&\, \textrm{else}
                  \end{aligned} \right\}
 \right] \, .
\end{equation}
For the $y$-integral over $(1)$ four cases are therefore relevant regarding the signs of $\alpha$ and $\beta$:
\begin{align}
 \alpha>0, \beta>0 :&\, \int\limits_{-\infty}^{+\infty} (1) \,\mathrm{d}y = \begin{cases}
                                                                             \frac{-2\pi}{\beta-\alpha r} &:\, r<0 \\
                                                                             0                            &:\, r>0
                                                                            \end{cases} \\
 \alpha>0, \beta<0 :&\, \int\limits_{-\infty}^{+\infty} (1) \,\mathrm{d}y = \begin{cases}
                                                                             0                            &:\, r<0 \\
                                                                             \frac{-2\pi}{\beta-\alpha r} &:\, r>0
                                                                            \end{cases} \\
 \alpha<0, \beta>0 :&\, \int\limits_{-\infty}^{+\infty} (1) \,\mathrm{d}y = \begin{cases}
                                                                             0                            &:\, r<0 \\
                                                                             \frac{ 2\pi}{\beta-\alpha r} &:\, r>0
                                                                            \end{cases} \\
 \alpha<0, \beta<0 :&\, \int\limits_{-\infty}^{+\infty} (1) \,\mathrm{d}y = \begin{cases}
                                                                             \frac{ 2\pi}{\beta-\alpha r} &:\, r<0 \\
                                                                             0                            &:\, r>0
                                                                            \end{cases} \, .
\end{align}
Next, consider $(2)$:
\begin{equation}
  \int\limits_{-\infty}^{+\infty}
    (2) \,\mathrm{d}y
= \oint_\mathcal{C}
    f_2(z) \,\mathrm{d}z \label{eqn:int_f2}
\end{equation}
with
\begin{align}
  f_2(z) =& \frac{1}{z + i \alpha}    \cdot \frac{1}{r z - i} \\
         =& \frac{1}{z - \frac{i}{r}} \cdot \frac{1}{r(z + i \alpha)} \, .
\end{align}
The residues of $f_2$ are:
\begin{align}
 \mathrm{res}_{z_1} f_2 =& \frac{1}{r(-i \alpha)  - i}
                        =  \frac{1}{  -i \alpha r - i}
                        =  \frac{-1}{i(1 + \alpha r)}
                        \textrm{ for } \alpha < 0 \\
 \mathrm{res}_{z_4} f_2 =& \frac{1}{r(\frac{i}{r} + i \alpha)}
                        =  \frac{1}{i + i \alpha r}
                        =  \frac{1}{i(1 + \alpha r)}
                        \textrm{ for } r > 0\, .
\end{align}
The integral in \eqn{int_f2} can be evaluated:
\begin{equation}
   \oint_\mathcal{C} f_2(z) \,\mathrm{d}z
 = 2 \pi i \left[ \left\{ \begin{aligned}
                   \frac{-1}{i(1 + \alpha r)} :&\, \alpha<0 \\
                   0                          :&\, \alpha>0
                  \end{aligned} \right\}
                 +\left\{ \begin{aligned}
                   0                         :&\, r < 0 \\
                   \frac{1}{i(1 + \alpha r)} :&\, r > 0
                  \end{aligned} \right\}
 \right] \, .
\end{equation}
For the $y$-integral over $(2)$ two cases are therefore relevant regarding the sign of $\alpha$:
\begin{align}
 \alpha > 0 :&\, \int\limits_{-\infty}^{+\infty} (2) \,\mathrm{d}y = \begin{cases}
                                                                      0                        &:\, r<0 \\
                                                                      \frac{ 2\pi}{1+\alpha r} &:\, r>0
                                                                     \end{cases} \\
 \alpha < 0 :&\, \int\limits_{-\infty}^{+\infty} (2) \,\mathrm{d}y = \begin{cases}
                                                                      \frac{-2\pi}{1+\alpha r} &:\, r<0 \\
                                                                      0                        &:\, r>0
                                                                     \end{cases} \, .
\end{align}
We continue with $(3)$:
\begin{equation}
  \int\limits_{-\infty}^{+\infty}
    (3) \,\mathrm{d}y
= \oint_\mathcal{C}
    f_3(z) \,\mathrm{d}z \label{eqn:int_f3}
\end{equation}
with
\begin{align}
  f_3(z) =& \frac{1}{z - i}    \cdot \frac{1}{r z + i \beta} \\
         =& \frac{1}{z + i \frac{\beta}{r}} \cdot \frac{1}{r(z - i)} \, .
\end{align}
The residues of $f_3$ are:
\begin{align}
 \mathrm{res}_{z_2} f_3 =& \frac{1}{r i + i \beta}
                        =  \frac{1}{i (r + \beta)} \\
 \mathrm{res}_{z_3} f_3 =& \frac{1}{r(-i\frac{\beta}{r} - i)}
                        =  \frac{1}{-i \beta - i r}
                        =  \frac{-1}{i(\beta + r)}
                        \textrm{ for } r>0, \beta< 0 \textrm{ or } r<0, \beta>0 \, .
\end{align}
The integral in \eqn{int_f3} can be evaluated:
\begin{equation}
   \oint_\mathcal{C} f_3(z) \,\mathrm{d}z
 = 2 \pi i \left[ \frac{1}{i(\beta + r)} \\
                 +\left\{ \begin{aligned}
                   \frac{-1}{i(\beta + r)} :&\, \left\{\begin{aligned}
                                                       r>0,&\, \beta<0 \textrm{ or} \\
                                                       r<0,&\, \beta>0
                                                      \end{aligned} \right. \\
                   0                           :&\, \textrm{else}
                  \end{aligned} \right\}
 \right] \, .
\end{equation}
For the $y$-integral over $(3)$ two cases are therefore relevant regarding the sign of $\beta$:
\begin{align}
 \beta > 0 :&\, \int\limits_{-\infty}^{+\infty} (3) \,\mathrm{d}y = \begin{cases}
                                                                      0                      &:\, r<0 \\
                                                                      \frac{2\pi}{\beta + r} &:\, r>0
                                                                    \end{cases} \\
 \beta < 0 :&\, \int\limits_{-\infty}^{+\infty} (3) \,\mathrm{d}y = \begin{cases}
                                                                      \frac{2\pi}{\beta + r} &:\, r<0 \\
                                                                      0                      &:\, r>0
                                                                    \end{cases} \, .
\end{align}
Finally we consider the $y$-integral over $(4)$:
\begin{equation}
  \int\limits_{-\infty}^{+\infty}
    (4) \,\mathrm{d}y
= \oint_\mathcal{C}
    f_4(z) \,\mathrm{d}z \label{eqn:int_f4}
\end{equation}
with
\begin{align}
  f_4(z) =& \frac{1}{z - i}           \cdot \frac{1}{r z - i} \\
         =& \frac{1}{z - \frac{i}{r}} \cdot \frac{1}{r(z - i)} \, .
\end{align}
The residues of $f_4$ are:
\begin{align}
 \mathrm{res}_{z_2} f_4 =& \frac{1}{r i -i}
                        =  \frac{1}{i (r - 1)} \\
 \mathrm{res}_{z_4} f_4 =& \frac{1}{r(\frac{i}{r} - i)}
                        =  \frac{1}{i(1 - r)}
                        =  \frac{-1}{i(r - 1)}
                        \textrm{ for } r>0 \, .
\end{align}
The integral in \eqn{int_f4} can be evaluated:
\begin{equation}
   \oint_\mathcal{C} f_4(z) \,\mathrm{d}z
 = 2 \pi i \left[ \frac{1}{i(r - 1)} \\
                 +\left\{ \begin{aligned}
                   0                   :&\, r < 0 \\
                   \frac{-1}{i(r - 1)} :&\, r > 0
                  \end{aligned} \right\}
 \right] \, .
\end{equation}
The $y$-integral over $(4)$ is thus:
\begin{equation}
   \int\limits_{-\infty}^{+\infty} (4) \,\mathrm{d}y
 = \begin{cases}
    \frac{2 \pi}{r-1} &:\, r<0 \\
    0                 &:\, r>0
   \end{cases} \, .
\end{equation}
The $y$-integrals are now complete. The results are inserted back into the remaining
integral over $r$ and the cases for different signs of $r$ are evaluated to limit the ranges of the integrals over $r$.
We need to consider $\alpha>0$, $\beta>0$ for compatibility with earlier assumptions
regarding $s$ and $t$ in \eqn{geom_series_s} and \eqn{geom_series_t}:
\begin{align}
 I&(s,t) = \frac{-1}{4 \pi}
           \int\limits_{-\infty}^{+\infty}
            \frac{\mathrm{d}r}
                 {\sqrt{a + 2 b r + c r^2}}  \nonumber \\
   ~     &  \left[
             \left\{\begin{aligned}
              \frac{-2\pi}{\beta-\alpha r} &:\, r<0 \\
              0                            &:\, r>0
             \end{aligned}\right\}
           - \left\{\begin{aligned}
              0                        &:\, r<0 \\
              \frac{ 2\pi}{1+\alpha r} &:\, r>0
             \end{aligned}\right\}
           - \left\{\begin{aligned}
               0                      &:\, r<0 \\
               \frac{2\pi}{\beta + r} &:\, r>0
             \end{aligned}\right\}
           + \left\{\begin{aligned}
              \frac{2 \pi}{r-1} &:\, r<0 \\
              0                 &:\, r>0
             \end{aligned}\right\}
           \right] \, . \nonumber \\
 =& \frac{-1}{4 \pi}
    \int\limits_{-\infty}^{0}
      \left[ \frac{-2 \pi}{\beta - \alpha r} + \frac{2 \pi}{r - 1} \right]
      \frac{\mathrm{d}r}
           {\sqrt{a + 2 b r + c r^2}}
  + \frac{1}{4 \pi}\int\limits_{0}^{+\infty}
      \left[ \frac{2 \pi}{1 + \alpha r} + \frac{2 \pi}{\beta + r} \right]
      \frac{\mathrm{d}r}
           {\sqrt{a + 2 b r + c r^2}} \nonumber \\
 =& \frac{1}{2}
    \int\limits_{0}^{-\infty}
      \left[ \frac{-1}{\beta - \alpha r} + \frac{1}{r - 1} \right]
      \frac{\mathrm{d}r}
           {\sqrt{a + 2 b r + c r^2}}
  + \frac{1}{2}\int\limits_{0}^{+\infty}
      \left[ \frac{1}{1 + \alpha r} + \frac{1}{\beta + r} \right]
      \frac{\mathrm{d}r}
           {\sqrt{a + 2 b r + c r^2}} \nonumber \\
 =& \frac{1}{2}
    \int\limits_{0}^{+\infty}
      \left[ \frac{-1}{\beta - \alpha (-r)} + \frac{1}{(-r) - 1} \right]
      \frac{-\mathrm{d}r}
           {\sqrt{a - 2 b r + c r^2}}
  + \frac{1}{2}\int\limits_{0}^{+\infty}
      \left[ \frac{1}{1 + \alpha r} + \frac{1}{\beta + r} \right]
      \frac{\mathrm{d}r}
           {\sqrt{a + 2 b r + c r^2}} \nonumber \\
 =& \frac{1}{2}
    \int\limits_{0}^{+\infty}
      \left[ \frac{1}{\beta + \alpha r} + \frac{1}{1 + r} \right]
      \frac{\mathrm{d}r}
           {\sqrt{a - 2 b r + c r^2}}
  + \frac{1}{2}\int\limits_{0}^{+\infty}
      \left[ \frac{1}{1 + \alpha r} + \frac{1}{\beta + r} \right]
      \frac{\mathrm{d}r}
           {\sqrt{a + 2 b r + c r^2}} \, . \label{eqn:int_I_split}
\end{align}
A change of variables is performed from $r$ to $x$ with
\begin{equation}
  x = \frac{r - 1}{r + 1} \, .
\end{equation}
Note that for this change of variables the Jacobian is:
\begin{equation}
 \frac{\mathrm{d}x}{\mathrm{d}r} = \frac{2}{(r+1)^2} ~ \Rightarrow ~ \mathrm{d}x = \frac{2 \,\mathrm{d}r}{(r+1)^2} \, .
\end{equation}
In \eqn{int_I_split} four relatively similar terms occur:
\begin{align}
 \frac{1}{\beta + \alpha r} ~& \qquad \qquad \qquad \quad (\textrm{the original}), \nonumber \\
 \frac{1}{1 + r}            =& \left. \frac{1}{\beta + \alpha r} \right|_{\alpha = 1,
                                                                          \beta  = 1} \, , \nonumber \\
 \frac{1}{1 + \alpha r}     =& \left. \frac{1}{\beta + \alpha r} \right|_{\beta  = 1} \qquad \textrm{and} \nonumber \\
 \frac{1}{\beta + r}        =& \left. \frac{1}{\beta + \alpha r} \right|_{\alpha = 1} \qquad . \nonumber
\end{align}
Therefore it is sufficient to consider the original term only and the others will follow from it.
Furthermore, $\alpha$ and $\beta$ are replaced again with their respective expressions in $s$ and $t$:
\begin{equation}
  s = \frac{1 - \alpha}{1 + \alpha} \quad \mathrm{and} \quad t = \frac{1 - \beta}{1 + \beta} \, ,
\end{equation}
leading to
\begin{align}
  \left( \beta + \alpha r \right)^{-1}
  =& \left( \frac{1-t}{1+t} + \frac{1-s}{1+s} \,r \right)^{-1}
  =  \left( \frac{(1-t)(1+s) + (1+t)(1-s) r}{(1+t)(1+s)} \right)^{-1} \nonumber \\
  =& \frac{(1+t)(1+s)}{1 +s-t -st +r -sr+tr -str}
  =  \frac{(1+t)(1+s)}{(r+1)-st(r+1)-(s-t)(r-1)} \label{eqn:beta_alpha_r} \\
  =& \frac{1}{r+1} \cdot \frac{(1+t)(1+s)}{1-st-(s-t)\underbrace{\frac{r-1}{r+1}}_{=\,x}}
  =  \frac{1}{r+1} \cdot \frac{(1+t)(1+s)}{1-st-(s-t)x} \nonumber
\end{align}
where $\alpha=1$ implies $s=0$ and $\beta=1$ implies $t=0$.
Next consider the argument of the inverse square roots in \eqn{int_I_split}.
New variables are introduced:
\begin{align}
 a^{+} =&\, a + 2b + c \nonumber \\
 a^{-} =&\, a - 2b + c \\
 d     =&\, c - a \nonumber
\end{align}
which can be solved for the old variables $a$, $b$, $c$ as follows:
\begin{align}
 b =&\, \frac{1}{4} \left( a^{+} - a^{-} \right) \nonumber \\
 a =&\, \frac{1}{4} \left( a^{+} + a^{-} - 2d \right) \\
 c =&\, \frac{1}{4} \left( a^{+} + a^{-} + 2d \right) \nonumber
\end{align}
Inserting these expressions into the inverse square root argument results in:
\begin{align}
 a \pm 2 b r + c r^2
 =&\, \frac{1}{4} \left[ \left( a^{+} + a^{-} - 2d \right) \pm 2 \left( a^{+} - a^{-} \right) r + \left( a^{+} + a^{-} + 2d \right) r^2 \right] \nonumber \\
 =&\, \frac{1}{4} \left[ a^{+} + a^{-} - 2d \pm 2ra^{+} \mp 2ra^{-} + r^2a^{+} + r^2a^{-} + 2r^2d \right] \nonumber \\
 =&\, \frac{1}{4} \left[ \left( 1 \pm 2r + r^2 \right) a^{+}  + 2 \left( r^2 - 1 \right) d + \left( 1 \mp 2r + r^2 \right) a^{-} \right] \nonumber \\
 =&\, \frac{1}{4} \left[ \left( r \pm 1 \right)^2 a^{+}  + 2 (r-1)(r+1) d + \left( r \mp 1 \right)^2 a^{-} \right] \label{eqn:a_pm_etc} \\
 =&\, \frac{1}{4} (r+1)^2 \Biggl[ a^{\pm}  + 2 \underbrace{\frac{r-1}{r+1}}_{=\,x} d + \Biggl( \underbrace{\frac{r-1}{r+1}}_{=\,x} \Biggr)^2 a^{\mp} \Biggr] \nonumber \\
 =&\, \frac{1}{4} (r+1)^2 \left( a^{\pm}  + 2 d x + a^{\mp} x^2 \right) \, . \nonumber
\end{align}

Now \eqn{beta_alpha_r} and \eqn{a_pm_etc} can be substitited back into \eqn{int_I_split}:
\begin{align}
 I(s,t)
 =& \phantom{+}\,
     \frac{1}{2} \int\limits_{0}^{+\infty}
     \frac{1}{r+1}
     \left(
         \frac{(1+t)(1+s)}{1-st-(s-t)x}
       + 1
     \right)
     \frac{\mathrm{d}r}{\frac{1}{2} (r+1) \sqrt{ a^{-}  + 2 d x + a^{+} x^2 }} \nonumber \\
 ~& +
     \frac{1}{2} \int\limits_{0}^{+\infty}
     \frac{1}{r+1}
     \left(
         \frac{1+s}{1-sx}
       + \frac{1+t}{1+tx}
     \right)
     \frac{\mathrm{d}r}{\frac{1}{2} (r+1) \sqrt{ a^{+}  + 2 d x + a^{-} x^2 }} \nonumber \\
 =& \phantom{+}\,
     \frac{1}{2} \int\limits_{0}^{+\infty}
     \underbrace{\frac{2 \,\mathrm{d}r}{(r+1)^2}}_{=\mathrm{d}x}
     \left(
         \frac{(1+t)(1+s)}{1-st-(s-t)x}
       + 1
     \right)
     \frac{1}{\sqrt{ a^{-}  + 2 d x + a^{+} x^2 }} \nonumber \\
 ~& +
     \frac{1}{2} \int\limits_{0}^{+\infty}
     \underbrace{\frac{2 \,\mathrm{d}r}{(r+1)^2}}_{=\mathrm{d}x}
     \left(
         \frac{1+s}{1-sx}
       + \frac{1+t}{1+tx}
     \right)
     \frac{1}{\sqrt{ a^{+}  + 2 d x + a^{-} x^2 }} \, .
\end{align}
With this it follows for $I(s,t)$:
\begin{align}
 I(s,t)
 =& \phantom{+}\,
     \frac{1}{2} \int\limits_{-1}^{1}
     \left(
         \frac{(1+t)(1+s)}{1-st-(s-t)x}
       + 1
     \right)
     \frac{\mathrm{d}x}{\sqrt{ a^{-}  + 2 d x + a^{+} x^2 }} \nonumber \\
 ~& +
     \frac{1}{2} \int\limits_{-1}^{1}
     \left(
         \frac{1+s}{1-sx}
       + \frac{1+t}{1+tx}
     \right)
     \frac{\mathrm{d}x}{\sqrt{ a^{+}  + 2 d x + a^{-} x^2 }} \, .
\end{align}
The following function $h^{\pm}(s,t)$ is introduced:
\begin{equation}
   h^{\pm}(s,t)
 = \frac{(1+s)(1+t)}{2}
   \int\limits_{-1}^{1}
     \frac{\mathrm{d}x}{\left( 1-st-(s-t)x \right)\sqrt{ a^{\mp}  + 2 d x + a^{\pm} x^2 }}
\end{equation}
and $I(s,t)$ can be expressed using $h^{\pm}(s,t)$ as follows:
\begin{equation}
 I(s,t) = h^{+}(s,t) + h^{+}(0,0) + h^{-}(s,0) + h^{-}(0,t) \, . \label{eqn:I_from_h}
\end{equation}
Now consider $h^{\pm}(s,t)$ in more detail:
\begin{align}
 h^{\pm}(s,t)
 =&\, \frac{(1+s)(1+t)}{2}
    \int\limits_{-1}^{1}
     \frac{\mathrm{d}x}{\underbrace{\left( 1-st-(s-t)x \right)}_{\begin{aligned}
            =& (1-st)\left(1- \frac{s-t}{1-st} x \right)
            \end{aligned}}\sqrt{ a^{\mp}  + 2 d x + a^{\pm} x^2 }} \nonumber \\
 =&\, \frac{1}{2}
    \frac{(1+s)(1+t)}{1-st}
    \int\limits_{-1}^{1}
    \underbrace{\frac{1}{1- \frac{s-t}{1-st} x}}_{\begin{aligned}
                      = \sum\limits_{l=0}^{\infty} \left( \frac{s-t}{1-st} \right)^l x^l
                                                          \end{aligned}}
     \frac{\mathrm{d}x}{\sqrt{ a^{\mp}  + 2 d x + a^{\pm} x^2 }} \nonumber \\
 =&\, \frac{1}{2}
    \frac{(1+s)(1+t)}{1-st}
    \int\limits_{-1}^{1}
      \sum\limits_{l=0}^{\infty} \left( \frac{s-t}{1-st} \right)^l
      \frac{x^l}{\sqrt{ a^{\mp}  + 2 d x + a^{\pm} x^2 }}
      \,\mathrm{d}x \nonumber \\
 =&\, \frac{1}{2}
    \frac{(1+s)(1+t)}{1-st}
    \sum\limits_{l=0}^{\infty} \left( \frac{s-t}{1-st} \right)^l
    \int\limits_{-1}^{1}
      \frac{x^l}{\sqrt{ a^{\mp}  + 2 d x + a^{\pm} x^2 }}
      \,\mathrm{d}x \nonumber \\
 =&\, \frac{1}{2}
    \frac{(1+s)(1+t)}{1-st}
    \sum\limits_{l=0}^{\infty} \left( \frac{s-t}{1-st} \right)^l
    T^{\pm}_l \label{eqn:h_pm_st}
\end{align}
with
\begin{equation}
 T^{\pm}_l
 = \int\limits_{-1}^{1}
      \frac{x^l}{\sqrt{ a^{\mp}  + 2 d x + a^{\pm} x^2 }}
      \,\mathrm{d}x \, . \label{eqn:T_pm_l}
\end{equation}
The last three terms of \eqn{I_from_h} can be simplified already:
\begin{align}
  h^{+}(0,0) =& \frac{1}{2} T^{+}_0 \label{eqn:h_p_00} \\
  h^{-}(s,0) =& \frac{1}{2} (1+s) \sum\limits_{l=0}^{+\infty} s^l T^{-}_l
             =  \frac{1}{2} \left[ \sum\limits_{l=0}^{+\infty} T^{-}_l s^l + \sum\limits_{l=0}^{+\infty} T^{-}_l s^{l+1} \right] \nonumber \\
      ~      =& \frac{1}{2} T^{-}_0 + \frac{1}{2} \sum\limits_{l=1}^{+\infty} \left( T^{-}_l + T^{-}_{l-1} \right) s^l \label{eqn:h_m_s0} \\
  h^{-}(0,t) =& \frac{1}{2} (1+t) \sum\limits_{l=0}^{+\infty} (-t)^l T^{-}_l
             =  \frac{1}{2} \left[ \sum\limits_{l=0}^{+\infty} (-1)^l T^{-}_l t^l + \sum\limits_{l=0}^{+\infty} (-1)^l T^{-}_l t^{l+1} \right] \nonumber \\
      ~      =& \frac{1}{2} T^{-}_0 + \frac{1}{2} \sum\limits_{l=1}^{+\infty} (-1)^l \left( T^{-}_l - T^{-}_{l-1} \right) t^l \label{eqn:h_m_0t} \, .
\end{align}
Equation~(\ref{eqn:h_pm_st}) can be further expanded for $h^{+}(s,t)$:
\begin{align}
 h^{+}(s,t)
 =& \frac{(1+s)(1+t)}{2}
    \sum\limits_{l=0}^{+\infty} (s-t)^l \frac{1}{(1-st)^{l+1}} T^{+}_l \nonumber \\
 =& \frac{(1+s)(1+t)}{2}
    \sum\limits_{l=0}^{+\infty}
      \left( \sum\limits_{j=0}^{l} (-1)^{j} {\genfrac(){0pt}{0}{l}{j}} s^{l-j} t^{j} \right)
      \left( \sum\limits_{k=0}^{+\infty} {\genfrac(){0pt}{0}{k+l}{k}} s^k t^k \right)
      \,T^{+}_l \nonumber \\
 =& \frac{(1+s)(1+t)}{2}
    \sum\limits_{k=0}^{+\infty}
    \sum\limits_{l=0}^{+\infty}
    \sum\limits_{j=0}^{l}
      (-1)^{j}
      \frac{\bcancel{l!}\,(k+l)!}{j!\,(l-j)!\,k!\,\bcancel{l!}}
      \,T^{+}_l
      s^{k+l-j} t^{k+j} \nonumber \\
 =& \frac{1}{2} (1+s+t+st)
    \sum\limits_{k=0}^{+\infty}
    \sum\limits_{l=0}^{+\infty}
    \sum\limits_{j=0}^{l}
      (-1)^{j}
      \frac{(k+l)!}{j!\,(l-j)!\,k!}
      \,T^{+}_l
      s^{k+l-j} t^{k+j} \label{eqn:h_p_st} \, .
\end{align}
The representation of $I(s,t)$ from \eqn{I_from_h} is equated with the representation from \eqn{I_gf}.
The goal now is to identify which terms from \eqn{h_p_00}, \eqn{h_m_s0}, \eqn{h_m_0t} and \eqn{h_p_st}
contribute to $I_{m,n}$ for any given powers $m$ and $n$ of $s$ and $t$, respectively.
The contributions from $h^{+}(0,0)$, $h^{-}(s,0)$ and $h^{-}(0,t)$ are easily related to $(m=0, n=0)$, $(m=0, n\geq1)$ and $(m\geq1, n=0)$.
The contributions from $h^{+}(s,t)$ are a little bit more involved
and thus, only the sums
\begin{equation}
  \sum\limits_{k=0}^{+\infty}
  \sum\limits_{l=0}^{+\infty}
  \sum\limits_{j=0}^{l}
    (-1)^{j}
    \frac{(k+l)!}{j!\,(l-j)!\,k!}
    \,T^{+}_l
    s^{k+l-j} t^{k+j} \label{eqn:sums_h_p_st}
\end{equation}
are considered for now.
For the following, assume $m\geq1$ and $n\geq1$ are given. The task is now to find all terms in \eqn{sums_h_p_st} with matching $s^m$ and $t^n$.
There exist two cases, namely $m \geq n$ and $n \geq m$.

Assume $m \geq n$ for the first part.
Then there are $(n+1)$ contributions to $s^m t^n$ with $m=k+l-j$ and $n=k+j$,
since $k \geq 0$ and $j \geq 0$, leading to the list of possible combinations of values of $j$ and $k$ shown in Tab.~(\ref{tab:from_j_k_to_n}).
\begin{table}[htbp]
  \centering
  \begin{tabular}{c|c}
    $j$ & $k$ \\
    \hline
    0 & $n$ \\
    1 & $n-1$ \\
    ... & ... \\
    $n-1$ & 1 \\
    $n$ & 0
  \end{tabular}
  \caption{$(n+1)$ possible combinations of $j \geq 0$ and $k \geq 0$ which sum up to a given $n=k+j$.}
  \label{tab:from_j_k_to_n}
\end{table}
For every row in Tab.~(\ref{tab:from_j_k_to_n}) (indexed by $j=0, 1, ..., n$), the appropriate value of $k$ is given by $k=n-j$.
Then, $l$ has to be chosen sufficiently large, so that $m=k+l-j$ is fulfilled.
For the given $m$ and a given $j$ and appropriately chosen $k~(=n-j$, see above), the matching value
of $l$ is $l=m-k+j$ and inserting the chosen value for $k$ into this leads to $l=m-n+2j$.

Now assume $n \geq m$ for the second part.
There are $(m+1)$ contributions to $s^m t^n$ with $m=k+l-j$ and $n=k+j$,
labeled $k=0, 1, ..., m$, as shown in Tab.~(\ref{tab:from_k_lmj_to_m}).
\begin{table}[htbp]
  \centering
  \begin{tabular}{c|c}
    $k$ & $l-j$ \\
    \hline
    0 & $m$ \\
    1 & $m-1$ \\
    ... & ... \\
    $m-1$ & 1 \\
    $m$ & 0
  \end{tabular}
  \caption{$(m+1)$ possible combinations of $k \geq 0$ and $l \geq 0$, $0 \leq j \leq l$ which sum up to a given $m=k+l-j$.}
  \label{tab:from_k_lmj_to_m}
\end{table}
For the given $n$, $j$ has to be chosen for $n=k+j$ as $j=n-k$.
Note that due to $n \geq m$ and $k \leq m$, $j \geq 0$ is guaranteed for above values of $k$.
Since $l \geq 0$ and $0 \leq j \leq l$ hold, it follows that $(l-j) \geq 0$ for any occuring values of $l$ and $j$.
Then, $l$ has to be found for the representation of $m$ from $m=k+l-j$.
Insert the chosen value for $j$ into $m$ as $m=k+l-(n-k) = k+l-n+k = l-n+2k$
and solve for $l$ as $l=m+n-2k$.
The value of $k$ takes on a maximum of $m$ and at this limit, the minimal value of $l$ occurs:
$l_\textrm{min} = m+n-2m = n-m$ and since $n \geq m$ was assumed, $l_\textrm{min} \geq 0$ is guaranteed.

For the two cases $m \geq n$ and $n \geq m$, the values of $k, l$ and $j,l$, respectively, are inserted into the summands in \eqn{sums_h_p_st}
and the contributions are summed up.
For the case of $m \geq n$, this leads to:
\begin{align}
     c^{+}_{m,n} s^m t^n
  :=& \frac{1}{2}
      \sum\limits_{j=0}^{n}
    (-1)^{j}
    \frac{(\bcancel{n}\bcancel{-j}+m\bcancel{-n}+\bcancel{2}j)!}{j!\,(m-n+\bcancel{2}j\bcancel{-j})!\,(n-j)!}
    \,T^{+}_{m-n+2j}
    s^{\bcancel{n}\bcancel{-j}+m\bcancel{-n}\bcancel{+2j}\bcancel{-j}} t^{n\bcancel{-j}\bcancel{+j}} \nonumber \\
 =& \frac{1}{2}
    \sum\limits_{j=0}^{n}
    (-1)^{j}
    \frac{(m+j)!}{j!\,(m-n+j)!\,(n-j)!}
    \,T^{+}_{m-n+2j}
    s^m t^n \, . \label{eqn:c_mn_mgeqn}
\end{align}
For the case of $n \geq m$, this leads to:
\begin{align}
     c^{+}_{m,n} s^m t^n
  :=& \frac{1}{2}
      \sum\limits_{k=0}^{m}
    (-1)^{n-k}
    \frac{(\bcancel{k}+m+n-\bcancel{2}k)!}{(n-k)!\,(m\bcancel{+n}-\bcancel{2}k\bcancel{-n}\bcancel{+k})!\,k!}
    \,T^{+}_{m+n-2k}
    s^{\bcancel{k}+m\bcancel{+n}\bcancel{-2k}\bcancel{-n}\bcancel{+k}} t^{\bcancel{k}+n\bcancel{-k}} \nonumber \\
  =& \frac{1}{2}
     \sum\limits_{k=0}^{m}
    (-1)^{n-k}
    \frac{(m+n-k)!}{(n-k)!\,(m-k)!\,k!}
    \,T^{+}_{m+n-2k}
    s^m t^n \nonumber \\
  \xrightarrow[k \rightarrow (m-k)]{(m-k) \rightarrow k} ~
  =& \frac{1}{2}
     \sum\limits_{k=0}^{m}
    (-1)^{k+n-m}
    \frac{(n+k)!}{(n-m+k)!\,k!\,(m-k)!}
    \,T^{+}_{n-m+2k}
    s^m t^n \, . \label{eqn:c_mn_ngeqm}
\end{align}
Comparing the two expressions from \eqn{c_mn_mgeqn} and from \eqn{c_mn_ngeqm} with either $m \geq n$  or $n \geq m$, respectively,
in mind, a generalization of the integer quantities in the $c^{+}_{m,n}$ can be found and is tabulated in Tab.~(\ref{tab:generalization_c_mn}).
\begin{table}[htbp]
  \centering
  \begin{tabular}{c | c | c | c}
    quantity                     & $m \geq n$ & $n \geq m$ & generally                 \\
    \hline
                     end of loop & $n$        & $m$        & $\mathrm{min}\,(m,n)$     \\
                 power of $(-1)$ & $j$        & $j+n-m$    & $\mathrm{max}\,(0, n-m)$  \\
          factorial in numerator & $j+m$      & $j+n$      & $j+\mathrm{max}\,(m,n)$   \\
    1st factorial in denominator & $n-j$      & $m-j$      & $\mathrm{min}\,(m,n) - j$ \\
    2nd factorial in denominator & $m-n+j$    & $n-m+j$    & $|m-n|+j$                 \\
                index of $T^{+}$ & $m-n+2j$   & $n-m+2j$   & $|m-n|+2j$
  \end{tabular}
  \caption{Generalization of the integer quantities in $c^{+}_{m,n}$.}
  \label{tab:generalization_c_mn}
\end{table}

The min/max functions can be expressed analytically:
\begin{align}
  \mathrm{min}\,(m,n)
  =& \frac{1}{2} (m+n-|m-n|) = \begin{cases}
                                 \frac{1}{2} (\bcancel{m}+n\bcancel{-m}+n) = n & ~:~ m \geq n \\
                                 \frac{1}{2} (m\bcancel{+n}\bcancel{-n}+m) = m & ~:~ n \geq m
                               \end{cases} \\
  \mathrm{max}\,(m,n)
  =& \frac{1}{2} (m+n+|m-n|) = \begin{cases}
                                 \frac{1}{2} (m\bcancel{+n}+m\bcancel{-n}) = m & ~:~ m \geq n \\
                                 \frac{1}{2} (\bcancel{m}+n+n\bcancel{-m}) = n & ~:~ n \geq m
                               \end{cases} \\
  \mathrm{max}\,(0,n-m)
  =& \frac{1}{2} (|m-n|-m+n) = \begin{cases}
                                 \frac{1}{2} (\bcancel{m}\bcancel{-n}\bcancel{-m}\bcancel{+n}) = 0 & ~:~ m \geq n \\
                                 \frac{1}{2} (n-m-m+n) = n-m & ~:~ n \geq m
                               \end{cases}
\end{align}
These expressions now can be inserted into $c^{+}_{m,n}$ and are valid for both $m \geq n$ and $n \geq m$:
\begin{equation}
    c^{+}_{m,n}
  = \frac{1}{2}
    \sum\limits_{j=0}^{\frac{m+n-|m-n|}{2}}
      (-1)^{j+\frac{|m-n|-m+n}{2}}
      \frac{\left(\frac{m+n+|m-n|}{2} +j \right)!}
           {\left(\frac{m+n-|m-n|}{2} -j \right)!\, (|m-n|+j)!\, j!}
      T^{+}_{|m-n|+2j} \, . \label{eqn:c_p_mn}
\end{equation}
The coefficients $c^{+}_{m,n}$ can now be used to express $h^{+}(s,t)$ from \eqn{h_p_st} again:
\begin{align}
 h^{+}&(s,t)
 = \, (1+s+t+st)
    \sum\limits_{m=0}^{+\infty}
    \sum\limits_{n=0}^{+\infty}
      c^{+}_{m,n}
      s^m t^n \nonumber \\
 =&   \sum\limits_{m=0}^{+\infty}
      \sum\limits_{n=0}^{+\infty}
        c^{+}_{m,n}
        s^m t^n
    + \sum\limits_{m=0}^{+\infty}
      \sum\limits_{n=0}^{+\infty}
        c^{+}_{m,n}
        s^{m+1} t^n
    + \sum\limits_{m=0}^{+\infty}
      \sum\limits_{n=0}^{+\infty}
        c^{+}_{m,n}
        s^m t^{n+1}
    + \sum\limits_{m=0}^{+\infty}
      \sum\limits_{n=0}^{+\infty}
        c^{+}_{m,n}
        s^{m+1} t^{n+1}            \nonumber \\
 =& \phantom{+} \,   c^{+}_{0,0}
                   + \sum\limits_{m=1}^{+\infty}
                       c^{+}_{m,0}
                       s^m
                   + \sum\limits_{n=1}^{+\infty}
                       c^{+}_{0,n}
                       t^n
                   + \sum\limits_{m=1}^{+\infty}
                     \sum\limits_{n=1}^{+\infty}
                       c^{+}_{m,n}
                       s^m t^n    \nonumber \\
 ~& + \sum\limits_{m=1}^{+\infty}
        c^{+}_{m-1,0}
        s^m
    + \sum\limits_{m=1}^{+\infty}
      \sum\limits_{n=1}^{+\infty}
        c^{+}_{m-1,n}
        s^m t^n                   \nonumber \\
 ~& + \sum\limits_{n=1}^{+\infty}
        c^{+}_{0,n-1}
        t^n
    + \sum\limits_{m=1}^{+\infty}
      \sum\limits_{n=1}^{+\infty}
        c^{+}_{m,n-1}
        s^m t^n                   \nonumber \\
 ~& + \sum\limits_{m=1}^{+\infty}
      \sum\limits_{n=1}^{+\infty}
        c^{+}_{m-1,n-1}
        s^m t^n                   \nonumber \\
 =& \phantom{+}\, \,c^{+}_{0,0}   \nonumber \\
 ~&          +  \sum\limits_{n=1}^{+\infty}
                  \left( c^{+}_{0,n} + c^{+}_{0,n-1} \right)
                  t^n             \nonumber \\
 ~&          +  \sum\limits_{m=1}^{+\infty}
                  \left( c^{+}_{m,0} + c^{+}_{m-1,0} \right)
                  s^m             \nonumber \\
 ~&          +  \sum\limits_{m=1}^{+\infty}
                \sum\limits_{n=1}^{+\infty}
                  \left( c^{+}_{m,n} + c^{+}_{m-1,n} + c^{+}_{m,n-1} + c^{+}_{m-1,n-1} \right)
                  s^m t^n \label{eqn:h_p_st_c_mn}
\end{align}
As it turns out, the expression for $c^{+}_{m,n}$ in \eqn{c_p_mn}
generalizes further to $c^{\pm}_{m,n}$ for $m \geq 0$ and $n \geq 0$:
\begin{equation}
    c^{\pm}_{m,n}
  := \frac{1}{2}
    \sum\limits_{j=0}^{\frac{m+n-|m-n|}{2}}
      (-1)^{j+\frac{|m-n|-m+n}{2}}
      \frac{\left(\frac{m+n+|m-n|}{2} +j \right)!}
           {\left(\frac{m+n-|m-n|}{2} -j \right)!\, (|m-n|+j)!\, j!}
      T^{\pm}_{|m-n|+2j} \, . \label{eqn:c_pm_mn}
\end{equation}
First, evaluate $c^{\pm}_{m,n}$ for $m=0$ and $n=0$:
\begin{equation}
  c^{\pm}_{0,0} = \frac{1}{2} T^{\pm}_0 \, .
\end{equation}
Next, consider the case of $m=0$ and $n \geq 1$:
\begin{equation}
  c^{\pm}_{0,n} = \frac{1}{2} (-1)^n \,T^{\pm}_n
\end{equation}
and similarly for $m \geq 1$ and $n=0$:
\begin{equation}
  c^{\pm}_{m,0} = \frac{1}{2} \,T^{\pm}_m \, .
\end{equation}
These terms can be found in \eqn{h_p_00}, \eqn{h_m_s0} and \eqn{h_m_0t}:
\begin{align}
  h^{+}(0,0) =& c^{+}_{0,0} \label{eqn:h_p_00_c_mn} \\
  h^{-}(s,0) =& c^{-}_{0,0} + \sum\limits_{m=1}^{+\infty} \left( c^{-}_{m,0} + c^{-}_{m-1,0} \right) s^m \label{eqn:h_m_s0_c_mn} \\
  h^{-}(0,t) =& c^{-}_{0,0} + \sum\limits_{n=1}^{+\infty} \left( c^{-}_{0,n} + c^{-}_{0,n-1} \right) t^n \label{eqn:h_m_0t_c_mn} \, .
\end{align}
Remember the two representations of $I(s,t)$ from \eqn{I_gf} and \eqn{I_from_h}:
\begin{equation*}
 I(s,t) = \sum\limits_{m=0}^{+\infty}
          \sum\limits_{n=0}^{+\infty}
            I_{m,n} s^m t^n
        = h^{+}(s,t) + h^{+}(0,0) + h^{-}(s,0) + h^{-}(0,t)
        \, .
\end{equation*}
Collecting terms with equal powers of $s$ and $t$
from \eqn{h_p_st_c_mn}, \eqn{h_p_00_c_mn}, \eqn{h_m_s0_c_mn} and \eqn{h_m_0t_c_mn}
finally leads to the following form of the $I_{m,n}$ coefficients:
\begin{equation}
 I_{m,n} = \left\{
           \begin{array}{lllll}
             c^{+}_{m,n} &+\, c^{+}_{m-1,n} &+\, c^{+}_{m,n-1} &+\, c^{+}_{m-1,n-1} &\textrm{for}~~ m \geq 1 , n \geq 1 \\
             c^{+}_{m,0} &+\, c^{+}_{m-1,0} &+\, c^{-}_{m,0}   &+\, c^{-}_{m-1,0}   &\textrm{for}~~ m \geq 1 , n = 0 \\
             c^{+}_{0,n} &+\, c^{+}_{0,n-1} &+\, c^{-}_{0,n}   &+\, c^{-}_{0,n-1}   &\textrm{for}~~ m = 0    , n \geq 1 \\
             c^{+}_{0,0} &+\, c^{+}_{0,0}   &+\, c^{-}_{0,0}   &+\, c^{-}_{0,0}     &\textrm{for}~~ m = 0    , n = 0 \, .
           \end{array} \right. \label{eqn:Imn}
\end{equation}

\FloatBarrier
\subsection{Closed Form Expressions for the Coefficients}
Next, find closed form expressions for the coefficients $T^{\pm}_l$ from \eqn{T_pm_l}.
The first case for $l=0$ is:
\begin{equation}
 T^{\pm}_0 = \int\limits_{-1}^{1} \frac{\mathrm{d}x}{\sqrt{ a^{\mp}  + 2 d x + a^{\pm} x^2 }}
\end{equation}
The corresponding antiderivative is tabulated in Ref.~\cite{gradshteyn_ryzhik} labeled 2.261.
The following shorthand notations are needed:
\begin{align}
      R =& {a'} + {b'} x + {c'} x^2 \label{eqn:def_R} \\
 \Delta =& 4 {a'} {c'} - {b'}^2 \label{eqn:def_delta}
\end{align}
with
\begin{align}
 {a'} =& a^{\mp} = a \mp 2 b + c \nonumber \\
 {b'} =& 2 d = 2 (c-a) \label{eqn:def_primes} \\
 {c'} =& a^{\pm} = a \pm 2 b + c \nonumber \, ,
\end{align}
where $a$, $b$ and $c$ are from \eqn{def_abc}:
\begin{equation}
 \begin{array}{llll}
 a &= {\mathbf{x}'}_{u'}^2 &= {\mathbf{x}'}_{u'} \cdot {\mathbf{x}'}_{u'} &= |{\mathbf{x}'}_{u'}|^2\\
 b &        ~              &= {\mathbf{x}'}_{u'} \cdot {\mathbf{x}'}_{v'} &= |{\mathbf{x}'}_{u'}| \cdot |{\mathbf{x}'}_{v'}| \cdot \cos(\theta) \\
 c &= {\mathbf{x}'}_{v'}^2 &= {\mathbf{x}'}_{v'} \cdot {\mathbf{x}'}_{v'} &= |{\mathbf{x}'}_{v'}|^2
 \end{array}
\end{equation}
where $\theta$ is the angle between ${\mathbf{x}'}_{u'}$ and ${\mathbf{x}'}_{v'}$.
The signs of $c'$ and $\Delta$ need to be considered for selecting which antiderivative can be used.
It follows for $c'$:
\begin{equation}
 {c'} = |{\mathbf{x}'}_{u'}|^2 \pm 2 \cdot |{\mathbf{x}'}_{u'}| \cdot |{\mathbf{x}'}_{v'}| \cdot \cos(\theta) + |{\mathbf{x}'}_{v'}|^2 \, .
\end{equation}
For the case of $(+)$, the minimal value is taken on for $\cos(\theta)=-1$ and for this it holds:
\begin{align}
 {c'} =& |{\mathbf{x}'}_{u'}|^2 + |{\mathbf{x}'}_{v'}|^2 + 2 \cdot |{\mathbf{x}'}_{u'}| \cdot |{\mathbf{x}'}_{v'}| \cdot (-1) \nonumber \\
   ~  =& |{\mathbf{x}'}_{u'}|^2 + |{\mathbf{x}'}_{v'}|^2 - 2 \cdot |{\mathbf{x}'}_{u'}| \cdot |{\mathbf{x}'}_{v'}|            \nonumber \\
   ~  =& \left( |{\mathbf{x}'}_{u'}| - |{\mathbf{x}'}_{v'}| \right)^2 > 0 \, , \label{eqn:check_c_prime}
\end{align}
since $|{\mathbf{x}'}_{u'}|=0$ occurs never simultaneously with $|{\mathbf{x}'}_{v'}|=0$ for any $(u', v') \in [0,1]^2$.
The minimal value of $c'$ occurs for the case of $(-)$ at $\cos(\theta)=+1$ and \eqn{check_c_prime} follows.
Thus, ${c'}>0$ is ensured for any $(u', v') \in [0,1]^2$.
Regarding $\Delta$, note that:
\begin{align}
 \Delta =& 4 \cdot (a \mp 2 b + c) \cdot (a \pm 2 b + c) - (2 (c-a))^2 \nonumber \\
    ~   =& 4 \left[ (a \mp 2 b + c) \cdot (a \pm 2 b + c) - (c-a)^2 \right] \nonumber \\
    ~   =& 4 \left[ a^2 \pm 2 a b + a c \mp 2 a b - 4 b^2 \mp 2 b c + a c \pm 2 b c + c^2 - (c^2 - 2 a c + a^2) \right] \nonumber \\
    ~   =& 4 \left[ \bcancel{a^2} \bcancel{\pm 2 a b} + a c \bcancel{\mp 2 a b} - 4 b^2 \bcancel{\mp 2 b c} + a c \bcancel{\pm 2 b c} \bcancel{+ c^2} \bcancel{- c^2} + 2 a c \bcancel{- a^2} \right] \nonumber \\
    ~   =& 4 \left[ 4 a c - 4 b^2 \right] = 16 (a c - b^2) \label{eqn:def_delta_simple} \\
    ~   =& 16 \left[ |{\mathbf{x}'}_{u'}|^2 \cdot |{\mathbf{x}'}_{v'}|^2 - \left( |{\mathbf{x}'}_{u'}| \cdot |{\mathbf{x}'}_{v'}| \cdot \cos(\theta) \right)^2 \right] \nonumber \\
    ~   =& 16 \cdot |{\mathbf{x}'}_{u'}|^2 \cdot |{\mathbf{x}'}_{v'}|^2 \cdot \underbrace{\left( 1 - \cos^2(\theta) \right)}_{\geq 0} \geq 0 \, .
\end{align}
Now the tabulated integrals can be considered:
\begin{equation}
 \int \frac{\mathrm{d}x}{\sqrt{R}} =
 \begin{cases}
  \frac{1}{\sqrt{c'}} \left[ \mathrm{arsinh}\left( \frac{2 c' x + b'}{\sqrt{\Delta}} \right) \right]_{-1}^{+1} & \textrm{ for } {c'}>0, \Delta>0 \\
  \frac{1}{\sqrt{c'}} \left[ \ln            \left(       2 c' x + b'                 \right) \right]_{-1}^{+1} & \textrm{ for } {c'}>0, \Delta=0
 \end{cases} \, . \label{eqn:int_oosr}
\end{equation}
The case for $\Delta>0$ will be considered first and it will be shown that the solution obtained this way is applicable also for $\Delta=0$.
It follows with \eqn{arsinh}:
\begin{align}
    \int\limits_{-1}^{+1} \frac{\mathrm{d}x}{\sqrt{R}}
 =& \frac{1}{\sqrt{c'}} \left[ \mathrm{arsinh}\left( \frac{2 c' x + b'}{\sqrt{\Delta}} \right) \right]_{-1}^{+1} \nonumber \\
 =& \frac{1}{\sqrt{c'}} \left[ \ln \left( \frac{2 c' x + b'}{\sqrt{\Delta}} + \sqrt{ \left(\frac{2 c' x + b'}{\sqrt{\Delta}}\right)^2 + 1} \right) \right]_{-1}^{+1} \nonumber \\
 =& \frac{1}{\sqrt{c'}} \left[ \ln \left( \frac{2 c' x + b'}{\sqrt{\Delta}} + \sqrt{ \frac{ \left(2 c' x + b'\right)^2}{\Delta} + 1} \right) \right]_{-1}^{+1} \nonumber \\
 =& \frac{1}{\sqrt{c'}} \left[ \ln \left( \frac{2 c' x + b'}{\sqrt{\Delta}} + \frac{1}{\sqrt{\Delta}}\sqrt{ \left(2 c' x + b'\right)^2 + \Delta} \right) \right]_{-1}^{+1} \nonumber \\
 =& \frac{1}{\sqrt{c'}} \left[ \ln \left( 2 c' x + b' + \sqrt{ \left(2 c' x + b'\right)^2 + \Delta} \right) - \frac{1}{2} \ln(\Delta) \right]_{-1}^{+1} \nonumber \\
 =& \frac{1}{\sqrt{c'}} \left[ \ln \left( 2 c' x + b' + \sqrt{ \left(2 c' x + b'\right)^2 + \Delta} \right) \right]_{-1}^{+1} \nonumber \\
 =& \frac{1}{\sqrt{c'}} \left[  \ln \left( b' + 2 c' + \sqrt{ \left(b' + 2 c'\right)^2 + \Delta} \right)
                              - \ln \left( b' - 2 c' + \sqrt{ \left(b' - 2 c'\right)^2 + \Delta} \right) \right] \nonumber \\
 =& \frac{1}{\sqrt{c'}} \ln \left( \frac{ \sqrt{ \left(b' + 2 c'\right)^2 + \Delta} + b' + 2 c' }
                                        { \sqrt{ \left(b' - 2 c'\right)^2 + \Delta} + b' - 2 c' } \right) \, .
\end{align}
The definition of $\Delta$ from \eqn{def_delta} is now inserted to simplify the equation further:
\begin{align}
 \int\limits_{-1}^{+1} \frac{\mathrm{d}x}{\sqrt{R}}
 =& \frac{1}{\sqrt{c'}} \ln \left( \frac{ \sqrt{ \left(b' + 2 c'\right)^2 + 4 a' c' - {b'}^2} + b' + 2 c' }
                                        { \sqrt{ \left(b' - 2 c'\right)^2 + 4 a' c' - {b'}^2} + b' - 2 c' } \right) \nonumber \\
 =& \frac{1}{\sqrt{c'}} \ln \left( \frac{ \sqrt{ \bcancel{{b'}^2} + 4 b' c' + 4 {c'}^2 + 4 a' c' \bcancel{- {b'}^2}} + b' + 2 c' }
                                        { \sqrt{ \bcancel{{b'}^2} - 4 b' c' + 4 {c'}^2 + 4 a' c' \bcancel{- {b'}^2}} + b' - 2 c' } \right) \nonumber \\
 =& \frac{1}{\sqrt{c'}} \ln \left( \frac{ 2 \sqrt{ (a' + b' + c') c' } + b' + 2 c' }
                                        { 2 \sqrt{ (a' - b' + c') c' } + b' - 2 c' } \right) \nonumber \\
 =& \frac{1}{\sqrt{c'}} \ln \left( \frac{ 2 \sqrt{ (a' + b' + c') c' } + b' + 2 c' }
                                        { 2 \sqrt{ (a' - b' + c') c' } + b' - 2 c' } \right) \, . \label{eqn:t0_almost}
\end{align}
Note that
\begin{align}
 R(+1) = a' + b' + c' =& a \pm 2 b + c + 2 (c - a) + a \mp 2 b + c =  4 c \label{eqn:four_c} \\
 R(-1) = a' - b' + c' =& a \pm 2 b + c - 2 (c - a) + a \mp 2 b + c =  4 a \label{eqn:four_a} \, .
\end{align}
Inserting the definitions for $a'$, $b'$ and $c'$ into \eqn{t0_almost} leads to:
\begin{align}
T^{\pm}_0 = \int\limits_{-1}^{+1} \frac{\mathrm{d}x}{\sqrt{R}}
 =& \frac{1}{\sqrt{a \pm 2 b + c}}
    \ln \left( \frac{ \bcancel{2} \sqrt{ 4 c (a \pm 2 b + c)} + \bcancel{2} (c-a) + \bcancel{2} (a \pm 2 b + c) }
                    { \bcancel{2} \sqrt{ 4 a (a \pm 2 b + c)} + \bcancel{2} (c-a) - \bcancel{2} (a \pm 2 b + c) } \right) \nonumber \\
 =& \frac{1}{\sqrt{a \pm 2 b + c}}
    \ln \left( \frac{ 2 \sqrt{ c (a \pm 2 b + c)}          +c \bcancel{-a} \bcancel{+a} \pm 2 b          +c  }
                    { 2 \sqrt{ a (a \pm 2 b + c)} \bcancel{+c}         -a           - a \mp 2 b \bcancel{-c} } \right) \nonumber \\
 =& \frac{1}{\sqrt{a \pm 2 b + c}}
    \ln \left( \frac{ \bcancel{2} \sqrt{ c (a \pm 2 b + c)} +\bcancel{2}c \pm \bcancel{2} b }
                    { \bcancel{2} \sqrt{ a (a \pm 2 b + c)} -\bcancel{2}a \mp \bcancel{2} b } \right) \nonumber \\
 =& \frac{1}{\sqrt{a \pm 2 b + c}}
    \ln \left( \frac{ \sqrt{ c (a \pm 2 b + c)} +c \pm b }
                    { \sqrt{ a (a \pm 2 b + c)} -a \mp b } \right) \, . \label{eqn:t_pm_0}
\end{align}
It remains to be shown that this results also works for $\Delta=0$.
Start at \eqn{t_pm_0} and insert $0=\Delta=16\left(ac-b^2\right) \Rightarrow ac = b^2$:
\begin{align}
    T^{\pm}_0
 =& \frac{1}{\sqrt{a \pm 2 b + c}}
    \ln \left( \frac{ \sqrt{ a c \pm 2 b c + c^2} +c \pm b }
                    { \sqrt{ a^2 \pm 2 a b + a c} -a \mp b } \right) \nonumber \\
 =& \frac{1}{\sqrt{a \pm 2 b + c}}
    \ln \left( \frac{ \sqrt{ b^2 \pm 2 b c + c^2} +c \pm b }
                    { \sqrt{ a^2 \pm 2 a b + b^2} -a \mp b } \right) \nonumber \\
 =& \frac{1}{\sqrt{a \pm 2 b + c}}
    \ln \left( \frac{ \sqrt{ \left(b \pm c \right)^2 } +c \pm b }
                    { \sqrt{ \left(a \pm b \right)^2 } -a \mp b } \right) \, .
\end{align}
In the numerator, select the case $c \pm b$.
In the denominator, the case $-a \mp b$ is selected.
This leads to:
\begin{align}
    T^{\pm}_0
 =& \frac{1}{\sqrt{a \pm 2 b + c}}
    \ln \left( \frac{ 2 (  c \pm b) }{ 2 ( -a \mp b) } \right) \nonumber \\
 =& \frac{1}{\sqrt{c'}}
    \ln \left( \frac{  c \pm 2b + c }{ -a -a \mp 2b } \right) \nonumber \\
 =& \frac{1}{\sqrt{c'}}
    \ln \left( \frac{c-a + a \pm 2b + c }{c-a - a \mp 2b - c } \right) \nonumber \\
 =& \frac{1}{\sqrt{c'}}
    \ln \left( \frac{\bcancel{2}(c-a) + \bcancel{2}(a \pm 2b + c)}{\bcancel{2}(c-a) - \bcancel{2}(a \pm 2b + c)} \right) \nonumber \\
 =& \frac{1}{\sqrt{c'}}
    \ln \left( \frac{b' + 2 c'}{b' - 2 c'} \right) \nonumber \\
 =& \frac{1}{\sqrt{c'}}
    \left[ \ln(b' + 2 c') - \ln(b' - 2 c') \right] \nonumber \\
 =& \frac{1}{\sqrt{c'}}
    \left[ \ln(2 c' x + b') \right]_{-1}^{+1}
\end{align}
which is the case for $\Delta=0$ in \eqn{int_oosr}.
This proves the universal applicability of \eqn{t_pm_0} for computing $T^{\pm}_0$.
The next item to consider is $T^{\pm}_1$:
\begin{equation}
 T^{\pm}_1 = \int\limits_{-1}^{+1} \frac{x}{\sqrt{a^{\mp} + 2 d x + a^{\pm} x^2}} \mathrm{d}x \, .
\end{equation}
From Ref.~\cite{gradshteyn_ryzhik} the formula 2.264.2 can be used:
\begin{equation}
   \int \frac{x}{\sqrt{R}} \mathrm{d}x
 = \frac{\sqrt{R}}{c'} - \frac{b'}{2 c'} \int \frac{\mathrm{d}x}{\sqrt{R}}
\end{equation}
with $R$, $a'$, $b'$ and $c'$ as defined in \eqn{def_R} and \eqn{def_primes}.
Application to this problem and re-use of \eqn{four_c} as well as \eqn{four_a} leads to:
\begin{align}
 T^{\pm}_1
 =& \int\limits_{-1}^{+1} \frac{x}{\sqrt{R}} \mathrm{d}x
  = \frac{1}{c'} \Biggl( \left[ \sqrt{R} \right]_{-1}^{+1} - \frac{b'}{2} \underbrace{\int\limits_{-1}^{+1} \frac{\mathrm{d}x}{\sqrt{R}}}_{=T^{\pm}_0} \Biggr) \nonumber \\
 =& \frac{1}{c'} \left( \left[ \sqrt{a' + b' x + c' x^2} \right]_{-1}^{+1} - \frac{b'}{2} T^{\pm}_0 \right) \nonumber \\
 =& \frac{1}{c'} \left( \left[ \sqrt{a' + b' + c'} - \sqrt{a' - b' + c'} \right] - \frac{b'}{2} T^{\pm}_0 \right) \nonumber \\
 =& \frac{1}{a \pm 2 b + c} \left( \left[ \sqrt{4 c} - \sqrt{4 a} \right] - \frac{\bcancel{2} (c-a)}{\bcancel{2}} T^{\pm}_0 \right) \nonumber \\
 =& \frac{1}{a \pm 2 b + c} \left( 2 \left(\sqrt{c} - \sqrt{a} \right) - (c-a) T^{\pm}_0 \right) \label{eqn:t_pm_1} \, .
\end{align}
The remaining $T^{\pm}_l$ for $l \ge 2$ is tabulated in Ref.~\cite{gradshteyn_ryzhik} as 2.263.1 with $m=l$ and $n=0$ there.
The resulting antiderivative reads then:
\begin{equation}
 \int \frac{x^l}{\sqrt{R}} \mathrm{d}x
 = \frac{x^{l-1}}{l c' \sqrt{R^{-1}}} - \frac{(2 l - 1)b'}{2 l c'} \int \frac{x^{l-1}}{\sqrt{R}} \mathrm{d}x - \frac{(l-1)a'}{l c'} \int \frac{x^{l-2}}{\sqrt{R}} \mathrm{d}x \, .
\end{equation}
Insert the limits of the integral in $T^{\pm}_l$:
\begin{align}
 T^{\pm}_l
 =& \int\limits_{-1}^{+1} \frac{x^l}{\sqrt{R}} \mathrm{d}x
 = \frac{1}{l c'} \Biggl\{
       \left[ x^{l-1} \sqrt{R} \right]_{-1}^{+1}
     - \frac{(2 l - 1)b'}{2} \underbrace{\int\limits_{-1}^{+1} \frac{x^{l-1}}{\sqrt{R}} \mathrm{d}x}_{=T^{\pm}_{l-1}}
     -   (l-1) a'            \underbrace{\int\limits_{-1}^{+1} \frac{x^{l-2}}{\sqrt{R}} \mathrm{d}x}_{=T^{\pm}_{l-2}}
   \Biggr\} \nonumber \\
 =& \frac{1}{l c'}
    \left[
      \sqrt{R(+1)} -(-1)^{l-1} \sqrt{R(-1)}
      - \bcancel{\frac{1}{2}} (2 l - 1) \bcancel{2} (c-a) T^{\pm}_{l-1}
      - (l-1) (a \mp 2 b + c) T^{\pm}_{l-2}
    \right] \nonumber \\
 =& \frac{1}{l (a \pm 2b + c)}
    \left[
      \sqrt{4 c} +(-1)^l \sqrt{4 a}
      - (2 l - 1) (c-a) T^{\pm}_{l-1}
      - (l-1) (a \mp 2 b + c) T^{\pm}_{l-2}
    \right] \nonumber \\
 =& \frac{1}{l (a \pm 2b + c)}
    \left[
      2 \left( \sqrt{c} +(-1)^l \sqrt{a} \right)
      - (2 l - 1) (c-a) T^{\pm}_{l-1}
      - (l-1) (a \mp 2 b + c) T^{\pm}_{l-2}
    \right] \, .
\end{align}
This is the general result for the $T^{\pm}_l$ with $l \ge 2$.

Consider the coefficients $K_{m,n}$ from \eqn{Kmn} next.
The partial derivatives only act on the factors $a$, $b$ and $c$ within $I_{m,n}$.
However, $I_{m,n}$ is constructed from $c_{m,n}^{\pm}$ via \eqn{Imn};
hence the partial derivatives act on the $c_{m,n}^{\pm}$.
These in turn only consist of a summation over $T^{\pm}_l$, which finally are dependent on $a$, $b$ and $c$.
New terms $S^{\pm}_l$ are introduced with:
\begin{equation}
  S^{\pm}_l = -2 \left( A \frac{\partial}{\partial a} + B \frac{\partial}{\partial b} + C \frac{\partial}{\partial c} \right) T^{\pm}_l \, . \label{eqn:def_S_pm_l}
\end{equation}
First consider the polynomial in the denominator of $T^{\pm}_l$:
\begin{align}
  a^{\mp}+2 d x + a^{\pm} x^2
  =& a \mp 2b +c +2 (c-a) x + (a \pm 2 b + c) x^2 \nonumber \\
  =& a \left( 1 -2 x + x^2 \right) + b \left( \mp 2 - \pm 2 x^2 \right) + c \left( 1 +2 x +x^2 \right)
\end{align}
and then the partial derivatives:
\begin{align}
  \frac{\partial}{\partial a} \left( a^{\mp}+2 d x + a^{\pm} x^2 \right)^{-\frac{1}{2}}
  =& -\frac{1}{2} \left( a^{\mp}+2 d x + a^{\pm} x^2 \right)^{-\frac{3}{2}} \left( 1 - 2 x + x^2 \right) \\
  \frac{\partial}{\partial b} \left( a^{\mp}+2 d x + a^{\pm} x^2 \right)^{-\frac{1}{2}}
  =& -\frac{1}{2} \left( a^{\mp}+2 d x + a^{\pm} x^2 \right)^{-\frac{3}{2}} \left( \mp 2 - \pm 2 x^2 \right) \\
  \frac{\partial}{\partial c} \left( a^{\mp}+2 d x + a^{\pm} x^2 \right)^{-\frac{1}{2}}
  =& -\frac{1}{2} \left( a^{\mp}+2 d x + a^{\pm} x^2 \right)^{-\frac{3}{2}} \left( 1 + 2 x + x^2 \right) \, .
\end{align}
These expressions can be used to formulate the partial derivatives of $T^{\pm}_l$:
\begin{equation}
    \frac{\partial}{\partial a} T^{\pm}_l
  = \int\limits_{-1}^{+1} \frac{\partial}{\partial a} \left( \frac{1}{\sqrt{a^{\mp}+2 d x + a^{\pm} x^2}} \right) x^l \,\mathrm{d}x
  = -\frac{1}{2} \int\limits_{-1}^{+1} \frac{1 - 2 x + x^2}{\left( a^{\mp}+2 d x + a^{\pm} x^2 \right)^{\frac{3}{2}}} x^l \,\mathrm{d}x
\end{equation}
and similarly:
\begin{align}
  \frac{\partial}{\partial b} T^{\pm}_l
  =& -\frac{1}{2} \int\limits_{-1}^{+1} \frac{ \mp 2 - \pm 2 x^2}{\left( a^{\mp}+2 d x + a^{\pm} x^2 \right)^{\frac{3}{2}}} x^l \,\mathrm{d}x \\
  \frac{\partial}{\partial c} T^{\pm}_l
  =& -\frac{1}{2} \int\limits_{-1}^{+1} \frac{1 + 2 x + x^2 }{\left( a^{\mp}+2 d x + a^{\pm} x^2 \right)^{\frac{3}{2}}} x^l \,\mathrm{d}x \, .
\end{align}
Above three expressions for the partial derivatives of $T^{\pm}_l$ are now inserted into \eqn{def_S_pm_l}:
\begin{equation}
  S^{\pm}_l = \int\limits_{-1}^{+1} \frac{A^{\mp} + 2 D x + A^{\pm} x^2}
              {\left( a^{\mp}+2 d x + a^{\pm} x^2 \right)^{\frac{3}{2}}} x^l \,\mathrm{d}x \label{eqn:int_S_pm_l}
\end{equation}
with
\begin{align}
  A^{\pm} =& A \pm 2B + C \\
  D =& C - A
\end{align}
and $A$, $B$ and $C$ from \eqn{def_ABC}.
Recall the definitions of $R$ and $\Delta$ from \eqn{def_R} and \eqn{def_delta_simple}:
\begin{align}
                        R(x)   =& a^{\mp}+2 d x + a^{\pm} x^2 \nonumber \\
  \textrm{ with } \sqrt{R(+1)} =& 2 \sqrt{c} \nonumber \\
  \textrm{ and }  \sqrt{R(-1)} =& 2 \sqrt{a} \nonumber \\
                        \Delta =& 16 (ac - b^2) \nonumber \, .
\end{align}
We continue by splitting up $S^{\pm}_l$:
\begin{equation}
    S^{\pm}_l
  = A^{\mp} \int\limits_{-1}^{+1}
              \frac{1}{R^{\frac{3}{2}}} x^l \,\mathrm{d}x
  + 2 D     \int\limits_{-1}^{+1}
              \frac{x}{R^{\frac{3}{2}}} x^l \,\mathrm{d}x
  + A^{\pm} \int\limits_{-1}^{+1}
              \frac{x^2}{R^{\frac{3}{2}}} x^l \,\mathrm{d}x \label{eqn:S_from_ints}
\end{equation}
These three contributions can be handled by partial integration.
Some antiderivatives are needed for this:
\begin{align}
  \int \frac{      \mathrm{d}x}{R^\frac{3}{2}} =&  \frac{2(2a^{\pm} x + 2 d)}{16\left(ac-b^2\right) \sqrt{R}} = \frac{a^{\pm}x + d}{4\left(ac-b^2\right)\sqrt{R}} =: f_1(x)\label{eqn:GR_2_264_5} \\
  \int \frac{x   \,\mathrm{d}x}{R^\frac{3}{2}} =& -\frac{2(2a^{\mp} + 2 d x)}{16\left(ac-b^2\right) \sqrt{R}} = \frac{-(a^{\mp} + d x)}{4\left(ac-b^2\right)\sqrt{R}} =: f_2(x) \label{eqn:GR_2_264_6} \\
  \int \frac{x^2 \,\mathrm{d}x}{R^\frac{3}{2}} =& -\frac{\left(16\left(ac-b^2\right)-(2d)^2\right)x-2a^{\mp}2d}{a^{\pm}16\left(ac-b^2\right)\sqrt{R}} +\frac{1}{a^{\pm}} \int \frac{\mathrm{d}x}{\sqrt{R}} \nonumber \\
   ~                                           =& -\frac{\left(4\left(ac-b^2\right)-d^2\right)x-a^{\mp}d}{4 a^{\pm}\left(ac-b^2\right)\sqrt{R}} +\frac{1}{a^{\pm}} \int \frac{\mathrm{d}x}{\sqrt{R}} \label{eqn:GR_2_264_7}
\end{align}
where \eqn{GR_2_264_5}, \eqn{GR_2_264_6} and \eqn{GR_2_264_7} are from Ref.~\cite{gradshteyn_ryzhik}
with the identifiers 2.264.5, 2.264.6 and 2.264.7, respectively.
Additionally, identity 2.263.1 from Ref.~\cite{gradshteyn_ryzhik} is needed for the cases with $l>0$:
\begin{equation}
    \int \frac{x^2}{R^\frac{3}{2}} x^l \,\mathrm{d}x
  = \int \frac{x^{l+2}}{R^\frac{3}{2}} \,\mathrm{d}x
  =   \frac{x^{l+1}}{l a^{\pm} \sqrt{R}}
    - \frac{(2l+1)d}{l a^{\pm}}      \int \underbrace{\frac{x}{R^\frac{3}{2}}}_{f_2'(x)} x^l \,\mathrm{d}x
    - \frac{(l+1)a^{\mp}}{l a^{\pm}} \int \underbrace{\frac{1}{R^\frac{3}{2}}}_{f_1'(x)} x^l \,\mathrm{d}x \, . \label{eqn:int_x2_r32}
\end{equation}
Note that for $l=0$, \eqn{GR_2_264_5}, \eqn{GR_2_264_6} and \eqn{GR_2_264_7}
can be used directly to express $S^{\pm}_0$ in \eqn{S_from_ints}.
Some reoccuring terms are evaluated now to have them at hand later on:
\begin{align}
  a^{\pm} + a^{\mp} =& a \bcancel{\pm 2b} + c + a \bcancel{\mp 2b} + c = 2(a+c) \\
  a^{\pm} - a^{\mp} =& \bcancel{a} \pm 2b \bcancel{+ c} \bcancel{- a} \pm 2b \bcancel{- c} = \pm 4 b \\
  a^{\pm} + d       =& \bcancel{a} \pm 2b + c + c \bcancel{-a} = 2 (c \pm b) \\
  a^{\pm} - d       =& a \pm 2b \bcancel{+c} \bcancel{-c} + a = 2 (a \pm b) \\
  a^{\mp} + d       =& \bcancel{a} \mp 2b + c + c \bcancel{-a} = 2 (c \mp b) \\
  a^{\mp} - d       =& a \mp 2b \bcancel{+c} \bcancel{-c} + a = 2 (a \mp b) \\
  a^{\pm} a^{\mp}   =& (a \pm 2b + c)(a \mp 2b + c) \nonumber \\
        ~           =& a^2 \bcancel{\pm 2 a b} + a c \bcancel{\mp 2 a b} - 4 b^2 \bcancel{\mp 2 b c} + a c \bcancel{\mp 2 b c} + c^2 \nonumber \\
        ~           =& a^2 + 2 a c + c^2 - 4 b^2 = (a + c)^2 - 4b^2 \, .
\end{align}
The derivation in this text starts with the case of $l>0$ and the $l=0$ case is handled afterwards.
For the partial integration to come, we need the function $g(x)$:
\begin{align}
              g    (x) =&\,      x^l \\
  \Rightarrow {g'}(x) =&\, l \, x^{l-1} \, .
\end{align}
For $l>0$, we can insert \eqn{int_x2_r32} for the third integral with $x^{l+2}$ in \eqn{S_from_ints}:
\begin{align}
    S^{\pm}_l
  =&\, A^{\mp} \int\limits_{-1}^{+1}
              \frac{1}{R^{\frac{3}{2}}} x^l \,\mathrm{d}x
  + 2 D     \int\limits_{-1}^{+1}
              \frac{x}{R^{\frac{3}{2}}} x^l \,\mathrm{d}x \nonumber \\
  ~& + A^{\pm} \left\{
         \left[ \frac{x^{l+1}}{l a^{\pm} \sqrt{R}} \right]_{-1}^{+1}
       - \frac{(2l+1)d}{l a^{\pm}}      \int\limits_{-1}^{+1} \frac{x}{R^\frac{3}{2}} x^l \,\mathrm{d}x
       - \frac{(l+1)a^{\mp}}{l a^{\pm}} \int\limits_{-1}^{+1} \frac{1}{R^\frac{3}{2}} x^l \,\mathrm{d}x
       \right\} \nonumber \\
  =&\, A^{\pm} \left[ \frac{x^{l+1}}{l a^{\pm} \sqrt{R}} \right]_{-1}^{+1}
  + \left( A^{\mp} - A^{\pm} \frac{(l+1)a^{\mp}}{l a^{\pm}} \right) \int\limits_{-1}^{+1} \frac{1}{R^\frac{3}{2}} x^l \,\mathrm{d}x
  + \left( 2 D     - A^{\pm} \frac{(2l+1)d}{l a^{\pm}}      \right) \int\limits_{-1}^{+1} \frac{x}{R^\frac{3}{2}} x^l \,\mathrm{d}x \, . \nonumber
\end{align}
Intermezzo:
\begin{equation}
    \left[ \frac{x^{l+1}}{l a^{\pm} \sqrt{R}} \right]_{-1}^{+1}
  = \frac{1}{l a^{\pm}}\left( \frac{1}{2 \sqrt{c}} - \frac{(-1)^{l+1}}{2 \sqrt{a}} \right)
  = \frac{1}{l a^{\pm}}\left( \frac{1}{2 \sqrt{c}} + \frac{(-1)^l    }{2 \sqrt{a}} \right)
\end{equation}
Now continue with above expression for $S^{\pm}_l$:
\begin{align}
  S^{\pm}_l
  =& \phantom{+}~ \frac{A^{\pm}}{l a^{\pm}}\left( \frac{1}{2 \sqrt{c}} + \frac{(-1)^l    }{2 \sqrt{a}} \right) \nonumber \\
  ~&          +   \left( A^{\mp} - A^{\pm} \frac{(l+1)a^{\mp}}{l a^{\pm}} \right) \int\limits_{-1}^{+1} \frac{1}{R^\frac{3}{2}} x^l \,\mathrm{d}x \nonumber \\
  ~&          +   \left( 2 D     - A^{\pm} \frac{(2l+1)d}{l a^{\pm}}      \right) \int\limits_{-1}^{+1} \frac{x}{R^\frac{3}{2}} x^l \,\mathrm{d}x \, . \label{eqn:S_pm_l_full}
\end{align}
The first integral in \eqn{S_pm_l_full} can be solved using partial integration:
\begin{align}
     \int\limits_{-1}^{+1}
       \underbrace{              \frac{1}{R^{\frac{3}{2}}}  }_{{f_1'}(x)}
       \underbrace{x^l \vphantom{\frac{1}{R^{\frac{3}{2}}}} }_{g(x)} \,\mathrm{d}x
  =&   \Biggl[ \underbrace{              \frac{a^{\pm}x + d}{4\left(ac-b^2\right)\sqrt{R}}  }_{f_1(x)}
               \underbrace{x^l \vphantom{\frac{a^{\pm}x + d}{4\left(ac-b^2\right)\sqrt{R}}} }_{g(x)}
       \Biggr]_{-1}^{+1}
     - \int\limits_{-1}^{+1}
         \underbrace{                       \frac{a^{\pm}x + d}{4\left(ac-b^2\right)\sqrt{R}}  }_{f_1(x)}
         \underbrace{l \, x^{l-1} \vphantom{\frac{a^{\pm}x + d}{4\left(ac-b^2\right)\sqrt{R}}} }_{{g'}(x)} \,\mathrm{d}x \nonumber \\
  =& \frac{1}{4\left(ac-b^2\right)} \Biggl[
       \frac{a^{\pm}     + d}{2\sqrt{c}}
     - \frac{a^{\pm}(-1) + d}{2\sqrt{a}} (-1)^l
     - l a^{\pm} \underbrace{\int\limits_{-1}^{+1} \frac{x^l    }{\sqrt{R}}\,\mathrm{d}x}_{T^{\pm}_l}
     - l d       \underbrace{\int\limits_{-1}^{+1} \frac{x^{l-1}}{\sqrt{R}}\,\mathrm{d}x}_{T^{\pm}_{l-1}}
     \Biggr] \nonumber \\
  =& \frac{1}{4\left(ac-b^2\right)} \left[
                \frac{a^{\pm} + d}{2\sqrt{c}}
       + (-1)^l \frac{a^{\pm} - d}{2\sqrt{a}}
       - l a^{\pm} T^{\pm}_l - l d T^{\pm}_{l-1} \right] \nonumber \\
  =& \frac{1}{4\left(ac-b^2\right)} \left[
                \frac{\bcancel{2}(c \pm b)}{\bcancel{2}\sqrt{c}}
       + (-1)^l \frac{\bcancel{2}(a \pm b)}{\bcancel{2}\sqrt{a}}
       - l a^{\pm} T^{\pm}_l - l d T^{\pm}_{l-1} \right] \nonumber \\
  =& \frac{1}{4\left(ac-b^2\right)} \left[
                \frac{c \pm b}{\sqrt{c}}
       + (-1)^l \frac{a \pm b}{\sqrt{a}}
       - l a^{\pm} T^{\pm}_l - l d T^{\pm}_{l-1} \right] \, .
\end{align}
Next comes the second integral in \eqn{S_pm_l_full}:
\begin{align}
     \int\limits_{-1}^{+1}
       \underbrace{              \frac{x}{R^{\frac{3}{2}}}  }_{{f_2'}(x)}
       \underbrace{x^l \vphantom{\frac{x}{R^{\frac{3}{2}}}} }_{g(x)} \,\mathrm{d}x
  =&   \Biggl[ \underbrace{              \frac{-(a^{\mp} + d x)}{4\left(ac-b^2\right)\sqrt{R}} }_{f_2(x)}
               \underbrace{x^l \vphantom{\frac{-(a^{\mp} + d x)}{4\left(ac-b^2\right)\sqrt{R}}}}_{g(x)}
       \Biggr]_{-1}^{+1}
     - \int\limits_{-1}^{+1}
         \underbrace{                       \frac{-(a^{\mp} + d x)}{4\left(ac-b^2\right)\sqrt{R}} }_{f_2(x)}
         \underbrace{l \, x^{l-1} \vphantom{\frac{-(a^{\mp} + d x)}{4\left(ac-b^2\right)\sqrt{R}}}}_{{g'}(x)} \,\mathrm{d}x \nonumber \\
  =& \frac{1}{4\left(ac-b^2\right)} \Biggl[
       \frac{-(a^{\mp} + d     )}{2\sqrt{c}}
     - \frac{-(a^{\mp} + d (-1))}{2\sqrt{a}} (-1)^l
     + l a^{\mp} \underbrace{\int\limits_{-1}^{+1} \frac{x^{l-1}}{\sqrt{R}}\,\mathrm{d}x}_{T^{\pm}_{l-1}}
     + l d       \underbrace{\int\limits_{-1}^{+1} \frac{x^l    }{\sqrt{R}}\,\mathrm{d}x}_{T^{\pm}_l}
     \Biggr] \nonumber \\
  =& \frac{1}{4\left(ac-b^2\right)} \left[
                \frac{-(a^{\mp} + d)}{2\sqrt{c}}
       - (-1)^l \frac{-(a^{\mp} - d)}{2\sqrt{a}}
       + l a^{\mp} T^{\pm}_{l-1} + l d T^{\pm}_l \right] \nonumber \\
  =& \frac{1}{4\left(ac-b^2\right)} \left[
                \frac{-\bcancel{2}(c \mp b)}{\bcancel{2}\sqrt{c}}
       + (-1)^l \frac{ \bcancel{2}(a \mp b)}{\bcancel{2}\sqrt{a}}
       + l a^{\mp} T^{\pm}_{l-1} + l d T^{\pm}_l \right] \nonumber \\
  =& \frac{1}{4\left(ac-b^2\right)} \left[
       -        \frac{c \mp b}{\sqrt{c}}
       + (-1)^l \frac{a \mp b}{\sqrt{a}}
       + l a^{\mp} T^{\pm}_{l-1} + l d T^{\pm}_l \right] \, .
\end{align}
The two solutions above can now be inserted into \eqn{S_pm_l_full}:
\begin{align}
  S^{\pm}_l
  =& \phantom{+}~ \frac{A^{\pm}}{l a^{\pm}}
                  \left( \frac{1}{2 \sqrt{c}} + \frac{(-1)^l    }{2 \sqrt{a}} \right) \nonumber \\
  ~&          +   \frac{1}{4\left(ac-b^2\right)}
                  \left( A^{\mp} - A^{\pm} \frac{(l+1)a^{\mp}}{l a^{\pm}} \right)
                  \left[
                             \frac{c \pm b}{\sqrt{c}}
                    + (-1)^l \frac{a \pm b}{\sqrt{a}}
                    - l a^{\pm} T^{\pm}_l - l d T^{\pm}_{l-1} \right] \nonumber \\
  ~&          +   \frac{1}{4\left(ac-b^2\right)}
                  \left( 2 D     - A^{\pm} \frac{(2l+1)d}{l a^{\pm}}      \right)
                  \left[
                     -        \frac{c \mp b}{\sqrt{c}}
                     + (-1)^l \frac{a \mp b}{\sqrt{a}}
                     + l a^{\mp} T^{\pm}_{l-1} + l d T^{\pm}_l \right] \, .
\end{align}
In order to proceed, some re-ordering is needed:
\begin{align}
  (ac&-b^2) S^{\pm}_l \nonumber \\
  =& \phantom{+}~ \left[
                    - \frac{l a^{\pm}}{4} \left( A^{\mp} - A^{\pm} \frac{(l+1)a^{\mp}}{l a^{\pm}} \right)
                    + \frac{l d      }{4} \left( 2 D     - A^{\pm} \frac{(2l+1)d}{l a^{\pm}}      \right)
                  \right] T^{\pm}_l \nonumber \\
  ~&          +    \left[
                    - \frac{l d      }{4} \left( A^{\mp} - A^{\pm} \frac{(l+1)a^{\mp}}{l a^{\pm}} \right)
                    + \frac{l a^{\mp}}{4} \left( 2 D     - A^{\pm} \frac{(2l+1)d}{l a^{\pm}}      \right)
                  \right] T^{\pm}_{l-1} \nonumber \\
  ~&          +    \left[
                      \frac{(c \pm b)}{4} \left( A^{\mp} - A^{\pm} \frac{(l+1)a^{\mp}}{l a^{\pm}} \right)
                    - \frac{(c \mp b)}{4} \left( 2 D     - A^{\pm} \frac{(2l+1)d}{l a^{\pm}}      \right)
                    + \frac{ac-b^2}{2 l a^{\pm}} A^{\pm}
                  \right] \frac{1}{\sqrt{c}} \nonumber \\
  ~&          +    \left[
                      \frac{(a \pm b)}{4} \left( A^{\mp} - A^{\pm} \frac{(l+1)a^{\mp}}{l a^{\pm}} \right)
                    + \frac{(a \mp b)}{4} \left( 2 D     - A^{\pm} \frac{(2l+1)d}{l a^{\pm}}      \right)
                    + \frac{ac-b^2}{2 l a^{\pm}} A^{\pm}
                  \right] \frac{(-1)^l}{\sqrt{a}} \nonumber
\end{align}
and furthermore:
\begin{align}
  ~& a^{\pm} \left(ac-b^2\right) S^{\pm}_l \nonumber \\
  =& \left[
       - \frac{l a^{\pm}}{4} \left(   a^{\pm} A^{\mp} - A^{\pm} \frac{(l+1)a^{\mp}}{l} \right)
       + \frac{l d      }{4} \left( 2 a^{\pm} D       - A^{\pm} \frac{(2l+1)d}{l}      \right)
     \right] T^{\pm}_l \nonumber \\
  +& \left[
       - \frac{l d      }{4} \left(   a^{\pm} A^{\mp} - A^{\pm} \frac{(l+1)a^{\mp}}{l} \right)
       + \frac{l a^{\mp}}{4} \left( 2 a^{\pm} D       - A^{\pm} \frac{(2l+1)d}{l}      \right)
     \right] T^{\pm}_{l-1} \nonumber \\
  +& \left[
         \frac{(c \pm b)}{4} \left(   a^{\pm} A^{\mp} - A^{\pm} \frac{(l+1)a^{\mp}}{l} \right)
       - \frac{(c \mp b)}{4} \left( 2 a^{\pm} D       - A^{\pm} \frac{(2l+1)d}{l}      \right)
       + \frac{ac-b^2}{2 l} A^{\pm}
     \right] \frac{1}{\sqrt{c}} \nonumber \\
  +& \left[
         \frac{(a \pm b)}{4} \left(   a^{\pm} A^{\mp} - A^{\pm} \frac{(l+1)a^{\mp}}{l} \right)
       + \frac{(a \mp b)}{4} \left( 2 a^{\pm} D       - A^{\pm} \frac{(2l+1)d}{l}      \right)
       + \frac{ac-b^2}{2 l} A^{\pm}
     \right] \frac{(-1)^l}{\sqrt{a}} \nonumber \\
  =& \left[
       - \frac{l a^{\pm}}{4} \left(   a^{\pm} A^{\mp} - \left(1+\frac{1}{l}\right) a^{\mp} A^{\pm} \right)
       + \frac{l d      }{4} \left( 2 a^{\pm} D       - \left(2+\frac{1}{l}\right) d       A^{\pm} \right)
     \right] T^{\pm}_l \nonumber \\
  +&  \left[
       - \frac{l d      }{4} \left(   a^{\pm} A^{\mp} - \left(1+\frac{1}{l}\right) a^{\mp} A^{\pm} \right)
       + \frac{l a^{\mp}}{4} \left( 2 a^{\pm} D       - \left(2+\frac{1}{l}\right) d       A^{\pm} \right)
     \right] T^{\pm}_{l-1} \nonumber \\
  +&  \left[
         \frac{(c \pm b)}{4} \left(   a^{\pm} A^{\mp} - \left(1+\frac{1}{l}\right) a^{\mp} A^{\pm} \right)
       - \frac{(c \mp b)}{4} \left( 2 a^{\pm} D       - \left(2+\frac{1}{l}\right) d       A^{\pm} \right)
       + \frac{ac-b^2}{2 l} A^{\pm}
     \right] \frac{1}{\sqrt{c}} \nonumber \\
  +&  \left[
         \frac{(a \pm b)}{4} \left(   a^{\pm} A^{\mp} - \left(1+\frac{1}{l}\right) a^{\mp} A^{\pm} \right)
       + \frac{(a \mp b)}{4} \left( 2 a^{\pm} D       - \left(2+\frac{1}{l}\right) d       A^{\pm} \right)
       + \frac{ac-b^2}{2 l} A^{\pm}
     \right] \frac{(-1)^l}{\sqrt{a}} \, . \nonumber
\end{align}
The next operation to be performed on above equation is to sort the terms by their power of $l$:
\begin{align}
  a^{\pm} & \left(ac-b^2\right) S^{\pm}_l \nonumber \\
  =& \phantom{+}~ l \left[
                    - \frac{a^{\pm}}{4} \left(   a^{\pm} A^{\mp} -   a^{\mp} A^{\pm} \right)
                    + \frac{d      }{4} \left( 2 a^{\pm} D       - 2 d       A^{\pm} \right)
                  \right] T^{\pm}_l
                  +
                  \left(
                      \frac{a^{\pm}}{4} a^{\mp} A^{\pm}
                    - \frac{d      }{4} d       A^{\pm}
                  \right) T^{\pm}_l \nonumber \\
  ~&          + l \left[
                    - \frac{d      }{4} \left(   a^{\pm} A^{\mp} -   a^{\mp} A^{\pm} \right)
                    + \frac{a^{\mp}}{4} \left( 2 a^{\pm} D       - 2 d       A^{\pm} \right)
                  \right] T^{\pm}_{l-1}
                  +
                  \underbrace{\left(
                      \frac{d      }{4} a^{\mp} A^{\pm}
                    - \frac{a^{\mp}}{4} d       A^{\pm}
                  \right)}_{=0} T^{\pm}_{l-1} \nonumber \\
  ~&          +   \left[
                      \frac{c \pm b}{4} \left(   a^{\pm} A^{\mp} -   a^{\mp} A^{\pm} \right)
                    - \frac{c \mp b}{4} \left( 2 a^{\pm} D       - 2 d       A^{\pm} \right)
                  \right] \frac{1}{\sqrt{c}} \nonumber \\
  ~&          +   \frac{1}{l} \left(
                    - \frac{c \pm b}{4} a^{\mp} A^{\pm}
                    + \frac{c \mp b}{4} d       A^{\pm}
                    + \frac{ac-b^2}{2} A^{\pm}
                  \right) \frac{1}{\sqrt{c}} \nonumber \\
  ~&          +   \left[
                      \frac{a \pm b}{4} \left(   a^{\pm} A^{\mp} -   a^{\mp} A^{\pm} \right)
                    + \frac{a \mp b}{4} \left( 2 a^{\pm} D       - 2 d       A^{\pm} \right)
                  \right] \frac{(-1)^l}{\sqrt{a}} \nonumber \\
  ~&          +   \frac{1}{l} \left(
                    - \frac{a \pm b}{4} a^{\mp} A^{\pm}
                    - \frac{a \mp b}{4} d       A^{\pm}
                    + \frac{ac-b^2}{2} A^{\pm}
                  \right) \frac{(-1)^l}{\sqrt{a}} \nonumber \\
  =& \phantom{+}~ \frac{l}{4} \left[
                    - a^{\pm} \left(   a^{\pm} A^{\mp} -   a^{\mp} A^{\pm} \right)
                    + d       \left( 2 a^{\pm} D       - 2 d       A^{\pm} \right)
                  \right] T^{\pm}_l
                  + \frac{1}{4}
                  \left(
                      a^{\pm} a^{\mp}
                    - d^2
                  \right)  A^{\pm} T^{\pm}_l \nonumber \\
  ~&          +   \frac{l}{4} \left[
                    - d       \left(   a^{\pm} A^{\mp} -   a^{\mp} A^{\pm} \right)
                    + a^{\mp} \left( 2 a^{\pm} D       - 2 d       A^{\pm} \right)
                  \right] T^{\pm}_{l-1} \nonumber \\
  ~&          +   \frac{1}{4} \left[
                      (c \pm b) \left(   a^{\pm} A^{\mp} -   a^{\mp} A^{\pm} \right)
                    - (c \mp b) \left( 2 a^{\pm} D       - 2 d       A^{\pm} \right)
                  \right] \frac{1}{\sqrt{c}} \nonumber \\
  ~&          +   \frac{1}{4 l} \left(
                    - (c \pm b) a^{\mp}
                    + (c \mp b) d
                    + 2 \left(ac-b^2\right)
                  \right) A^{\pm} \frac{1}{\sqrt{c}} \nonumber \\
  ~&          +   \frac{1}{4} \left[
                      (a \pm b) \left(   a^{\pm} A^{\mp} -   a^{\mp} A^{\pm} \right)
                    + (a \mp b) \left( 2 a^{\pm} D       - 2 d       A^{\pm} \right)
                  \right] \frac{(-1)^l}{\sqrt{a}} \nonumber \\
  ~&          +   \frac{1}{4 l} \left(
                    - (a \pm b) a^{\mp}
                    - (a \mp b) d
                    + 2\left(ac-b^2\right)
                  \right) A^{\pm} \frac{(-1)^l}{\sqrt{a}} \, . \label{eqn:S_pm_almost_there}
\end{align}
Now have a closer look on the $1/l$ terms for $1/\sqrt{c}$ and $1/\sqrt{a}$:
\begin{align}
  -(c \pm b) a^{\mp} + (c \mp b) d + 2 \left(ac-b^2\right)
  =& -c \underbrace{(a^{\mp} - d)}_{=2(a\mp b)} \mp b \underbrace{(a^{\mp} + d)}_{=2(c\mp b)} + 2ac - 2b^2 \nonumber \\
  =& 2 \left(-ac \pm bc \mp bc +b^2 +ac -b^2 \right) = 0 \\
  -(a \pm b) a^{\mp} - (a \mp b) d + 2\left(ac-b^2\right)
  =& -a\underbrace{(a^{\mp} + d)}_{=2(c\mp b)} \mp b \underbrace{(a^{\mp} - d)}_{=2(a\mp b)} + 2ac-2b^2 \nonumber \\
  =& 2 \left(-ac \pm ab \mp ab +b^2 +ac -b^2 \right) = 0
\end{align}
and the $l$-independent term for $T^{\pm}_l$:
\begin{align}
  a^{\pm} a^{\mp} - d^2
  =& (a+c)^2 - 4b^2 - (c-a)^2 \nonumber \\
  =& \bcancel{a^2} + 2ac \bcancel{+c^2} - 4b^2 \bcancel{-c^2} +2ac \bcancel{-a^2} = 4\left(ac-b^2\right) \, .
\end{align}
Insert above findings into \eqn{S_pm_almost_there}:
\begin{align}
  a^{\pm} & \left(ac-b^2\right) S^{\pm}_l \nonumber \\
  =& \phantom{+}~ \frac{l}{4} \left[
                    - a^{\pm} \left(   a^{\pm} A^{\mp} -   a^{\mp} A^{\pm} \right)
                    + d       \left( 2 a^{\pm} D       - 2 d       A^{\pm} \right)
                  \right] T^{\pm}_l
              +   \left(ac-b^2\right)  A^{\pm} T^{\pm}_l \nonumber \\
  ~&          +   \frac{l}{4} \left[
                    - d       \left(   a^{\pm} A^{\mp} -   a^{\mp} A^{\pm} \right)
                    + a^{\mp} \left( 2 a^{\pm} D       - 2 d       A^{\pm} \right)
                  \right] T^{\pm}_{l-1} \nonumber \\
  ~&          +   \frac{1}{4} \left[
                      (c \pm b) \left(   a^{\pm} A^{\mp} -   a^{\mp} A^{\pm} \right)
                    - (c \mp b) \left( 2 a^{\pm} D       - 2 d       A^{\pm} \right)
                  \right] \frac{1}{\sqrt{c}} \nonumber \\
  ~&          +   \frac{1}{4} \left[
                      (a \pm b) \left(   a^{\pm} A^{\mp} -   a^{\mp} A^{\pm} \right)
                    + (a \mp b) \left( 2 a^{\pm} D       - 2 d       A^{\pm} \right)
                  \right] \frac{(-1)^l}{\sqrt{a}} \, . \label{eqn:S_pm_l_final}
\end{align}
This can be simplified by introducing two constants $k_1$ and $k_2$ with:
\begin{align}
  k_1 =&\,          -       \frac{ a^{\pm} A^{\mp} - a^{\mp} A^{\pm} }{4} \label{eqn:k1_original} \\
  k_2 =&\, \phantom{-}~\,   \frac{ a^{\pm} D       - d       A^{\pm} }{2} \label{eqn:k2_original}
\end{align}
which leads to:
\begin{align}
  a^{\pm} \left(ac-b^2\right) S^{\pm}_l
  =& \left[ l \left( a^{\pm}   k_1 +     d     k_2 \right)
                +                        \left(ac-b^2\right)  A^{\pm} \right] T^{\pm}_l
                +   l \left(     d     k_1 +  a^{\mp}  k_2 \right) T^{\pm}_{l-1} \nonumber \\
  ~&          -        \frac{ (c \pm b) k_1 + (c \mp b) k_2 }{\sqrt{c}}
              - (-1)^l \frac{ (a \pm b) k_1 - (a \mp b) k_2 }{\sqrt{a}} \, . \label{eqn:S_pm_l_Merkel}
\end{align}
Some simplifications can be made for $k_1$ and $k_2$:
\begin{align}
  k_1 =& -\frac{1}{4} \left( a^{\pm} (A \mp 2 B + C) - a^{\mp} (A \pm 2B + C) \right) \nonumber \\
  ~   =& -\frac{1}{4} \Bigl( (A+C) \underbrace{(a^{\pm} - a^{\mp})}_{=\pm 4b} \mp 2B \underbrace{(a^{\pm} + a^{\mp})}_{=2(a+c)} \Bigr) \nonumber \\
  ~   =& \pm (a+c) B \mp b(A+C) \label{eqn:k1_result} \\
  k_2 =& \frac{1}{2} \left( a^{\pm} (C-A) - d (A \pm 2B + C) \right) \nonumber \\
  ~   =& \frac{1}{2} \Bigl(-A \underbrace{(a^{\pm}+d)}_{=2(c \pm b)} \mp 2B d + C\underbrace{(a^{\pm}-d)}_{2(a \pm b)}) \Bigr) \nonumber \\
  ~   =& C (a \pm b) \mp B (c-a) - A (c \pm b) \, .
\end{align}
This concludes the derivation of the $S^{\pm}_l$.
It remains to consider the $l=0$ term $S^{\pm}_0$.
In \eqn{S_pm_l_final}, the quantity $T^{\pm}_{l-1}$ is not defined for $l=0$.
However, $T^{\pm}_{l-1}$ is multiplied by $l$ and if one puts aside concerns about undefined quantites, whatever value $T^{\pm}_{-1}$ might have,
it does not enter into $S^{\pm}_0$ anyway.
It will be shown below that $S^{\pm}_0$ is equal to the result of inserting $l=0$ into $S^{\pm}_l$.
We start again with \eqn{S_from_ints} and insert $l=0$, leading to:
\begin{equation}
  S^{\pm}_0
  = A^{\mp} \int\limits_{-1}^{+1}
              \frac{1}{R^{\frac{3}{2}}} \,\mathrm{d}x
  + 2 D     \int\limits_{-1}^{+1}
              \frac{x}{R^{\frac{3}{2}}} \,\mathrm{d}x
  + A^{\pm} \int\limits_{-1}^{+1}
              \frac{x^2}{R^{\frac{3}{2}}} \,\mathrm{d}x \, .
\end{equation}
\eqn{GR_2_264_5}, \eqn{GR_2_264_6} and \eqn{GR_2_264_7}
can be inserted directly and no partial integration is needed for this:
\begin{align}
   S^{\pm}_0 % \nonumber \\
  =& \phantom{+}~ A^{\mp} \left[ \frac{  a^{\pm}x + d   }{4\left(ac-b^2\right)\sqrt{R}} \right]_{-1}^{+1}
              +     2 D       \left[ \frac{-(a^{\mp}  + d x)}{4\left(ac-b^2\right)\sqrt{R}} \right]_{-1}^{+1} \nonumber \\
  ~& +                A^{\pm} \left[ -\frac{\left(4\left(ac-b^2\right)-d^2\right)x-a^{\mp}d}{4 a^{\pm}\left(ac-b^2\right)\sqrt{R}} \right]_{-1}^{+1}
              + \frac{A^{\pm}}{a^{\pm}} \underbrace{\int\limits_{-1}^{+1} \frac{\mathrm{d}x}{\sqrt{R}}}_{=T^{\pm}_0} \nonumber \\
  =& \phantom{+}~\,   \frac{A^{\mp}}{4\left(ac-b^2\right)} \left[
                      \frac{a^{\pm}     + d}{2\sqrt{c}}
                    - \frac{a^{\pm}(-1) + d}{2\sqrt{a}} \right] %\nonumber \\
  +                 \frac{2 D}{4\left(ac-b^2\right)} \left[
                      \frac{-(a^{\mp} + d     )}{2\sqrt{c}}
                    - \frac{-(a^{\mp} + d (-1))}{2\sqrt{a}} \right] \nonumber \\
  ~&+           \frac{A^{\pm}}{4 a^{\pm}\left(ac-b^2\right)} \left[ - \frac{      4\left(ac-b^2\right)-d^2            -a^{\mp}d}{2\sqrt{c}}
                                                         + \frac{\left(4\left(ac-b^2\right)-d^2\right)(-1) -a^{\mp}d}{2\sqrt{a}} \right] %\nonumber \\
    +         \frac{A^{\pm}}{a^{\pm}} T^{\pm}_0 \nonumber \\
  =& \phantom{+}~\,  \frac{A^{\mp}}{4\left(ac-b^2\right)} \left[
       \frac{a^{\pm} + d}{2\sqrt{c}}
     + \frac{a^{\pm} - d}{2\sqrt{a}} \right]
  +    \frac{2 D}{4\left(ac-b^2\right)} \left[
     - \frac{a^{\mp} + d}{2\sqrt{c}}
     + \frac{a^{\mp} - d}{2\sqrt{a}} \right] \nonumber \\
  ~& +   \frac{A^{\pm}}{4 a^{\pm}\left(ac-b^2\right)} \left[   \frac{ d^2 + a^{\mp}d -4\left(ac-b^2\right) }{2\sqrt{c}}
                                                           + \frac{ d^2 - a^{\mp}d -4\left(ac-b^2\right) }{2\sqrt{a}} \right]
     + \frac{A^{\pm}}{a^{\pm}} T^{\pm}_0 \nonumber \\
  =& \frac{1}{4 a^{\pm}\left(ac-b^2\right)} \Biggl\{ \nonumber \\
  ~& \phantom{+}\,
     \left[
           a^{\pm} \left( a^{\pm} + d               \right) A^{\mp}
       - 2 a^{\pm} \left( a^{\mp} + d               \right) D
       +           \left( d^2 + a^{\mp}d -4\left(ac-b^2\right) \right) A^{\pm}
     \right] \frac{1}{2\sqrt{c}} \nonumber \\
  ~& +
     \left[
           a^{\pm} \left( a^{\pm} - d               \right) A^{\mp}
       + 2 a^{\pm} \left( a^{\mp} - d               \right) D
       +           \left( d^2 - a^{\mp}d -4\left(ac-b^2\right) \right) A^{\pm}
     \right] \frac{1}{2\sqrt{a}}
     \Biggr\} % \nonumber \\
   + \frac{A^{\pm}}{a^{\pm}} T^{\pm}_0 \, . \label{eqn:S0_almost}
\end{align}
The coefficient for $1/\sqrt{c}$ can be simplified:
\begin{align}
 ~& \frac{1}{2} \left[
        a^{\pm} \left( a^{\pm} + d               \right) A^{\mp}
    - 2 a^{\pm} \left( a^{\mp} + d               \right) D
    +           \left( d^2 + a^{\mp}d -4\left(ac-b^2\right) \right) A^{\pm} \right] \nonumber \\
 =& \frac{1}{2} \Bigl[
                  \left( a^{\pm} + d \right) a^{\pm} A^{\mp}
    \underbrace{- \left( a^{\pm} + d \right) a^{\mp} A^{\pm}
                + \left( a^{\pm} + d \right) a^{\mp} A^{\pm}}_{=0} \nonumber \\
 ~& \phantom{\frac{1}{2} \Bigl[}
    - 2 a^{\pm} \left( a^{\mp} + d               \right) D
    +           \left( d^2 + a^{\mp}d -4\left(ac-b^2\right) \right) A^{\pm} \Bigr] \nonumber \\
 =& \frac{1}{2} \Bigl[
                \underbrace{\left( a^{\pm} + d \right)}_{=2(c\pm b)} \left( a^{\pm} A^{\mp} - a^{\mp} A^{\pm} \right) \nonumber \\
 ~& \phantom{\frac{1}{2} \Bigl[}
    +           \left( a^{\pm} + d \right) a^{\mp} A^{\pm}
    - 2 a^{\pm} \left( a^{\mp} + d               \right) D
    +           \Bigl( \underbrace{d^2 + a^{\mp}d}_{=d(d + a^{\mp})} -4\left(ac-b^2\right) \Bigr) A^{\pm} \Bigr] \nonumber \\
 =& (c \pm b) \left( a^{\pm} A^{\mp} - a^{\mp} A^{\pm} \right) \nonumber \\
 ~& + \frac{1}{2} \Bigl[
        - \underbrace{\left( a^{\mp} + d \right)}_{=2(c \mp b)}
            \left( 2 a^{\pm} D - 2 d A^{\pm} \right)
        \underbrace{- \left( a^{\mp} + d \right) d       A^{\pm}
                    + \left( a^{\pm} + d \right) a^{\mp} A^{\pm}}_{=\left(\bcancel{-a^{\mp}d} - d^2 + a^{\pm}a^{\mp} \bcancel{+d a^{\mp}}\right)A^{\pm}}
        - 4\left(ac-b^2\right) A^{\pm} \Bigr] \nonumber \\
 =&   (c \pm b) \left(   a^{\pm} A^{\mp} -   a^{\mp} A^{\pm} \right)
    - (c \mp b) \left( 2 a^{\pm} D       - 2 d       A^{\pm} \right)
    + \frac{1}{2} \left( a^{\pm}a^{\mp} - d^2 - 4\left(ac-b^2\right) \right) A^{\pm} \nonumber
\end{align}
and with
\begin{align}
     a^{\pm}a^{\mp} - d^2 - 4\left(ac-b^2\right)
  =& (a+c)^2 - 4b^2 - (c-a)^2 - 4ac+ 4b^2  \nonumber \\
  =& a^2 +2ac + c^2 - 4b^2 - c^2 +2ac - a^2 - 4ac + 4b^2  = 0 \label{eqn:apm_amp_d2_zero}
\end{align}
we arrive at:
\begin{align}
  \frac{1}{2} \Bigl[ &
        a^{\pm} \left( a^{\pm} + d               \right) A^{\mp}
    - 2 a^{\pm} \left( a^{\mp} + d               \right) D
    +           \left( d^2 + a^{\mp}d -4\left(ac-b^2\right) \right) A^{\pm} \Bigr] \nonumber \\
 =&   (c \pm b) \left(   a^{\pm} A^{\mp} -   a^{\mp} A^{\pm} \right)
    - (c \mp b) \left( 2 a^{\pm} D       - 2 d       A^{\pm} \right) \, . \label{eqn:factor_S0_sqrt_c}
\end{align}
The coefficient for $1/\sqrt{a}$ can be simplified as well:
\begin{align}
 ~& \frac{1}{2} \left[
          a^{\pm} \left( a^{\pm} - d               \right) A^{\mp}
      + 2 a^{\pm} \left( a^{\mp} - d               \right) D
      +           \left( d^2 - a^{\mp}d -4\left(ac-b^2\right) \right) A^{\pm} \right] \nonumber \\
 =& \frac{1}{2} \Bigl[
                  \left( a^{\pm} - d \right) a^{\pm} A^{\mp}
    \underbrace{- \left( a^{\pm} - d \right) a^{\mp} A^{\pm}
                + \left( a^{\pm} - d \right) a^{\mp} A^{\pm}}_{=0} \nonumber \\
 ~& \phantom{\frac{1}{2} \Bigl[}
    + 2 a^{\pm} \left( a^{\mp} - d               \right) D
    +           \left( d^2 - a^{\mp}d -4\left(ac-b^2\right) \right) A^{\pm} \Bigr] \nonumber \\
 =& \frac{1}{2} \Bigl[
    \underbrace{\left( a^{\pm} - d \right)}_{=2(a\pm b)} \left( a^{\pm} A^{\mp} - a^{\mp} A^{\pm} \right)
                + \left( a^{\mp} - d \right) 2 a^{\pm} D
    \underbrace{- \left( a^{\mp} - d \right) 2 d       A^{\pm}
                + \left( a^{\mp} - d \right) 2 d       A^{\pm}}_{=0} \nonumber \\
 ~& \phantom{\frac{1}{2} \Bigl[}
    + \left( a^{\pm} - d \right) a^{\mp} A^{\pm}
    + \Bigl( \underbrace{d^2 - a^{\mp}d}_{=-d(a^{\mp}-d)} -4\left(ac-b^2\right) \Bigr) A^{\pm} \Bigr] \nonumber \\
 =& (a\pm b) \left( a^{\pm} A^{\mp} - a^{\mp} A^{\pm} \right)
    + \frac{1}{2} \Bigl[
    \phantom{+}\,
    \underbrace{\left( a^{\mp} - d \right)}_{=2(a\mp b)} \left( 2 a^{\pm} D - 2 d A^{\pm} \right) \nonumber \\
 ~& \phantom{
      (a\pm b) \left( a^{\pm} A^{\mp} - a^{\mp} A^{\pm} \right)
      + \frac{1}{2} \Bigl[
    }
    + \left( a^{\mp} - d \right) d A^{\pm}
    + \left( a^{\pm} - d \right) a^{\mp} A^{\pm}
    -4\left(ac-b^2\right) A^{\pm} \Bigr] \nonumber \\
 =&   (a\pm b) \left( a^{\pm} A^{\mp} - a^{\mp} A^{\pm} \right)
    + (a\mp b) \left( 2 a^{\pm} D - 2 d A^{\pm} \right) \nonumber \\
 ~& + \frac{1}{2} \Bigl[
        \underbrace{
            \left( a^{\mp} - d \right) d
        +   \left( a^{\pm} - d \right) a^{\mp}
        }_{=\bcancel{a^{\mp} d} - d^2 + a^{\pm}a^{\mp} \bcancel{- d a^{\mp}} }
        - 4 \left(ac-b^2\right) \Bigr]  A^{\pm}
\end{align}
and with \eqn{apm_amp_d2_zero} we arrive at:
\begin{align}
 \frac{1}{2} \Bigl[ &
          a^{\pm} \left( a^{\pm} - d               \right) A^{\mp}
      + 2 a^{\pm} \left( a^{\mp} - d               \right) D
      +           \left( d^2 - a^{\mp}d -4\left(ac-b^2\right) \right) A^{\pm} \Bigr] \nonumber \\
 =&   (a\pm b) \left( a^{\pm} A^{\mp} - a^{\mp} A^{\pm} \right)
    + (a\mp b) \left( 2 a^{\pm} D - 2 d A^{\pm} \right) \, . \label{eqn:factor_S0_sqrt_a}
\end{align}
\eqn{factor_S0_sqrt_c} and \eqn{factor_S0_sqrt_a} are inserted into \eqn{S0_almost} to arrive at:
\begin{align}
  S^{\pm}_0
  = \frac{1}{4 a^{\pm}\left(ac-b^2\right)} \Biggl\{
       & \phantom{+}\,
       \left[
           (c \pm b) \left(   a^{\pm} A^{\mp} -   a^{\mp} A^{\pm} \right)
         - (c \mp b) \left( 2 a^{\pm} D       - 2 d       A^{\pm} \right)
       \right] \frac{1}{\sqrt{c}} \nonumber \\
  ~&   +
       \left[
           (a\pm b) \left( a^{\pm} A^{\mp} - a^{\mp} A^{\pm} \right)
         + (a\mp b) \left( 2 a^{\pm} D - 2 d A^{\pm} \right)
       \right] \frac{1}{\sqrt{a}}
     \Biggr\}
     + \frac{A^{\pm}}{a^{\pm}} T^{\pm}_0 \nonumber \\
\Leftrightarrow
  a^{\pm} \left(ac-b^2\right) S^{\pm}_0
  =& \phantom{+}\,
       \left(ac-b^2\right) A^{\pm} T^{\pm}_0 \nonumber \\
  ~& \phantom{+}~
       \frac{1}{4}
       \left[
           (a\pm b) \left( a^{\pm} A^{\mp} - a^{\mp} A^{\pm} \right)
         + (a\mp b) \left( 2 a^{\pm} D - 2 d A^{\pm} \right)
       \right] \frac{1}{\sqrt{a}} \nonumber \\
  ~& + \frac{1}{4}
       \left[
           (c \pm b) \left(   a^{\pm} A^{\mp} -   a^{\mp} A^{\pm} \right)
         - (c \mp b) \left( 2 a^{\pm} D       - 2 d       A^{\pm} \right)
       \right] \frac{1}{\sqrt{c}} \, .
\end{align}
This is equivalent to \eqn{S_pm_l_final} for $l=0$ when ignoring the term $T^{\pm}_{l-1}$,
which is anyway multiplied by $l=0$ and does not enter the value of $S^{\pm}_0$.

% \FloatBarrier
% \subsection{Relation to the 1986 NESTOR article}
% \eqn{t_pm_0} is the result for $T^{\pm}_0$ Merkel published in his article on NESTOR.
% \eqn{S_pm_l_Merkel} is the formulation of the $S^{\pm}_l$ Merkel used in his article.

% Merkel published a later re-write of the original NESTOR article in Ref.~\cite{merkel_1988}.
% Some typos were present in the original NESTOR article and these have been corrected in the re-write.
% The following list summarizes the typos in the original NESTOR article and references the corresponding equations in this text.
% \begin{itemize}
%   \item \eqn{t_pm_1} features a square root in the first factor, which is not present in the actual solution.
%   \item ...
% \end{itemize}

\FloatBarrier
\section{Numerical Implementation}

\FloatBarrier
\subsection{Coefficients involving Factorials}
The $I_{m,n}$ have to be re-evaluated for every change of the surface geometry in the free-boundary problem,
since the $T^{\pm}_l$ depend on the geometry of the surface via $a$, $b$ and $c$.
In order to save computational workload, the coefficients of the $T^{\pm}_l$ can be cached.
The coefficients are re-ordered to yield a linear index $l$ for $T^{\pm}_l$ with $l=|m-n|+2j$,
where $j$ is the summation index in \eqn{c_pm_mn}.
Then, $j=(l-|m-n|)/2$ and the limits of the summation
are (for $j=0$)             $l_\textrm{min} = |m-n|$
and (for $j=(m+n-|m-n|)/2$) $l_\textrm{max} = m+n$.
The summation variable $l$ has to be incremented in steps of $2$.
\eqn{c_pm_mn} then can be re-formulated as follows with the coefficients $\mathtt{cmn}(l,m,n)$:
\begin{equation}
    c^{\pm}_{m,n}
  = \frac{1}{2}
    \sum\limits_{\substack{l=|m-n|\\l+=2}}^{m+n}
     \underbrace{
      (-1)^{\frac{l-m+n}{2}}
      \frac{\left(\frac{m+n+l}{2}\right)!}
           {\left(\frac{m+n-l}{2}\right)!\,
            \left(\frac{l+|m-n|}{2}\right)!\,
            \left(\frac{l-|m-n|}{2}\right)!}
     }_{=: \mathtt{cmn}(l,m,n)}
      T^{\pm}_l \, . \label{eqn:c_pm_mn_l}
\end{equation}
If either both $m$ and $n$ are even or both $m$ and $n$ are odd, $|m-n|$ and $m+n$ are even
and also $l$ is even and so the factorials are evaluated only for integer arguments.
If one of $m$ or $n$ is even and the other one is odd, both $|m-n|$ and $m+n$ are odd
and $l$ is odd; the terms similar to, e.g., $m+n+l$ occuring in \eqn{c_pm_mn_l}
are therefore even again and so the factorials are also in this case evaluated only for integer arguments.
Some shorthand notation is introduced:
\begin{align}
 j_{m,n} &\equiv m+n \nonumber \\
 i_{m,n} &\equiv m-n \nonumber \\
 k_{m,n} &\equiv |i_{m,n}| = |m-n| \nonumber \\
 s_{m,n} &\equiv (j_{m,n}+k_{m,n})/2 = \frac{m+n+|m-n|}{2} \nonumber \, .
\end{align}
Note that since $j_{m,n}$ and $k_{m,n}$ are either both even or both odd,
their sum is always even and $s_{m,n}$ is therefore always an integer.
The factorials are split up as follows:
\begin{equation}
   \mathtt{cmn}(l,m,n)
 = (-1)^\frac{l-i_{m,n}}{2}
   \underbrace{ \frac{\left( \frac{j_{m,n}+l}{2} \right)!}
                     {\left( \frac{j_{m,n}-l}{2} \right)!} }_{=:f_1^{(l)}}
   \cdot
   \frac{1}{
     \underbrace{ \left( \frac{l+k_{m,n}}{2} \right)! }_{=:f_2^{(l)}} \,
     \underbrace{ \left( \frac{l-k_{m,n}}{2} \right)! }_{=:f_3^{(l)}} }
\end{equation}
where the superscripts of the $f$-factors denote the corresponding value of $l$.
Note that with summation index $j$ from \eqn{c_pm_mn} it holds:
\begin{align}
 f_1^{(j)} &= \frac{\left(s_{m,n}         + j \right)!}
                   {\left(s_{m,n}-k_{m,n} - j \right)!} \\
 f_2^{(j)} &= (k_{m,n}+j)! \\
 f_3^{(j)} &= j! \, .
\end{align}
The $\mathtt{cmn}(l,m,n)$ are computed iteratively.
The first term in the sum of \eqn{c_pm_mn_l} occurs at $j=0$ corresponding to $l=|m-n|=k_{m,n}$.
The value of $f_1$ is there:
\begin{equation}
   f_1^{(l=k_{m,n})} = f_1^{(j=0)}
 = \frac{      s_{m,n}                !}
        {\left(s_{m,n}-k_{m,n} \right)!} \, .
\end{equation}
This can be expressed explicitly by examining the term $s_{m,n}!$ in more detail:
\begin{align}
 s_{m,n}! &= s_{m,n} \cdot (s_{m,n}-1) \cdot ... \cdot (s_{m,n}-(k_{m,n}-1))
  \cdot \underbrace{(s_{m,n}-k_{m,n}) \cdot ... \cdot 1}_{=(s_{m,n}-k_{m,n})!} \nonumber \\
\Rightarrow
 \frac{s_{m,n}!}{(s_{m,n}-k_{m,n})!} &=
   \underbrace{s_{m,n} \cdot (s_{m,n}-1) \cdot ... \cdot (s_{m,n}-(k_{m,n}-1))}_{k_{m,n} \textrm{ terms}} \nonumber \\
\Rightarrow
 f_1^{(l=k_{m,n})} = f_1^{(j=0)} &= \prod_{i=1}^{k_{m,n}} (s_{m,n} - (i-1)) \, . \label{eqn:f_1_start}
\end{align}
Furthermore, for $j=0$, it holds:
\begin{align}
 f_2^{(j=0)} &= k_{m,n}! = \prod_{i=1}^{k_{m,n}} i \label{eqn:f_2_start} \\
 f_3^{(j=0)} &= 0! = 1 \, . \label{eqn:f_3_start}
\end{align}
Note that the same product ranges occur in \eqn{f_1_start} and \eqn{f_2_start}.
Using these values of $f_1^{(j=0)}$, $f_2^{(j=0)}$ and $f_3^{(j=0)}$ for $l=k_{m,n}$
allows to compute the coefficient $\mathtt{cmn}(l,m,n)$ as follows:
\begin{equation}
 \mathtt{cmn}(l,m,n) = \frac{f_1^{(l)}}{f_2^{(l)} f_3^{(l)}} (-1)^{\frac{l-i_{m,n}}{2}} \, .
\end{equation}
The next values of $f_1^{(l)}$, $f_2^{(l)}$ and $f_3^{(l)}$ for $l \rightarrow l+2$ can now be computed
iteratively based on the available values for $l=k_{m,n}$.
First look at the factorials that occur in $f_1^{(l+2)}$:
\begin{align}
 \left( \frac{j_{m,n}+l+2}{2} \right)! &= \left( \frac{j_{m,n}+l+2}{2} \right) \cdot \left( \frac{j_{m,n}+l  }{2} \right)! \nonumber \\
 \left( \frac{j_{m,n}-l  }{2} \right)! &= \left( \frac{j_{m,n}-l  }{2} \right) \cdot \left( \frac{j_{m,n}-l-2}{2} \right)! \nonumber
\end{align}
and thus
\begin{align}
 f_1^{(l+2)} &= \frac{\left( \frac{j_{m,n}+(l+2)}{2} \right)!}
                     {\left( \frac{j_{m,n}-(l+2)}{2} \right)!} \nonumber \\
    ~        &= \left(\frac{j_{m,n}+l+2}{2} \right) \cdot \left( \frac{j_{m,n}-l  }{2} \right) \cdot
                \frac{\left( \frac{j_{m,n}+l}{2} \right)!}
                     {\left( \frac{j_{m,n}-l}{2} \right)!} \nonumber \\
    ~        &= f_1^{(l)} \cdot \frac{1}{4} \cdot (j_{m,n}+l+2) \cdot (j_{m,n}-l) \, .
\end{align}
The other two factorials $f_2^{(l+2)}$ and $f_3^{(l+2)}$ are simpler to compute:
\begin{align}
 f_2^{(l+2)} &= f_2^{(l)} \cdot (l+k_{m,n})/2 \\
 f_3^{(l+2)} &= f_3^{(l)} \cdot (l-k_{m,n})/2 \, .
\end{align}
Note that the factors $f_1^{(l)}$, $f_2^{(l)}$ and $f_3^{(l)}$ remain integers throughout the computation of all $\mathtt{cmn}(l,m,n)$
and only multiplication, shifting (division by powers of 2) and summation of integers is needed to compute them.
With above tools at hand, Algorithm~(\ref{alg:cmn}) can be constructed to efficiently compute the coefficients $\mathtt{cmn}(l,m,n)$
for given $m$ and $n$.
This algorithm is used to fill a three-dimensional matrix $\mathtt{cmn}(l,m,n)$ for all mode number combinations
$0 \leq m \leq m_\mathrm{f}$, $0 \leq n \leq n_\mathrm{f}$ and $0 \leq l \leq m_\mathrm{f}+n_\mathrm{f}$.
Note that for any given $(m,n)$, only every second entry along the $l$ dimension of the matrix $\mathtt{cmn}(l,m,n)$
between $l_\mathrm{min}=|m-n|$ and $l_\mathrm{max}=m+n$ is assigned a finite value. The remainder of the matrix remains initialized to zero.
\begin{algorithm}
\caption{Calculate $\mathtt{cmn}(l,m,n)$}
\begin{algorithmic}
\Require $m \geq 0, n \geq 0$
\State $j_{m,n} \gets m + n$
\State $i_{m,n} \gets m - n$
\State $k_{m,n} \gets |i_{m,n}|$
\State $s_{m,n} \gets (j_{m,n} + k_{m,n})/2$
\State $f_1 \gets 1$
\State $f_2 \gets 1$
\State $f_3 \gets 1$
\State $\mathtt{cmn}(:,m,n) \gets 0$
\For{$(i \gets 1 ; i \leq k_{m,n} ; i \gets i+1)$}
  \State $f_1 \gets f_1 \cdot (s_{m,n} - i + 1)$
  \State $f_2 \gets f_2 \cdot i$
\EndFor
\For{$(l \gets k_{m,n} ; l \leq j_{m,n} ; l \gets l+2)$}
  \State $\mathtt{cmn}(l,m,n) \gets f_1/(f_2 \cdot f_3) \cdot (-1)^{(l-i_{m,n})/2}$
  \State $f_1 \gets f_1 \cdot (l+2+j_{m,n}) \cdot (j_{m,n}-l) / 4$
  \State $f_2 \gets f_2 \cdot (l+2+k_{m,n})/2$
  \State $f_3 \gets f_3 \cdot (l+2-k_{m,n})/2$
\EndFor
\end{algorithmic}
\label{alg:cmn}
\end{algorithm}
\FloatBarrier
Now that the $\mathtt{cmn}(l,m,n)$ are available, the coefficients $c^{\pm}_{m,n}$ can be evaluated as follows:
\begin{equation}
   c^{\pm}_{m,n}
  = \frac{1}{2}
    \sum\limits_{\substack{l=|m-n|\\l+=2}}^{m+n}
     \mathtt{cmn}(l,m,n)\, T^{\pm}_l \, . \label{eqn:c_pm_mn_short}
\end{equation}
These coefficients are not used directly in NESTOR,
but only through their appearance in the coefficients $I_{m,n}$ as defined in \eqn{Imn}.
For $I_{0,0}$, only $c^{\pm}_{0,0}$ is required.
The limits of the summation in \eqn{c_pm_mn_short} are then
$l_\mathrm{min} = |0-0| = 0$ and $l_\mathrm{max} = 0+0 = 0$ and thus only $\mathtt{cmn}(0,0,0)$ is used in $c^{\pm}_{0,0}$:
\begin{equation}
 c^{\pm}_{0,0} = \frac{1}{2} \,\mathtt{cmn}(0,0,0)\, T^{\pm}_0 \, .
\end{equation}
From this $I_{0,0}$ can be built:
\begin{align}
  I_{0,0} =& c^{+}_{0,0} + c^{+}_{0,0} + c^{-}_{0,0} + c^{-}_{0,0} \nonumber \\
          =& \mathtt{cmn}(0,0,0) \left( T^{+}_0 + T^{-}_0 \right) \, .
\end{align}
Remember that only the entry for $l=0$ in $\mathtt{cmn}(l,0,0)$ is non-zero
and above equation for $I_{0,0}$ can be re-formulated as follows:
\begin{equation}
  I_{0,0} = \sum_{l=0}^{m_\mathrm{f} + n_\mathrm{f}}
    \frac{1}{2} \left( \mathtt{cmn}(l,0,0) + \mathtt{cmn}(l,0,0) \right)
    \left( T^{+}_l + T^{-}_l \right) \label{eqn:I_00}
\end{equation}
where the summation is now over all values of $l$ in steps of $1$, that is along the whole $l$ dimension of the $\mathtt{cmn}$ matrix.
Next consider the coefficients $I_{0,n}$ and $I_{m,0}$ where one of $(m,n)$ is zero and the other is equal to or greater than $1$.
For $I_{0,n}$ with $n \geq 1$ we know from \eqn{Imn}:
\begin{equation}
  I_{0,n} = c^{+}_{0,n} + c^{+}_{0,n-1} + c^{-}_{0,n} + c^{-}_{0,n-1} \, .
\end{equation}
The limits of the summation in the coefficients $c^{\pm}_{0,n}$ for $n\geq 0$
are $l_\mathrm{min} = |0-n| = n$ and $l_\mathrm{max} = 0+n = n$, leading to
\begin{equation}
  c^{\pm}_{0,n} = \frac{1}{2} \,\mathtt{cmn}(n,0,n)\, T^{\pm}_n \, .
\end{equation}
This form of the $c^{\pm}_{0,n}$ can be used to express $I_{0,n}$ as
\begin{equation}
  I_{0,n} = \frac{1}{2} \left[
      \mathtt{cmn}(n,0,n)     \left( T^{+}_n + T^{-}_n \right)
    + \mathtt{cmn}(n-1,0,n-1) \left( T^{+}_{n-1} + T^{-}_{n-1} \right)
    \right]
\end{equation}
and this can be reformulated to
\begin{equation}
  I_{0,n} = \sum_{l=0}^{m_\mathrm{f} + n_\mathrm{f}}
    \frac{1}{2} \left( \mathtt{cmn}(l,0,n) + \mathtt{cmn}(l,0,n-1) \right)
    \left( T^{+}_l + T^{-}_l \right) \label{eqn:I_0n}
\end{equation}
by recognizing again that the matrix $\mathtt{cmn}(l,m,n)$ contains non-zero values only between $|m-n|$ and $m+n$
in the $l$-dimension for any given $(m,n)$.
For $I_{m,0}$ with $m \geq 1$ we know from \eqn{Imn}:
\begin{equation}
  I_{m,0} = c^{+}_{m,0} + c^{+}_{m-1,0} + c^{-}_{m,0} + c^{-}_{m-1,0} \, .
\end{equation}
The limits of the summation in the coefficients $c^{\pm}_{m,0}$ for $m\geq 0$
are $l_\mathrm{min} = |m-0| = m$ and $l_\mathrm{max} = m+0 = m$, leading to
\begin{equation}
  c^{\pm}_{m,0} = \frac{1}{2} \,\mathtt{cmn}(m,m,0)\, T^{\pm}_m \, .
\end{equation}
This form of the $c^{\pm}_{m,0}$ can be used to express $I_{m,0}$ as
\begin{equation}
  I_{m,0} = \frac{1}{2} \left[
      \mathtt{cmn}(m,m,0)     \left( T^{+}_m + T^{-}_m \right)
    + \mathtt{cmn}(m-1,m-1,0) \left( T^{+}_{m-1} + T^{-}_{m-1} \right)
    \right]
\end{equation}
and this can be reformulated to
\begin{equation}
  I_{m,0} = \sum_{l=0}^{m_\mathrm{f} + n_\mathrm{f}}
    \frac{1}{2} \left( \mathtt{cmn}(l,m,0) + \mathtt{cmn}(l,m-1,0) \right)
    \left( T^{+}_l + T^{-}_l \right) \label{eqn:I_m0}
\end{equation}
by recognizing again that the matrix $\mathtt{cmn}(l,m,n)$ contains non-zero values only between $|m-n|$ and $m+n$
in the $l$-dimension for any given $(m,n)$.
Finally for $m \geq 1$ and $n \geq 1$, the coefficients $I_{m,n}$ are defined as
\begin{equation}
  I_{m,n} = c^{+}_{m,n} + c^{+}_{m-1,n} + c^{+}_{m,n-1} + c^{+}_{m-1,n-1} \, .
\end{equation}
These can be reformulated as follows (again, only for $m \geq 1$ and $n \geq 1$):
\begin{align}
  ~& I_{m,n} \\
  =& \phantom{+}~ \frac{1}{2} \sum_{\substack{l=|m-n|\\l+=2}}^{m+n}     \mathtt{cmn}(l,m,n) T^{+}_l
              +   \frac{1}{2} \sum_{\substack{l=|m-n-1|\\l+=2}}^{m+n-1} \mathtt{cmn}(l,m-1,n) T^{+}_l \nonumber \\
           ~& +   \frac{1}{2} \sum_{\substack{l=|m-n+1|\\l+=2}}^{m+n-1} \mathtt{cmn}(l,m,n-1) T^{+}_l
              +   \frac{1}{2} \sum_{\substack{l=|m-n|\\l+=2}}^{m+n-2}   \mathtt{cmn}(l,m-1,n-1) T^{+}_l \\
  =& \sum_{l=0}^{m_\mathrm{f}+n_\mathrm{f}}
     \frac{1}{2} \left( \mathtt{cmn}(l,m,n) + \mathtt{cmn}(l,m-1,n) + \mathtt{cmn}(l,m,n-1) + \mathtt{cmn}(l,m-1,n-1) \right) T^{+}_l \, . \label{eqn:I_mn}
\end{align}
The re-formulated expressions for $I_{m,n}$ from \eqn{I_00}, \eqn{I_0n}, \eqn{I_m0} and \eqn{I_mn}
can be combined and abstracted by introducing a new coefficient matrix $\mathtt{cmns}$ with:
\begin{equation}
  \mathtt{cmns}(l,m,n) = \frac{1}{2}
  \begin{cases}
    \mathtt{cmn}(l,0,0) + \mathtt{cmn}(l,0,0)                                               & \textrm{ for } m=0, n=0 \\
    \mathtt{cmn}(l,0,n) + \mathtt{cmn}(l,0,n-1)                                             & \textrm{ for } m=0, n \geq 1 \\
    \mathtt{cmn}(l,m,0) + \mathtt{cmn}(l,m-1,0)                                             & \textrm{ for } m \geq 1, n=0 \\
    \phantom{+} \mathtt{cmn}(l,m,n \phantom{-1}~\,) + \mathtt{cmn}(l,m-1,n \phantom{-1}~\,) & ~ \\
             +  \mathtt{cmn}(l,m,n          -1    ) + \mathtt{cmn}(l,m-1,n          -1    ) & \textrm{ for } m \geq 1, n \geq 1
  \end{cases}
\end{equation}
and where $0 \leq l \leq m_\mathrm{f}+n_\mathrm{f}$.
Then:
\begin{equation}
  I_{m,n}(u',v') = \sum_{l=0}^{m_\mathrm{f}+n_\mathrm{f}} \mathtt{cmns}(l,m,n)
  \begin{cases}
    T^{+}_l(u',v') & \textrm{ for } m \geq 1, n \geq 1 \\
    T_l    (u',v') & \textrm{ else}
  \end{cases}
\end{equation}
with $T_l(u',v') = T^{+}_l(u',v') + T^{-}_l(u',v')$.
Similarly, the coefficients $K_{m,n}(u',v')$ can then be expressed as follows:
\begin{equation}
  K_{m,n}(u',v') = \sum_{l=0}^{m_\mathrm{f}+n_\mathrm{f}} \mathtt{cmns}(l,m,n)
  \begin{cases}
    S^{+}_l(u',v') & \textrm{ for } m \geq 1, n \geq 1 \\
    S_l    (u',v') & \textrm{ else}
  \end{cases}
\end{equation}
with $S_l(u',v') = S^{+}_l(u',v') + S^{-}_l(u',v')$.

\FloatBarrier
\subsection{Surface-dependent Coefficients}
Coefficients are introduced for $S^{\pm}_l$ based on \eqn{S_pm_l_Merkel}:
\begin{align}
  R^{\pm}_0              =&\, \frac{k_1 d       + k_2 a^{\mp}}{a^{\pm}(ac-b^2)} \\
  R^{\pm}_1              =&\, \frac{k_1 a^{\pm} + k_2 d      }{a^{\pm}(ac-b^2)} \\
  R^{\pm}_{\mathrm{a},1} =&\, \frac{A^{\pm}}{a^{\pm}}
\end{align}
Note that:
\begin{align}
  R^{\pm}_0 + R^{\pm}_1 =& \frac{k_1  d + k_2  a^{\mp} + k_1 a^{\pm} + k_2 d }{a^{\pm}(ac-b^2)}
                        =  \frac{k_1 (d + a^{\pm}) + k_2 (a^{\mp} + d) }{a^{\pm}(ac-b^2)} \nonumber \\
                        =& \frac{k_1 2 (c \pm b) + k_2 2 (c \mp b) }{a^{\pm}(ac-b^2)}
                        =  2 \frac{k_1 (c \pm b) + k_2 (c \mp b) }{a^{\pm}(ac-b^2)} \\
  R^{\pm}_0 - R^{\pm}_1 =& \frac{k_1  d + k_2  a^{\mp} - k_1 a^{\pm} - k_2 d }{a^{\pm}(ac-b^2)}
                        =  \frac{k_1 (d - a^{\pm}) + k_2 (a^{\mp} - d) }{a^{\pm}(ac-b^2)} \nonumber \\
                        =& \frac{k_1 (-2) (a \pm b) + k_2 2 (a \mp b) }{a^{\pm}(ac-b^2)}
                        = -2 \frac{k_1 (a \pm b) - k_2 (a \mp b) }{a^{\pm}(ac-b^2)} \, .
\end{align}
Using these new coefficients, the $S^{\pm}_l$ can be formulated as follows:
\begin{equation}
  S^{\pm}_l =   \left(l R^{\pm}_1 + R^{\pm}_{\mathrm{a},1} \right) T^{\pm}_l + l R^{\pm}_0 T^{\pm}_{l-1}
              -        \frac{R^{\pm}_0 + R^{\pm}_1}{2 \sqrt{c}}
              + (-1)^l \frac{R^{\pm}_0 - R^{\pm}_1}{2 \sqrt{a}} \, .
\end{equation}
Note that:
\begin{align}
  4(ac-b^2)
  =& 4(ac-b^2)+d^2-d^2 \nonumber \\
  =& 4ac-4b^2+(c-a)^2-d^2 \nonumber \\
  =& 4ac-4b^2+c^2-2ac+a^2-d^2 \nonumber \\
  =& c^2+2ac+a^2 - 4b^2 - d^2 \nonumber \\
  =& (a+c)^2-4b^2 - d^2 \nonumber \\
  =& a^{\pm}a^{\mp} - d^2 \, .
\end{align}
The implementation of the coefficient $R^{\pm}_0$ is then done as follows
by using the definition of $k_1$ from \eqn{k1_original}
and the definition of $k_2$ from \eqn{k2_original}:
\begin{align}
  R^{\pm}_0
  =&\, \frac{k_1 d + k_2 a^{\mp}}{a^{\pm}(ac-b^2)} \nonumber \\
  =&\, \frac{-d\left( a^{\pm} A^{\mp} - a^{\mp} A^{\pm} \right) + a^{\mp} \left( 2 a^{\pm} D - 2 d A^{\pm} \right) }{4 a^{\pm}(ac-b^2)} \nonumber \\
  =&\, \frac{-d\left( a^{\pm} A^{\mp} - a^{\mp} A^{\pm} \right) + a^{\mp} \left( 2 a^{\pm} D - 2 d A^{\pm} \right) }{a^{\pm} \left( a^{\pm}a^{\mp} - d^2 \right)} \nonumber \\
  =&\, \frac{-d  a^{\pm} A^{\mp} - d a^{\mp} A^{\pm} + 2 a^{\mp} a^{\pm} D \bcancel{- 2 a^{\mp} d A^{\pm}} \bcancel{+ 2 a^{\mp} d A^{\pm}} }{a^{\pm} \left( a^{\pm}a^{\mp} - d^2 \right)} \nonumber \\
  =&\, \frac{- d a^{\mp} A^{\pm}/a^{\pm} -d  A^{\mp}  + 2 a^{\mp} D  }{ a^{\pm}a^{\mp} - d^2 }
\end{align}
and similarly for $R^{\pm}_1$:
\begin{align}
  R^{\pm}_1
  =&\, \frac{k_1 a^{\pm} + k_2 d}{a^{\pm}(ac-b^2)} \nonumber \\
  =&\, \frac{-a^{\pm} \left( a^{\pm} A^{\mp} - a^{\mp} A^{\pm} \right) + d \left( 2 a^{\pm} D - 2 d A^{\pm} \right) }{4 a^{\pm}(ac-b^2)} \nonumber \\
  =&\, \frac{-a^{\pm} \left( a^{\pm} A^{\mp} - a^{\mp} A^{\pm} \right) + d \left( 2 a^{\pm} D - 2 d A^{\pm} \right) }{a^{\pm} \left( a^{\pm}a^{\mp} - d^2 \right)} \nonumber \\
  =&\, \frac{-a^{\pm} a^{\pm} A^{\mp} + a^{\pm} a^{\mp} A^{\pm} + 2 d a^{\pm} D - 2 d^2 A^{\pm} }{a^{\pm} \left( a^{\pm}a^{\mp} - d^2 \right)} \nonumber \\
  =&\, \frac{A^{\pm} \left( a^{\pm} a^{\mp} - 2 d^2 \right)/a^{\pm} - a^{\pm} A^{\mp} + 2 d D }{ a^{\pm}a^{\mp} - d^2 } \, .
\end{align}
The mapping between quantities in this analytical derivation
and the variables in the Fortran implementation of NESTOR is shown in Tab.~(\ref{tab:T_pm_S_pm}).
\begin{table}[htb]
  \centering
  \begin{tabular}[t]{ c | c }
    quantity & variable \\
    \hline
    $a$  & \texttt{guu\_b} \\
    $2b$ & \texttt{guv\_b} \\
    $c$  & \texttt{gvv\_b} \\
    \hline
    $a^{+}$ & \texttt{adp} \\
    $a^{-}$ & \texttt{adm} \\
    $d$     & \texttt{cma} \\
    $\Delta/4 = a^{+}a^{-}-d^2$ & \texttt{delt1u} \\
    \hline
    $2\sqrt{c}$  & \texttt{sqrtc} \\
    $2\sqrt{a}$  & \texttt{sqrta} \\
    $\sqrt{a^{+}}$ & \texttt{sqad1u} \\
    $\sqrt{a^{-}}$ & \texttt{sqad2u} \\
    \hline
    $A$  & \texttt{auu} \\
    $2B$ & \texttt{auv} \\
    $C$  & \texttt{avv} \\
    \hline
    $A^{+}$  & \texttt{azp1u} \\
    $A^{-}$  & \texttt{azm1u} \\
    $D$  & \texttt{cma11u}
    \end{tabular}
    \hspace{2cm}
    \begin{tabular}[t]{ c | c }
    quantity & variable \\
    \hline
    $R^{+}_0$ & \texttt{r0p} \\
    $R^{-}_0$ & \texttt{r0m} \\
    $R^{+}_1$ & \texttt{r1p} \\
    $R^{-}_1$ & \texttt{r1m} \\
    $R^{+}_{a,1}$ & \texttt{ra1p} \\
    $R^{-}_{a,1}$ & \texttt{ra1m} \\
    \hline
    $T^{+}_l$ & \texttt{tlp} \\
    $T^{-}_l$ & \texttt{tlm} \\
    $T_l$ & \texttt{tlpm} \\
    $T^{+}_{l-1}$ & \texttt{tlp1} \\
    $T^{-}_{l-1}$ & \texttt{tlm1} \\
    $T^{+}_{l-2}$ & \texttt{tlp2} \\
    $T^{-}_{l-2}$ & \texttt{tlm2} \\
    \hline
    $S^{+}_l$ & \texttt{slp} \\
    $S^{-}_l$ & \texttt{slm} \\
    $S_l$ & \texttt{slpm}
  \end{tabular}
  \caption{Mapping of the derived quantites $T^{\pm}_l$ and $S^{\pm}_l$ to the corresponding Fortran variables.}
  \label{tab:T_pm_S_pm}
\end{table}

These variables are initialized in the subroutine \texttt{surface}.
The computation of the coefficients $T^{\pm}_l$ and $S^{\pm}_l$ is then done in the subroutine \texttt{analyt}.
With these definitions at hand, the algorithms for computing the $T^{\pm}_l$ and $S^{\pm}_l$ can be formulated.
The following algorithm~(\ref{alg:const_initial_T_pm_0}) calculates some constants and the initial values of $T^{\pm}_l$ for $l=0$.
\begin{algorithm}
\caption{Calculate constants needed for and initial values of $T^{\pm}_l$}
\label{alg:const_initial_T_pm_0}
\begin{algorithmic}
\Require $a, 2b, c, A, 2B, C$
\State Initialize constants:
\State $a^{+} \gets a + 2b + c$
\State $a^{-} \gets a - 2b + c$
\State $d \gets c-a$
\State $2\sqrt{c} \gets 2 \cdot \sqrt{c}$
\State $2\sqrt{a} \gets 2 \cdot \sqrt{a}$
\State $\Delta/4 \gets a^{+} \cdot a^{-} - d \cdot d$
\State $A^{+} \gets A + 2B + C$
\State $A^{-} \gets A - 2B + C$
\State $D \gets C - A$
\State $R^{+}_1 \gets \left[  A^{+} \cdot (\Delta/4 - d \cdot d)/a^{+} - A^{-} \cdot a^{+} + 2 \cdot D \cdot d \right]/(\Delta/4)$
\State $R^{-}_1 \gets \left[  A^{-} \cdot (\Delta/4 - d \cdot d)/a^{-} - A^{+} \cdot a^{-} + 2 \cdot D \cdot d \right]/(\Delta/4)$
\State $R^{+}_0 \gets \left( -A^{+} \cdot a^{-} \cdot d / a^{+} - A^{-} \cdot d + 2 \cdot D \cdot a^{-} \right)/(\Delta/4)$
\State $R^{-}_0 \gets \left( -A^{-} \cdot a^{+} \cdot d / a^{-} - A^{+} \cdot d + 2 \cdot D \cdot a^{+} \right)/(\Delta/4)$
\State $R^{+}_{\mathrm{a},1} \gets A^{+} / a^{+}$
\State $R^{-}_{\mathrm{a},1} \gets A^{-} / a^{-}$
\State ~
\State Compute initial values for $T^{\pm}_{l-1}$ and $T^{\pm}_l$:
\State Compute $\sqrt{a^{+}}$
\State Compute $\sqrt{a^{-}}$
\State $T^{+}_{l-1} \gets 0$
\State $T^{-}_{l-1} \gets 0$
\State $T^{+}_l \gets 1/\sqrt{a^{+}} \cdot \log\left( \left[ \sqrt{a^{+}} \cdot 2\sqrt{c} + a^{+} + d \right]/\left[ \sqrt{a^{+}} \cdot 2\sqrt{a} - a^{+} + d \right] \right)$
\State $T^{-}_l \gets 1/\sqrt{a^{-}} \cdot \log\left( \left[ \sqrt{a^{-}} \cdot 2\sqrt{c} + a^{-} + d \right]/\left[ \sqrt{a^{-}} \cdot 2\sqrt{a} - a^{-} + d \right] \right)$
\State $T_l \gets T^{+}_l + T^{-}_l$
\end{algorithmic}
\end{algorithm}

Algorithm~(\ref{alg:T_pm_l_S_pm_l}) then iterates over the values of $l$ and computes the $T^{\pm}_l$ and $S^{\pm}_l$ along the way.
\begin{algorithm}
\caption{Calculate $T^{\pm}_l$ and $S^{\pm}_l$ for all $l$}
\label{alg:T_pm_l_S_pm_l}
\begin{algorithmic}
\Require results of Alg.~(\ref{alg:const_initial_T_pm_0}) and $\mathtt{mf}, \mathtt{nf}$
\State $\mathtt{sgn} \gets 1$
\For{$(l \gets 0 ; l \leq \mathtt{mf}+\mathtt{nf} ; l \gets l+1)$}
  \State Compute $S^{+}_l$, $S^{-}_l$, $S_l$:
  \State $         S^{+}_l \gets     \left( R^{+}_1 \cdot l + R^{+}_{\mathrm{a},1} \right) \cdot T^{+}_l + R^{+}_0 \cdot l \cdot T^{+}_{l-1}$
  \State $\phantom{S^{+}_l \gets } - \left( R^{+}_1 + R^{+}_0 \right)/\left(2\sqrt{c}\right) + \mathtt{sgn} \cdot \left( R^{+}_0 - R^{+}_1 \right)/\left(2\sqrt{a}\right)$
  \State $         S^{-}_l \gets     \left( R^{-}_1 \cdot l + R^{-}_{\mathrm{a},1} \right) \cdot T^{-}_l + R^{-}_0 \cdot l \cdot T^{-}_{l-1}$
  \State $\phantom{S^{-}_l \gets } - \left( R^{-}_1 + R^{-}_0 \right)/\left(2\sqrt{c}\right) + \mathtt{sgn} \cdot \left( R^{-}_0 - R^{-}_1 \right)/\left(2\sqrt{a}\right)$
  \State $S_l \gets S^{+}_l + S^{-}_l$
  \State ~
  \State Use $T^{\pm}_l$ and $S^{\pm}_l$ in mode number loop over \texttt{analysum} and \texttt{analysum2}.
  \State ~
  \State Update $T^{\pm}_l$ and $T^{\pm}_{l-1}$ for next $l$:
  \State $\mathtt{sgn} \gets -\mathtt{sgn}$
  \State $T^{+}_{l-2} \gets T^{+}_{l-1}$
  \State $T^{-}_{l-2} \gets T^{-}_{l-1}$
  \State $T^{+}_{l-1} \gets T^{+}_l$
  \State $T^{-}_{l-1} \gets T^{-}_l$
  \State $T^{+}_l \gets \left[ \left( 2\sqrt{c} + \mathtt{sgn} \cdot 2\sqrt{a} \right) - (2\cdot (l+1)-1) \cdot d \cdot T^{+}_{l-1} - l \cdot a^{-} T^{+}_{l-2} \right]/\left( a^{+} \cdot (l+1) \right)$
  \State $T^{-}_l \gets \left[ \left( 2\sqrt{c} + \mathtt{sgn} \cdot 2\sqrt{a} \right) - (2\cdot (l+1)-1) \cdot d \cdot T^{-}_{l-1} - l \cdot a^{+} T^{-}_{l-2} \right]/\left( a^{-} \cdot (l+1) \right)$
  \State $T_l \gets T^{+}_l + T^{-}_l$
\EndFor
\end{algorithmic}
\end{algorithm}
Note that in above algorithm,
\texttt{analysum} and \texttt{analysum2} perform the Fourier transform introduced in \eqn{imn_ft}.
The formulation of the algorithm as stated in this way
has the advantage that the coefficients $T^\pm_l$ and $S^\pm_l$ are used on-the-fly
and thus do not need to be stored for each $l$ before performing the Fourier transform.

\FloatBarrier
\subsection{External Magnetic Field}
The external magnetic field is composed of two contributions.
The obvious one is the field from the main confinement coils, $\mathbf{B}^\textrm{coils}$.
The second contribution comes from the net toroidal plasma current,
which is modeled in the \texttt{vac1} version of NESTOR as a line current along the magnetic axis.
This second contribution is denoted~$\mathbf{B}^\textrm{I}$.
The cylindrical components of the full external magnetic field are then given by:
\begin{align}
  B^\textrm{ext}_R       =&\, B^\textrm{coils}_R       + B^\textrm{I}_R       \\
  B^\textrm{ext}_Z       =&\, B^\textrm{coils}_Z       + B^\textrm{I}_Z       \\
  B^\textrm{ext}_\phi =&\, B^\textrm{coils}_\phi + B^\textrm{I}_\phi \, .
\end{align}
The covariant components~$(B^\textrm{ext}_\theta, B^\textrm{ext}_\zeta)$ of the external magnetic field
are computed from the cylindrical components~$(B^\textrm{ext}_R, B^\textrm{ext}_Z, B^\textrm{ext}_\phi)$
of the external magnetic field as follows:
\begin{align}
 B^{\textrm{ext}}_\theta =&\, R_\theta B^\textrm{ext}_R + Z_\theta B^\textrm{ext}_Z \\
 B^{\textrm{ext}}_\zeta  =&\, R_\zeta  B^\textrm{ext}_R + Z_\zeta  B^\textrm{ext}_Z + R B^\textrm{ext}_\phi \, .
\end{align}
With N from 3.71 we have: % TODO: fix reference !!!
\begin{equation}
 F = - \mathbf{B}^\textrm{ext} \cdot \mathbf{N}
\end{equation}
and $F$ is computed as follows:
\begin{equation}
 F = - \left[
     B^\textrm{ext}_R       \cdot N^R
   + B^\textrm{ext}_Z       \cdot N^Z
   + B^\textrm{ext}_\phi \cdot N^\phi
 \right] \, .
\end{equation}

\FloatBarrier
\subsection{Regularized Surface Quantities}
The Green's function and its derivative are regularized
and evaluated on the four-dimensional surface grid.
This is the computation of $h^\textrm{reg}$ and $g^\textrm{reg}$, respectively.
These functions take in contributions from all toroidal field periods,
as is noted in Eqn. 6.22 and Eqn. 6.23. % TODO: fix references !!!

The toroidal symmetry of the various field periods
implies that the Fourier representation of a quantity
is uniquely defined by its values in a single (i.e., the first) toroidal module.

Furthermore, the singularity in the Green's function surface integrals
only appears if $(\zeta_k, \theta_l)$ and $(\zeta_{k'}, \theta_{l'})$ are in the same toroidal module.
The subtraction method therefore only has to be applied for contributions
from the first toroidal module.
The contributions from all other toroidal modules
can be added without the need to explicitly cancel the singular behaviour of the integrand.

For an axisymmetric case (Tokamak),
there is still the need to compute an integral along the toroidal direction.
The singularity then appears in each of the poloidal planes
where the poloidal coordinates $\theta_l$ and $\theta_{l'}$ coincide.

Field period-invariant parts of geometric quantities can be pre-computed and need only to be extended by
the field period-dependent parts when summing over toroidal modules.
This is done for the distance between $\mathbf{x}$ and $\mathbf{x}'$ as follows:
\begin{align}
     |\mathbf{x} - \mathbf{x}'|^2
  =& (x-x')^2 + (y-y')^2 + (z-z')^2 \nonumber \\
  =& x^2 - 2 x x' + {x'}^2  +  y^2 - 2 y y' + {y'}^2  +  z^2 - 2 z z' + {z'}^2 \nonumber \\
  =& \underbrace{(x^2 + y^2)}_{=r^2} + z^2 + \underbrace{({x'}^2 + {y'}^2)}_{={r'}^2} + {z'}^2 - 2 z z' - 2(x x' + y y') \nonumber \\
  =& r^2 + z^2 + {r'}^2 + {z'}^2 - 2 z z' - 2(x x' + y y') \, .
\end{align}
Only the last term in above equation is dependent on the toroidal angle.
One can now introduce the toroidally-invariant part $g_\mathrm{save}$ with:
\begin{equation}
  g_\mathrm{save} = r^2 + z^2 + {r'}^2 + {z'}^2 - 2 z z' \label{eqn:gsave}
\end{equation}
and arrives at:
\begin{equation}
  |\mathbf{x} - \mathbf{x}'|^2 = g_\mathrm{save}- 2(x x' + y y') \, .
\end{equation}
Similarly for the distance vector projected onto the surface normal:
\begin{align}
     (\mathbf{x} - \mathbf{x}') \cdot \mathbf{N}'
  =& (x-x') {N^x}' + (y-y') {N^y}' + (z-z') {N^z}' \nonumber \\
  =& \left( x {N^x}' + y {N^y}' \right) - \underbrace{\left( x' {N^x}' + y' {N^y}' \right)}_{=R' {N^R}'} + (z-z') {N^z}' \, .
\end{align}
Again the toroidally-invariant part can be identified and is denoted $d_\mathrm{save}$ in this case:
\begin{equation}
    d_\mathrm{save}
  = -R' {N^R}' + (z-z') {N^z}'
  = - \left(R' {N^R}' + z' {N^z}' \right) + z {N^z}' \label{eqn:dsave} \, .
\end{equation}
For the quantitiy of interest it follows:
\begin{equation}
  (\mathbf{x} - \mathbf{x}') \cdot \mathbf{N}' = x {N^x}' + y {N^y}' + d_\mathrm{save} \, .
\end{equation}

First only consider the three-dimensional stellarator case
in the following.
It holds:
\begin{align}
 h^\textrm{reg} = h^{(0)} - h^\textrm{sing} + \sum\limits_{p=1}^{n_\textrm{fp}-1} h^{(p)} \\
 g^\textrm{reg} = g^{(0)} - g^\textrm{sing} + \sum\limits_{p=1}^{n_\textrm{fp}-1} g^{(p)}
\end{align}
where $p$ is the toroidal module index (ranging from 0 for the first module to $n_\textrm{fp}-1$ for the last one)
and the four-dimensional indexing $(\zeta_k, \theta_l, \zeta_{k'}, \theta_{l'})$ has been left our for brevity.

The inverse distance quantites are computed as follows:
\begin{align}
 f_\textrm{temp} =&\, \frac{1}{g_\textrm{save} - 2 (x x' + y y')} \\
 h_\textrm{temp} =&\, \sqrt{f_\textrm{temp}}
\end{align}
Then, we have:
\begin{align}
 \frac{1}{      |\mathbf{x} - \mathbf{x}'|^{3/2}} =&\, h_\textrm{temp} \cdot f_\textrm{temp} \\
 \frac{1}{\sqrt{|\mathbf{x} - \mathbf{x}'|}     } =&\, h_\textrm{temp} \, .
\end{align}

The Green's function kernel is then given by:
\begin{equation}
 h^{(0)} = h_\textrm{temp}
\end{equation}
and the Green's function derivative is given by:
\begin{equation}
 g^{(0)} = h_\textrm{temp} \cdot f_\textrm{temp} \cdot \left( x N^{x'} + y N^{y'} + d_\textrm{save} \right) \, .
\end{equation}

The equivalently-singular kernels are computed next.
Note:
\begin{align}
 \delta u =&\, \frac{1}{\pi} \tan(\pi (u - u')) \\
 \delta v =&\, \frac{1}{\pi} \tan(\pi (v - v')) \, .
\end{align}
Using these, define $g_1$ and $g_2$:
\begin{align}
 g_1 =&\, g_{uu} (\delta u)^2 + 2 g_{uv} \delta u \delta v + g_{vv} (\delta v)^2 \\
 g_2 =&\, a_{uu} (\delta u)^2 + 2 a_{uv} \delta u \delta v + a_{vv} (\delta v)^2 \, .
\end{align}
Then perform the following operations in-place:
\begin{align}
 g_2 \leftarrow&\, \frac{g_2}{g_1} \\
 g_1 \leftarrow&\, \frac{1}{\sqrt{g_1}} \, .
\end{align}
This leads to:
\begin{align}
 g_1 =&\, \frac{1}
               {\sqrt{ g_{uu} (\delta u)^2 + 2 g_{uv} \delta u \delta v + g_{vv} (\delta v)^2 }} \\
 g_2 =&\, \frac{       a_{uu} (\delta u)^2 + 2 a_{uv} \delta u \delta v + a_{vv} (\delta v)^2  }
               {       g_{uu} (\delta u)^2 + 2 g_{uv} \delta u \delta v + g_{vv} (\delta v)^2  } \, .
\end{align}
Now, $h^\textrm{sing}$ and $g^\textrm{sing}$ can be formed:
\begin{align}
 h^\textrm{sing} =&\, g_1 \\
 g^\textrm{sing} =&\, g_1 \cdot g_2 \, ,
\end{align}
which effectively leads to:
\begin{align}
 h^\textrm{sing} =&\, \frac{1}
                           {\sqrt{ g_{uu} (\delta u)^2 + 2 g_{uv} \delta u \delta v + g_{vv} (\delta v)^2 }}  \\
 g^\textrm{sing} =&\, \frac{       a_{uu} (\delta u)^2 + 2 a_{uv} \delta u \delta v + a_{vv} (\delta v)^2               }
                           { \left[g_{uu} (\delta u)^2 + 2 g_{uv} \delta u \delta v + g_{vv} (\delta v)^2 \right]^{3/2} }  \, ,
\end{align}

\FloatBarrier
\subsection{Discrete Fourier Transforms}
The source term~$f$ is computed as follows (see 6.59): % TODO: fix reference !!!
\begin{equation}
 f(\theta_l, \zeta_k)
 = \frac{(2 \pi)^2}{n_\theta n_\zeta}
   \sum\limits_{k', l'}
     F(\theta_{l'}, \zeta_{k'}) h^\textrm{reg}(\theta_l, \zeta_k, \theta_{l'}, \zeta_{k'}) \, .
\end{equation}
The Green's function is computed by summing $K^\textrm{sin}_{m,n}(\theta_l, \zeta_k)$
and the discrete Fourier transform of the Green's function derivative:
\begin{align}
 ~&\, \tilde{K}^\textrm{sin}_{m,n}(\theta_l, \zeta_k)
 = K^\textrm{sin}_{m,n}(\theta_l, \zeta_k) + \nonumber \\
 ~&\, \frac{(2 \pi)^2}{n_\theta n_\zeta}
   \sum\limits_{k', l'}
     \left[   g^\textrm{reg}(\theta_l, \zeta_k, \theta_{        l'}, \zeta_{        k'})
            - g^\textrm{reg}(\theta_l, \zeta_k, \theta_{2 \pi - l'}, \zeta_{2 \pi - k'}) \right]
     \sin(m' \theta_{l'} - n' \zeta_{k'}) \, .
\end{align}
In case of an asymmetric computation, the cosine terms are computed as well:
\begin{align}
 ~&\, \tilde{K}^\textrm{cos}_{m,n}(\theta_l, \zeta_k)
 = K^\textrm{cos}_{m,n}(\theta_l, \zeta_k) + \nonumber \\
 ~&\, \frac{(2 \pi)^2}{n_\theta n_\zeta}
   \sum\limits_{k', l'}
     \left[   g^\textrm{reg}(\theta_l, \zeta_k, \theta_{        l'}, \zeta_{        k'})
            + g^\textrm{reg}(\theta_l, \zeta_k, \theta_{2 \pi - l'}, \zeta_{2 \pi - k'}) \right]
     \cos(m' \theta_{l'} - n' \zeta_{k'}) \, .
\end{align}

The right-hand side is compute by summing $I_{m,n}$
and the discrete Fourier transform of the Green's function:
\begin{equation}
 \tilde{I}^\textrm{sin}_{m,n}
 = I^\textrm{sin}_{m,n}
   + \frac{1}{2 n_\textrm{fp}} \hat{\mathbf{b}}^\textrm{sin}_{m, n}
\end{equation}
where
\begin{equation}
 \hat{\mathbf{b}}^\textrm{sin}_{m, n}
 = 2 n_\textrm{fp} \frac{(2 \pi)^2}{n_\theta n_\zeta n_\textrm{fp}}
   \sum\limits_{k, l}
     \left[ f(\theta_l, \zeta_k) - f(2 \pi - \theta_l, 2 \pi - \zeta_k) \right] \sin(m \theta_l - n \zeta_k) \, .
\end{equation}
In case of an asymmetric case, the following terms are needed as well:
\begin{equation}
 \tilde{I}^\textrm{cos}_{m,n}
 = I^\textrm{cos}_{m,n}
   + \frac{1}{2 n_\textrm{fp}} \hat{\mathbf{b}}^\textrm{cos}_{m, n}
\end{equation}
where
\begin{equation}
 \hat{\mathbf{b}}^\textrm{cos}_{m, n}
 = 2 n_\textrm{fp} \frac{(2 \pi)^2}{n_\theta n_\zeta n_\textrm{fp}}
   \sum\limits_{k, l}
     \left[ f(\theta_l, \zeta_k) + f(2 \pi - \theta_l, 2 \pi - \zeta_k) \right] \cos(m \theta_l - n \zeta_k) \, .
\end{equation}

The matrix is computed by a discrete Fourier transform
of $\tilde{K}_{m,n}(\theta, \zeta)$:
\begin{equation}
 \left( \tilde{\mathbf{A}}^\textrm{sin}_{\textrm{sin}'} \right)_{m, n, m', n'}
 = \frac{(2 \pi)^2}{n_\theta n_\zeta}
   \sum\limits_{k, l}
     \tilde{K}^\textrm{sin}_{m,n}(\theta_l, \zeta_k) sin(m' \theta_l - n' \zeta_k) \, .
\end{equation}
For an asymmetric case, additional matrices are needed:
\begin{align}
 \left( \tilde{\mathbf{A}}^\textrm{sin}_{\textrm{cos}'} \right)_{m, n, m', n'}
 =&\, \frac{(2 \pi)^2}{n_\theta n_\zeta}
      \sum\limits_{k, l}
        \tilde{K}^\textrm{sin}_{m,n}(\theta_l, \zeta_k) cos(m' \theta_l - n' \zeta_k) \\
 \left( \tilde{\mathbf{A}}^\textrm{cos}_{\textrm{sin}'} \right)_{m, n, m', n'}
 =&\, \frac{(2 \pi)^2}{n_\theta n_\zeta}
      \sum\limits_{k, l}
        \tilde{K}^\textrm{cos}_{m,n}(\theta_l, \zeta_k) sin(m' \theta_l - n' \zeta_k) \\
 \left( \tilde{\mathbf{A}}^\textrm{cos}_{\textrm{cos}'} \right)_{m, n, m', n'}
 =&\, \frac{(2 \pi)^2}{n_\theta n_\zeta}
      \sum\limits_{k, l}
        \tilde{K}^\textrm{cos}_{m,n}(\theta_l, \zeta_k) cos(m' \theta_l - n' \zeta_k) \, .
\end{align}

\FloatBarrier
\subsection{Solution of Linear System}
NESTOR solves for~$\Phi$ in tangential Fourier space.
The right-hand side~$\mathbf{b}$ is the sum of
the regularized discrete Fourier transform of the (external magnetic field)-part in the Green's function and
the analytical singular integrals~$I_{m,n}$.
The matrix~$\mathbf{A}$ is given by the sum of
the regularized discrete Fourier transform of the Green's function and
the analytical singular integrals~$K_{m,n}$.
The linear system of equations to be solved is thus:
\begin{equation}
 \mathbf{A} \mathbf{x} = \mathbf{b}
\end{equation}
where the matrix is given by:
\begin{align}
  \mathbf{A}
  = \begin{pmatrix}
      \mathbf{A}^\textrm{sin}_{\textrm{sin}'} & \mathbf{A}^\textrm{sin}_{\textrm{cos}'} \\[2ex]
      \mathbf{A}^\textrm{cos}_{\textrm{sin}'} & \mathbf{A}^\textrm{cos}_{\textrm{cos}'}
    \end{pmatrix}
\end{align}
and the solution and right-hand-side vectors are
$\mathbf{x} = \{ \hat{\Phi}^\textrm{sin}_{m,n}, \hat{\Phi}^\textrm{cos}_{m,n} \}$ and
$\mathbf{b} = \{ \tilde{I}^\textrm{sin}_{m,n}, \tilde{I}^\textrm{cos}_{m,n} \}$, respectively.
The matrix is square and its row-count, its column-count and the length of the vectors
is a linear index along the two-dimensional mode numbers $m$ and $n$.
In the stellarator-symmetric case, only $\mathbf{A}^\textrm{sin}_{\textrm{sin}'}$ is non-zero.

% TODO: fix equation references here !!! !!!
The diagonal elements (see Eqn. 6.36) are incorporated as follows:
\begin{align}
 \mathbf{A}^\textrm{sin}_{\textrm{sin}'}
 =&\, \tilde{\mathbf{A}}^\textrm{sin}_{\textrm{sin}'} + (2 \pi)^2 \pi \cdot \mathbb{1} \\
 \mathbf{A}^\textrm{cos}_{\textrm{cos}'}
 =&\, \tilde{\mathbf{A}}^\textrm{cos}_{\textrm{cos}'} + (2 \pi)^2 \pi \cdot \mathbb{1}
\end{align}
and the (0,0)-mode needs a factor of 2 (see notes on orthogonality of Fourier basis in Eqn. 2.14):
\begin{equation}
 \left( \mathbf{A}^\textrm{cos}_{\textrm{cos}'} \right)_{0, 0, 0, 0}
 = \left( \tilde{\mathbf{A}}^\textrm{cos}_{\textrm{cos}'} \right)_{0, 0, 0, 0} + (2 \pi)^2 \pi \cdot 2 \, .
\end{equation}
Furthermore, the $(0,n)$-modes for $n < 0$ in the matrix and right-hand side are set to zero
in order to force the solution to not need those modes (for compatibility with VMEC, it is assumed).
In \texttt{vac1}, this is solved by the LAPACK method \texttt{dgesv} (LU decomposition with partial pivoting).
In \texttt{vac2}, this is solved by the LAPACK methods \texttt{dpotrf} and \texttt{dpotrs} (Cholesky decomposition).
The use of the Cholesky decomposition in \texttt{vac2} indicates that the problem to be solved
can be formulated using a symmetric matrix (which is planned to be ported to \texttt{vac1} now as well).

\FloatBarrier
\subsection{Vacuum Magnetic Pressure}
The tangential derivatives of the scalar magnetic potential
are computed during the inverse Fourier transform
from the Fourier coefficients~$\hat{\Phi}^\textrm{sin}_{mn}$ back to real space:
\begin{align}
 \frac{\partial \Phi}{\partial \theta} (\theta_l, \varphi_k)
 =&\, \sum\limits_{n=-n_\textrm{f}}^{n_\textrm{f}} \sum\limits_{m=0}^{m_\textrm{f}}
       m               \hat{\Phi}^\textrm{sin}_{m,n} \cos(m \theta_l - n n_\textrm{fp} \varphi_k) \\
 \frac{\partial \Phi}{\partial \varphi} (\theta_l, \varphi_k)
 =&\, \sum\limits_{n=-n_\textrm{f}}^{n_\textrm{f}} \sum\limits_{m=0}^{m_\textrm{f}}
      -n n_\textrm{fp} \hat{\Phi}^\textrm{sin}_{m,n} \cos(m \theta_l - n n_\textrm{fp} \varphi_k) \, .
\end{align}
This is for the stellarator-symmetric case.
For the asymmetric case, cosine-terms are included as well.

The covariant components of the vacuum magnetic field
are computed from the scalar magnetic potential~$\Phi$
and the external magnetic field~$\mathbf{B}^{\textrm{ext}}$ due to coils (from the \texttt{mgrid} file):
\begin{align}
 B^{\textrm{vac}}_\theta  =&\, \frac{\partial \Phi}{\partial \theta } + B^\textrm{ext}_\theta  \\
 B^{\textrm{vac}}_\varphi =&\, \frac{\partial \Phi}{\partial \varphi} + B^\textrm{ext}_\varphi \, .
\end{align}

The contravariant components of the vacuum magnetic field
are computed from the covariant components and the metric elements.
Note that the covariant magnetic field components
can be computed as follows from the contravariant ones
(see \eqn{bsubtheta_from_contra} and \eqn{bsubzeta_from_contra}):
\begin{align}
  B^{\textrm{vac}}_\theta  =&\, g_{\theta \theta } B^{\textrm{vac},\theta} + g_{\theta  \varphi} B^{\textrm{vac},\varphi} \\
  B^{\textrm{vac}}_\varphi =&\, g_{\theta \varphi} B^{\textrm{vac},\theta} + g_{\varphi \varphi} B^{\textrm{vac},\varphi} \, .
\end{align}
This is a linear system of two equations:
\begin{equation}
  \begin{pmatrix}
    B^{\textrm{vac}}_\theta \\
    B^{\textrm{vac}}_\varphi
  \end{pmatrix}
  =
  \begin{pmatrix}
    g_{\theta \theta } & g_{\theta  \varphi} \\
    g_{\theta \varphi} & g_{\varphi \varphi}
  \end{pmatrix}
  \begin{pmatrix}
    B^{\textrm{vac},\theta } \\
    B^{\textrm{vac},\varphi}
  \end{pmatrix} \, .
\end{equation}
This linear system can be solved by inverting the matrix of metric coefficients:
\begin{equation}
  \begin{pmatrix}
    B^{\textrm{vac},\theta } \\
    B^{\textrm{vac},\varphi}
  \end{pmatrix}
  =
  \mathbf{G}
  \begin{pmatrix}
    B^{\textrm{vac}}_\theta \\
    B^{\textrm{vac}}_\varphi
  \end{pmatrix}
\end{equation}
with
\begin{equation}
  \mathbf{G} =
  \frac{1}{ g_{\theta \theta} g_{\varphi \varphi} - g_{\theta \varphi}^2 }
  \begin{pmatrix}
    \phantom{-} g_{\varphi \varphi} &          -  g_{\theta \varphi} \\
             -  g_{\theta  \varphi} & \phantom{-} g_{\theta \theta }
  \end{pmatrix} \, .
\end{equation}
The magnetic pressure on the boundary from the vacuum region is then computed as follows:
\begin{align}
 \frac{|\mathbf{B}^{\textrm{vac}}|^2}{2}
 = \frac{1}{2} \left(   B^{\textrm{vac}}_\theta  B^{\textrm{vac}, \theta }
                      + B^{\textrm{vac}}_\varphi B^{\textrm{vac}, \varphi} \right) \, .
\end{align}

\section{NESTOR for Tokamaks}
This is a variant of NESTOR, which makes explicit use of axisymmetry
to carry out the toroidal integrals analytically.

The grid steps are:
\begin{align}
 \Delta_u      =&\, \frac{1}{n_{\theta,1}} \\
 \Delta_\theta =&\, 2 \pi \Delta_u \, .
\end{align}
The Fourier basis functions can be evaluated
for a fixed grid in advance.
Here are the regular basis functions:
\begin{align}
 c^{lm} =&\, \cos(\Delta_\theta (m+1) l) \Delta_u \\
 s^{lm} =&\, \sin(\Delta_\theta (m+1) l) \Delta_u \, .
\end{align}
Here are the poloidal derivative basis functions:
\begin{align}
 c_m^{lm} =&\, 2 \pi (m+1) \cos(\Delta_\theta (m+1) l) \Delta_u \\
 s_m^{lm} =&\, 2 \pi (m+1) \sin(\Delta_\theta (m+1) l) \Delta_u \, .
\end{align}
The Fourier basis functions are used
to compute the realspace geometry
of the last closed flux surface
as provided by VMEC:
\begin{align}
 R(\theta_l) = \sum\limits_{m = 0}^{M}
   \hat{R}^\textrm{cos}_m \cos(m l \Delta_\theta) \\
 Z(\theta_l) = \sum\limits_{m = 0}^{M}
   \hat{Z}^\textrm{cos}_m \sin(m l \Delta_\theta)
\end{align}
as well as the poloidal derivatives:
\begin{align}
 R_u(\theta_l) = \sum\limits_{m = 0}^{M}
   (- 2 \pi m) \hat{R}^\textrm{cos}_m \sin(m l \Delta_\theta) \\
 Z_u(\theta_l) = \sum\limits_{m = 0}^{M}
   (  2 \pi m) \hat{Z}^\textrm{cos}_m \cos(m l \Delta_\theta)
\end{align}
plus, in case of a non-stellarator-symmetric
(respectively non-up/down-symmetric) case,
the other sin/cos terms.
The metric element~$g_{\theta \theta}$ is given by:
\begin{equation}
 g_{\theta \theta} = R_u^2 + Z_u^2 \, .
\end{equation}
The cylindrical components of the external magnetic field
$B^R(\theta_l)$, $B^\varphi(\theta_l)$ and $B^Z(\theta_l)$
are interpolated from the \texttt{mgrid} file
at the circumference of the plasma boundary.
The normal component of the magnetic field is given by:
\begin{equation}
 (\mathbf{B} \cdot \mathbf{n})(\theta_l) =
 2 \pi R(\theta_l) \left[
   Z_u(\theta_l) B^R(\theta_l) - R_u(\theta_l) B^Z(\theta_l)
 \right] \, .
\end{equation}
The total poloidal current~$G$
is furthermore computed as:
\begin{equation}
 G = \frac{2 \pi}{n_\theta} \sum\limits_{l = 0}^{n_\theta - 1}
   R(\theta_l) B^\varphi(\theta_l)
\end{equation}
which is simply an average over the plasma boundary:
\begin{equation}
 G = 2 \pi \langle R B^\varphi \rangle \, .
\end{equation}
The normal magnetic field component
is Fourier-transformed:
\begin{align}
 \hat{B}_{n,\textrm{ext}}^s =&\,
   \sum\limits_{l = 0}^{n_\theta - 1}
     (\mathbf{B} \cdot \mathbf{n})(\theta_l) s^{l m} \\
 \hat{B}_{n,\textrm{ext}}^c =&\,
   \sum\limits_{l = 0}^{n_\theta - 1}
     (\mathbf{B} \cdot \mathbf{n})(\theta_l) c^{l m} \, .
\end{align}
Define the following shorthand notation:
\begin{align}
 \sigma^R_{l l'} =&\, R(\theta_l) + R(\theta_{l'}) \\
 \delta^R_{l l'} =&\, R(\theta_l) - R(\theta_{l'}) \\
 \delta^R_{l l'} =&\, Z(\theta_l) - Z(\theta_{l'})
\end{align}
as well as:
\begin{align}
 a^{+}_{l l'} =&\, \sqrt{(\sigma^R_{l l'})^2 + (\delta^Z_{l l'})^2} \\
 a^{-}_{l l'} =&\, \sqrt{(\delta^R_{l l'})^2 + (\delta^Z_{l l'})^2} \, .
\end{align}
Then define
\begin{equation}
 w_{l l'} = \frac{a^{+}_{l l'} - a^{-}_{l l'}}{a^{+}_{l l'} + a^{-}_{l l'}}
\end{equation}
which leads to
\begin{align}
 x_{l l'} =&\, 1 - \frac{4 w_{l l'}}{(1 + w_{l l'})^2} \\
          =&\, \frac{(1 - w_{l l'})^2}{(1 + w_{l l'})^2} \, .
\end{align}
This is used as argument to complete elliptic integrals to compute:
\begin{equation}
 E_{l l'} = \frac{1}{\pi w_{l l'}} \left[
   \frac{1 + w_{l l'}^2}{1 + w_{l l'}} \mathcal{K}(x_{l l'})
   - (1 + w_{l l'}) \mathcal{E}(x_{l l'})
 \right] \, .
\end{equation}
Moreover, it is computed:
\begin{equation}
 a_\textrm{sing}(l - l')
 = \log\left( \frac{1}{\pi} \sin\left( \frac{\pi}{n_\theta} (l - l') \right) \right) \, .
\end{equation}
Furthermore, we compute:
\begin{equation}
 A = \frac{1}{2} \left( a^{+}_{l l'} + a^{-}_{l l'} \right) \, .
\end{equation}
The off-diagonal elements of the interaction matrix in realspace
can now be computed as:
\begin{equation}
 G_{l,l',\textrm{reg}}
 =
 \pi R(\theta_l) R(\theta_{l'}) \frac{E_{l l'}}{A_{l l'}}
 + \frac{1}{2} \left[ R(\theta_l) + R(\theta_{l'}) \right] a_\textrm{sing}(l - l') \, .
\end{equation}
The diagonal elements of the interaction matrix in realspace are computed as:
\begin{equation}
 G_{l,l,\textrm{reg}}
 =
 R(\theta_l) \left(
   3 \log(2) - 2 - \log(\frac{\sqrt{g_{\theta \theta}}}{R(\theta_l)})
 \right) \, .
\end{equation}
The matrix is symmetric
and therefore, the upper triangle computed above
is mirrored into the lower triangle.
The Fourier transforms for forming the matrix blocks are computed now:
\begin{align}
  f^\texttt{cc}_{m,m'} =&\, f^\texttt{cc}_{m,m',\textrm{reg}} -  f^\texttt{cc}_{m,m',\textrm{sing}} \\
  f^\texttt{cs}_{m,m'} =&\, f^\texttt{cs}_{m,m',\textrm{reg}} -  f^\texttt{cs}_{m,m',\textrm{sing}} \\
  f^\texttt{ss}_{m,m'} =&\, f^\texttt{ss}_{m,m',\textrm{reg}} -  f^\texttt{ss}_{m,m',\textrm{sing}} \, .
\end{align}
The regular matrix block contributions are:
\begin{align}
 f^\texttt{cc}_{m,m',\textrm{reg}}
 =&\, \sum\limits_{l=0}^{n_\theta-1} \sum\limits_{l'=0}^{n_\theta-1}
   c_m^{lm} G_{l',l,\textrm{reg}} c_m^{l' m'} \\
 f^\texttt{cs}_{m,m',\textrm{reg}}
 =&\, \sum\limits_{l=0}^{n_\theta-1} \sum\limits_{l'=0}^{n_\theta-1}
   c_m^{lm} G_{l',l,\textrm{reg}} s_m^{l' m'} \\
 f^\texttt{ss}_{m,m',\textrm{reg}}
 =&\, \sum\limits_{l=0}^{n_\theta-1} \sum\limits_{l'=0}^{n_\theta-1}
   s_m^{lm} G_{l',l,\textrm{reg}} s_m^{l' m'} \, .
\end{align}
The singular matrix kernel is
\begin{equation}
 g^s_m = - \frac{n_\theta}{2 (m + 1)}
\end{equation}
and this is used to form:
\begin{equation}
 \hat{g}^s_{m m'} = \frac{1}{2} \left(g^s_m + g^s_{m'} \right) \, .
\end{equation}
This is used to form the singular matrix blocks:
\begin{align}
 f^\texttt{cc}_{m,m',\textrm{sing}}
 =&\, \sum\limits_{l=0}^{n_\theta-1}
   c_m^{lm} R(\theta_l) \hat{g}^s_{m m'} c_m^{l' m'} \\
 f^\texttt{cs}_{m,m',\textrm{sing}}
 =&\, \sum\limits_{l=0}^{n_\theta-1}
   c_m^{lm} R(\theta_l) \hat{g}^s_{m m'} s_m^{l' m'} \\
 f^\texttt{ss}_{m,m',\textrm{sing}}
 =&\, \sum\limits_{l=0}^{n_\theta-1}
   s_m^{lm} R(\theta_l) \hat{g}^s_{m m'} s_m^{l' m'} \, .
\end{align}
The right-hand-side vector of the linear system~$\mathbf{g}$
is computed as follows:
\begin{equation}
 \mathbf{g} =
 \begin{pmatrix}
  \hat{g}^s \\
  \hat{g}^c
 \end{pmatrix}
\end{equation}
with
\begin{align}
 \hat{g}^s =&\, \hat{g}_\textrm{reg}^s + \hat{g}_\textrm{sing}^s \\
 \hat{g}^c =&\, \hat{g}_\textrm{reg}^c - \hat{g}_\textrm{sing}^c \, .
\end{align}
In there, we use:
\begin{align}
 \hat{g}_{m,\textrm{reg}}^s
 =&\,
 \sum\limits_{l = 0}^{n_\theta}
   \sum\limits_{l' = 0}^{n_\theta}
     G_{l',l,\textrm{reg}} c_m^{lm} \Delta_u \\
 \hat{g}_{m,\textrm{reg}}^c
 =&\, -
 \sum\limits_{l = 0}^{n_\theta}
   \sum\limits_{l' = 0}^{n_\theta}
     G_{l',l,\textrm{reg}} s_m^{lm} \Delta_u \, .
\end{align}
The singular part is compute as follows:
\begin{align}
 \hat{g}_{m,\textrm{sing}}^s
 =&\,
 \sum\limits_{l = 0}^{n_\theta}
   \alpha_m R(\theta_l) s^{lm} \\
 \hat{g}_{m,\textrm{sing}}^c
 =&\,
 \sum\limits_{l = 0}^{n_\theta}
   \alpha_m R(\theta_l) c^{lm}
\end{align}
with
\begin{equation}
 \alpha_m =
   \frac{1}{2}
   \left[
     \pi + 2 \pi (m+1) \log(2 \pi)
   \right] \, .
\end{equation}
The matrix of the linear system~$\mathcal{H}$ is
now constructed from the blocks:
\begin{equation}
 \mathcal{H} =
 \begin{pmatrix}
   f^\texttt{cc}_{m,m'} & -f^\texttt{cs}_{m,m'} \\
  -f^\textrm{cs}_{m',m} &  f^\textrm{ss}_{m,m'}
 \end{pmatrix}
\end{equation}
The inverse of the matrix is obtained
by computing the Choleskey decomposition of the matrix
and then solving for a set of right-hand-side vectors
forming the equally-sized identity matrix.
The solution to the linear system is then
computed as follows:
\begin{equation}
 \begin{pmatrix}
  \hat{\Phi}^s \\
  \hat{\Phi}^c
 \end{pmatrix}
 =
 \mathcal{H}^{-1}
 \left[
 I_\textrm{tor}
 \begin{pmatrix}
  \hat{g}^s \\
  \hat{g}^c
 \end{pmatrix}
 +
 \begin{pmatrix}
  \hat{B}_{n,\textrm{ext}}^s \\
  \hat{B}_{n,\textrm{ext}}^c
 \end{pmatrix}
 \right]
\end{equation}
The potential derivative~$\Phi_u$ is computed
directly from the Fourier coefficients of the potential:
\begin{equation}
 \Phi_u(\theta_l)
 =
 I_\textrm{tor}
 - n_{\theta,1} \sum\limits_{m = 0}^{m_f}
     \left[
       \hat{\Phi}^s_m c_m^{lm} - \hat{\Phi}^c_m s_m^{lm}
     \right] \, .
\end{equation}
The vacuum magnetic pressure~$|\mathbf{B}_\textrm{vac}|^2/2$
is computed as follows:
\begin{equation}
 \frac{\vert\mathbf{B}_\textrm{vac}\vert^2}{2}
 =
 \frac{1}{2} \left[
   \frac{G^2}{(2 \pi R)^2}
    +
    \frac{\Phi_u^2}{g_{\theta \theta}^2}
 \right] \, .
\end{equation}

\chapter{VMEC Inputs}

\section{Numerical Resolution and Symmetry Assumptions}

\begin{itemize}
 \item \texttt{lasym} \\
       Type: \texttt{bool} \\
       Default: \texttt{false} \\
       This flag controls the assumption of stellarator symmetry in the computation.
       If set to \texttt{true}, the additional Fourier coefficients in \eqn{r_full} and \eqn{z_full}
       are assumed throughout the computation instead of the reduced sets in \eqn{r_symm} and \eqn{z_symm}.

 \item \texttt{nfp} \\
       Type: \texttt{int} \\
       Range: $>0$ \\
       Default: 1 \\
       This is the number of field periods, $n_\textrm{fp}$, to make use of toroidal symmetry.

 \item \texttt{mpol} \\
       Type: \texttt{int} \\
       Range: $\geq 2$ \\
       Default: 6 \\
       This controls the poloidal resolution of the computation in Fourier space.
       The set of poloidal mode numbers taken into account is
       $\{0, 1, ..., (\texttt{mpol} - 1)\}$.
       The number of poloidal Fourier modes taken into account is thus \texttt{mpol}.

 \item \texttt{ntor} \\
       Type: \texttt{int} \\
       Range: $\geq 0$ \\
       Default: 0 \\
       This controls the toroidal resolution of the computation in Fourier space.
       The set of toroidal mode numbers taken into account is
       $\{-\texttt{ntor}, ..., -1, 0, 1, ..., \texttt{ntor}\}$.
       The number of toroidal Fourier modes taken into account is thus $2 \times \texttt{ntor} + 1$.
       If $0$ is specified, an axisymmetric (Tokamak) run is done.

 \item \texttt{ntheta} \\
       Type: \texttt{int} \\
       Range: $\geq 0$ \\
       Default: 0 \\
       This controls the poloidal resolution of the computation in real space,
       i.e., the number of grid points in the poloidal direction.
       A value of $0$ implies that it is automatically chosen as the minimally allowed number of poloidal grid points.
       Any given value is checked against the minimally allowed number of poloidal grid points,
       which is $2 \times \texttt{mpol} + 6$.

 \item \texttt{nzeta} \\
       Type: \texttt{int} \\
       Range: $\geq 0$ \\
       Default: 0 \\
       This controls the toroidal resolution of the computation in real space,
       i.e., the number of grid points per field period in the toroidal direction.
       A value of $0$ implies that it is automatically chosen as the minimally allowed number of toroidal grid points per period.
       Any given value is checked against the minimally allowed number of toroidal grid points,
       which is $2 \times \texttt{ntor} + 4$.
       In case of an axisymmetric calculation ($\texttt{ntor} = 0$),
       a value of $0$ is replaced with $1$, but a different positive number can also be taken into account.
       Think of this as being able to average over many poloidal cutplanes for an axisymmetric run.

\end{itemize}

\section{Multi-Grid Steps}

\begin{itemize}

 \item \texttt{ns\_array} \\
       Type: \texttt{vector<int>} \\
       Range: each entry $\ge 3$, each entry $\geq$ the previous one \\
       Default: $\{ 31 \}$ \\
       This controls the radial resolution, i.e., the number of flux surfaces,
       in a series of consecutive resolution steps.
       The equilibrium is solved for the first entry in \texttt{ns\_array},
       then interpolated onto the next radial resolution grid, solved there, and so on.

 \item \texttt{ftol\_array} \\
       Type: \texttt{vector<double>} \\
       Range: each entry $> 0$ \\
       Default: $\{ 10^{-10} \}$ \\
       This controls the convergence criterion for each multi-grid step.
       The sum of the raw, un-preconditioned, force residuals has to fall below
       the given entry in \texttt{ftol\_array} for that multi-grid step to be considered converged.

 \item \texttt{niter\_array} \\
       Type: \texttt{vector<int>} \\
       Range: each entry $> 0$ \\
       Default: $\{ 100 \}$ \\
       This controls the number of allowed iterations for each multi-grid step
       for trying to achieve the requested force tolerance levels in \texttt{ftol\_array}.

\end{itemize}

\section{Global Physics Parameters}

\begin{itemize}

 \item \texttt{phiedge} \\
       Type: \texttt{double} \\
       Default: $1.0$ \\
       This specifies the toroidal magnetic flux enclosed by the boundary.
       In a fixed-boundary computation, this implicitly defines the magnetic field strength.
       In a free-boundary computation, this mainly influences the area of the poloidal cross section of the boundary
       and thus sets the volume of the plasma.

 \item \texttt{ncurr} \\
       Type: \texttt{int} \\
       Range: 0 or 1 \\
       Default: 0 \\
       This flag selects whether the rotational transform profile ($\texttt{ncurr} = 0$)
       or the toroidal current profile ($\texttt{ncurr} = 1$)
       are used as the second radial profile (next to the mass or pressure profile)
       to define the equilibrium to compute.

\end{itemize}

\section{Profile of Mass or Pressure}

\begin{itemize}

 \item \texttt{pmass\_type} \\
       Type: \texttt{string} \\
       Range: one of $\{ \texttt{power\_series},
                         \texttt{gauss\_trunc},
                         \texttt{two\_lorentz},
                         \texttt{two\_power},
                         \texttt{two\_power\_gs}, \\
                         \texttt{akima\_spline},
                         \texttt{cubic\_spline},
                         \texttt{pedestal},
                         \texttt{rational},
                         \texttt{line\_segment} \}$ \\
       Default: \texttt{power\_series} \\
       This defines the parameterization used for specifying the mass or pressure profile.

 \item \texttt{am} \\
       Type: \texttt{vector<double>} \\
       Default: $\{ \}$ \\
       These are the expansion coefficients for a parametric mass or pressure profile.
       Their interpretation changes depending on the value of \texttt{pmass\_type}.
       The pressure profile would be specified in Pascals.

 \item \texttt{am\_aux\_s} \\
       Type: \texttt{vector<double>} \\
       Default: $\{ \}$ \\
       These are the knot locations for a spline-parameterized specification
       of the mass or pressure profile.
       The first entry should be $0.0$ and the last entry should be $1.0$.
       At least 4 entries need to be provided.
       Equal number of entries need to be specified here and in \texttt{am\_aux\_f}.

 \item \texttt{am\_aux\_f} \\
       Type: \texttt{vector<double>} \\
       Default: $\{ \}$ \\
       These are the knot values for a spline-parameterized specification
       of the mass or pressure profile.
       At least 4 entries need to be provided.
       Equal number of entries need to be specified here and in \texttt{am\_aux\_s}.
       The pressure profile would be specified in Pascals.

 \item \texttt{pres\_scale} \\
       Type: \texttt{double} \\
       Range: $\geq 0$ \\
       Default: $1.0$ \\
       This is a global scaling factor for the mass or pressure profile in Pascals.

 \item \texttt{gamma} \\
       Type: \texttt{double} \\
       Range: $\geq 0$ \\
       Default: $0.0$ \\
       This is the adiabatic index (also ratio of specific heats) $\gamma$.
       Specifying $0$ implies that the pressure profile is specified.
       For all other values, the mass profile is specified.

 \item \texttt{spres\_ped} \\
       Type: \texttt{double} \\
       Range: $0 < \texttt{spres\_ped} \leq 1$ \\
       Default: $1.0$ \\
       This specifies the radial position in $s$ of a pedestal in the pressure profile.
       Radially outside of this location, the pressure is assumed to be constant.

\end{itemize}

\section{(Initial Guess for) Rotational Transform Profile}

Note that the rotational transform profile can be specified in the input file
even in case of a constrained-toroidal-current computation
(where $\texttt{ncurr} = 1$ and the rotational transform profile is thus computed on-the-fly internally).
The rotational transform profile specified in the input file
is then only used as an initial guess for computing the radial profiles of the magnetic fluxes
in the first iteration.
The key takeaway here is that a weird rotational transform profile in the input file
can mess up a VMEC++ computation even if the input file specifies
to run VMEC++ in constrained-toroidal-current mode
(and the user thus expects not having to care about the rotational transform profile).
The suggested workflow here is to only specify the rotational transform profile in the input file
if a constrained-rotational-transform computation is supposed to be done (with $\texttt{ncurr} = 0$).
The option to specify an initial guess for the rotational transform profile
in case of a constrained-toroidal-current computation is then reserved for the case
where the default initial guess for the rotational transform profil ($\iota = 1$) does not work.

\begin{itemize}

 \item \texttt{piota\_type} \\
       Type: \texttt{string} \\
       Range: one of $\{ \texttt{power\_series},
                         \texttt{sum\_atan},
                         \texttt{akima\_spline},
                         \texttt{cubic\_spline}, \\
                         \texttt{rational},
                         \texttt{line\_segment},
                         \texttt{nice\_quadratic} \}$ \\
       Default: \texttt{power\_series} \\
       This defines the parameterization used for specifying the rotational transform profile.

 \item \texttt{ai} \\
       Type: \texttt{vector<double>} \\
       Default: $\{ \}$ \\
       These are the expansion coefficients for a parametric rotational transform profile.
       Their interpretation changes depending on the value of \texttt{piota\_type}.

 \item \texttt{ai\_aux\_s} \\
       Type: \texttt{vector<double>} \\
       Default: $\{ \}$ \\
       These are the knot locations for a spline-parameterized specification
       of the rotational transform profile.
       The first entry should be $0.0$ and the last entry should be $1.0$.
       At least 4 entries need to be provided.
       Equal number of entries need to be specified here and in \texttt{ai\_aux\_f}.

 \item \texttt{ai\_aux\_f} \\
       Type: \texttt{vector<double>} \\
       Default: $\{ \}$ \\
       These are the knot values for a spline-parameterized specification
       of the rotational transform profile.
       At least 4 entries need to be provided.
       Equal number of entries need to be specified here and in \texttt{ai\_aux\_s}.

\end{itemize}

\section{(Initial Guess for) Toroidal Current Profile}

\begin{itemize}

 \item \texttt{pcurr\_type} \\
       Type: \texttt{string} \\
       Range: one of $\{ \texttt{power\_series},
                         \texttt{power\_series\_i},
                         \texttt{gauss\_trunc},
                         \texttt{sum\_atan}, \\
                         \texttt{two\_power},
                         \texttt{two\_power\_gs},
                         \texttt{akima\_spline\_i},
                         \texttt{akima\_spline\_ip}, \\
                         \texttt{cubic\_spline\_i},
                         \texttt{cubic\_spline\_ip},
                         \texttt{pedestal},
                         \texttt{rational}, \\
                         \texttt{line\_segment\_i},
                         \texttt{line\_segment\_ip},
                         \texttt{sum\_cossq\_s},
                         \texttt{sum\_cossq\_sqrts},
                         \texttt{sum\_cossq\_s\_free} \}$ \\
       Default: \texttt{power\_series} \\
       This defines the parameterization used for specifying the toroidal current profile.
       Traditionally, the profile of the radial derivative of the enclosed toroidal current
       was specified. This is referred to as the ``I-prime'' profile.
       The radial profile of the enclosed toroidal current was then computed by numerical quadrature.
       Later, additional profile types were implemented that allow to specify the radial profile
       of the enclosed toroidal current directly.
       These are distinguished between with the suffixes \texttt{\_i} (enclosed current)
       and \texttt{\_ip} (radial derivative of enclosed current)
       in the names of the profile parameterizations.

 \item \texttt{ac} \\
       Type: \texttt{vector<double>} \\
       Default: $\{ \}$ \\
       These are the expansion coefficients for a parametric toroidal current profile.
       Their interpretation changes depending on the value of \texttt{pcurr\_type}.

 \item \texttt{ac\_aux\_s} \\
       Type: \texttt{vector<double>} \\
       Default: $\{ \}$ \\
       These are the knot locations for a spline-parameterized specification
       of the toroidal current profile.
       The first entry should be $0.0$ and the last entry should be $1.0$.
       At least 4 entries need to be provided.
       Equal number of entries need to be specified here and in \texttt{ac\_aux\_f}.

 \item \texttt{ac\_aux\_f} \\
       Type: \texttt{vector<double>} \\
       Default: $\{ \}$ \\
       These are the knot values for a spline-parameterized specification
       of the toroidal current profile.
       At least 4 entries need to be provided.
       Equal number of entries need to be specified here and in \texttt{ac\_aux\_s}.

 \item \texttt{curtor} \\
       Type: \texttt{double} \\
       Default: $0.0$ \\
       This is the net toroidal current in Amperes.
       The toroidal current profile is scaled to yield a total net toroidal current as given here.

 \item \texttt{bloat} \\
       Type: \texttt{double} \\
       Range: $\texttt{bloat} > 1$ \\
       Default: $1.0$ \\
       This is a scalar factor to re-scale the radial scale for the toroidal current profile.
       This factor can only deviate from $1.0$ in case of a constrained-current computation,
       i.e., when $\texttt{ncurr} = 1$.

\end{itemize}

\section{Free-Boundary Parameters}

\begin{itemize}

 \item \texttt{lfreeb} \\
       Type: \texttt{bool} \\
       Default: \texttt{false} \\
       This flag controls whether VMEC++ is run in fixed-boundary mode (\texttt{false}) or in free-boundary mode (\texttt{true}).

 \item \texttt{mgrid\_file} \\
       Type: \texttt{string} \\
       Default: \texttt{"NONE"} \\
       This is the filename to the MGRID file (in NetCDF format) that contains the magnetic field response factors for the external coils.

 \item \texttt{extcur} \\
       Type: \texttt{vector<double>} \\
       Default: $\{ \}$ \\
       This is the array of coil currents that determines the net external magnetic field as produced by the coil contributions in the MGRID file.

 \item \texttt{nvacskip} \\
       Type: \texttt{int} \\
       Range: $1 \leq \texttt{nvacskip} \leq 10$ \\
       Default: 1 \\
       This is the interval for performing a full update of the free-boundary computation.
       Values greater than 1 mean that the matrix in the Nestor module is re-used for the next \texttt{nvacskip} iterations
       after the one that computed it (such as the first iteration that the free-boundary computation was activated).

 \item \texttt{free\_boundary\_method} \\
       Type: \texttt{string} \\
       Default: \texttt{"nestor"} \\
       This enum controls which method is used in VMEC++
       to compute the free-boundary force contribution.
       Options are \texttt{"nestor"} for the standard
       NEumann Solver for TORoidal systems
       and \texttt{"only\_coils"} for using only the vacuum field due to external coils.
       The option \texttt{"only\_coils"} only works for zero pressure (\texttt{pres\_scale = 0})
       and zero net toroidal current (\texttt{curtor = 0}).

\end{itemize}

\section{Tweaking Parameters}

\begin{itemize}

 \item \texttt{nstep} \\
       Type: \texttt{int} \\
       Range: $\geq 1$ \\
       Default: 10 \\
       This is the interval of iterations at which the convergence progress (force residuals, etc.) are printed to the screen.

 \item \texttt{aphi} \\
       Type: \texttt{vector<double>} \\
       Default: $\{ 1.0 \}$ \\
       This is a power series expansion, without a constant offset,
       of the remapping profile for changing the radial location of the flux surfaces
       as a function of the normalized toroidal flux.

 \item \texttt{delt} \\
       Type: \texttt{double} \\
       Range: $0 < \texttt{delt} \leq 1$ \\
       Default: $1.0$ \\
       This is the initial value for the artificial time step used in the iterative solver in VMEC++.
       The time step used along the computation is automatically reduced on-the-fly
       if a too-large time step lead to crossing of the flux surfaces.

 \item \texttt{tcon0} \\
       Type: \texttt{double} \\
       Range: $0 \leq \texttt{tcon0} \leq 1$ \\
       Default: $1.0$ \\
       This is a scaling factor that can be used to tune down the spectral condensation force contribution.

 \item \texttt{lforbal} \\
       Type: \texttt{bool} \\
       Default: \texttt{false} \\
       This is a flag to use a non-variational form of the MHD forces
       for the innermost flux surface (the first one outside the magnetic axis).
       It can sometimes help convergence to enable this option.

\end{itemize}

\section{Initial Guess for Magnetic Axis Geometry}

\begin{itemize}

 \item \texttt{raxis\_c} \\
       Type: \texttt{vector<double>} \\
       Range: At least 1 value needs to be given, at most entries up to $n = \texttt{ntor}$ are considered. \\
       Default: $\{ \}$ \\
       These are the stellarator-symmetric (cosine) expansion coefficients $\hat{R}_{n}^{\cos}$ for the $R$ coordinate of the magnetic axis.

 \item \texttt{zaxis\_s} \\
       Type: \texttt{vector<double>} \\
       Range: At most entries up to $n = \texttt{ntor}$ are considered. The first entry ($n = 0$) is ignored. \\
       Default: $\{ \}$ \\
       These are the stellarator-symmetric (sine) expansion coefficients $\hat{Z}_{n}^{\sin}$ for the $Z$ coordinate of the magnetic axis.

 \item \texttt{raxis\_s} \\
       Type: \texttt{vector<double>} \\
       Range: At most entries up to $n = \texttt{ntor}$ are considered. The first entry ($n = 0$) is ignored. \\
       Default: $\{ \}$ \\
       These are the non-stellarator-symmetric (sine) expansion coefficients $\hat{R}_{n}^{\sin}$ for the $R$ coordinate of the magnetic axis.

 \item \texttt{zaxis\_c} \\
       Type: \texttt{vector<double>} \\
       Range: At least 1 value needs to be given, at most entries up to $n = \texttt{ntor}$ are considered. \\
       Default: $\{ \}$ \\
       These are the non-stellarator-symmetric (cosine) expansion coefficients $\hat{Z}_{n}^{\cos}$ for the $Z$ coordinate of the magnetic axis.

\end{itemize}

\section{(Initial Guess for) Boundary Geometry}

\begin{itemize}

 \item \texttt{rbc} \\
       Type: \texttt{vector<double>[]} \\
       Default: $\{ \}$ \\
       These are the stellarator-symmetric (cosine) expansion coefficients $\hat{R}_{m, n}^{\cos}$ for the $R$ coordinate of the (initial guess for the) boundary.

 \item \texttt{zbs} \\
       Type: \texttt{vector<double>[]} \\
       Default: $\{ \}$ \\
       These are the stellarator-symmetric (sine) expansion coefficients $\hat{Z}_{m, n}^{\sin}$ for the $Z$ coordinate of the (initial guess for the) boundary.

 \item \texttt{rbs} \\
       Type: \texttt{vector<double>[]} \\
       Default: $\{ \}$ \\
       These are the non-stellarator-symmetric (sine) expansion coefficients $\hat{R}_{m, n}^{\sin}$ for the $R$ coordinate of the (initial guess for the) boundary.

 \item \texttt{zbc} \\
       Type: \texttt{vector<double>[]} \\
       Default: $\{ \}$ \\
       These are the non-stellarator-symmetric (cosine) expansion coefficients $\hat{Z}_{m, n}^{\cos}$ for the $Z$ coordinate of the (initial guess for the) boundary.

\end{itemize}

\chapter{VMEC Outputs}

Legacy Fortran VMEC produces four main output files:
\texttt{wout},
\texttt{jxbout},
\texttt{mercier} and
\texttt{threed1}.
These are merged into a single HDF5 file in VMEC++.
HDF5 is chosen, because it allows (in principle)
to adjust the other tools in the VMEC toolchain
to adopt the new VMEC++ output file format,
since a well-established HDF5 Fortran interface exists
and is maintained by the HDF group.
An intermediate conversion program from the single HDF5
into the legacy output files is available as part of the Python code vendored with VMEC++,
since most of the VMEC toolchain relies on the legacy output files.

The HDF5 file produced by VMEC++ has the following structure:
\begin{itemize}
 \item \texttt{/indata} - This is a 1:1 duplicate of the input data used to create this VMEC++ run.
 \item \texttt{/jxbout} - This is the equivalent of the \texttt{jxbout} output file produced by Fortran VMEC.
 \item \texttt{/mercier} - This is the equivalent of the \texttt{mercier} output file produced by Fortran VMEC.
 \item \texttt{/threed1} - This is the equivalent of the \texttt{threed1} output file produced by Fortran VMEC.
 \item \texttt{/wout} - This is the equivalent of the \texttt{wout} output file produced by Fortran VMEC.
\end{itemize}

\section{\texttt{indata}}
The \texttt{indata} members are specified and described above.
Their definitions are therefore not repeated here.

\section{\texttt{wout}}

The first set of members in the \texttt{wout} group
contains the following members:
\begin{itemize}
 \item \texttt{version} \\
       Type: \texttt{string} \\
       This is the version number of VMEC++, e.g.,
       \texttt{"8.52"} for the initial release based on serial VMEC 8.52 from STELLOPT's \texttt{v251} branch,
       on which \texttt{educational\_VMEC} is based, on which in turn VMEC++ is based.
 \item \texttt{sign\_of\_jacobian} \\
       Type: \texttt{int} \\
       This is the sign of the Jacobian of the coordintate transform between flux coordinates
       and cylindrical coordinates. It encodes the handedness of the coordinate system.
       As in Fortran VMEC (where this is called \texttt{signgs}), this is fixed to $-1$ as of now.
 \item \texttt{mgrid\_mode} \\
       Type: \texttt{string} \\
       In case of a free-boundary run,
       this indicates if the mgrid file was normalized to unit currents~(\texttt{"S"}) or not~(\texttt{"R"}).
       For a given mgrid file, this changes the interpretation of the coil currents.
 \item \texttt{ier\_flag} \\
       Type: \texttt{int} \\
       This is the status code indicating success of or problems during this VMEC++ run.
\end{itemize}
The next set of members in the \texttt{wout} group
indicates the convergence of this VMEC++ run:
\begin{itemize}
 \item \texttt{fsqt} \\
       Type: \texttt{double[]} \\
       Evolution of the total force residual along the run.
       Not implemented yet.
 \item \texttt{wdot} \\
       Type: \texttt{double[]} \\
       Evolution of the MHD energy decay along the run.
       Not implemented yet.
 \item \texttt{nfev} \\
       Type: \texttt{int} \\
       Number of evaluations of the ideal MHD forward model during this run.
 \item \texttt{fsqr} \\
       Type: \texttt{double} \\
       Invariant force residual of the force on $R$ at end of the run.
 \item \texttt{fsqz} \\
       Type: \texttt{double} \\
       Invariant force residual of the force on $Z$ at end of the run.
 \item \texttt{fsql} \\
       Type: \texttt{double} \\
       Invariant force residual of the force on $\lambda$ at end of the run.
\end{itemize}
A number of scalar plasma properties are available
in the next set of members in the \texttt{wout} group:
\begin{itemize}
 \item \texttt{wb} \\
       Type: \texttt{double} \\
       Magnetic energy: volume integral of $\vert\mathbf{B}\vert^2/(2 \mu_0)$.
 \item \texttt{wp} \\
       Type: \texttt{double} \\
       Kinetic energy: volume integral of $p$.
 \item \texttt{Rmajor\_p} \\
       Type: \texttt{double} \\
       Major radius of the plasma.
 \item \texttt{Aminor\_p} \\
       Type: \texttt{double} \\
       Minor radius of the plasma.
 \item \texttt{volume\_p} \\
       Type: \texttt{double} \\
       Plasma volume.
 \item \texttt{b0} \\
       Type: \texttt{double} \\
       Toroidal magnetic flux density from poloidal current and magnetic axis position at $\varphi=0$:
       \begin{equation}
        \texttt{b0} = F_\psi(0) / R(s=0, \theta=0, \varphi=0)
       \end{equation}
       with the poloidal current at the magnetic axis~$F_\psi(0)$
       (obtained by linear extrapolation from the innermost two half-grid points):
       \begin{equation}
        F_\psi(0) = \frac{3}{2} \langle B_\varphi \rangle(s_{0.5}) - \frac{1}{2} \langle B_\varphi \rangle(s_{1.5})
       \end{equation}
       and in there the covariant magnetic field component~$B_\varphi$ averaged over flux surfaces:
       \begin{equation}
        \langle B_\varphi \rangle (s)= \int\limits_0^{2 \pi} \int\limits_0^{2 \pi} B_\varphi(s ,\theta, \varphi) \,\mathrm{d}\theta \,\mathrm{d}\varphi
       \end{equation}
       (\texttt{bvcoH} in VMEC++, on half-grid).
       One can think of this as approximating the toroidal field as originating from a line current along the $z$ axis,
       which combines all poloidal currents in the coils and the plasma,
       and then asking what magnetic field at the radius of the magnetic axis at $\varphi=0$ this would create.
       Because the quantity is divided by $R$ at $\varphi = 0$,
       it is not a toroidal average and should not be used when, e.g., scaling equilibria based on magnetic field strength
       \footnote{Thanks to Damien Huet for stressing this important point.}.
 \item \texttt{rmax\_surf} \\
       Type: \texttt{double} \\
       Maximum of $R$ on the plasma boundary over all grid points.
 \item \texttt{rmin\_surf} \\
       Type: \texttt{double} \\
       Minimum of $R$ on the plasma boundary over all grid points.
 \item \texttt{zmax\_surf} \\
       Type: \texttt{double} \\
       Maximum of $Z$ on the plasma boundary over all grid points.
 \item \texttt{aspect} \\
       Type: \texttt{double} \\
       Aspect ratio, i.e., ratio of major radius over minor radius, of the plasma boundary.
 \item \texttt{betatotal} \\
       Type: \texttt{double} \\
       Total plasma beta.
 \item \texttt{betapol} \\
       Type: \texttt{double} \\
       Polodial plasma beta.
 \item \texttt{betator} \\
       Type: \texttt{double} \\
       Torodial plasma beta.
 \item \texttt{betaxis} \\
       Type: \texttt{double} \\
       Plasma beta on the magnetic axis.
 \item \texttt{volavgB} \\
       Type: \texttt{double} \\
       Volume-averaged magnetic field strength.
 \item \texttt{IonLarmor} \\
       Type: \texttt{double} \\
       Larmor radius of plasma ions.
 \item \texttt{ctor} \\
       Type: \texttt{double} \\
       Net toroidal plasma current.
 \item \texttt{rbtor} \\
       Type: \texttt{double} \\
       Poloidal ribbon current in the plasma at the boundary,
       i.e., sum of all (assumed, in case of a fixed-boundary run) coil currents
       plus the net poloidal plasma current.
 \item \texttt{rbtor0} \\
       Type: \texttt{double} \\
       Poloidal ribbon current in the plasma at the axis,
       i.e., sum of all (assumed, in case of a fixed-boundary run) coil currents,
       but without the poloidal plasma current.
\end{itemize}
A number of radial profiles of plasma parameters are available
in the next set of members in the \texttt{wout} group:
\begin{itemize}
 \item \texttt{iotaf} \\
       Type: \texttt{double[]} \\
       Dimensions: \texttt{[ns]} \\
       Rotational transform~$\iota$ on the full-grid.
 \item \texttt{q\_factor} \\
       Type: \texttt{double[]} \\
       Dimensions: \texttt{[ns]} \\
       Safety factor~$1/\iota$ on the full-grid.
 \item \texttt{presf} \\
       Type: \texttt{double[]} \\
       Dimensions: \texttt{[ns]} \\
       Kinetic pressure~$p$ on the full-grid.
 \item \texttt{phi} \\
       Type: \texttt{double[]} \\
       Dimensions: \texttt{[ns]} \\
       Enclosed toroidal magnetic flux~$\phi$ on the full-grid.
 \item \texttt{phipf} \\
       Type: \texttt{double[]} \\
       Dimensions: \texttt{[ns]} \\
       Radial derivative of enclosed toroidal magnetic flux~$\phi'$ on the full-grid.
 \item \texttt{chi} \\
       Type: \texttt{double[]} \\
       Dimensions: \texttt{[ns]} \\
       Enclosed poloidal magnetic flux~$\chi$ on the full-grid.
 \item \texttt{chipf} \\
       Type: \texttt{double[]} \\
       Dimensions: \texttt{[ns]} \\
       Radial derivative of enclosed toroidal magnetic flux~$\chi'$ on the full-grid.
 \item \texttt{jcuru} \\
       Type: \texttt{double[]} \\
       Dimensions: \texttt{[ns]} \\
       Radial derivative of enclosed poloidal current on full-grid.
 \item \texttt{jcurv} \\
       Type: \texttt{double[]} \\
       Dimensions: \texttt{[ns]} \\
       Radial derivative of enclosed toroidal current on full-grid.
 \item \texttt{iotas} \\
       Type: \texttt{double[]} \\
       Dimensions: \texttt{[ns - 1]} \\
       Rotational transform~$\iota$ on the half-grid.
 \item \texttt{mass} \\
       Type: \texttt{double[]} \\
       Dimensions: \texttt{[ns - 1]} \\
       Plasma mass profile~$m$ on half-grid.
 \item \texttt{pres} \\
       Type: \texttt{double[]} \\
       Dimensions: \texttt{[ns - 1]} \\
       Kinetic pressure~$p$ on the half-grid.
 \item \texttt{beta\_vol} \\
       Type: \texttt{double[]} \\
       Dimensions: \texttt{[ns - 1]} \\
       Flux-surface averaged plasma beta on half-grid.
 \item \texttt{buco} \\
       Type: \texttt{double[]} \\
       Dimensions: \texttt{[ns - 1]} \\
       Profile of encosed toroidal current~$I$ on half-grid.
 \item \texttt{bvco} \\
       Type: \texttt{double[]} \\
       Dimensions: \texttt{[ns - 1]} \\
       Profile of encosed poloidal ribbon current~$G$ on half-grid.
 \item \texttt{vp} \\
       Type: \texttt{double[]} \\
       Dimensions: \texttt{[ns - 1]} \\
       Differential volume~$V'$ on half-grid.
 \item \texttt{equif} \\
       Type: \texttt{double[]} \\
       Dimensions: \texttt{[ns]} \\
       Radial force balance residual on full-grid.
 \item \texttt{specw} \\
       Type: \texttt{double[]} \\
       Dimensions: \texttt{[ns]} \\
       Spectral width~$M$ on full-grid.
 \item \texttt{phips} \\
       Type: \texttt{double[]} \\
       Dimensions: \texttt{[ns - 1]} \\
       Radial derivative of enclosed toroidal magnetic flux~$\phi'$ on the half-grid.
 \item \texttt{over\_r} \\
       Type: \texttt{double[]} \\
       Dimensions: \texttt{[ns - 1]} \\
       $\langle \tau / R \rangle / V'$ on half-grid
 \item \texttt{jdotb} \\
       Type: \texttt{double[]} \\
       Dimensions: \texttt{[ns]} \\
       $\mathbf{j} \cdot \mathbf{B}$ on full-grid.
 \item \texttt{bdotb} \\
       Type: \texttt{double[]} \\
       Dimensions: \texttt{[ns]} \\
       $\mathbf{B} \cdot \mathbf{B}$ on full-grid.
 \item \texttt{bdotgradv} \\
       Type: \texttt{double[]} \\
       Dimensions: \texttt{[ns]} \\
       Flux-surface-averaged toroidal magnetic field component $\mathbf{B} \cdot \nabla \zeta$ on full-grid.
 \item \texttt{DMerc} \\
       Type: \texttt{double[]} \\
       Dimensions: \texttt{[ns]} \\
       Full Mercier stability criterion on the full-grid.
 \item \texttt{DShear} \\
       Type: \texttt{double[]} \\
       Dimensions: \texttt{[ns]} \\
       Mercier stability criterion contribution due to magnetic shear on the full-grid.
 \item \texttt{DWell} \\
       Type: \texttt{double[]} \\
       Dimensions: \texttt{[ns]} \\
       Mercier stability criterion contribution due to magnetic well on the full-grid.
 \item \texttt{DCurr} \\
       Type: \texttt{double[]} \\
       Dimensions: \texttt{[ns]} \\
       Mercier stability criterion contribution due to plasma currents on the full-grid.
 \item \texttt{DGeod} \\
       Type: \texttt{double[]} \\
       Dimensions: \texttt{[ns]} \\
       Mercier stability criterion contribution due to geodesic curvature on the full-grid.
\end{itemize}
The Fourier representation of the magnetic axis geometry
is the next set of members in the \texttt{wout} group:
\begin{itemize}
 \item \texttt{raxis\_cc}, \texttt{raxis\_cs} \\
       Type: \texttt{double[]} \\
       Dimensions: \texttt{[ntor + 1]} \\
       Fourier coefficients of~$R(\zeta)$ of the magnetic axis geometry.
 \item \texttt{zaxis\_cs}, \texttt{zaxis\_cc} \\
       Type: \texttt{double[]} \\
       Dimensions: \texttt{[ntor + 1]} \\
       Fourier coefficients of~$Z(\zeta)$ of the magnetic axis geometry.
\end{itemize}
The state vector if VMEC++, i.e., the Fourier representation of the flux surface geometry ($R$ and $Z$)
and the stream function~$\lambda$
form the next set of members in the \texttt{wout} group:
\begin{itemize}
 \item \texttt{mnmax} \\
       Type: \texttt{int} \\
       Number of Fourier coefficients for the state vector.
 \item \texttt{xm} \\
       Type: \texttt{int[]} \\
       Dimensions: \texttt{[mnmax]} \\
       Poloidal mode numbers~$m$ for the Fourier coefficients in the state vector.
 \item \texttt{xn} \\
       Type: \texttt{int[]} \\
       Dimensions: \texttt{[mnmax]} \\
       Toroidal mode numbers times number of toroidal field periods~$n * n_\textrm{fp}$
       for the Fourier coefficients in the state vector.
 \item \texttt{rmnc}, \texttt{rmns} \\
       Type: \texttt{double[][]} \\
       Dimensions: \texttt{[ns][mnmax]} \\
       $R$ of the geometry of the flux surfaces on the full-grid.
 \item \texttt{zmns}, \texttt{zmnc} \\
       Type: \texttt{double[][]} \\
       Dimensions: \texttt{[ns][mnmax]} \\
       $Z$ of the geometry of the flux surfaces on the full-grid.
 \item \texttt{lmns\_full}, \texttt{lmnc\_full} \\
       Type: \texttt{double[][]} \\
       Dimensions: \texttt{[ns][mnmax]} \\
       $\lambda$ stream function on the full-grid.
 \item \texttt{lmns}, \texttt{lmnc} \\
       Type: \texttt{double[][]} \\
       Dimensions: \texttt{[ns][mnmax]} \\
       $\lambda$ stream function on the half-grid (for backwards-compatibility with ancient versions of Fortran VMEC).
\end{itemize}
The Fourier representation of
the Jacobian and the magnetic field on the flux surfaces
form the next set of members in the \texttt{wout} group:
\begin{itemize}
 \item \texttt{mnyq} \\
       Type: \texttt{int} \\
       Number of poloidal Fourier modes used to describe the Nyquist-quantites,
       i.e., the Jacobian, the magnetic field strength, and the magnetic field components.
 \item \texttt{nnyq} \\
       Type: \texttt{int} \\
       Number of toroidal Fourier modes used to describe the Nyquist-quantites,
       i.e., the Jacobian, the magnetic field strength, and the magnetic field components.
 \item \texttt{mnmax\_nyq} \\
       Type: \texttt{int} \\
       Number of Fourier coefficients for the Nyquist-quantites,
       i.e., the Jacobian, the magnetic field strength, and the magnetic field components.
 \item \texttt{xm} \\
       Type: \texttt{int[]} \\
       Dimensions: \texttt{[mnmax\_nyq]} \\
       Poloidal mode numbers~$m$ for the Fourier coefficients in the Nyquist-quantites,
       i.e., the Jacobian, the magnetic field strength, and the magnetic field components.
 \item \texttt{xn} \\
       Type: \texttt{int[]} \\
       Dimensions: \texttt{[mnmax\_nyq]} \\
       Toroidal mode numbers times number of toroidal field periods~$n * n_\textrm{fp}$
       for the Fourier coefficients in the Nyquist-quantites,
       i.e., the Jacobian, the magnetic field strength, and the magnetic field components.
 \item \texttt{gmnc}, \texttt{gmns} \\
       Type: \texttt{double[][]} \\
       Dimensions: \texttt{[ns - 1][mnmax\_nyq]} \\
       Fourier coefficients of the Jacobian~$\sqrt{g}$ of the coordinate transform between flux coordinates
       and cylindrical coordinates on the half-grid.
 \item \texttt{bmnc}, \texttt{bmns} \\
       Type: \texttt{double[][]} \\
       Dimensions: \texttt{[ns - 1][mnmax\_nyq]} \\
       Fourier coefficients of the magnetic field strength~$\vert\mathbf{B}\vert$ on the half-grid.
 \item \texttt{bsubumnc}, \texttt{bsubumns} \\
       Type: \texttt{double[][]} \\
       Dimensions: \texttt{[ns - 1][mnmax\_nyq]} \\
       Fourier coefficients of the covariant magnetic field component~$B_\theta$ on the half-grid.
 \item \texttt{bsubvmnc}, \texttt{bsubvmns} \\
       Type: \texttt{double[][]} \\
       Dimensions: \texttt{[ns - 1][mnmax\_nyq]} \\
       Fourier coefficients of the covariant magnetic field component~$B_\zeta$ on the half-grid.
 \item \texttt{bsubsmns}, \texttt{bsubsmnc} \\
       Type: \texttt{double[][]} \\
       Dimensions: \texttt{[ns][mnmax\_nyq]} \\
       Fourier coefficients of the covariant magnetic field component~$B_s$ on the full-grid.
 \item \texttt{bsupumnc}, \texttt{bsupumns} \\
       Type: \texttt{double[][]} \\
       Dimensions: \texttt{[ns - 1][mnmax\_nyq]} \\
       Fourier coefficients of the contravariant magnetic field component~$B^\theta$ on the half-grid.
 \item \texttt{bsupvmnc}, \texttt{bsupvmns} \\
       Type: \texttt{double[][]} \\
       Dimensions: \texttt{[ns - 1][mnmax\_nyq]} \\
       Fourier coefficients of the contravariant magnetic field component~$B^\zeta$ on the half-grid.
 \item \texttt{currumnc}, \texttt{currumns} \\
       Type: \texttt{double[][]} \\
       Dimensions: \texttt{[ns][mnmax\_nyq]} \\
       Fourier coefficients of the contravariant current density component~$j^\theta$ on the full-grid.
       Not implemented yet.
 \item \texttt{currvmnc}, \texttt{currvmns} \\
       Type: \texttt{double[][]} \\
       Dimensions: \texttt{[ns - 1][mnmax\_nyq]} \\
       Fourier coefficients of the contravariant current density component~$j^\zeta$ on the full-grid.
       Not implemented yet.
\end{itemize}
In case of a free-boundary run,
the Fourier representation of
the vacuum scalar magnetic potential on the plasma boundary
is the last set of members in the \texttt{wout} group:
\begin{itemize}
 \item \texttt{mnmaxpot} \\
       Type: \texttt{int} \\
       Number of Fourier modes used to represent
       the vacuum scalar magnetic potential on the boundary.
       Not implemented yet.
 \item \texttt{xmpot} \\
       Type: \texttt{int[]} \\
       Dimensions: \texttt{[mnmaxpot]} \\
       Poloidal mode numbers~$m$ for the Fourier coefficients
       of the vacuum scalar magnetic potential.
       Not implemented yet.
 \item \texttt{xnpot} \\
       Type: \texttt{int[]} \\
       Dimensions: \texttt{[mnmaxpot]} \\
       Toroidal mode numbers times number of toroidal field periods~$n * n_\textrm{fp}$
       for the Fourier coefficients
       of the vacuum scalar magnetic potential.
       Not implemented yet.
 \item \texttt{potsin}, \texttt{potcos} \\
       Type: \texttt{double[]} \\
       Dimensions: \texttt{[mnmaxpot]} \\
       Fourier coefficients of the vacuum scalar magnetic potential on the plasma boundary.
       Not implemented yet.
\end{itemize}

\chapter{Using VMEC}

\section{Representation of Flux Surface Geometry}
An inverse representation is used to describe the flux surface geometry in VMEC++.
This means that the (cylindrical) coordinates of the flux surface geometry
are expressed as functions of the flux coordinates.
\\
The three flux coordinates are:
\begin{itemize}
 \item $s$ is the normalized toroidal flux enclosed in a given flux surface.
       Its value is $0$ at the magnetic axis (in the core of the plasma) and $1$ at the last closed flux surface.
 \item $\theta$ is the poloidal angle-like coordinate.
       It increases by $2 \pi$ once around the torus the short way, i.e., wristband-like.
 \item $\varphi$ is the toroidal angle
       and it coincides with the angle coordinate in a cylindrical coordinate system.
       It increases by $2 \pi$ once around the torus the long way.
\end{itemize}
The full toroidal angle $\varphi$ is assumed to be made up of $n_\textrm{fp}$ identical, rotated, field periods of the plasma.
The toroidal coordinate in each field period is then
\begin{equation}
 \zeta = n_\textrm{fp} \varphi \, ,
\end{equation}
such that $\zeta$ goes from $0$ to $2 \pi$ over a single field period.

Having the flux coordinates $(s, \theta, \zeta)$, the cylindrical coordinates~$(R, \varphi, Z)$ of points on the flux surfaces
are given by:
\begin{align}
 R       =&\, R(s, \theta, \zeta) \nonumber \\
 \varphi =&\, \zeta / n_\textrm{fp} \\
 Z       =&\, Z(s, \theta, \zeta) \nonumber \, .
\end{align}
A set of discrete flux surfaces is considered in VMEC++.
This is reflected in the radial coordinate $s$, which only takes on discrete values in VMEC++.
The number of flux surfaces is an input parameter, which is called \texttt{ns}.
The individual flux surfaces are then (usually) located at equal increments in $s$:
\begin{equation}
 s_j = \frac{j}{\texttt{ns} - 1} \textrm{ for } j \in \{0, 1, ..., \texttt{ns} - 1\} \, .
\end{equation}
Toroidal symmetry is taken into account for the number of field periods, $n_\textrm{fp}$.

For each discrete flux surface, the coordinates $R$ and $Z$ on it are represented as two-dimensional Fourier series:
\begin{align}
 R(s_j, \theta, \zeta) =&\, \sum\limits_{m, n}
   \left[ \hat{R}_{j, m, n}^{\cos} \cos(m \theta - n \zeta) +
          \hat{R}_{j, m, n}^{\sin} \sin(m \theta - n \zeta) \right] \label{eqn:r_full} \\
 Z(s_j, \theta, \zeta) =&\, \sum\limits_{m, n}
   \left[ \hat{Z}_{j, m, n}^{\cos} \cos(m \theta - n \zeta) +
          \hat{Z}_{j, m, n}^{\sin} \sin(m \theta - n \zeta) \right] \label{eqn:z_full} \, .
\end{align}
In case of stellarator symmetry, half of the Fourier coefficients can be omitted
and the flux surface geometry is given by:
\begin{align}
 R(s_j, \theta, \zeta) =&\, \sum\limits_{m, n}
   \hat{R}_{j, m, n}^{\cos} \cos(m \theta - n \zeta) \label{eqn:r_symm} \\
 Z(s_j, \theta, \zeta) =&\, \sum\limits_{m, n}
   \hat{Z}_{j, m, n}^{\sin} \sin(m \theta - n \zeta) \textsl{}\label{eqn:z_symm} \, .
\end{align}
The summation over $m$ and $n$ goes as follows.
The poloidal mode number $m$ goes from $0$ to $\texttt{mpol} - 1$.
For $m = 0$, only non-negative toroidal mode numbers are taken into account,
thus $n \in \{0, 1, ..., \texttt{ntor}\}$.
For $m > 0$, all toroidal mode numbers are taken into account,
thus $n \in \{-\texttt{ntor}, ..., -1, 0, 1, ..., \texttt{ntor}\}$.
This scheme serves to remove the redundancy (and associated consistency requirement) among the Fourier coefficients with $m = 0$.

The geometry of the magnetic axis is represented in a similar functional form.
The axis is a line along the toroidal direction and thus has to variation in the poloidal direction.
Therefore, it is only parameterized in the toroidal direction.

The Fourier series for $R$ and $Z$ of the flux surface geometry
are typically implemented as follows:
\begin{align}
 R(s_j, \theta, \varphi) =& \sum_{\texttt{mn} = 0}^{\texttt{mnmax}}
   \left[   R_\texttt{mn}^\mathrm{cos}(s_j) \cos(\texttt{xm[mn]} \theta - \texttt{xn[mn]} \varphi)
          + R_\texttt{mn}^\mathrm{sin}(s_j) \sin(\texttt{xm[mn]} \theta - \texttt{xn[mn]} \varphi) \right] \\
 Z(s_j, \theta, \varphi) =& \sum_{\texttt{mn} = 0}^{\texttt{mnmax}}
   \left[   Z_\texttt{mn}^\mathrm{sin}(s_j) \sin(\texttt{xm[mn]} \theta - \texttt{xn[mn]} \varphi)
          + Z_\texttt{mn}^\mathrm{cos}(s_j) \cos(\texttt{xm[mn]} \theta - \texttt{xn[mn]} \varphi) \right]
\end{align}
on, for example, the $j$-th flux surface with flux surface label (i.e., radial coordinate)~$s_j$.
The dataset names of these Fourier coefficients in the \texttt{wout} group
of the main VMEC++ output file are listed in Tab.~\ref{tab:fc_rz}.
\begin{table}[h]
 \centering
 \begin{tabular}{|c|c|c|}
  \hline
  symbol & dataset name & stellarator-symmetric? \\
  \hline
  $R_\texttt{mn}^\mathrm{cos}$ & \texttt{rmnc} & y \\
  $Z_\texttt{mn}^\mathrm{sin}$ & \texttt{zmns} & y \\
  \hline
  $R_\texttt{mn}^\mathrm{sin}$ & \texttt{rmns} & n \\
  $Z_\texttt{mn}^\mathrm{cos}$ & \texttt{zmnc} & n \\
  \hline
 \end{tabular}
 \caption{Dataset names for the Fourier coefficient arrays in the \texttt{wout} group
          that describe the cylindrical coordinates~$R$ and $Z$ of the flux-surface geometry.}
 \label{tab:fc_rz}
\end{table}

\section{Representation of Magnetic Field}
The magnetic field~$\mathbf{B}$ in the VMEC++ outputs
is available in both a contravariant form
and a covariant form.
The contravariant form is defined as follows:
\begin{equation}
 \mathbf{B} = B^\theta \hat{\mathbf{e}}_\theta + B^\zeta \hat{\mathbf{e}}_\zeta
\end{equation}
and the covariant form is defined as follows:
\begin{equation}
 \mathbf{B} = B_s \nabla s + B_\theta \nabla \theta + B_\zeta \nabla \zeta \, .
\end{equation}
The components $B^\theta$, $B^\zeta$, $B_\theta$ and $B_\zeta$
are available on the radial half-grid.
The component $B_s$ is available on the radial full-grid.
All these components are expressed as two-dimensional Fourier series
to represent their scalar, real values on a discrete set of flux surfaces:
\begin{equation}
 X(s_j, \theta, \varphi) =
 \sum\limits_{m} \sum\limits_{n}
   \left[   \hat{X}_{mn}^\textrm{cos}(s_j) \cos(m \theta - n n_\textrm{fp} \varphi)
          + \hat{X}_{mn}^\textrm{sin}(s_j) \sin(m \theta - n n_\textrm{fp} \varphi) \right]
\end{equation}
where $X$ is one of $\{B^\theta, B^\zeta, B_s, B_\theta, B_\zeta\}$
and $\hat{X}_{mn}^\textrm{cos}$ ($\hat{X}_{mn}^\textrm{sin}$)
are the Fourier coefficients for the cosine (sine) basis functions, respectively,
on the $j$-th flux surface with flux surface label (i.e., radial coordinate)~$s_j$.
The number of toroidal field periods is denoted by~$n_\textrm{fp}$ here.
The poloidal angle-like coordinate~$\theta$ spans from $0$ to $2 \pi$
the short way around the torus.
The toroidal angle (indentical with the cylindrical angle)~$\varphi$
spans from $0$ to $2 \pi$ once around the machine.

The summation over $m$ and $n$ is to be taken over all relevant mode number combinates
needed to represent the magnetic field components.
Requiring that the components are real-valued
allows to make use of certain symmetries, reducing the number of required mode number
combinations to take into account.
In VMEC++, a linear mode index is used in conjunction with mode number arrays
for convenient one-dimensional summation:
\begin{align}
 X(s_j, \theta, \varphi) =
 \sum\limits_{\texttt{mn}=0}^{\texttt{mnmax\_nyq} - 1}
   \Bigl[ &\phantom{+}~ \hat{X}_\texttt{mn}^\textrm{cos}(s_j) \cos(\texttt{xm}[\texttt{mn}] \theta - \texttt{xn}[\texttt{mn}] \varphi) \nonumber \\
         ~&         +   \hat{X}_\texttt{mn}^\textrm{sin}(s_j) \sin(\texttt{xm}[\texttt{mn}] \theta - \texttt{xn}[\texttt{mn}] \varphi) \Bigr] \, .
\end{align}
It is noted that a larger number of Fourier coefficients~($\texttt{mnmax\_nyq}$)
is used to describe the magnetic field components,
than is used to describe the flux surface geometry~($\texttt{mnmax}$).
The dataset names of these Fourier coefficients in the \texttt{wout} group
of the main VMEC++ output file are listed in Tab.~\ref{tab:fc_gb}.
Note that the notation~$B = \vert\mathbf{B}\vert$
has been used for the magnetic field strength.
\begin{table}[h]
 \centering
 \begin{tabular}{|c|c|c|}
  \hline
  symbol & dataset name & stellarator-symmetric? \\
  \hline
  $\sqrt{g}_\texttt{mn}^\mathrm{cos}$   & \texttt{gmnc}     & y \\
  $B_\texttt{mn}^\mathrm{cos}$          & \texttt{bmnc}     & y \\
  $B_{\theta,\texttt{mn}}^\mathrm{cos}$ & \texttt{bsubumnc} & y \\
  $B_{\zeta,\texttt{mn}}^\mathrm{cos}$  & \texttt{bsubvmnc} & y \\
  $B_{s,\texttt{mn}}^\mathrm{sin}$      & \texttt{bsubsmns} & y \\
  $B_\texttt{mn}^{\theta,\mathrm{cos}}$ & \texttt{bsupumnc} & y \\
  $B_\texttt{mn}^{\zeta,\mathrm{cos}}$  & \texttt{bsupvmnc} & y \\
  \hline
  $\sqrt{g}_\texttt{mn}^\mathrm{sin}$   & \texttt{gmns}     & n \\
  $B_\texttt{mn}^\mathrm{sin}$          & \texttt{bmns}     & n \\
  $B_{\theta,\texttt{mn}}^\mathrm{sin}$ & \texttt{bsubumns} & n \\
  $B_{\zeta,\texttt{mn}}^\mathrm{sin}$  & \texttt{bsubvmns} & n \\
  $B_{s,\texttt{mn}}^\mathrm{cos}$      & \texttt{bsubsmnc} & n \\
  $B_\texttt{mn}^{\theta,\mathrm{sin}}$ & \texttt{bsupumns} & n \\
  $B_\texttt{mn}^{\zeta,\mathrm{sin}}$  & \texttt{bsupvmns} & n \\
  \hline
 \end{tabular}
 \caption{Dataset names for the Fourier coefficient arrays in the \texttt{wout} group
          that describe the Jacobian, the magnetic field strength,
          and the co- and contravariant magnetic field components.}
 \label{tab:fc_gb}
\end{table}

\section{Re-Computing the Rotational Transform Profile}
The contravariant magnetic field components are defined as follows in VMEC~\cite{lee_1988}:
\begin{align}
 B^{\theta}  =&\, \frac{\Phi'}{\sqrt{g}} \left( \iota - \frac{\partial \lambda}{\partial \varphi} \right) \label{eqn:bsupu} \\
 B^{\varphi} =&\, \frac{\Phi'}{\sqrt{g}} \left(    1  + \frac{\partial \lambda}{\partial \theta } \right) \label{eqn:bsupv} \, .
\end{align}
The covariant magnetic field components can be computed from these using the metric elements:
\begin{align}
 B_{\theta } =&\, g_{\theta  \theta} B^{\theta} + g_{\theta  \varphi} B^{\varphi} \label{eqn:bsubu} \\
 B_{\varphi} =&\, g_{\theta \varphi} B^{\theta} + g_{\varphi \varphi} B^{\varphi} \label{eqn:bsubv} \, .
\end{align}
The radial profile of enclosed net toroidal current is given by:
\begin{align}
 \frac{\mu_0}{2 \pi} I_\zeta
 =&\, \langle B_\theta \rangle \label{eqn:curtor} \, .
\end{align}
where $\langle . \rangle$ is an average over the corresponding flux surface:
\begin{equation}
 \langle B_\theta \rangle = \frac{1}{(2 \pi)^2} \int\limits_{0}^{2 \pi} \int\limits_{0}^{2 \pi}
        B_\theta(s, \theta, \varphi) \,\mathrm{d}\theta \,\mathrm{d}\varphi \, .
\end{equation}
Now, \eqn{bsupu} and \eqn{bsupv} are inserted into \eqn{bsubu}:
\begin{align}
 B_{\theta } =&\, \frac{\Phi'}{\sqrt{g}} \left[
       g_{\theta  \theta}  \left( \iota - \frac{\partial \lambda}{\partial \varphi} \right)
     + g_{\theta  \varphi} \left(    1  + \frac{\partial \lambda}{\partial \theta } \right) \right] \, .
\end{align}
This in turn is inserted into \eqn{curtor}:
\begin{equation}
 \frac{\mu_0}{2 \pi} I_\zeta
 = \left\langle
     \frac{\Phi'}{\sqrt{g}} \left[
       g_{\theta  \theta}  \left( \iota - \frac{\partial \lambda}{\partial \varphi} \right)
     + g_{\theta  \varphi} \left(    1  + \frac{\partial \lambda}{\partial \theta } \right) \right] \right\rangle \, .
\end{equation}
The linearity of the integral allows to split this term into several parts.
Also, the quantities $\Phi'$ and $\iota$ only depend on the radial coordinate
and can be pulled out of the surface-averaging operator.
It thus follows:
\begin{equation}
 \frac{\mu_0}{2 \pi} \frac{I_\zeta}{\Phi'}
 =
 \iota \left\langle \frac{g_{\theta  \theta}}{\sqrt{g}} \right\rangle
 + \left\langle - \frac{g_{\theta  \theta }}{\sqrt{g}} \frac{\partial \lambda}{\partial \varphi}
                + \frac{g_{\theta  \varphi}}{\sqrt{g}} \frac{\partial \lambda}{\partial \theta } \right\rangle
 + \left\langle \frac{g_{\theta  \varphi}}{\sqrt{g}} \right\rangle \, .
\end{equation}
This can be re-arranged for the iota profile:
\begin{equation}
 \iota \left\langle \frac{g_{\theta  \theta}}{\sqrt{g}} \right\rangle
 =
 \frac{\mu_0}{2 \pi} \frac{I_\zeta}{\Phi'}
 - \left\langle - \frac{g_{\theta  \theta }}{\sqrt{g}} \frac{\partial \lambda}{\partial \varphi}
                + \frac{g_{\theta  \varphi}}{\sqrt{g}} \frac{\partial \lambda}{\partial \theta } \right\rangle
 - \left\langle \frac{g_{\theta  \varphi}}{\sqrt{g}} \right\rangle \label{eqn:iota_eqn} \, .
\end{equation}
The quantity $\lambda$ is represented as a Fourier series in VMEC
(with only $\sin$-coefficients assuming stellarator-symmetry):
\begin{equation}
 \lambda(s, \theta, \varphi)
 =
 \sum\limits_{m,n}
   \hat{\lambda}^{\sin}_{mn}(s)
   \sin\left(m \theta - n n_\textrm{fp} \varphi \right) \, .
\end{equation}
This yields analytical derivatives in both tangential directions:
\begin{align}
 \frac{\partial \lambda}{\partial \theta}
 =&\, \sum\limits_{m,n}
        m
        \hat{\lambda}^{\sin}_{mn}
        \cos\left(m \theta - n n_\textrm{fp} \varphi \right) \label{eqn:lu} \\
 \frac{\partial \lambda}{\partial \varphi}
 =&\, \sum\limits_{m,n}
        (-n n_\textrm{fp})
        \hat{\lambda}^{\sin}_{mn}
        \cos\left(m \theta - n n_\textrm{fp} \varphi \right) \label{eqn:lv} \, .
\end{align}
\eqn{lu} and \eqn{lv} are now inserted into \eqn{iota_eqn}.
The Fourier coefficients can be regarded as functions
of only the radial coordinate $s$ and can therefore be pulled out of the surface averaging operator.
\begin{equation}
 \iota \left\langle \frac{g_{\theta  \theta}}{\sqrt{g}} \right\rangle
 =
 - \sum\limits_{m,n} \left\langle \left(- \frac{g_{\theta  \theta }}{\sqrt{g}} (-n n_\textrm{fp})
                                        + \frac{g_{\theta  \varphi}}{\sqrt{g}}   m
                                  \right) \cos\left(m \theta - n n_\textrm{fp} \varphi \right)
                      \right\rangle
                      \hat{\lambda}^{\sin}_{mn}
 - \left\langle \frac{g_{\theta  \varphi}}{\sqrt{g}} \right\rangle
 + \frac{\mu_0}{2 \pi} \frac{I_\zeta}{\Phi'} \label{eqn:full_iota} \, .
\end{equation}
A matrix $\mathbf{A}$ with entries $a_{mn}$ and vectors $b$ and $c$ are now introduced with
\begin{equation}
 a_{mn} = \left\langle \left(  \frac{g_{\theta  \theta }}{\sqrt{g}} n n_\textrm{fp}
                                   + \frac{g_{\theta  \varphi}}{\sqrt{g}} m
                             \right) \cos\left(m \theta - n n_\textrm{fp} \varphi \right) \right\rangle
          /
          \left\langle \frac{g_{\theta  \theta}}{\sqrt{g}} \right\rangle \label{eqn:a_mat}
\end{equation}
as well as
\begin{equation}
 b = \left\langle \frac{g_{\theta  \varphi}}{\sqrt{g}} \right\rangle
     /
     \left\langle \frac{g_{\theta  \theta}}{\sqrt{g}} \right\rangle \label{eqn:b_vec} \, .
\end{equation}
and
\begin{equation}
 c = \frac{\mu_0}{2 \pi} \frac{I_\zeta}{\Phi'}
     /
     \left\langle \frac{g_{\theta  \theta}}{\sqrt{g}} \right\rangle \label{eqn:c_vec} \, .
\end{equation}
This allows to write \eqn{full_iota} as a linear system of equations:
\begin{equation}
 \iota = -\sum\limits_{m,n} a_{mn} \hat{\lambda}^{\sin}_{mn} - b + c \label{eqn:iota_from_linear_eqns} \, .
\end{equation}

\subsection{Numerical Implementation}
The objective of this section is go detail out the computation of
the quantites from \eqn{a_mat}, \eqn{b_vec}, \eqn{c_vec} for evaluation of $\iota$
from \eqn{iota_from_linear_eqns}.
The rotational transform profile is on the half-grid in VMEC.
Also, the metric elements, the Jacobian and the enclosed toroidal current as well as the toroidal flux derivative
are available on the half-grid in VMEC.
The state variable $\lambda$ is on the full-grid in VMEC
and needs to be averaged onto the half-grid before being used in \eqn{iota_from_linear_eqns}.

\subsection{Comparison with Mercier's Expression}
Mercier's expression for the rotational transform on the magnetic axis
of a stellarator is as follows:
\begin{equation}
 \iota =
 \frac{1}{2 \pi} \int\limits_{0}^{L}
   \frac{1}{\cosh \eta} \left[
       \frac{\mu_0 J}{2 B_0}
     - \left(\cosh \eta - 1 \right) d'
     - \tau
  \right]  \,\mathrm{d}l - N
\end{equation}
The three contributions in this expression
are due to the toroidal current,
a poloidal rotation of the poloidal cross section of the flux surfaces close to the axis
and torsion of the magnetic axis.
$\lambda$ is identically zero on the magnetic axis,
because the magnetic field is purely toroidal there and has no variation over the surface
of the axis, because the axis has no surface.
It is expected that the expression in \eqn{iota_from_linear_eqns} should approach Mercier's result
when moving towards the magnetic axis.
Here, $\mu_0 J$ is the current density on the magnetic axis.

\section{Current Density from VMEC}
The current density from VMEC is computed using Ampere's law:
\begin{equation}
 \mathbf{j} = \frac{1}{\mu_0} \nabla \times \mathbf{B} \, .
\end{equation}
If the curl of a covariant representation of a field is computed,
this results in a contravariant representation of the resulting field and vice versa~\cite{dHaseleer}.
Therefore, the covariant magnetic field components $B_i$ are needed
to compute the contravariant current density components $j^i$.
The covariant magnetic field components~$(B_s, B_\theta, B_\varphi)$ represent the equilibrium magnetic field $\mathbf{B}$ as follows:
\begin{equation}
 \mathbf{B} = B_s \nabla s + B_\theta \nabla \theta + B_\zeta \nabla \zeta \, .
\end{equation}
They are available in the output of VMEC as two-dimensional Fourier series on the half-grid:
\begin{align}
 B_s      (s_{j+0.5}, \theta, \varphi) =& \sum_{m,n} \left[  \hat{B}_{s,      mn}^\mathrm{cos} (s_{j+0.5}) \,\mathrm{cos}(m \theta - n n_\mathrm{fp} \varphi)
                                                            + \hat{B}_{s,      mn}^\mathrm{sin} (s_{j+0.5}) \,\mathrm{sin}(m \theta - n n_\mathrm{fp} \varphi) \right] \nonumber \\
 B_\theta (s_{j+0.5}, \theta, \varphi) =& \sum_{m,n} \left[  \hat{B}_{\theta, mn}^\mathrm{cos} (s_{j+0.5}) \,\mathrm{cos}(m \theta - n n_\mathrm{fp} \varphi)
                                                            + \hat{B}_{\theta, mn}^\mathrm{sin} (s_{j+0.5}) \,\mathrm{sin}(m \theta - n n_\mathrm{fp} \varphi) \right] \\
 B_\varphi  (s_{j+0.5}, \theta, \varphi) =& \sum_{m,n} \left[  \hat{B}_{\varphi,  mn}^\mathrm{cos} (s_{j+0.5}) \,\mathrm{cos}(m \theta - n n_\mathrm{fp} \varphi)
                                                            + \hat{B}_{\varphi,  mn}^\mathrm{sin} (s_{j+0.5}) \,\mathrm{sin}(m \theta - n n_\mathrm{fp} \varphi) \right] \nonumber
\end{align}
with the $m,n$ summation as detailed in Sec.~\ref{sec:torcoords_numbering},
albeit more Fourier coefficients are retained in this expansion
in order to satisfy the Nyquist sampling theorem.
Note that, in contrast to the contravariant magnetic field representation,
the radial component $B_s \neq 0$ in the covariant representation.
Given the covariant magnetic field representation, the curl of it takes a particurarly simple form~\cite{dHaseleer}:
\begin{equation}
 \mu_0 \mathbf{j}^i = \frac{1}{\sqrt{g}} \sum_k \left(  \frac{\partial B_j}{\partial i}
                                             - \frac{\partial B_i}{\partial j} \right)
                                       \hat{\mathbf{e}}_k, \quad (i,j,k) ~ \mathrm{cyc.} ~ (s, \theta, \varphi)
\end{equation}
with $\sqrt{g}$ being the Jacobian of the coordinate transform between flux coordinates and cylindrical coordinates.
The current density is, in practise, often computed to act as a source term in the Biot-Savart volume integrals.
There, the Jacobian~$\sqrt{g}$ is required in the integral.
It is therefore customary to compute~$\mathbf{j} \sqrt{g}$:
\begin{equation}
 j^i \sqrt{g} = \frac{1}{\mu_0} \left( \frac{B_k}{\partial u^j} - \frac{B_j}{\partial u^k} \right) , \quad (i,j,k) \textrm{ cyc. } (s,\theta,\zeta) \, .
\end{equation}
The concrete contravariant components of the current density are thus:
\begin{align}
 j^s      \sqrt{g} =&\, \frac{1}{\mu_0} \left( \frac{\partial B_\zeta }{\partial \theta} - \frac{\partial B_\theta}{\partial \zeta } \right) \nonumber \\
 j^\theta \sqrt{g} =&\, \frac{1}{\mu_0} \left( \frac{\partial B_s     }{\partial \zeta } - \frac{\partial B_\zeta }{\partial s     } \right) \label{eqn:currx} \\
 j^\zeta  \sqrt{g} =&\, \frac{1}{\mu_0} \left( \frac{\partial B_\theta}{\partial s     } - \frac{\partial B_s     }{\partial \theta} \right) \nonumber
\end{align}
The current density vector $\mathbf{j}$ is constrained to be tangential to flux surfaces in VMEC++.
Therefore, the current density can only have non-zero components in the tangential $\theta$ and $\varphi$ directions and $j^s = 0$ is assumed.
This allows to formulate a magnetic differential equation:
\begin{align}
 \frac{\partial B_\zeta }{\partial \theta} - \frac{\partial B_\theta}{\partial \zeta } = 0 \, .
\end{align}
We only consider $j^\theta$ and $j^\zeta$ in the following.
The two-dimensional Fourier series representation of the covariant magnetic field components
allows to perform the tangential derivatives in \eqn{currx} analytically:
\begin{align}
 \frac{\partial B_i}{\partial \theta}(s, \theta, \zeta)
 =&\, \sum\limits_{m,n} \left[
   \hat{B}^\textrm{cos}_{i,m,n}(s) m (-\sin)(m \theta - n \zeta) +
   \hat{B}^\textrm{sin}_{i,m,n}(s) m   \cos (m \theta - n \zeta)
 \right] \nonumber \\ ~
 =&\, \sum\limits_{m,n} \left[
   \left(-m \hat{B}^\textrm{cos}_{i,m,n}(s)\right) \sin(m \theta - n \zeta) +
   \left( m \hat{B}^\textrm{sin}_{i,m,n}(s)\right) \cos(m \theta - n \zeta)
 \right] \\
 \frac{\partial B_i}{\partial \zeta}(s, \theta, \zeta)
 =&\, \sum\limits_{m,n} \left[
   \hat{B}^\textrm{cos}_{i,m,n}(s) (-n) (-\sin)(m \theta - n \zeta) +
   \hat{B}^\textrm{sin}_{i,m,n}(s) (-n) \cos(m \theta - n \zeta)
 \right] \nonumber \\ ~
 =&\, \sum\limits_{m,n} \left[
   \left( n \hat{B}^\textrm{cos}_{i,m,n}(s)\right) \sin(m \theta - n \zeta) +
   \left(-n \hat{B}^\textrm{sin}_{i,m,n}(s)\right) \cos(m \theta - n \zeta)
 \right] \, .
\end{align}
The quantities $j^i \sqrt{g} \equiv J^i$ are also Fourier-decomposed:
\begin{equation}
 J^i(s, \theta, \zeta) = \sum\limits_{m,n} \left[
   \hat{J}^{i,\textrm{cos}}_{m,n}(s) \cos(m \theta - n \zeta) +
   \hat{J}^{i,\textrm{sin}}_{m,n}(s) \sin(m \theta - n \zeta)
 \right] \, .
\end{equation}
The orthogonality of the Fourier basis now allows to express \eqn{currx}
for each of the coefficients independently:
\begin{align}
 \hat{J}^{\theta,\textrm{cos}}_{m,n}(s) =&\, \frac{1}{\mu_0} \left(
   -n \hat{B}^\textrm{sin}_{s,m,n}(s) - \frac{\partial \hat{B}^\textrm{cos}_{\zeta,m,n}}{\partial s}
 \right) \\
 \hat{J}^{\theta,\textrm{sin}}_{m,n}(s) =&\, \frac{1}{\mu_0} \left(
   ~ \,\, n \hat{B}^\textrm{cos}_{s,m,n}(s) - \frac{\partial \hat{B}^\textrm{sin}_{\zeta,m,n}}{\partial s}
 \right) \\
 \hat{J}^{\zeta,\textrm{cos}}_{m,n}(s) =&\, \frac{1}{\mu_0} \left(
   \frac{\partial \hat{B}^\textrm{cos}_{\theta,m,n}}{\partial s} - m \hat{B}^\textrm{sin}_{s,m,n}(s)
 \right) \\
 \hat{J}^{\zeta,\textrm{sin}}_{m,n}(s) =&\, \frac{1}{\mu_0} \left(
   \frac{\partial \hat{B}^\textrm{sin}_{\theta,m,n}}{\partial s} + m \hat{B}^\textrm{cos}_{s,m,n}(s)
 \right)
\end{align}
If the current density components are desired to be computed on the full-grid,
the radial derivatives can make use of the fact that $B_\theta$ and $B_\zeta$
are computed on the half-grid in VMEC++.
Derivatives with respect to $s$ are computed as finite differences across the discrete flux surfaces.
Regularization terms have to be included in order to
reduce the impact of catastrophic cancellation near the magnetic axis~\cite{hirshman_vmec_currents}.
Only the stellarator-symmetric current density components~$\hat{j}_{mn}^{\theta, \mathrm{cos}}$
and~$\hat{j}_{mn}^{\varphi,  \mathrm{cos}}$ are considered in the following.
The non-stellarator-symmetric components follow analogously.
The radial derivatives~$\partial \hat{B}_{\theta, mn}^\mathrm{cos} / \partial s$
and~$\partial \hat{B}_{\varphi, mn}^\mathrm{cos} / \partial s$ are computed as follows:
\begin{align}
\frac{\partial \hat{B}_{\theta, mn}^\mathrm{cos}}{\partial s} (s_j)
 =&\, \begin{cases}
        \frac{1}{\Delta s} \left( \hat{B}_{\theta, mn}^\mathrm{cos}(s_{j+0.5}) - \hat{B}_{\theta, mn}^\mathrm{cos}(s_{j-0.5}) \right) &:\, \textrm{ even } m \\
        \frac{\sqrt{s_j}}{\Delta s} \left(
            \hat{b}_{\theta, mn}^\mathrm{cos}(s_{j+0.5})
          - \hat{b}_{\theta, mn}^\mathrm{cos}(s_{j-0.5}) \right) \\
        ~ + \frac{1}{4 \sqrt{s_j}} \left(
              \hat{b}_{\theta, mn}^\mathrm{cos}(s_{j+0.5})
            + \hat{b}_{\theta, mn}^\mathrm{cos}(s_{j-0.5}) \right)
          &:\, \textrm{ odd } m
      \end{cases} \\
 \frac{\partial \hat{B}_{\varphi, mn}^\mathrm{cos}}{\partial s} (s_j)
 =&\, \begin{cases}
        \frac{1}{\Delta s} \left( \hat{B}_{\varphi, mn}^\mathrm{cos}(s_{j+0.5}) - \hat{B}_{\varphi, mn}^\mathrm{cos}(s_{j-0.5}) \right) &:\, \textrm{ even } m \\
        \frac{\sqrt{s_j}}{\Delta s} \left(
            \hat{b}_{\varphi, mn}^\mathrm{cos}(s_{j+0.5})
          - \hat{b}_{\varphi, mn}^\mathrm{cos}(s_{j-0.5}) \right) \\
        ~ + \frac{1}{4 \sqrt{s_j}} \left(
              \hat{b}_{\varphi, mn}^\mathrm{cos}(s_{j+0.5})
            + \hat{b}_{\varphi, mn}^\mathrm{cos}(s_{j-0.5}) \right)
          &:\, \textrm{ odd } m
      \end{cases}
\end{align}
with
\begin{align}
 \hat{b}_{\theta, mn}^\mathrm{cos}(s_{j+0.5}) =&\, \frac{\hat{B}_{\theta, mn}^\mathrm{cos}(s_{j+0.5})}{\sqrt{s_{j+0.5}}} \\
 \hat{b}_{\varphi, mn}^\mathrm{cos}(s_{j+0.5}) =&\, \frac{\hat{B}_{\varphi, mn}^\mathrm{cos}(s_{j+0.5})}{\sqrt{s_{j+0.5}}} \, .
\end{align}
This is the differencing scheme introduced in Ref.~\cite{hirshman_schwenn_nuehrenberg_1990}.
The regularized interpolation of~$\hat{B}_{s,mn}^\mathrm{sin}$ is done as follows:
\begin{align}
 \hat{B}_{s,mn}^\mathrm{sin}
 = \begin{cases}
    \frac{1}{2} \left( \hat{B}_{s, mn}^\mathrm{sin}(s_{j+0.5}) + \hat{B}_{s, mn}^\mathrm{sin}(s_{j-0.5}) \right) &:\, \textrm{ even } m \\
    \frac{1}{2 \sqrt{s_j}} \left(
        \hat{b}_{s, mn}^\mathrm{cos}(s_{j+0.5})
      + \hat{b}_{s, mn}^\mathrm{cos}(s_{j-0.5}) \right) &:\, \textrm{ odd } m
   \end{cases}
\end{align}
with
\begin{align}
 \hat{b}_{s, mn}^\mathrm{cos}(s_{j+0.5})
 = \sqrt{s_{j+0.5}} \hat{B}_{s,mn}^\mathrm{sin}(s_{j+0.5}) \, .
\end{align}
The mapping between intermediate quantities occuring throughout the computation of the current density in VMEC
and the corresponding variables in the Fortran VMEC implementation is found in Tab.~\ref{tab:compute_currents_vars}.
\begin{table}[tb]
 \centering
 \begin{tabular}{|c|c|}
  \hline
  Fortran Variable name & Physical quantity \\
  \hline
  \texttt{bu0} & $\hat{b}_{\theta, mn}^\mathrm{cos}(s_{j-0.5})$ \\
  \texttt{bu1} & $\hat{b}_{\theta, mn}^\mathrm{cos}(s_{j+0.5})$ \\
  \texttt{bv0} & $\hat{b}_{\varphi, mn}^\mathrm{cos}(s_{j-0.5})$ \\
  \texttt{bv1} & $\hat{b}_{\varphi, mn}^\mathrm{cos}(s_{j+0.5})$ \\
  \hline
  \texttt{t1} & $\hat{B}_{s,mn}^\mathrm{sin} (s_j)$ \\
  \texttt{t2} & $\frac{\partial \hat{B}_{\theta, mn}^\mathrm{cos}}{\partial s} (s_j)$ \\
  \texttt{t3} & $\frac{\partial \hat{B}_{\varphi, mn}^\mathrm{cos}}{\partial s} (s_j)$ \\
  \hline
 \end{tabular}
 \caption{Fortran variable names for the various physical quantities
          occuring throughout the computation of the current density in VMEC.}
 \label{tab:compute_currents_vars}
\end{table}
The \texttt{b\{u,v\}\{0,1\}} terms are used to compute the \texttt{t\{2,3\}} terms as follows:
\begin{align}
 \texttt{t2} =&\, \begin{cases}
                   \frac{   1      }{\Delta s} \left( \hat{B}_{\theta, mn}^\mathrm{cos}(s_{j+0.5}) - \hat{B}_{\theta, mn}^\mathrm{cos}(s_{j-0.5}) \right)     &:\, \textrm{even } m \\
                   \frac{\sqrt{s_j}}{\Delta s} \left( \texttt{bu1} - \texttt{bu0} \right) + \frac{1}{4 \sqrt{s_j}} \left( \texttt{bu1} + \texttt{bu0} \right) &:\, \textrm{odd  } m
                  \end{cases} \\
 \texttt{t3} =&\, \begin{cases}
                   \frac{   1      }{\Delta s} \left( \hat{B}_{\varphi, mn}^\mathrm{cos}(s_{j+0.5}) - \hat{B}_{\varphi, mn}^\mathrm{cos}(s_{j-0.5}) \right)       &:\, \textrm{even } m \\
                   \frac{\sqrt{s_j}}{\Delta s} \left( \texttt{bv1} - \texttt{bv0} \right) + \frac{1}{4 \sqrt{s_j}} \left( \texttt{bv1} + \texttt{bv0} \right) &:\, \textrm{odd  } m  \, .
                  \end{cases}
\end{align}
Using these definitions, the Fourier coefficients of the current density (times the Jacobian)
are computed as follows in the VMEC subroutine \texttt{Compute\_Currents}:
\begin{align}
 \texttt{currumnc} = (\sqrt{g} j^\theta)^{\textrm{cos}}_{mn}
   =&\, \frac{1}{\mu_0} \left( -n \cdot \texttt{t1} - \texttt{t3} \right) \\
 \texttt{currvmnc} = (\sqrt{g} j^\varphi )^{\textrm{cos}}_{mn}
   =&\, \frac{1}{\mu_0} \left( -m \cdot \texttt{t1} + \texttt{t2} \right) \, .
\end{align}
Linear extrapolation is used to compute the current density Fourier coefficents with $m \leq 1$ at the magnetic axis,
with all other Fourier harmonics set to zero at the magnetic axis.
Linear extrapolation is also used to compute the current density coefficients at the LCFS for all Fourier harmonics.
The source term in the Biot-Savart integral over the plasma current density,
namely the cylindrical components of the current density~$(j^{R}, j^{\varphi}, j^{Z})$ times the Jacobian,
are then computed from the contravariant representation as follows
for the remaining $j^R$ and $j^Z$:
\begin{align}
 \sqrt{g} j^{R}       =&\,   \sqrt{g} j^\theta \frac{\partial R}{\partial \theta} + \sqrt{g} j^\varphi \frac{\partial R}{\partial \varphi} \\
 \sqrt{g} j^{Z}       =&\,   \sqrt{g} j^\theta \frac{\partial Z}{\partial \theta} + \sqrt{g} j^\varphi \frac{\partial Z}{\partial \varphi} \, .
\end{align}

\section{Field Line Following in VMEC}
Straight-fieldline-coordinate are $(s, \theta^*, \zeta)$ in VMEC, where
\begin{equation}
 \theta^* = \theta + \lambda(s,\theta,\zeta) \label{eqn:theta_star}
\end{equation}
and $(s,\theta,\zeta)$ are the VMEC-internal coordinates
that lead to spectrally-condensed Fourier representations for $R$ and $Z$.
The stream function $\lambda$ is expressed as a Fourier series as well
(considering only the stellarator-symmetric case for now):
\begin{equation}
 \lambda(s,\theta,\zeta) = \sum\limits_{m,n} \hat{\lambda}^\textrm{sin}_{m,n}(s) \sin(m \theta - n \zeta) \, .
\end{equation}
Assume a certain location $(s,\theta,\zeta)$ is to be mapped along the magnetic field lines
to another toroidal location $\zeta'$.
In order to achieve this, we need to:
\begin{enumerate}
 \item compute $\theta^*_\textrm{target} = \theta^*(s,\theta,\zeta)$ and
 \item find $\theta'$ such that $\theta^*(s,\theta',\zeta') = \theta^*_\textrm{target}$.
\end{enumerate}
The first point is done by simply evaluating \eqn{theta_star}.
For the second step, a non-linear root-finding method can be employed to
find the root of the following objective function:
\begin{equation}
 f(\theta') = \theta^*(s,\theta',\zeta') - \theta^*_\textrm{target} \, . \label{eqn:theta_star_objective}
\end{equation}
It can be safely assumed that there exists only one solution
and furthermore, $\theta^*$ can be assumed to be monotonic with respect to $\theta$.
This allows to use a Newton method for solving \eqn{theta_star_objective}.
The Newton method needs derivate information, which can be computed analytically:
\begin{equation}
 \frac{\partial f}{\partial \theta}(s,\theta',\zeta')
 =
 \frac{\partial \theta^*}{\partial \theta}(s,\theta',\zeta')
 =
 1 + \frac{\partial \lambda}{\partial \theta}(s,\theta',\zeta') \, . \label{eqn:df_dthetap}
\end{equation}
The derivative of $\lambda$ can be computed analytically as well:
\begin{equation}
 \frac{\partial \lambda}{\partial \theta}(s,\theta',\zeta')
 =
 \sum\limits_{m,n} m \hat{\lambda}^\textrm{sin}_{m,n}(s) \cos(m \theta' - n \zeta') \, .
\end{equation}
It is expected that the objective function $f$ needs to be evaluated multiple times
for different values of $\theta'$, but always at the given values of $s$ and $\zeta'$
during the root-finding procedure.
Thus, a transformed Fourier representation for $\lambda$ and $\partial \lambda/\partial \theta$
can be derived:
\begin{align}
 \lambda\vert_{\zeta'}(\theta')
 =&\,
 \sum\limits_{m,n} \hat{\lambda}^\textrm{sin}_{m,n}(s) \sin(m \theta' - n \zeta') \nonumber \\ ~
 =&\,
 \sum\limits_{m} \sum\limits_{n} \hat{\lambda}^\textrm{sin}_{m,n}(s) \left[
   \sin(m \theta') \cos(n \zeta') - \cos(m \theta') \sin(n \zeta')
 \right] \nonumber \\ ~
 =&\,
 \sum\limits_{m} \left[
   \Bigl(\sum\limits_{n} \hat{\lambda}^\textrm{sin}_{m,n}(s) \cos(n \zeta') \Bigr) \sin(m \theta') -
   \Bigl(\sum\limits_{n} \hat{\lambda}^\textrm{sin}_{m,n}(s) \sin(n \zeta') \Bigr) \cos(m \theta')
 \right] \nonumber \\ ~
 =&\,
 \sum\limits_{m} \left[
   \hat{\lambda}^\textrm{sin}_{m,\zeta'}(s) \sin(m \theta') -
   \hat{\lambda}^\textrm{cos}_{m,\zeta'}(s) \cos(m \theta')
 \right] \label{eqn:lambda_at_zetap}
\end{align}
with
\begin{align}
 \hat{\lambda}^\textrm{sin}_{m,\zeta'}(s)
 =&\, \sum\limits_{n} \hat{\lambda}^\textrm{sin}_{m,n}(s) \cos(n \zeta') \label{eqn:lmszeta} \\
 \hat{\lambda}^\textrm{cos}_{m,\zeta'}(s)
 =&\, \sum\limits_{n} \hat{\lambda}^\textrm{sin}_{m,n}(s) \sin(n \zeta') \label{eqn:lmczeta} \, .
\end{align}
The coefficients $\hat{\lambda}^\textrm{sin}_{m,\zeta'}(s)$ and $\hat{\lambda}^\textrm{cos}_{m,\zeta'}(s)$
can be computed once for the given values of $s$ and $\zeta'$
and then only the summation over $m$ in \eqn{lambda_at_zetap} has to be performed
in every iteration of the root-finding algorithm.
Similarly, the computation of $\partial \lambda/\partial \theta$ can be streamlined,
now already based on \eqn{lambda_at_zetap}:
\begin{equation}
 \frac{\partial \lambda}{\partial \theta}\vert_{\zeta'}(\theta')
 =
 \sum\limits_{m} \left[
   m \hat{\lambda}^\textrm{sin}_{m,\zeta'}(s) \cos(m \theta') +
   m \hat{\lambda}^\textrm{cos}_{m,\zeta'}(s) \sin(m \theta')
 \right] \label{eqn:lu_at_zetap}
\end{equation}
Thus, the following algorithm can be used for field-line following on VMEC equilibria:
\begin{enumerate}
 \item Assume that $s$, $\theta$, $\zeta$ and $\zeta'$ are given.
       Assume that the Fourier coefficients $\hat{\lambda}^\textrm{sin}_{m,n}(s)$ are given for all $m$ and $n$.
       The desired tolerance is $\epsilon$.
 \item Compute $\theta^*_\textrm{target}$ using \eqn{theta_star}.
 \item For each $m$, pre-compute \eqn{lmszeta} and \eqn{lmczeta}.
 \item Initialize $\theta' \leftarrow \theta$ as an initial guess.
 \item Evaluate the objective function \eqn{theta_star_objective}:
       \begin{enumerate}
         \item Compute $\lambda\vert_{\zeta'}(\theta')$ using \eqn{lambda_at_zetap}.
         \item Compute $\theta^*(\theta')$ using \eqn{theta_star}.
         \item Compute $f(\theta')$ using \eqn{theta_star_objective}.
       \end{enumerate}
 \item Consider converged if $f(\theta') < \epsilon$ and exit if converged.
 \item Compute the derivative of the objective function \eqn{df_dthetap}:
       \begin{enumerate}
         \item Compute $\partial \lambda / \partial \theta \vert_{\zeta'}(\theta')$ using \eqn{lu_at_zetap}.
         \item Compute $\partial f / \partial \theta\vert_{\zeta'}(\theta')$ using \eqn{df_dthetap}.
       \end{enumerate}
 \item Perform the Newton step:
       \begin{equation}
         \theta' \leftarrow \theta' - \frac{f(\theta')}{\partial f / \partial \theta (\theta')} \, . \nonumber
       \end{equation}
 \item Go back to step 5.
\end{enumerate}
The result of this algorithm is a set of VMEC coordinates $(s,\theta',\zeta')$,
which are on the same field line as the VMEC coordinates $(s,\theta,\zeta)$.

\section{Inverse Coordinate Transform}
Assume flux surface shapes are given in VMEC(-like) coordinates $R(s, \theta, \varphi)$ and $Z(s, \theta, \varphi)$
as function of normalized toroidal flux $s$ with radius-like radial coordinate $\rho = \sqrt{s}$
and angle-like poloidal coordinate $\theta$ and toroidal coordinate $\varphi$,
equal to the cylindrical coordinates angle.
The representation is assumed to be as Fourier series for a number of discrete flux surfaces:
\begin{align}
 R(s_j, \theta, \varphi) =&\, \sum_{m,n} R^\textrm{cos}_{mn}(s_j) \cos(m \theta - n n_\textrm{fp} \varphi) \label{eqn:r_of_flux_coords} \\
 Z(s_j, \theta, \varphi) =&\, \sum_{m,n} Z^\textrm{sin}_{mn}(s_j) \sin(m \theta - n n_\textrm{fp} \varphi) \label{eqn:z_of_flux_coords}
\end{align}
with the number of toroidal field periods $n_\textrm{fp}$
at the $j$-th flux surface, identified by radial coordinate $s_j$.
A point in space (inside or outside the plasma) is identified by cylindrical coordinates $(R_0, \varphi_0, Z_0)$.
The goal of the method described here is to find VMEC coordinates $(s, \theta, \varphi)$,
for a given equilibrium and for now, assuming that the point under consideration is within the plasma volume,
corresponding to the point's real-space coordinates $(R_0, \varphi_0, Z_0)$.

\FloatBarrier
\subsection{Basic Approach}
It is noted that there is no direct inverse of the coordinate parameterization
in \eqn{r_of_flux_coords} and \eqn{z_of_flux_coords}.
In turn, the approach used here is to formulate this inverse coordinate transform as a non-linear minimization problem:
\begin{align}
 \min_{0 \leq s \leq 1, 0 \leq \theta < 2 \pi} \delta(s, \theta)
\end{align}
with the real-space distance~$\delta$ between the trial point $(R(s, \theta, \varphi_0), \varphi_0, Z(s, \theta, \varphi_0))$
and the given point $(R_0, \varphi_0, Z_0)$, whose VMEC coordinates~$(s, \theta)$ are to be inferred:
\begin{equation}
 \delta^2 (s, \theta) = \left[ R(s, \theta, \varphi_0) - R_0 \right]^2 + \left[ Z(s, \theta, \varphi_0) - Z_0 \right]^2 \, .
\end{equation}
This implies that the minimization has to be carried out only in the poloidal cutplane over $(s, \theta)$
at a fixed toroidal angle $\varphi_0$ for the given point.

\FloatBarrier
\subsection{Initial Guess}
A good initial guess $(s^{(-1)}, \theta^{(-1)})$ can greatly accelerate convergence of this iterative method.
Assume the magnetic axis location at toroidal angle $\varphi_0$ is given by $(R_\textrm{axis}(\varphi_0), Z_\textrm{axis}(\varphi_0))$ (see below).
Then, the distance to the target point is given by $d$:
\begin{align}
 \Delta R_\textrm{axis} =&\, R_0 - R_\textrm{axis}(\varphi_0) \\
 \Delta Z_\textrm{axis} =&\, Z_0 - Z_\textrm{axis}(\varphi_0) \\
 d =&\, \sqrt{ \Delta R_\textrm{axis}^2 + \Delta Z_\textrm{axis}^2 } \, .
\end{align}
Assume the minor radius of the plasma is approximated by $a_\textrm{eff}$ (see below).
Then, the initial guess for the radial coordinate can be approximated as $\rho^{(-1)} = d / a_\textrm{eff}$.
The initial guess for the poloidal coordinate~$\theta^{(-1)}$ is computed
by computing the simple poloidal angle of the point in question
with respect to the magnetic axis:
\begin{align}
 \theta^{(-1)} = \texttt{atan2}(\Delta Z_\textrm{axis}, \Delta R_\textrm{axis}) \, .
\end{align}

\subsubsection{Magnetic Axis Geometry}
The magnetic axis is the central flux surface, which collapses to a filament.
Its geometry is given as a purely toroidal Fourier series,
since the magnetic axis filament has no poloidal dependence:
\begin{align}
 R_\textrm{axis}(\varphi) = \sum_{n=0}^{\texttt{ntor}} R^\textrm{cos}_{0n}(0) \cos(n n_\textrm{fp} \varphi) \label{eqn:axis_r} \\
 Z_\textrm{axis}(\varphi) = \sum_{n=0}^{\texttt{ntor}} Z^\textrm{sin}_{0n}(0) \sin(n n_\textrm{fp} \varphi) \label{eqn:axis_z} \, .
\end{align}

\subsubsection{Minor Radius Approximation}
In the simplest case, where only the $(m=1,n=0)$ harmonic is non‐zero,
the cross‐section is an ellipse in the poloidal plane:
\begin{align}
 R(\theta) =&\, R^\textrm{cos}_{0,0} + R^\textrm{cos}_{1,0} \cos(\theta) \\
 Z(\theta) =&\,                        Z^\textrm{sin}_{1,0} \sin(\theta)
\end{align}
If $R^\textrm{cos}_{1,0} = Z^\textrm{sin}_{1,0} = a$, that is a perfect circle of radius $a$.
If $R^\textrm{cos}_{1,0} \neq Z^\textrm{sin}_{1,0}$, you get an ellipse with semi-axes $R^\textrm{cos}_{1,0}$ and $Z^\textrm{sin}_{1,0}$.
The geometric mean of those semi-axes
\begin{equation}
 a_\textrm{eff} = \sqrt{R^\textrm{cos}_{1,0} \cdot Z^\textrm{sin}_{1,0}} \label{eqn:minor_radius_approximation}
\end{equation}
is exactly the radius of a circle having the same area~$A$ as that ellipse
($A = \pi R^\textrm{cos}_{1,0} Z^\textrm{sin}_{1,0}$, so $\pi a_\textrm{eff}^2 = A$).
In plasma physics, the minor radius is often taken to be the radius of a circle of equal cross‐sectional area
as the flux surface under consideration.

\FloatBarrier
\subsection{Iterative Algorithm}
A Newton-type iteration algorithm is employed in this context~\cite{attenberger_1987}.
For brevity, we consider in the following only $R(s,\theta)$ and $Z(s,\theta)$.
A Taylor expansion can be used to approximate $R_0$ and $Z_0$ around $s^{(k)}$ and $\theta^{(k)}$
in the $k$-th iteration:
\begin{align}
 R_0 =&\, R(s^{(k)},\theta^{(k)}) +
          (     s -      s^{(k)}) \frac{\partial R}{\partial      s}(s^{(k)}, \theta^{(k)}) +
          (\theta - \theta^{(k)}) \frac{\partial R}{\partial \theta}(s^{(k)}, \theta^{(k)}) + ... \label{eqn:approx_r} \\
 Z_0 =&\, Z(s^{(k)},\theta^{(k)}) +
          (     s -      s^{(k)}) \frac{\partial Z}{\partial      s}(s^{(k)}, \theta^{(k)}) +
          (\theta - \theta^{(k)}) \frac{\partial Z}{\partial \theta}(s^{(k)}, \theta^{(k)}) + ... \label{eqn:approx_z} \, .
\end{align}
The following abbreviations are introduced:
\begin{align}
 R^{(k)}        \equiv&\, R(s^{(k)},\theta^{(k)}) \\
 R^{(k)}_s      \equiv&\, \frac{\partial R}{\partial      s}(s^{(k)}, \theta^{(k)}) \\
 R^{(k)}_\theta \equiv&\, \frac{\partial R}{\partial \theta}(s^{(k)}, \theta^{(k)})
\end{align}
and equivalently for $Z$.
If the Taylor expansions in \eqn{approx_r} and \eqn{approx_z} are terminated after the shown terms,
these expressions form a rank-2 linear system of equations:
\begin{equation}
  \begin{pmatrix}
    R_0 - R^{(k)} \\
    Z_0 - Z^{(k)}
  \end{pmatrix}
  =
  \begin{pmatrix}
    R^{(k)}_s & R^{(k)}_\theta \\
    Z^{(k)}_s & Z^{(k)}_\theta
  \end{pmatrix}
  \begin{pmatrix}
    s - s^{(k)} \\
    \theta - \theta^{(k)}
  \end{pmatrix} \label{eqn:approx_rz} \, .
\end{equation}
This system can be solved for $(s - s^{(k)})$ and $(\theta - \theta^{(k)})$.
The determinant of the matrix in \eqn{approx_rz} is:
\begin{equation}
 \tau^{(k)} = R^{(k)}_s Z^{(k)}_\theta - R^{(k)}_\theta Z^{(k)}_s \, .
\end{equation}
Then, the solution is given by:
\begin{equation}
 \begin{pmatrix}
         s \\
    \theta
  \end{pmatrix}
  =
 \begin{pmatrix}
         s^{(k)} \\
    \theta^{(k)}
  \end{pmatrix}
  +
  \frac{1}{\tau^{(k)}}
  \begin{pmatrix}
     Z^{(k)}_\theta & -R^{(k)}_\theta \\
    -Z^{(k)}_s      &  R^{(k)}_s
  \end{pmatrix}
  \begin{pmatrix}
    R_0 - R^{(k)} \\
    Z_0 - Z^{(k)}
  \end{pmatrix} \, .
\end{equation}
This can be used as an iteration equation for $s$ and $\theta$,
where the values for the $(k+1)$-th iteration are given by:
\begin{align}
 s^{(k+1)} =&\, s^{(k)} +
   \frac{ Z^{(k)}_\theta \left( R_0 - R^{(k)} \right) -
          R^{(k)}_\theta \left( Z_0 - Z^{(k)} \right) }{\tau^{(k)}} \\
 \theta^{(k+1)} =&\, \theta^{(k)} +
   \frac{ R^{(k)}_s \left( Z_0 - Z^{(k)} \right) -
          Z^{(k)}_s \left( R_0 - R^{(k)} \right) }{\tau^{(k)}} \, .
\end{align}
The inverse coordinate transform is considered converged
if the distance between the targeted point $(R_0, Z_0)$ and the model values $(R^{(k)},Z^{(k)})$
is smaller than a prescribed tolerance~$\epsilon$:
\begin{equation}
 \left( R_0 - R^{(k)} \right)^2 + \left( Z_0 - Z^{(k)} \right)^2 < \epsilon^2 \, .
\end{equation}
In case the iteration gives a worse result than the previous iterate,
the step size is halved
and an averaged value of the Jacobian is used:
\begin{align}
 s^{(k+2)} =&\, s^{(k)} + \frac{1}{2}
   \frac{ Z^{(k)}_\theta \left( R_0 - R^{(k)} \right) -
          R^{(k)}_\theta \left( Z_0 - Z^{(k)} \right) }{0.75 \tau^{(k)} + 0.25 \tau^{(k+1)}} \label{eqn:adaptive_step_s} \\
 \theta^{(k+2)} =&\, \theta^{(k)} + \frac{1}{2}
   \frac{ R^{(k)}_s \left( Z_0 - Z^{(k)} \right) -
          Z^{(k)}_s \left( R_0 - R^{(k)} \right) }{0.75 \tau^{(k)} + 0.25 \tau^{(k+1)}} \label{eqn:adaptive_step_u} \, .
\end{align}
This process can be repeated as necessary
until a new iterate for $s$ and $\theta$ is found
which reduces the objective.

\FloatBarrier
\subsection{Radial Derivative of VMEC Flux Surface Geometry}
One key point to consider when working with flux surface geometries from VMEC
is the interpolation of the flux surface geometry and quantities on them
in the radial direction.
This is because VMEC solves only for discrete flux surfaces
and uses finite differences in the radial direction,
which leaves it to the user to choose how to interpolate quantities between flux surfaces.
Here, an interpolation approach is described, which was found to work very well in general,
and in particular for the coordinate transform use case considered here.

The interpolation is set up over $\rho = \sqrt{s}$ as the radial coordinate.
Cubic splines are used with natural boundary conditions to set up the actual interpolation.
The interpolation and associated analytical derivatives are set up
for the radial dependence of each Fourier coefficient in the representations
\eqn{r_of_flux_coords} and \eqn{z_of_flux_coords} individually.
The boundary condition at the plasma boundary is the natural boundary condition provided by the spline implementation.
The boundary condition at the magnetic axis, i.e., at the center of the plasma,
is adjusted to satisfy the poloidal parity of the respective Fourier coefficients to interpolate:
\begin{align}
 X_{m, n}(-s) =&\, \phantom{-} ~ X_{m, n}(s) \textrm{ if } m \textrm{ even } \\
 X_{m, n}(-s) =&\,          -    X_{m, n}(s) \textrm{ if } m \textrm{ odd }
\end{align}
for $X \in \{ R^\textrm{cos}, Z^\textrm{sin} \}$.
Importantly, note that the parity of the interpolation at the magnetic axis
only depends on the poloidal mode number and not on the phase (sin or cos)
of the function to interpolate.
This regularization is particularly relevant for obtaining reliable
inverse coordinate transforms near the magnetic axis, i.e.,
inside of the first flux surface of the VMEC equilibrium under consideration.

The geometry of the flux surfaces in VMEC is represented on the ``full'' radial grid:
\begin{equation}
 s_j = \frac{j}{\texttt{ns} - 1} \textrm{ for } j = 0, 1, ..., (\texttt{ns} - 1)
\end{equation}
where $j=0$ represents the magnetic axis, i.e., the core of the plasma
and $j=\texttt{ns} - 1$ represents the plasma boundary.

The even and odd symmetry boundary conditions at the left end of the range
in the radial direction (at the magnetic axis) can be enforced in some spline
interpolation implementations.
Some spline interpolation implementations
feature only natural boundary conditions at both ends.
One can mirror the data according to the desired symmetry
and interpolate over that extended dataset
and thereby still establish symmetry with the desired parity at the magnetic axis
using such a spline implementation.
The following concrete formulation is used
for even symmetry at the magnetic axis when interpolating Fourier coefficients on the full-grid:
\begin{align}
 \rho[\texttt{ns} - 1 + j] =&\, \sqrt{ \frac{j}{\texttt{ns} - 1} } \textrm{ for } 0 \leq j < \texttt{ns} \\
 f[\texttt{ns} - 1 + j]    =&\, X[j]                               \textrm{ for } 0 \leq j < \texttt{ns} \\
 \textrm{ and }             & \textrm{ then } \nonumber \\
 \rho[\texttt{ns} - 1 - j] =&\, -\rho[\texttt{ns} - 1 + j]         \textrm{ for } 1 \leq j < \texttt{ns} \\
 f[\texttt{ns} - 1 - j]    =&\, f[\texttt{ns} - 1 + j]             \textrm{ for } 1 \leq j < \texttt{ns}
\end{align}
where $\rho$ and $f$ are the arrays of length $2 \cdot \texttt{ns} - 1$ to be handed to the spline interpolation
and $X$ is the Fourier coefficient array on the full grid to be interpolated.
For enforcing odd parity at the magnetic axis,
the following formulation is used instead:
\begin{align}
 \rho[\texttt{ns} - 1 + j] =&\, \sqrt{ \frac{j}{\texttt{ns} - 1} } \textrm{ for } 0 \leq j < \texttt{ns} \\
 f[\texttt{ns} - 1 + j]    =&\, X[j]                               \textrm{ for } 0 \leq j < \texttt{ns} \\
 \textrm{ and }             & \textrm{ then } \nonumber \\
 \rho[\texttt{ns} - 1 - j] =&\, -\rho[\texttt{ns} - 1 + j]         \textrm{ for } 1 \leq j < \texttt{ns} \\
 f[\texttt{ns} - 1 - j]    =&\, -f[\texttt{ns} - 1 + j]            \textrm{ for } 1 \leq j < \texttt{ns} \label{eqn:mirror_odd}
\end{align}
where the only difference to the formulation for even parity
is the negative sign in mirroring the data in \eqn{mirror_odd}.

\FloatBarrier
\subsection{Extrapolation beyond the Plasma Boundary}
VMEC only provides the flux surface geometry up to the plasma boundary.
Extrapolation of the geometry can be set up in a way that makes
the extrapolated flux surface geometry to tend towards ellipses far away from the plasma.
This works for most plasma boundary shapes, but note that it is possible
that the extrapolation scheme outlined below leads to extrapolated flux surfaces overlapping
with the VMEC-provided flux surfaces, in particular close to the plasma boundary.
The extrapolation method is described in Ref.~\cite{attenberger_1987}.
The Fourier coefficients of the flux surface geometry are extrapolated
to $\rho > 1$ as follows:
\begin{align}
 X_{mn}(\rho) =&\, \begin{cases}
                    X_{mn}(1)          \cdot \rho  & \textrm{ for } (m,n)=(1,0) \\
                    X_{mn}(1) \phantom{\cdot \rho} & \textrm{ else }
                   \end{cases}
\end{align}
for $X \in \{ R^\textrm{cos}, Z^\textrm{sin} \}$.
Therefore, the radial derivatives are extrapolated as follows
outside of the plasma boundary at $\rho > 1$:
\begin{align}
 \frac{\partial X_{mn}}{\partial \rho}(\rho) =&\, \begin{cases}
                                                   X_{mn}(1) & \textrm{ for } (m,n)=(1,0) \\
                                                   0         & \textrm{ else. }
                                                  \end{cases}
\end{align}

\FloatBarrier
\subsection{Actual Implementation of Inverse Coordinate Transform}
The actual implementation of the inverse coordinate transform for VMEC coordinates
is suggested to always use the extrapolation for $\rho > 1$ introduced above.
If points outside the plasma boundary shall be marked as such,
one can identify cases where $\rho > 1$ was found as radial VMEC coordinate
for a given real-space point, and replace the inferred coordinates
with $\hat{\rho} = \infty$ and $\hat{\theta} = \textrm{NaN}$, $\hat{\varphi} = \textrm{NaN}$.
Note that the actual implementation uses $\rho$ as the radial coordinate,
since this was found to work better with the interpolation
of flux surface geometry near the magnetic axis.
The actual implementation is presented in Algorithm~\ref{alg:inverse_coordinate_transform}.
\begin{algorithm}
  \caption{Inverse Coordinate Transform $(R_0, \varphi_0, Z_0) \to (\rho, \theta, \varphi_0)$ with extrapolation}
  \begin{algorithmic}[1]
    \Require
      $R_0, \varphi_0, Z_0$ ; $\epsilon > 0$ ; $n_\textrm{max} \in \mathbb{N}$

    % Initialize squared tolerance
    \State $\epsilon^2 \gets \epsilon \cdot \epsilon$

    % Compute magnetic axis and distance to axis
    \State $\{ R_\textrm{axis}(\varphi_0), Z_\textrm{axis}(\varphi_0) \} \gets \texttt{GetMagneticAxis}(\varphi_0)$
    \State $\Delta R_\textrm{axis} \gets R_0 - R_\textrm{axis}(\varphi_0)$ ; $\Delta Z_\textrm{axis} \gets Z_0 - Z_\textrm{axis}(\varphi_0)$
    \State $d \gets \sqrt{\Delta R_\textrm{axis}^2 + \Delta Z_\textrm{axis}^2}$

    % Approximate minor radius and initial guess
    \State $r_0 \gets \texttt{MinorRadiusApproximation}()$
    \State $ \rho \gets d / r_0$
    \State $\theta \gets \texttt{atan2}(\Delta Z_\textrm{axis}, \Delta R_\textrm{axis})$
    \If{$\theta < 0$}
      \State $\theta \gets \theta + 2\pi$
    \EndIf

    % Initialize iteration variables
    \State $\delta^2 \gets +\infty$ ; ${\delta^2}^{(-1)} \gets +\infty$
    \State $\beta \gets 1$
    \State $\tau^{(-1)} \gets \textrm{NaN}$

    \For{$i = 1$ to $n_\textrm{max}$}

      % Evaluate forward map and residuals
      \State Evaluate $R$, $\partial R / \partial \rho$ and $\partial R / \partial \theta$ at $(\rho, \theta)$ and analogously for $Z$
      \State $\Delta R \gets R(\rho, \theta) - R_0$ ; $\Delta Z \gets Z(\rho, \theta) - Z_0$
      \State ${\delta^2}^{(-1)} \gets \delta^2$ ; $\delta^2 \gets \Delta R^2 + \Delta Z^2$
      \If{$\delta^2 < \epsilon^2$}
        \State \textbf{break}
      \EndIf

      % Compute Jacobian determinant
      \State $\tau^{(-1)} \gets \tau$
      \State $\tau \gets \partial R / \partial \rho (\rho, \theta) \cdot \partial Z / \partial \theta (\rho, \theta) - \partial R / \partial \theta (\rho, \theta) \cdot \partial Z / \partial \rho (\rho, \theta)$
      \If{$i=1$}
        \State $\tau^{(-1)} \gets \tau$
      \EndIf

      % Handle case of un-successful step - distance to target increased -> mix Jacobian with previous Jacobian and reduce step length
      \If{$\delta^2 > {\delta^2}^{(-1)}$}
        \State $\tau \gets 0.75\,\tau^{(-1)} + 0.25\,\tau$ ; $\beta \gets \beta/2$
      \Else
        % Restore original step length in case of a successful step
        \State $\beta \gets 1$
      \EndIf

      % Compute next iterate via Newton-like update
      \State $\rho   \gets \rho   - \beta \left[\partial Z / \partial \theta (\rho, \theta) \cdot \Delta R - \partial R / \partial \theta (\rho, \theta) \cdot \Delta Z \right] / \tau$
      \State $\theta \gets \theta - \beta \left[\partial R / \partial \rho   (\rho, \theta) \cdot \Delta Z - \partial Z / \partial \rho   (\rho, \theta) \cdot \Delta R \right] / \tau$

      % Handle crossing over axis
      \If{$\rho < 0$}
        \State $\theta \gets \theta + \pi$ ; $\rho \gets -\rho$
      \EndIf

      % Make sure \theta stays within [0, 2 pi]
      \While{$\theta < 0$}
        \State $\theta \gets \theta + 2 \pi$
      \EndWhile
      \State $\theta \gets \theta \bmod 2 \pi$

    \EndFor

    \State \Return $(\rho, \theta, \varphi_0)$
  \end{algorithmic}
  \label{alg:inverse_coordinate_transform}
\end{algorithm}
Note that the step is taken in the ``reverse'' direction,
which is probably required to incorporate the negative Jacobian used in the VMEC coordinate system.
The method $\texttt{GetMagneticAxis}(\varphi_0)$ implements \eqn{axis_r} and \eqn{axis_z}.
The method $\texttt{MinorRadiusApproximation}()$ implements \eqn{minor_radius_approximation}.

\FloatBarrier
\subsection{Testing Strategy}
The functionality provided by this inverse coordinate algorithm is actually quite fundamental
to stellarator modeling and data analysis.
It is therefore used in quite a number of places of the overall stellarator toolchain.
Therefore, this algorithm deserves significant, robust testing
in order to get an implementation ready for production use.

This section outlines a testing strategy that has proven sufficiently thorough
to get an implementation ready for large-scale use in data analysis for Wendelstein~7-X.
Credit for thinking out this testing strategy goes to Oliver Ford at IPP Greifswald.
This algorithm should be tested on a number of different VMEC equilibria,
which capture typical plasma shapes occuring in the context
where this coordinate transform algorithm is supposed to be used.
Random number generation is used to come up with test point coordinates.
It is important for debuggability to ensure proper random seed management
in the testing harness in order to be able to debug errors which occur also very late in a testing cycle,
which would be otherwise very hard to reproduce.
The tests are carried out it batches of points,
which each are generated with their own random seed,
which is drawn from a random number generator themselves (seeded with the ``seed of seeds'').
This multi-layer random number generation allows to reproduce errors
in a given batch of test points quicker than if all points are tested in a single sequence,
since the random number generator does not have to go through generating
all point coordinates starting from a given seed,
but only needs to go through number in given batch
to reproduce an error that occured in that batch.
Unit testing of the implementation is done in 10 batches of 1000 points each.
Extended testing to catch rare errors and really ensure that an implementation
is ready for production use should run at least 10000 batches of 5000 test points each.

Radial derivatives of the flux surface geometry are inaccurate near the plasma boundary,
due to the (assumed chosen) natural boundary condition for the spline there.
Therefore, the radial displacement of test points near the plasma boundary
is done via the tangent vector in poloidal direction,
whose components $(\partial R / \partial \theta, \partial Z / \partial \theta)$
can be computed analytically from the Fourier series representation of $R$ and $Z$.
This tangent vector is then rotated by 90 degrees into the perpendicular orientation
by swapping its components and inverting one of them (see below).

The inverse coordinate transform algorithm presented here
is an iterative method with a user-prescribed (but usually chosen statically for a given application) convergence tolerance.
This means that it cannot be expected that the real space point
corresponding to inferred VMEC coordinates is closer to the target point in distance
than the prescribed convergence tolerance.
It is therefore imperative to avoid false negatives during testing of this algorithm
to compute the real-space distance between the point corresponding to the inferred coordinates
and the target point.
More importantly, if a test is made to check if a point is tested that was supposed to be inside the plasma boundary
actually is inferred to be inside the plasma boundary,
the test point has to be placed at least as far away from the plasma boundary towards the inside of the plasma
as the convergence tolerance distance is specified in the coordinate transform.
Otherwise, the inferred point may be within the convergence tolerance radius
around the target point, but fall just outside the plasma boundary,
even though the original point was inside the plasma boundary,
and an unsuitably-setup test would fail.

The inverse coordinate transform algorithm is tested then on a few different cases:
\begin{enumerate}
 \item points equally distributed in $(\rho, \theta, \varphi)$ space, i.e., anywhere within the plasma volume;
 \item points close to the magnetic axis
       - this particularly stresses the coordinate interpolation near the magnetic axis;
 \item points close the plasma boundary, but still predictably within the plasma boundary; and
 \item points close the plasma boundary, but already predictably outside of the plasma boundary
       - this particularly tests the coordinate extrapolation.
\end{enumerate}
For each of these cases, real-space cylindrical coordinates of test points~$[R_0, \varphi_0, Z_0]$
are made available. For points sampled close to the plasma boundary (cases 3 and 4),
an additional flag~\texttt{outside\_flag} is returned, which indicates if a given point is supposed
to be located inside the plasma boundary or not.

The testing procedure then is orchestrated as follows:
\begin{enumerate}
 \item Sample random test point coordinates~$[R_0, \varphi_0, Z_0]$ according to a method randomly selected from above list of test cases.
 \item Use the inverse coordinate transform algorithm under test to obtain the corresponding VMEC coordinates~$[\rho, \theta, \varphi]$.
 \item Transform the obtained VMEC coordinates back into cylindrical coordinates~$[R, \varphi, Z]$.
 \item Test if the returned value for $\varphi$ matches $\varphi_0$ - they should be bit-identical.
 \item If the extrapolation is inhibited by replacing the VMEC coordinates of points outside of the plasma boundary
       with $\rho = +\infty$, it is checked that a provided \texttt{outside\_flag} (see below) is consistent with the point
       actually being inferred as inside or outside the plasma boundary.
 \item Compute the Eucledian distance between $(R_0, Z_0)$ and $(R, Z)$ and make sure it is below the convergence tolerance $\epsilon$.
       If not, consider the test failed.
 \item Repeat all steps above for every test case in the current batch.
 \item Repeat all steps above for every batch in the whole set of test points.
\end{enumerate}
The random generation of test points for each of the four cases is presented next:
\subsubsection{Points Equally Distributed in the Plasma Volume}
The points equally distributed in the plasma volume are generated as follows:
\begin{align}
  %
  % sample normalized flux coordinate in [0,1]
  s &\sim \mathcal{U}(0,1)\\
  %
  % sample poloidal angle in [0,2\pi]
  \theta &\sim \mathcal{U}(0,2\pi)\\
  %
  % sample toroidal angle in [0,2\pi]
  \varphi_0 &\sim \mathcal{U}(0,2\pi)\\
  %
  % map to cylindrical coordinates via the inverse transform
  (R_0, \varphi_0, Z_0) &= \texttt{ToCylinderCoordinates}\bigl(s, \theta, \varphi_0 \bigr)
\end{align}
where \texttt{ToCylinderCoordinates} implements \eqn{r_of_flux_coords} and \eqn{z_of_flux_coords}.

\subsubsection{Points Close to the Magnetic Axis}
Test points close to the magnetic axis are generated as follows:
\begin{align}
  %
  % approximate minor radius
  a_\textrm{eff} &= \texttt{MinorRadiusApproximation}() \\
  %
  % small reference distance along minor radius
  d &= \frac{a_\textrm{eff}}{10} \\
  %
  % random radial and vertical offsets in [0,d]
  dr &\sim \mathcal{U}(0,d) \\
  dz &\sim \mathcal{U}(0,d) \\
  %
  % random toroidal angle in [0,2π]
  \varphi_0 &\sim \mathcal{U}(0,2\pi) \\
  %
  % magnetic axis coordinates at this toroidal angle
  \bigl[R_{\mathrm{axis}}(\varphi_0),\,Z_{\mathrm{axis}}(\varphi_0)\bigr] &= \texttt{GetMagneticAxis}(\varphi_0) \\
  %
  % construct test point near the axis
  R_{0} &= R_{\mathrm{axis}}(\varphi) + dr \\
  Z_{0} &= Z_{\mathrm{axis}}(\varphi) + dz \, .
\end{align}

\subsubsection{Points Close to the Plasma Boundary}

\begin{align}
%
% random poloidal angle on boundary
\theta_0 &\sim \mathcal{U}(0,2\pi)\\
%
% random toroidal angle on boundary
\varphi_0 &\sim \mathcal{U}(0,2\pi)\\
%
% signed offset distance from boundary (≥ requested accuracy)
d & = \epsilon \cdot \bigl(\mathcal{U}(0,1) + 1\bigr)\\
%
% cylindrical coords on boundary surface s=1
\bigl[R_{b},\,\varphi_{b},\,Z_{b}\bigr] &= \texttt{ToCylinderCoordinates}\bigl(1, \theta_0, \varphi_0 \bigr)\\
%
% arc‐length element along boundary curve
\ell & = \sqrt{\bigl(dR/d\theta\bigr)^{2} + \bigl(dZ/d\theta\bigr)^{2}}\\
%
% perpendicular displacements of magnitude d
dR_{\perp} & = \frac{dZ/d\theta}{\ell}\,d\\
dZ_{\perp} & = -\,\frac{dR/d\theta}{\ell}\,d\\
%
% final test point and outside flag (50\% inside, 50\% outside)
\bigl[R_{0},\,\varphi_{0},\,Z_{0},\,\texttt{outside\_flag}\bigr] &=
\begin{cases}
\bigl[R_{b}-dR_{\perp},\,\varphi_{b},\,Z_{b}-dZ_{\perp},\, \texttt{false} \bigr] \textrm{ in 50 \% of cases }  \\
\bigl[R_{b}+dR_{\perp},\,\varphi_{b},\,Z_{b}+dZ_{\perp},\, \texttt{true } \bigr] \textrm{ in 50 \% of cases }
\end{cases}
\end{align}
where $(dR_{\perp}, dZ_{\perp})$ is the outward-facing unit vector on the plasma boundary
and \texttt{outside\_flag} indicates to the test harness if a given test point
is supposed to be inside the plasma boundary or outside of it.

% \section{Computing the Magnetic Field}
% This section introduces how to compute the magnetic field from a VMEC equilibrium.
% Several methods are described here:
% \begin{enumerate}
%  \item Computation of the contravariant magnetic field components
%        from $\lambda$, $\iota$, $\sqrt{g}$ and $\Phi'$
%  \item Computation of the cylindrical magnetic field components
%        from the contravariant magnetic field components
%  \item Computation of the Cartesian magnetic field components
%        from the cylindrical magnetic field components.
% \end{enumerate}

% {\color{red} TODO:
% R * Btor towards axis - is constant? Attenberger check...
% }

\FloatBarrier
\section{Calculation of \texorpdfstring{$\nabla\alpha$}{grad-alpha} from VMEC equilibria}
But how do we get $\nabla\alpha$ or $\mathbf{\hat{e}}^\alpha$,
if not through the easily computable magnetic field direction and flux surface normal?
We can go via the angle definition, $\alpha = \theta - \iota\phi$ and get~\cite{plunk_collisionless_2014}:
\begin{align}
 \nabla\alpha = \nabla\theta - \iota\nabla\phi - \iota'\phi\nabla u_r \, ,
\end{align}
where $u_r$ is any (radial) coordinate and $\iota' = d\iota/du_r$.
Unfortunately, this is only true if the poloidal and toroidal angles are straight-fieldline-coordinates
(i.e. coordinates in which field lines are straight lines),
which is not the case for VMEC coordinates.
To transform to according coordinates, a stream function $\lambda(s,\theta_v, \phi)$ is necessary~\cite{stella_doc}:
\begin{align}
 \theta = \theta_v + \lambda(s,\theta_v, \phi) \, ,
\end{align}
where $\theta_v$ is the VMEC poloidal angle and
$\theta$ is a straight-fieldline-coordinate together with $\phi$ and $s$,
which are unchanged.
We then get $\alpha = \theta_v + \lambda - \iota\phi$ and accordingly~\cite{stella_doc}:
\begin{align}
    \nabla\alpha &= \nabla(\theta_v + \lambda - \iota\phi) \\
    &=   \bigg(\frac{\partial\lambda}{\partial s}
       - \phi\frac{d\iota}{ds}\bigg)\nabla s
       + \bigg(1 + \frac{\partial\lambda}{\partial\theta_v}\bigg)\nabla\theta_v
       + \bigg(-\iota + \frac{\partial\lambda}{\partial\phi}\bigg)\nabla\phi \, .
\end{align}
Find the relevant part in stella diagnostic code
\href{https://github.com/stellaGK/stella/blob/master/POST_PROCESSING/stellapy/data/geometry/calculate_geometricQuantitiesVMEC.py#L738}{here}.

The stream function $\lambda$ is expressed as a Fourier series:
\begin{align}
 \lambda(s, \theta_v, \phi) = \sum_{m=0}^M\sum_{n=-N}^{N}\hat{\lambda}_{m,n}^{\sin}(s)\sin{(m\theta - nN_\mathrm{fp}\phi)} \, ,
\end{align}
where $\hat{\lambda}_{m,n}^{\sin}(s)$ are the Fourier coefficients given in the VMEC output on a $(n_s, M\times 2N)$ grid.
From this, we get analytical expressions for the partial derivatives with respect to $\theta_v$ and $\phi$,
which can directly be calculated from the Fourier coefficients~\cite{stella_doc}:
\begin{align}
 \frac{\partial\lambda}{\partial\theta_v}(s, \theta_v, \phi) &=
   \sum_{m=0}^M\sum_{n=-N}^{N}
     m\hat{\lambda}_{m,n}^{\sin}(s)\cos{(m\theta - nN_\mathrm{fp}\phi)} \\
 \frac{\partial\lambda}{\partial\phi}(s, \theta_v, \phi) &=
   \sum_{m=0}^M\sum_{n=-N}^{N}
     -nN_\mathrm{fp}\hat{\lambda}_{m,n}^{\sin}(s)\cos{(m\theta - nN_\mathrm{fp}\phi)} \, .
\end{align}
The derivative with respect to $s$ can be pulled into the sum,
but needs be to calculated numerically based on the available $n_s$ grid of Fourier coefficients:
\begin{align}
 \frac{\partial\lambda}{\partial s}(s, \theta_v, \phi) &=
   \sum_{m=0}^M\sum_{n=-N}^{N}
     \bigg[ \frac{\partial}{\partial s}\hat{\lambda}_{m,n}^{\sin}(s) \bigg] \sin{(m\theta - nN_\mathrm{fp}\phi)} \\
 \frac{\partial}{\partial s}\hat{\lambda}_{m,n}^{\sin}(s) &\approx
   \frac{\hat{\lambda}_{m,n}^{\sin}[i_s +1] - \hat{\lambda}_{m,n}^{\sin}[i_s]}{\Delta s} \, .
\end{align}

\FloatBarrier
\section{Summary of Coordinate Sign Conventions} \label{sec:sign_conventions}
In VMEC, the toroidal angle, whether it is $\phi$ or $\zeta$
(the latter is simply the rescaling of $\phi$ by a positive integer, see \eqn{zeta}),
increases counter-clockwise when the plasma is viewed from above,
which results in a right-handed cylindrical coordinate system $(R, \phi, Z)$.

The poloidal angle $\theta$ increases from the outboard to the inboard side over the top of the plasma,
resulting in a left-handed toroidal coordinate system $(s, \theta, \phi)$.

Due to the different handedness, the Jacobian of the transformation between the toroidal flux
coordinates and cylindrical coordinates is negative;
this is reflected in the output variable \texttt{signgs}, which is fixed to $-1$~\cite{landreman_vmec_signs}.

A positive value of \texttt{phiedge}, the toroidal flux enclosed by the LCFS,
corresponds to the magnetic field pointing in the $\nabla\zeta$ direction,
i.e., counter-clockwise when viewed from above.
In fixed-boundary mode, the sign of \texttt{phiedge} determines the orientation of the magnetic field.
In free-boundary mode, the sign of \texttt{phiedge} should be chosen according to the orientation
of the vacuum magnetic field produced by the coils.
For instance, coils with counter-clockwise currents, which produce a toroidal magnetic field
in the clockwise direction (viewed from above), correspond to a negative \texttt{phiedge}. In VMEC++ free-boundary, the wrong input sign will trigger the error message:
``\texttt{rBtor} and \texttt{bSubVVac} must have the same sign
- maybe flip the sign of \texttt{phiedge} or the sign of the coil currents''.
The sign of \texttt{rBtor}, which is the total poloidal current at the LCFS, follows the sign of \texttt{phiedge},
while the sign of \texttt{bSubVVac}, which is the vacuum-side poloidal current contribution, is determined by the coils.

The signs of the output quantities \texttt{phi} and \texttt{phipf} are the same as \texttt{phiedge},
while the sign of the output toroidal flux quantity \texttt{phips} differs from \texttt{phiedge}
by the factor \texttt{signgs}, i.e., it has the opposite sign~\cite{landreman_vmec_signs}.

\chapter{BNORM}

\FloatBarrier
\section{Introduction}
Every current is surrounded by an accompanying magnetic field.
A toroidal plasma cannot confine itself by its own currents.
Thus, external confinement coils are required to produce the (mostly toroidal) confinement magnetic field.
In a Stellarator, these confinement coils are required (in VMEC: assumed)
to produce good flux surfaces in the confimenent region.
The diamagnetic current~$\mathbf{j}_\perp$ inside the plasma arises from a finite plasma pressure:
\begin{equation}
 \nabla p = \mathbf{j}_\perp \times \mathbf{B} \, .
\end{equation}
The current paths inside the plasma are closed: $\nabla \cdot \mathbf{j} = 0$.
This imples $\nabla \cdot \mathbf{j}_\perp = -\nabla \cdot \mathbf{j}_{||}$,
where $\mathbf{j} = \mathbf{j}_\perp + \mathbf{j}_{||}$ is the total current density in the plasma
and $\mathbf{j}_{||}$ is the Pfirsch-Schlüter current density.
The diamagnetic current is a current flowing inside the plasma itself.
It has its own magnetic field associated with it,
which is superposed to the confinement magnetic field,
thereby contributing to the total magnetic field inside the plasma domain.
VMEC can be used to compute the total magnetic field inside the plasma domain
such that flux surfaces are in force balance.

It is a non-trivial problem to separate the individual contributions (plasma current and coil currents)
to the total magnetic field from each other.
The magnetic field due to the confinement coils (``external'' magnetic field)
is typically computed using the Biot-Savart law
for a given geometry of the coils and given coil currents along those current paths.
It can be computed in all of space; both inside the plasma domain
as well as outside of it and also on the boundary.

The total magnetic field at the plasma boundary is assumed to be purely tangential to the boundary.
This implies that the normal component of the magnetic field contribution from the coils
is cancelled by the normal component of the magnetic field due to plasma currents.
A Stellarator coil configuration must have flux surfaces in vacuum alone;
otherwise, it is not a suitable Stellarator magnetic configuration.
The magnetic field on each of these vacuum flux surfaces
is purely tangential to the flux surfaces by definition.
However, if a plasma is created now inside the flux surface region
and pressure builds up inside the plasma,
the magnetic field due to coils inside the plasma region is expelled from the plasma region.
This is the so-called diamagnetic behavior of the plasma.
Good flux surfaces are assumed to persist even in case of a finite plasma pressure in VMEC.
In fact, stochastization of the magnetic field inside the plasma (breaking up of flux surfaces)
poses a limit on the achievable plasma pressure. SPEC can be used to compute these limits,
but this is not the topic of this particular note.
The buildup of plasma pressure implies that currents start to flow inside the plasma.
These currents necessarily change the magnetic field structure inside the plasma as well as outside of it.
Assuming that good flux surfaces still exist,
the flux surfaces change shape and shift around in space (e.g. the Shafranov shift is one of these outcomes).
Suddenly, the vacuum flux surfaces (on which the external magnetic field is purely tangential)
do not coincide with the plasma flux surfaces anymore.
This implies that a non-zero normal component of the plasma magnetic field is present
and that (because of the assumption on the continued existence of flux surfaces in the plasma)
only the new total magnetic field is purely tangential to the new plasma boundary.

Inside the plasma domain, the magnetic field due to plasma currents
can be obtained by subtracting the coil field (available from Biot-Savart)
from the total magnetic field (available from VMEC).
The currents inside the plasma can be computed using Ampere's law:
\begin{equation}
 \mu_0 \mathbf{j} = \nabla \times \mathbf{B} \, .
\end{equation}
A covariant representation of the total magnetic field inside the plasma is available in VMEC
and this can be used to compute the contravariant representation of the current density inside the plasma.
This can then be used to compute the magnetic field due to plasma currents anywhere in space.
Of particular interest is the computation of the magnetic field due to plasma currents
outside the plasma domain, where it is not available trivially.
Here, a Biot-Savart (three-dimensional) volume integral over the plasma current density
can be used to compute the magnetic field outside the plasma domain.
This approach is commonly used to predict magnetic diagnostic signals.

The virtual casing principle can be used to reduce the volume integral
to a surface integral over the plasma boundary
by replacing the volume current density inside the plasma domain
with a surface current density flowing inside the plasma boundary surface,
which generates the same magnetic field as the volume plasma currents
outside the plasma domain.
This method is available in, e.g., DIAGNO.
The surface integral is nearly-singular when evaluating the external field due to plasma currents
close to the surface. This is the reason why an adaptive integration routine
is used in DIAGNO to compute the predictions for the magnetic diagnostics
(which are typically located quite close to the LCFS).

BNORM computes the normal component of the external magnetic field via the virtual casing integral \emph{on} the LCFS.
This is the normal component of the magnetic field due to plasma currents alone.
The integral is singular in this case, since the evaluation location is located
in the same surface as where the surface current density (the source term of the Biot-Savart integral) is defined.
A regularization method is therefore required.

\FloatBarrier
\section{Theoretical Foundation}
Assume an arbitrary magnetic field $\mathbf{B}_0$ is given in a toroidal domain.
In case of a VMEC computation, this is the total magnetic field
due to both external confinement coils and the plasma currents.
Furthermore, $\mathbf{B}_0$ shall be purely tangential to the boundary of the domain,
such that $\mathbf{B}_0 \cdot \mathbf{n} = 0$, where $\mathbf{n}$ is the exterior normal of the boundary surface.
The virtual-casing principle allows to define a surface current $\mathbf{j}$ flowing on the boundary
that is given by:
\begin{equation}
  \mu_0 \mathbf{j} = \mathbf{B}_0 \times \mathbf{n}
\end{equation}
which leads to the same magnetic field outside the toroidal domain
as the current source terms inside the volume of the domain.
When computing the magnetic field outside the toroidal domain,
one can replace the Biot-Savart volume integral over the current within the domain
by a surface integral over above surface current density.
The magnetic vector potential $\mathbf{A}(\mathbf{x})$ outside the domain
due to the surface current is given by the Biot-Savart law,
which takes on the form of a two-dimensional surface integral in this case:
\begin{equation}
 \mathbf{A}(\mathbf{x}) = \frac{\mu_0}{4 \pi} \iint \frac{\mathbf{j}(\mathbf{x}')}{|\mathbf{x} - \mathbf{x}'|} \,\mathrm{d}^2\mathbf{x}' \, .
\end{equation}
The magnetic field outside the toroidal domain
is then given by $\mathbf{B} = \nabla \times \mathbf{A}$.
The normal component of this magnetic field~$B_\mathrm{n}$ is then given by
\begin{equation}
 B_\mathrm{n} = -\mathbf{n} \cdot \mathbf{B} = -\mathbf{n} \cdot (\nabla \times \mathbf{A}) \label{eqn:b_norm_def} \, .
\end{equation}
This is the main quantity of interest in this computation.
The outward-directed surface normal~$\mathbf{n}$ is given by:
\begin{equation}
 \mathbf{n} = - \frac{\mathbf{x}_u \times \mathbf{x}_v}{|\mathbf{x}_u \times \mathbf{x}_v|} \, .
\end{equation}
Inserting this into Eqn.~(\ref{eqn:b_norm_def}), it follows:
\begin{align}
 B_\mathrm{n}
 =&\ \frac{1}{|\mathbf{x}_u \times \mathbf{x}_v|}
     \left( \mathbf{x}_u \times \mathbf{x}_v \right) \cdot (\nabla \times \mathbf{A}) \nonumber \\
 =&\ \frac{1}{|\mathbf{x}_u \times \mathbf{x}_v|}
     \left(  \mathbf{x}_v \cdot \frac{\partial \mathbf{A}}{\partial u}
           - \mathbf{x}_u \cdot \frac{\partial \mathbf{A}}{\partial v} \right) \nonumber \\
 =&\ \frac{1}{|\mathbf{x}_u \times \mathbf{x}_v|}
     \left[  \frac{\partial}{\partial u} \left(\mathbf{x}_v \cdot \mathbf{A} \right)
           - \frac{\partial}{\partial v} \left(\mathbf{x}_u \cdot \mathbf{A} \right) \right] \, .
\end{align}
In above formula, the following trick was used:
\begin{align}
 ~&\,   \frac{\partial}{\partial u} \left(\mathbf{x}_v \cdot \mathbf{A} \right)
      - \frac{\partial}{\partial v} \left(\mathbf{x}_u \cdot \mathbf{A} \right) \nonumber \\
 =&\,   \mathbf{x}_v \cdot \frac{\partial \mathbf{A}}{\partial u} \bcancel{+ \mathbf{x}_{uv} \cdot \mathbf{A}}
      - \mathbf{x}_u \cdot \frac{\partial \mathbf{A}}{\partial v} \bcancel{- \mathbf{x}_{uv} \cdot \mathbf{A}} \nonumber \\
 =&\,   \mathbf{x}_v \cdot \frac{\partial \mathbf{A}}{\partial u}
      - \mathbf{x}_u \cdot \frac{\partial \mathbf{A}}{\partial v} \, .
\end{align}
The derivatives are conveniently computed by Fourier-transforming
$\mathbf{x}_v \cdot \mathbf{A}$ and $\mathbf{x}_u \cdot \mathbf{A}$
and then computing the tangential derivatives analytically.
The remainder of this note deals with the computation of the vector potential.
Have a closer look at the current density:
\begin{align}
 \mu_0 |\mathbf{x}_u \times \mathbf{x}_v| \,\mathbf{j}
 =&\, \mathbf{B}_0 \times \left( \mathbf{x}_u \times \mathbf{x}_v \right) \nonumber \\
 =&\,   \mathbf{x}_u \cdot \left( \mathbf{B}_0 \cdot \mathbf{x}_v \right)
      - \mathbf{x}_v \cdot \left( \mathbf{B}_0 \cdot \mathbf{x}_u \right)
\end{align}
according to the BAC-CAB rule.
The Jacobian from realspace to normalized angle-like coordinates on the surface is:
\begin{equation}
 \mathrm{d}^2\mathbf{x} = |\mathbf{x}_u \times \mathbf{x}_v| \,\mathrm{d}u \,\mathrm{d}v \, .
\end{equation}
It follows for the vector potential~$\mathbf{A}$:
\begin{equation}
 \mathbf{A}(u, v) = \frac{1}{4 \pi} \iint
    \frac{  \mathbf{x}_{u'} \cdot \left( \mathbf{B}_0' \cdot \mathbf{x}_{v'} \right)
          - \mathbf{x}_{v'} \cdot \left( \mathbf{B}_0' \cdot \mathbf{x}_{u'} \right) }
         {|\mathbf{x}(u,v) - \mathbf{x}(u',v')|}
    \,\mathrm{d}u' \,\mathrm{d}v' \, .
\end{equation}
This function is evaluated on a grid in $u$ and $v$,
which is the same grid on which the integral over $u'$ and $v'$ is computed,
since this allows to re-use certain geometric quantities.
However, this approach introduces a singularity in the integrand.
The distance $|\mathbf{x}(u,v) - \mathbf{x}(u',v')|$ in the denominator of the integrand
vanishes when $u = v'$ simulatenous with $v = v'$, leading to a $1/x$-type singularity in the integrand.
A subtraction method is applied to solve this issue.
An function with the same singularity, which can be Fourier-transformed analytically,
is subtracted and its analytical Fourier transform is added back in afterwards.
The regularizing function approximates the actual integrand:
\begin{equation}
  \frac{  \mathbf{x}_{u'} \cdot \left( \mathbf{B}_0' \cdot \mathbf{x}_{v'} \right)
        - \mathbf{x}_{v'} \cdot \left( \mathbf{B}_0' \cdot \mathbf{x}_{u'} \right) }
       {|\mathbf{x}(u,v) - \mathbf{x}(u',v')|}
  \approx
  \frac{  \mathbf{x}_{u} \cdot \left( \mathbf{B}_0 \cdot \mathbf{x}_{v} \right)
        - \mathbf{x}_{v} \cdot \left( \mathbf{B}_0 \cdot \mathbf{x}_{u} \right)}
       {\left[ g_{uu} \delta u^2 + 2 g_{uv} \delta u \delta v + g_{vv} \delta v^2 \right]^{1/2} }
\end{equation}
where
\begin{align}
 g_{uu} =&\, \mathbf{x}_{u} \cdot \mathbf{x}_{u} \\
 g_{uv} =&\, \mathbf{x}_{u} \cdot \mathbf{x}_{v} \\
 g_{vv} =&\, \mathbf{x}_{v} \cdot \mathbf{x}_{v} \\
 \delta u =&\, \frac{1}{\pi} \tan\left(\pi (u - u') \right) \\
 \delta v =&\, \frac{1}{\pi} \tan\left(\pi (v - v') \right) \, .
\end{align}
The analytical Fourier transform of this equivalently-singular function is:
\begin{equation}
 I(a,b,c) = \pi \int\limits_0^1 \int\limits_0^1
   \frac{\mathrm{d}u \,\mathrm{d}v}
        {\left[ a \tan^2(\pi u) + 2 b \tan(\pi u) \tan(\pi v) + c \tan^2(\pi v) \right]^{1/2} } \, .
\end{equation}
The analytical Fourier transform results in:
\begin{equation}
 I(a,b,c) = T_{0}^{+} + T_{0}^{-}
\end{equation}
with
\begin{equation}
 T_{0}^{\pm} = \frac{1}{\sqrt{a \pm 2b + c}} \log\left(
   \frac{\sqrt{c \left(a \pm 2b + c\right)} + c \pm b}
        {\sqrt{a \left(a \pm 2b + c\right)} - a \mp b} \right) \, .
\end{equation}
The vector potential is then computed as follows:
\begin{equation}
 \mathbf{A}(u,v) = \mathbf{A}_\textrm{reg}(u,v) + \mathbf{A}_\textrm{sing}(u,v)
\end{equation}
with
\begin{align}
 \mathbf{A}_\textrm{reg}(u,v)
 =&\, \frac{1}{4 \pi} \iint \,\mathrm{d}u' \,\mathrm{d}v' \Biggl\{
      \frac{  \mathbf{x}_{u'} \cdot \left( \mathbf{B}_0' \cdot \mathbf{x}_{v'} \right)
            - \mathbf{x}_{v'} \cdot \left( \mathbf{B}_0' \cdot \mathbf{x}_{u'} \right) }
           {|\mathbf{x}(u,v) - \mathbf{x}(u',v')|} \nonumber \\
 ~&\,   - \frac{\pi \left[  \mathbf{x}_{u} \cdot \left( \mathbf{B}_0 \cdot \mathbf{x}_{v} \right)
                      - \mathbf{x}_{v} \cdot \left( \mathbf{B}_0 \cdot \mathbf{x}_{u} \right) \right] }
           {\left[ g_{uu} \delta u^2 + 2 g_{uv} \delta u \delta v + g_{vv} \delta v^2 \right]^{1/2} }
         \Biggr\} \label{eqn:a_sing}
\end{align}
and
\begin{equation}
 \mathbf{A}_\textrm{sing}(u,v)
 = \left[   \mathbf{x}_{u} \cdot \left( \mathbf{B}_0 \cdot \mathbf{x}_{v} \right)
          - \mathbf{x}_{v} \cdot \left( \mathbf{B}_0 \cdot \mathbf{x}_{u} \right) \right]
   I(g_{uu}, g_{uv}, g_{vv}) \, .
\end{equation}
The discrete Fourier transform in Eqn.~(\ref{eqn:a_sing}) is performed
numerically, where the singular point is not considered in the summation.

% \FloatBarrier
% \section{Numerical Implementation}

\FloatBarrier
\section{Further Considerations}
The source term of the normal component of the magnetic field computed by BNORM
is the diamagnetic current density, which arises due to a finite plasma pressure.
Thus, for a vacuum computation, the resulting normal field should vanish,
since the source term is missing in case of a zero-pressure/vacuum computation.

% TODO: run VMEC for a vanishing-pressure case and run BNORM on the output; expected outcome: B.n vanishes as well.

BNORM computes B.n due to plasma currents;
NESTOR computes $|B|^2/2$ due to plasma currents.

BNORM plays a central role in the classical two-stage Stellarator optimization approach.
The plasma is optimized subject to physics objective functions using fixed-boundary VMEC.
BNORM is then used to compute the normal component of the magnetic field due to plasma currents
on the LCFS. This components needs to be cancelled by the external magnetic field.
A jump can be present in the tangential magnetic field components across the LCFS
between vacuum and plasma, as long as the total magnitude of the magnetic field
is conserved between plasma and vacuum.
The normal component of the magnetic field due to plasma currents
is therefore (along with the surface geometry) input to the coil design code
and defines the target to match/cancel.

% Workflow:
% \begin{enumerate}
%  \item fixed-boundary optimization using fixed-bdy VMEC
%  \item BNORM: get normal component of magnetic field on LCFS
%  \item NESCOIL/... to design coils for given BNORM output and LCFS geometry from VMEC
% \end{enumerate}

\chapter{Solov'ev Analytical Tokamak Equilibrium}
Here, the analytical Solov'ev Tokamak equilibrium
presented in Ref.~\cite{hirshman_whitson_1983}
is investigated and converted to VMEC inputs.
This set of notes is based on the investigations
on this topic by J. D. Hanson (Auburn University),
which were distributed as Mathematica Notebooks,
and TeX-ed here for further work.

\section{Setup}

\subsection{Information Survey from Sec. VI.}
Here, we collect all information
related to the Solov'ev equilibrium
available in the section ``I.~Introduction''
of the original article.

\begin{itemize}

\item
In the axisymmetric case, can integrate $F_\beta = 0$ to yield:
\begin{equation}
 B_\zeta = F(\rho) \label{eqn:bsubv_f}
\end{equation}

\item
Some metric elements vanish:
\begin{equation}
 g_{\theta \zeta} = g_{\rho \zeta} = 0
\end{equation}

\item
axisymmetry implies:
\begin{equation}
 \frac{\partial \lambda}{\partial \zeta} = 0
\end{equation}

\item
Contravariant magnetic field component:
\begin{equation}
 B^\zeta = \frac{B_\zeta}{g_{\zeta \zeta}}
\end{equation}
with
\begin{equation}
 g_{\zeta \zeta} = R^2
\end{equation}
where $R$ is the major radius

\item
now find the inverse Grad-Shafranov equation:
\begin{equation}
 F_\rho = \frac{\chi'}{\mu_0 \sqrt{g}}
 \left[
   \frac{\partial}{\partial \rho  } \left( \frac{\chi' g_{\theta \theta}}{\sqrt{g}} \right) -
   \frac{\partial}{\partial \theta} \left( \frac{\chi' g_{\rho   \theta}}{\sqrt{g}} \right)
 \right]
 + \frac{F F'}{\mu_0 R^2}
 + p'
\end{equation}

\item
We can also write Eqn.~\ref{eqn:bsubv_f} as:
\begin{equation}
 F(\rho) = \frac{\phi' R^2}{\sqrt{g}} \left( 1 + \frac{\partial \lambda}{\partial \theta} \right)
\end{equation}

\item
This yields
\begin{equation}
 \phi'(\rho) = \langle \frac{\sqrt{g}}{R^2} \rangle F(\rho)
\end{equation}

\item
and
\begin{equation}
 \frac{\partial \lambda}{\partial \theta}
 = \frac{\sqrt{g} / R^2}{\langle \sqrt{g} / R^2 \rangle} - 1
\end{equation}
where the $\langle . \rangle$ denotes a normalized $\theta$ average.

\end{itemize}

\subsection{Information Survey from Sec. VI.}
Here, we collect all information
related to the Solov'ev equilibrium
available in the section ``VI.~Moment Analysis of Solov'ev Equilibrium''
of the original article.

\begin{itemize}

\item
The magnetic field is represented as
\begin{equation}
 \mathbf{B} = \chi' \nabla \zeta \times \nabla \rho + F(\rho) \nabla \zeta \, .
\end{equation}

\item
A solution to the axisymmetric Grad-Shafranov equation,
with the above representation of the magnetic field, is:
\begin{equation}
 \rho^2 = \frac{\beta_1}{\chi_0^2}
 \left[
   Z^2 R_0^2 + \left( \frac{\beta_0}{8 \beta_1} \right) \left( R^2 - R_m^2 \right)^2
 \right]
\end{equation}

\item
where
\begin{equation}
 \chi' = 2 \rho \chi_0 \label{eqn:chip}
\end{equation}

\item
$\rho^2$ is the normalized poloidal flux $0 \leq \rho \leq 1$

\item
The pressure profile is
\begin{equation}
 p(\rho) = \beta_0 \left( 1 - \rho^2 \right)
\end{equation}

\item
The current flux function~$F$ is
\begin{equation}
 F^2 = R_0^2 \left( 1 - 4 \beta_1 \rho^2 \right)
\end{equation}

\item
The toroidal field is normalized to unity,
if $R_0$ is identified with the mean major radius;
this says something about \texttt{phiedge}.

\item
Note that $R = R_m$ is the magnetic axis,
which is determined by the boundary curve $\rho = 1$.
This could either be a typo (the axis is at $\rho = 0$),
or it is meant to say that the boundary geometry determines the magnetic axis position,
and the magnetic axis geometry is not something the user can freely choose.

\item
The spectral analysis is trivial in terms of the variables
$u = R^2$ and $Z$, yielding:
\begin{align}
 u =
 R^2 =&\, R_m^2 - \chi_0 \sqrt{\frac{8}{\beta_0}} \rho \cos(\theta) \\
 Z   =&\, \frac{\chi_0}{R_0} \frac{1}{\sqrt{\beta_1}} \rho \sin(\theta)
\end{align}

\item
Here, $\theta$ is a geometric angle
yielding a rapidly converging Fourier expansion for $R$ and $Z$,
which is not equal to the magnetic angle $\theta^*$,
in which magnetic field lines are straigt.

\item
The Jacobian is
\begin{equation}
 \sqrt{g} = \frac{\chi_0^2}{R_0} \sqrt{\frac{2}{\beta_0 \beta_1}} \rho
\end{equation}

\item
It is easy to evaluate $\lambda$ explicitly:
\begin{equation}
 \lambda_\theta = \frac{\sqrt{1 - a^2}}{1 - a \cos(\theta)} - 1
\end{equation}
where
\begin{equation}
 a(\rho) = \frac{\chi_0}{R_m^2} \sqrt{\frac{8}{\beta_0}} \rho < 1 \label{eqn:def_a}
\end{equation}

\item
Thus:
\begin{equation}
 \lambda = \sum_{m = 1} \hat{\lambda}_m \sin(m \theta)
\end{equation}
with
\begin{equation}
 \hat{\lambda}_m = \frac{2}{m} \left[ \frac{1}{a} \left( 1 - \sqrt{1 - a^2} \right) \right]^m
\end{equation}

% TODO: What is the correct asymptotic behaviour towards the magnetic axis?
%       It seems that $a$ goes to zero as $\rho$ goes to zero,
%       and $\hat{\lambda}_m$ blows up because it depends on $1/a$?

\item
Note that for $a^2 < 1$, $\hat{\lambda}_m$ decays exponentially with $m$.

\item
The magnetic flux profile is found to be
\begin{equation}
 \frac{\phi'}{\chi'} \equiv q(\rho)
 = \frac{\chi_0}{R_m^2}
 \sqrt{
   \frac{1}{2 \beta_0 \beta_1} \left( \frac{1 - 4 \beta_1 \rho^2}{1 - a^2} \right)
 }
\end{equation}

\end{itemize}

\subsection{Information Survey from Sec. VI.}
Here, we collect all information
related to the Solov'ev equilibrium
available in the section ``IX. Numerical Examples''
of the original article.

\begin{itemize}

\item
The value of $R_0(0)$ should be $4$.

\item
The caption of Figure 2 lists
a few more parameters:

\item
$R$ of the boundary geometry:
\begin{equation}
 \left(\frac{R}{4}\right)^2 = 1 - \frac{1}{2} \rho \cos(\theta)
\end{equation}

\item
$Z$ of the boundary geometry:
\begin{equation}
 Z = \frac{\sqrt{10}}{2} {\color{red} \rho} \sin(\theta)
\end{equation}
Note that the $\rho$ was probably forgotten in the article.

\item
The pressure profile is:
\begin{equation}
 p(\rho) = \frac{1}{8} (1 - \rho^2)
\end{equation}

\item
The poloidal flux profile is:
\begin{equation}
 \chi = \rho^2
\end{equation}

\end{itemize}

\section{Derived Quantities}

\subsection{Rotational Transform Profile}
One can simply compute $\iota = 1 / q$ as follows:
\begin{align}
 \iota(\rho) = \frac{1}{q(\rho)}
 =&\, \frac{1}{
  \frac{\chi_0}{R_m^2}
 \sqrt{
   \frac{1}{2 \beta_0 \beta_1} \left( \frac{1 - 4 \beta_1 \rho^2}{1 - a^2} \right)
 }
           } \nonumber \\
 =&\,
 \frac{R_m^2}{\chi_0}
 \sqrt{2 \beta_0 \beta_1 \left( \frac{1 - a^2}{1 - 4 \beta_1 \rho^2} \right)}
\end{align}
and with $a$ from Eqn.~\ref{eqn:def_a} we find:
\begin{align}
 \iota(\rho)
 =&\, \frac{R_m^2}{\chi_0} \sqrt{2 \beta_0 \beta_1 \left( \frac{1 - \left[ \frac{\chi_0}{R_m^2} \sqrt{\frac{8}{\beta_0}} \rho \right]^2}{1 - 4 \beta_1 \rho^2} \right)} \nonumber \\
 =&\, \frac{R_m^2}{\chi_0} \sqrt{2 \beta_0 \beta_1 \left( \frac{1 - \frac{\chi_0^2}{R_m^4} \frac{8}{\beta_0} \rho^2}{1 - 4 \beta_1 \rho^2} \right)}
\end{align}

\subsection{Toroidal Flux}
VMEC profiles are specified in terms of the toroidal magnetic flux.
We know:
\begin{equation}
 \iota = \frac{\chi'}{\phi'} \equiv \frac{\partial \chi}{\partial \rho} / \frac{\partial \phi}{\partial \rho}
\end{equation}
for some radial coordinate~$\rho$.
This can be reformulated as:
\begin{equation}
 \frac{\partial \phi}{\partial \rho} = \frac{1}{\iota} \frac{\partial \chi}{\partial \rho}
\end{equation}
and we can integrate radially to get the toroidal flux profile:
\begin{equation}
 \phi(\rho) = \int\limits_{0}^{\rho} \frac{1}{\iota} \frac{\partial \chi}{\partial \rho'} \,\mathrm{d}\rho'
\end{equation}
assuming that $\phi(0) = 0$ at the magnetic axis.
We know $\chi'$ already from Eqn.~\ref{eqn:chip}:
\begin{equation}
 \phi(\rho)
 = \int\limits_{0}^{\rho} \frac{2 \rho' \chi_0}{\iota} \,\mathrm{d}\rho'
 = \int\limits_{0}^{\rho} 2 \rho' \chi_0 q(\rho') \,\mathrm{d}\rho'
\end{equation}
and using the expression for the safety factor, we get:
\begin{align}
 \phi(\rho) =&\, 2 \chi_0 \int\limits_{0}^{\rho}
 \rho'
 \frac{\chi_0}{R_m^2}
 \sqrt{
   \frac{1}{2 \beta_0 \beta_1} \left( \frac{1 - 4 \beta_1 {\rho'}^2}{1 - a^2(\rho')} \right)
 }
 \,\mathrm{d}\rho' \nonumber \\
 =&\, 2 \frac{\chi_0^2}{R_m^2} \sqrt{\frac{1}{2 \beta_0 \beta_1}}
   \int\limits_{0}^{\rho}
     \rho'
     \sqrt{\frac{1 - 4 \beta_1 {\rho'}^2}{1 - \beta_2^2 {\rho'}^2}}
     \,\mathrm{d}\rho'
\end{align}
The abstract integral used here can be solved by Wolfram Alpha:
\begin{align}
 \int x \sqrt{\frac{1 - a x^2}{1 - b x^2}} \,\mathrm{d}x
 =&\,
 \frac{1}{2 a b^{3/2} (a x^2 - 1)}
 \sqrt{b - a}
 \sqrt{\frac{a x^2 - 1}{b x^2 - 1}}
 \sqrt{\frac{a - a b x^2}{a - b}}
 \Biggl[
 \nonumber \\
~& \phantom{+}~ \sqrt{b}
   \sqrt{b - a}
   (a x^2 - 1)
   \sqrt{\frac{a (b x^2 - 1)}{b - a}}
   \nonumber \\
~& + (a - b)
   \sqrt{a x^2 - 1}
   \,\mathrm{sinh}^{-1} \left(
     \frac{\sqrt{b} \sqrt{a x^2 - 1}}{\sqrt{b - a}}
   \right)
 \Biggr]
\end{align}
A direct inverse of this function is not obvious,
which is why finding the value of $\rho$ for a given value of $\phi$
will be done numerically for the time being.

One could also think about using VMEC in \texttt{lRFP} mode,
where the poloidal flux is constrained and used as radial variable.

\subsection{Jacobian}

The VMEC-like Jacobian is given by:
\begin{equation}
 \sqrt{g} = R \tau
\end{equation}
with
\begin{align}
 \tau = R_\theta Z_s - R_s Z_\theta \, .
\end{align}

The radial derivatives with respect to $s$ are computed as follows.
The analytical model uses $\rho$ as the radial coordinate.
Since $\iota$ is constant, we have $\rho = \sqrt{s}$
with the normalized toroidal flux $s = \phi / \phi_\textrm{LCFS}$.
This implies for the radial derivatives:
\begin{align}
 R_s = \frac{\mathrm{d}R}{\mathrm{d}s}
 = \frac{\mathrm{d}R}{\mathrm{d}\rho} \frac{\mathrm{d}\rho}{\mathrm{d}s}
 = R_\rho \frac{\mathrm{d}\rho}{\mathrm{d}s}
\end{align}
and with the definition of $\rho$ we have
\begin{align}
 \frac{\mathrm{d}\rho}{\mathrm{d}s}
 = \frac{\partial}{\partial s} \left( \sqrt{s} \right)
 = \frac{1}{2 \sqrt{s}} = \frac{1}{2 \rho}
\end{align}
It thus follows:
\begin{align}
 R_s = R_\rho \frac{1}{2 \rho}
\end{align}

First order finite-differences are used for the radial derivatives:
\begin{align}
 R_s(s_{j+0.5}, \theta_l) =&\,
   \frac{R(s_{j+1}, \theta_l) - R(s_{j}, \theta_l)}{s_{j+1} - s_{j}}
\end{align}

Linear interpolation is used for the poloidal derivatives:
\begin{align}
 R_\theta(s_{j+0.5}, \theta_l) =&\,
   \frac{R_\theta(s_{j+1}, \theta_l) + R_\theta(s_{j}, \theta_l)}{2}
\end{align}

\subsection{Fourier Representation of the Boundary}

The expressions for the Fourier coefficients of the flux surface geometry are:
\begin{align}
 \hat{R}^\textrm{cos}_m(\rho) =&\,
   \frac{2 - \delta_{m,0}}{2 \pi}
   \int_{0}^{2 \pi}
     R(\rho, \theta) \cos(m \theta) \,\mathrm{d}\theta \\
 \hat{Z}^\textrm{sin}_m(\rho) =&\,
   \frac{2 - \delta_{m,0}}{2 \pi}
   \int_{0}^{2 \pi}
     Z(\rho, \theta) \sin(m \theta) \,\mathrm{d}\theta
\end{align}
Note that $Z$ only has a single Fourier coefficient:
\begin{align}
 Z =&\, \frac{\chi_0}{R_0} \frac{1}{\sqrt{\beta_1}} \rho \sin(\theta) \\
 \Rightarrow
 \hat{Z}^\textrm{sin}_1(\rho) =&\, \frac{\chi_0}{R_0} \frac{1}{\sqrt{\beta_1}} \rho
\end{align}
However, $R$ has more Fourier coefficients.
Note that:
\begin{align}
  R^2 =&\, R_m^2 - \chi_0 \sqrt{\frac{8}{\beta_0}} \cos(\theta) \\
  \Leftrightarrow
  \left( \frac{R}{R_m} \right)^2
    =&\, 1 - \underbrace{\frac{\chi_0}{R_m^2} \sqrt{\frac{8}{\beta_0}}}_{\equiv \beta_2} \rho \cos(\theta)
\end{align}
and hence
\begin{align}
 \frac{R}{R_m} = \sqrt{ 1 - \beta_2 \rho \cos(\theta) }
\end{align}
For $m=0$, it thus follows:
\begin{align}
 \hat{R}^\textrm{cos}_0(\rho)
%  =&\,
%    \frac{1}{2 \pi}
%    \int_{0}^{2 \pi}
%      R(\rho, \theta)
%      \,\mathrm{d}\theta \\
 =&\,
   \frac{R_m}{2 \pi}
   \int_{0}^{2 \pi}
     \sqrt{ 1 - \beta_2 \rho \cos(\theta) }
     \,\mathrm{d}\theta
\end{align}
This is a complete elliptic integral
and can therefore be evaluated analytically.
Gradshteyn-Ryshik can help us out here with its 3.671.4:
\begin{align}
 \int\limits_0^\pi \sqrt{a + b \cos(x)} \,\textrm{d}x
 =
 2 \sqrt{a + b} \,\mathbf{E}\left( \sqrt{\frac{2 b}{a + b}} \right)
 \textrm{ for }
 a > b > 0
\end{align}
with the complete elliptic integral of the second kind~$\mathbf{E}$.
For $m>0$, we have:
\begin{align}
 \hat{R}^\textrm{cos}_m(\rho)
 =&\,
   \frac{R_m}{\pi}
   \int_{0}^{2 \pi}
     \sqrt{ 1 - \beta_2 \rho \cos(\theta) }
     \cos(m \theta)
     \,\mathrm{d}\theta \\
\end{align}

Adaptive arbitrary-precision quadrature was used to compute the Fourier coefficients
of $R$ for a number of poloidal modes.
\begin{figure}[h]
 \centering
 \includegraphics[width=0.5\textwidth]{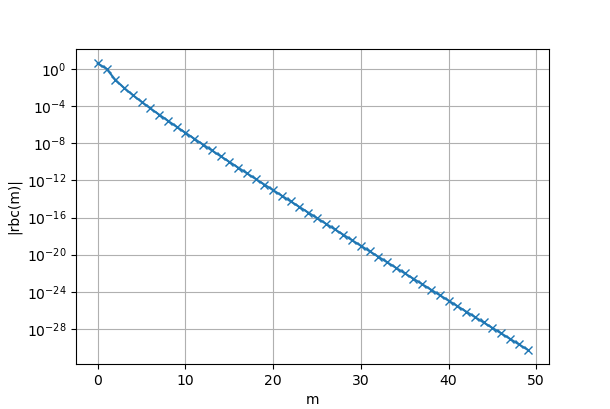}
 \caption{Decay of the boundary geometry coefficients for $R$ with poloidal mode number $m$.}
\end{figure}

% \subsection{Scan of $\beta_1$}
% The value of $\beta_1$ is $1/16$ in case of a flat profile of $\iota$.
% This is assumed to be case that is published in the 1983 article.
% This value has been scanned over the values $1/32$ (blue), $1/16$ (orange) and $1/8$ (green) in the following plots.

% \begin{figure}[h]
%  \centering
%  \includegraphics[width=0.5\textwidth]{img/solovev/boundary_shapes.png}
%  \caption{Boundary and inner flux surfaces shapes.
%           $\beta_1 = 1/32$ (blue), $1/16$ (orange) and $1/8$ (green).}
% \end{figure}

% \begin{figure}[h]
%  \centering
%  \includegraphics[width=0.5\textwidth]{img/solovev/iota_profiles.png}
%  \caption{Rotational transform profile.
%           $\beta_1 = 1/32$ (blue), $1/16$ (orange) and $1/8$ (green).}
% \end{figure}

% \begin{figure}[h]
%  \centering
%  \includegraphics[width=0.5\textwidth]{img/solovev/phi_profiles.png}
%  \caption{Toroidal flux function $\phi$ profile.
%           $\beta_1 = 1/32$ (blue), $1/16$ (orange) and $1/8$ (green).}
% \end{figure}

% \begin{figure}[h]
%  \centering
%  \includegraphics[width=0.5\textwidth]{img/solovev/iprime_profiles.png}
%  \caption{Profile of radial derivative of enclosed toroidal current (I-prime).
%           $\beta_1 = 1/32$ (blue), $1/16$ (orange) and $1/8$ (green).}
% \end{figure}

\chapter{Acknowledgements}
\begin{itemize}
 \item Philipp Jurašić has taken over product ownership for VMEC++ at Proxima Fusion, which includes the maintenance and further development of VMEC++, as of 2026. Thanks a lot and all the best for this role!
 \item Thanks to Mark Cianciosa for managing and providing the VMEC source code and setting up the new \texttt{cmake} build system.
 \item Thanks to Matt Landreman for contributing an independent solution to the analytical math and code structure of NESTOR,
       which was very helpful in deriving the NESTOR equations.
 \item Thanks to
       Jim Hanson,
       Mark Cianciosa,
       Sam Lazerson,
       Joachim Geiger,
       John Schmitt,
       Florian Hindenlang,
       Omar Maj,
       Dharya Malhotra,
       Antoine Cerfon,
       Antoine Baillod,
       Joaquim Loizu,
       Stuart Hudson,
       Ksenia Aleynikova,
       Chris Smiet,
       Andrew Ware,
       Caoxiang Zhu,
       Zhisong Qu,
       David Pfefferle,
       Bharat Medasani,
       Rory Conlin,
       Daniel Dudt,
       Dario Panici,
       Egemen Kolemen,
       Philipp Jurasic,
       Sophia Henneberg,
       Carolin Nührenberg,
       Oliver Ford,
       and
       Gavin Weir
       for the many and sometimes lengthy discussions
       related to VMEC which made it at all possible for me to get this deeply involved.
 \item Thanks to Erika Strumberger for providing the NEMEC source code and additional derivations related
       to the second implementation of NESTOR by Peter Merkel.
 \item Thanks to Enrico Guiraud and Veronika Siska for substantially contributing to the successful implementation of VMEC++.
 \item Thanks to Jan-Peter Bähner for providing the section on the calculation of $\nabla \alpha$.
 \item Thanks to Camilla Lazzati for providing the section on coordinate sign conventions,
       adapting Matt Landreman's notes on Fortran VMEC and manually verifying them for VMEC++.
 \item Special thanks to Klara Höfler for many insightful discussions on math and wording
       as well as independent solutions to some critical parts of the derivation of the analytical parts of NESTOR.
 \end{itemize}

\printbibliography

\end{document}